\documentclass[a4paper, 10pt, twoside]{book}
\usepackage[top=2cm,left=1.5cm,right=1.5cm]{geometry}
\usepackage[numbers,square]{natbib}
\usepackage[english]{babel}
\usepackage{epsfig}
\usepackage{amssymb}
\usepackage{amsfonts}
\usepackage{epsf}
\usepackage{subfigure}
\usepackage{rotating}
\usepackage{graphicx}
\usepackage{amsmath}
\usepackage{fancyhdr}
\usepackage{lineno}
\usepackage{graphics}
\usepackage{pstricks}
\usepackage{color}
\usepackage{tocbibind}

\usepackage{hyperref}

\hypersetup{
    colorlinks,
    citecolor=green,
    filecolor=black,
    linkcolor=blue,
    urlcolor=black
}

\let\a   = \alpha     \let\b = \beta    \let\g = \gamma    \let\d = \delta
       \let\h = \eta      
      \let\k = \kappa      
                     \let\r = \rho
\let\s   = \sigma            
            
\newcommand{\pd}{\partial}

\newcommand{\muu}{\mu_{1}}
\newcommand{\nuu}{\nu_{1}}
\newcommand{\ku}{k_{1}}
\newcommand{\xu}{x_{1}}

\newcommand{\mud}{\mu_{2}}
\newcommand{\nud}{\nu_{2}}
\newcommand{\kd}{k_{2}}
\newcommand{\xd}{x_{2}}

\newcommand{\mut}{\mu_{3}}
\newcommand{\nut}{\nu_{3}}
\newcommand{\kt}{k_{3}}
\newcommand{\xt}{x_{3}}

\newcommand{\muq}{\mu_{4}}
\newcommand{\nuq}{\nu_{4}}
\newcommand{\kq}{k_{4}}
\newcommand{\xq}{x_{4}}

\newcommand{\kc}{k_{5}}

\newcommand{\ks}{k_{6}}

\newcommand{\beq}{\begin{equation}}
\newcommand{\eeq}{\end{equation}}
\newcommand{\bea}{\begin{eqnarray}}
\newcommand{\eea}{\end{eqnarray}}
\newcommand{\bann}{\begin{eqnarray*}}
\newcommand{\eann}{\end{eqnarray*}}
\newcommand{\bmi}{\begin{minipage}}
\newcommand{\emi}{\end{minipage}}

\newcommand{\spu}{\,\,\,}
\newcommand{\nn}{\nonumber}

\newcommand{\llangle}{\left\langle}
\newcommand{\rrangle}{\right\rangle}

\newcommand{\ksl}{k \! \! \!  /}

\makeatletter
\def\cleardoublepage{\clearpage\if@twoside
\ifodd\c@page
\else\hbox{}\thispagestyle{empty}\newpage
\if@twocolumn\hbox{}\newpage\fi\fi\fi}
\makeatother

\begin{document}
\thispagestyle{empty}
\mbox{}
\vspace{-1.5cm}
\begin{center}
\begin{figure}[h]
\begin{center}
\includegraphics[scale=0.2]{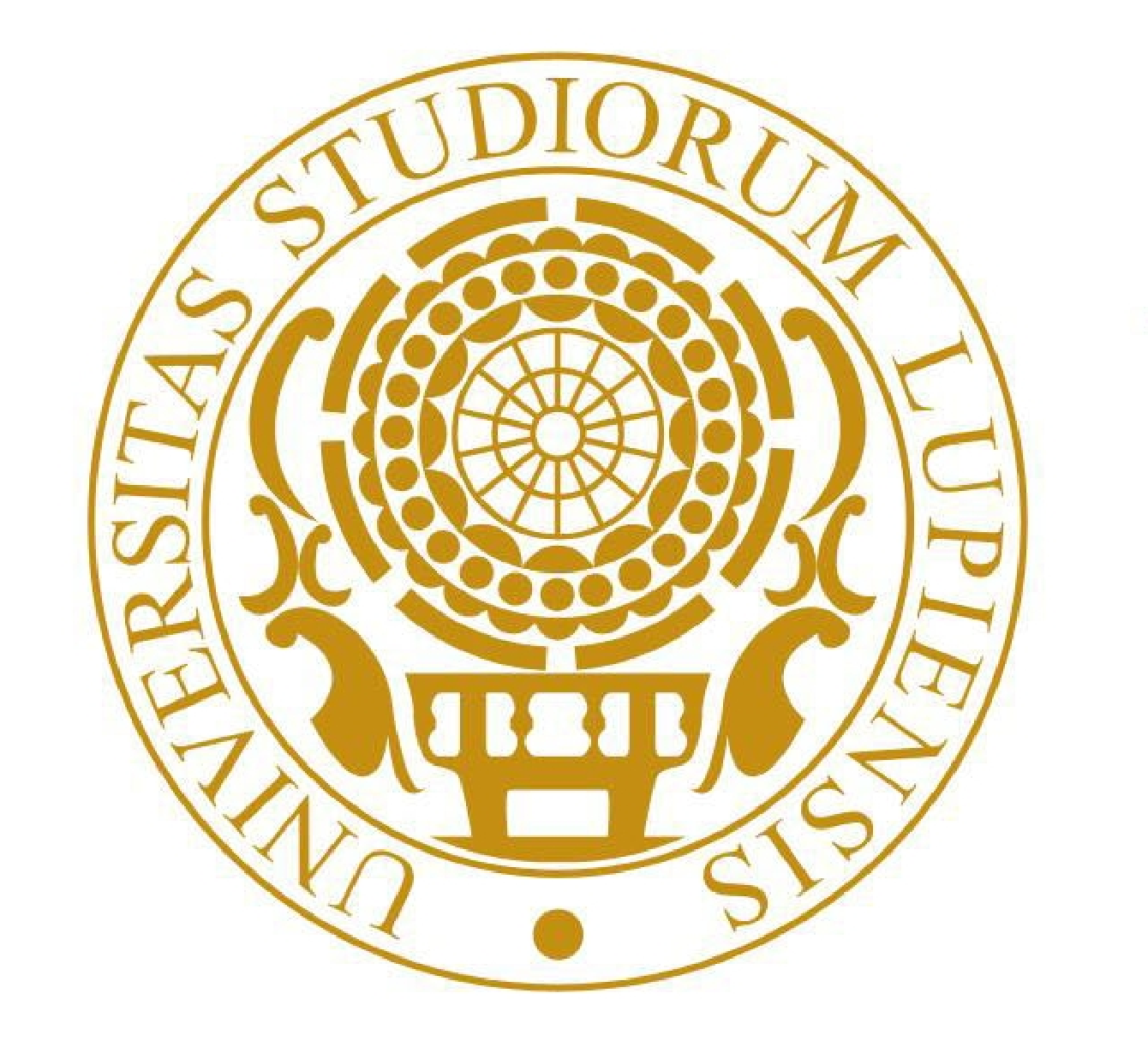}
\end{center}
\end{figure}
\begin{tabular*}{17cm}{c}
\hline
\end{tabular*}
{\LARGE \mbox{{Universit\`a Del Salento}}}\\
\vspace{0.5cm}
{\large {Facolt\`a di Scienze Matematiche, Fisiche e Naturali }}\\
\vspace{0.5cm}
{\large {Dipartimento di Matematica e Fisica ``Ennio De Giorgi''}}
\begin{tabular*}{17cm}{c}
\hline
\end{tabular*}
\vspace{2mm}
\vspace{2.5cm}
\\
{\huge {\textbf {Conformal Anomaly Actions \\ and Dilaton Interactions \\ 
}}} 
\end{center}
\vspace{3.0cm}
\begin{large}
\begin{flushleft}
\hspace{0.8cm}\emph{Supervisor}\\
\hspace{0.8cm}Prof. Claudio Corian\`{o}\\
\hspace{15cm}\emph{Candidate} \\
\hspace{15cm}Mirko Serino
\end{flushleft}
\end{large}
%
%
%
\vspace{3.0cm}
\begin{center}
\begin{tabular*}{17cm}{c}
\hline
\end{tabular*}
\end{center}
\vspace{0.1cm}
\begin{center}
{\sc \large {Tesi di Dottorato In Fisica - XXVI ciclo}}\\
\vspace{0.2 cm}
{\sc \large {Anno Accademico 2013 - 2014}}\\
\end{center}

\frontmatter

\newpage

\null \vspace {\stretch {1}}
\begin{flushright}
\vspace{0.5cm}
\emph{pi\`{u} diventa tutto inutile \\
             e pi\`{u} credi che sia vero \\
						e il giorno della Fine \\
						non ti servir\`{a} l'inglese \\
\vspace{0.3cm}
Franco Battiato}
\end{flushright}
\vspace {\stretch {2}}\null

\clearpage{\pagestyle{empty}\cleardoublepage}

\mbox{}
\vspace{4cm}
\begin{flushleft}
{\Huge{\textbf{Aknowledgements}}}
\end{flushleft}
\vspace{0.5cm}

It is a pleasure to acknowledge all the people who made this work possible.
First, I want to thank my advisor, Claudio Corian\`o, for his teachings and his relentless encouragement through all these years, 
ever since I was an undergraduate student.
Obviously, next come my colleagues: Luigi Delle Rose, for his unceasing support and collaboration, 
Carlo Marzo, for his friendship, his humour and all of his integrations by parts (!!!); 
Antonio Costantini, Perla Tedesco and Annalisa De Lorenzis, for being part of our group; 
Roberta Armillis, Antonio Mariano and Antonio Quintavalle, whom I always remember with pleasure; 
and all the people in our department with whom I have shared something important, who are by far too many to be listed here. \\
I also want to thank professor Pietro Colangelo and professor Emil Mottola, for their support and the collaborations we had.

Very special warm thanks and a huge hug are for all the members of my family, 
for being by my side at every step along my path, especially in my most difficult moments. \\

Last but not least, I want to dedicate this work to the most extraordinary woman I have ever met, loved and been with:
to Ilaria, wherever you are, from Here-And-Now to Eternity...

\clearpage{\pagestyle{empty}\cleardoublepage}

\tableofcontents%

\newpage

\clearpage{\pagestyle{empty}\cleardoublepage}%

\vspace{3cm}

\chapter[List of publications]{List of publications}

\vspace{0.5cm}%

The chapters of this thesis are based on the following research papers:
\begin{itemize}
\item C.~Corian\`o, L.~Delle~Rose, E.~Mottola, and M.~Serino \\
\newblock{\em Graviton vertices and the mapping of anomalous correlators to momentum space for a general conformal field theory} \\
\newblock {\em JHEP} 1208 (2012) 147   [arXiv:1203.1339 [hep-th]]
\item C.~Corian\`o, L.~Delle Rose, A.~Quintavalle, M.~Serino \\
\newblock {\em Dilaton interactions and the anomalous breaking of scale invariance in the Standard Model} \\
\newblock {\em JHEP} 1306 (2013) 077, [arXiv:1206.0590 [hep-ph]]
\item C.~Corian\`o, L.~Delle Rose, C.~Marzo, M.~Serino \\
\newblock {\em Higher order dilaton interactions in the nearly conformal limit of the Standard Model} \\
\newblock {\em Phys.Lett.} B717  182-187 (2012), [ arXiv:1207.2930 [hep-ph]]
\item C.~Corian\`o, L.~Delle Rose, C.~Marzo, M.~Serino \\
\newblock {\em Conformal trace relations from the dilaton Wess-Zumino action} \\
\newblock {\em Phys.Lett. } B726 (2013) 896-905, [ arXiv:1306.4248 [hep-th]]
\item C.~Corian\`o, C.~Marzo, L.~Delle Rose, M.~Serino \\
\newblock {\em The dilaton Wess-Zumino action in 6 dimensions from Weyl-gauging: local anomalies and trace relations} \\
\newblock {\em Class. Quantum Grav.} 31 (2014) 105009,  [arXiv:1311.1804  [hep-th]]
\end{itemize}

\vspace{0.5cm}

The following papers are related to the topics presented in this thesis but are not discussed in detail:

\begin{itemize}
\item C.~Corian\`{o}, L.~Delle~Rose, A.~Quintavalle and M.~Serino \\
\newblock {\em The conformal anomaly and the neutral currents sector of the Standard Model} \\
\newblock {\em Phys.Lett.},  B700 (2011) 29-38 [arXiv:1101.1624 [hep-ph]]
\item C.~Corian\`{o}, L.~Delle~Rose, and M.~Serino \\
\newblock {\em Gravity and the neutral currents: effective interactions from the trace anomaly} \\
\newblock {\em Phys.Rev.}, D83 (2011) 125028 [arXiv:1102.4558 [hep-ph]]
\item C.~Corian\`o, L.~Delle Rose, M.~Serino \\
\newblock {\em Three and four point functions of stress energy tensors in D=3 for the analysis of cosmological non-gaussianities } \\
\newblock {\em JHEP} 1212 (2012) 090  [ arXiv:1210.0136 [hep-th]]
\item C.~Corian\`o, L.~Delle Rose, E.~Mottola, M.~Serino \\
\newblock {\em Solving the conformal constraints for scalar operators in momentum space and the evaluation of Feynman's master integrals} \\
\newblock {\em JHEP} 1307 (2013) 011  [arXiv:1304.6944 [hep-th]]
\item C.~Corian\`o, A.~Costantini, L.~Delle Rose, M.~Serino \\
\newblock {\em Sum rules and the effective action of the composite Axion/Dilaton/Axino supermultiplet in N = 1 theories }
\newblock {\em JHEP} 1406 (2014) 136   [arXiv:1402.6369  [hep-ph]]
\end{itemize}
%

\vspace{1cm}%

{\Large \bfseries Proceedings}

\begin{itemize}
\item L.~Delle Rose, M.~Serino
\newblock {\em Dilaton interactions in QCD and in the electroweak sector of the Standard Model}\\
\newblock {\em AIP Conf.Proc.} 1492 (2012) 210-213  [arXiv:1208.6432] 
\item L.~Delle Rose, M.~Serino
\newblock {\em Massless scalar degrees of freedom in QCD and in the electroweak sector from the trace anomaly  } \\
\newblock {\em AIP Conf.Proc.} 1492 (2012) 205-209 [arXiv:1208.6425]
\end{itemize}

\clearpage{\pagestyle{empty}\cleardoublepage}

\chapter[Introduction]{Introduction}

The first  keystone of our present understanding of the fundamental interactions is the gauge symmetry.
In fact, the structure of the Standard Model of electroweak interactions is based on the symmetry group $SU(2)_L \times U(1)_Y$,
where $L$ stands for the weak isospin, which characterizes left-handed fermions, and $Y$ is the hypercharge quantum number. Similarly, the theory of strong interactions, QCD, is entirely based on the non abelian group describing a gauged colour symmetry, $SU(3)_C$.

The symmetry of these groups dictates the structure of the fermion multiplets which couple to the corresponding gauge currents and predicts the existence of four massless gauge bosons.
By incorporating the spontaneous breaking of the gauge symmetry via the Higgs mechanism, the Standard Model accounts for a gauge-invariant generation of the masses both of the fermions and of the $W^{\pm}/ Z$ bosons mediating 
weak interactions, while keeping the photon massless.

Beside the symmetry principles, the second fundamental pillar of a consistent quantum field theory is renormalizablity,
which is invoked in order to remove the infinities plaguing the perturbative computations in order to obtain predictive results.
In the case of quantum electrodynamics or QED, the problem of renormalizability, for instance, was solved in the late $40$'s by Feynman 
\cite{Feynman:1949hz,Feynman:1948fi,Feynman:1949zx,Feynman:1950ir,Feynman:1948km,Feynman:1951gn}, 
Schwinger \cite{Schwinger:1948iu,Schwinger:1948yk,Schwinger:1948yj,Schwinger:1949ra,Schwinger:1949zz, Schwinger:1951xk,Schwinger:1953tb}, Tomonaga \cite{Tomonaga:1946zz} and Dyson \cite{Dyson:1949bp,Dyson:1949ha}.
A similar result in the case of non abelian gauge theories was presented only 20 years later, after that Yang-Mills theories 
were recognized as a possible description of the fundamental interactions. In this case, the proof of renormalizability was given by 't Hooft in 1971 \cite{THooft:1971fh,THooft:1971rn}.

Together, gauge symmetry and renormalizability allow to successfully account for three of the four fundamental interactions observed in nature, i.e. electromagnetism, weak and strong interactions, at least up to the highest energy which can be experimentally tested.
Nonetheless, this picture is far from being complete. In fact, gravitation is still out of this framework, and has so far defied any attempt to a proper quantization, consistent with perturbative unitarity and renormalizability. 
Its formalism is obtained by gauging the $10$-parameter Poincar\'{e} group of rigid translations, boosts and rotations
acting in Minkowski space, which turns into the group of general coordinate transformations, also called diffeomorphisms.

Despite the fact that General Relativity is a gauge theory which has successfully passed all the experimental tests performed so far
and that almost a century has passed since 1916, when it was formulated for the first time,
a consistent quantum description of gravitation is still lacking. 
This happens because no renormalization program is viable for the quantized Einstein's theory.
In fact, due to well known power counting arguments, a necessary condition for a field theory to be renormalizable 
is that the coupling constants in the Lagrangian must not have negative mass dimensions.
The Einstein-Hilbert action contains one such parameter, namely Newton's constant $G$, whose dimension is $\left[ M \right]^{-2}$ and therefore the attempt to quantize General Relativity inevitably results in a non renormalizable theory.
This implies that the description of the dynamics of the gravitational field needs to be modified at very high energies,
i.e. near the Planck energy $E_{P} = (\hbar\, c^5/G)^{\frac{1}{2}}  \approx 1.22\,\times\, 10^{19}$ GeV, where it is expected 
to play a decisive role in the dynamics of the early universe. 

At present, nobody has ever formulated a consistent quantum theory of gravitation, but there are two reasons why one can draw
really significant insights by coupling a relativistic field theory to a purely classical background metric.

First, the value of $E_P$ is bigger by fifteen orders of magnitude than the highest energy at which all the known quantum field theories 
can be experimentally tested so far, i.e. $14$ TeV's at the Large Hadron Collider. This means that the description of
any scattering process will not be affected at all by the quantum fluctuations in the gravitational field.

Second, and more importantly for the purposes of this work, beside the symmetry under general coordinate transformations, 
which is enjoyed by any system embedded in curved space according to the formalism of General Relativity, there is another one,
which is typical of a subset of theories containing no dimensioful parameters, i.e. \emph{Weyl symmetry}.
A theory is said to be Weyl-symmetric if its action is invariant under the local rescaling of the metric tensor
$g_{\mu\nu}\rightarrow e^{2\sigma(x)}g_{\mu\nu}$, where $\sigma(x)$ is a well-behaved function of the coordinates.
It is trivial to see that the condition expressing Weyl invariance is the tracelessness of the energy-momentum tensor,
${T^\mu}_\mu = 0$. \\
On the other hand, it is known that the possibility to define a traceless energy-momentum tensor, for instance by introducing
proper terms of improvement, is the condition obeyed in Minkowski space by field theories which are invariant under the conformal group 
\cite{Polchinski:1987dy}.We remark that the conformal constraints in Minkowski space are implemented through rather nontrivial operators, 
whereas Weyl symmetry is much more straightforward to study. In particular, the tracelessness condition on the stress-energy tensor, derived in curved space as a consequence of 
Weyl symmetry, obviously remains valid in the flat limit as well. \\
Under some mild assumptions, it is possible to provide a simple algebraic criterion in curved space which allows to establish whether
a classical field theory is conformally invariant \cite{Iorio:1996ad} or not.
There is a great advantage in studying field theories through their formal embedding in a curved space-time background. In fact many aspects of conformal invariance can be studied much more easily than in Minkowski space, 
by analysing the constraints of the abelian Weyl group.
This issue is reviewed in chapter \ref{Weyl}, to set the stage for all the subsequent work presented in this thesis.
So far for classical conformal field theories.

\vspace{0.5cm}
{\large \bfseries Conformal anomalies and the low-energy effective actions for gravity}
\vspace{0.5cm}

It is known that, due to renormalization effects, the naive tracelessness condition of the energy-momentum tensor that conformal field theories 
enjoy at the classical level is not inherited by the vacuum expectation value of the corresponding quantum operator 
\cite{Capper:1974ic,Capper:1973mv,Capper:1974ed,Duff:1977ay,Duff:1993wm,Birrell:1982ix}.
Terms violating the classical Weyl symmetry due to quantum effects are known as \emph{conformal} or \emph{trace anomalies}
and are of two kinds.
Terms of the first kind are proportional to the beta functions of the theory and depend on the background gauge fields, 
so that they vanish only at the renormalization group fixed points for theories containing interactions, 
whereas they do not appear at all in free field theories.
Terms of the second kind are c-number contributions depending on the background metric tensor and are always present in the trace of the 
energy-momentum tensor of any field theory. 
It goes without saying that for theories which are not conformally invariant at the classical level, these terms appear in the
trace of vaccum expectation value of the energy-momentum tensor together with the vacuum expectation value (vev) of the operators which classically break conformal symmetry,
e.g. the mass terms.

An outstanding feature of the conformal anomalies is that, whether they are seen as a consequence of renormalization of UV singularities \cite{Duff:1977ay}
or as an infrared effect \cite{Giannotti:2008cv}, they affect the physics of the system at all energy scales.
This circumstance is generally true for chiral anomalies and  has led 't Hooft to formulate his famous anomaly matching conditions 
\cite{THooft:1979bh}, which are a powerful constraint for chiral theories describing the low-energy limit of QCD.

For conformal anomalies, as we have mentioned, there are contributions which depend on the beta-functions of the theory, and hence on the energy scale, 
but there are also pure c-number contributions, built out of the metric tensor, just as in the chiral case. \\
The existence of such scale-independent terms allows to look at the connection between conformal symmetry and General Relativity also in other 
ways. In fact, since anomalies are a quantum effect that does not depend on the energy scale and are built out of the metric background field, 
they affect the description of gravitation in the infrared regime of the quantum theory, providing the first corrections to the 
classical Einstein-Hilbert action \cite{Mottola:2006ew}.

The dynamics of a quantum system in the infrared can be effectively described by a proper low-energy action encoding 
the symmetries of the theory and describing the interactions of the degrees of freedom surviving in the infrared regime \cite{Wilson:1973jj}, 
after integrating out the ultraviolet modes. If the theory is anomalous, the effective action is modified and can always be thought as a sum of the anomalous and of the ordinary (non anomalous) one, which is homogeneous under the action of the anomalous symmetry transformation. \\
Historically, the first example of this kind is the Wess-Zumino effective action for the $SU(3)_{L}\times SU(3)_{R}$ 
flavour symmetry of low-energy QCD, describing the pion dynamics \cite{Wess:1971yu} and incorporating the effects of the chiral anomalies. 
In this case, pions appear as (pseudo-) Goldstone bosons introduced specifically to solve the variational problem defined by the anomalous 
Ward identities. \\
It is important to notice that anomalous effective actions solving the chiral constraints can also be defined without introducing 
additional scalar fields, but in this case they are non-local (see e.g. \cite{Coriano:2008pg} for an overview). \\
Of course, any variational solution of the anomaly equation cannot account for the non anomalous part part of the effective action
and is always defined modulo homogeneous terms. Determining such contributions requires a separate effort. For example, 
in the case of the one-particle irreducible effective action, it is necessary to explicitly evaluate the Feynman diagrams in the perturbative series,
taking into account all the degrees of freedom in the fundamental Lagrangian.

On the perturbative side, a signature of scalar degrees of freedom is present in computations of Feynman diagrams as well.
The chiral anomaly was discovered by Schwinger \cite{Schwinger:1951nm} and, later, by Adler, Bell and Jackiw 
\cite{Adler:1969gk,Bell:1969ts} in the $AVV$ diagram describing the decay of an axial-vector current ($A$) into two vector currents ($V$).
Soon after that, it was pointed out by Dolgov and Zakharov that a salient feature of this diagram is the presence of a one-particle massless pole
\cite{Dolgov:1971ri}. Finally, this massless pole was interpreted as a signature of a scalar degree of freedom, identified with the pion, 
interpolating between the axial and the vector currents. This interpretation is encoded in the modified PCAC relation connecting the divergence
of the axial current to the pion field plus the chiral anomaly term \cite{Adler:1969gk}, which successfully accounts for the experimental 
pion decay rate into two photons.

The situation for conformal anomalies has shown to be quite similar.
A non-local effective action for the trace anomaly in $4$ dimensions was proposed for the first time in a paper by Deser, Duff and Isham
\cite{Deser:1976yx}, but this action has a rather complicated form, as it contains logarithmic terms such as 
$\log\left((\Box + R)/\mu^2\right)$, where $\mu$ is a renormalization scale and $R$ the Ricci scalar, which are hard to expand
around the flat limit $g_{\mu\nu}= \eta_{\mu\nu}$. Moreover, the possibility of the existence of such logarithmic terms in the anomalous
effective action was subsequently ruled out through cohomological arguments in 
\cite{Mazur:2001aa,Boulanger:2007ab,Boulanger:2007st}.

An action which provides the minimal variational solution of the anomaly equation in a general curved background and can be 
easily expanded around  flat space was found by Riegert in 1987 \cite{Riegert:1987kt}. \\
A salient feature of Riegert's effective action is that, just like the non-local action for chiral anomalies, it predicts the existence of 
massless scalar poles coupled to the anomaly, as shown in \cite{Giannotti:2008cv}, so that, in order to complete the correspondence 
between chiral and conformal anomalies, this pole should be found in perturbative computations. \\
Given the structure of the trace anomaly (see chapter \ref{TTTVertex}), the easiest way to look for such an interpolating scalar
state in perturbation theory is to evaluate explicitly the $TVV$ correlator in flat space, where $T$ is the energy-momentum tensor and $V$ 
stands for a vector current.
This Green function, at $1$-loop, gives the next-to-leading order contribution to the interaction between a graviton and two gauge bosons
and is affected by the trace anomaly. In \cite{Giannotti:2008cv,Armillis:2009pq} 
its computation was performed in QED and the pole predicted by Riegert's action was found.
As Riegert's action holds for non abelian gauge theories as well, the computation was subsequently performed in QCD and in the Standard Model 
\cite{Armillis:2010qk,Coriano:2011zk}, confirming the presence of this contribution in all cases.

A similar analysis of anomaly poles in three point functions was performed in a supersymmetric context for $\mathcal{N}=1$ super
Yang-Mills theory in \cite{Coriano:2014gja}. In this work, it is shown that anomaly poles appear in the $\mathcal{JVV}$ correlator, with
$\mathcal{J}$ the Ferrara-Zumino hypercurrent and $\mathcal{V}$ the vector supercurrent.

The next thing to accomplish, in order to complete this program, is to look for anomaly poles in the contributions to the trace anomaly
depending only on the metric tensor. This requires the study of correlation functions of the energy-momentum tensor alone, 
as this is the quantum
operator sourced by the metric tensor. Actually, a trace anomaly already affects the $2$-point function of the energy-momentum tensor,
but it depends on the renormalization scheme, so that the first correlator where scheme-independent contributions can be found 
is the $TTT$ vertex.

Chapter \ref{TTTVertex} of this thesis presents a complete $1$-loop computation of the three graviton vertex in $4$ dimensions,
in three different free field theories which are conformally invariant, namely the scalar field with a proper term of improvement, 
the Dirac fermion and the abelian gauge boson. 
The computation is performed in the off-shell kinematic configuration, by evaluating all the diagrams in the perturbative expansion
with the Passarino-Veltman tensor-reduction technique, implemented in a symbolic manipulation program.
The results are tested by checking the general covariance and the trace Ward identities which descend from the well known master equations
for the conservation and the anomalous trace of the energy-momentum tensor. Renormalization is performed in the $\overline{MS}$ scheme.

The general result is given in terms of a set of $499$ scalar coefficients multiplying a corresponding basis of rank-$6$ tensors.
Due to the size of the general result, explicit coefficients are provided only in the limit in which two of the three gravitons are on the mass-shell, for which we expand all the correlator on a basis of only $13$ tensors.

In the end, Riegert's effective action for the conformal anomaly is explicitly introduced and we briefly review one of its two possible
local formulations, i.e. in terms of two auxiliary scalar fields, which is discussed, for instance, in \cite{Mottola:2006ew}. 
In this paper the infrared effective action for gravity is studied, with special focus on the terms induced by the anomaly,
which are shown to be relevant in the infrared and, as such, provide the first quantum corrections to General Relativity.
The possible appearance of scalar poles interpolating between the gravitons and the anomalous contributions to the $TTT$ vertex 
is briefly discussed as a suggestion for further investigation. \\

\vspace{0.5cm}
{\large \bfseries Conformal symmetry in position and in momentum space}
\vspace{0.5cm}

In chapter \ref{Mapping}, we temporarily turn away from the discussion of the conformal anomaly effective actions and exploit our computation
of the $TTT$ correlator to elaborate on the connection between conformal invariance in position and momentum space. \\
The implications of the constraints of conformal invariance have been worked out mostly in position space, as for instance in
\cite{Osborn:1993cr, Erdmenger:1996yc}, where the structure of various important $3$-point correlators was established,
modulo a small set of constants.

On the other hand, explicit evaluations of correlators in specific conformally invariant field theories are performed through the usual Feynman 
expansion, which is commonly and most easily implemented in momentum space.
This discrepancy is likely to hinder the comparison of results found in the two ways,
especially when it comes to such complicated correlators as the $TTT$, whose first explicit perturbative computation was performed 
for the first time in \cite{Coriano:2012wp} and is discussed here in chapter \ref{TTTVertex}.
Motivated by the search for a clear-cut way to map the results obtained with these different approaches into each other,
we develop two methods of comparison to which chapter \ref{Mapping} is devoted.

The first method is called the inverse mapping procedure. It starts from the integral expression in momentum space of the $1$-loop diagrams 
defining the correlators and proceeds with their (inverse) Fourier transform. This allows to set a precise correspondence between 
Feynman diagrams with specific different topologies and the non-local terms in the position space expressions for 
$3$-point functions which are provided in \cite{Osborn:1993cr}. 
Counterterms, which correspond to local terms in position space, are considered separately, 
as they have to be added by hand in most cases.

Clearly, this method makes sense only with theories for which a specific Lagrangian formulation exists and that are clearly defined in momentum
space. Nevertheless, the implications of conformal constraints are much more general, as they do not rely on a specific Lagrangian.
In this sense, after using extensively the inverse-mapping procedure to compare perturbative results with the constructions in
\cite{Osborn:1993cr}, we turn to the development of a second method, which works in the opposite direction. 
It is a general algorithm to Fourier-transform position space results, finding their expressions in terms of integrals in momentum space.
In order to deal with non Fourier-integrable expressions, the use of an intermediate regulator is required, in the spirit of differential regularization
\cite{Freedman:1991tk}. An interesting result of this analysis is a criterion to establish whether the conformal correlator which is studied
can be realized in the framework of a Lagrangian theory or not. 
In fact, the mapping procedure can end on momentum integrals containing logarithmic terms, 
which clearly cannot be generated from any Lagrangian theory. 
So, if the expressions of the transformed correlators contain combinations of such terms which cannot be re-expressed as 
ordinary Feynman integrals, then it is established that the underlying conformal field theory cannot be formulated in terms of a local Lagrangian. \\

\vspace{0.5cm}
{\large \bfseries Dilatons and effective actions for conformal anomalies}
\vspace{0.5cm}

Coming back to the discussion of the effective actions for conformal anomalies, we have already mentioned that Riegert's nonlocal action can be expressed in a local form
at the cost of introducing two auxiliary scalar fields \cite{Mottola:2006ew}. The presence of two such fields is the consequence
of the existence of $2$ independent cocycles for the Weyl group in $4$  dimensions 
(see \cite{Mazur:2001aa} and the discussion in chapter \ref{Recursive} of this work for details). \\
However, as the Weyl group is abelian, application of the general method of Wess and Zumino to the trace anomaly 
\cite{Schwimmer:2010za} implies that a local effective action can be built in terms of one single (pseudo-)Goldstone boson, 
which is usually called \emph{dilaton}. The construction of this effective action is discussed extensively in chapter \ref{Recursive}.

From chapter \ref{Effective} onwards, this thesis deals with dilaton interactions. \\
Dilaton states may be either fundamental or composite scalars. In the first case, they result from the compactification
of extra dimensions (graviscalars) or, on the other hand, they may appear as effective degrees of freedom of a more fundamental field theory, similar Nambu-Goldstone (NG) modes of a broken symmetry. While massless NG modes are always present in the case of a spontaneously broken global symmetry, for radiative breakings their massless nature is not 
necessarily guaranteed. In fact, non perturbative effects may contribute with a mass term and shift the position of the massless poles encountered in the 1-particle irreducible (1PI) anomaly action. 

If the dilaton is not a fundamental field, then, in close analogy with the pion case, it can be thought of as an effective state mediating the coupling of matter to the trace anomaly, according to the interactions derived from the Wess-Zumino action. \\
On the perturbative side, the pole identified in the $TVV$ correlator in  \cite{Giannotti:2008cv}
suggests that there might be such effective state interacting with matter via the trace anomaly.
Moreover, as the pion is a composite state of fermions, it is quite natural to elaborate on the idea that the dilaton
is a composite state of particles belonging to a strongly interacting sector which might be accessible in the near future at high energy colliders.
This possibility was suggested, for instance, in \cite{Goldberger:2007zk}.
In this sense, the Wess-Zumino action for the conformal anomaly could describe the low-energy limit of a theory
whose more fundamental components might be revealed at energies higher than those probed so far at the LHC. 
Then the dilaton could prove to be an effective degree of freedom surviving at energies lower than the scale $\Lambda$
at which conformal symmetry is broken.

Chapter \ref{Effective}, which is based on \cite{Coriano:2012nm}, discusses this scenario and presents complete $1$-loop
computations of the interactions of a graviscalar particle, derived from the compactification of large extra dimensions, with the
neutral gauge currents of the Standard Model. \\
Then we turn, in the same chapter, to a discussion of scale invariant extensions of the Standard Model and to the possibility that the anomaly
poles found in the $TVV$ in these theories \cite{Giannotti:2008cv,Armillis:2009pq,Armillis:2010qk,Coriano:2011zk} 
might be describing the emergence of an effective dilaton.
We must mention that a very similar scenario shows up in the $\mathcal{N}=1$ super Yang-Mills theory,
where anomaly poles appearing the triangle correlator of the Ferrara hypercurrent and two vector supercurrents can be interpreted
as a signal of the exchange of a composite dilaton/axion/dilatino multiplet in the effective Lagrangian \cite{Coriano:2014gja}.

Chapter \ref{Traced4T}, based on \cite{Coriano:2012dg}, extends the results of chapter \ref{Effective} 
by working in the conformal limit of the Standard Model, in which all the masses are set to zero. 
In this regime, we present the computation of $3$- and $4$-point traced correlators of the energy-momentum
tensor, which exactly match dilaton self-interactions in the on-shell limit.
Techniques presented in chapter \ref{Mapping}, relying on the connection between the trace anomaly and the
gravitational counterterms for conformal field theories, are used to secure the correctness of the result.

Finally, chapter \ref{Recursive} deals with the Wess-Zumino action for the geometric sector of the conformal anomaly 
both in $4$ and $6$ dimensions, putting together the results presented in \cite{Coriano:2013xua,Coriano:2013nja}.
The anomalous effective action is explicitly built by using the most general renormalization scheme, exploiting a cohomological
method presented for the first time in \cite{Mazur:2001aa}. Possible kinetic terms for the dilaton, which obviously cannot be derived by
the analysis of the anomalous constraints, are systematically reviewed.
After the derivation of the most general anomalous effective action, an interesting result is proven.
First, the Wess-Zumino anomalous effective action is written in another way, 
i.e. as a perturbative functional expansion with respect to the dilaton field,
which is also a power series in the inverse conformal breaking scale, $1/\Lambda$.
Each term of this power series, in turn, is a well defined and simple combination of traced Green functions of the energy-momentum tensor.
For consistency, then, one can require that each term in the perturbative expansion must match the term proportional to the same power
of $1/\Lambda$ in the explicit expression of the anomalous effective action which is derived in the first part with cohomological methods.
Imposing this consistency condition results in an infinite set of recurrence relations,
which allow to compute traced correlators of the energy-momentum tensor to an arbitrarily high order.

\mainmatter

\clearpage{\pagestyle{empty}\cleardoublepage}

\chapter{Conformal symmetry and Weyl symmetry}\label{Weyl}

\section{Introduction}

In this introductory chapter, we set the stage for all the results to be presented in the rest of the thesis.
The computations that we present are performed by embedding the quantum field theories that we are going to investigate 
in a curved metric background $g_{\mu\nu}$.
Hereafter, the term \emph{matter fields} will be used by us to refer to any fundamental field except for the metric tensor.
In particular we are going to present a short introduction on the concept of Weyl symmetry and its various realization in a curved background. 
This will be useful for the analysis presented in the later chapters.
Weyl invariance in a curved background allows to address the issue of conformal invariance in any free-falling frame.
We recall that conformal invariance implies the possibility to define a traceless energy-momentum tensor,
as discussed, for instance, in \cite{Polchinski:1987dy}. The search for theories which exhibit scale invariance but not conformal invariance has 
been at the center of several recent studies, as reviewed in \cite{Nakayama:2013is}. The advantage of dealing with a Weyl invariant theory in a 
curved background respect to a conformal symmetric theory in flat background is in the different character of the two symmetry groups, as 
the first one is abelian. 
Therefore, one can derive specific implications for a certain conformal invariant theory by starting from a simpler Weyl invariant 
theory on a curved background and the specialising the result to a local free falling frame \cite{Iorio:1996ad}.

In the following sections, we first introduce the conformal group, then proceed to discuss Weyl invariance in curved spacetime background. 
Finally, we review the argument presented in \cite{Iorio:1996ad}. Our attention will be limited to the case of any spacetime dimensions except for the case of $d=2$ since in this case the conformal group is infinite dimensional.

\section{The conformal group}

We present a brief review, in $d > 2$ dimensions and euclidean space, of the transformations which identify the conformal group $SO(2,d)$.
These may be defined as the transformations $x_\mu \rightarrow x'_\mu(x)$ that preserve the infinitesimal length up to a local factor 
\bea
d x_\mu d x^\mu \rightarrow d x'_\mu d x'^\mu = \Omega(x)^{-2} d x_\mu d x^\mu \, .
\label{ConformalMeasure}
\eea
In the infinitesimal form, the conformal transformations are given by
\bea
\label{xtransf}
x'_\mu(x) = x_\mu + a_\mu + \omega_{\mu\nu}\, x^\nu + \sigma\, x_\mu + b_\mu\, x^2 - 2\, b \cdot x\, x_\mu\ , ,
\eea
with
\bea
\Omega(x) = 1 -\lambda(x) \qquad \mbox{and} \qquad \lambda(x) = \sigma - 2 b \cdot x \, .
\label{Omega}
\eea
The transformation in eq. (\ref{xtransf}) is defined by translations ($a_\mu$), 
boosts and rotations ($\omega_{\mu\nu} = - \omega_{\nu\mu}$), 
dilatations ($\sigma$) and special conformal transformations ($b_\mu$). 
The first two define the Poincar\'{e} subgroup which leaves invariant the infinitesimal length and for which $\Omega(x) = 1$.
If we also consider the inversion
\bea
x_\mu \rightarrow x'_\mu = \frac{x_\mu}{x^2} \, , \qquad \qquad \Omega(x) = x^2 \, ,
\eea 
we can enlarge the conformal group to $O(2,d)$. 
Special conformal transformations can be realized by a translation preceded and followed by an inversion. \\

The Poincar\'{e} subgroup containts the basic set of symmetries for any relativistic system.
Its albegra is given by
\bea
i\, \left[ J^{\mu\nu}, J^{\rho\sigma}\right] 
&=& 
\delta^{\nu\rho}\, J^{\mu\sigma} - \delta^{\mu\rho}\, J^{\nu\sigma}
- \delta^{\mu\sigma}\, J^{\rho\nu} +   \delta^{\nu\sigma}\, J^{\rho\mu} \, ,\nn \\
i\, \left[P^\mu, J^{\rho\sigma} \right] 
&=&  
\delta^{\mu\rho}\,P^\sigma - \delta^{\mu\sigma}\, P^\rho \, ,\nn \\
\left[P^\mu,P^\nu\right] 
&=& 0 \, ,
\eea
where the $J$'s are the generators of the Lorentz group and the four components of the momentum $P^\mu$ generate rigid translations.
For scale invariant theories, which contain no dimensionful paramters, the Poincar\'{e} group can be extended by including the dilatation
generator $D$, corresponding to the fourth terms in the coordinate transformations in (\ref{xtransf}), 
whose commutation relations with the other generators are
\bea
\left[  P^\mu, D \right]             &=&    i\, P^\mu\,  , \nn \\
\left[ J^{\mu\nu}, D \right]   &=&   0\, .
\label{Dilatation}
\eea
Finally, it is possible to further extend this group so as to include the four special conformal transformations, 
whose generators we call $K^\mu$, extending the algebra as
\bea
\left [ K^\mu,D \right] &=& -i\, K^\mu   \, , \nn \\
\left [ P^\mu, K^\nu \right] &=& 2\,i\,\delta^{\mu\nu}\,D + 2\,i\,J^{\mu\nu}\, , \nn \\
\left [ K^\mu,K^\nu \right] &=& 0 \, , \nn \\
\left [ J^{\rho\sigma}, K^\mu \right] &=& i\,\delta^{\mu\rho}\,K^\sigma - i\,\delta^{\mu\sigma}\,K^	\rho \, .
\label{Conformality}
\eea
By the way, it is clear from (\ref{Dilatation}) and (\ref{Conformality}) that scale invariance does not require conformal invariance, 
but conformal invariance necessarily implies scale invariance.

Having specified the elements of the conformal group, we can define a quasi primary field $\mathcal O^i(x)$, where the index $i$ runs over the 
representation of the group to which the field belongs, through the transformation property under a conformal transformation $g$
\bea
\mathcal O^i(x) \stackrel{g}{\rightarrow} \mathcal O'^i(x') = \Omega(x)^\eta D^i_j(g) \mathcal O^j(x) \,,
\eea
where $\eta$ is the scaling dimension of the field and $D^i_j(g)$ denotes the representation of $O(1,d-1)$.
In the infinitesimal form we have
\bea
\delta_g \mathcal O^i(x) = 
- (L_g \mathcal O)^i(x) \,, \qquad \mbox{with} \qquad  L_g = v \cdot \partial + \eta \, \lambda + \frac{1}{2}
 \partial_{[ \mu} v_{\nu ]} \Sigma^{\mu\nu} \,,
\eea
where the vector $v_\mu$ is the infinitesimal coordinate variation $v_\mu = \delta_g x_\mu = x'_\mu(x) - x_\mu$ and 
$(\Sigma_{\mu\nu})^i_j$ are the generators of $O(1,d-1)$ in the representation of the field $\mathcal O^i$. 
The explicit form of the operator $L_g$ can be obtained from eq. (\ref{xtransf}) and eq. (\ref{Omega}) and is given by
\begin{align}
& \mbox{translations:} && L_g = a^\mu \partial_\mu \,, \nn \\
& \mbox{rotations:} && L_g = \frac{\omega^{\mu\nu}}{2} \left[ x_\nu \partial_\mu - x_\mu \partial_\nu - \Sigma_{\mu\nu} \right] \,, 
\nn \\
& \mbox{scale transformations :}    && L_g = \sigma \left[ x \cdot \partial +  \eta \right] \,, \nn  \\
& \mbox{special conformal transformations. :}    && L_g = b^\mu \left[ x^2 \partial_\mu - 2 x_\mu \, x \cdot \partial - 2 \eta \, x_\mu - 2 x_\nu 
{\Sigma_{\mu}}^{\nu} \right] \,.
\end{align}
As already remarked, invariance of a matter system under the conformal group implies the possibility to define a 
traceless energy-momentum tensor $T_I^{\mu\nu}$ \cite{Polchinski:1987dy},
\beq
{T_I^\mu}_\mu = 0\, ,
\eeq
where the subscript $I$ indicates that possible improvement terms have been added to the minimal energy-momentum tensor
which is obtained from the sole requirement of invariance under the Poincar\'{e} group.

\section{Weyl symmetry and its connection to conformal invariance}

Now that conformal symmetry in flat space has been introduced, we can move on to curved space and discuss Weyl symmetry.
We assume that the reader is familiar with basic General Relativity, including the Vielbein formalism which is necessary to embed
fermions in a gravitational field, for which we refer to \cite{Birrell:1982ix,Weinberg:1972gc}.
In the following, we follow the discussion in \cite{Iorio:1996ad}.

Let us suppose that our theory in flat space is described by an action functional
\beq
\mathcal S = \int d^d x \, \mathcal{L}(\Phi,\pd_\mu\Phi)\, ,
\eeq
depending on the matter fields $\Phi$ and their first derivatives. 
This theory can be easily embedded in curved space, replacing ordinary derivatives with diffeomorphism-invariant ones
\beq
\mathcal S = \int d^dx\, \sqrt{g}\, \mathcal{L}(\Phi,\nabla_\mu\Phi)\, ,
\eeq
The energy-momentum tensor of the theory is defined as the source of the gravitational 
field appearing in Einstein's equations and it is given by
\beq
T^{\mu\nu} = \frac{2}{\sqrt{g}}\, \frac{\delta \mathcal{S}}{\delta g_{\mu\nu}}\, .
\eeq
For a Lagrangian in flat space written in a diffeomorphic invariant form, scale invariance is equivalent to global Weyl invariance. 
The equivalence can be shown quite straightforwardly by rewriting a scale transformation acting on the coordinates of flat space 
and the matter fields $\Phi$,
\bea \label{ChIntFlatScaling}
x^\mu &\to& {x'}^{\mu}=e^\sigma x^\mu \, , \nn\\
\Phi(x)&\to& \Phi'(x')=e^{-d_\Phi \sigma}\Phi(x) \, ,
\eea
in terms of a rescaling of the metric tensor, the Vielbein and the matter fields
\bea \label{ChIntCurvedScaling}
g_{\mu\nu}(x) &\to&  e^{2\sigma}\, g_{\mu\nu}(x)\, , \nn\\
{V}_{a\,\rho}(x) &\to&  e^{\sigma}\, V_{a\,\rho}(x)\, , \nn\\
\Phi(x)&\to& e^{-d_\Phi \sigma}\Phi(x)\, ,
\eea
leaving the coordinates $x$ of the manifold invariant. We have denoted with $d_\Phi$ the field scaling dimension, 
which is deduced by an ordinary dimensional analysis of the Lagrangian density.
The reason why (\ref{ChIntFlatScaling}) can be traded for (\ref{ChIntCurvedScaling}) is that metric tensors
and the Vielbein always appear in order to contract derivative terms to obtain diffeomorphic scalars.

Once we move to a curved metric background, it is natural to promote the global scaling parameter $\sigma$ 
to a local function, so that the transformation laws of the metric, Vielbein and matter fields are
\bea 
\label{ChIntWeylTransf}
{g'}_{\mu\nu}(x)  &=&  e^{2\sigma(x)}\, g_{\mu\nu}(x)\, , \nn \\
{V'}_{a\,\rho}(x)  &=&  e^{\sigma(x)}\, V_{a\,\rho}(x)\, , \nn \\
\Phi'(x)           &=&  e^{-d_{\Phi}\, \sigma(x)}\, \Phi(x)\, .
\eea
The transformations of metric, Vielbein and matter fields shown in (\ref{ChIntWeylTransf}) define the abelian Weyl group.
It is natural to ask whether is possible to modify the theory in such a way that (\ref{ChIntWeylTransf}) leave the action functional invariant. 
Historically, the scale symmetry was the first whose gauging was systematically studied, in an attempt, made by Weyl, to connect 
electromagnetism with geometry. For this reason, this procedure was named Weyl-gauging.
It can be implemented in the same way as for QED, introducing an appropriate new field which takes a role similar to the vector potential.
This allows to define a new Lagrangian which is diffeomorphic and Weyl invariant at the same time.

For instance, for a free scalar theory described by the action
\beq
\mathcal{S}_{\phi} = \frac{1}{2}\, \int d^d x\, \sqrt{g} \, g^{\mu\nu} \, \partial_{\mu}\phi\,  \partial_\nu \phi , 
\label{ChIntkin}
\eeq
the derivative terms are modified according to
\beq \label{ChIntDerTransf}
\pd_\mu \rightarrow \pd^W_\mu = \pd_\mu - d_{\phi}\, W_{\mu}\, ,
\eeq
where $W_{\mu}$ is a vector gauge field that shifts under a Weyl trasnformation as
\beq \label{ChIntWeylGaugeField}
W_{\mu} \rightarrow W_{\mu} - \pd_\mu \sigma \, .
\eeq
In the case of a covariant derivative acting on higher spin fields, such as a spin-$1$ field $v_\mu$, the Weyl-gauging has to be supplemented
with a prescription to render the generally covariant derivative Weyl invariant, which is to add to (\ref{ChIntDerTransf}) the modified 
Christoffel connection
\beq \label{ChIntModChristoffel}
\hat\Gamma^\lambda_{\mu\nu} =
\Gamma^\lambda_{\mu\nu} + {\delta_\mu}^\lambda\, W_\nu + {\delta_\nu}^\lambda\, W_\mu - g_{\mu\nu}\, W^\lambda \, .
\eeq
It is easy to check that this Christoffel symbol is Weyl invariant. 
So, pursuing closely the analogy with the gauging of a typical abelian theory, 
we can define the Weyl covariant derivatives acting on vector fields as
\bea \label{ChIntWeylChristoffel}
\nabla^W_\mu v_\nu &=& \pd_\mu v_\nu - d_v\, W_\mu v_\nu - \hat\Gamma^\lambda_{\mu\nu} v_\lambda \, , \nn\\
\nabla^W_\mu v_\nu  &\rightarrow& e^{- d_v\sigma(x)}\, \nabla^W_\mu v_\nu \, ,
\eea
with an obvious generalisation to tensors of arbitrary rank. \\
Of course, the extension of such a derivative to the fermion case requires the Vielbein formalism and is obtained by the relation
\beq \label{ChIntWeylSpinConnection}
\nabla_{\mu} \rightarrow \nabla^W_\mu = \nabla_{\mu} - d_\psi\, W_{\mu} + 2\, {\Sigma_\mu}^\nu\, W_\nu\, , \quad 
\Sigma^{\mu\nu} \equiv V^{a\mu}\, V^{b\nu} \Sigma_{ab}\, ,
\eeq
where we have denoted with $d_\psi$ the scaling dimension of the spinor field ($\psi$) and with 
$\Sigma_{ab}$ the spinor generators of the Lorentz group. 

If we Weyl-gauge the scalar action (\ref{ChIntkin}) according to the prescriptions in (\ref{ChIntDerTransf}) 
and (\ref{ChIntWeylGaugeField}), we obtain
\bea \label{ChIntWeylGaugeFieldScalar}
S_{\phi,W}
&=& 
\frac{1}{2}\, \int d^dx\, \sqrt{g}\,  g^{\mu\nu}\, \pd^W_\mu \phi\, \pd^W_\nu \phi
= \frac{1}{2}\, \int d^dx\, \sqrt{g}\,  g^{\mu\nu}\,\bigg( \pd_\mu - \frac{d-2}{2}\, W_\mu \bigg)\,\phi\,
\bigg( \pd_\nu - \frac{d-2}{2}\, W_\nu \bigg)\, \phi \nn \\
&=&
\frac{1}{2}\, \int d^dx\, \sqrt{g}\,  g^{\mu\nu} \bigg\{ \pd_\mu \phi\, \pd_\nu\phi
- \frac{d-2}{2}\, \bigg( \phi\,W_\mu\,\pd_\nu\phi + \phi\,W_\nu\,\pd_\mu \phi - \frac{d-2}{2}\, W_\mu\,W_\nu\,\phi^2  \bigg)
\bigg\}
\eea
which, using $\phi\, \pd_{\mu}\phi = 1/2\, \pd_\mu \phi^2$ and integrating by parts, can be written in the form
\beq
\mathcal S_{\phi,W} = 
\frac{1}{2}\, \int d^dx\, \sqrt{g}\,  g^{\mu\nu}\, \bigg(\pd_\mu \phi\, \pd_\nu\phi 
                                                          + \phi^2\, \frac{d-2}{2}\, \Omega_{\mu\nu}(W) \bigg)\, ,
\label{ChIntgauged}
\eeq
where we have introduced the quantity 
\beq \label{ChIntOmega}
\Omega_{\mu\nu}(W) = \nabla_\mu W_\nu - W_\mu\,W_\nu + \frac{1}{2}\,g_{\mu\nu}\, W^2 \, .
\eeq
The result of this procedure is a Weyl invariant Lagrangian in which the Weyl variation of the
ordinary kinetic term of $\phi$ is balanced by the variation of the $\Omega$ term. 

This term plays a prominent role in the subsequent discussion, as we are going to ask whether it is possible to
enhance scale invariance to Weyl invariance without introducing the new degree of freedom $W_\mu$.
To this aim, we notice that there are two second rank tensors which can be built out of $W_\mu$ and its first covariant derivative, 
namely $W_\mu W_\nu$ and $\nabla_\mu W_\nu$. \\
From  (\ref{ChIntWeylGaugeField}), we can infer how they transform under a finite Weyl scaling,
\beq
\Delta (W_\mu W_\nu) = \sigma_\mu\sigma_\nu -(W_\mu\sigma_\nu+W_\nu\sigma_\mu)\, ,
\label{DeltaW2}
\eeq
where we habve introduced $ \sigma_\mu \equiv \pd_\mu \sigma $ to keep the notation easy.
To compute the finite Weyl-variation of $\Delta (\nabla_\mu W_\nu)$ we recall the 
transformation rule for the Christoffel connection, $\Gamma_{\mu\nu}^\lambda$
\beq
\Delta \Gamma_{\mu\nu}^\lambda = g^{\lambda\sigma}
\bigl(g_{\mu\sigma}\sigma_\nu+
g_{\nu\sigma}\sigma_\mu-
g_{\mu\nu}\sigma_\sigma\bigr),  \label{5.2}
\eeq             
which implies
\beq
\Delta (\nabla_\mu W_\nu)= - \nabla_\mu\sigma_\nu - g_{\mu\nu}\, \sigma \cdot \sigma 
+ 2\, \sigma_\mu\sigma_\nu + g_{\mu\nu}\, W \cdot  \sigma - \left(W_\mu\, \sigma_\nu + W_\nu\, \sigma_\mu \right)\, .
\label{DeltaNablaW}
\eeq
If we notice that contracting (\ref{DeltaW2}) we obtain
\beq
\Delta (g_{\mu\nu}W \cdot  W)= g_{\mu\nu}\left(\sigma \cdot  \sigma - 2\, W \cdot \sigma \right) \, ,
\label{5.4}
\eeq
then we immediately conclude that the variation under a finite Weyl shift of $\Omega_{\mu\nu}[W]$
is independent of $W_\mu$ and symmetric. More precisely, it is given by
\beq
\Delta\Omega_{\mu\nu}[W] = 
- \left(\nabla_\mu\sigma_\nu- \sigma_\mu\sigma_\nu 
+ \frac{1}{2}\,g_{\mu\nu}\;\sigma\!\cdot\!  \sigma \right) = - \Omega_{\mu\nu}[\sigma] \, . 
\label{5.6}
\eeq
As  $\Omega_{\mu\nu}[\sigma]$ depends only on the scaling parameter $\sigma$, 
one can argue that it may be related to some purely geometrical object. 
In fact, the variation of the Ricci tensor under a local scale trasnformation is given by
\beq
\Delta R_{\mu\nu}= R_{\mu\nu}[e^{2\sigma} g_{\mu\nu}]-R_{\mu\nu}[g_{\mu\nu}] 
= g_{\mu\nu}\nabla^2 \sigma
+(n-2)\Bigl(\nabla_\mu \sigma_\nu-\sigma_\mu \sigma_\nu+g_{\mu\nu}\;\sigma\!\cdot\!  \sigma\Bigr)\, .
\label{RicciVarInt}
\eeq
From (\ref{RicciVarInt}) we see that the tensor
\beq
S_{\mu\nu}= R_{\mu\nu}- \frac{g_{\mu\nu}}{2(n-1)}\, R
\label{DefineSInt}
\eeq
transforms under Weyl-scalings in the same way as $\Omega_{\mu\nu}$, i.e.
\beq
\Delta S_{\mu\nu}=(n-2)\Omega_{\mu\nu}[\sigma] \, .  
\label{DeltaSInt}
\eeq
Now it is clear when Weyl-gauging can be replaced by a non-minimal coupling to the curvature.
Since, according to (\ref{DefineSInt}) and (\ref{DeltaSInt}), the Weyl variation of $\Omega_{\mu\nu}[W]$ 
is proportional to the variation of $S_{\mu\nu}$, we see that whenever $W_\mu$ appears in the action 
only in the combination $\Omega_{\mu\nu}[W]$, it can be replaced by $S_{\mu\nu}$.
Replacing terms depending on $W_\mu$ with a non minimal couplig to the Ricci tensor is a procedure referred to as
\emph{Ricci gauging}.

We still have to explore under what conditions, in an action which is Weyl-gauged, the terms depending on $W_\mu$
appear only in the combination $\Omega_{\mu\nu}$.
We are going to see that the sufficient condition for Ricci gauging to be possible is preciely conformal invariance in flat space.

We can start our discussion by representing a conformal transformation of the metric in the form of a diffeomorphism
\beq
\frac{\pd x^\mu}{\pd x'^\alpha} \frac{\pd x^\nu}{\pd x'^\beta}g_{\mu\nu}(x) = 
\hat g_{\alpha\beta}(x') \quad \text{and} \quad \hat g_{\alpha\beta}(x)=  e^{\hat \sigma(x)}g_{\alpha\beta}(x) \, .
\label{6a.1}
\eeq
The functions $\hat\sigma$ in (\ref{6a.1}) form a subgroup of the group of local Weyl transformations that is induced 
by conformal transformations, that we call the conformal Weyl group. 
Given   (\ref{6a.1}), we can characterize the functions $\hat \sigma$ via the condition
\beq 
\frac{\pd x^\mu}{\pd x'^\alpha}\frac{\pd x^\nu}{\pd x'^\beta}R_{\mu\nu}(x)=
\hat{R}_{\alpha\beta}(y)= R_{\alpha\beta}[e^{2 \hat{\sigma}}g_{\mu\nu}](x')\, , 
\label{6a.2}
\eeq
which leads, in virtue of the transformation law of the tensor $S_{\mu\nu}$ under Weyl shifts, given by
(\ref{RicciVarInt}) and (\ref{DeltaSInt}), to the differential equation
\beq
(n-2)\sigma_{\alpha\beta}[\hat\sigma] = \hat S_{\mu\nu}-S_{\mu\nu}\, .
\label{6a.3}
\eeq
The existence of global solutions of (\ref{6a.3}) is a non-trivial problem in general. 
Now, as in the flat-space limit $S_{\alpha\beta}$ vanishes, the condition (\ref{6a.3}) reduces to
\beq
\pd_\nu \hat \sigma_\mu - \hat\sigma_\mu\hat\sigma_\nu + \frac{g_{\mu\nu}}{2} \;\hat\sigma\!\cdot\!\hat\sigma = 0 \, . 
\label{ChapIntDiffCondition}
\eeq
We find that the general solution of (\ref{ChapIntDiffCondition}) identifying the subset of the $\hat\sigma$ functions 
defining the conformal group of flat space is simply
\beq
\hat\sigma(x)= \log\left( 	\frac{1}{1- 2 \, b \cdot x + b^2 \, x^2} \right) \, ,
\label{ChIntHatSigma}
\eeq
where $b$ is any constant vector.  We see that (\ref{ChIntHatSigma}), for $b$ infinitesimal, 
corresponds exactly to eq.  (\ref{Omega}). When exponentiated as in (\ref{6a.1}), 
it gives a metric tensor producing the variation of the line interval defined in (\ref{ConformalMeasure}). \\

Now suppose that an action $\mathcal{S}$ admits Ricci gauging, so that the gauged action satisfies the condition 
\beq
\mathcal{S}(\Phi, S_{\mu\nu})= \mathcal{S}(\Phi', S_{\mu\nu}+ (n-2)\, \Omega_{\mu\nu}[\sigma])\, ,
\label{ca}
\eeq
where $\Phi'$ denotes the Weyl-transformed fields. 
It follows in particular that, if $\Omega_{\mu\nu}[\sigma]=0$, the action is invariant even without gauging. 
But the condition $\Omega_{\mu\nu}[\sigma]=0$ defines the conformal Weyl group in flat space, 
as shown in eqs.  (\ref{6a.3})-(\ref{ChIntHatSigma}); Ricci gauging is equivalent to the identity transformation, 
as the Riemann tensor always vanishes in flat space \cite{Weinberg:1972gc}.
This proves that, \emph{if an action admits Ricci gauging, it is conformally invariant in flat space}. \\

All that remains to be proved is that conformal invariance in flat space is also sufficient for the action to admit Ricci gauging.
This result will be proved \emph{for actions which which contain only first derivatives of the conformally variant fields}. 
Suppose that an action $\mathcal{S}_0$ is conformally invariant in flat space.
For infinitesimal conformal transformations we then have the conservation law
\beq
\delta \mathcal{S}_0 = \int d^dx \hat\sigma_{\mu}j^{\mu} = c_{\mu}\int j^{\mu}=0 \, ,  
\label{VirialInt}
\eeq
where $j^\mu$ is the virial current \cite{Polchinski:1987dy,Callan:1970ze}, defined as
\beq
j^\mu = \pi_\nu \left( d_{\Phi} \delta^{\mu\nu} + 2\, \Sigma^{\mu\nu} \right) \Phi \, , 
\quad 
\pi^\mu = \frac{\delta \mathcal{S}}{\delta (\pd_\mu\Phi)} \, ,
\label{VirialDefInt}
\eeq
where we remind that the $\Sigma$'s are the generators of the Lorentz group in the representation of the field $\Phi$
and $\hat\sigma = b_\nu\,x^\nu$. 
Eq.  (\ref{VirialInt}) holds for an arbitrary vector $b_\mu$, so it implies that $j^\mu$ is the divergence of a second rank tensor,
\beq
j^{\mu}= \pd_\nu J^{\mu\nu}\, .
\label{TotDivInt}
\eeq
As the actions we are considering contain only first derivatives of the conformally variant fields, the same must be true for $j^\mu$.
Therefore (\ref{TotDivInt}) is telling us that the tensor $J^{\mu\nu}$ depends only on the conformally variant fields 
and not on their derivatives, for otherwise $j^\mu$ would contain higher derivatives of the same fields.
This implies that $j^\mu$ is at most linear in the derivatives of the conformally variant fields. 
So, given the definition of the virial current (\ref{VirialDefInt}), 
the action $\mathcal{S}_{0}$ must be at most quadratic in the same derivatives. \\
We have got the intermediate result that \emph{as far as we consider actions containing only first derivatives of the conformally variant fields, 
conformal invariance is allowed only for those which are at most quadratic in such derivatives}. \\
In the case where the action is linear in the first derivatives the same argument implies that the virial current is identically zero. 
This happens, for instance, for the Dirac fermion, which does vary under coformal transformations. 
On the other hand, the argument does not impose any conditons on the conformally invariant fields. \\

Next we consider finite conformal transformations. 
As our actions are at most quadratic in the first derivatives of the conformally variant fields 
and there are no higher derivatives of any field whatsoever, their variation under finite conformal shifts is simply
\beq
\Delta \mathcal{S}_0 = \int d^dx\, \left(\hat\sigma_\mu j^\mu + \hat\sigma_\mu \hat\sigma_\nu K^{\mu\nu}\right)\, ,  
\label{6.6}
\eeq
where $K^{\mu\nu}$ does not contain derivatives of the fields.
Using the relation (\ref{TotDivInt}) , we can integrate by parts and write (\ref{6.6}) as
\beq
\Delta \mathcal{S}_0 = \int d^dx\, \left( - J^{\mu\nu}\pd_\mu \hat \sigma_\nu +\hat\sigma_\mu\hat\sigma_\nu K^{\mu\nu}\right)\, .  
\label{6.7}
\eeq
We can use (\ref{ChapIntDiffCondition}) to recast (\ref{6.7}) in the form
\beq
\Delta \mathcal{S}_0 = \int d^dx\, \hat\sigma_\mu \hat\sigma_\nu\, 
\left(K^{\mu\nu}- J^{\mu\nu}+\frac{\delta^{\mu\nu}}{2}{J^\lambda}_\lambda \right)\, .  
\label{6.8}
\eeq
The function $\hat\sigma$ is given in (\ref{ChIntHatSigma}): 
it has a specific form but otherwise depends on an arbitrary four-vector $b_\mu$.
This is sufficient
to conclude that the integrand in (\ref{6.8}) must vanish identically, so that
\beq
K^{\mu\nu}= J^{\mu\nu}- \frac{\delta^{\mu\nu}}{2}J
\qquad \text{or} \qquad 
J^{\mu\nu}= K^{\mu\nu}-\frac{\delta^{\mu\nu}}{n-2} K\, ,
\label{6.9}
\eeq
where we have denoted with $K$ and $J$ the traces.
Hence, invariance under finite conformal transformations implies that the virial tensor $J^{\mu\nu}$ 
is a specific linear function of the tensor $K^{\mu\nu}$ which appears in the quadratic expansion. \\
This allows to construct the Ricci-gauged action, for which we return to curved space, 
where the results just derived can be written as
\beq
j^\mu = \nabla_\nu J^{\mu\nu}\, ,
\quad  \text{with} \quad  
J^{\mu\nu}= K^{\mu\nu}- \frac{g^{\mu\nu}}{n-2} T  \, .   
\label{8.9}
\eeq
From the discussion in the preceding section, we learn that the Weyl gauge field is introduces only to compensate the variation of the 
derivatives of the fields which change under Weyl transformations, according to eq.  (\ref{ChIntWeylSpinConnection}). 
As the action is at most quadratic in the derivatives of the conformally variant fields, we can Weyl-gauge it by adding terms that are at most
quadratic in the Weyl field $W_\mu$, namely
\beq
\mathcal{S}= \mathcal{S}_0 + \int d^dx\, \sqrt{g}\, \left(W_\mu j^\mu + W_\mu W_\nu K^{\mu\nu}\right) \, .
\label{6.10}
\eeq
The form of the first term follows from eqs.  (\ref{ChIntWeylSpinConnection}) and (\ref{VirialDefInt}), 
while the tensor in the quadratic term is necessarily the same as in (\ref{6.6}), because derivative terms are the only ones that have
non trivial properties under both Weyl and conformal transformations.
By the same integration by part as above, this time in curved space, we find that
\beq
\mathcal{S} = \mathcal{S}_0 + 
\int d^dx \, \sqrt{g}\, \left(- J^{\mu\nu}\, \nabla_\mu W_\nu + W_\mu W_\nu\, K^{\mu\nu}\right) \, , 
\label{6.11}
\eeq
which, using (\ref{6.9}), can be written as
\beq
\mathcal{S} =\mathcal{S}_0 - \int d^dx\, \sqrt{g}\, J^{\mu\nu}\, \Omega_{\mu\nu}[W] \, .
\label{6.12}
\eeq
This shows that, for theories which are conformally invariant in the flat limit, the supplementary terms brought by the Weyl-gauging
appear only in the form $\Omega_{\mu\nu}[W]$, which is exactly the condition for Ricci gauging.  

Therefore we have proved that \emph{a necessary and sufficient condition for a scale invariant action $\mathcal{S}$ to allow
for Ricci gauging is that the flat-space limit of the ungauged action $\mathcal{S}_0$ is conformally invariant}.
The Ricci-gauging is achieved by (\ref{6.12}).

All the discussion so far is purely classical. At the quantum level, Weyl symmetry is violated by the trace anomaly,
which will be introduced in chapter \ref{TTTVertex}. \\

\clearpage{\pagestyle{empty}\cleardoublepage}

\chapter{The Three Graviton Vertex}\label{TTTVertex}

\section{Introduction}

In several recent works \cite{Giannotti:2008cv,Armillis:2009pq,Armillis:2010qk} certain correlation functions describing the 
interaction between a gauge theory and gravity with massless fields in the internal loop and related therefore to the trace 
anomalies in these theories have been analysed. The interesting property that such anomalous amplitudes contain massless poles in 
2-particle intermediate states has been exposed in these investigations. In particular, this has been demonstrated in the $TVV$ amplitude 
in QED, characterized by the insertion of the energy-momentum tensor ($T$) on 2-point functions of vector gauge currents ($V$). 
As long as the gravitational field is kept as a classical external source, which will always be the case in our treatment,
this amplitude gives the $1$-loop contribution to the interaction between a gauge theory and gravity, 
a part of which is mediated by the trace anomaly.

The complete evaluation of this amplitude in QCD and in the Standard Model \cite{Armillis:2010qk,Coriano:2011zk} confirms the 
conclusion of \cite{Giannotti:2008cv}, namely the presence of an effective massless scalar, ``dilaton-like" degree of freedom in 
intermediate 2-particle states, that is intimately connected with the trace anomaly, in the sense that the non-zero residue of the pole is 
necessarily proportional to the coefficient of the anomaly. The perturbative results of 
\cite{Giannotti:2008cv,Armillis:2009pq,Armillis:2010qk} 
are also in agreement with the anomaly-induced gravitational effective action in 4 dimensions whose non-local form was found in 
\cite{Riegert:1984kt}. It had been argued in \cite{Mottola:2006ew,Mottola:2010gp} that
the local covariant form of this anomalous effective action necessarily implies effective massless scalar degrees of freedom . 

This is the 4-dimensional analogue of the anomaly-induced action in 2-dimensional CFT's coupled to a background metric generated 
by the 2-dimensional trace anomaly and related to the central term in the infinite dimensional Virasoro algebra \cite{Polyakov:1981}. 
The anomaly-induced scalar in the 2-dimensional case is the Liouville mode of 
non-critical string theory on the 2-dimensional world sheet of the string.

In even dimensions greater than 2 it is important to recognize that the anomaly-induced effective action discussed in 
\cite{Mottola:2006ew,Mazur:2001aa,Riegert:1984kt,Mottola:2010gp} is determined only up to Weyl invariant terms. The full quantum 
effective action is not determined by the trace anomaly alone, and hence only when certain anomalous contributions to the $TVV$ or 
other amplitudes are isolated from their non-anomalous parts should any comparison with the anomaly-induced effective action be made.
The non-anomalous components are dependent upon additional Weyl invariant terms in the quantum effective action,
corresponding to traceless parts of the Green functions of the theory.

While \cite{Giannotti:2008cv,Armillis:2009pq,Armillis:2010qk,Coriano:2011zk} are focused on the search for signatures 
of massless scalar degrees of freedom in correlators describing the interaction of gauge fields with the background gravitational field,
exploiting the connection with Riegert's anomalous effective action, no such study has been attempted for the gravitational
field self-interactions, until recently. The simplest Green function accounting for such interaction, in the limit in which
the gravitational field is kept classical, is the $3$-graviton correlator, whose explicit evaluation is technically quite demanding. \\
In this chapter we present the first explicit perturbative $1$-loop computation of the three graviton vertex in $4$ dimensions.
The computation was performed in momentum space in a completely off-shell configuration, 
but the remarkable complexity of the general result allows us to present here, in a compact form, only the expression with two of three gravitons 
on the external lines in an on-shell configuration. \\
We discuss the derivation of general covariance and anomalous trace Ward identities for the correlator, 
all of which were explicitly checked in order to secure the correctness of the result.
The computation of the necessary one loop tensor integrals with three denominators and up to rank $6$ that are necessary 
for the evaluation of the correlator was performed with the Passarino-Veltman technique.

Though the original motivation for the explicit evaluation of the $\langle TTT \rangle$ Green function was the search of anomaly poles
in the geometric sector of the trace anomaly, we did not not attempt in this work to address the issue of their presence in the $TTT$ correlator.
Although this is an important motivation for initiating this study, the actual demonstration of the existence of such poles
requires a considerable additional effort, due to the extreme complexity of the result.
We expect to address this final point in a related work making use of the technical framework 
and building upon the results of the present study.

\section{Conventions and the trace anomaly equation}

Before beginning the discussion of the $TTT$ correlator investigated in our work, we introduce our definitions and conventions.

We recall that the ordinary definition of the energy-momentum tensor (which we will address as EMT from now on)
in a classical theory described by an action $\mathcal S$, which is embedded in curved space, is
\beq \label{Ch1EMT}
T^{\mu\nu}(z) = -\frac{2}{\sqrt{g_z}}\frac{\delta \mathcal{S}}{\delta g_{\mu\nu}(z)} = 
g^{\mu\alpha}(z)\,g^{\nu\beta}(z)\,\frac{2}{\sqrt{g_z}}\,\frac{\delta\mathcal S}{\delta g^{\alpha\beta}(z)} \, ,
\eeq
with $\textrm{det}\, g_{\mu\nu}(z)\equiv g_z$.

Now we introduce the generating functional of the theory in euclidean conventions, which we call $\mathcal W$,
\beq\label{Ch1Generating}
\mathcal W = \frac{1}{\mathcal{N}} \, \int \, \mathcal D\Phi \, 
e^{- \mathcal S - \int d^4x\, \sqrt{g}\,  A^a_\mu \,V^{a\,\mu} }\, ,
\eeq
where $\mathcal{N}$ is a normalization constant, $\Phi$ stands generally for all the quantum fields 
of the theory and we have explicitly added the coupling of vector currents to background gauge fields $A^a_\mu$.
Given (\ref{Ch1EMT}), the vacuum expectation value (vev) of the EMT is given, in terms of $\mathcal W$, by
\beq \label{Ch1VEVEMT}
\llangle T^{\mu\nu}(z) \rrangle_s = \frac{2}{\sqrt{g_z}}\frac{\delta\, \mathcal W}{\delta\, g_{\mu\nu}(z)}\,,
\eeq
with the subscript $s$ meaning that the background fields are kept turned on.
From now on the dependence on coordinates will be dropped when not strictly necessary.

As for conformally invariant field theories, which will be the subject of most of this work, the trace of the EMT is zero at the classical
level, ${T^\mu}_\mu = 0$, one would naively expect this to be true also for the vev of the EMT,
\beq
g_{\mu\nu} \llangle T^{\mu\nu} \rrangle_s = 0\, .
\eeq
But this is known not to be true, as the quantum theory shows non vanishing terms in this trace.
In particular, when the matter system which is classically conformal invariant at the classical level is embedded in a background of
gauge fields $A^a_\mu$ and a gravitational field described by the metric tensor $g_{\mu\nu}$, it is found, in $4$ dimensions,
that the traced vev is given by \cite{Duff:1977ay,Duff:1993wm,Birrell:1982ix}
\beq \label{Ch1TraceAnomaly}
g_{\mu\nu}\llangle T^{\mu\nu} \rrangle_s  \equiv \mathcal{A}[g, A]  =
\sum_{I=f,s,V}n_I \, \bigg[ \beta_a(I)\, F + \beta_b(I)\, G + \beta_c(I)\,\Box R + \beta_d(I)\, R^2  \bigg]
- \frac{\kappa}{4} \, n_V \, F^{a\,\mu\nu}\, F^a_{\mu\nu} \, ,
\eeq
where $g$ and $A^a$ are short-hand notations respectively for the background metric and gauge field and
the coefficients $\beta_a$, $\beta_b$, $\beta_c$ and $\beta_d$ depend on the field content of the Lagrangian theory
(e.g. fermions, scalars, vector bosons) and we have a multiplicity factor $n_I$ for each particle species\footnote{Equivalent and more popular 
notations are $c\equiv 16 \pi^2 \beta_a$ and $a\equiv -16 \pi^2 \beta_b$}. Actually the coefficient of $R^2$ must vanish identically
\beq
\beta_d \equiv 0
\eeq
since a non-zero $R^2$ term does not satisfy the Wess-Zumino consistency condition for conformal anomalies
\cite{Bonora:1983ff,AntMazMott:1992}. In addition, the value of $\beta_c$ is regularization dependent, corresponding to 
the fact that it can be changed by the addition of an arbitrary local term in the effective action proportional
to the integral of $R^2$.
In particular, the values for $\beta_c$ reported in table \ref{Ch1AnomalyCoeff} hold in dimensional regularization,
for which one finds the constraint \cite{Duff:1977ay,Birrell:1982ix}
\beq\label{Ch1constraints}
\beta_c = -\frac{2}{3}\,\beta_a \, .
\eeq
Thus only $\beta_a$, $\beta_b$ and $\kappa$ correspond to true anomalies in the trace of the stress tensor.
For the purpose of this study, the gauge field sector of the trace anomaly is not concerned, so that from now on
we will assume that there is no background gauge field, implying that the last term on the r.h.s in eq.  (\ref{Ch1TraceAnomaly}),
proportional to the squared field-strengths $F^a_{\mu\nu}$, is absent, so that 
the trace anomaly functional depends only on the metric, $\mathcal A \equiv \mathcal{A}[g]$.
In table \ref{Ch1AnomalyCoeff} we list the values of the coefficients for the three conformal free field theories with 
spin $0, \frac{1}{2}, 1$, in which the computations described in this work were performed.
\begin{table}
$$
\begin{array}{|c|c|c|c|}\hline
I & \beta_a(I)\times 2880\,\pi^2 & \beta_b(I)\times 2880\,\pi^2 & \beta_c(I)\times 2880\,\pi^2
\\ \hline\hline
S & \frac{3}{2} & -\frac{1}{2} & -1
\\ \hline
F & 9 & -\frac{11}{2} & -6
\\ \hline
V & 18 & -31 & -12
\\
\hline
\end{array}
$$
\caption{Anomaly coefficients for a conformally coupled scalar, a Dirac Fermion and a vector boson}
\label{Ch1AnomalyCoeff}
\end{table}

$\mathcal{A}[g]$ contains the diffeomorphism-invariants built out of the Riemann tensor,
${R^{\alpha}}_{\beta\gamma\delta}$, as well as the Ricci tensor $R_{\alpha\beta}$ and the scalar curvature $R$.
$G$ and $F$ in eq.  (\ref{Ch1TraceAnomaly}) are the Euler density and the square of the Weyl tensor respectively.
All our conventions are listed in appendix \ref{Geometrical}.

Eq.  (\ref{Ch1TraceAnomaly}) plays the role of a generating functional for the anomalous Ward identities of any underlying field theory. 
These conditions are not necessarily linked to any Lagrangian, since the solution of these and of the other (non anomalous) 
Ward identities - which typically constrain a given correlator - are based on generic requirements of conformal invariance.
Nevertheless, for our purposes, all these identities can be extracted from an ordinary generating functional, defined in terms 
of a generic Lagrangian $\mathcal{L}$, which offers a convenient device to identify such relations.

Inserting these definitions in (\ref{Ch1TraceAnomaly}) and multiplying both sides by $\sqrt{g}$  we obtain
\beq
\label{Ch1TraceAnomalySymm}
2 \, g_{\mu\nu}\frac{\delta\, \mathcal{W}}{\delta \, g_{\mu\nu}}=
\sqrt{g} \, \mathcal A[g] \, .
\eeq
From (\ref{Ch1TraceAnomaly}) and (\ref{Ch1TraceAnomalySymm}) we can extract identities for the anomaly of 
correlators involving $n$ insertions of energy-momentum tensors, just by taking $n$ functional derivatives of both sides 
with respect to the metric  of  (\ref{Ch1TraceAnomalySymm}) and setting $g_{\mu\nu}=\delta_{\mu\nu}$ at the end.
For vertices with multiple insertions of gravitons, such as the $TTT$ vertex, which are really involved,
a successful test of the anomalous Ward identities is crucial in order to secure the correctness of the result of the perturbative computation.

\section{Definitions for the $TTT$ Amplitude}
\label{Ch1DiagTTT}

For a multi-graviton vertex, it is convenient to define the corresponding correlation function as the n-th
functional variation with respect to the metric of the generating functional $\mathcal W$ evaluated in the flat-space limit
\bea \label{Ch1NPF}
\llangle T^{\mu_1\nu_1}(x_1)...T^{\mu_n\nu_n}(x_n) \rrangle 
&=&
\bigg[\frac{2}{\sqrt{g_{x_1}}}...\frac{2}{\sqrt{g_{x_n}}} \,
\frac{\delta^n \mathcal{W}}{\delta g_{\mu_1\nu_1}(x_1)...\delta g_{\mu_n\nu_n}(x_n)}\bigg]
\bigg|_{g_{\mu\nu} = \delta_{\mu\nu}} \nonumber \\ 
&=&  
2^n\, \frac{\delta^n \mathcal{W}}{\delta g_{\mu_1\nu_1}(x_1)...\delta g_{\mu_n\nu_n}(x_n)}\bigg|_{g_{\mu\nu} = 
\delta_{\mu\nu}} \, ,
\eea
so that it is explicitly symmetric with respect to the exchange of any couple of metric tensors.
As we are going to deal with correlation functions evaluated in the flat-space limit all through the work,
we will omit to specify it from now on, so as to keep our notation easy.

\begin{figure}[t]
\centering
\includegraphics[scale=0.8]{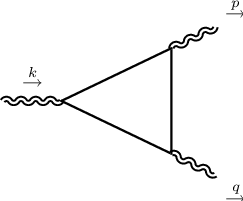}
\hspace{5mm}
\includegraphics[scale=0.8]{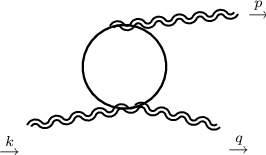}
\hspace{5mm}
\includegraphics[scale=0.8]{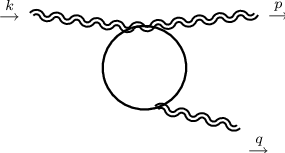}
\hspace{5mm}
\includegraphics[scale=0.8]{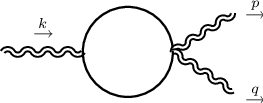}
\hspace{5mm}
\caption{One loop expansion of the 3-graviton vertex in terms of the triangle and the self-energy-type contributions.
Shown here are the diagrams for the scalar particle. The expansion for the other two CFT's investigated can be obtained by replacing 
the scalar by a fermion or a photon in the loops. In the former case one has to consider, for the triangle case, two inequivalent contributions,
distinguished by the direction of flow of the momentum flow of the fermion; for the latter, ghost corrections follow the same topologies.}
\label{Ch1Fig.diagramsTTT}
\end{figure}

The 3-point function we are interested in studying is found by evaluating (\ref{Ch1NPF}) for $n=3$,
\bea\label{Ch13PF}
\llangle T^{\mu\nu}(x_1)T^{\rho\sigma}(x_2)T^{\alpha\beta}(x_3)\rrangle
&=&
8 \, \bigg[ - \, \llangle \frac{\delta \mathcal S}{\delta g_{\mu\nu}(x_1)}\frac{\delta \mathcal S}{\delta g_{\rho\sigma}(x_2)}
\frac{\delta \mathcal S}{\delta g_{\alpha\beta}(x_3)}\rrangle \nonumber \\
&& \hspace{-15mm}
+ \, \llangle \frac{\delta^2 \mathcal S }{\delta g_{\alpha\beta}(x_3)\delta g_{\mu\nu}(x_1)}
\frac{\delta \mathcal S }{\delta g_{\rho\sigma}(x_2)} \rrangle
 + \llangle \frac{\delta^2 \mathcal S }{\delta g_{\rho\sigma}(x_2)\delta g_{\mu\nu}(x_1)}
\frac{\delta \mathcal S }{\delta g_{\alpha\beta}(x_3)} \rrangle
\nonumber \\
&&\hspace{-15mm}
+ \, \llangle \frac{\delta^2 \mathcal S}{\delta g_{\rho\sigma}(x_2)\delta g_{\alpha\beta}(x_3)}
\frac{\delta \mathcal S }{\delta g_{\mu\nu}(x_1)}\rrangle
- \,\llangle \frac{\delta^3 \mathcal S}{\delta g_{\rho\sigma}(x_2)\delta g_{\alpha\beta}(x_3)
\delta g_{\mu\nu}(x_1)}\rrangle \bigg] \, .
\nonumber \\
\eea
The last term is identically zero in dimensional regularization, being proportional to a massless tadpole.
The Green function
\beq
\llangle \frac{\delta\mathcal S}{\delta g_{\mu\nu}(x_1)}
            \frac{\delta\mathcal S}{\delta g_{\rho\sigma}(x_2)}
            \frac{\delta\mathcal S}{\delta g_{\alpha\beta}(x_3)}
            \rrangle \, 
\label{Ch1Triangle}
\eeq
has the diagrammatic representation of a triangle topology, while the contributions
\bea \label{Ch1bubbles}
\llangle \frac{\d^2 \mathcal S}{\delta g_{\rho\sigma}(x_2)\delta g_{\alpha\beta}(x_3)}
            \frac{\delta\mathcal S}{\delta g_{\mu\nu}(x_1)} \rrangle \, ,
\quad
\llangle \frac{\d^2 \mathcal S}{\delta g_{\alpha\beta}(x_3)\delta g_{\mu\nu}(x_1)}
            \frac{\delta\mathcal S}{\delta g_{\rho\sigma}(x_2)} \rrangle \, ,
\quad
\llangle \frac{\d^2 \mathcal S}{\delta g_{\rho\sigma}(x_2)\delta g_{\mu\nu}(x_1)}
            \frac{\delta\mathcal S}{\delta g_{\alpha\beta}(x_3)} \rrangle
\eea
have the topology of 2-point functions and are traditionally named, in perturbative analysis, as ``bubbles''. 
We decide to call them "$k$", "$q$" and "$p$" bubbles respectively, naming each one after the momentum
flowing into or out of the single graviton vertex.
The diagrammatic representation of the four contributions is show in fig. \ref{Ch1Fig.diagramsTTT}.

We convey to choose a dependence on the momenta such that $k$ is incoming at the point $x_1$ 
and $q$ and $p$ are outgoing at $x_2$ and $x_3$ respectively. 
These conventions are summarized by the Fourier transform
\beq
\int \, d^4x_1\,d^4x_2\,d^4x_3\, 
\llangle T^{\mu\nu}(x_1)T^{\rho\sigma}(x_2)T^{\alpha\beta}(x_3)\rrangle \,
e^{-i(k\cdot x_1 - q\cdot x_2 - p\cdot x_3)} =
(2\pi)^4\,\delta^{(4)}(k-p-q)\, \llangle T^{\mu\nu}T^{\rho\sigma}T^{\alpha\beta}\rrangle(q,p)\, .
\label{Ch13PFMom}
\eeq
Of course, for a $2$-point function we have
\beq\label{Ch12PFMom}
\int \, d^4x_2 \, d^4x_3 \, \llangle T^{\rho\sigma}(x_2)T^{\alpha\beta}(x_3) \rrangle\, e^{-i(q\cdot x_2 - p\cdot x_3)} =
(2\pi)^4\,\delta^{(4)}(p-q)\, \llangle T^{\rho\sigma}T^{\alpha\beta} \rrangle(p) \, .
\eeq

It proves particularly useful to introduce a specific notation for the flat limit of functional derivatives with respect to the metric,
\beq \label{Ch1Flat}
\left[\mathcal F\right]^{\muu\nuu\mud\nud\dots\mu_{n}\nu_{n}}(\xu,\xd,\dots,x_n) \equiv
\frac{\delta^n\, \mathcal F}{\delta g_{\muu\nuu}(\xu)\,\delta g_{\mud\nud}(\xd)\dots\delta g_{\mu_n\nu_n}(x_{n})} 
\bigg|_{g_{\mu\nu}=\delta_{\mu\nu}} \, , 
\eeq
for any functional (or function) $\mathcal F$ which depends on the background field $g_{\mu\nu}(x)$. 

\subsection{General covariance Ward identities for the $TTT$}
\label{Ch1DiagTTTWardCov}

The requirement of general covariance for the generating functional $\mathcal W$
immediately leads to the master Ward identity for the conservation of the energy-momentum tensor.
Assuming that the integration measure is invariant under diffeomorphisms $\mathcal D \Phi' = \mathcal D \Phi'$,
standard manipulations lead to the master equation
\beq \label{Ch1masterWI0}
\nabla_\nu \llangle T^{\mu\nu}(x_1) \rrangle = 
\nabla_\nu \bigg(\frac{2}{\sqrt{g_{x_1}}}\frac{\delta\mathcal W}{\delta g_{\mu\nu}(x_1)}\bigg) = 0 \, ,
\eeq
which, expanding the covariant derivative, can be written as
\beq
\frac{2}{\sqrt{g_{x_1}}}\bigg(\pd_{\nu}\frac{\delta\mathcal W}{\delta g_{\mu\nu}(x_1)}
- \Gamma^\lambda_{\lambda\nu}(x_1)\frac{\delta\mathcal W}{\delta g_{\mu\nu}(x_1)}
+ \Gamma^\mu_{\kappa\nu}(x_1)\frac{\delta\mathcal W}{\delta g_{\kappa\nu}(x_1)}
+ \Gamma^\nu_{\kappa\nu}(x_1)\frac{\delta\mathcal W}{\delta g_{\mu\kappa}(x_1)}\bigg) = 0 \, ,
\eeq
where the first of the three Christoffel symbols (for their definition see appendix \ref{Sign}) is generated by differentiation 
of ${1}/{\sqrt{g_{x_1}}}$ in the definition of $T_{\mu\nu}(\xu)$ in (\ref{Ch1masterWI0}) together with
\beq
\Gamma^\alpha_{\alpha\beta}= \frac{1}{2} \, g^{\alpha\gamma} \, \pd_\beta \, g_{\alpha\gamma} \,
\eeq
or, equivalently, as
\beq
2 \, \bigg(\pd_{\nu}\frac{\delta\mathcal W}{\delta g_{\mu\nu}(x_1)}
+ \Gamma^\mu_{\kappa\nu}(x_1)\frac{\delta\mathcal W}{\delta g_{\kappa\nu}(x_1)}\bigg) = 0\, .
\label{Ch1masterWI}
\eeq
By taking one and two functional derivatives of (\ref{Ch1masterWI}) with respect to  $g_{\rho\sigma}(x_2)$ and
$g_{\rho\sigma}(x_2)$ and $g_{\alpha\beta}(x_3)$ respectively, one gets, in curved spacetime,
\bea
&&
4 \, \bigg[ \pd_\nu \frac{\delta^2\mathcal W}{\delta g_{\rho\sigma}(x_2)\delta g_{\mu\nu}(x_1)}
+ \frac{\delta \Gamma^\mu_{\kappa\nu}(x_1)}{\delta g_{\rho\sigma}(x_2)}
\frac{\delta\mathcal W}{\delta g_{\kappa\nu}(x_1)}
+ \Gamma^\mu_{\kappa\nu}(x_1) \frac{\delta^2 \mathcal W}{\delta g_{\mu\nu}(x_1)\delta g_{\rho\sigma}(x_2)}\bigg]
= 0 \, , \label{Ch1WI2PFCoordinate} \\
&&
8 \bigg[\pd_\nu\frac{\delta^3\mathcal W}{\delta g_{\alpha\beta}(x_3)\delta g_{\rho\sigma}(x_2)\delta g_{\mu\nu}(x_1)}
+ \frac{\delta \Gamma^\mu_{\kappa\nu}(x_1)}{\delta g_{\rho\sigma}(x_2)}
  \frac{\delta^2 \mathcal W}{\delta g_{\alpha\beta}(x_3)\delta g_{\kappa\nu}(x_1)}
+ \frac{\delta \Gamma^\mu_{\kappa\nu}(x_1)}{\delta g_{\alpha\beta}(x_3)}
  \frac{\delta^2 \mathcal W}{\delta g_{\rho\sigma}(x_2)\delta g_{\kappa\nu}(x_3)} 
\nonumber \\
&&
+ \frac{\delta^2\Gamma^\mu_{\kappa\nu}(x_1)}{\delta g_{\rho\sigma}(x_2)\delta g_{\alpha\beta}(x_3)}
  \frac{\delta\mathcal W}{\delta g_{\mu\nu}(x_1)}
+ \Gamma^\mu_{\kappa\nu}(x_1)
  \frac{\delta^3\mathcal W}{\delta g_{\rho\sigma}(x_2)\delta g_{\alpha\beta}(x_2)\delta g_{\kappa\nu}(x_1)}\bigg] = 0 \, .
\label{Ch1WI3PF} 
\eea
As we are interested in the flat spacetime limit, we must evaluate (\ref{Ch1WI2PFCoordinate}) 
and (\ref{Ch1WI3PF}) by letting the Christoffel symbols go to zero.
Another simplification is obtained by noticing that the Green's functions
\beq
\llangle \frac{\delta \mathcal S}{\delta g_{\mu\nu}(x_1)}\rrangle = - \frac{\delta \mathcal W}{\delta g_{\mu\nu}(x_1)}
\eeq
and
\beq\label{Ch1Tadpole2PF}
\llangle \frac{\delta^2 \mathcal S}{\delta g_{\mu\nu}(x_1)\delta g_{\alpha\beta}(x_3)}\rrangle
\eeq
are proportional to massless tadpoles, so that we can ignore them in the expression of the $2$-point function,
\beq
\frac{\d^2 \mathcal W}{\delta g_{\alpha\beta}(x_3) \delta g_{\mu\nu}(z)}
= \llangle \frac{\delta \mathcal S}{\delta g_{\mu\nu}(x_1)}
\frac{\delta \mathcal S}{\delta g_{\alpha\beta}(x_3)}\rrangle
- \llangle \frac{\d^2 \mathcal S}{\d g_{\alpha\beta}(x_3)\delta g_{\mu\nu}(x_1)} \rrangle
= \llangle \frac{\delta \mathcal S}{\delta g_{\mu\nu}(x_1)}
\frac{\delta \mathcal S}{\delta g_{\alpha\beta}(x_3)}\rrangle \, .
\eeq
Thus the Ward identity for the $2$-point function in flat coordinates is immediately seen to be
\beq
\pd_{\nu} \llangle T^{\mu\nu}(x_1) T^{\rho\sigma}(x_2) \rrangle = 0 \, ,
\eeq
where, due to the vanishing of (\ref{Ch1Tadpole2PF}), we have set
\beq \label{Ch12PF}
\llangle T^{\mu\nu}(x_1) T^{\rho\sigma}(x_2) \rrangle
\equiv
4\, \llangle \frac{\delta\mathcal S}{\delta g_{\mu\nu}(x_1)}
\frac{\delta\mathcal S}{\delta g_{\rho\sigma}(x_2)} \rrangle \, .
\eeq
Obviously, its form in momentum space, exploiting the Fourier-transform (\ref{Ch12PFMom}), is
\beq \label{Ch1WI2PFMom}
p_{\mu}\llangle T^{\mu\nu}T^{\rho\sigma} \rrangle(p) = 0 \, .
\eeq
The terms surviving in (\ref{Ch1WI3PF}) are those in the first line.
In order to make them explicit, we evaluate the functional derivative of the Christoffel symbols using
the rules in appendix \ref{Sign}, namely (\ref{Ch1Christoffel}), (\ref{Ch1Tricks}) and (\ref{Ch1Tricks2}), finding
\bea\label{Ch1GammaDerivatives1}
\frac{\delta\Gamma^\mu_{\kappa\nu}(x_1)}{\delta g_{\rho\sigma}(x_2)}
&=&
\frac{1}{2}\delta^{\mu\alpha}\bigg[-s^{\rho\sigma}_{\,\,\,\,\,\,\kappa\nu}\pd_\alpha
+ s^{\rho\sigma}_{\,\,\,\,\,\,\alpha\nu}\pd_\k + s^{\rho\sigma}_{\,\,\,\,\,\,\alpha\kappa} \pd_\nu\bigg]\delta^{4}(x_{12})\, ,
\eea
where we establish the convention $\delta^{4}(x_{12})\equiv \delta(x_1-x_2)$ and so on for the other couples of points.
Plugging this into (\ref{Ch1WI3PF}) and using (\ref{Ch12PF}), the second term becomes
\bea
8 \, \frac{\delta\Gamma^\mu_{\k\nu}(x_1)}{\delta g_{\r\s}(x_2)}
\frac{\delta^2 \mathcal W}{\delta g_{\alpha\beta}(x_3)\delta g_{\kappa\nu}(x_1)} 
&=&
\bigg[\delta^{\mu\r}\llangle T^{\nu\sigma}(x_1)T^{\alpha\beta}(x_3)\rrangle \, \pd_\nu +
\delta^{\mu\s}\llangle T^{\nu\rho}(x_1)T^{\alpha\beta}(x_3)\rrangle \, \pd_\nu 
\nonumber\\
&& \hspace{5mm}
- \, \llangle T^{\rho\sigma}(x_1)T^{\alpha\beta}(x_3)\rrangle \, \pd^\mu\bigg]\delta^{4}(x_{12}) \, .
\eea
A completely analogous relation holds for the exchanged term
$\big(g_{\alpha\beta}(x_3)\leftrightarrow g_{\rho\sigma}(x_2)\big)$. \\
Finally, we can recast the Ward identity (\ref{Ch1WI3PF}) in the form
\bea \label{Ch1WI3PFcoordinate}
\pd_\nu\llangle T^{\mu\nu}(x_1)T^{\rho\sigma}(x_2)T^{\alpha\beta}(x_3) \rrangle
&=&
\bigg[\llangle T^{\rho\sigma}(x_1)T^{\alpha\beta}(x_3)\rrangle\pd^\mu\delta^{4}(x_{12}) + 
\llangle T^{\alpha\beta}(x_1)T^{\rho\sigma}(x_2)\rrangle\pd^\mu\delta^{4}(x_{13}) \bigg]\nn \\
&-&
\bigg[\delta^{\mu\rho}\llangle T^{\nu\sigma}(x_1)T^{\alpha\beta}(x_3)\rrangle
+     \delta^{\mu\sigma}\llangle T^{\nu\rho}(x_1)T^{\alpha\beta}(x_3)\rrangle\bigg]\pd_\nu\delta^{4}(x_{12})\nn\\
&-&
\bigg[\delta^{\mu\alpha}\llangle T^{\nu\beta}(x_1)T^{\rho\sigma}(x_2)\rrangle
+ \delta^{\mu\beta}\llangle T^{\nu\alpha}(x_1)T^{\rho\sigma}(x_2)\rrangle\bigg]\pd_\nu\delta^{4}(x_{13})\, ,\nn\\
\eea
having used the definitions (\ref{Ch1NPF}) and (\ref{Ch13PF}).\\
Fourier-transforming according to (\ref{Ch13PFMom}) and (\ref{Ch12PFMom}), 
we get the Ward identity in momentum space that we need, i.e.
\bea\label{Ch1WI3PFmomenta2a}
&& k_\nu \llangle T^{\mu\nu}T^{\rho\sigma}T^{\alpha\beta} \rrangle(q,p) =
   p^\mu \llangle T^{\rho\sigma}T^{\alpha\beta}\rrangle(q) 
 + q^\mu \llangle T^{\rho\sigma}T^{\alpha\beta}\rrangle(p) \nn \\
&&
-  p_\nu \bigg[\delta^{\mu\beta} \llangle T^{\nu\alpha}T^{\rho\sigma} \rrangle(q)
             + \delta^{\mu\alpha}\llangle T^{\nu\beta}T^{\rho\sigma}  \rrangle(q)\bigg]
-  q_\nu \bigg[\delta^{\mu\sigma}\llangle T^{\nu\rho}T^{\alpha\beta}  \rrangle(p)
+              \delta^{\mu\rho}  \llangle T^{\nu\sigma}T^{\alpha\beta}\rrangle(p)\bigg].
\eea
Similar Ward identities can be obtained when we contract with the momenta of the other lines. These are going to be essential in 
order to test the correctness of the computation once we turn to perturbation theory.

\subsection{The anomalous Ward identities for the TTT}
\label{Ch1DiagTTTWardAnom} 

The anomalous Ward identities for the 3-graviton vertex is obtained performing two functional variations 
of (\ref{Ch1TraceAnomalySymm}) and taking the flat-space limit, thereby obtaining
\bea
\delta_{\mu\nu}\llangle T^{\mu\nu}T^{\rho\sigma}T^{\alpha\beta} \rrangle(q,p)
&=&
4 \, \left[\mathcal A[g]\right]^{\rho\sigma\alpha\beta}(q,p)
- 2 \, \llangle T^{\rho\sigma} T^{\alpha\beta} \rrangle(q) - 2 \, \llangle T^{\rho\sigma}T^{\alpha\beta} \rrangle(p)\nn\\
&=&
4 \, \bigg[ \beta_a\,\big(\big[F\big]^{\rho\sigma\alpha\beta}(q,p)
- \frac{2}{3} \big[\sqrt{g}\Box\,R\big]^{\rho\sigma\alpha\beta}(q,p)\big)
+ \beta_b\, \big[G\big]^{\rho\sigma\alpha\beta}(q,p) \bigg]\nn\\
&-&
2 \, \llangle T^{\rho\sigma}T^{\alpha\beta} \rrangle(q) 
- 2 \, \llangle T^{\rho\sigma}T^{\alpha\beta} \rrangle(p) \, ,\label{Ch1munu3PFanomaly}
\eea
where $\left[ \dots \right]^{\alpha\beta\rho\sigma}(p,q)$ are contributions generated by functional derivatives of the anomaly,
according to the notation introduced in (\ref{Ch1Flat}).
We remark, if not obvious, that all the contractions with the metric tensor in the flat spacetime limit ($\delta_{\mu\nu}$) 
should be understood as being 4-dimensional. This is the case for all the anomaly equations.
The various contributions to the trace anomaly are given in terms of
the functional derivatives of quadratic invariants in appendix \ref{Ch1Functionals}. Analogous anomalous 
Ward identities can be obtained by tracing the other two pairs of indices. 

\section{Three free field theory realizations of conformal symmetry}
\label{Ch1lags}

At this point we illustrate our perturbative computation in momentum space. 
It was performed within the context of three free field theories, namely a conformally coupled (improven) scalar,
a Dirac fermion and the free Maxwell field.

The actions for the scalar and the fermion fields are respectively
\bea\label{Ch1}
\mathcal{S}^s
&=&
\frac{1}{2} \, \int d^4 x \, \sqrt{g}\,
\bigg[g^{\mu\nu}\,\nabla_\mu\phi\,\nabla_\nu\phi - \chi\,R\,\phi^2 \bigg]\, ,\label{Ch1scalarAction}\\
\mathcal{S}^f
&=&
\frac{1}{2} \, \int d^4 x \, V \, {V_\alpha}^\rho\,
\bigg[\bar{\psi}\,\gamma^\alpha\,(\mathcal{D}_\rho\,\psi) - (\mathcal{D}_\rho\,\bar{\psi}) \, \gamma^\alpha\,\psi \bigg]  \, . 
\label{Ch1DiracAction}
\eea
Here $\chi$ is the parameter corresponding to the ``improvement term" one must add to the action of the free scalar field 
so as to obtain a Weyl-invariant action; in particular, its value in $4$ dimensions has to be $\chi = 1/6$;
the symbol ${V_\alpha}^\rho$ is instead the Vielbein and $V(= \sqrt{g})$ its determinant, 
needed to embed fermions in the curved background, with its covariant derivative $\mathcal{D}_\mu$ as
\beq
\mathcal{D}_\mu = \pd_\mu + \Gamma_\mu =
\pd_\mu + \frac{1}{2} \, \Sigma^{\alpha\beta} \, {V_\alpha}^\sigma \, \nabla_\mu\,V_{\beta\sigma} \, .
\eeq
The $\Sigma^{\alpha\beta}$ are the generators of the Lorentz group  in the case of a spin $1/2$-field. 
For more details on embedding classical fields into a curved background, we refer to \cite{Birrell:1982ix,Weinberg:1972gc}. \\
The action $\mathcal{S}_V$ for the photon field is given by the sum of three terms
\beq
\mathcal{S}_V = \mathcal{S}_M + \mathcal{S}_{gf} + \mathcal{S}_{gh}\, ,
\eeq
where the superscript $V$ stands for vector boson and the three contributions are the Maxwell action, 
the gauge fixing contribution and the ghost action, that must be taken into account as well, 
as gravity couples to any field with the same strength,
\bea
\mathcal{S}_M    &=&   \frac{1}{4} \, \int d^4 x \, \sqrt{g} \, F^{\a\b} F_{\a\b}\, ,\\
\mathcal{S}_{gf} &=&   \frac{1}{2 \xi} \, \int d^4 x \, \sqrt{g} \, \left( \nabla_{\alpha}A^\alpha \right)^2\, ,\\
\mathcal{S}_{gh} &=& - \int d^4 x \, \sqrt{g}\, \partial^\a \bar{c} \, \partial_\a c\, .
\eea
When performing formal manipulations with the \emph{Vielbein}, one has to correspondingly modify the definition of 
the functional derivative with respect to the background metric, so that the EMT defined in (\ref{Ch1EMT})becomes,
in the fermion case,
\beq \label{Ch1FerTEI}
T^{\mu\nu} = - \frac{1}{V} \, V^{\alpha\mu} \, \frac{\delta\mathcal{S}}{\delta {V^\alpha}_\nu}\, .
\eeq
This tensor is not symmetric in general, but its antisymmetric part does not contribute to our calculations, 
so that, for our purposes, we can adopt the symmetric definition
\beq\label{Ch1SymmFerTEI}
T^{\mu\nu}
\stackrel{def}{\equiv} - \frac{1}{2\,V}\bigg(
V^{\alpha\mu} \, \frac{\delta}{\delta {V^\alpha}_\nu} + 
V^{\alpha\nu}\, \frac{\delta}{\delta {V^\alpha}_\mu}\bigg) \, \mathcal S \, .
\eeq
The EMT's for the scalar and the fermion are then
\bea
T^{s\,\mu\nu}
&=&
\nabla^\mu \phi \, \nabla^\nu\phi - \frac{1}{2} \, g^{\mu\nu}\,g^{\alpha\beta}\,\nabla_\alpha \phi \, \nabla_\beta \phi
+ \chi \bigg[g^{\mu\nu} \Box - \nabla^\mu\,\nabla^\nu - \frac{1}{2}\,g^{\mu\nu}\,R + R^{\mu\nu} \bigg]\, \phi^2 \\
T^{f\,\mu\nu}
&=&
\frac{1}{4}\,
\bigg[ g^{\mu\rho}\,{V_\alpha}^\nu  + g^{\nu\rho}\,{V_\alpha}^\mu - 2\,g^{\mu\nu}\,{V_\alpha}^\rho \bigg]
\bigg[\bar{\psi} \, \gamma^{\alpha} \, \left(\mathcal{D}_\rho \,\psi\right) -
\left(\mathcal{D}_\rho \, \bar{\psi}\right) \, \gamma^{\alpha} \, \psi \bigg],
\eea
while the EMT for the photon field is given by the sum of three terms
\beq
T_V^{\mu\nu} = T^{\mu\nu}_M + T^{\mu\nu}_{gf} + T^{\mu\nu}_{gh}\, ,
\eeq
with
\bea
T^{\mu\nu}_M
&=&
F^{\mu\a}{F^\nu}_{\a}  - \frac{1}{4}g^{\mu\nu}F^{\a\b}F_{\a\b} \, ,
\label{Ch1TEI M}
\\
T^{\mu\nu}_{gf}
&=&
\frac{1}{\xi}\left\{ A^\mu\nabla^\nu(\nabla_\rho A^\rho) + A^\nu\nabla^\mu(\nabla_\rho A^\rho ) -g^{\mu\nu}\, 
\left[ A^\rho \nabla_\rho(\nabla_\sigma A^\sigma) + \frac{1}{2}(\nabla_\rho A^\rho)^2 \right]\right\}\, ,
\label{Ch1TEI g.f.}
\\
T^{\mu\nu}_{gh}
&=&
- \left(\pd^\mu\bar{c}\, \pd^\nu c + \pd^\nu\bar{c}\, \pd^\mu c - g^{\mu\nu}\pd^{\r}\bar{c}\,\pd_{\r}c \right)
\label{Ch1TEI gh}\, .
\eea
Now, to perform the explicit one loop computation, one must write down the integrals corresponding to the Feynman diagrams drawn in
fig. \ref{Ch1Fig.diagramsTTT}. Since the vev's of the third order derivatives correspond to massless tadpoles, 
which can be consistently set to zero in dimensional regularization,
in order to write down the perturbative expansion vertices with no more than to gravitons are needed; 
these require in turn to perform two variations of the action with respect to the background metric.
These results, together with the euclidean propagators, are listed in appendix 	\ref{Ch1Vertices}.
Once one has found them, tensor integrals must be evaluated. As mentioned in the introduction, they have been computed
by implementing in a symbolic manipulation program the Passarino-Veltman reduction technique.
Given the complexity of the result and to avoid any error, we have checked that all the expressions obtained 
satisfy the corresponding Ward identities derived above.  \\
Finally, we must remark that, as the $3$-graviton correlator does not have any gauge field on the external lines, one would expect it
not to depend on the gauge-fixing procedure, which enters only in the virtualities running in the loop.
This is known to happen in the case of the gauge-field $2$-point function and we explicitly checked that it is the case also for our
Green function, as expected. In other words, the computation of the correlator with the vertices listed in appendix \ref{Ch1Vertices}
is completely equivalent to the same computation performed by sending $1/\xi \rightarrow 0$ and omitting the diagrams with ghost loops.
This is another non trivial check of our computation in the gauge boson sector. \\

We end this section by providing the expressions of the $2$-point functions, which are necessary to test our computations.
Their general structure before renormalization is
\bea
\llangle T^{\mu\nu} \, T^{\alpha\beta} \rrangle(p)
&=&
C_1(p)\, \bigg[ \frac{1}{2} \, \bigg( \Theta^{\mu\alpha}(p) \, \Theta^{\nu\beta}(p) + 
\Theta^{\mu\beta}(p) \, \Theta^{\nu\alpha}(p)\bigg)
- \frac{1}{3} \, \Theta^{\mu\nu}(p) \, \Theta^{\alpha\beta}(p)\bigg]
+ \frac{C_2}{3}\,\Theta^{\mu\nu}(p) \, \Theta^{\alpha\beta}(p) \,  , \nn \\
\eea
where the transverse tensor $\Theta$ is 
\beq \label{DefineTheta}
\Theta^{\alpha\beta}(p) = \delta^{\alpha\beta}(p)\, p^2 - p^\alpha p^\beta \, .
\eeq
The values of the form factors for the three free field theories at hand are respectively given by
\bea
C^s_1(p)
&=&
\frac{16 + 15 \, \mathcal{B}_0(p^2)}{14400\,\pi^2}\, ,
\qquad \hspace{4mm}
C^s_2 = - \frac{1}{1440\,\pi^2} \, , \\
C^f_1(p)
&=&
\frac{2 + 5 \,\mathcal{B}_0(p^2)}{800\,\pi^2} \, ,
\qquad \hspace{6mm}
C^f_2 = - \frac{1}{240\,\pi^2}\, , \\
C^V_1(p)
&=&
\frac{-11 + 10\, \mathcal{B}_0(p^2)}{800\,\pi^2} \, ,
\qquad 
C^V_2 = - \frac{1}{120\,\pi^2}\, .
\eea
Here, $\mathcal{B}_0(p^2)$ is the $2$-point scalar integral, the only one that can appear in a massless $2$-point correlator
in the flat limit, as no scale is present and tadpoles can be consistently set to zero.
Its expression is
\beq
\mathcal{B}_0(p^2) =
\frac {1}{\pi^2} \int d^dl\, \frac{1}{l^2\,(l + p)^2} = 
 \frac{2}{\epsilon} - \gamma + \log \pi + 2 + \ln\left(\frac{\mu^2}{p^2}\right) \, ,
\label{BareB0}
\eeq
with $\gamma$ being the Euler constant. \\
The Ward identities discussed so far were tested before performing renormalization, but of course they hold even after the subtraction 
of the ultraviolet singularities is performed. 
For diffeomorphism invariance Ward identities this is easily understood, as the master equation
from which these constraints descends, (\ref{Ch1masterWI0}), can be very easily derived requiring the bare 
generating functional to be invariant under general coordinate transformations.
Every correlator in such a Ward identity, for instance (\ref{Ch1WI3PFmomenta2a}), can be split into a finite part and an infinite one,
with the latter featuring a $1/\epsilon$ pole, so that the coefficients of $1/\epsilon$ and the finite (i.e. renormalized) 
contributions on both sides can be separately equated.

The situation is rather different for trace Ward identities, because the master equation from which they descend, 
(\ref{Ch1TraceAnomaly}), does not hold for the bare generating functional, but only for the renormalized one.
A thorough, detailed discussion of the last point, using the method of $\zeta$-function regularization and holding also for the
more general case of non conformal field theories, can be found in \cite{Birrell:1982ix}.
A more direct approach for CFT's, due to Duff \cite{Duff:1977ay} and enlightening 
the relation between anomalies and counterterms, will be reviewed in chapter \ref{Mapping}.

\section{Renormalization of the $TTT$}
\label{Ch1Renormalization}

In this section we address the problem of the renormalization of the $3$-graviton vertex.
In \cite{Freedman:1974gs,Freedman:1974ze} it was shown that for scalar and gauge field theories,
with and without spontaneous symmetry breaking, the counterterms for the theory in flat space 
are sufficient to renormalize Green functions with one insertion of the EMT and an arbitrary number of matter fields.

But for correlators including more than one insertion of the EMT, this is known not to be true.
In fact, the theories we are dealing with are not renormalizable, i.e. there are not enough constants in the Lagrangians
which could be split into a finite and an infinite part so as to produce the contributions needed to subtract infinities from
correlators involving multiple insertions of the EMT.
Actually, power-counting arguments can immediately show that the number of divergent correlators is infinite 
when gravitational interactions are present, as remarked in the Introduction.

It is then necessary to introduce counterterms by hand. 
For $d=4$  with dimensional regularization and in the $\overline{MS}$ renormalization scheme, it is known that the contribution 
that must be added to the generating functional in order to remove $1$-loop divergences is \cite{Duff:1977ay,Duff:1993wm}
\beq\label{Ch1CounterAction}
S_{counter} = 
- \frac{\mu^{-\epsilon}}{\bar\epsilon}\sum_{I=f,s,V}n_I \int d^d x \sqrt{g} \bigg( \beta_a(I) F + \beta_b(I) G\bigg) \, ,
\qquad \frac{2}{\bar\epsilon} = \frac{2}{\epsilon} - \gamma - \log \pi \, .
\eeq
Notice that one of the two objects that appear in the counterterm integral, $G$, is a total divergence 
in 4 but not in $d$ dimensions. In particular, $G$ generates a counterterm which is effectively a projector on the extra $(4-d)$-dimensional 
space and, as such, gives a contribution which needs to be included in order to perform a correct renormalization of the vertex, 
in contrast with assertions that can be found in much of the literature on the subject. 
For instance, in \cite{Deser:1993yx} the famous distinction between $A$ and $B$-type anomalies 
for general (even) dimension was established, where the $B$-type are defined as the anomalies to which conformal invariants built out of 
the Weyl tensor contribute, whereas $A$-type ones are those proportional to the Euler density contribution.
In \cite{Deser:1993yx} it was also claimed that the $A$-type anomalies are not associated with poles in $1/\epsilon$.
Such a result cannot be tested at the level of the $2$-point function, as the Euler anomaly does not contribute to it.
To the extent of our knowledge, the fact tha this point is not true was first pointed out in \cite{Osborn:1993cr} and has been confirmed
by our explicit computation in dimensional regularization in our approach.
Another way to show the presence of $1/\epsilon$ poles associated to type-$A$ anomalies relies on cohomological arguments,
which are used to build the Wess-Zumino effective action for conformal anomalies in dimensional regularization \cite{Mazur:2001aa}.
This approach is a starting point for the results presented in chapter \ref{Recursive}, where it is briefly reviewed.
The intimate relation between anomalies and counterterms will be further explored in chapter \ref{Mapping}.

\begin{figure}[t]
\centering
\includegraphics[scale=0.7]{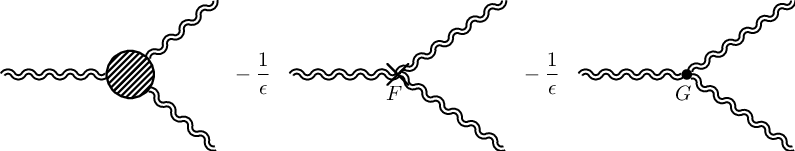}
\hspace{5mm}
\caption{$TTT$ and its counterterms generated with the choice of the square of the Weyl ($F$) 
tensor in 4 dimensions and the Euler density ($G$).}
\label{Ch1Fig.diagrams3grav}
\end{figure}

We have used the 4-dimensional realization of $F$
\beq\label{Ch1RecallF}
F =
R^{\alpha\beta\gamma\delta}R_{\alpha\beta\gamma\delta} - 2\,R^{\alpha\beta}R_{\alpha\beta} + \frac{1}{3} \, R^2 \, .
\eeq
As remarked, $G$ does not contribute to every correlator. For instance, in the case of the $TT$, the counterterm is obtained by
functional differentiation twice of $S_{counter}$, but one can easily check (see eq.  (\ref{Ch1Magic})) that the second variation of
$G$ vanishes in the flat limit. Hence, the only counterterm is given by
\beq
\label{Ch12PFCounterterm}
D_F^{\alpha\beta\rho\sigma}(x_1,x_2) =
4 \, \frac{\delta^2}{\delta g_{\alpha\beta}(x_1)\delta g_{\rho\sigma}(x_2)} \, \int\,d^d w\,\sqrt{g} \, F
\bigg|_{g_{\mu\nu}= \delta_{\mu\nu}}   \, ,
\eeq
whose form in momentum space is
\bea
D_F^{\alpha\beta\rho\sigma}(p) 
&=& 
4 \, \bigg[ \frac{1}{2}\,
\left( \Theta^{\mu\alpha}(p)\, \Theta^{\nu\beta}(p) + \Theta^{\mu\beta}(p)\, \Theta^{\nu\alpha}(p) \right)
- \frac{1}{3}\, \Theta^{\mu\nu}(p)\, \Theta^{\alpha\beta}(p)\bigg] \, ,
\label{Ch12PFCountertermMom}
\eea
where $\Theta$ was defined in (\ref{DefineTheta}).
We obtain the renormalized $2$-point function  by adding it according to (\ref{Ch1CounterAction}), i.e.
\beq\label{Ch1Ren2PF2}
\llangle T^{\alpha\beta}\,T^{\rho\sigma} \rrangle_{ren}(p) = 
\llangle T^{\alpha\beta}\,T^{\rho\sigma} \rrangle(p) - \frac{\beta_a}{\,{\bar\epsilon}} \, D_F^{\alpha\beta\rho\sigma}(p)\, .
\eeq
In the case of the $3$-graviton vertex the counterterm action (\ref{Ch1CounterAction}) generates the vertices 
\beq
-\frac{\mu^{-\epsilon}}{\bar\epsilon}\, \bigg(\beta_a D_F^{\mu\nu\rho\sigma\alpha\beta}(x_1,x_2,x_3)
+ \beta_b \, D_G^{\mu\nu\rho\sigma\alpha\beta}(x_1,x_2,x_3)\bigg)\, ,
\eeq
where
\bea
D_F^{\mu\nu\rho\sigma\alpha\beta}(x_1,x_2,x_3)
&=&
8 \, \frac{\delta^3}{\delta g_{\mu\nu}(x_1) \delta g_{\rho\sigma}(x_2)\delta g_{\alpha\beta}(x_3)}
\int\,d^d w\,\sqrt{g}\, F\bigg|_{g_{\mu\nu}=\delta_{\mu\nu}} \, , 
\label{Ch1DF}\\
D_G^{\mu\nu\rho\sigma\alpha\beta}(z,x,y)
&=&
8 \, \frac{\delta^3}{\delta g_{\mu\nu}(x_1) \delta g_{\rho\sigma}(x_2)\delta g_{\alpha\beta}(x_3)}
\int\,d^d w\,\sqrt{g}\, G\bigg|_{g_{\mu\nu}=\delta_{\mu\nu}}  \, .
\label{Ch1DG}
\eea
The explicit form of (\ref{Ch1DF}) and (\ref{Ch1DG}) is derived by functionally deriving three times the general functional
\beq
\mathcal{I}(a,b,c) \equiv \int\,d^4 x\,\sqrt{g}\,
\big(a\,R^{abcd}R_{abcd} + b\,R^{ab}R_{ab} + c\, R^2 \big)\, ,
\eeq
with respect to the metric for appropriate $a, b$ and $c$, i.e.
\bea
&&
a = 1 \, ,\quad b = -2 \, ,\quad c = \frac{1}{3} \, , \nn \\
&&
a = 1 \, ,\quad b = -4 \, ,\quad c = 1 \, . \nonumber
\eea
For convenience, the computations leading to the general result are reproduced in appendix \ref{Ch1FunctionalIntegral}.
The renormalized correlator is represented in fig. \ref{Ch1Fig.diagrams3grav}. 
It goes without saying that the counterterms for the $3$-point function 
in momentum space are given by the transform (\ref{Ch13PFMom}). \\

It is known that $D_G^{\mu\nu\rho\sigma\alpha\beta}(q,p)$ is found to vanish identically in four dimensions.
In fact, its explicit form is
\beq\label{Ch1ExplicitDG}
D_G^{\mu\nu\rho\sigma\alpha\beta}(q,p) =
-240 \big(E^{\mu\sigma\alpha\gamma\kappa,\nu\rho\beta\delta\lambda} +
E^{\mu\rho\alpha\gamma\kappa,\nu\sigma\beta\delta\lambda} + \alpha\leftrightarrow \beta\big)
\,q_\gamma\, q_\delta\, p_\kappa\, p_\lambda\, ,
\eeq
where $E^{\mu\sigma\alpha\gamma\kappa,\nu\rho\beta\delta\lambda}$ is a projector onto
completely antisymmetric tensors with five indices, so that it would yield zero in four dimensions,
reflecting the fact that the integral of the Euler density is a topological invariant in integer dimensions.
We have explicitly checked that, given the structure of the counterterm Lagrangian in (\ref{Ch1CounterAction}), one necessarily needs
to include the contribution from the $G$ part of the functional, in the form given by $D_G$, in order to remove all
the divergences. 


The fully renormalized 3-point correlator in momentum space can be written down as
\beq\label{Ch1Ren3PF}
\llangle T^{\mu\nu}T^{\rho\sigma}T^{\alpha\beta} \rrangle_{ren}(q,p) =
\llangle T^{\mu\nu}T^{\rho\sigma}T^{\alpha\beta} \rrangle_{bare}(q,p) -
\frac{\mu^{-\epsilon}}{{\bar\epsilon}}\,\bigg(\beta_a\,D_F^{\mu\nu\rho\sigma\alpha\beta}(q,p)
+ \beta_b\,D_G^{\mu\nu\rho\sigma\alpha\beta}(q,p)\bigg)\,
\eeq
and the goal is to proceed with an identification both of $D_F$ and $D_G$ from the
diagrammatic expansion in momentum space.
The cancellation of all of the ultraviolet poles, for suitable expressions of $D_F$ and $D_G$, has been thoroughly checked
from our explicit results. 

At this point, it is necessary to comment about the difference between our approach and that followed in \cite{Osborn:1993cr}, 
where the choice of $F$ is slightly different from ours, since the authors essentially define
a counterterm which is given by an integral of the form
\beq
\mathcal{\tilde{S}}_{counter} = 
- \frac{\mu^{-\epsilon}}{\bar\epsilon} \int d^4x \sqrt{g} \, \big( \beta_a \, F^{d} + \beta_b \, G \big) \, ,
\label{Ch1newren}
\eeq
based on the $d$-dimensional expression of the squared of the Weyl tensor ($F_d$).
Such a choice does not generate the anomaly contribution proportional to $\square R$ (sometimes refferred to as ``local anomaly'').
In fact, the authors choose to work with $\beta_c=0$ from the very beginning, since the inclusion of the local anomaly contribution amounts just 
to a finite renormalization with respect to (\ref{Ch1newren}). 
We briefly comment on the connection between the choice of the counterterm in (\ref{Ch1newren}) and the quoted finite
renormalization, discussed for the first time in \cite{Capper:1974ic}.
Notice that in $d$ dimensions, if we take the trace of the functional derivative in (\ref{Ch1Magic}) 
for $a=1$, $b = -{4}/({d-2})$, $c = {2}/((d-1)(d-2))$, which are the $d$-dimensional coefficients appearing in $F_d$, 
we can explicitly check that the contribution proportional to $\square R$ in the anomalous trace cancels.
For this purpose we can expand the integrand of (\ref{Ch1newren}) around $d=4$ (in $\epsilon = 4 - d$) up to $O(\epsilon)$,
obtaining that the counterterm action can be separated in a polar plus a finite part, i.e.
\beq
\mathcal{\tilde{S}}_{counter} =
\mathcal S_{counter} + \mathcal S_{fin.\,ren.} = \mathcal S_{counter} 
+ \beta_a \, \int d^4x\, \sqrt{g} \, \bigg(R^{\alpha\beta}\,R_{\alpha\beta} - \frac{5}{18} \, R^2\bigg)  + O(\epsilon)\, .
\eeq
Recalling the definition (\ref{Ch1VEVEMT}) and using (\ref{Ch1Magic}),
we see that the contribution of this finite part to the vev of the EMT is
\beq
g_{\mu\nu} \llangle \, T^{\mu\nu} \, \rrangle_{fin.ren.} = -\beta_c \, \square R \, .
\eeq

\begin{figure}
\begin{center}
\includegraphics[scale=0.7]{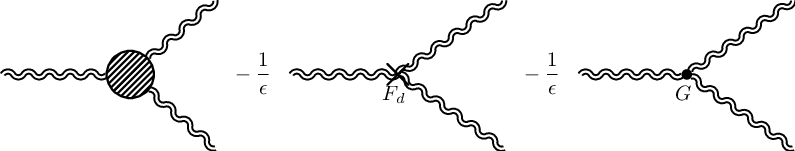}
\hspace{5mm}
\caption{The contributions to the renormalized $TTT$ vertex from the square of the Weyl tensor in $d$-dimensions ($F_dd$) 
and the Euler density ($G$).}
\label{Ch1Fig.diagrams3gravd}
\includegraphics[scale=0.7]{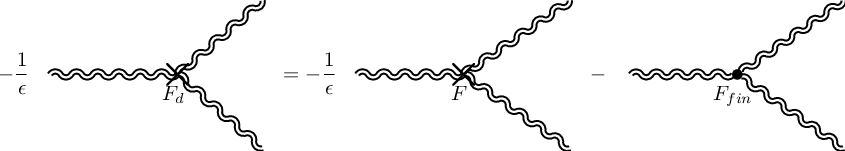}
\hspace{5mm}
\caption{The relation between the counterterm generated by $F_d$ and the same obtained from $F$. The difference is a finite
renormalization ($F_{fin}$) term in the counterterm Lagrangian, which generates the local contribution to the trace anomaly.}
\end{center}
\end{figure}

Comparing this with (\ref{Ch1TraceAnomaly}), we see that this extra contribution will cancel the local anomaly. \\
It is immediately realized that this approach is equivalent, for what concerns the anomaly, to supplying the action of the theory with
the finite renormalization usually met in the literature, i.e.
\beq
\mathcal S^{(2)}_{fin.\,ren.} \equiv - \frac{\beta_c}{12} \, \int d^4x\, \sqrt{g} \, R^2\, ,
\eeq
which is known to cancel the local anomaly, due to the similar relation
\beq
g_{\mu\nu} \, \frac{2}{\sqrt{g}} \, \frac{\delta \mathcal S^{(2)}_{fin.\,ren.}}{\delta g_{\mu\nu}} = 
- \beta_c \,\square R\, ,
\eeq
which can be checked using (\ref{Ch1Magic}) once again .

\section{The renormalized 3-graviton vertex with two lines on-shell}
\label{Ch1ExplicitTTT}

In general, a rank-$6$ tensor depending on $2$ momenta can be expanded on a basis made up of $499$ tensors
built out of the $2$ momenta and the metric tensor $\delta_{\mu\nu}$.
The problem is that the scalar coefficients multiplying such tensors are really complicated, in the most general kinematic
configuration and it is no point reporting them explicitly.

Nevertheless, we have found that, in all of the three cases explicitly examined, if we go on-shell on the two outgoing gravitons,
the $TTT$ vertex can be expanded on a basis made up of just thirteen tensors.
This amounts to contracting the amplitude with polarization tensors $e^s_{\lambda\kappa}(p)$  which are transverse  and traceless
\beq\label{Ch1helicity}
p^\lambda\,e^s_{\lambda\kappa}(p) = 0 \, , \quad  {(e^s)^\lambda}_\lambda(p) = 0 \, ,
\eeq
where the superscript denotes the helicity state of the graviton carrying the momentum $p$.

Given the assignment of the momenta established in (\ref{Ch13PFMom}), it is immediate
to see that the contraction of the amplitude with the polarization tensors with the properties (\ref{Ch1helicity})
for the two outgoing gravitons is equivalent to the replacements
\beq
p^2       \rightarrow 0  \, , \quad q^2      \rightarrow 0 \, , \quad
p^\alpha  \rightarrow 0  \, , \quad p^\beta  \rightarrow 0 \, , \quad
q^\rho    \rightarrow 0  \, , \quad q^\sigma \rightarrow 0 \, ,
\label{Ch1OnShell}
\eeq
so that these conditions can be used to select the tensors that are non-vanishing after the on-shell limit is taken.
We are going to express our result for the amplitude in terms of such tensors.

The expansion of our Green's function for a theory with $n_S$ scalars, $n_F$ fermions and $n_V$ vector bosons can be generally written as
\beq
\llangle T^{\mu\nu}T^{\rho\sigma}T^{\alpha\beta} \rrangle(q,p)\bigg|_{On-Shell} \equiv 
\sum_{n_I=n_S,n_F,n_V} n_I \, \sum_{i=1}^{13} \Omega^I_i(s) \, t_i^{\mu\nu\rho\sigma\alpha\beta}(q,p)\, ,
\quad s \equiv k^2 = 2 \, p \cdot q \, .
\eeq
The form factors  for the three theories that we consider in this chapter are listed in table \ref{Ch1FormFactors}.
\begin{table}
$$
\begin{array}{|c|c|c|c|}
\hline
i
&   \Omega^S_{i}(s) 
&   \Omega^F_{i}(s) 
&   \Omega^V_{i}(s)
\\
\hline 
\hline 
1
&  - \frac{1}{720\,\pi^2}\times \frac{1}{2 \, s}
&  - \frac{1}{240\,\pi^2}\times \frac{1}{s}
&    \frac{1}{1152\,\pi^2}\times \frac{72}{5 \, s}  \\
\hline
2
&  - \frac{1}{720\,\pi^2} \times \frac{1}{s}
&  - \frac{1}{240\,\pi^2} \times \frac{1}{3 \, s}
&    \frac{1}{1152\,\pi^2} \times \frac{64}{5 \, s}
\\
\hline
3
&  - \frac{1}{720\,\pi^2}\times \frac{7 + 30 \, \mathcal{B}_0(s)}{120}
&    \frac{1}{240\,\pi^2}\times \frac{13 - 30 \, \mathcal{B}_0(s)}{60}
&    \frac{1}{1152\,\pi^2}\times \frac{82 - 120 \, \mathcal{B}_0(s)}{25}
\\
\hline
4
&  -  \frac{1}{720\,\pi^2}\times \frac{2 + 5 \, \mathcal{B}_0(s)}{10}
&     \frac{1}{240\,\pi^2}\times \frac{7 - 70 \, \mathcal{B}_0(s)}{120}
&     \frac{1}{1152\,\pi^2}\times \frac{2 \, (482 + 130 \, \mathcal{B}_0(s))}{25}
\\
\hline
5
&     \frac{1}{720\,\pi^2}\times \frac{1}{6}
&   - \frac{1}{240\,\pi^2}\times \frac{-1 + 10 \, \mathcal{B}_0(s)}{48}
&   - \frac{1}{1152\,\pi^2}\times \frac{79 + 50 \, \mathcal{B}_0(s)}{5}
\\
\hline
6
&    \frac{1}{720\,\pi^2}\times \frac{23 + 20 \, \mathcal{B}_0(s)}{20}
&    \frac{1}{240\,\pi^2}\times \frac{33 + 70 \, \mathcal{B}_0(s)}{60}
&  - \frac{1}{1152\,\pi^2}\times \frac{104 \, (22 + 5 \, \mathcal{B}_0(s))}{25}
\\
\hline
7
&  - \frac{1}{720\,\pi^2}\times \frac{s\,(16 + 15 \, \mathcal{B}_0(s))}{20}
&  - \frac{1}{240\,\pi^2}\times \frac{3\,s\,(2 + 5 \, \mathcal{B}_0(s))}{10}
&  - \frac{1}{1152\,\pi^2}\times \frac{s\,(-11 + 10 \, \mathcal{B}_0(s))}{80}
\\
\hline
8
&  - \frac{1}{720\,\pi^2}\times \frac{s\,(47 + 30 \, \mathcal{B}_0(s))}{80}
&  - \frac{1}{240\,\pi^2}\times \frac{3\, s\,(9 + 10 \, \mathcal{B}_0(s))}{40}
&    \frac{1}{1152\,\pi^2}\times \frac{s\,(2 + 5 \, \mathcal{B}_0(s))}{40}
\\
\hline
9
&    \frac{1}{720\,\pi^2}\times \frac{s\,(2 + 5 \, \mathcal{B}_0(s))}{40}
&  - \frac{1}{240\,\pi^2}\times \frac{7 s\,(1 - 10 \, \mathcal{B}_0(s))}{480}
&  - \frac{1}{1152\,\pi^2}\times \frac{s\,(487 + 130 \, \mathcal{B}_0(s))}{50}
\\
\hline
10
&    \frac{1}{720\,\pi^2}\times \frac{s \, (9 + 10 \, \mathcal{B}_0(s))}{20}
&    \frac{1}{240\,\pi^2}\times \frac{s \, (137 + 430\,  \mathcal{B}_0(s))}{480}
&  - \frac{1}{1152\,\pi^2}\times \frac{s \, (883 - 230 \, \mathcal{B}_0(s))}{50}
\\
\hline
11
&  - \frac{1}{720\,\pi^2}\times \frac{s \, (7 + 5 \, \mathcal{B}_0(s))}{20}
&  - \frac{1}{240\,\pi^2}\times \frac{7 \, s \, (9 + 10 \, \mathcal{B}_0(s))}{240}
&    \frac{1}{1152\,\pi^2}\times \frac{s \, (467 + 130 \, \mathcal{B}_0(s))}{25}
\\
\hline
12
&  - \frac{1}{720\,\pi^2}\times \frac{s \, (121 + 90 \, \mathcal{B}_0(s))}{240}
&  - \frac{1}{240\,\pi^2}\times \frac{s \, (97 + 130 \, \mathcal{B}_0(s))}{240}
&    \frac{1}{1152\,\pi^2}\times \frac{2 \, s \, (299 + 35 \, \mathcal{B}_0(s))}{25}
\\
\hline
13
&   \frac{1}{720\,\pi^2}\times \frac{5 \, s^2 \, (3 + 2  \, \mathcal{B}_0(s))}{32}
&   \frac{1}{240\,\pi^2}\times \frac{5 \, s^2 \, (9 + 10 \, \mathcal{B}_0(s))}{96}
&  -\frac{s^2 \, (13 - \mathcal{B}_0(s))}{1152\,\pi^2}
\\
\hline
\end{array}
$$
\caption{Form factors for the vertex $\llangle T^{\mu\nu}T^{\rho\sigma}T^{\alpha\beta} \rrangle (q,p)$ with the
graviton lines $(\alpha,\beta,p)$ and $(\rho,\sigma,q)$ on the mass shell.}
\label{Ch1FormFactors}
\end{table}

Below we give the explicit form of the 13 tensors $t_i^{\mu\nu\alpha\beta\rho\sigma}(p,q)$,
\bea
t_1^{\mu \nu \rho \sigma\alpha \beta}(q,p)
&=&
\big( p^{\mu} p^{\nu} + q^{\mu} q^{\nu}\big) \, p^{\rho} p^{\sigma} q^{\alpha} q^{\beta} \nn\\
t_2^{\mu \nu \rho \sigma\alpha \beta}(q,p)
&=&
\big( p^{\mu} q^{\nu} + p^{\nu} q^{\mu} \big) \, p^{\rho} p^{\sigma} q^{\alpha} q^{\beta}
\nn \\
t_3^{\mu\nu\rho\sigma\alpha\beta}(q,p)
&=&
\big(p^{\mu} p^{\nu}  + q^{\mu} q^{\nu} \big) \,
\big( p^{\sigma} q^{\beta} \delta^{\alpha \rho} + p^{\sigma} q^{\alpha} \delta^{\beta \rho}
+ p^{\rho} q^{\beta} \delta^{\alpha \sigma} + p^{\rho} q^{\alpha}\delta^{\beta \sigma} \big) \nn \\
t_4^{\mu\nu\rho\sigma\alpha\beta}(q,p)
&=&
p^{\rho} p^{\sigma} \, \big(
q^{\beta} q^{\nu} \delta^{\alpha \mu} + q^{\beta} q^{\mu} \delta^{\alpha \nu}
+ q^{\alpha} q^{\nu} \delta^{\beta \mu} + q^{\alpha} q^{\mu} \delta^{\beta \nu}
\big) \nn \\
&+&
q^{\alpha} q^{\beta} \, \big(
p^{\nu} p^{\sigma} \delta^{\mu \rho} + p^{\nu} p^{\rho} \delta^{\mu \sigma}
+ p^{\mu} p^{\sigma} \delta^{\nu \rho} + p^{\mu} p^{\rho} \delta^{\nu \sigma}
\big)
\nn \\
t_5^{\mu\nu\rho\sigma\alpha\beta}(q,p)
&=&
\big(p^{\mu} q^{\nu} + q^{\mu} p^{\nu}\big) \, \bigg(
  p^{\rho}   \big(q^{\alpha} \delta^{\beta\sigma} + q^{\beta} \delta^{\alpha\sigma} \big)
+ p^{\sigma} \big(q^{\alpha}\delta^{\beta \rho}   + q^{\beta} \delta^{\alpha\rho} \big) \bigg) \nn \\
t_6^{\mu\nu\rho\sigma\alpha\beta}(q,p)
&=&
\delta^{\mu \nu} p^{\rho} p^{\sigma} q^{\alpha} q^{\beta}
\nn \\
t_7^{\mu\nu\rho\sigma\alpha\beta}(q,p)
&=&
  p^{\rho}p^{\sigma}  \, \big(\delta^{\mu\alpha}\delta^{\nu\beta} + \delta^{\mu\beta}\delta^{\nu\alpha}\big)
+ q^{\alpha}q^{\beta} \, \big(\delta^{\mu\rho}\delta^{\nu\sigma} + \delta^{\mu\sigma}\delta^{\nu\rho}\big) \nn\\
&-&
\frac{1}{2} \, \bigg(
  p^{\mu}p^{\rho}   \big( \delta^{\alpha\sigma}\delta^{\nu\beta} + \delta^{\beta\sigma}\delta^{\nu\alpha} \big)
+ p^{\nu}p^{\rho}   \big( \delta^{\alpha\sigma}\delta^{\mu\beta} + \delta^{\beta\sigma}\delta^{\mu\alpha} \big) \nn \\
&+&
  p^{\mu}p^{\sigma} \big( \delta^{\alpha\rho}\delta^{\nu\beta}   + \delta^{\beta\rho}\delta^{\nu\alpha} \big)
+ p^{\nu}p^{\sigma} \big( \delta^{\alpha\rho}\delta^{\mu\beta}   + \delta^{\beta\rho}\delta^{\mu\alpha}  \big) \nn \\
&+&
  q^{\mu}q^{\alpha} \big( \delta^{\beta\sigma}\delta^{\nu\rho}   + \delta^{\beta\rho}\delta^{\nu\sigma}  \big)
+ q^{\nu}q^{\alpha} \big( \delta^{\beta\sigma}\delta^{\mu\rho}   + \delta^{\beta\rho}\delta^{\mu\sigma}  \big)\nn \\
&+&
  q^{\mu}q^{\beta}  \big( \delta^{\alpha\sigma}\delta^{\nu\rho}  + \delta^{\alpha\rho}\delta^{\nu\sigma}  \big)
+ q^{\nu}q^{\beta}  \big( \delta^{\alpha\sigma}\delta^{\mu\rho}  + \delta^{\alpha\rho}\delta^{\mu\sigma} \big)
\bigg) \nn \\
t_8^{\mu\nu\rho\sigma\alpha\beta}(q,p)
&=&
\big(p^{\mu} p^{\nu}+q^{\mu} q^{\nu}\big) \,
\big(\delta^{\alpha \sigma} \delta^{\beta \rho} + \delta^{\alpha \rho} \delta^{\beta \sigma})
\nn \\
t_9^{\mu\nu\rho\sigma\alpha\beta}(q,p)
&=&
  p^{\rho} \, \bigg(
  q^{\mu} (\delta^{\alpha \sigma} \delta^{\beta \nu}+\delta^{\alpha \nu} \delta^{\beta \sigma})
+ q^{\nu} (\delta^{\alpha \sigma} \delta^{\beta \mu}+\delta^{\alpha \mu} \delta^{\beta \sigma})
\bigg) \nn \\
&+&
p^{\sigma} \, \bigg(
   q^{\mu} (\delta^{\alpha \rho} \delta^{\beta \nu}+\delta^{\alpha \nu} \delta^{\beta \rho})
+  q^{\nu} (\delta^{\alpha \rho} \delta^{\beta \mu} + \delta^{\alpha \mu} \delta^{\beta \rho})
\bigg) \nn \\
&+&
q^{\alpha} \, \bigg(
  p^{\mu}  (\delta^{\beta \sigma} \delta^{\nu \rho} +  \delta^{\beta \rho} \delta^{\nu \sigma})
+ p^{\nu}  (\delta^{\beta \sigma} \delta^{\mu\rho}  +  \delta^{\beta \rho} \delta^{\mu \sigma})
\bigg) \nn \\
&+&
q^{\beta} \, \bigg(
  p^{\mu}  ( \delta^{\alpha\sigma}\delta^{\nu\rho} + \delta^{\alpha\rho} \delta^{\nu\sigma})
+ p^{\nu}  ( \delta^{\alpha\sigma}\delta^{\mu\rho} + \delta^{\alpha\rho}\delta^{\mu\sigma})
\bigg)
\nn
\eea
\bea
t_{10}^{\mu\nu\rho\sigma\alpha\beta}(q,p)
&=&
p^{\rho} \, \bigg(
  q^{\alpha} (\delta^{\beta \nu} \delta^{\mu \sigma} + \delta^{\beta \mu} \delta^{\nu \sigma})
+ q^{\beta}  (\delta^{\alpha \nu} \delta^{\mu \sigma} + \delta^{\alpha \mu} \delta^{\nu\sigma})
\big) \nn \\
&+&
p^{\sigma}  \, \bigg(
   q^{\alpha} (\delta^{\beta \nu} \delta^{\mu\rho} + \delta^{\beta \mu} \delta^{\nu \rho})
+  q^{\beta}  (\delta^{\alpha \nu} \delta^{\mu \rho} + \delta^{\alpha \mu} \delta^{\nu \rho}) \bigg) \nn \\
&-& p.q \, \bigg(
  \delta^{\alpha \rho}    (\delta^{\beta \nu} \delta^{\mu \sigma} + \delta^{\beta \mu} \delta^{\nu \sigma})
+ \delta^{\alpha \nu}     (\delta^{\beta \sigma} \delta^{\mu \rho} + \delta^{\beta \rho} \delta^{\mu \sigma}) \nn \\
&+&
  \delta^{\alpha \mu}     (\delta^{\beta \sigma} \delta^{\nu \rho} + \delta^{\beta \rho} \delta^{\nu \sigma} )
+ \delta^{\alpha \sigma}  (\delta^{\beta \nu} \delta^{\mu \rho}+\delta^{\beta \mu} \delta^{\nu\rho})
\bigg) \nn \\
t_{11}^{\mu\nu\rho\sigma\alpha\beta}(q,p)
&=&
\big(p^{\nu} q^{\mu} + p^{\mu} q^{\nu}\big) \,
\big(\delta^{\alpha \sigma} \delta^{\beta \rho} + \delta^{\alpha \rho} \delta^{\beta \sigma}\big)
\nn \\
t_{12}^{\mu\nu\rho\sigma\alpha\beta}(q,p)
&=&
 \delta^{\mu \nu} \,
\bigg( p^{\rho} \big( q^{\beta} \delta^{\alpha \sigma} + q^{\alpha} \delta^{\beta\sigma} \big)
+ p^{\sigma}  \big( q^{\beta} \delta^{\alpha \rho} + q^{\alpha} \delta^{\beta \rho} \big)
 \bigg)
\nn\\
t_{13}^{\mu\nu\rho\sigma\alpha\beta}(q,p)
&=&
\delta^{\mu \nu} \, \big(\delta^{\alpha \sigma} \delta^{\beta \rho}+\delta^{\alpha \rho} \delta^{\beta \sigma}\big).
\eea
In this limit, the correlator is affected by ultraviolet divergences coming from the $2$-point integrals $\mathcal{B}_0(s)$ .
This is true in the off-shell case too, as all the other contributions to the scalar coefficients
of its tensor expansion are made up of the three invariants $p^2$, $q^2$ and $p \cdot q$ plus the scalar 3-point integral
\beq
\mathcal {C}_0(s,s_1,s_2) =
\frac {1}{\pi^2} \int d^dl\, \frac{1}{l^2\,(l + p_1)^2\, (l + p_2)^2}\, , \quad s = (p_1+p_2)^2 \, , s_i= p_i^2\, , \quad i=1,2 \, ,
\eeq
which is finite for $d=4$. 

The bare $2$-point integral in $4$ dimensions is defined in eq. (\ref{BareB0}).
After using the renormalization procedure discussed in the previous section in the $\overline{MS}$ scheme, it becomes 
\beq
B_0^{\overline{MS}}(p^2) = 2 + \log\left(\frac{\mu^2}{p^2}\right)\, ,
\label{RenB0}
\eeq

For the sake of completeness, we report that we have checked that by taking the trace of these 13 tensors one reproduces the Weyl, 
Euler and local contributions to the trace anomaly satisfied by the vertex, which in this on-shell case are given by
\bea
\delta_{\mu\nu}\llangle T^{\mu\nu}T^{\rho\sigma}T^{\alpha\beta}\rrangle(q,p)\bigg|_{On-Shell}
&=&
4 \, \bigg\{\beta_a\,\bigg( \big[F\big]^{\rho\sigma\alpha\beta}(q,p)
- \frac{2}{3} \big[ \sqrt{g}\Box\,R\big]^{\rho\sigma\alpha\beta}(q,p) \bigg) 
\nonumber\\
&+& \beta_b\, \big[G\big]^{\rho\sigma\alpha\beta}(q,p)\bigg\}\bigg|_{On-Shell} \, ,
\label{Ch1munu3PFanomaly1}
\eea
\bea
\delta_{\alpha\beta}\llangle T^{\mu\nu}T^{\rho\sigma}T^{\alpha\beta}\rrangle(q,p)\bigg|_{On-Shell}
&=&
4 \, \bigg\{\beta_a\,\bigg(\big[F\big]^{\mu\nu\rho\sigma}(-k,q)
- \frac{2}{3} \big[\sqrt{g}\Box\,R\big]^{\mu\nu\rho\sigma}(-k,q)\bigg)
\nonumber\\
&+& 
\beta_b\, \big[G\big]^{\mu\nu\rho\sigma}(-k,q)
- \frac{1}{2}\,\llangle T^{\mu\nu}T^{\rho\sigma}\rrangle(k)\bigg\}\bigg|_{On-Shell} \, ,
\label{Ch1albe3PFanomaly1}
\eea
\bea
\delta_{\rho\sigma}\llangle T^{\mu\nu}T^{\rho\sigma}T^{\alpha\beta}\rrangle(q,p)\bigg|_{On-Shell}
&=&
4 \, \bigg\{\beta_a\,\bigg(\big[F\big]^{\mu\nu\alpha\beta}(-k,p)
- \frac{2}{3} \big[\sqrt{g}\Box\,R\big]^{\mu\nu\alpha\beta}(-k,p)\bigg)
\nonumber\\
&+& 
\beta_b\, \big[G\big]^{\mu\nu\alpha\beta}(-k,p) 
- \frac{1}{2} \,\llangle T^{\mu\nu}T^{\alpha\beta}\rrangle(k)\bigg\}\bigg|_{On-Shell}\, ,
\nonumber\\
\label{Ch1rosi3PFanomaly1}
\eea
with
\bea
\left[F\right]^{\rho\sigma\alpha\beta}(q,p)\bigg|_{On-Shell}
&=&
2\, p^{\rho} \, p^{\sigma} q^{\alpha} q^{\beta}
- p\cdot q\,\bigg( p^{\sigma}  q^{\beta} \delta^{\alpha\rho} - p^{\rho} q^{\beta} \delta^{\alpha\sigma}
- p^{\sigma} q^{\alpha} \delta^{\beta\rho} - p^{\rho} q^{\alpha} \delta^{\beta\sigma}\bigg)
\nonumber \\
&+&
(p\cdot q)^2\, \bigg(\delta^{\alpha\sigma} \delta^{\beta\rho} + \delta^{\alpha\rho} \delta^{\beta\sigma}\bigg) \, ,
\\
\left[G\right]^{\rho\sigma\alpha\beta}(q,p)\bigg|_{On-Shell}
&=&
2\, p^{\rho} p^{\sigma} q^{\alpha} q^{\beta}
- p\cdot q\,\bigg( p^{\sigma} q^{\beta} \delta^{\alpha\rho} -  p^{\rho} q^{\beta} \delta^{\alpha \sigma}
- p^{\sigma} q^{\alpha} \delta^{\beta\rho} - p^{\rho} q^{\alpha} \delta^{\beta \sigma}\bigg)
\nonumber\\
&+&
(p\cdot q)^2\, \bigg(\delta^{\alpha \sigma} \delta^{\beta \rho} + \delta^{\alpha \rho} \delta^{\beta \sigma}\bigg) \, ,
\\
\left[\sqrt{g}\,\square R\right]^{\rho\sigma\alpha\beta}(q,p)\bigg|_{On-Shell}
&=&
\frac{1}{2}\, p\cdot q\,\bigg(p^{\sigma} q^{\beta} \delta^{\alpha\rho}   + p^{\rho} q^{\beta} \delta^{\alpha \sigma}
+ p^{\sigma} q^{\alpha} \delta^{\beta\rho} + p^{\rho} q^{\alpha} \delta^{\beta \sigma}\bigg)
\nonumber\\
&-&
\frac{3}{2}\, (p\cdot q)^2 \bigg(g^{\alpha \sigma} \delta^{\beta \rho} - \delta^{\alpha \rho} \delta^{\beta \sigma}\bigg)\, .
\eea

\section{Conclusions and perspectives: the integrated anomaly and the nonlocal action}

Before coming to our conclusions, we offer here a brief discussion of the possible extensions of our analysis in the context 
of the emergence of massless degrees of freedom in the computation of correlators of the form $TVV$ and $TTT$, 
as predicted by Riegert's non local solution \cite{Riegert:1984kt} of the anomaly equation.

We recall that an action that formally solves the anomaly equation (\ref{Ch1TraceAnomaly}) takes the form 
\bea
S_{anom}[g,A] 
&=& 
\frac {1}{8}\int d^4x\sqrt{g}\int d^4x'\sqrt{-g'} \left(G + \frac{2}{3} \square R\right)_x\, 
G_4(x,x')\,\left[ 2\,\beta_a\, F + \beta_b\, \left(G + \frac{2}{3} \square R\right) 
- 2\, \frac{\kappa}{4}\, F_{\mu\nu}F^{\mu\nu}\right]_{x'}
\nn \\
&& 
-\, \frac{\beta_a+\beta_b}{18}\, \int d^4x\,\sqrt{g}\, R^2
\label{Ch1Riegertactions}
\eea
where we keep the sum on all the contributions coming from different particle species implicit. \\
The notation $G_4(x,x')$ denotes the Green's function of the differential operator defined by
\beq
\Delta_4 \equiv  \nabla_\mu\left(\nabla^\mu\nabla^\nu - 2 R^{\mu\nu} + \frac{2}{3}\, g^{\mu\nu}\, R \right)
\nabla_\nu = \square^2 - 2\, R^{\mu\nu}\nabla_\mu \nabla_\nu -\frac{1}{3}\, (\nabla^\mu R)\, \nabla_\mu 
+ \frac{2}{3}\,  \square R \, ,
\label{Ch1operator4}
\eeq
which is the only order-$4$ differential operator enjoying the invariance property under Weyl transformations
\beq
\sqrt{g}\, G_4(x,x')\rightarrow \sqrt{g}\, G_4(x,x')\, , \quad \text{for} \quad 
g_{\mu\nu} \rightarrow e^{2\,\sigma(x)}\, g_{\mu\nu}(x)\, ,
\eeq
whereas the combination $\left( G + 2/3 \square R\right)$ transforms as
\beq
\left( G + \frac{2}{3}\, \square R \right) \rightarrow  
\left( G + \frac{2}{3}\, \square R \right) + 4\, \Delta_4 \sigma\, , \quad \text{for} \quad 
g_{\mu\nu} \rightarrow e^{2\,\sigma(x)}\, g_{\mu\nu}(x)\, ,
\eeq
which immediately helps understanding how the trace of the EMT derived from (\ref{Ch1operator4}) 
reproduces the trace anomaly in (\ref{Ch1TraceAnomaly}). 

As shown in \cite{Giannotti:2008cv,Mottola:2006ew}, performing repeated variations of the "anomaly-induced" action 
(\ref{Ch1Riegertactions}) with respect to the background metric $g_{\mu\nu}$ and to the  $A_{\alpha}$ gauge field, 
one can reproduce the anomalous contribution of correlators with multiple insertions of the EMT or of gauge currents. 
Of course, this action does not reproduce the homogeneous contributions to the anomalous trace Ward identity (no variational solution
could). Because these require an independent computation in order to be identified, 
such as the perturbative $1$-loop analysis undertaken in this work.

The action can be reformulated in such a way that its interactions become local \cite{Mottola:2006ew}, 
by introducing two auxiliary scalar fields. 
After some manipulations, one can show that the apparently double pole in $G_4(x,x')$ reduces to a single pole
and the anomaly-induced action near a flat background takes the simpler form
\beq
S_{anom}[g,A]  \rightarrow  
- \frac{\kappa}{24}\int d^4x\sqrt{g}\int d^4x'\sqrt{-g'}\, R_x\, 
\square^{-1}_{x,x'}\, [F_{\alpha\beta}F^{\alpha\beta}]_{x'} \, .
\label{Ch1SSimple}
 \eeq
Notice that this action is valid to first order in metric variations around flat space. Its local expression is given by
\beq
S_{anom} [g,A;\varphi,\psi'] = \int\,d^4x\,\sqrt{g}\,
\left[ -\psi'\square\,\varphi - \frac{R}{3}\, \psi'  + \frac{c}{2} F_{\alpha\beta}F^{\alpha\beta} \varphi\right]  \, .
\label{Ch1effact}
\eeq
with $\psi'$ and $\varphi$ defined as in \cite{Giannotti:2008cv}.
$R_x$, in the equations above, is the linearised version of the Ricci scalar
\beq
 R_x \equiv \partial^x_\mu\, \partial^x_\nu \, h^{\mu\nu} - \square \,  h, \qquad h=\eta_{\mu\nu} \, h^{\mu\nu} \, .
 \eeq
Eq.  (\ref{Ch1effact}) shows the appearance of coupled massless degrees of freedom, whose interpretation was been offered in 
\cite{Giannotti:2008cv}, to which we refer for further details, using the approach of dispersion relations.

This analysis, so far, has been limited to the $TVV$ correlator and could be extended, with a lot of additional effort, 
to the case of the $TTT$ vertex whose explicit computation has been discussed in this work. 
In particular, this analysis could test directly if the pole structure present in the 
expression of the $TTT$ vertex will match the prediction of the same vertex once this is computed using (\ref{Ch1Riegertactions}) 
by functional differentiation with respect to the metric. 
This point is technically very involved, since it requires a comparison between the off-shell result of a direct computation of the $TTT$
in perturbative field theory, as done in this work, with the anomalous part of the same correlator computed from Riegert' s variational solution.
We hope to come back to discuss this point in a related work.

\clearpage{\pagestyle{empty}\cleardoublepage}

\chapter{Conformal correlators in position and momentum space}\label{Mapping}

 \section{Introduction}

The analysis of correlation functions in $d$-dimensional quantum field theory possessing conformal invariance has found widespread 
interest over the years (see \cite{Fradkin:1997df} for an overview). Given the infinite dimensional character of the conformal algebra in $2$ 
dimensions, conformal field theories (CFT's from now on) for $d=2$ have received much more attention than their 
$4$-dimensional counterparts.

In $d$ dimensional CFT's the structure of generic conformal correlators is not entirely fixed just by conformal symmetry, 
but for $2$- and $3$-point functions the situation is rather special and these can be significantly constrained, up to a small number of constants.
From the CFT side, some important information, mainly due to \cite{Osborn:1993cr,Erdmenger:1996yc}, is available. 
These results  concern the $TOO$ - with $O$ denoting a generic scalar operator - $TVV$ and $TTT$ vertices, which are determined by 
implementing the conformal constraints in position space. In the analysis of \cite{Osborn:1993cr}, in particular, it was shown for the first time 
that some of these vertices are expressible in terms of few linearly independent tensor structures: specifically, their numbers are
$1$ for the $TOO$ vertex, $2$ for the $TVV$ and $3$ for the $TTT$. \\
Imposing the conformal Ward identities and identifying these tensor structures directly in momentum space turns out the be technically quite
involved. The main goal of the present chapter is to present the result of a systematic study, initiated in \cite{Coriano:2012wp}, 
enabling comparison of general results of d-dimensional CFT's based on position space analyses, such as those in
\cite{Osborn:1993cr, Erdmenger:1996yc}, with explicit realizations of anomalous $3$-point vertices in free CFT's, 
most commonly expressed in momentum space. 
Recent results of studies of $3$- and $4$-point functions in  $d=3$ in the context of the $ADS/CFT$ correspondence
are contained in  \cite{Bzowski:2011ab,Maldacena:2011nz,Raju:2012zs}.


Another significant difference between the position-space approach and the computations in momentum space is that 
conformal anomalies necessarily arise quite differently in the two contexts.
In the former case, in fact, anomalous terms show up as ultra-local divergences proportional to delta functions or derivatives thereof
at coincident spatial points. Thus a very careful regularization procedure is required to
determine these anomalous ultra-local contributions which are absent for any finite point separation.
The special strategy followed in determining these anomalous ultra-local contributions in position space, developed in \cite{Osborn:1993cr}, 
deserves some comments. In this approach, the diffeomorphism and trace Ward identities are solved 
in each case by combining a completely homogeneous solution, which is built on the ground of the requirements of Lorentz and 
conformal invariance and is non local - i.e. obtained  keeping the three points separate - with inhomogeneous terms.
The inhomogeneous terms are of two kinds: terms of the first kind are semi-local (two coincident points out of three), 
are identified via the Operator Product Expansion of correlators and account for ordinary contact terms in the Ward identities; 
the terms of the second kind are ultra-local (all three points coincident). In particular, the latter contributions 
arise from the need to subtract ultraviolet singularities appearing in the ultra-local limit and are responsible for the corresponding trace anomalies.

It is clear that such a separation, based on the distinction of terms according to the separation/coincidence of points
at which operators are evaluated, does not make sense in momentum space. Here anomalies are thought either 
as a remnant of the renormalization of ultraviolet divergences, that breaks conformal invariance via the introduction of a mass scale $\mu$
(see for example \cite{Duff:1977ay,Duff:1993wm,Birrell:1982ix}) or as an infrared effect, as can be seen in the dispersive approach
of \cite{Giannotti:2008cv}, where the trace anomaly is shown originating from the imposition of all  the non anomalous Ward identities
and the spectral representation of the amplitude. 
At first glance this appears to be quite different than the ultra-local delta function terms obtained in the position 
space approach of \cite{Osborn:1993cr, Erdmenger:1996yc}. 
Thus, the relationship between this approaches requires some clarification, and this is a principal motivation for the present work.

A necessary comment is that the eventual agreement of the two approaches may seem less surprising if it is remembered that coincident point 
singularities in euclidean position space become light cone singularities in Minkowski spacetime, and these light cone singularities are associated 
with the propagation of massless fields, which generally have long range infrared effects. \\

This chapter is composed of two main parts. 

In the first part, building on the results of \cite{Osborn:1993cr, Erdmenger:1996yc}, 
we analyse the structure of the $3$-point correlators in configuration and in momentum space for a general CFT. 
In particular we generalize previous studies of the $TVV$ correlator, perturbatively evaluated in $4$ dimensions in
\cite{Armillis:2009pq, Armillis:2010qk,Coriano:2011zk} in QED, QCD and the Standard Model, to $d$ dimensions.
We also study the $TTT$ vertex, whose computation in $4$ dimensions was presented in the previous chapter, 
and perform a complete investigation of this correlator by the same approach. 
We give particular emphasis to the discussion of the connection between the general approach of \cite{Osborn:1993cr} 
and the perturbative picture. In particular, we give a diagrammatic interpretation of the various contact terms introduced 
by Osborn and Petkou in order to solve the Ward identities for generic positions of the points of the correlators. 
This allows to close a gap between their bootstrap method, previous investigations of the $TVV$
\cite{Giannotti:2008cv,Armillis:2009pq, Armillis:2010qk}, and the recent computation of the $TTT$ vertex.
We show that the perturbative analysis in momentum space in dimensional regularization is in complete agreement with their results. \\
It should be remarked that, in general, the momentum space formulation of the correlators of a CFT has remained largely unexplored
until the publication of \cite{Coriano:2013jba,Bzowski:2013sza}, where conformal constraints in momentum space for $3$-point functions are
systematically explored, proving, as expected, much more difficult to implement with respect to the corresponding position space constraints.
The lack of this investigation for such a long time is mainly due to the fact that momentum space is ideally suited for perturbative computations, 
which in turn always stem from a Lagrangian formulation. This Lagrangian is often missing for CFT's, 
which can be defined on the sole ground of symmetry principles.

This brings us to the second part of the chapter, contained in section \ref{Ch1direct},  where we discuss a general procedure to map 
to momentum space any massless correlator given in position space and not necessarily related to a Lagrangian description.
The investigation of these correlators in momentum space reveals, in general, some specific facts, such as the presence of single and 
multi-logarithmic integrands which, in general, cannot be re-expressed in terms of ordinary master integrals, typical of the Feynman
expansion. In particular, we conclude that, whenever the mentioned logarithmic integrals do not cancel, the theory does not possess 
a Lagrangian formulation, because otherwise no such integral would appear in the perturbative expansion. \\
To address these points, one has to formulate an alternative and general approach to perform the transforms, 
not directly linked to the Lagrangian realization, since in this case such representation may not exist.
The method that is proposed relies on a $d$-dimensional version of differential regularization, similar to the approach suggested in
\cite{Erdmenger:1996yc}. We use the standard technique of \emph{pulling out derivatives via partial integration}
in singular correlators in such a way to make them Fourier-integrable, i.e. expressible as integrals in momentum space.
This is combined with {\em the method of uniqueness} \cite{Kazakov:1986mu},  here generalized to tensor structures, 
in order to formulate a complete and self-consistent procedure.
As in \cite{Osborn:1993cr, Erdmenger:1996yc} we need an extra regulator ($\omega$), 
unrelated to the dimensional regularization parameter $\epsilon$. 
Our approach is defined as a generic algorithm which can handle rather straightforwardly any massless correlator 
written in configuration space. 
The algorithm has been implemented in a symbolic manipulation program and can handle, in principle, correlators of any rank. \\
The aim of the method is to test the Fourier-integrability of a given correlator, by checking the cancellation of the singularities 
in the extra regulator $\omega$ directly in momentum space, and to provide us with the direct expression of the transform.

\section{The correlators and the corresponding Ward identities}

We provide the basic definitions of the correlators that we are going to investigate. 
We suppose that the theory admits an euclidean generating functional $\mathcal W$, in analogy with (\ref{Ch1Generating}),
which depends on the background metric $g_{\mu\nu}$, acting as source of the EMT $T^{\mu\nu}$,
on gauge fields $A^a_\mu$ coupled to the gauge currents $V^a_\mu$, and a source $J$ for each scalar operator $\mathcal{O}$ 
of the spectrum (for the sake of simplicity, we do not distinguish them ).
Thus, if the classical theory is described by an action $\mathcal S$, we are embedding this into a curved space via the coupling 
to the metric $g$ and supplying it with additional source terms,
\beq
\mathcal{W}[g,A,J]= \int \mathcal{D}\Phi\, 
e^{- \mathcal S - \int d^dx\, \sqrt{g}\, \left( A^a_\mu \,V^{a\,\mu} + J \, \mathcal O  \right)}\, .
\label{GenPlusSources}
\eeq
Then, the functional averages of $\mathcal O$, $V$ and $T$ are obtained by differentiating 
the generating functional with respect to the corresponding sources, i.e.
\beq
\llangle O(x) \rrangle_s = - \frac{1}{\sqrt{g_x}} \frac{\delta \mathcal{W}}{\delta J(x)}\, , 
\quad
\llangle V^{a \, \mu}(x) \rrangle_s = -\frac{1}{\sqrt{g_x}} \frac{\delta \mathcal{W}}{\delta A^a_\mu(x)}\, ,
\quad
\llangle T^{\mu\nu}(x) \rrangle_s = \frac{2}{\sqrt{g_x}}\frac{\delta\, \mathcal W}{\delta\, g_{\mu\nu}(x)}\, .
\eeq
The construction of the correlators is straightforward.
If the scalar operator $O$ is coupled to the source $J$, the three point function is defined via a triple functional derivative
with respect to $g_{\mu\nu}$ once and to the scalar source $J$ twice, evaluated switching off the sources at the end, i.e.
\bea
\llangle T^{\mu\nu}(x_1) O (x_2)O(x_3) \rrangle 
&=&  
\bigg\{ \frac{\delta^2}{\delta J(x_2)  \delta J(x_3)} \bigg[  
\frac{2}{\sqrt{g_{x_1}}} \frac{\delta \mathcal{W}}{\delta g_{\mu\nu}(x_1)} \bigg]_{g=\delta}\bigg\}_{J=0} \nn \\
&& \hspace{-20mm}
=\, \llangle T^{\mu\nu}(x_1) O(x_2) O(x_3) \rrangle_{J=0}  + \llangle \frac{\delta T^{\mu\nu}[J](x_1)}{
\delta J(x_2)}  O(x_3) 
\rrangle_{J= 0}  + \llangle \frac{\delta T^{\mu\nu}[J](x_1)}{\delta J(x_3)}  O(x_2) \rrangle_{J= 0}\, ,
\label{DefineTOO}
\eea
The second correlator that we will analyse will be the $VVV$ vertex, which is defined by the third functional derivative of the
generating functional with respect to the source gauge field $A^a_{\mu}(x)$
\bea
\llangle V^{a\,\mu}(x_1) V^{b\,\nu}(x_2) V^{c\,\rho}(x_3) \rrangle = -
\frac{\delta^3 \mathcal{W}|_{g=\delta}}{\delta A^a_\mu(x_1) \delta A^b_\nu(x_2) \delta A^c_\rho(x_3)}\bigg |_{A=0}\, .
\eea
We remark that, due to Furry's theorem, the gauge theory has to be non abelian 
in order to define a non vanishing $VVV$ correlator. 

To derive the $TVV$ correlator, we can first perform a functional derivative with respect to the metric 
and then insert the vector currents by functionally differentiating with respect to the gauge field sources $A$, specifically
\bea
\llangle T^{\mu\nu}(x_1)\, V^{a \, \alpha} (x_2)\, V^{b \, \beta} (x_3) \rrangle 
&=&  
\bigg\{\frac{\delta^2}{\delta A^a_\alpha(x_2) \delta A^b_\beta(x_3)} 
\bigg[\frac{2}{\sqrt{g_{x_1}}} \frac{\delta \mathcal{W}}{\delta g_{\mu\nu}(x_1)} \bigg]_{g=\delta} \bigg\}_{A=0} 
\nn \\
&=&
\llangle T^{\mu\nu}(x_1)\, V^{a\,\alpha}(x_2)\, V^{b\,\beta}(x_3) \rrangle_{A=0} \nn \\
&&  
+\, \llangle \frac{\delta T^{\mu\nu}(x_1)}{\delta A^a_\alpha(x_2)} V^{b\,\beta}(x_3) \rrangle_{A= 0} +
\llangle \frac{\delta T^{\mu\nu}(x_1)}{\delta A^b_\beta(x_3)}  V^{a\,\alpha}(x_2) \rrangle_{A= 0}
\label{DefineTVV}
\eea
where $T_{\mu\nu}$ is calculated in the presence of the background source $A_\mu^a$. The first term in 
the previous expression represents the insertion of the three operators, while the last two are contact terms, with the topology of 
$2$-point functions, exploiting the linear dependence of the EMT from the source field $A$. 

Finally, for the definition of the $TTT$ Green function, which obviously does not change after including the additional sources 
in the generating functional, we refer to section (\ref{Ch1DiagTTT}).


Now we turn to the derivation of non-anomalous Ward identities, which hold for general dimensions, 
by which we mean away from the (even) values of the space dimension for which the trace Ward identities become anomalous
\cite{Birrell:1982ix}. \\
We assume that the generating functional  $W[g,A,J]$ is invariant under diffeomorphisms,
\bea
\mathcal{W}[g',A',J'] = \mathcal{W}[g,A,J]  \, ,
\eea
where $g'$ and $A'$ and $J'$ are transformed metric, gauge field and scalar source under the general infinitesimal 
coordinate transformation $x^\mu \rightarrow {x'}^\mu = x^\mu + \epsilon^\mu(x)$, under which they change according to
\beq
\delta g_{\mu\nu} = \nabla_{\mu} \epsilon_{\nu}  +  \nabla_{\nu} \epsilon_{\mu} \, , 
\qquad 
\delta A^{a}_{\mu}= 
- \left( \epsilon^{\lambda} \nabla_{\lambda} A^a_{\mu} + A^{a \, \lambda} \nabla_{\mu} \epsilon_{\lambda} \right) \, ,
\qquad
\delta J = - \epsilon^\lambda \pd_\lambda J \, .
\eeq
Diffeomorphism invariance and gauge invariance of the generating functional respectively imply the relations
\bea
&& \qquad 
\nabla_{\mu} \llangle T^{\mu\nu} \rrangle + \nabla^{\nu} A^a_\mu \llangle V^{a \, \mu} \rrangle + 
\nabla_{\mu} \left( A^{a\,\nu} \llangle V^{a \, \mu}\rrangle \right)+ \pd_\nu J\, \llangle\mathcal{O}\rrangle = 0 \, ,  \nn \\
&&  \qquad 
\nabla_{\mu} \llangle V^{a \, \mu} \rrangle + f^{abc} A^b_{\mu} \llangle V^{c \mu}\rrangle = 0 \, ,
\label{Ch1BaiscWard}
\eea
where $f^{abc}$ are the structure constants of the gauge group. \\
Naive conformal invariance gives the tracelessness condition
\beq \label{Ch1NaiveScaleWI}
g_{\mu\nu} \llangle T^{\mu\nu} \rrangle_s + 	\left( d - \eta \right)\, J\, \llangle \mathcal{O}\rrangle_s = 0 \, .
\eeq
This last Ward identity is naive, due to the appearance of an anomaly at quantum level, after renormalization of the correlator, 
for even dimensions. However, it is the correct identity away from $d=2\,k$, for integer $k$.
In this respect, the functional differentiation of (\ref{Ch1BaiscWard}) and (\ref{Ch1NaiveScaleWI}) 
allows to derive ordinary Ward identities for the various correlators. \\
If we want to include anomalies in $4$-dimensional space, then remembering eq.  (\ref{Ch1TraceAnomaly}) 
and the conditions on the $\beta$ coefficients discussed below it, we can write, in dimensional regularization,
\beq
g_{\mu\nu}\llangle T^{\mu\nu} \rrangle_s =
\sum_{I=f,s,V}n_I\, \bigg[ \beta_a(I)\, \bigg( F - \frac{2}{3}\, \square R \bigg)+ \beta_b(I)\, G \bigg]
- \frac{\kappa}{4} \, n_V\, F^{a\,\mu\nu}\,F^a_{\mu\nu} + F[J] \, ,
\eeq
where by $F[J]$ we have denoted a possible functional of the background source for the scalar operators,
whose form is not unique, but depends on the dimensions of the corresponding operators.
For example, in $4$ dimensions and for the operator $\mathcal{O} = \phi^2 $, where $\phi$ 
is the standard elementary scalar field, then $F[J] = \frac{p}{2}\, J^2$, with $p$ a c-number.

Now let us list the Ward identities implied by (\ref{Ch1BaiscWard}) for the various correlators.
In the case of the $TOO$ vertex one has the equation
\beq
\pd^{x_1}_\mu \llangle T^{\mu\nu}(x_1)\mathcal{O}(x_2)\mathcal{O}(x_3) \rrangle =
\pd^{x_1}_\nu \delta^{(d)}(x_{12}) \llangle \mathcal{O}(x_1)\mathcal{O}(x_3)\rrangle +
\pd^{x_1}_\nu \delta^{(d)}(x_{13}) \llangle \mathcal{O}(x_1)\mathcal{O}(x_2)\rrangle \, .
\label{DiffWardTOO}
\eeq
For the $VVV$ the conservation equation is
\beq
\pd^{x_1}_\mu \llangle V^{a\,\mu}(x_1)V^{b\,\nu}(x_2)V^{c\,\rho}(x_3) \rrangle = 
f^{abd}\, \delta^{(d)}(x_{12}) \llangle V^{d\,\nu}(x_1)V^{c\,\rho}(x_3) \rrangle - 
f^{acd}\, \delta^{(d)}(x_{13}) \llangle V^{d\,\rho}(x_1)V^{c\,\nu}(x_2) \rrangle  \, ,
\label{DiffWardVVV}
\eeq
Finally, for the case of the $TVV$ we obtain
\bea
\partial_{\mu}^{x_1} \llangle T^{\mu\nu}(x_1) V^{a\,\alpha}(x_2) V^{b\,\beta}(x_3) \rrangle  
&=&
\partial^{\nu}_{x_1} \delta^d(x_{12}) \llangle V^{a\,\alpha}(x_1) V^{b\,\beta}(x_3) \rrangle 
+ \partial^{\nu}_{x_1} \delta^d(x_{31}) \llangle V^{a\,\alpha}(x_2) V^{b\,\beta}(x_1) \rrangle \nn \\
&-& 
\delta^{\nu\alpha} \partial^{x_1\,}_{\mu} \left( \delta^d(x_{12})  \llangle V^{a\,\mu}(x_1) V^{b\,\beta}(x_3) 
\rrangle \right) - 
\delta^{\nu\beta} \partial^{x_1\,}_{\mu} \left( \delta^d(x_{31})  \llangle V^{a\,\alpha}(x_2) V^{b\,\mu}(x_1) 
\rrangle \right) \, , \nn \\
\label{TVVWardCoord}
\eea
together with the vector current Ward identities, following from gauge invariance,
\bea
\partial_{\alpha}^{x_2} \llangle T^{\mu\nu}(x_1) V^{a\,\alpha}(x_2) V^{b\,\beta}(x_3) \rrangle  &=& 0 \,, \qquad
\partial_{\beta}^{x_3} \llangle T^{\mu\nu}(x_1) V^{a\,\alpha}(x_2) V^{b\,\beta}(x_3) \rrangle  = 0 \, .
\label{GaugeInvTVV}
\eea
The general covariance Ward identity for the $TTT$ vertex was already given in section \ref{Ch1DiagTTTWardCov}.

Now let us move to trace Ward identities for both general $d$ and $4$ dimensions.
Discarding the  $VVV$,  the naive identity (\ref{Ch1NaiveScaleWI}) gives the non-anomalous conditions, for $d$ dimensions
\bea
\delta_{\mu\nu} \, \llangle T^{\mu\nu}(x_1) \mathcal{O}(x_2)\mathcal{O}(x_2) \rrangle  
&=& 
\left( d-\eta \right)\, \left( \delta^{(d)}(x_{12})\,\llangle \mathcal{O}(x_1)\mathcal{O}(x_3)\rrangle +
\delta^{(d)}(x_{13})\,\llangle \mathcal{O}(x_1)\mathcal{O}(x_2)\rrangle \right) \, , 
\label{TraceWardTOO} \\
\delta_{\mu\nu} \, \llangle T^{\mu\nu}(x_1) V^{a\,\alpha}(x_2) V^{b\,\beta}(x_3) \rrangle  
&=& 
0 \, ,
\label{TraceWardTVV}
\eea 
which become, when anomalies are properly included,
\bea
\delta_{\mu\nu} \, \llangle T^{\mu\nu}(x_1) \mathcal{O}(x_2)\mathcal{O}(x_2) \rrangle  
&=& 
\left( 4-\eta \right)\, \left( \delta^{(d)}(x_{12})\,\llangle \mathcal{O}(x_1)\mathcal{O}(x_3)\rrangle +
\delta^{(d)}(x_{13})\,\llangle \mathcal{O}(x_1)\mathcal{O}(x_2)\rrangle \right) \nn 
+ \frac{\delta^2 F[J](x_1)}{\delta J(x_2)\delta J(x_3)} \, ,   
\label{AnTraceWardTOO}
\\
\delta_{\mu\nu} \, \llangle T^{\mu\nu}(x_1) V^{a\,\alpha}(x_2) V^{b\,\beta}(x_3) \rrangle 
&=& 
\delta^{ab}\, \kappa\, \left( \pd^\beta\delta^{(4)}(x_{12})\pd^\alpha\delta^{(4)}(x_{13}) - 
\delta^{\alpha\beta}\, \pd^\lambda\delta^{(4)}(x_{12})\pd_\lambda\delta^{(4)}(x_{13})   \right) \,  .
\label{AnTraceWardTVV}
\eea
The trace Ward identity for the $TTT$ was given in section \ref{Ch1DiagTTTWardAnom} in momentum space in 4 dimensions.
Here we report the coordinate-space versions of the non anomalous identity, holding for general $d$.
\beq
\delta_{\mu\nu}\llangle T^{\mu\nu}(x_1)T^{\alpha\beta}(x_2)T^{\rho\sigma}(x_3) \rrangle  =
-2\, \bigg( \delta^{(4)}(x_{12}) + \delta^{(4)}(x_{13} ) \bigg)\,
\llangle T^{\alpha\beta}(x_2)T^{\rho\sigma}(x_3)\rrangle \, ,
\label{NoAnTraceTTT}
\eeq
and of the anomalous identity, valid only for $d=4$
\bea
\delta_{\mu\nu}\llangle T^{\mu\nu}(x_1)T^{\alpha\beta}(x_2)T^{\rho\sigma}(x_3) \rrangle  
&=&
4\, \bigg( \beta_a\, \left[ F(x_1) - \frac{2}{3}\,\Box R(x_1) \right]^{\rho\sigma\alpha\beta}(x_2,x_3)
+ \beta_b\, \left[ G(x_1)\right]^{\rho\sigma\alpha\beta}(x_2,x_3) \bigg) \nn \\
&&
-\, 2\, \bigg( \delta^{(4)}(x_{12}) + \delta^{(4)}(x_{13} ) \bigg)\,
\llangle T^{\alpha\beta}(x_2)T^{\rho\sigma}(x_3)\rrangle   \, ,
\label{AnTraceTTT}
\eea
with the compact notation for functional derivatives in the flat space limit introduced in (\ref{Ch1Flat}).

\section{Inverse mappings: correlators in position space from the momentum space Feynman expansion}

Having by now defined all the fundamental (anomalous and regular) Ward identities which allow to test the correctness
of the correlators we are interested in studying, we turn to compare the expressions of these correlators in position space
with their perturbative realizations in free field theories in momentum space.

We remind that an important result of \cite{Osborn:1993cr} is the identification of the solution of the Ward identities in terms of a set of
constants and of certain linearly independent tensor structures in euclidean position space. Consistency requires that the Fourier transforms 
of these tensor structures must occur in direct computations of the same vertex functions in free field theories in momentum space,
which are defined, in turn, via certain $1$-loop integrals that can be computed according to a well-defined set of Feynman rules,
once the Lagrangian theory has been specified.
This implies that, after establishing the combination of these integrals defining the Green function in momentum space,
we can use them to infer what those tensor structures must be, and find the exact correspondence between CFT amplitudes 
in position space and momentum space {\it a posteriori}.
Obviously, this is only possible provided that we have enough linearly independent vertex functions for different free theories to determine
the linear combinations uniquely. We call this procedure an \emph{inverse mapping}, as it allows to re-express the correlators 
of  \cite{Osborn:1993cr} in such a form that their Fourier-integrability is explicit.
By \emph{integrable} we mean,  in this context, a function of coordinates whose Fourier transform is not divergent and contains 
no additional regulator, in a sense made precise in section \ref{Ch1pull}.
This result is obtained by pulling out  derivatives of the corresponding diagrams, in the spirit of differential regularization \cite{Freedman:1991tk}
in such a way that integrability becomes trivial. 

More technical details on the inverse mapping method are explicitly provided in appendix \ref{Ch1InverseTTT}, where the $TTT$
correlator is used as an illustrative example.

\subsection{The TOO case}

The first correlator that we are going to investigate is the $TOO$. In the perspective of comparing the coordinate space results of
\cite{Osborn:1993cr} with momentum space perturbative expansions, it is also the most ambiguous, as such a perturbative
expansion requires establishing the scaling dimensions of the scalar operators once for all.
For this reason, we are going to perform such a test only in one specific case, i.e. for $\mathcal O = \phi^2$.

The general structure of this Green function in coordinate space - for non coincident points -  is
\beq
\label{Ch1TOOPO}
\llangle T_{\mu\nu}(x_1) \, O(x_2) \, O(x_3) \rrangle =
\frac{a}{(x^2_{12})^{d/2} \, (x^2_{23})^{\eta-d/2} \, (x^2_{31})^{d/2}} \, h^1_{\mu\nu}(\hat{X}_{23}) \, ,
\eeq
where $a$ is a constant, $\eta$ the scaling dimension of the scalar operator $O $  and where
\bea   \label{Ch1TOOstructures}
\hat{X}_\mu = \frac{X_\mu}{\sqrt{X^2}} \, , \nn \qquad
h^1_{\mu\nu}(\hat X) = \hat{X}_\mu \, \hat{X}_\nu - \frac{1}{d} \, \delta_{\mu\nu} \, ,
\eea
where
\bea
x_{ij} \equiv x_i - x_j \, , \qquad  X_{ij} = - X_{ji} \equiv
\frac{x_{ik}}{x^2_{ik}} - \frac{x_{jk}}{x^2_{jk}} \, ,  \quad  i,j,k = 1,2,3 \, , \quad k\neq i \, , k \neq j \, .
\eea
In the short-distance limits of its external points, this vertex is singular for $\eta \rightarrow d/2$ and needs regularization. 
In \cite{Osborn:1993cr} the Ward identities are solved through the analysis of the short distance limits of (\ref{Ch1TOOPO}),
by which we mean the limits $x_1 \rightarrow x_2$, $x_1 \rightarrow x_3$. 
Some singular terms are found and thus the authors are forced to regularize them 
with the method of differential regularization \cite{Freedman:1991tk}, which finally gives the modified expression
\bea
\llangle T_{\mu\nu}(x_1) \, O(x_2) \, O(x_3) \rrangle 
&=&
\frac{a}{(x^2_{12})^{d/2} \, (x^2_{23})^{\eta-d/2} \, (x^2_{31})^{d/2}} \, h^1_{\mu\nu}(\hat{X}_{23}) 
\nn \\
&&
+\,  \left[ \hat A_{\mu\nu}(x_{12}) - A_{\mu\nu}(x_{12}) + 
             \hat A_{\mu\nu}(x_{31}) - A_{\mu\nu}(x_{31})\right]  \, \frac{N}{(x_{23}^2)^{\eta}} \, ,
\label{CompleteTOO}
\eea
where we have introduced the structures
\bea
A_{\mu\nu}(s) = \frac{a}{N}\, \frac{1}{s^d}\, 
                                  \left( \frac{s_\mu s_\nu}{s^2} - \frac{1}{d} \delta_{\mu\nu} \right) \, , 
\qquad
\hat A_{\mu\nu}(s) = \frac{a}{N\,d} \left( \frac{\partial_{\mu}\partial_{\nu}}{d-2} \, \frac{1}{s^{d-2}} + 
\frac{\eta -d +1}{\eta}\, S_d \, \delta_{\mu\nu} \,  \delta^d(s) \right) \, ,
\label{Ahat}
\eea
and $N$ is defined as the normalization constant of the $2$-point function of the scalar operator
\beq
\llangle \mathcal{O}(x_1)\mathcal{O}(x_2) \rrangle = \frac{N}{(x^2_{12})^\eta}\, .
\eeq
It is important to make some comments at this point, as the $TOO$ is the simplest function, between those addressed in this chapter,
to require such a regularization procedure in order to account for the inhomogeneous terms in the Ward identities.
Notice that, in general, one can define a differentially regularized tensor $\hat{A}$  as
\beq
\hat{A}_{\mu\nu}(s) =  \frac{a}{N\,d}\, 
\left( \frac{1}{d-2}\pd_\mu\pd_\nu \frac{2}{s^{d-2}} + C\,\delta_{\mu\nu}\,S_d \delta^{(d)}(s)  \right) \, ,
\eeq
which \emph{exactly coincides  with $A_{\mu\nu}(s)$ for $s \neq 0$} and where the only difference between them is in the
$\delta$-function term, whose coefficient is not fixed \emph{a priori}, reflecting the arbitrariness typical of any regularization. 
It is precisely this kind of term that discerns (\ref{CompleteTOO}) from (\ref{Ch1TOOPO}).
The ambiguity in its coefficient is solved by requiring the Ward identities to be satisfied by (\ref{CompleteTOO}).
In this way, the terms $\hat{A}- A$ are the contact contributions (sometimes called \emph{semi-local} in the literature) 
which consistently account for the r.h.s. of (\ref{DiffWardTOO}) and (\ref{TraceWardTOO}).
Essentially the same argument holds for the more complicated $3$-point functions we are going to discuss in the next sections:
only the formulas are more complicated, due to the increasing number of tensor structures.

In the expression above $S_d$ is the "volume" of the sphere in $d$-dimensions, 
\beq
S_d = 2\,\pi^{\frac{d}{2}}/\Gamma(d/2) \, .
\eeq
Introducing (\ref{Ahat}) into (\ref{CompleteTOO}), we find the explicit expression
\bea
\llangle T_{\mu\nu}(x_1) \, O(x_2) \, O(x_3) \rrangle
&=&
\frac{a}{(d-2)^2}\bigg\{
( \partial_{\mu}^{12} \partial_{\nu}^{31} + \partial_{\nu}^{12} \partial_{\mu}^{31}  ) +
\frac{d-2}{d} ( \partial_{\mu \nu}^{12}  + \partial_{\mu\nu}^{31} )
\bigg\} \frac{1}{(x^2_{12})^{d/2-1}  (x^2_{23})^{\eta-d/2+1}  (x^2_{31})^{d/2-1}} \nn \\
&+&
a\, \frac{ x^2_{12}  x^2_{23} + x^2_{31} x^2_{23} -
(x^2_{23})^2}{(x^2_{12})^{d/2}  (x^2_{23})^{\eta-d/2+1}  (x^2_{31})^{d/2}} \frac{\delta_{\mu\nu}}{d}+ 
a\, \frac{\eta -d+1}{d\, \eta} \, S_d \, \delta_{\mu\nu} \, \frac{\delta^d(x_{12}) +  
\delta^d(x_{31}) }{(x_{23}^2)^\eta}\, ,
\label{ExpCompleteTOO}
\eea
where, from now on, we set $\pd^{12}_\mu \equiv \frac{\pd}{\pd x_{12\,\mu}}$ and 
$\pd^{12}_{\mu\nu} \equiv \frac{\pd}{\pd x_{12\,\mu}}\frac{\pd}{\pd x_{12\,\nu}}$.
Notice that the first term of the second line proportional to $\delta_{\mu\nu}$ is not manifestly integrable, 
but one can use identities such as 
$x_{12}^2+ x_{13}^2 - x_{23}^2=2 x_{12}\cdot x_{13}$ in order to rewrite it in the form
\beq
\frac{ x^2_{12} \, x^2_{23} + x^2_{31} \, x^2_{23} - (x_{23}^2)^2}{(x^2_{12})^{d/2} \, (x^2_{23})^{d/2}\, 
(x^2_{31})^{d/2}}=
\frac{2 }{(d-2)^2} \partial^{12}_\mu \partial^{31 \,\mu} 
\frac{1}{ (x_{12}^2)^{d/2-1} (x_{31}^2)^{d/2-1} (x_{23}^2)^{\eta-d/2+1}}
\eeq
which shows its integrability when $\eta < d-1$.

Now, in order to test the consistency of the result (\ref{Ch1TOOPO}) obtained from the application of the conformal Ward identities 
for the $TOO$, we can consider a particular scalar free field theory. 
We suppose for instance that the scalar operator $\mathcal O$ is given by $\mathcal O = \phi^2$ with dimensions $\eta = d-2$, whose EMT is
\bea
T_{\mu\nu}
&=&
\pd_\mu \phi \, \pd_\nu\phi - \frac{1}{2} \, \delta_{\mu\nu}\,\pd_\alpha \phi \, \pd^\alpha \phi
+ \frac{1}{4}\,\frac{d-2}{d-1}\, \bigg[\delta_{\mu\nu} \pd^2 - \pd_\mu\,\pd_\nu\bigg]\, \phi^2 
\eea
which is conserved and traceless in $d$ dimensions. \\
Using the Feynman rules in momentum space together with the expression of a scalar propagator,
after applying the inverse mapping procedure detailed in appendix \ref{Ch1InverseTTT},  
we obtain the $T\phi^2\phi^2$ correlation function in $d$ dimensions
\bea
\label{Ch1TphiphiDer}
\llangle T_{\mu\nu}(x_1) \phi^2 (x_2) \phi^2 (x_3) \rrangle 
&=&  
\frac{2 a (d-1)}{d (d-2)^2} \bigg[ \partial_{\mu}^{12} 
\partial_{\nu}^{31} + \partial_{\nu}^{12} \partial_{\mu}^{31}  - \delta_{\mu\nu} \partial^{12} \cdot \partial^{31}  - 
\frac{d-2}{2(d-1)} \bigg( - \partial_{\mu\nu}^{12} - \partial_{\mu\nu}^{31}  + \partial_{\mu}^{12} \partial_{\nu}^{31} 
\nn \\
&+& 
\partial_{\nu}^{12} \partial_{\mu}^{31} + \delta_{\mu\nu} \left(\partial^2_{12} + \partial^2_{31} - 2 \partial^{12} \cdot 
\partial^{31} \right)  \bigg) \bigg]    \frac{1}{(x_{12}^2)^{d/2-1} (x_{23}^2)^{d/2-1} (x_{31}^2)^{d/2-1}} \nn \\
&-& 
a \frac{d-1}{d(d-2)} S_d \delta_{\mu\nu} \frac{\delta^d(x_{12}) +  \delta^d(x_{31}) }{(x_{23}^2)^{d-2}} \,.
\eea
The equivalence of this expression with the solution given in (\ref{ExpCompleteTOO})  can be explicitly checked.
We remark that (\ref{Ch1TphiphiDer}) is clearly integrable and does not require any intermediate regularization. 
The first term in the previous expression comes from the triangle topology diagram while the last two, 
proportional to the delta functions, are contact terms with $2$-point topology (see eq.  (\ref{DefineTOO})).

\subsection{The $VVV$ case}

The $VVV$ vertex function is pretty easy to handle with the inverse mapping procedure.
\begin{figure}[t]
\begin{center}
\includegraphics[scale=0.7]{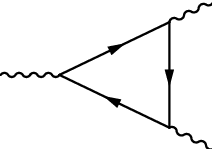}\qquad\qquad
\includegraphics[scale=0.7]{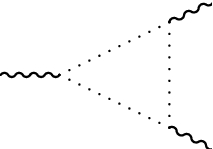}
\caption{The fermion and the scalar sectors contributing to the conformal VVV vertex in any dimension.}
\label{Ch1VVV}
\end{center}
\end{figure}
In  \cite{Osborn:1993cr} the general CFT requirements fix the structure of the $VVV$ to be
\bea \label{Ch1VVVcoord}
\llangle V_{\mu}^a(x_1) V_{\nu}^b (x_2) V_{\rho}^c(x_3) \rrangle
&=&
\frac{f^{abc}}{(x^2_{12})^{d/2-1}\,(x^2_{23})^{d/2-1}\,(x^2_{31})^{d/2-1}}  \, 
\bigg\{ (a - 2b) \, X_{23\,\mu}\,X_{31\,\nu}\,X_{12\,\rho} \nonumber \\
&-&
b \left[\frac{1}{x^2_{23}} \, X_{23\,\mu} \, I_{\nu\rho}(x_{23})  + 
\frac{1}{x^2_{31}} \, X_{31\,\nu} \, I_{\mu\rho}(x_{31})+ 
\frac{1}{x^2_{12}} \, X_{12\,\rho}\,I_{\mu\nu}(x_{12} ) \right]  \bigg\} \, ,
\eea
where $I_{\mu\nu}(x)$ is the inversion operator defined as
\beq
I^{\mu\nu}(x)=
\delta^{\mu\nu} - 2 \frac{x^\mu x^\nu}{x^2} \, .
\eeq
The correlator is Fourier-integrable, although this is not immediately evident from (\ref{Ch1VVVcoord}). The simplest way to prove this point 
consists in showing that (\ref{Ch1VVVcoord}) can be reproduced in $d$-dimensions by the combination of the scalar and the fermion 
sectors of a free field theory.
For this purpose we use two realizations of the vector current $V_{\mu}^a$, using scalar and fermion fields
\bea
V_{\mu}^a = \phi^{*} t^a \left(\partial_{\mu} \phi \right) - \left( \partial_{\mu} \phi^{*}\right) t^a \phi \, ,  
\qquad  
V_{\mu}^a = \bar \psi \, t^a \gamma_{\mu} \psi \, .
\label{SAndF}
\eea
The diagrammatic expansion of this correlator consists of two triangle diagrams, the direct and the exchanged, both in the scalar and fermion 
sectors. These two types of diagrams are shown in fig. \ref{Ch1VVV} and it is well known how to write down their expressions
in momentum space, using the corresponding Feynman rules. \\
Performing the inverse mapping procedure, we find the result
\bea
\llangle V_{\mu}^a(x_1) V_{\nu}^b (x_2) V_{\rho}^c(x_3) \rrangle^{f} &=&
- \frac{ c_f \, f^{a b c} }{(d-2)^3} \Delta_{\mu \alpha \nu \beta \rho \gamma} \partial^{\alpha}_{12}\partial^{\beta}_{23}
\partial^{\gamma}_{31}  \frac{1}{(x_{12}^2)^{d/2-1} (x_{23}^2)^{d/2-1} (x_{31}^2)^{d/2-1}} \, ,  \\
\llangle V_{\mu}^a(x_1) V_{\nu}^b (x_2) V_{\rho}^c(x_3) \rrangle^{s} 
&=&
\frac{c_s \, f^{a b c}}{(d-2)^2}   \left( \partial_{\mu}^{12} + \partial_{\mu}^{31} \right) \left( \partial_{\nu}^{23} + 
\partial_{\nu}^{12} \right) \left( \partial_{\rho}^{31} + 
\partial_{\rho}^{23} \right) \frac{1}{(x_{12}^2)^{d/2-1} (x_{23}^2)^{d/2-1} (x_{31}^2)^{d/2-1}}  \nn \\
\eea
for the fermion and the scalar sector respectively, where we have introduced the operator
\bea
\label{Ch1Delta6}
\Delta_{\mu \alpha \nu \beta \rho \gamma} = \frac{1}{4}\, 
Tr \left[ \gamma_{\mu} \gamma_{\alpha} \gamma_{\nu} \gamma_{\beta} \gamma_{\rho} \gamma_{\gamma} \right]  \, ,
\eea
and $c_f, c_s$ are normalization constants whose numerical values are irrelevant here.
Written in this form, with derivatives pulled out, the two expressions are manifestly integrable.
 
Tracing over the $\gamma$ matrices and applying the derivatives over all the denominators, 
we generate the result of  \cite{Osborn:1993cr} by taking a linear combination of these two sectors
\bea
\llangle V_{\mu}^a(x_1) V_{\nu}^b (x_2) V_{\rho}^c(x_3) \rrangle = 
\bigg( a \, t^a_{\mu\nu\rho} + b \, t^b_{\mu\nu\rho}  \bigg) 
\frac{ f^{abc}}{(x_{12}^2)^{d/2-1} (x_{23}^2)^{d/2-1} (x_{31}^2)^{d/2-1}} \, ,
\eea
where
\bea
t^a_{\mu\nu\rho} &=& \frac{1}{d(d-2)^2}   \left( \partial_{\mu}^{12} + 
\partial_{\mu}^{31} \right) \left( \partial_{\nu}^{23} + \partial_{\nu}^{12} \right) \left( \partial_{\rho}^{31} + 
\partial_{\rho}^{23} \right) - \frac{1}{d} t^b_{\mu\nu\rho} \, , \\
t^b_{\mu\nu\rho} &=&   - \frac{1}{(d-2)^3}\, 
\Delta_{\mu \alpha \nu \beta \rho \gamma}\partial^{\alpha}_{12}\partial^{\beta}_{23}\partial^{\gamma}_{31} \, .
\eea
We have explicitly checked the equivalence between this expression and eq.  (\ref{Ch1VVVcoord}).
No additional term is required, in position space, to account for the general covariance Ward identity (\ref{DiffWardVVV}).

\subsection{The $TVV$ case}

The next correlator that we are going to discuss is the $TVV$, for which, together with the $TTT$ vertex, 
the analysis required to confirm the correspondence between the position space solutions given in \cite{Osborn:1993cr}
and perturbative computations in momentum space is much more involved, due to the growth of the number of tensors structures.

We begin with the expression of the $TVV$ in position space at separate points, which is
\beq
\llangle T_{\mu\nu}(x_1) V^a_\alpha(x_2)  V^b_\beta(x_3) \rrangle =
\frac{\delta^{ab}}{(x^2_{12})^{d/2} \, (x^2_{31})^{d/2} \, 
(x^2_{23})^{d/2-1}} \, I_{\alpha\sigma}(x_{12}) \, I_{\beta\rho}(x_{31}) \, t_{\mu\nu\rho\sigma}(X_{23}) \, ,
\label{TVVSepCoord}
\eeq
where the structure $t_{\mu\nu\rho\sigma}(X)$ is given by the following combination of $4$-indices structures,
which are traceless with respect to $(\mu,\nu)$,
\beq
t_{\mu\nu\rho\sigma}(X) = 
a\, h^1_{\mu\nu}(\hat X)\, \delta_{\rho\sigma} + 
b\, h^{1}_{\mu\nu}(\hat X)\,h^{1}_{\rho\sigma}(X) +
c\, h^{2}_{\mu\nu\rho\sigma}(\hat X) + 
e\, h^{3}_{\mu\nu\rho\sigma}\, ,
\eeq
with $h^1_{\mu\nu}$ already introduced in (\ref{Ch1TOOstructures}), whereas
\bea
h^{2}_{\mu\nu\rho\sigma}(\hat X)  
&=& 
\hat{X}_\mu\, \hat{X}_\rho\, \delta_{\nu\sigma} + 
\hat{X}_\nu\, \hat{X}_\sigma\, \delta_{\mu\sigma} + 
\left( \rho \leftrightarrow \sigma \right) - 
\frac{4}{d}\, \hat{X}_\mu\, \hat{X}_\nu\, \delta_{\rho\sigma} -
\frac{4}{d}\, \hat{X}_\rho\, \hat{X}_\sigma\, \delta_{\mu\nu} +
\frac{4}{d^2}\, \delta_{\mu\nu}\, \delta_{\rho\sigma} \, ,
\nn \\
h^{3}_{\mu\nu\rho\sigma} 
&=& 
\delta_{\mu\rho}\delta_{\nu\sigma}  + \delta_{\mu\sigma}\delta_{\nu\rho} - 
\frac{2}{d}\, \delta_{\mu\nu}\delta_{\rho\sigma} \, .
\eea
On the other hand, if one considers the Ward identity (\ref{TVVWardCoord}) at separate points, i.e. with vanishing r.h.s. ,
the four coefficients $a$,$b$,$c$ and $e$ are found to be constrained by
\beq
d\,a -2\,b +2\,(d-2)\,c = 0 \, , \quad b= d\,(d-2)\,e\, ,
\label{ConstraintsTVV}
\eeq
so that there are two independent contributions to the $TVV$ vertex for general dimensions.

As usual, the next step in the analysis of \cite{Osborn:1993cr} is to study the $2$-point coincidence limits 
$x_1\rightarrow x_2$, $x_1\rightarrow x_3$ in order to identify the terms which are responsible 
for the r.h.s. of the Ward identity (\ref{TVVWardCoord}).
Unlike the case of the $VVV$ and similarly to the case of the $TOO$, to which we refer for the details concerning the regularization procedure
of the short-distance singularities, some of the terms appearing in this limit on the r.h.s. of (\ref{TVVWardCoord}) 
are found to need regularization. Then differential regularization is used to pull out derivatives 
and find some regularization-dependent terms proportional to $\delta$-functions.
The solution obtained is connected to the normalization constant of the vector current (unrenormalized) $2$-point function, which is
\beq
\llangle V^{a}_\mu (x_1)V^{b}_\nu (x_2) \rrangle = 
C_V \, \frac{I_{\mu\nu}(x_{12})}{(x^2_{12})^{d-1}} \, .
\label{VVPoistion}
\eeq
so that the complete, unrenormalized $TVV$ correlator is given by
\bea
\llangle T_{\mu\nu}(x_1) V^a_\alpha(x_2)  V^b_\beta(x_3) \rrangle 
&=&
\frac{\delta^{ab}}{(x^2_{12})^{d/2} \, (x^2_{31})^{d/2} \, 
(x^2_{23})^{d/2-1}} \, I_{\alpha\sigma}(x_{12}) \, I_{\beta\rho}(x_{31}) \, t_{\mu\nu\rho\sigma}(X_{23}) \nn \\
&&
+ \delta^{ab}\,C_V\, \left[ \hat A_{\mu\nu\alpha\rho}(x_{12}) - A_{\mu\nu\alpha\rho}(x_{12}) \right] 
\frac{I_{\rho\beta}(x_{23})}{(x_{23}^2)^{d-1}} \nn\\
&&
+ \delta^{ab}\,C_V	\, \left[ \hat A_{\mu\nu\sigma\beta}(x_{31}) - A_{\mu\nu\sigma\beta}(x_{31}) \right] 
\frac{I_{\sigma\alpha}(x_{23})}{(x_{23}^2)^{d-1}}\, .
\label{Ch1TVVcoord}
\eea
Here, the structures $A$ and $\hat{A}$ are respectively
\bea
C_V\, A_{\mu\nu\rho\sigma}(s) 
&=& 
\frac{1}{s^d}\, I_{\rho\alpha}(s)\, t_{\mu\nu\alpha\sigma}(s) \, , \nn \\
C_V\, \hat{A}_{\mu\nu\rho\sigma}(s) 
&=&
\bigg[ 
\frac{2e}{d(d-2)}\delta_{\mu\nu}\pd_\rho\pd_\sigma - \frac{2c}{d(d-2)}\,\delta_{\rho\sigma}  \pd_\mu\pd_\nu
- \frac{c+d e}{d(d-2)}\left(\delta_{\nu\rho}\pd_\mu\pd_\sigma +  \delta_{\mu\rho}\pd_\nu\pd_\sigma \right) \nn \\
&&
+\, \frac{c-(d-2)e}{d(d-2)}\left( \delta_{\nu\sigma}\pd_\mu\pd_\rho + \delta_{\mu\sigma}\pd_\nu\pd_\rho  \right) 
\bigg]\, \frac{1}{s^{d-2}}
-\, \frac{e}{d(d-4)}\,\pd_\mu\pd_\nu\pd_\rho\pd_\sigma \frac{1}{s^{d-4}} \nn \\
&&
+\, \frac{1}{d}\, \bigg[ 2e\, \delta_{\mu\nu}\delta_{\rho\sigma}- \left(c + d e\right)
\left(\delta_{\mu\sigma}\delta_{\nu\rho}+\delta_{\mu\rho}\delta_{\nu\sigma}\right)  \bigg]\,S_d\, \delta^{(d)}(s)\, .
\label{hatATVV}
\eea
%
%
%
Again, $\hat{A}_{\mu\nu\rho\sigma}(s) = {A}_{\mu\nu\rho\sigma}(s) $ for $s\neq 0$ and the coefficients in front
of the $\delta$-function terms in the last line of (\ref{hatATVV}) are determined only after imposing the Ward identities. 

\begin{figure}[t]
\begin{center}
\includegraphics[scale=0.7]{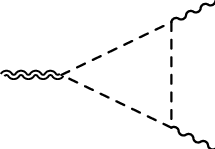}\qquad
\includegraphics[scale=0.7]{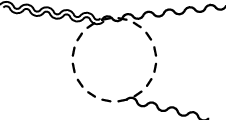}\qquad
\includegraphics[scale=0.7]{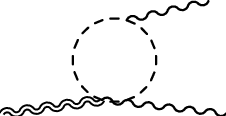}
\caption{The fermion/scalar sectors in the TVV vertex.}
\label{Ch1TVVfigure}
\end{center}
\end{figure}

Now we are ready to check the correspondence between the complete expression of the unrenormalized $TVV$ in position space, 
(\ref{Ch1TVVcoord}), and the inverse-mapped momentum space $1$-loop computations.
We have drawn the diagrammatic structure of the $TVV$ in fig. \ref{Ch1TVVfigure}.
Using the information that the most general $TVV$ is parametrized by just two independent constants, we conclude that,
in any dimension, it can be fully constructed as a linear combination of two contributions coming from independent free theories, 
the fermion and the scalar. \\
Therefore we can write
\bea
\llangle T_{\mu\nu}(x_1) V^a_{\alpha} (x_2) V^b_{\beta} (x_3) \rrangle
&=&
\sum_{I=s,f}n_I\, \bigg( \llangle T_{\mu\nu}(x_1) V^a_\alpha(x_2) V^b_\beta(x_3) \rrangle^I_{A=0}  \nn \\
&&
+\, \llangle \frac{\delta T_{\mu\nu}(x_1)}{\delta A^{a\,\alpha}(x_2)} V^b_\beta(x_3) \rrangle^I_{A= 0}
+   \llangle \frac{\delta T_{\mu\nu}(x_1)}{\delta A^{b \,\beta}(x_3)} V^a_\alpha(x_2) \rrangle^I_{A= 0} \bigg)
\label{Ch1contactTVV}
\eea
where the sum is over the same scalar (s) and fermion (f) sectors introduced for the $VVV$ and $n_I$ stands for the number 
of corresponding fields. 

For  the diagrammatic interpretation of  the various contributions to this correlator, 
(except for the counterterm, which will be addressed in the next section),
among the terms above, the first one corresponds to the triangle topology, while the remaining two are the two bubbles 
(see fig. \ref{Ch1TVVfigure}).

Using the Feynman rules in momentum space and going through the inverse-mapping procedure,
one can obtain the following parametrization of the triangle contribution to the $TVV$ vertex for scalars within the loop,
\bea
\llangle T_{\mu\nu}(x_1) V^a_{\alpha} (x_2) V^b_{\beta} (x_3) \rrangle_{A=0}^{s} = 
\nn
c  \, \delta^{ab} \frac{2(d-1)}{d (d-2)^3} \bigg[ \partial_{\mu}^{12} \partial_{\nu}^{31} +  
\partial_{\nu}^{12} \partial_{\mu}^{31}  - \delta_{\mu\nu} \partial^{12} \cdot \partial^{31}  
\nn \\
 - \frac{d-2}{2(d-1)} \bigg( - \partial_{\mu\nu}^{12} - \partial_{\mu\nu}^{31}
 + \partial_{\mu}^{12} \partial_{\nu}^{31} + \partial_{\nu}^{12} \partial_{\mu}^{31} + 
\delta_{\mu\nu} \left(\partial^2_{12} + \partial^2_{31} - 2 \partial^{12} \cdot \partial^{31} \right)  \bigg) \bigg] \times 
\nn \\
\times  \left( \partial_{\alpha}^{12} + \partial_{\alpha}^{23}\right)
\left( \partial_{\beta}^{31} + \partial_{\beta}^{23}\right)   
\frac{1}{(x_{12}^2)^{d/2-1} (x_{23}^2)^{d/2-1} (x_{31}^2)^{d/2-1}}\, .
\label{Ch1scalarsector}
\eea
whereas for fermions we have
\bea
\llangle T_{\mu\nu}(x_1) V^a_{\alpha} (x_2) V^b_{\beta} (x_3) \rrangle_{A=0}^{f} =  
\frac{c \, \delta^{ab}}{d(d-2)^3} \, A_{\mu\nu\xi\eta}\, \Delta_{\xi\rho\alpha\sigma\beta\lambda}\,
(\partial_{\eta}^{12} + \partial_{\eta}^{31} ) \, \partial^{\rho}_{12} \partial^{\sigma}_{23} \partial^{\lambda}_{31} 
\nn\\
\times \frac{1}{(x_{12}^2)^{d/2-1} (x_{23}^2)^{d/2-1} (x_{31}^2)^{d/2-1}} \, ,
\label{Ch1fermionsector}
\eea
where $ \Delta_{\mu\rho\alpha\sigma\beta\lambda}$ is defined in eq.  (\ref{Ch1Delta6}) 
and $A_{\mu\nu\rho\sigma}$ in appendix \ref{Ch1Vertices}. \\
In (\ref{Ch1scalarsector})-(\ref{Ch1fermionsector}) $c$ is a normalization constant and
these terms can be seen to exactly correspond to the expression (\ref{TVVSepCoord}), holding for separate points, if one sets
\bea
&&
e = \frac{c}{d-2} \, , \quad c = \frac{1}{S_d^3}\, \frac{d}{2(d-1)}\, ,
\nn \\
&&
e=0 \, ,  \quad \hspace{7mm} c = \frac{1}{S_d^3}\, \frac{d\,2^{\frac{d}{2}}}{2}\, ,
\eea
in the scalar and fermion sector respectively, the values of the other coefficients following from (\ref{ConstraintsTVV}).

The only subtle difference to notice is in the scalar sector, where the $\partial^2_{12}$ and $\partial^2_{31}$ terms,
which are proportional to $\delta_{\mu\nu}$, vanish in the non-coincident point limit and their sum is given by
\bea
&& 
- \frac{c \, \delta^{ab} }{d (d-2)^2} \delta_{\mu\nu} \left(\partial^2_{12} + \partial^2_{31} \right) \left( 
\partial_{\alpha}^{12} + \partial_{\alpha}^{23}\right)  \left( \partial_{\beta}^{31} + \partial_{\beta}^{23}\right)  
\frac{1}{(x_{12}^2)^{d/2-1} (x_{23}^2)^{d/2-1} (x_{31}^2)^{d/2-1}} \nn \\
&& 
= 
\frac{2 c \, \delta^{ab} }{d (d-2)} S_d \delta_{\mu\nu}   \bigg[\partial_{\alpha}^{23}  \left( \partial_{\beta}^{31} + 
\partial_{\beta}^{23}\right) \frac{\delta^d(x_{12}) }{ (x_{23}^2)^{d/2-1} (x_{31}^2)^{d/2-1}} + 
\partial_{\beta}^{23} \left( \partial_{\alpha}^{12} 
+ \partial_{\alpha}^{23}\right)  \frac{ \delta^d(x_{31})}{(x_{12}^2)^{d/2-1} (x_{23}^2)^{d/2-1} } \bigg]\,. 
\label{Ch1top2a}
\eea
They have the topology of $2$-point functions and must be carefully summed to those arising from the bubble diagrams
in order to reproduce exactly the contributions identified as $\hat A - A$  in (\ref{Ch1TVVcoord}).
The bubble contributions are
\bea
\llangle \frac{\delta T_{\mu\nu}(x_1)}{\delta A^{a\,\alpha}(x_2)}  V^b_\beta(x_3) \rrangle_{A= 0}^{s}
&=& 
\frac{c \, \delta^{ab} (d-1)}{d (d-2)^2} S_d \delta^d(x_{12}) 
\left((\partial^{23}_{\mu} + \partial^{31}_{\mu} ) \delta_{\nu\alpha} + 
(\partial^{23}_{\nu} + \partial^{31}_{\nu} ) \delta_{\mu \alpha}  
- \delta_{\mu \nu}  (\partial^{23}_{\alpha} + \partial^{31}_{\alpha}) \right)\times \nn \\
&& 
\times  (\partial^{23}_{\beta} + \partial^{31}_{\beta} ) 
\frac{1}{(x_{31}^2)^{d/2-1} (x_{23}^2)^{d/2-1}} \label{Ch1top2b} \, , \\
\llangle \frac{\delta T_{\mu\nu}(x_1)}{\delta A^{b \,\beta}(x_3)}  V^a_\alpha(x_2) \rrangle_{A= 0}^{s}
&=&
\frac{c \, \delta^{ab} (d-1)}{d (d-2)^2} S_d \delta^d(x_{31}) 
\left((\partial^{23}_{\mu} + \partial^{12}_{\mu} ) \delta_{\nu \beta} + (\partial^{23}_{\nu} + \partial^{12}_{\nu} ) 
\delta_{\mu \beta}  - \delta_{\mu \nu}  (\partial^{23}_{\alpha} + \partial^{12}_{\beta}) \right)\times  
\nn \\
&& 
\times  (\partial^{23}_{\alpha} + \partial^{12}_{\alpha} ) 
\frac{1}{(x_{12}^2)^{d/2-1} (x_{23}^2)^{d/2-1}} \label{Ch1top2c}\, ,
\eea
and we thoroughly checked that the sum of (\ref{Ch1top2a}), (\ref{Ch1top2b}) and (\ref{Ch1top2c}) corresponds
to the solution of the Ward identities (\ref{Ch1TVVcoord}) for $e=c/(d-2)$ and applying (\ref{ConstraintsTVV}). \\
Similarly, the contact terms for the fermion sector in the diagrammatic expansion are found to be, after the inverse mapping,
\bea
\llangle \frac{\delta T_{\mu\nu}(x_1)}{\delta A^{a\,\alpha}(x_2)}  V^b_\beta(x_3) \rrangle_{A= 0}^{f}
&= & 
\frac{c \, \delta^{ab}}{d(d-2)^2} S_d \delta^d(x_{12}) 
\Delta^{(2)}_{\mu \nu \alpha \beta \rho \sigma}  \partial^{\rho}_{31} 
\frac{1}{(x_{31}^2)^{d/2-1}} \partial^{\sigma}_{31} \frac{1}{(x_{31}^2)^{d/2-1}} \, , \\
\llangle \frac{\delta T_{\mu\nu}(x_1)}{\delta A^{b \,\beta}(x_3)}  V^a_\alpha(x_2) \rrangle_{A= 0}^{f}
&=& 
\frac{c \, \delta^{ab}}{d(d-2)^2} S_d \delta^d(x_{31}) 
\Delta^{(2)}_{\mu \nu \beta \alpha \rho \sigma}  \partial^{\rho}_{12} 
\frac{1}{(x_{12}^2)^{d/2-1}} \partial^{\sigma}_{31} \frac{1}{(x_{12}^2)^{d/2-1}}\, ,
\eea
with $\Delta^{(2)}$ defined by
\bea
\Delta^{(2)}_{\mu \nu \alpha \beta \rho \sigma} 
&=& 
  \delta_{\alpha\nu } \delta_{\beta\sigma}\delta_{\mu\rho }
+ \delta_{\alpha\mu } \delta_{\beta\sigma}\delta_{\nu\rho }
+ \delta_{\alpha\nu } \delta_{\beta\rho}  \delta_{\mu\sigma }
+ \delta_{\alpha\mu } \delta_{\beta\rho}  \delta_{\nu\sigma } 
- \delta_{\alpha\nu } \delta_{\beta\mu}   \delta_{\rho\sigma }
- \delta_{\alpha\mu } \delta_{\beta\nu}   \delta_{\rho\sigma } \nn \\
&-& 
2 \, \delta_{\mu\nu} \, \left(\delta_{\alpha\rho}\delta_{\beta\sigma} + \delta_{\alpha\sigma}\delta_{\beta\rho}
- \delta_{\alpha\beta}\delta_{\rho\sigma} \right)\, .
\eea
These two contributions exactly match the $\hat A - A$ terms in (\ref{Ch1TVVcoord}) for $e=0$ and applying (\ref{ConstraintsTVV}),
so that we can conclude that our check for the agreement between the position space solution and the momentum space
perturbative computation is successful also for the $TVV$. \\

To conclude with the $TVV$, we notice that the Green functions discussed so far are unrenormalized.
The issue of renormalization will be addressed separately in section \ref{Counterterms}.


\subsection{The $TTT$ case}\label{Ch1InverseMappingTTT}

Now we are ready to turn to the analysis of the $3$-graviton vertex, whose perturbative computation was presented in chapter
\ref{TTTVertex}.
The general structure of the $TTT$ correlator in position space for separate points is \cite{Osborn:1993cr}
\beq \label{Ch1bareTTT}
\llangle T^{\mu\nu}(x_1) \, T^{\rho\sigma}(x_2) \, T^{\alpha\beta}(x_3) \rrangle =
\frac{1}{(x^2_{12})^{d/2} \, (x^2_{23})^{d/2} \, (x^2_{31})^{d/2}} \,
\mathcal{I}^{\mu\nu,\mu'\nu'}\, \mathcal{I}^{\rho\sigma,\rho'\sigma'} \, t^{\mu'\nu'\rho'\sigma'\alpha\beta}(X_{12}) \, , 
\eeq
\beq \label{Ch1Inversion}
\mathcal{I}^{\mu \nu,\alpha \beta} (s) =
I^{\mu\rho}(s)I^{\nu\sigma}(s) \epsilon_{T}^{\rho\sigma,\alpha\beta} \, ,
\quad s=  x - y \, ,
\eeq
where the tensor
\beq\label{Ch1epislon}
\epsilon_{T}^{\mu\nu,\alpha\beta} =
\frac{1}{2} \, (\delta^{\mu\alpha} \delta^{\nu \beta} + \delta^{\mu \beta} \delta^{\nu \alpha} \bigl)
- \frac{1}{d} \, \delta^{\mu \nu} \delta^{\alpha \beta}
\eeq
is the projector onto the space of symmetric traceless tensors. \\
If we introduce
\bea
h^{4\,\mu\nu\rho\sigma\alpha\beta}(X) 
&=& 
h^{3\,\mu\nu\alpha\rho} \hat{X}^{\sigma}\hat{X}^{\beta} + h^{3\,\mu\nu\alpha\sigma} \hat{X}^{\rho}\hat{X}^{\beta}
+ (\alpha \leftrightarrow \beta) \nn \\
&&
-\, \frac{2}{d}\, \delta^{\rho\sigma}\, h^{2\,\mu\nu\alpha\beta}(\hat{X})
- \frac{2}{d}\, \delta^{\alpha\beta}\, h^{2\,\mu\nu\rho\sigma}(\hat{X}) 
- \frac{8}{d^2}\, \delta^{\rho\sigma}\, \delta^{\alpha\beta}\, h^{1\,\mu\nu}(\hat{X})\, , \nn \\
h^{5\,\mu\nu\rho\sigma\alpha\beta}(\hat{X}) 
&=& 
\bigg[ \left( \delta^{\mu\rho}\delta^{\nu\alpha}\delta^{\rho\beta} + \delta^{\nu\rho}\delta^{\mu\alpha}\delta^{\rho\beta} +
\left( \rho \leftrightarrow \sigma \right) \bigg] + \bigg[ \alpha \leftrightarrow \beta \right] 
\nn \\
&&
-\, \frac{4}{d}\,\delta^{\mu\nu}\,h^{3\,\rho\sigma\alpha\beta} -
\frac{4}{d}\,\delta^{\rho\sigma}\,h^{3\,\mu\nu\alpha\beta} -
\frac{4}{d}\,\delta^{\alpha\beta}\,h^{3\,\mu\nu\rho\sigma} -
\frac{8}{d^2}\, \delta^{\mu\nu}\delta^{\rho\sigma}\delta^{\alpha\beta}\, ,
\eea
then the rank-$6$ tensor $t^{\mu\nu\rho\sigma\alpha\beta}(X)$ is written as
\bea
t^{\mu\nu\rho\sigma\alpha\beta}(X)
&=& a\, h^{5\,\mu\nu\rho\sigma\alpha\beta} + b\, h^{4\,\alpha\beta\mu\nu\rho\sigma}(\hat X) + 
b'\, \bigg(h^{4\, \mu\nu\rho\sigma\alpha\beta}(\hat X) + h^{4\,\rho\sigma\mu\nu\alpha\beta}(\hat X)\bigg)
\nn \\
&&
+\, c\, h^{3\,\mu\nu\rho\sigma}h^{1\,\alpha\beta}(\hat X) + 
c' \, \bigg (h^{3\, \rho\sigma\alpha\beta}h^{1\,\mu\nu}(\hat X) + 
h^{3\, \mu\nu\alpha\beta}h^{1\,\rho\sigma}(\hat X) \bigg) \nn \\
&&
+\, e\, h^{2\,\mu\nu\rho\sigma}(\hat X)h^{1\,\alpha\beta}(\hat X) + e'\,  \bigg(
h^{2\, \rho\sigma\alpha\beta}(\hat X)h^{1\,\mu\nu}(\hat X) + h^{2\, \mu\nu\alpha\beta}(\hat X)h^{1\, \rho\sigma}(\hat X)\bigg)
\nn \\
&& 
+\, f \, h^{1\,\mu\nu}(\hat X)\, h^{1\,\rho\sigma}(\hat X)\, h^{1\,\alpha\beta}(\hat X) \, ,
\eea
with the constraints
\bea
&&
b+b' = -2\,a\, , \quad c' = c\, , \quad e + e' = -4\,b' - 2\,c  \nn \\
&&
d^2\,a + 2\,(b+b') - (d-2)\,b' - d\,c + e' = 0 \, , \nn \\
&&
d\,(d+2)(2\,b'+c)+4\,(e+e') + f = 0\, ,
\eea
leaving only three unconstrained coefficients, say $a$, $b$ and $c$.

The $TTT$ needs regularization in the coincidence limits too, so we introduce the $2$-point function for the energy-momentum tensor,
\beq
\llangle T^{\mu\nu}(x_1)T^{\alpha\beta}(x_2) \rrangle = 
C_T\, \frac{\mathcal{I}^{\mu\nu\alpha\beta}(x_{12})}{(x_{12}^2)^{d/2}}\, .
\label{TTPoistion}
\eeq
Following arguments similar to those for the $TOO$ and $TVV$ vertices, the authors in \cite{Osborn:1993cr} arrive at the 
unrenormalized expression
\bea
\llangle T^{\mu\nu}(x_1) \, T^{\rho\sigma}(x_2) \, T^{\alpha\beta}(x_3) \rrangle 
&=&
\frac{1}{(x^2_{12})^{d/2} \, (x^2_{23})^{d/2} \, (x^2_{31})^{d/2}} \,
\mathcal{I}^{\mu\nu,\mu'\nu'}\, \mathcal{I}^{\rho\sigma,\rho'\sigma'} \, t^{\mu'\nu'\rho'\sigma'\alpha\beta}(X_{12}) \nn \\
&& \hspace{-5mm}
+\, \frac{C_T}{2}\, \bigg[ 
\hat{A}^{\mu\nu\rho\sigma\gamma\delta}(x_{12}) - A^{\mu\nu\rho\sigma\gamma\delta}(x_{12}) \bigg]\, 
\bigg( \frac{\mathcal{I}^{\gamma\delta\alpha\beta}(x_{23})}{(x^2_{23})^{d/2}} + 
\frac{\mathcal{I}^{\gamma\delta\alpha\beta}(x_{13})}{(x^2_{13})^{d/2}} \bigg) \nn \\
&& \hspace{-5mm}
+\, \frac{C_T}{2}\, \bigg[ 
\hat{A}^{\mu\nu\alpha\beta\gamma\delta}(x_{13}) - A^{\mu\nu\alpha\beta\gamma\delta}(x_{13}) \bigg]\, 
\bigg( \frac{\mathcal{I}^{\gamma\delta\rho\sigma}(x_{32})}{(x^2_{32})^{d/2}} + 
\frac{\mathcal{I}^{\gamma\delta\rho\sigma}(x_{12})}{(x^2_{12})^{d/2}} \bigg) \nn \\
&& \hspace{-5mm}
+\, \frac{C_T}{2}\, \bigg[ 
\hat{A}^{\rho\sigma\alpha\beta\gamma\delta}(x_{23}) - A^{\rho\sigma\alpha\beta\gamma\delta}(x_{23}) \bigg]\, 
\bigg( \frac{\mathcal{I}^{\gamma\delta\mu\nu}(x_{31})}{(x^2_{31})^{d/2}} + 
\frac{\mathcal{I}^{\gamma\delta\mu\nu}(x_{21})}{(x^2_{21})^{d/2}} \bigg)\, ,
\label{HatATTT}
\eea
where the $A$ tensor is found to be, by computing the short distance limit of (\ref{Ch1bareTTT}),
\beq
A^{\mu\nu\rho\sigma\alpha\beta}(s) = \frac{1}{(s^2)^{d/2}}\, t^{\mu\nu\rho\sigma\alpha\beta}(s)\, ,
\label{UnregATTT}
\eeq
whereas its regularized counterpart is given by
\bea
\hat{A}^{\mu\nu\rho\sigma\alpha\beta}(s) 
&=& 
\mathcal{D.\,R.}\left[A^{\mu\nu\rho\sigma\alpha\beta}(s)\right] + 
\bigg( C\, h^{5\, \mu\nu\rho\sigma\alpha\beta}(s) + D\, \left( \delta^{\mu\nu}     \, h^{3\,\rho\sigma\alpha\beta} + 
\delta^{\rho\sigma}\, h^{3\,\mu\nu\alpha\beta} \right) \bigg)\, S_d\, \delta^{(d)}(s) \, ,
\eea
where the short-hand notation on the r.h.s. stands for the differentially regularized version of $A$, obtained, just as in the
case of the $TOO$ and the $TVV$ vertices, by re-expressing it as a (lengthy) combination of derivatives of powers of $s^2$
which are lower than $d$ (compare with eq.   (\ref{UnregATTT})) which is no point reporting here explicitly
and can be found in eqs.  (6.37)-(6.38) of \cite{Osborn:1993cr}. \\
Next comes the imposition of the Ward identities: by requiring the general covariance (\ref{Ch1WI3PFcoordinate}) and  the non anomalous
trace (\ref{NoAnTraceTTT}) constraints to be satisfied, it is found that the values of the so far arbitrary coefficients $C$ and $D$ are
\beq
C = \frac{(d-2)(2\,a+b)-d\,c}{d\,(d+2)} \, , \quad D = \frac{C_T}{d} \, .
\eeq
As the general solution of the Ward identities, for any CFT, is parametrized by $3$ independent constants,
we conclude that computing the correlator in the $3$ independent free field theories considered in chapter \ref{TTTVertex}
is enough to account for the complete unrenormalized result in $4$ dimensions, whereas for $d\neq 4$ the spin $1$ sector is not conformally 
invariant and we cannot build the general expression just by superposing the scalar and the fermion sectors.
However, the combination of the scalar and fermion sectors corresponds to a special solution for $d=3$,
where the $TTT$ is parametrized just by $2$ independent constants, whereas in $d=2$ there is just $1$ such constant
\cite{Osborn:1993cr, Erdmenger:1996yc}.


As done before for the $TOO$, $VVV$ and $TVV$ correlators, here we check the result (\ref{Ch1bareTTT}) building explicitly 
the correlator in position space anti-transforming the diagrammatic expansion in free field theory.
This allows to come up with an expression for this vertex which is manifestly integrable.
We will be using the Feynman rules obtained from the Lagrangian descriptions for scalars, fermions and spin $1$
in configuration space, given in section \ref{Ch1lags}.
We start by testing the non-coincident case, for which we can omit the contact terms.
This corresponds only to the diagrams with triangle topology.
We give the expressions in $d$ dimensions for the scalar and the fermion cases, while - as already remarked -
we have to limit our analysis to $d=4$ for the spin-$1$ vector boson. Moreover, in the vector case the gauge-fixing and ghost parts 
of the amplitude have to cancel since the vertex is gauge invariant, as explained in section \ref{Ch1lags}, so that, performing our inverse 
mapping, we include in the interactions vertices for the vector field only the Maxwell contributions.

We have
\bea
\llangle \frac{\delta\mathcal S}{\delta g_{\mu\nu}(x_1)}\,\frac{\delta\mathcal S}{\delta g_{\rho\sigma}(x_2)}
\frac{\delta\mathcal S}{\delta g_{\alpha\beta}(x_3)} \rrangle^{s}
&=&
C^{s}_{TTT} V_{\mathcal{S}\phi\phi}^{\mu\nu}(i\,\partial^{12},-i\,\partial^{31})\,
V_{\mathcal{S}\phi\phi}^{\rho\sigma}(i\,\partial^{23},-i\,\partial^{12})\,
V_{\mathcal{S}\phi\phi}^{\alpha\beta}(i\,\partial^{31},-i\,\partial^{23})
\nonumber\\
&\times&
\frac{1}{(x_{12}^2)^{d/2-1}\,(x_{23}^2)^{d/2-1}\,(x_{31}^2)^{d/2-1}}
\, , \label{Ch1ScalarTriangle}\\
\llangle \frac{\delta\mathcal S}{\delta g_{\mu\nu}(x_1)}\,\frac{\delta\mathcal S}{\delta g_{\rho\sigma}(x_2)}
\frac{\delta\mathcal S}{\delta g_{\alpha\beta}(x_3)} \rrangle^{f}
&=&\nn
\eea
\bea
C^{f}_{TTT}\,(-1)\,\,\bigg( \textrm{Tr}\big[V_{\mathcal{S}\bar\psi \psi}^{\mu\nu}(i\,\partial^{12},-i\,\partial^{31})\,
\,i\,\gamma\cdot\partial^{12}\,V_{\mathcal{S}\bar\psi \psi}^{\rho\sigma}(i\,\partial^{23},-i\,\partial^{12})\,
\,i\,\gamma\cdot\partial^{23}\,V_{\mathcal{S}\bar\psi \psi}^{\alpha\beta}(i\,\partial^{31},-i\,\partial^{23})\,
i\,\gamma\cdot\partial^{31}\big]
&&\,
\nonumber\\
\hspace{-10mm}
+\textrm{Tr}\big[ V_{\mathcal{S}\bar\psi \psi}^{\mu\nu}(i\,\partial^{31},-i\,\partial^{12})\,
i\,\gamma\cdot\,\partial^{31}\,V_{\mathcal{S}\bar\psi \psi}^{\alpha\beta}(i\,\partial^{23},-i\,\partial^{31})\,
i\,\gamma\cdot\,\partial^{12}\,V_{\mathcal{S}\bar\psi \psi}^{\rho\sigma}(i\,\partial^{12},-i\,\partial^{23})\,
i\,\gamma\cdot\partial^{12}\big]\bigg)\,
\nonumber\\
\times
\frac{1}{(x_{12}^2)^{d/2-1}\,(x_{23}^2)^{d/2-1}\,(x_{31}^2)^{d/2-1}} \, , &&
\nonumber \\
\label{Ch1FermionTriangle}
\eea
\bea
\llangle \frac{\delta\mathcal S}{\delta g_{\mu\nu}(x_1)}\,\frac{\delta\mathcal S}{\delta g_{\rho\sigma}(x_2)}
\frac{\delta\mathcal S}{\delta g_{\alpha\beta}(x_3)} \rrangle^V
&=&
C^{V}_{TTT}\,(-1)^3\,
V_{\mathcal{S}AA}^{\mu\nu\gamma\delta}(i\,\partial^{12},-i\,\partial^{31})\,
V_{\mathcal{S}AA}^{\rho\sigma\zeta\xi}(i\,\partial^{23},-i\,\partial^{12})\,
V_{\mathcal{S}AA}^{\alpha\beta\chi\omega}(i\,\partial^{31},-i\,\partial^{23})\,
\nonumber\\
&\times&
\frac{\delta_{\gamma\xi}\,\delta_{\delta\chi}\delta_{\zeta\omega}}{x_{12}^2\,x_{23}^2\,x_{31}^2}
\bigg|_{\frac{1}{\xi}\rightarrow 0}\, .
\label{Ch1VectorTriangle}
\eea
Due to the complexity of the expressions, we have chosen an implicit notation in which the dependences of the vertices on the coordinates 
are obtained by replacing the momenta of the vertices in appendix \ref{Ch1Vertices} with appropriate derivatives with respect to the external
position variables. For instance  
\beq
V^{\mu\nu}_{\mathcal{S}\phi\phi}(p,q)\to V^{\mu\nu}_{\mathcal{S}\phi\phi}(\hat{p},\hat{q})= 
V^{\mu\nu}_{\mathcal{S}\phi\phi}(i\, \partial^{12},- i\, \partial^{23}) \, ,
\eeq
with 
\beq
\hat{p}\to i\, \partial^{12} \qquad\qquad \hat{q}\to - i\, \partial^{23} \, .
\eeq
Explicitly
\beq
V^{\mu\nu}_{\mathcal{S}\phi\phi}(i\, \partial^{12},- i\, \partial^{23}) =
\frac{1}{2}\,(i\,\partial_{12\,\alpha}) \, (- i\,\partial_{23\,\beta}) \, C^{\mu\nu\alpha\beta} +
\chi \bigg( \delta^{\mu\nu} \left( i\,\partial_{12} - i\,\partial_{23} \right)^2 
- \left( i\,\partial_{12}^{\mu} - i\,\partial_{23}^{\mu}\right)\,
\left( i\,\partial_{12}^{\nu} - i\,\partial_{23}^{\nu} \right) \bigg) \, .
\eeq
The replacements of $p,q$ and $l$ by the operator expressions $\hat{p},\hat{q}$ and $\hat{l}$ are specific for each vertex. 
In appendix \ref{Ch1InverseTTT} we provide some more details on this procedure. Notice that we have chosen the coupling parameter 
for the scalar field in $d$ dimensions at the corresponding conformal value $\chi = (d-2)/4(d-1)$.

Expanding the derivatives contained in each vertex, the expression given in (\ref{Ch1bareTTT}) is recovered by setting
\beq
C^{s}_{TTT} = -\frac{8}{S_d^3\,(d-2)^3}\, ,
\quad C^{f}_{TTT} = \frac{2^{d/2+1}}{S_d^3\,(d-2)^3}\, ,
\quad C^{V}_{TTT} = \frac{1}{S_4^3} \, .
\label{OverAllTTT}
\eeq
It was explicitly checked that the results in (\ref{Ch1ScalarTriangle})-(\ref{Ch1VectorTriangle}) with 
overall coefficients (\ref{OverAllTTT}) match the result for separate points presented in (\ref{Ch1bareTTT}),
with $a$, $b$ and $c$ assuming the values corresponding to the respective theories, as listed in \cite{Osborn:1993cr}:
\bea
&&
a = \frac{1}{8\,S_d^3}\, \frac{d^3}{(d-1)^3}\, , \quad 
b = - \frac{1}{8\,S_d^3}\, \frac{d^4}{(d-1)^3}\, , \quad
c = - \frac{1}{8\,S_d^3}\, \frac{d^2\,(d-2)^2}{(d-1)^3}\, , \quad \text{for the scalar}
\nn \\
&&
a = 0 \, , \quad 
b = -\frac{1}{16\,S_d^3}\, d^2\,2^{\frac{d}{2}} \, , \quad
c = -\frac{1}{8\,S_d^3}\, d^2\,2^{\frac{d}{2}} \, , \quad \text{for the fermion}
\nn \\
&&
a = - \frac{16}{S_4^3} \, , \quad 
b = 0 \, , \quad
c = - \frac{64}{S_4^3}\, , \quad \text{for the photon} \, .
\eea
Next we compute the contributions with the topology of $2$-point functions, 
which are needed to account for the behaviour of the vertex in the short distance limits of two coincident points.
In coordinate space we can write them in a manifestly integrable form by pulling out derivatives in the same way as
for the triangle diagram. We replace the momenta with derivatives with respect to the corresponding
coordinates acting on propagators, obtaining very compact expressions for the vertex. 
As already mentioned, more details on this computation can be found in appendix \ref{Ch1InverseTTT}, 
whereas here we just quote the results.

In the scalar case we have
\bea
\llangle \frac{\delta^2\mathcal{S}}{\delta g_{\mu\nu}(x_1)\delta g_{\alpha\beta}(x_3)}\,
\frac{\delta\mathcal{S}}{\delta g_{\rho\sigma}(x_2)} \rrangle^{s}
&=&
\frac{C^{s}_{TTT}}{2}\,
V^{\rho\sigma}_{\mathcal{S}\phi\phi}(i\,\partial^{23},-i\,\partial^{12})\,
V^{\mu\nu\alpha\beta}_{\mathcal{S}\mathcal{S}\phi\phi}(i\,\partial^{12},-i\,\partial^{23},i\,\partial^{23}-i\,\partial^{31})\, \nn\\
&& \times \frac{\delta^{(d)}(x_{31})}{(x^2_{12})^{d/2-1}(x^2_{23})^{d/2-1}} \, ,
\nonumber
\eea
\bea
\llangle \frac{\delta^2\mathcal{S}}{\delta g_{\mu\nu}(x_1)\delta g_{\rho\sigma}(x_2)}\,
\frac{\delta\mathcal{S}}{\delta g_{\alpha\beta}(x_3)} \rrangle^{s}
&=&
\frac{C^{s}_{TTT}}{2}\,
V^{\alpha\beta}_{\mathcal{S}\phi\phi}(i\,\partial^{31},-i\,\partial^{23})\,
V^{\mu\nu\alpha\beta}_{\mathcal{S}\mathcal{S}\phi\phi}(i\,\partial^{23},-i\,\partial^{31},-i\,\partial^{23}+i\,\partial^{12})
\nn\\
&&\times \frac{\delta^{(d)}(x_{12})}{(x^2_{23})^{d/2-1}(x^2_{31})^{d/2-1}} \, ,
\nonumber
\eea
\bea
\llangle \frac{\delta^2\mathcal{S}}{\delta g_{\alpha\beta}(x_3)\delta g_{\rho\sigma}(x_2)}\,
\frac{\delta\mathcal{S}}{\delta g_{\mu\nu}(x_1)} \rrangle^{s}
&=&
\frac{C^{s}_{TTT}}{2}\,
V^{\mu\nu}_{\mathcal{S}\phi\phi}(i\,\partial^{12},-i\,\partial^{31})\,
V^{\alpha\beta\rho\sigma}_{\mathcal{S}\mathcal{S}\phi\phi}
(i\,\partial^{31},-i\,\partial^{12},i\,\partial^{12}-i\,\partial^{23})\, \nn \\
&& \times \frac{\delta^{(d)}(x_{23})}{(x^2_{12})^{d/2-1}(x^2_{31})^{d/2-1}}\, .
\label{Ch1ScalarKBubble}
\eea
Notice that in the three contributions above, the $p,q,$ and $l$ dependence of the vertices correspond to mappings 
onto $\hat{p}, \hat{q}$ and $\hat{l}$ which are specific for each bubble. 
Similarly, in the fermion sector we obtain
\bea
\llangle \frac{\delta^2\mathcal{S}}{\delta g_{\mu\nu}(x_1)\delta g_{\alpha\beta}(x_3)}\,
\frac{\delta\mathcal{S}}{\delta g_{\rho\sigma}(x_2)} \rrangle^{f}
&=&
- C^{f}_{TTT}\,\delta^{(d)}(x_{31})\, \textrm{tr}\,
\big[
V^{\mu\nu\alpha\beta}_{\mathcal{S}\mathcal{S}\bar\psi \psi}(i\,\partial^{12},-i\,\partial^{23})\,
i\,\gamma\cdot\partial^{12}
V^{\rho\sigma}_{\mathcal{S}\bar{\psi }\psi }(i\,\partial^{23},-i\,\partial^{12})\, i\, \gamma\cdot\partial^{23}\big]  \nn \\
&\times&
\frac{1}{(x^2_{23})^{d/2-1}(x^2_{12})^{d/2-1}} \, ,
\eea
and similar expressions for the $k-$ and $p$-bubbles.
Finally, for the spin-1 vector field we have
\bea
\llangle \frac{\delta^2\mathcal{S}}{\delta g_{\mu\nu}(x_1)\delta g_{\alpha\beta}(x_3)}\,
\frac{\delta\mathcal{S}}{\delta g_{\rho\sigma}(x_2)} \rrangle^V
&=&
\frac{C^{V}_{TTT}}{2}\, \,\delta^{(d)}(x_{31})\,
V^{\mu\nu\rho\alpha\beta\chi}_{\mathcal{S}\mathcal{S}AA}(i\,\partial^{12},-i\,\partial^{23})\,
V^{\rho\sigma\tau\omega}_{\mathcal{S}AA}(i\,\partial^{23},-i\,\partial^{12})
\,\frac{\delta_{\zeta\tau}\,\delta_{\chi\omega}}{x^2_{12}\,x^2_{23}}\bigg|_{\frac{1}{\xi}\rightarrow 0} \, , \nn \\
\eea
and similarly for the other bubble-type contributions.
Again, we find that these results are in exact correspondence with the contact terms $\hat A - A$ given in (\ref{HatATTT}),
which completes our check successfully.

The complete structure of the $TTT$ vertex in $4$ dimensions and in position space is thus obtained by combining the 
triangle and the ``k",``p" and ``q"-bubble topologies in the form
\bea
\llangle T^{\mu\nu}(x_1)\,T^{\rho\sigma}(x_2)\,T^{\alpha\beta}(x_3) \rrangle
&=&
\sum_{I=s,f,V} 8 \, C^{I}_{TTT}\, 
\bigg[- \llangle \frac{\delta \mathcal{S}}{\delta g_{\mu\nu}(x_1)}\,
\frac{\delta \mathcal{S}}{\delta g_{\sigma\rho}(x_3)}\,
\frac{\delta \mathcal{S}}{\delta g_{\alpha\beta}(x_2)} \rrangle^I
\nonumber\\
&&\hspace{-45mm}
+ \, \llangle \frac{\delta^2\mathcal{S}}{\delta g_{\mu\nu}(x_1)\,\delta g_{\alpha\beta}(x_3)}\,
\frac{\delta\mathcal{S}}{\delta g_{\rho\sigma}(x_2)}\rrangle^I  +
\llangle \frac{\delta^2\mathcal{S}}{\delta g_{\mu\nu}(x_1)\,\delta g_{\rho\sigma}(x_2)}
\frac{\delta\mathcal{S}}{\delta g_{\alpha\beta}(x_3)}\rrangle^I 
+ \llangle \frac{\delta^2\mathcal{S}}{\delta g_{\alpha\beta}(x_3)\,\delta g_{\rho\sigma}(x_2)}\,
\frac{\delta\mathcal{S}}{\delta g_{\mu\nu}(x_1)}\rrangle^I \bigg] \, .
\nonumber \\
\eea
This expression is in agreement with the form of the unrenormalized energy-momentum tensor
three point function given in \cite{Osborn:1993cr}. The integrability of this result is manifest, due to the $(d/2-1)$ exponent of 
each propagator in position space, which corresponds, generically, to a $1/l^2$ behaviour in momentum space. \\

We are now ready to discuss the renormalization of the correlators discussed so far, elaborating on the meaning of the counterterms
and their relation to the trace anomaly.

\section{Counterterms and their relation to the trace anomaly}
\label{Counterterms}

So far, in comparing the position-space results of \cite{Osborn:1993cr} with perturbative computations in momentum space,
we discarded the issue of renormalization of the divergent correlators, particularly of the $TVV$ and $TTT$ vertices 
(the $VVV$ is finite, whereas the ultraviolet behaviour of the $TOO$ is of no particular interest and 
and will not be considered any longer).
The reason was simply that the inverse mapping procedure we used naturally allows to establish a direct correspondence
between $3$- and $2$-point function topologies and solutions of the conformal constraints at separate points or in the
coincidence limits in which $2$ out of $3$ points are pinched.
On the other hand, divergent contributions, corresponding to poles in $1/\epsilon$ in the dimensional regularization scheme,
are found, in position space, in the limit in which all the three points coincide $x_1 \approx x_2 \approx x_3$, which come
from the high-momentum region in the loop integrals defining our correlators in the perturbative picture
(see, in particular, eq.  (8.13) of \cite{Osborn:1993cr}).
 
In this section we discuss the structure of counterterms for conformal $2$- and $3$-point functions 
and show how one can derive them by imposing Ward identities on the renormalized vertices and telling the divergent 
contributions apart from the finite ones we already treated in the previous section. 
We also comment on the relation between the traces of the counterterms in the analytically continued dimension $d$
and the trace anomaly of the corresponding correlators.

The results of this discussion are complementary with the ones of the previous section and, together, they complete the 
study of the correspondence between position and momentum space results for conformal correlators, which was
the first goal of this chapter.

\subsection{The counterterms for $2$-point functions}

As we are going to see, the interpretation of the anomaly and of its origin, 
in the process of renormalization, can be different, depending on the way the correlator is represented.
In fact, the anomaly can be attributed either to the renormalized amplitude in $4$ dimensions or, alternatively, 
to the specific structure of the counterterm in dimensional regularization, which violates conformal invariance in $d$ dimensions, 
while being traceless for $d=4$.
In the first case the anomaly emerges as a feature of the $d=4$ renormalized amplitude and, 
specifically, of its 4-dimensional trace (in different even dimensions there will be a similar mechanism at work).

We start by illustrating the case of the $TT$, which allows to discuss both the renormalization of a $2$-point function
and the connection of counterterms and trace anomalies.
Together with the discussion of the counterterms for the $VV$, which follows, this part is a warm-up exercise in view of the analysis 
of the counterterms for $3$-point functions that we will discuss afterwards. 

In the $TT$ case conformal symmetry fixes this correlator up to constant. \\
Recalling eq.  (\ref{TTPoistion}), the conformal EMT $2$-point function is given by
\beq\label{Ch12PFOsborn}
\llangle T^{\mu \nu}(x) \, T^{\alpha \beta} (0)\rrangle =
\frac{C_T}{x^{2d}} \, \mathcal{I}^{\mu\nu ,\alpha\beta}(x) \, ,
\eeq
with $\mathcal{I}^{\mu\nu,\alpha\beta}(s)$ defined in (\ref{Ch1Inversion}) and (\ref{Ch1epislon}). \\
In order to move in the framework of differential regularization,  we pull out some derivatives and rewrite our correlator as
\beq
\label{Ch12nd2PFOsborn}
\llangle T^{\mu\nu}(x) \, T^{\alpha\beta}(0) \rrangle
= \frac{C_T}{4\,d\,(d-2)^2\,  (d+1)} \, \hat{\Delta}^{(d)\,\mu\nu\alpha\beta}
\frac{1}{(x^2)^{d - 2}} \, ,
\eeq
where the differential operator $\hat\Delta^{d\,\mu\nu\alpha\beta}$ is defined as
\bea 
\hat{\Delta}^{(d)\,\mu\nu\alpha\beta}
&=&
\frac{1}{2}\left( \hat{\Theta}^{\mu\alpha} \hat{\Theta}^{\nu\beta} +
\hat{ \Theta}^{\mu\beta}\hat{ \Theta}^{\nu\alpha} \right)
- \frac{1}{d-1}\hat{ \Theta^{\mu\nu}} \hat{\Theta}^{\alpha\beta}\, ,
\quad
\text{with}
\quad
\hat{\Theta}^{\mu\nu} = \pd^\mu  \pd^\nu - \delta^{\mu\nu} \, \Box  \, ,
\nn \\
\pd_\mu \, \hat{\Delta}^{(d)\,\mu\nu\alpha\beta}
&=&
0 \, , \quad
\delta^{(d)}_{\mu\nu} \, \hat{\Delta}^{(d)\,\mu\nu\alpha\beta} = 0   \, .
\label{Ch1TransverseDeltaCoord}
\eea
This form of the $TT$ correlator is Fourier-integrable (again, for the meaning of integrability see section \ref{Ch1direct},
in particular eq.  (\ref{Ch1fund})). It is also characterized by a UV divergence in the limit $x\to 0$.
To move to momentum space we can split the ${1}/{(x^2)^{d-2}}$ term
into the product of two ${1}/{(x^2)^{d/2-1}}$ factors and apply straightforwardly the fundamental transform
in eq.  (\ref{Ch1fund}), obtaining
\bea \label{Ch1TransTT}
\llangle T^{\mu\nu} \, T^{\alpha\beta} \rrangle (p)
&\equiv&
\int \, d^d x \, \llangle T^{\mu\nu}(x) T^{\alpha\beta}(0) \rrangle
e^{- i\, p \cdot x}  \nn\\
&=&\frac{C_T}{4\,d\,(d-2)^2\,  (d+1)} \, \int \, d^d x \, e^{- i\, p \cdot x}
\,\hat{ \Delta}^{(d)\,\mu\nu\alpha\beta}\, \frac{1}{(x^2)^{d/2-1}} \, \frac{1}{(x^2)^{d/2-1}} \nonumber \\
&=&
\frac{ (2\pi)^d \,  C(d/2 - 1)^2 \,  C_T}{4\,d\,(d-2)^2\,  (d+1)}  \,
\Delta^{(d)\,\mu\nu\alpha\beta}(p) \, \int \, d^d l  \, \frac{1}{l^2 (l+p)^2} \, ,
\eea
where we use the momentum space counterparts of the operators introduced in (\ref{Ch1TransverseDeltaCoord}),
\bea \label{Ch1TransverseDeltaMom}
\Theta^{\mu\nu}(p)
&=&
\delta^{\mu\nu} \, p^2 - p^\mu \, p^\nu \, , \nn \\
\Delta^{(d)\,\mu\nu\alpha\beta}(p)
&=&
\frac{1}{2} \, \bigg(\Theta^{\mu\alpha}(p) \, \Theta^{\nu\beta}(p) + \Theta^{\mu\beta}(p) \, \Theta^{\nu\alpha}(p) \bigg)
- \frac{1}{d-1} \, \Theta^{\mu\nu}(p) \, \Theta^{\alpha\beta}(p) \,  .
\eea
%
Of course, due to conformal invariance, in $d$ dimensions the $TT$ correlator is anomaly-free,
as apparent from (\ref{Ch1TransTT}),
\beq
\delta_{\mu\nu} \llangle T^{\mu\nu} \, T^{\alpha\beta} \rrangle \, (p) =
\delta_{\alpha\beta}  \llangle T^{\mu\nu} \, T^{\alpha\beta} \rrangle \, (p) = 0 \, .
\eeq
Now, as we move to $d=4$ the correlator in momentum space has a UV singularity, coming from the $2$-point integral, shown in
eqs.  (\ref{BareB0}) and (\ref{RenB0}).
%
%
Then one has to plug (\ref{BareB0}) into (\ref{Ch1TransTT}), \emph{expand all the result},
including the $1/(d-1)$ factor in $\Delta^{(d)\,\mu\nu\alpha\beta}(p)$, around $d=4$ 
and discard the terms that are $O(\epsilon)$, so as to end with the general expression of the bare $2$-point correlator,
already met in eq.  (\ref{BareB0}),
\bea \label{Ch12PFp}
\llangle T^{\mu\nu} \, T^{\alpha\beta} \rrangle_{bare}(p)
&=&
C_1(p)\, \bigg[ \frac{1}{2} \, \bigg( \Theta^{\mu\alpha}(p) \, \Theta^{\nu\beta}(p) + 
\Theta^{\mu\beta}(p) \, \Theta^{\nu\alpha}(p)\bigg)
- \frac{1}{3} \, \Theta^{\mu\nu}(p) \, \Theta^{\alpha\beta}(p)\bigg]
+ \frac{C_2}{3}\,\Theta^{\mu\nu}(p) \, \Theta^{\alpha\beta}(p) \,  \nn\\
&\equiv&
 C_1(p)\, \Delta^{(4)\,\mu\nu\alpha\beta}(p) +
\frac{C_2}{3}\Theta^{\mu\nu}(p) \, \Theta^{\alpha\beta}(p)  \, .
\eea
It must be pointed out that, whereas both the contributions to the unrenormalized correlator in the last line of (\ref{Ch12PFp})
separately respect the energy-momentum conservation Ward identity for the $2$-point function (\ref{Ch1WI2PFMom}), 
only the first one, proportional to $\Delta^{(4)\,\alpha\beta\rho\sigma}(p)$ and carrying the divergence, is traceless in $d=4$,
while tracing the second, finite term we obtain the anomalous relation
\beq
\delta^{(4)}_{\mu\nu} \llangle T^{\mu\nu}T^{\alpha\beta} \rrangle(p) = 
\frac{C_2}{3}\, \delta^{(4)}_{\mu\nu}\, \Theta^{\mu\nu}(p)\,\Theta^{\alpha\beta}(p) = 
C_2\, p^2 \, \Theta^{\alpha\beta}(p) = 2 \, \beta_c\,\big[\Box R \big]^{\alpha\beta}(p) \, .
\eeq
The last equality can be checked directly from eq.  (\ref{Ch1TraceAnomaly}),  by computing the first functional derivative of its r.h.s. 
around flat space, which leaves $ \Box R $ as the only contribution to the $TT$ anomaly.
The superscript $4$ on the Kronecker $\delta$ means that the metric is $4$-dimensional, as usual. \\
The singular contribution in eq.  (\ref{Ch12PFp}) can be eliminated by the ordinary renormalization procedure 
in the $\overline{MS}$ scheme, leaving a result that is finite and whose trace can be taken {\em directly in 4 dimensions}.
The last two equations allow to fix the final structure of the fully renormalized correlator in the form
\beq\label{Ch1Ren2PF1}
\llangle T^{\mu\nu} \, T^{\alpha\beta} \rrangle_{ren}(p) = 
\llangle T^{\mu\nu}\,T^{\alpha\beta}\rrangle_{bare}(p)
+ 6 \, \beta_c\, \frac{\mu^{-\epsilon}}{\bar{\epsilon}} \, \Delta^{(4)\,\mu\nu\alpha\beta}(p)=
\llangle T^{\mu\nu} \, T^{\alpha\beta}\rrangle_{bare}(p)
- 4 \, \beta_a\, \frac{\mu^{-\epsilon}}{\bar{\epsilon}} \, \Delta^{(4)\,\mu\nu\alpha\beta}(p)\, ,
\eeq
where we have used eq.  (\ref{Ch1constraints}) in the last step. \\
So far, the anomaly can be unambiguously attributed to the regularization procedure, not to the counterterm, which is traceless
in the physical dimension where traces are taken. \\

Now we want to explore a second approach to the problem, which is the one exploited in \cite{Osborn:1993cr} and that is particularly
suited to renormalization in position space. \\
To explain how to switch over to this point of view, let us write the renormalized correlator \emph{around the physical dimension} $d=4$, 
but without doing any series expansion. Its form is
\beq
\llangle T^{\mu\nu} \, T^{\alpha\beta} \rrangle(p)
= \frac{C_T}{4\,d\,(d-2)^2\, (d+1)} \, \hat{\Delta}^{(d)\,\mu\nu\alpha\beta}(p)\, \mathcal{B}_0(p^2)
- 4\, \beta_a\, \frac{\mu^{-\epsilon}}{\bar{\epsilon}} \, \Delta^{(4)\,\mu\nu\alpha\beta}(p)  \,  ,
\label{Ch1super}
\eeq
where the counterterm is meant to remove the ultraviolet singularity as $d \rightarrow 4$ and it is implicitly meant
that the values of $C_T$ and $\beta_a$ depend on the field content of the theory. \\
In other words, one keeps everything $d$-dimensional and subtracts from it the $4$-dimensional ultraviolet divergence, 
which does not exist away from $d=4$. \\
As noticed above, the counterterm is traceless for $d=4$ (i.e. contracting the indices with a 4-dimensional metric) 
but not in general dimensions $d$. 
The key observation is that, as the correlator is written in $d$ dimensions, it is natural to compute its trace by contracting it with the 
$d$-dimensional metric. \\
We are free to split the  $\delta^{(d)}_{\mu\nu}$ into a direct sum ($\oplus$)
of a 4-dimensional ($\delta_{\mu\nu}\equiv \delta^{(4)}_{\mu\nu}$) and of a $(d-4)$-dimensional metrics acting 
on the subspaces $E_4$ and $E_{d-4}$ of $d$-dimensional euclidean space $E_d$  
\bea
E_d = E_4 + \oplus E_{d-4} \, , \nn \\
\delta^{(d)}_{\mu\nu} =  \delta^{(4)}_{\mu\nu} + \delta^{(d-4)}_{\mu\nu} \, .
\eea
Then, by taking the trace we obtain
\beq \label{Ch1BrokenTrace}
\delta^{(d)}_{\mu\nu} \Delta^{(4)\mu\nu\alpha\beta}(p) = 
\delta^{(4)}_{\mu\nu}\Delta^{(4)\mu\nu\alpha\beta}(p) + 
\delta^{(d-4)}_{\mu\nu}\Delta^{(4)\mu\nu\alpha\beta}(p) = 
\delta^{(d-4)}_{\mu\nu} \Delta^{(4)\mu\nu\alpha\beta}(p) \, .
\eeq
To arrive at (\ref{Ch1BrokenTrace}) we have used the tracelessness property
\beq
\delta^{(4)}_{\mu\nu} \Delta^{(4)\mu\nu\alpha\beta}(p) =0 \, .
\eeq
Thus, we find that the $d$-dimensional trace of $\Delta^{(4)}$ is $O(\epsilon)$
\beq
\delta^{(d)}_{\mu\nu}\, \hat{\Delta}^{(4)\,\mu\nu\alpha\beta} = 
\frac{\epsilon}{3}\, p^2 \, \Theta^{\alpha\beta}(p) \, .
\eeq
It is then apparent that the trace of the renormalized $TT$ correlator around the physical dimension gives the correct anomaly. 
In particular, the trace operation cancels the ${1}/{\epsilon}$ pole of the counterterm
\beq
\delta^{(d)}_{\mu\nu}\llangle T^{\mu\nu} T^{\alpha\beta} \rrangle(p) =
-4 \, \frac{\beta_a }{\bar{\epsilon}} \, \delta^{(d-4)}_{\mu\nu} \, \Delta^{(4)\,\mu\nu\alpha\beta}(p)
=  2\, \beta_c\, \, p^2\, \Theta^{\alpha\beta}(p) + O(\epsilon)\, ,
\label{Ch1count}
\eeq
which is finite as $\epsilon\to 0$ and reproduces the expected anomaly.
From this point of view, the anomaly can be attributed to the counterterm. \\
Here the $TT$ case is used only as an illustrative example. This procedure is very general and can be applied to any correlator.
In the following, we present simple relations that allow to extend this argument to arbitrary correlators involving
vector currents and/or EMT's on their external lines.
This will complete our discussion of the mapping between momentum and position space solutions of the conformal Ward identities
and open the way to the computations presented in chapters \ref{Traced4T} and \ref{Recursive}, 
for which the relation between counterterms of CFT's and conformal anomalies is of paramount importance. \\

Now we explain the reason why this approach is ideally suited for trace anomalies in position space CFT's. \\
As already pointed out, in position space ultraviolet singularities appear, in $2$ as in $3$-point functions, in the form of $1/\epsilon$ poles,
only for completely coincident points (see, in particular, eq.  (8.13) of \cite{Osborn:1993cr}), i.e. \emph{1-loop divergences are local}.
This allows to write down the solution of the Ward identities as a sum of three pieces. 
The first piece is built on the grounds of conformal invariance constraints for general $d$ dimensions, keeping
all the points separate. It respects naive Ward identities. \\
The second term contains $delta$-functions forcing no more than two points to coincide and is obtained by regularizing
the terms appearing in the short-distance limits $x_1 \rightarrow x_2$ and $x_1 \rightarrow x_3$. 
These are needed to satisfy the (yet unrenormalized) Ward identities.
All this was reviewed in the previous section and, so far, neither ultraviolet singularities nor trace anomalies appear,
as everything is computed in general $d$ dimensions. \\
In the limit in which all the three points coincide, one has to add a counterterm to remove ultraviolet singularities.
This term is thus proportional to $1/\epsilon$ times $\delta$-functions and derivatives thereof enforcing all the three points to coincide.
As all the rest of the correlator is kept $d$-dimensional, if the Green function contains EMT's, 
the trace anomaly cannot descend but \emph{from the trace of the counterterm, taken in $d$ dimensions}. 

For example, let us write this down the renormalized $TT$ in position space,
\beq
\llangle T^{\mu\nu}(x) \, T^{\alpha\beta}(0) \rrangle
= \frac{C_T}{4\,d\,(d-2)^2 \,(d+1)} \, \hat{\Delta}^{(d)\,\mu\nu\alpha\beta}
\frac{1}{x^{2d - 4}} -4\, \beta_a\, \frac{\mu^{-\epsilon}}{\bar{\epsilon}} \, 
\hat{\Delta}^{(4)\,\mu\nu\alpha\beta} \, \delta^d(x-y) \, ,
\label{Ch1SuperPoistion}
\eeq
from which the local structure of the counterterm is manifest. \\

To prepare for the discussion of the  $TVV$ counterterm, we end by briefly recalling the structure of the famous vector $2$-point function,
given in (\ref{VVPoistion}) in position space for separate points and becoming, in momentum space
\beq
\llangle V^a_\alpha \, V^b_\beta \rrangle(p) =
\delta^{ab}\, C_V\, \Theta_{\alpha\beta}(p)\, B_0(p^2) \, ,
\label{VVd}
\eeq
the coefficient $C_V$ depends on the nature of the vector current $V$.
For the cases of the complex scalar and the fermion field of eq.  (\ref{SAndF}), this coefficient is
\beq
C^s_V = - \frac{1}{2^{d+1}\, \pi^{d-2}\,(d-1)}\, , \qquad
C^f_V = - \frac{d-2}{2^{d-1}\, \pi^{d-2}\,(d-1)} \, .
\eeq
eq.  (\ref{VVd}) can be expanded around $d=4$ to give
\beq
\llangle V^a_\alpha \, V^b_\beta \rrangle_{bare}(p) = 
\delta^{ab}\, \Theta_{\alpha\beta} \bigg( c_1 + \frac{c_2}{\bar\epsilon} \bigg)\, ,
\label{ExpandVVd}
\eeq
where the values of $c_1$ and $c_2$ do not matter and which shows that both the bare part of the $VV$ and the counterterms 
that has to be added, being both proportional to $\Theta_{\alpha\beta}(p)$, separately satisfy the gauge invariance constraint, 
\beq
p_{\alpha}\llangle V^a_\alpha \, V^b_\beta \rrangle_{bare}(p) = 0 \, .
\eeq
%

\subsection{Connection between counterterms and trace anomalies}\label{CountAnom}

We review a method to derive the trace anomaly for CFT's. 
The method is originally due to Duncan and Duff \cite{Duff:1977ay,Duncan:1976pv} and allows to generalize the discussion 
of the previous section to arbitrary correlators of EMT's and vector currents (we discard scalar operators). \\
As we have seen above, an argument in position space shows that the trace of the counterterms in $d$ dimensions has to yield 
the trace anomaly of the corresponding correlator.

Let us consider a general euclidean CFT described by a generating functional (\ref{GenPlusSources}), 
depending on the background gauge fields $A^a_\mu$ and metric $g_{\mu\nu}$. 
For some values of the space dimension, for instance $d=4$, the generating functional is affected by ultraviolet singularities.
In the framework of dimensional regularization, $1$-loop divergences are parametrized by a pole in $\epsilon = 4-d$.
For free field theories there are no higher order contributions and the generating functional is completely determined by 
the sum of the bare part (\ref{GenPlusSources}) plus the $1$-loop counterterms.
These are strongly constrained by the requirement of Weyl invariance for $d=4$, which implies that the dimensionally continued
generating functional must consist of a combination of contributions,  say $\mathcal{C}$,  enjoying the following properties:
\begin{itemize}
\item they must depend only on the background gauge fields $A^a_\mu$ and metric $g_{\mu\nu}$; \\
\item they must be invariant under gauge and general coordinate transformations ; \\
\item as there are no dimensionful constants in the bare theory, their mass dimension must be $4$ ; \\ 
\item they must be Weyl invariant in 4 dimensions, i.e.
\beq
\lim_{d \rightarrow 4} \bigg\{ \frac{2}{\sqrt{g}}\, g_{\mu\nu} \frac{\delta\mathcal{C} }{\delta g_{\mu\nu}}\bigg\}  = 0 \, .
\eeq
\end{itemize}
The set of these terms is well known in $4$ dimensions, where the only one depending on the gauge fields is the squared field-strength,
$F^{a\,\mu\nu}\,F^a_{\mu\nu}$, whereas the other two possible contributions were studied in chapter \ref{TTTVertex}
and are the Weyl tensor squared $F$ and the Euler density in $4$ dimensions, $G$. \\

We conclude that the renormalized generating functional for a CFT is written, in the $\overline{MS}$ scheme, in the form
\beq
\mathcal{W}_{ren}[g,A] \equiv \mathcal{W}[g,A] - \frac{\mu^{-\epsilon}}{\bar\epsilon}\mathcal{W}_{Ct}[g,A]  = 
\mathcal{W}[g,A]  - \frac{\mu^{-\epsilon}}{\bar\epsilon}\,
\int d^dx\, \sqrt{g}\, \bigg( c_F \, F + c_g\, G + c_A\, F^{a\,\mu\nu}\,F^a_{\mu\nu}  \bigg) \, .
\label{RenW}
\eeq
From this renormalized generating functional we can derive the vacuum expectation value of the EMT, using its definition 
(\ref{Ch1VEVEMT}), and thus we can infer its trace. As (\ref{RenW}) is written in dimensional regularization,
it is clear that the trace must be taken in $d$ dimensions.
Moreover, the bare action always enjoys conformal invariance in the limit $d\rightarrow 4$
(for scalars and fermions this is true already in $d$ dimensions) as we have seen in the example of the EMT $2$-point function,
\beq
\lim_{d\rightarrow 4} \bigg[ \frac{2}{\sqrt{g}}\, g_{\mu\nu}\, \frac{\delta \mathcal W[g,A] }{\delta g_{\mu\nu}}\bigg]  = 0 \, ,
\eeq
so it does not contribute to the trace anomaly. \\
We have to compute the contribution of the counterterm, taking the limit $d\rightarrow 4$ after tracing, i.e.
\beq
g_{\mu\nu} \llangle T^{\mu\nu} \rrangle_s \equiv 
\lim_{d\rightarrow 4} \bigg\{ g_{\mu\nu}\, \frac{2}{\sqrt{g}}\, g_{\mu\nu}\, \frac{\delta}{\delta g_{\mu\nu}} \,
 \bigg[ -\frac{\mu^{-\epsilon}}{\bar\epsilon}\, \int d^dx\, \sqrt{-g}\, 
\bigg( c_F\, F + c_G\, G + c_A\, F^{a\,\mu\nu}\,F^a_{\mu\nu}  \bigg) \bigg] \bigg\} \, .
\eeq
It is easy to show, following the procedure illustrated in appendix \ref{Ch1FunctionalIntegral}, that the following relations hold
\bea
&&
\frac{2}{\sqrt{g}}\, g_{\mu\nu}\, \frac{\delta}{\delta g_{\mu\nu}}\, \int d^d x\,\sqrt{g}\, F  =
- \epsilon \, \left(F - \frac{2}{3}\, \Box R\right)\, ,  \nn \\
&&
\frac{2}{\sqrt{g}}\, g_{\mu\nu}\, \frac{\delta}{\delta g_{\mu\nu}}\, \int d^d x\, \sqrt{g}\, G = 
- \epsilon \, G \, , \nn \\
&&
\frac{2}{\sqrt{g}}\, g_{\mu\nu}\, \frac{\delta}{\delta g_{\mu\nu}}\, \int d^d x\, \sqrt{g}\, F^{a\,\mu\nu}\,F^a_{\mu\nu} = 
- \epsilon \, F^{a\,\mu\nu}\, F^a_{\mu\nu} \, .
\label{CTVariations}
\eea
We see that, in this approach, the trace anomaly is intimately connected to the counterterms, as the $O(\epsilon)$ contributions
in (\ref{CTVariations}) cancel the $\epsilon$-pole  in the $1$-loop counterterm, yielding the finite result (\ref{Ch1TraceAnomaly}) 
if one sets the values of the coefficients
\beq
c_F = \beta_a \, , \qquad c_G = \beta_b \, , \qquad c_A = -\frac{\kappa}{4} \, .
\eeq
Thus, we conclude that an equivalent form of (\ref{Ch1TraceAnomaly}) is
\beq
\frac{2}{\sqrt{g}}\, g_{\mu\nu}\, \frac{\delta \mathcal{W}_{Ct}[g,A]}{\delta g_{\mu\nu}} = 
- \epsilon\, \mathcal{A}[g,A] \, ,
\label{dTraceCt}
\eeq
where the trace is meant to be $d$-dimensional, due to the presence of the counterterm $\mathcal{W}_{Ct}$,
computed in dimensional regularization. Of course we are neglecting $O(\epsilon^2)$ terms due to the difference between
$\epsilon$ and $\bar\epsilon$. This is the master equation  generating the trace Ward identities satisfied by the counterterms
in $d$ dimensions and extensively used in \cite{Osborn:1993cr}. We have checked that the counterterms for the $TVV$ 
and $TTT$ correlators studied in this chapter satisfy such identities, as we are going do discuss below. \\

Finally, we still have to remark that, for interacting CFT's, as Yang-Mills gauge bosons in $d=4$, divergences that are higher order 
than $1/\epsilon$ may exist, in general, so that the renormalized generating functional might be of the form
\bea
\mathcal{W}_{ren}[g,A] =  
\mathcal{W}[g,A] - \frac{\mu^{-\epsilon}}{\epsilon}\, \mathcal{W}^{(1)}_{Ct}[g,A]
- \frac{\mu^{-\epsilon}}{\epsilon^2}\, \mathcal{W}^{(2)}_{Ct}[g,A] + \dots  \, ,
\eea
where the superscripts stand for the order of the divergence in $1/\epsilon$.
Now, as the vev of the renormalized EMT is finite, so has to be its trace, then the condition
\beq
g_{\mu\nu}\, \frac{\delta \mathcal{W}^{(n)}_{Ct}[g,A]}{\delta g_{\mu\nu}} = O(\epsilon^n)\, , \quad n=1,2,3,\dots
\eeq
should hold.
To our knowledge, there are no conformal invariants depending only on the metric whose Weyl variation vanishes faster than linearly for $d
\rightarrow 4$ \cite{Mazur:2001aa}. Moreover, concerning the gauge sector of the trace anomaly, the same form as in
(\ref{Ch1TraceAnomaly}) of the gauge field contribution to the trace anomaly was derived in \cite{Adler:1976zt} for QED, 
whereas in \cite{Collins:1976yq} it was shown that it also holds for non abelian gauge theories after proper resummations 
are performed before taking the limit $d\rightarrow 4$. These results imply that the trace anomaly is completely determined 
by the $1$-loop contributions to the counterterm and is thus given by (\ref{Ch1TraceAnomaly}).

\subsection{The counterterm for the $TVV$}

We now turn to the issue of the renormalization of the $TVV$ in $4$ dimensions, which will complete the test of the correspondence
between the position space solution of \cite{Osborn:1993cr} and diagrammatic momentum space computations in dimensional
regularization. \\
The renormalized $3$-point function has to satisfy the requirement  of general covariance (\ref{TVVWardCoord}) 
as well as the anomalous Ward identity (\ref{AnTraceWardTVV}).
As explained at the end of section \ref{Ch1lags}, the implications of general covariance for the counterterms immediately descend
from the Ward identities for the corresponding Green functions. 
Specifically, we write the renormalized $VV$ and $TVV$ correlator as
\bea
\langle V^a_{\alpha}V^b_{\beta} \rangle_{ren}(p) 
&=&
\langle V^a_{\alpha}V^b_{\beta} \rangle_{bare}(p)
+ \frac{\mu^{-\epsilon}}{\bar\epsilon}\,\frac{\kappa}{4}\, D_{\alpha\beta}(p) \, ,
\nn \\
\langle T_{\mu\nu}V^a_{\alpha}V^b_{\beta} \rangle_{ren}(p,q) 
&=&
\langle T_{\mu\nu}V^a_{\alpha}V^b_{\beta} \rangle_{bare}(p,q) +
\frac{\mu^{-\epsilon}}{\bar\epsilon}\,\frac{\kappa}{4}\, D_{\mu\nu\alpha\beta}(p,q)\, , 
\label{RenTVVAndVV}
\eea
where the counterterms are derived according to the results of the previous section,
applying the definitions  of the $TVV$ and the $VV$ vertices to the renormalized generating functional (\ref{RenW}).
Their expressions in position space are found to be
\bea
D_{\alpha\beta}(x_1,x_2) 
&=&  
\frac{\delta^2 F^{c\,\gamma\delta}F^c_{\gamma\delta}}{\delta A^{a\,\alpha}(x_2)\delta A^{b\,\beta}(x_3)}\bigg|_{A=0} \, ,
\nn \\
D_{\mu\nu\alpha\beta}(x_1,x_2,x_3)
&=&
\frac{\delta^2}{\delta A^{a\,\alpha}(x_2) \delta A^{b\,\beta}(x_3)} \bigg[ \frac{2}{\sqrt{g_{x_1}}}
\frac{\delta F^{c\,\gamma\delta}F^c_{\gamma\delta}}{\delta g_{\mu\nu}(x_1)}\bigg]_{g=\delta} \bigg|_{A=0} \, ,
\label{PoistionTVVCt}
\eea
and we denote with $D_{\alpha\beta}(p)$, $D_{\mu\nu\alpha\beta}(p,q)$ their Fourier transforms. \\
We find that (\ref{PoistionTVVCt}) match the expressions presented in \cite{Osborn:1993cr} and satisfy the
constraints that are obtained by plugging (\ref{RenTVVAndVV}) into the general covariance and gauge invariance Ward identities, 
(\ref{TVVWardCoord}) and (\ref{GaugeInvTVV}), then passing to momentum space and isolating the divergent parts, i.e.
\bea    \label{Ch1DivWardTVV}
(p + q)^\mu \,  D_{\mu\nu\alpha\beta} (p,q)
&=&
q_\nu \, D_{\alpha\beta}(p) - \delta_{\nu\beta} \, q^\mu \, D_{\mu\alpha}(p) +
p_\nu \,  D_{\alpha\beta}(q) - \delta_{\nu\alpha} \,  p^\mu \,  D_{\mu\beta}(q)  \, ,  
\nn \\
p^\alpha \, D_{\mu\nu\alpha\beta}(p,q)
&=&
q^\beta \, D_{\mu\nu\alpha\beta}(p,q)  = 0 \, .
\eea
We provide the explicit form of the counterterms in momentum space,
\bea
D_{\alpha\beta}(p) 
&=& 
\Theta_{\alpha\beta}(p)\, , \nn \\
D_{\mu\nu\alpha\beta}(p,q)
&=&
\delta_{\alpha\beta} \,  ( p_\mu \, q_\nu + q_\mu \, p_\nu )
- (\delta_{\beta\nu} \, p_\mu + \delta_{\beta\mu} \,  p_\nu ) \,  q_\alpha -
( \delta_{\mu\alpha} \, q_\nu + \delta_{\alpha\nu} \, q_\mu ) \, p_\beta
 \nonumber \\
&&
+\, p \cdot q \,  (\delta_{\mu\beta}\, \delta_{\nu\alpha}  + \delta_{\mu\alpha} \, \delta_{\nu\beta} ) 
- \delta_{\mu\nu} \, ( p \cdot q \, \delta_{\alpha\beta} -   q_\alpha \,  p_\beta )  \, .
\eea
All that is left to check and is easily done is that the counterterm for the $TVV$ satisfies the $d$-dimensional trace relation discussed  
in \ref{CountAnom} and encoded in eq.  (\ref{dTraceCt}), which is
\beq
\delta^{(d)\, \mu\nu} \, D_{\mu\nu\alpha\beta}(p,q) =
- \epsilon  \, ( q_\alpha \, p_\beta - p \cdot q \, \delta_{\alpha\beta} ) \, ,
\eeq
reproducing the correct anomaly. \\

We know that the identification of the divergent parts of the $TVV$ correlator can be performed diagrammatically.
We just mention that the general form of the $TVV$ amplitude can be expanded in a basis of $13$ tensor structures 
defined in \cite{Giannotti:2008cv}.
A complete perturbative analysis shows that there is only $1$ tensor structure which is affected by the renormalization
procedure, which coincides with the $D_{\mu\nu\alpha\beta}(p,q)$ counterterm introduced above.
As found by direct computations in \cite{Giannotti:2008cv,Armillis:2009pq} for QED, in \cite{, Armillis:2010qk} for QCD 
and in \cite{Coriano:2011zk} for the electroweak sector, renormalization of the $TVV$ vertex always affects only this tensor structure.
Given the complexity of the computations and the wide difference between the general CFT 
approach and the ordinary diagrammatic one, this agreement is obviously non trivial. 

\subsection{The counterterms for the $TTT$}

For the case of the $TTT$, the discussion of the derivation of the counterterms was already done in chapter \ref{TTTVertex},
so here we do not report it. The definitions of the counterterms for the EMT $2$- and $3$-point functions (\ref{Ch12PFCounterterm}), 
(\ref{Ch1DF})  and (\ref{Ch1DG}) are identical, so that it is no point repeating the discussion. 

Nevertheless, it is instructive to see how one can derive the analogue of the $TTT$ counterterm (\ref{Ch1Ren3PF})
by using the Ward identities to constrain the scalar coefficients and knowing the counterterm just for the $2$-point function. 
In this case we are bound to introduce the generic counterterms to the $TTT$ vertex
\beq \label{Ch1Ren3PFansatz}
\llangle T^{\mu\nu}T^{\rho\sigma}T^{\alpha\beta} \rrangle_{ren}(p,q) =
\llangle T^{\mu\nu}T^{\rho\sigma}T^{\alpha\beta} \rrangle_{bare}(p,q)
+ \frac{1}{\bar\epsilon} \,
\bigg(C_F \, D_F^{\mu\nu\alpha\beta\rho\sigma}(p,q) + C_G \, D_G^{\mu\nu\alpha\beta\rho\sigma}(p,q)\bigg), \,
\eeq
written in terms of arbitrary coefficients $C_F$ and $C_G$ that one cannot know \emph{a priori}.
With the addition of the counterterms, it is clear that the renormalized vertex must satisfy (\ref{Ch1WI3PFcoordinate}) 
and two similar identities which follow by exchanging indices and momenta properly. In fact, 
renormalization has to preserve general covariance.
One can check that $D_G^{\mu\nu\alpha\beta\rho\sigma}(p,q)$ is transverse, as (\ref{Ch1ExplicitDG}) shows clearly,
\beq \label{Ch1DGConstraints1}
k_{\nu}D_G^{\mu\nu\alpha\beta\rho\sigma}(p,q) = 0 \, , \quad
p_{\alpha}D_G^{\mu\nu\alpha\beta\rho\sigma}(p,q) = 0 \, \quad
q_\sigma D_G^{\mu\nu\alpha\beta\rho\sigma}(p,q) =0\, ,
\eeq
so that, by inserting the expressions (\ref{Ch1Ren2PF2}) and (\ref{Ch1Ren3PFansatz}) into these Ward identities and taking
(\ref{Ch1DGConstraints1}) into account, one obtains three conditions on the F-contribution to the counterterm, the first being
\bea \label{Ch1DFConstraints1}
&&
C_F \, k_\nu D_F^{\mu\nu\alpha\beta\rho\sigma}(p,q) =
- \beta_a \bigg\{q^\mu D_F^{\rho\sigma\alpha\beta}(p) + p^\mu D_F^{\alpha\beta\rho\sigma}(q) \nn\\
&&
-  q_\nu \bigg[\delta^{\mu\rho}D_F^{\nu\sigma\alpha\beta}(p)
+ \delta^{\mu\sigma}D_F^{\nu\rho\alpha\beta}(p)\bigg]
-  p_\nu \bigg[\delta^{\mu\alpha}D_F^{\nu\beta\rho\sigma}(q)
+ \delta^{\mu\beta}D_F^{\nu\alpha\rho\sigma}(q)\bigg]\bigg\}\, .
\eea
and the other two coming from a permutation of the indices and of the momenta.
In (\ref{Ch1DFConstraints1}) we have used the expression (\ref{Ch1Ren2PF2}) for the renormalized $2$-point function.
These three identities are easily seen to be satisfied if $C_F = - \beta_a$, as one can check with a symbolic calculus program. \\
Once the first coefficient is fixed, the same argument can be applied to the three anomalous trace identities in $d = 4 - \epsilon$
dimensions, which can be exploited in order to fix $C_G$. 
These identities descend from the double functional derivative of (\ref{dTraceCt}) with respect to other $2$ metric tensors
and are
\bea\label{Ch1CTTraces}
\delta_{\mu\nu}D_F^{\mu\nu\alpha\beta\rho\sigma}(p,q)
&=&
-4 \,\epsilon \,\bigg( \big[F\big]^{\alpha\beta\rho\sigma}(p,q)
- \frac{2}{3} \big[\sqrt{g}\Box\,R\big]^{\alpha\beta\rho\sigma}(p,q) \bigg)
-  2 \, \beta_a\, \bigg( D_F^{\alpha\beta\rho\sigma}(p) + D_F^{\alpha\beta\rho\sigma}(q)\bigg) \, , \nn\\
\label{Ch1DFConstraints2}\,
\delta_{\mu\nu}D_G^{\mu\nu\alpha\beta\rho\sigma}(p,q)
&=&
- 4 \, \epsilon \,\big[G\big]^{\alpha\beta\rho\sigma}(p,q) \, .
\label{Ch1DGConstraints2}
\eea
According to the previously established notation, $\big[F\big]^{\alpha\beta\rho\sigma}(p,q)$ and
$\big[G\big]^{\alpha\beta\rho\sigma}(p,q)$ are the Fourier-transformed second functional derivatives
of the squared Weyl tensor and the Euler density respectively.
Requiring (\ref{Ch1munu3PFanomaly}) to be satisfied by the renormalized 2 and 3-point correlators we get
\bea \label{Ch1Trace}
\delta_{\mu\nu} \, \bigg( - \beta_a \,\ D_F^{\mu\nu\alpha\beta\rho\sigma}(p,q)
+ C_G \, D_G^{\mu\nu\alpha\beta\rho\sigma}(p,q) \bigg)
&=&
4 \, \epsilon \,
\bigg[ \beta_a \, \bigg(\left[ F \right]^{\alpha\beta\rho\sigma}(p,q)
           - \frac{2}{3}\left[ \sqrt{g}\Box\,R \right]^{\alpha\beta\rho\sigma}(p,q) \bigg)
\nn \\
&+&
\beta_b \, \left[ G \right]^{\alpha\beta\rho\sigma}(p,q) \bigg]
- 2 \, \beta_a\, \bigg(D_F^{\alpha\beta\rho\sigma}(p) + D_F^{\alpha\beta\rho\sigma}(q)\bigg)\, ,\nn\\
\eea
and other two similar equations, obtained by shuffling indices and momenta as for the general covariance Ward identities.
Solving these conditions allows one to obtain the relation $C_G = -{\beta_b}$, as expected.

\section{Handling massless correlators: a direct approach for general dimensions} \label{Ch1direct}

In the previous chapters we have tried to compare perturbative results in free field theory with general ones coming from imposing
conformal symmetry requirements on certain correlators. We have also seen that in this case one can work backward from
the explicit free field theory representation of these correlators in momentum space and match them with the general solutions 
given by the conformal constraints in position space.
This is the case of the $TOO$, $VVV$ and $TVV$ correlators in general dimensions, while for the $TTT$
the $4$-dimensional solution of the Ward identities is completely matched by a combination of scalar, vector and fermion sectors.
As we consider the same $3$-graviton vertex in $d$ dimensions, the vector contribution is not conformally invariant, 
and therefore the combination of the scalar and the fermion sectors does not match the most general $d$-dimensional solution. 
This raises the issue if there is, in general, a free field theory that can reproduce a given CFT correlator, and there is no simple answer. 
The goal of CFT, in fact, is to bootstrap certain correlation functions bypassing, if necessary, a Lagrangian formulation.

In fact, one of the main features of the standard CFT approach in the identification of the correlators is to work in position space 
with no reference to a Lagrangian. The finiteness of the Fourier transform is a necessary requirement in order to proceed with the identification, 
if this turns out to exist, of the corresponding Lagrangian field theory, since this could always be defined in momentum space.

Checking the finiteness (in  momentum space) of a general solution given in position space is not an obvious step, 
since a correlator in position space such as the $TTT$ contains several hundreds of terms, 
most of them characterized by a divergent Fourier expression in momentum space.
For this reason here we are going to illustrate a very general algorithm that allows to transform correlators of such a complexity using
a direct approach. Our analysis will be formulated in general but illustrated with few examples up to correlators of rank-$4$.
For obvious reasons, we will be choosing, as a test of our approach, some of the Green functions defined in the previous sections.
These, in fact, as we have seen, can be deduced from a Lagrangian formulation and therefore their expressions in momentum space
are well defined. \\ 
Obviously, we need some intermediate regularization of the integrands (in position space) of these correlators in order
to proceed with the definition of the Fourier transform of each individual term.
This is obtained by introducing a power-like regulator $(\omega)$ which is the analogous of the 
$\epsilon$ regulator of ordinary dimensional regularization but otherwise completely unrelated to it. 

The algorithm implements a sequence of integration by parts before proceeding with the identification of the 
$\omega$-regulated transforms. As a consistency condition, the correlators that we investigate have finite Fourier expressions, 
as expected, and we check the direct cancellation of all their Fourier singularities, which appear as poles (double and single) in $1/\omega$.

The finite products of the procedure, which correspond to the Fourier space integrands, manifest specific logarithmic terms.
These, in general, are a new feature of the momentum space form of a given CFT correlator. 
They are expected to appear once we rewrite any CFT correlation function from position to momentum space. 
In some cases, these log terms can be rewritten as ordinary (non-logarithmic) integrals, 
while in other cases this may not be possible, and we can think of the log-integrals, in all these second cases, 
as of new irreducible contributions. 

In the correlators that we explicitly investigate, obviously, we know beforehand that they have to be matched by free field theories. 
In this case, a brute force application of the algorithm would produce log-integrals which are, therefore, reducible to ordinary 
(non logarithmic) ones. When the $\omega$ singularities cancel, which indicates that it is possible to recollect the terms in position space 
(and using integration by parts) in such a way that the Fourier expression is manifestly finite, the logarithmic terms are absent.  
The use of the previous (Fourier-integrable) vertices allows to test this approach and show its consistency. \\

Before proceeding with an explicit discussion of the method, we list here, for definiteness, the steps that have to be followed in
order to transform the expression of any given CFT correlator in position space to momentum space:
\begin{enumerate}
\item expansion of the correlator into its single tensor components; \\
\item rewriting of each component in terms of some ``R-substitutions", that we will define below; \\
\item application of the dimensional shift $d\to d- 2 \omega$ which can be performed generically in the expression resulting from point $2$ ;  \\
\item implementation of the transform. The transform is implemented by eq.  (\ref{Ch1fund}) for each single difference $x_{ij}= x_i - x_j$.
\end{enumerate}
As we are going to describe below, this method and the regularization imposed by the dimensional shift allows to test quite
straightforwardly the integrability of any correlator,  a point already emphasized in \cite{Erdmenger:1996yc} 
where this regularization has been first introduced. 
The transform can be applied in several independent ways. These features share some similarities with the so called 
``method of uniqueness" (see for instance \cite{Kazakov:1986mu}) used for massless integrals in momentum or in configuration space.

\subsection{Pulling out derivatives}\label{Ch1pull}

One of the main steps that we will follow in the computation of the transform of the position space expression of the correlators
consists in rewriting a given position space tensor in terms of derivatives of other terms. We call this rule a ``derivative
relation." It allows one to reduce the degree of singularity of a given tensor structure, when the variables are coincident, in the
spirit of differential regularization. Differently from the standard approach of differential regularization, which is $4$-dimensional, 
we will be working in $d$ dimensions. We will be using the term ``integrable" to refer to expressions
for which the Fourier transform exists and that are well defined in $d$ dimensions, although they may be singular for $d=4$.
Derivative relations, combined with the basic transform
\bea \label{Ch1fund}
\frac{1}{(x^2)^\alpha}
&=& 
\frac{1}{4^\alpha \pi^{d/2}}\, \frac{\Gamma(d/2 - \alpha)}{\Gamma(\alpha)}\,
\int d^d l \, \frac{e^{il\cdot x}}{(l^2)^{d/2 - \alpha}}
\equiv C(\alpha) \, \int d^d l \, \frac{e^{i l\cdot x}}{(l^2)^{d/2 - \alpha}} \nn \\
C(\alpha) 
&=&
\frac{1}{4^{\alpha}\,\pi^{d/2}} \frac{\Gamma(d/2 - \alpha)}{\Gamma(\alpha)}
\eea
allow one to perform a direct mapping of these correlators to momentum space.
We proceed with a few examples to show how the lowering of the singularity takes place.

We start from tensors of rank $1$.
We use the relation
\beq
\frac{x^\mu}{(x^2)^\alpha} =
-\frac{1}{2 (\alpha -1)} \partial_\mu \frac{1}{(x^2)^{\alpha -1}} =
- \frac{i}{2^{2\alpha-1} \pi^{d/2}} \, \frac{\Gamma(d/2 +1 - \alpha) }{\Gamma(\alpha) } \, 
\int d^d l \, e^{i l\cdot x}\frac{l_\mu}{(l^2)^{d/2 - \alpha +1}} 
\eeq
to extract the derivative, where in the last step we have used (\ref{Ch1fund}).
Notice that by using (\ref{Ch1fund}) with $\alpha=d/2-1$ one can immediately obtain the equation
\beq
\Box \frac{1}{(x^2)^{d/2-1}}= - \frac{4\,\pi^{d/2 }}{\Gamma(d/2-1)} \, \delta^{(d)}(x)\, ,
\label{Ch1deltaeq}
\eeq
which otherwise needs Gauss' theorem to be derived.

Scalar $2$-point functions describing loops in position space are next in difficulty.
As an illustration, consider the generalized $2$-point function
\beq
\frac{1}{[(x-y)^2]^\alpha [(x-y)^2]^\beta}\, .
\eeq
Using (\ref{Ch1fund}) separately for the $1/[(x-y)^2]^\alpha$ and the $1/[(x-y)^2]^\beta$ factors,
the Fourier transform ($\mathcal{F T}$) of this expression is found to be
\bea\label{Ch1FourierB0General}
\mathcal{FT} \left[ \frac{1}{[(x-y)^2]^\alpha [(x-y)^2]^\beta} \right]
&\equiv&
\int\, d^d x \, d^d y \, \frac{e^{- i ( p\cdot x + q \cdot y )}}{[(x-y)^2]^\alpha [(x-y)^2]^\beta}
\nonumber \\
&=&
(2\pi)^{2d} \, C(\alpha) \, C(\beta)\, \int d^d l \, \frac{1}{[l^2]^\alpha [( l+p)^2 ]^\beta}\, .
\eea
The requirement of uniqueness for the transform allows to reformulate it by combining 
the powers of the propagators into a single factor,
\beq
\mathcal{FT}\left[\frac{1}{[(x-y)^2]^{\alpha + \beta}}\right] =
(2\pi)^{2d} \, \frac{C(\alpha + \beta)}{(p^2)^{d/2 - \alpha - \beta}}\, ,
\label{Ch1unsplit}
\eeq
giving, for consistency, a functional relation for the integral in (\ref{Ch1FourierB0General})
\bea \label{Ch1B0general}
\int\,d^dl \, \frac{1}{[l^2]^\alpha[(l+p)^2]^\beta} =
\frac{C(\alpha + \beta)}{C(\alpha) \, C(\beta)} \frac{1}{(p^2)^{d/2 - \alpha - \beta}} =
\pi^{d/2}\, \frac{\Gamma(d/2-\alpha)\Gamma(d/2-\beta)\Gamma(\alpha+\beta-d/2)}{\Gamma(\alpha)\Gamma(\beta)
\Gamma(d-\alpha-\beta)}\, \frac{1}{(p^2)^{\alpha+\beta-d/2}}\, . 
\nn \\
\eea
In the $TT$ and $TVV$ cases, position space expressions such as $x^{\mu_1}...x^{\mu_n}/ (x^2)^\alpha$ up to rank $4$ are common, 
and the use of derivative relations - before proceeding with their final transform to momentum space - can be done in several ways.
Also in this case, as for the scalar functions, uniqueness shows that the result does not depend on the way we combine the factors
at the denominators with the corresponding numerators.

In order to deal with tensor expressions in position space, we introduce some notation.
We denote by
\beq
{R^n}_{\mu_1\dots \mu_n}(x,\alpha)
\equiv
\frac{x_{\mu_1} \dots x_{\mu_n}}{(x^2)^\alpha} \, ,
\eeq
the ratio between a generic tensor monomial in the vector $x$ and a power of $x^2$.
This notation is meant to help us denote in a compact way the tensor structures appearing in the expansion of any tensor correlator.
We call these expressions ``R-terms". \\
After some differential and algebraic manipulation we can easily derive the first four R-terms,
\bea
{R^1}_\mu(x,\alpha)
&=&
-\frac{1}{2 \, (\alpha-1)} \, \pd_\mu \, \frac{1}{(x^2)^{\alpha-1}} \, , \nonumber \\
{R^2}_{\mu\nu}(x,\alpha)
&=&
\frac{1}{4 \, (\alpha-2) \, (\alpha-1)} \, \pd_\mu\,\pd_\nu \, \frac{1}{(x^2)^{\alpha-2}}
+ \frac{\delta_{\mu\nu}}{2\,(\alpha-1)} \,
\frac{1}{(x^2)^{\alpha-1}} \, , \nonumber \\
{R^3}_{\mu\nu\rho}(x,\alpha)
&=&
- \frac{1}{8 (\alpha-3)(\alpha-2)(\alpha-1)} \,\pd_\mu\,\pd_\nu \, \pd_\rho \frac{1}{(x^2)^{\alpha-3}} +
\frac{1}{2 (\alpha-1)}\, \big[ \delta_{\mu\nu}{R^1}_\rho
+ \delta_{\mu\rho}{R^1}_\nu + \delta_{\nu\rho}{R^1}_\mu \big](x,\alpha-1)\, , \nonumber \\
{R^4}_{\mu\nu\rho\sigma}(x,\alpha)
&=&
\frac{1}{16(\alpha-4)(\alpha-3)(\alpha-2)(\alpha-1)} \,
\pd_\mu \, \pd_\nu \, \pd_\rho \, \pd_\sigma \, \frac{1}{(x^2)^{\alpha-4}} \nonumber \\
&+&
\frac{1}{2(\alpha-1)} \, \big[
\delta_{\mu\nu} {R^2}_{\rho\sigma} + \delta_{\rho\sigma} {R^2}_{\mu\nu} +
\delta_{\mu\rho} {R^2}_{\nu\sigma} + \delta_{\nu\sigma} {R^2}_{\mu\rho} +
\delta_{\mu\sigma} {R^2}_{\nu\rho} + \delta_{\nu\rho} {R^2}_{\mu\sigma}
\big](x,\alpha-1)
\nonumber \\
&-&
\frac{1}{4(\alpha-2)(\alpha-1)} \, (\delta_{\mu\nu}\delta_{\rho\sigma} + \delta_{\mu\rho}\delta_{\nu\sigma}
+ \delta_{\mu\sigma}\delta_{\nu\rho})\, \frac{1}{(x^2)^{\alpha-2}} \label{Ch1Rterms} \, .
\eea
The use of R-terms allows to extract immediately the leading singularities of the correlators, as we show below.
One can use several different forms of R-substitutions for a given tensor component.
For example, a rank $2$ tensor can be rewritten in R-form in several ways
\bea
\frac{(x -y)_\mu(x-y)_\nu}{[(x-y)^2]^{d+1}}
&=&
{R^2}_{\mu\nu}(x-y,d+1)
\nonumber \\
&=&
{R^1}_\mu(x-y,d/2+1)\, {R^1}_\nu(x-y,d/2)
\nonumber \\
&=&
\frac{1}{(x-y)^2} \, {R^1}_\mu(x-y,d/2) \, {R^1}_\nu(x-y,d/2) \, .
\eea
The derivative relations in the three cases shown above are obviously different, but the transform is unique.
One can also artificially rewrite the numerators at will by introducing trivial identities in position space,
without affecting the final expression of the mapping.
We will be using this method in order to extract some of the logarithmic integrals generated by this procedure.
Obviously, this is possible only if we guarantee an intermediate regularization.
We implement it by a dimensional shift of the exponents of the propagators.
The regulator will allow to smooth out the singularity of the correlators around
the value $\alpha=d/2$, which is the critical value beyond which a function such as $1/[x^2]^{\alpha}$ is not integrable,
according to (\ref{Ch1fund}).

The structure of the singularities in position space of the corresponding scalars and tensor correlators can be
identified using the basic transform. For instance, using (\ref{Ch1fund}) for $\alpha=d/2$ one encounters a pole
in the expression of the transform. For this reason we regulate dimensionally in position space
such a singularity by shifting $d\rightarrow d- 2 \omega$. At the same time we compensate with a regularization scale $\mu$ to
preserve the dimension of the redefined correlator. A similar approach has been discussed in \cite{Dunne:1992ws}, in an attempt to relate 
differential and dimensional regularization. However, in our case as in \cite{Erdmenger:1996yc} $\omega$ is an independent regulator which 
serves to test integrability in momentum space, and for this reason is combined with a fundamental transform which is given by
\beq
\frac{\mu^{2 \omega}}{[x^2]^{d/2- \omega}} =
\frac{\mu^{2 \omega}}{4^{d/2- \omega} \pi^{d/2}} \, \frac{\Gamma(\omega)}{\Gamma(d/2-\omega)} \, \int d^d l \,
\frac{e^{i l\cdot x}}{[l^2]^{\omega}} \, ,
\eeq
that we can expand around $\omega \sim 0$ to obtain
\beq \label{Ch1second}
\frac{\mu^{2\omega}}{[x^2]^{d/2- \omega}} = 
\frac{\pi^{d/2}}{\Gamma(d/2)} \, \delta^{(d)}(x) \, \left[\frac{1}{\omega} - \gamma  + \log 4 + \psi (d/2) \right] -
\frac{1}{(4\pi )^{d/2 } \, \Gamma(d/2)} \, \int d^d l \,{e^{i l\cdot x}} \log\left(\frac{l^2}{\mu^2}\right) + O(\omega) \, .
\eeq
The subtraction of this pole in $d$ dimensions is obviously related to the need of redefining correlators which are not integrable, 
in analogy with the approach followed in differential regularization.
The most popular example is $1/[x^2]^2$, which has no transform for $d=4$, but is rewritten in the derivative form as \cite{Freedman:1991tk}
\beq \label{Ch1defin}
\frac{1}{x^4}=\Box \, G(x^2)\, ,
\eeq
where $G(x^2)$ is defined by
\beq
G(x^2)= \frac{\log x^2 M^2}{x^2} + c \, ,
\eeq
with $c$ being a constant. This second approach can be easily generalized to $d$ dimensions.
One can use derivative relations such as
\beq
\frac{1}{[x^2]^\alpha}= \frac{1}{2(\alpha -1)(2 \alpha - d)} \, \Box \, \frac{1}{[x^2]^{\alpha -1}}
\label{Ch1squareeq}
\eeq
which is correct as far as $\alpha\neq d/2$.
For $\alpha=d/2 $ this relation misses the singularity at $x=0$, which is apparent from (\ref{Ch1deltaeq}).
For this reason, as far as $\alpha=d/2 - \omega$ eq.  (\ref{Ch1squareeq}) remains valid and it can be used
together with (\ref{Ch1deltaeq}) and an expansion in $\omega$ to give
\bea \label{Ch1third}
\frac{\mu^{2 \omega}}{[x^2]^{d/2- \omega}}
&=&
- \frac{1}{2\,\omega}\frac{\mu^{2 \omega}}{d - 2 - 2 \omega} \, \Box
\frac{1}{[x^2]^{d/2-1-\omega}}
\nonumber \\
&=&
\frac{1}{4 - 2\,d}\left(\frac{1}{\omega} + \frac{2}{d-2} \right)\Box \frac{1}{[x^2]^{d/2-1}}-
\frac{1}{2 (d-2)}\Box\frac{ \log (\mu^2 x^2)}{[x^2]^{d/2-1}} \nonumber \\
&=&
\frac{\pi^{d/2}}{\Gamma(d/2)} \, \left(\frac{1}{\omega}
+ \frac{2}{d-2}\right) \, \delta^{(d)}(x) - \frac{1}{2(d-2)} \, \Box\frac{\log (\mu^2 x^2)}{(x^2)^{d/2-1}}.
\eea
The $d$-dimensional version of differential regularization can be obtained by requiring the subtraction of all
the terms in (\ref{Ch1third}) which are proportional to $\delta^d(x)$, giving
\beq
\frac{1}{[x^2]^{d/2}} \equiv -\frac{1}{2 (d-2)}\Box\frac{\log (\mu^2 x^2)}{(x^2)^{d/2-1}}\, .
\eeq
This procedure clearly agrees with the traditional version of differential regularization in $d=4$ \cite{Freedman:1991tk},
\beq
\frac{1}{x^4}\equiv-\frac{1}{4}\Box\frac{\log(x^2 \mu^2)}{x^2}  \, .
\eeq
Notice that this analysis shows that, according to (\ref{Ch1third}), 
the logarithmic integral in (\ref{Ch1second}) is given by
\bea    \label{Ch1LogInt}
\int d^d l e^{i l \cdot x}\log \left(\frac{l^2}{\mu^2}\right)
&=&
(2\pi)^d \, \left[ - \gamma + \log 4 + \psi (d/2) - \frac{2}{d-2}  \right] \, \delta^{(d)}(x)
+ \frac{(4\pi)^{d/2}}{2(d-2)}\Gamma(d/2) \, \Box \, \frac{ \log(\mu^2 x^2)}{[x^2]^{d/2-1}}
\nonumber \\
&=&
\frac{(4\pi)^{d/2}}{2(d-2)}\Gamma(d/2) \, \Box \, \frac{ \log(\bar{\mu}^2 x^2)}{[x^2]^{d/2-1}} \, ,
\eea
having redefined the regularization scale properly
\beq \label{Ch1massscale}
\log \bar{\mu}^2 = \log \mu^2  + \gamma - \log 4 - \psi (d/2) + \frac{2}{d-2} \, .
\eeq
Notice that also regulated (but singular) correlators can be mapped in several ways to momentum space, with identical results,
exactly as for no singular correlators. For instance, we can take $1/[x^2]^{d/2}$ and use on it eq.  (\ref{Ch1fund}) once
\bea \label{Ch1transf}
\int d^d x \, e^{i k \cdot x}\frac{1}{[x^2]^{d/2}}
\to
\int d^d x \, e^{i k \cdot x}\frac{\mu^{2\omega}}{[x^2]^{d/2-\omega}} 
&=&
\frac{1}{4^{d/2-\omega}\, \pi^{d/2}}\, \frac{\Gamma(\omega)}{\Gamma(d/2- \omega)}\,
\int d^d x\, d^d l\, e^{i (k + l) \cdot x}\frac{\mu^{2\omega}}{[l^2]^{\omega}}
\nonumber \\
&=& 
4^{\omega}\, \pi^{d/2}\, \frac{\Gamma(\omega)}{\Gamma(d/2- \omega)} \, \frac{\mu^{2\omega}}{[k^2]^{\omega}}\, ,
\eea
twice
\bea
\int d^dx \, \frac{\mu^{2\omega}}{x^2 [x^2]^{d/2-1-\omega}}
&=& 
\frac{1}{4^{d/2-\omega} \pi^d}\, \frac{\Gamma(d/2 - 1)\, \Gamma(1+\omega)}{\Gamma(d/2-1-\omega)}
\int d^dx\, d^d l_1\, d^d l_2\,  e^{i (k+ l_1 + l_2) \cdot x} \, \frac{\mu^{2\omega}}{[l_1^2]^{d/2-1} [l_2^2]^{1+\omega}}
\nonumber \\
&=& 
4^{\omega}\, \pi^{d/2}\, \frac{\Gamma(\omega)}{\Gamma(d/2- \omega)}\,
\frac{\mu^{2\omega}}{[k^2]^{\omega}}\, ,
\eea
(where in the last step (\ref{Ch1B0general}) was used) or any number of times, obtaining the same transform.

As one can easily work out, the use of the dimensional regulator generates, after a Laurent expansion in $\omega$, some
logarithmic integrals in momentum space. As we shall show, if the $1/\omega$ poles cancel, then these integrals can be avoided, in the sense
that it will be possible to rewrite the correlator in such a way that they are absent. 
This means that in this case one has to go back and try to rewrite the correlator in such a way that it takes an explicitly finite form already
in position space. In this case the mapping of the correlators onto momentum space is similar to the usual Feynman expansion 
typical of perturbation theory. The condition of Fourier transformability is obviously necessary in order to have, eventually,
a Lagrangian description of the correlator.
On the other hand, if the same poles do not cancel, then the logarithms are a significant aspect of the correlator which,
for sure, cannot be reproduced by a local field theory Lagrangian anyhow, in particular not by a free field theory.
We have left to appendix \ref{Ch1Distributional} a few more examples on the correct handling of these distributional identities.

\subsection{Regularization of tensors}

The regularization of other tensor contributions using this extension of differential regularization can be handled
in a similar and straightforward way.
The use of the derivative relations on the R-terms, that map the tensor structures into derivative of less singular terms,
combined at the last stage with the basic transform, allows to get full control of any correlator and guarantees
its consistent mapping onto momentum space.
We provide a few examples to illustrate the procedure.

Consider, for instance, the tensor structure
\beq
t_{\mu} = \frac{(x-y)_\mu}{[(x-y)^2]^{d/2 +1}} \, ,
\eeq
whose R-form is, trivially,
\beq
t_{\mu}
=
{R^1}_\mu\left( x-y,\frac{d}{2}+1\right)
=
- \frac{1}{d} \, \pd_\mu \, \frac{1}{[(x-y)^2]^{d/2}} \, ,
\eeq
where the derivative is intended with respect to $x-y$.
Now we send $d \to d - 2\omega$ in the exponent of the denominator, introducing the proper mass scale, 
since $d/2$ is a critical value for integrability.
This allows us to use the basic transform (\ref{Ch1fund}), getting
\beq
t_\mu(\omega) =
- \frac{i \, \mu^{2\omega}}{(d- 2\omega) \, 4^{d/2-\omega} \, \pi^{d/2}} \, 
\frac{\Gamma(\omega)}{\Gamma(d/2-\omega)} \, \int \, d^d l \, \frac{l_\mu}{[l^2]^\omega} \, e^{i l\cdot (x-y)} \, .
\eeq
We can expand in $\omega$ obtaining
\bea
t_\mu(\omega)
&=&
\frac{i}{d \, 2^d \pi^{d/2} \, \Gamma(d/2)}
\bigg[
- \bigg(\frac{1}{\omega} + \frac{2}{d} - \gamma + \log 4 + \psi (d/2)\bigg) \,
\int d^d l\, e^{i l\cdot (x-y)}\,  l_{\mu}
\nonumber\\
&&\hspace{25mm}
+ \int d^d l\, e^{i l\cdot (x-y)} \, l_{\mu}\, \log\left(\frac{l^2}{\mu^2}\right)
\bigg]
+ O(\omega) \nonumber \\
&=& \frac{\pi^{d/2}}{d\,\Gamma(d/2)} \, \pd_\mu
\bigg[
- \bigg( \frac{1}{\omega} + \frac{4(d-1)}{d(d-2)} \bigg) \,
\delta^{(d)}(x-y)
+ \frac{\Gamma(d/2)}{2(d-2)\pi^{d/2}} \, \Box \frac{\log(\bar{\mu}^2(x-y)^2)}{[(x-y)^2]^{d/2-1}}
\bigg] \, ,
\eea
where in the last step we have used (\ref{Ch1LogInt}).
Notice that the strength of the singularity has increased from $\delta(x)/\omega$ to $\partial_\mu\delta(x)/\omega$,
due to the higher power $(d/2)$ of the denominator in position space. It is clear that for finite correlators
these singular contributions must cancel.
In general, the introduction of the regulator $\omega$ allows to perform algorithmically the transform of any lengthy
expression, leaving its implementation to a symbolic calculus program. Obviously, for finite correlators this approach might
look redundant, but it can be extremely useful in order to check the cancellation of all the multiple and single
pole singularities in a very efficient way. We will present more examples of this approach in the next sections.

A more involved example is given by
\beq
t_{\mu\nu} = \frac{(x-y)_\mu (x-y)_\nu}{[(x-y)^2]^{d/2+1}} \, ,
\eeq
to which corresponds the regulated expression
\beq
t_{\mu\nu}(\omega)= \frac{\mu^{2\omega}(x-y)_\mu (x-y)_\nu}{[(x-y)^2]^{d/2+1 - \omega}}
\eeq
and a minimal R-form given by
\beq
t_{\mu\nu}(\omega) = \mu^{2\omega} \, {R^2}_{\mu\nu}\left(x - y,\frac{d}{2} + 1 - \omega\right) \, .
\eeq
Using the list of replacements given in (\ref{Ch1Rterms}), the derivative form of $t_{\mu\nu}$ is given by
\beq
t_{\mu\nu }(\omega)= \frac{\mu^{2\omega}}{(d - 2 - 2 \omega) \, (d - 2\omega)} \, \pd_\mu \, \pd_\nu \,
\frac{1}{[(x-y)^2]^{d/2 - \omega -1}}
+ \frac{\delta_{\mu\nu}}{d + 2 - 2\omega} \, \frac{\mu^{2\omega}}{[(x-y)^2]^{d/2 - \omega}} \, ,
\eeq
whose singularities are all contained in the second term, whose Fourier transform is given by
\beq
\mathcal{FT}\bigg[\frac{\delta_{\mu\nu}}{d + 2 - 2\omega} \, \frac{\mu^{2\omega}}{[(x-y)^2]^{d/2 - \omega}}\bigg]
= \frac{1}{\omega} \, \frac{\delta_{\mu\nu}}{2^d \, \pi^{d/2} \, (d+2) \, \Gamma(d/2)}  + O(\omega^0) \, ,
\eeq
where we have omitted the regular terms. The procedure therefore allows to identify quite straightforwardly
the leading singularities of any tensor in position space, giving, in this specific case,
\beq
\frac{(x-y)_\mu (x-y)_\nu}{[(x-y)^2]^{d/2+1 - \omega}} \sim
\frac{1}{\omega} \, \frac{\delta_{\mu\nu}}{2^d \, \pi^{d/2} \, (d+2) \, \Gamma(d/2)} \, .
\eeq
We can repeat the procedure for correlators of higher rank. The singularities, after performing all the substitutions,
are proportional to the non-derivative terms isolated by the repeated replacement of eq.  (\ref{Ch1Rterms}).

\subsection{Regularization of $3$-point functions}

In the case of $3$-point functions, the analysis of the corresponding singularities can be extracted quite simply. \\
Let us consider, for instance, the identity
\bea \label{Ch1fundFor3}
&&
\mathcal{FT}\bigg[\frac{1}{[(x-y)^2]^{\alpha_1}[(z-x)^2]^{\alpha_2}[(y-z)^2]^{\alpha_3 }}\bigg]
\equiv
\int \, d^d x \, d^d y \, d^d z \,
\frac{e^{- i(k\cdot z+  p\cdot x + q\cdot y)}}{[(x-y)^2]^{\alpha_1}[(z-x)^2]^{\alpha_2}[(y-z)^2]^{\alpha_3 }}
\nonumber \\
&=&
(2\pi)^{3d} \, \prod_{i=1}^{3}\left( \frac{\Gamma(d/2 - \alpha_i)}{4^{\alpha_i} \pi^{d /2}\Gamma(\alpha_i)}\right) \,
\delta^{(d)}(k+p+q)\,
\int \frac{d^dl}{[l^2]^{d/2-\alpha_1} [(l+p)^2]^{d/2-\alpha_2} [(l-q)^2]^{d/2-\alpha_3}}\, ,\nonumber \\
\eea
obtained using the fundamental transform (\ref{Ch1fund}), where all the physical momenta $(k,p,q)$ are treated as incoming.
The convention for matching the momenta in (\ref{Ch1fund}) with the couples of coordinate is
\beq
l_1 \leftrightarrow x-y\, , \qquad l_2 \leftrightarrow z-x\, , \qquad l_3 \leftrightarrow y-z \, ,
\eeq
and the shift $ l \rightarrow l-q $ (which is always possible in a regularized expression) has been performed at the end. \\
It is clear that the pre-factor on the r.h.s. of this relation has poles for $\alpha_i = d/2 + n$, with $n \geq 0$.
At the same time the loop integral is asymptotically divergent if $d = \sum_i \alpha_i $, where it develops a logarithmic singularity. 
In dimensional regularization such a singularity corresponds to a single pole in $\epsilon= d -\sum_i \alpha_i$. 
One can be more specific by discussing further examples of typical $3$-point functions. \\
For instance, consider the tensor structure
\beq
{\mathcal{Q}^1}_{\alpha\beta\mu\nu} =
\frac{(y-z)_\alpha \, (y-z)_\beta \, (y-z)_\mu \, (y-z)_\nu}
{[(x-y)^2]^{d/2+1} \, [(z-x)^2]^{d/2-1} \, [(y-z)^2]^{d/2+1}} \, ,
\eeq
which appears in the $TVV$ correlator and can be reduced to its R-form in several ways.
We use a minimal substitution and have
\beq
{\mathcal{Q}^1}_{\alpha\beta\mu\nu} = \frac{1}{[(x-y)^2]^{d/2+1}} \, \frac{1}{[(z-x)^2]^{d/2-1}} \,
{R^4}_{\alpha\beta\mu\nu}\left(y-z,\frac{d}{2}+1\right) \, ,
\eeq
after which an application of the derivative reductions in (\ref{Ch1Rterms}) gives
\bea
{\mathcal{Q}^1}_{\alpha\beta\mu\nu}
&=&
\frac{1}{(d-6)\,(d-4)\,(d-2)\,d}\, \frac{1}{[(x-y)^2]^{d/2+1}} \, \frac{1}{[(x-z)^2]^{d/2 - 1}}
\nonumber \\
&\times&
\bigg\{
\pd_\alpha \, \pd_\beta \, \pd_\mu \, \pd_\nu \, \frac{1}{[(y-z)^2]^{d/2 - 3}}
+ (d-6)\,(d-4) \,
\frac{\delta_{\mu\nu}\,\delta_{\alpha\beta} + \delta_{\mu\alpha}\,\delta_{\nu\beta} + \delta_{\mu\beta}\,\delta_{\nu\alpha}}
{[(y - z)^2]^{d/2 - 1}}
\nonumber \\
&+&
(d-6)  \, \left(
  \delta_{\mu\nu}    \, \pd_\alpha \,  \pd_\beta  + \delta_{\alpha\beta} \, \pd_\mu  \, \pd_\nu
+ \delta_{\mu\alpha} \, \pd_\nu    \,  \pd_\beta  + \delta_{\nu\beta}    \, \pd_\mu  \, \pd_\alpha
+ \delta_{\nu\alpha} \, \pd_\mu    \,  \pd_\beta  + \delta_{\mu\beta}    \, \pd_\nu  \, \pd_\alpha
\right) \, \frac{1}{[(y-z)^2]^{d/2-2}}
\bigg\}  \, .
\nonumber\\
\eea
Before moving to momentum space, a quick glance at this equation shows that its transform does not exist.
This appears obvious from the presence of the overall factor $1/[(x-y)^2]^{d/2+1}$ which needs regularization.
The mapping can be performed using the rules defined above, which give, for instance, for the coefficient of
$\delta_{\mu\nu}\, \delta_{\alpha\beta} + \delta_{\mu\alpha}\, \delta_{\nu\beta} + \delta_{\mu\beta}\, \delta_{\nu\alpha}$,
\bea
&&
\mathcal{F T}
\bigg[
\frac{1}{d\,(d-2)} \,
\frac{\mu^{2\omega}}{[(x-y)^2]^{d/2 + 1 - \omega} \, [(z-x)^2]^{d/2 - 1} \, [(y-z)^2]^{d/2 - 1}}
\bigg]
\nonumber \\
&=&
\frac{(2\pi)^{3d} \, \delta^{(d)}(k+p+q)}{d\,(d-2)} \, \frac{4^{1+\omega}}{(4\pi)^{3d/2}}\,
\frac{\Gamma(\omega-1)}{\Gamma(d/2-1)^2 \,\Gamma(d/2-1-\omega)} \,
\int \, d^d l \, \frac{\mu^{2\omega}}{(l^2)^{\omega-1}\,(l+p)^2\,(l-q)^2}\nonumber
\eea
\bea
&=&
\frac{\delta^{(d)}(k+p+q)}{d(d-2)}\, \frac{4\,\pi^{3d/2}}{\Gamma(d/2-1)^3}\,
\bigg[
-\frac{1}{\omega}  \int d^dl\, \frac{l^2}{(l+p)^2(l-q)^2}
+\,   \int d^dl\, \frac{l^2\, \log\left(l^2/\bar{\mu}^2\right)}{(l+p)^2(l-q)^2}
\bigg]
+  O(\omega). \nonumber \\
\eea
In a similar way, the Fourier transform of the first term is
\bea
&&
\mathcal{F T}\bigg[\frac{1}{(d-6)\,(d-4)\,(d-2)\,d}\, \frac{\mu^{2\omega}}{[(x-y)^2]^{d/2+1-\omega}} \,
\frac{1}{[(z-x)^2]^{d/2-1}} \, \pd_\mu\,\pd_\nu\,\pd_\alpha\,\pd_\beta \, \frac{1}{[(y-z)^2]^{d/2-3}}\bigg] \nonumber \\
&=&
\frac{(2\pi)^{3d}\,\delta^{(d)(k+p+q)}}{(d-6)\,(d-4)\,(d-2)\,d}\, \frac{4^{3+\omega}}{(4\pi)^{3d/2}}\,
\frac{2\,\Gamma(\omega-1)}{\Gamma(d/2-3)\,\Gamma(d/2-1)\,\Gamma(d/2+1-\omega)}
\nonumber \\
&\times&
\int d^dl\, \frac{(l-q)_\alpha\,(l-q)_\beta\,(l-q)_\mu\,(l-q)_\nu}{(l^2)^{\omega-1}\,(l+p)^2\,[(l-q)]^3}
\nonumber \\
&=&
\frac{\delta^{(d)(k+p+q)}}{d\,(d-2)}\,\frac{32\,\pi^{3d/2}}{\Gamma(d/2-1)^3}\,
\bigg[
- \frac{1}{\omega}\, \int d^dl\, \frac{l^2\,(l-q)_\alpha\,(l-q)_\beta\,(l-q)_\mu\,(l-q)_\nu}{(l+p)^2[(l-q)^2]^3}
\bigg]
\nonumber\\
&& \hspace{40mm}
+\, \int d^dl\, \frac{\log\left(l^2/\bar{\mu}^2\right)\, (l-q)_\alpha\,(l-q)_\beta\,(l-q)_\mu\,(l-q)_\nu}
{(l+p)^2[(l-q)^2]^3} + O(\omega)\, ,
\eea
illustrating quite clearly how the general procedure can be implemented.

Of course, the regularization can be performed by sending $d\to d - 2 \omega$ - with no distinction among the 
various terms - or, alternatively, one can regulate only the non integrable terms. The two approaches, in a generic
computation, will differ only at $O(\omega)$ and as such they are equivalent.

Another important comment concerns the possibility of performing an explicit computation of the logarithmic integrals.
They are indeed calculable in terms of generalized hypergeometric functions (for general $\omega$), but the small $\omega$
expansion of these functions is rather difficult to re-express as a combination of ordinary functions and polylogs.
This is due to the need of performing a double expansion (in $\epsilon$ and in $\omega$) if we move to $d=4$ and insist, as we
should, on the use of dimensional regularization in the computation of the momentum integrals. This difficulty is attributed to
the absence of simple expansions of hypergeometric functions (ordinary and generalized) about non integer (real) values of their indices.
However, if the $1/\omega$ terms for a combination of terms similar to those shown above cancel, 
there are some steps which can be taken in order to simplify this final part of the computation.

\subsection{Application to the $VVV$ case}

To illustrate the general procedure through a specific example, we reconsider the $VVV$ case, that we know to be integrable. 
We expand the position space correlator and perform the R-substitutions (\ref{Ch1Rterms}). The direct algorithm gives
an expression which is not immediately recognized as being integrable and is
\bea f^{abc} \, \bigg\{\frac{(a - 2 \, b)}{(d-2)^3} \,
&\times&
\bigg[ \pd^{31}_\mu \, \frac{1}{(x^2_{31})^{d/2-1}}
\, \pd^{12}_\nu \, \frac{1}{(x^2_{12})^{d/2-1}} \, \pd^{23}_\rho \, \frac{1}{(x^2_{23})^{d/2-1}} \nonumber \\
&+&
\pd^{12}_\mu \, \frac{1}{(x^2_{12})^{d/2-1}} \, \pd^{23}_\nu \, \frac{1}{(x^2_{23})^{d/2-1}} \, \pd^{31}_\rho \,
\frac{1}{(x^2_{31})^{d/2-1}}    \bigg] \nonumber \\
+  \frac{a}{d \, (d-2)^2} \,
&\times&
\bigg[ \frac{1}{(x^2_{12})^{d/2-1}} \left(
\pd^{31}_\mu \, \frac{1}{(x^2_{31})^{d/2-1}} \, \pd^{23}_\nu \, \pd^{23}_\rho \, \frac{1}{(x^2_{23})^{d/2-1}}  +
\pd^{23}_\nu \, \frac{1}{(x^2_{23})^{d/2-1}} \, \pd^{31}_\mu \, \pd^{31}_\rho \, \frac{1}{(x^2_{31})^{d/2-1}}  \right)
\nonumber \\
&+&
\frac{1}{(x^2_{23})^{d/2-1}} \left(
\pd^{31}_\rho \, \frac{1}{(x^2_{31})^{d/2-1}} \, \pd^{12}_\mu \, \pd^{12}_\nu \, \frac{1}{(x^2_{12})^{d/2-1}} +
\pd^{12}_\nu \, \frac{1}{(x^2_{12})^{d/2-1}} \, \pd^{31}_\mu \, \pd^{31}_\rho \, \frac{1}{(x^2_{31})^{d/2-1}}  
\right)
\nonumber \\ &+&
\frac{1}{(x^2_{31})^{d/2-1}} \left(
\pd^{23}_\rho \, \frac{1}{(x^2_{23})^{d/2-1}} \, \pd^{12}_\mu \, \pd^{12}_\nu \, \frac{1}{(x^2_{12})^{d/2-1}} +
\pd^{12}_\mu \, \frac{1}{(x^2_{12})^{d/2-1}} \, \pd^{23}_\nu \, \pd^{23}_\rho \, \frac{1}{(x^2_{23})^{d/2-1}}  
\right) \bigg]
\nonumber \\
- \frac{1}{d-2} \, \left( b - \frac{a}{d+2} \right) \,
&\times&
\bigg[  \frac{1}{(x^2_{31})^{d/2-1}} \, \left(
\frac{\delta_{\mu\nu}}{(x^2_{12})^{d/2}} \, \pd^{23}_\rho \, \frac{1}{(x^2_{23})^{d/2-1}} +
\frac{ \delta_{\nu\rho}}{(x^2_{23})^{d/2}} \, \pd^{12}_\mu\, \frac{1}{(x^2_{12})^{d/2-1}}\right) \nonumber \\
&+&
\frac{1}{(x^2_{23})^{d/2-1}} \, \left(
\frac{\delta_{\mu\nu}}{(x^2_{12})^{d/2}} \, \pd^{31}_\rho \, \frac{1}{(x^2_{31})^{d/2-1}} +
\frac{\delta_{\mu\rho}}{(x^2_{31})^{d/2}} \, \pd^{12}_\nu\, \frac{1}{(x^2_{12})^{d/2-1}} \right) \nonumber \\ &+&
\frac{1}{(x^2_{12})^{d/2-1}} \, \left(
\frac{\delta_{\mu\rho}}{(x^2_{31})^{d/2}} \, \pd^{23}_\nu \, \frac{1}{(x^2_{23})^{d/2-1}} +
\frac{\delta_{\nu\rho}}{(x^2_{23})^{d/2}} \, \pd^{31}_\mu\, \frac{1}{(x^2_{31})^{d/2-1}}\right)  \bigg] \bigg\} \, . 
\eea
The apparent non-integrability is due to terms of the form ${1}/{(x^2_{ij})^{d/2}}$ in the last addend. 
For this reason, ignoring any further information, to test the approach we proceed with a regularization of the non-integrable terms.
The expression in momentum space is obtained by sending $ d \to d - 2\omega$ in all the terms of the form 
${1}/{(x^2_{ij})^{d/2}}$. Expanding in $\omega$ the result, one can show that, as expected, the $1/\omega$ terms cancel, 
proving its integrability. We fill in few more details to clarify this point.
A typical not manifestly integrable term in $VVV$ is
\beq
\frac{1}{(x^2_{31})^{d/2-1}} \,
\frac{1}{(x^2_{12})^{d/2}} \, \pd^{23}_\rho \, \frac{1}{(x^2_{23})^{d/2-1}} +
\frac{1}{(x^2_{23})^{d/2-1}} \,
\frac{1}{(x^2_{12})^{d/2}} \, \pd^{31}_\rho \, \frac{1}{(x^2_{31})^{d/2-1}} \, ,
\eeq
which in momentum space after $\omega$ regularization gives (omitting an irrelevant constant)
\beq
\mu^{2 \omega}\, \Gamma(\omega)\, \int d^d l\, \frac{ 2 l^\rho - q^\rho}{(l^2) (l-q)^2 [(l+p)^2]^{\omega}}.
\eeq
Expanding in $\omega$, the residue of the pole is given by the integral
\beq
\int d^d l\frac{2 l^\rho - q^\rho}{l^2 (l-q)^2}
\eeq
which vanishes in dimensional regularization.
The finite term is logarithmic,
\beq
\int d^d l\frac{\log \left((l+p)^2/\mu^2\right)\, (2 l^\rho - q^\rho)}{l^2 (l-q)^2} \, .
\eeq
The scale dependence also disappears, since the $\log\mu^2$ term is also multiplied by the same vanishing integral. Obviously, the
non trivial part of the computation is in the appearance of a finite logarithmic integral which, due to the finiteness of the
correlator, has to be re-expressed in terms of other non-logarithmic contributions, i.e. of ordinary Feynman integrals.
There is no simple way to relate one single integral to an ordinary non-logarithmic contribution unless one performs the entire
computation and expresses the result in terms of special polylogarithmic functions, using consistency.  For correlators which are 
integrable, however, it is possible to relate two log integrals to regular Feynman integrals.
Single log integrals, at least in this case, can also be evaluated explicitly, as we illustrate in appendix \ref{Ch1Distributional}.

By applying the algorithm we get
\bea
&&
\llangle {V^a}_{\mu} \, {V^b}_{\nu} \, {V^c}_{\rho} \rrangle \, (p,q) =
(2\pi)^{3d} \, \delta^{(d)}(k+p+q) \, i \, f^{abc}
\nn \\
&\times&
\bigg\{
C(d/2-1)^3 \,
\bigg[
\frac{a\,(6-4d)+2\,b\,d}{d(d-2)^3}
\bigg(2 \, J_{\mu\nu\rho}(p,-q) + (p+q)_{\mu} \, J_{\nu\rho}(p,-q) + p_{\nu} \, J_{\mu\rho}(p,-q)
\nonumber \\
&& \hspace{50mm}
- \, q_{\rho} \, J_{\mu\nu}(p,-q) - p_{\nu}\,q_{\mu}\,J_{\rho}(p,-q) - p_{\mu}\,q_{\rho}\,J_{\nu}(p,-q)\bigg)
\bigg]
\nonumber \\
&+&
\frac{a}{d(d - 2)^2} \,
\bigg(- 2 \, \left(p_\mu +  q_\mu\right) \, \big( p_\nu\,J_\rho(p,-q) + q_\rho\,J_\nu(p,-q)\big)
+ q_\rho p_\nu \big(2 \, J_\mu(p,-q) +  (p-q)_\mu \, J(p,-q) \big) \bigg)
\nonumber \\
&-&
\frac{C(d/2-1)^2}{(4\pi)^{d/2}\,\Gamma(d/2)\,(d-2)} \,
\bigg(\frac{a}{d+2} - b \bigg) \,
\bigg[
\delta_{\mu\nu}  \, \bigg( 2 \, IL_\rho(p,0,-q) - q_\rho \, IL(p,0,-q)\bigg) \nonumber \\
&&
\hspace{54mm}
+ \, \delta_{\mu\rho} \, \bigg( 2 \, IL_\nu(-q,0,p)  + p_\nu  \, IL(-q,0,p)\bigg)
\nonumber \\
&&
\hspace{54mm}
+ \, \delta_{\nu\rho} \, \bigg( 2 \, IL_\mu(q,0,k)  + k_\mu  \, IL(q,0,k)\bigg)
\bigg]
\bigg\} \, .
\eea
The notations introduced for the momentum space integrals here and in the following point are explained in appendix \ref{Ch1Distributional}.
One can easily show the scale independence of the result, which is related to the finiteness of the expressions
and to the fact that the logarithmic contributions, in this case, are an artefact of the approach.
For this reason, when the scale independence of the regulated expressions has been proven, then one can go back and try to
rewrite the correlator in such a way that it is manifestly integrable. Obviously this may not be straightforward,
especially if the expression in position space is given by hundreds of terms. If, even after proving the finiteness of the expression, 
one is unable to rewrite it in an integrable form, one can always apply the algorithm that we have presented, generating the logarithmic integrals. 
Pairs of log integrals can be related to ordinary Feynman integrals by applying appropriate tricks.
We have illustrated in an appendix an example where we discuss the computation of the single log-integral appearing in $VVV$. 
In the case of the $TOO$ one encounters both single and double-log integrals.

\subsection{Application to the $TOO$ case and double logs}

A similar analysis can be pursued in the $TOO$ case. Also for this correlator we can apply a direct approach in order to show
the way to proceed in the test of its regularity. Using our basic transform (\ref{Ch1fundFor3}) and introducing the regulator
$\omega$ to regulate the intermediate singularities, we can easily transform it to momentum space
\bea
&&
\mathcal{FT}\bigg[ \llangle T_{\mu\nu}(x_1) \, O(x_2) \, O(x_3) \rrangle  \bigg]
\equiv
\llangle T_{\mu\nu} \, O \, O  \rrangle(p,q) =
(2\pi)^{3d} \, \delta^{(d)}(k+p+q) \, a
\nonumber\\
&\times&
\bigg\{
\frac{C(d/2-1)^3}{d\,(d-2)^2}\,
\bigg[
- 4\, (d-1)\, J_{\mu\nu}(p,-q) - 2 \, (d-1) \, \bigg( (q_\nu - p_\nu)\, J_\mu(p,-q)  + (q_\mu - p_\mu)\, J_\nu(p,-q) \bigg)
\nonumber \\
&+&
\bigg(d\,(p_\mu q_\nu + p_\nu q_\mu ) - (d-2)\, (p_\mu p_\nu + q_\mu q_\nu) \bigg) \, J(p,-q)
\bigg]
\nonumber \\
&+&
\frac{C(d/2-1)^2\, C(d/2-\omega)}{d}\, \delta_{\mu\nu}\,
\bigg(\int\,d^dl\,\frac{\mu^{2\omega}}{l^2[(l+p)^2]^{\omega}(l-q)^2}
    + \int\,d^dl\,\frac{\mu^{2\omega}}{l^2(l+p)^2[(l-q)^2]^{\omega}} \bigg)
\nonumber \\
&-&
\frac{C(d/2-1)\, C(d/2-\omega)^2}{d}\, \delta_{\mu\nu}\,
\int\,d^dl\,\frac{(\mu^{2\omega})^2}{[l^2]^2[(l+p)^2]^{\omega}[(l-q)^2]^{\omega}}
\bigg\} \, .
\eea
Once we perform an expansion in $\omega$, the expression above is affected by double and single poles,
which are expected to vanish so as to guarantee a finite result.

The coefficient of the double pole is easily seen to take the form
\beq
- \delta_{\mu\nu}\, \frac{a\, (2\,\pi^2)^d\, d\, C(d/2-1)}{\Gamma(d/2)^2} \, I(0) \, ,
\eeq
where the integral vanishes in dimensional regularization, being a massless tadpole. \\
The coefficient of the simple pole is instead given by
\bea \label{Ch1simple}
&&
\delta_{\mu\nu}\, \frac{4^d\, \pi^{5d/2}\, C(d/2-1)^2}{d\, \Gamma(d/2)}\,
\bigg\{
\frac{1}{\Gamma(d/2-2) \, \Gamma(d/2)^2} \,
\bigg[
2 \, \bigg( \gamma - \log 4 - \psi (d/2) \bigg) \, I(0)
\nonumber \\
&+&
\bigg( IL(p,0,0) + IL(-q,0,0)  \bigg) \bigg]  +
\frac{1}{\Gamma(d/2-1)^2 \, \Gamma(d/2)} \,
\bigg[
I(p) + I(q)
\bigg]
\bigg\} \, .
\eea
The first term of (\ref{Ch1simple}) vanishes as in the case of the double pole, while for the remaining contributions we use the relation
\bea\label{Ch1IntegralTrick}
IL(p,0,0) = \int d^dl\,\frac{\log\left(\frac{(l+p)^2}{\mu^2}\right)}{[l^2]^2} =
- \frac{\pd}{\pd\omega} \, \int d^d l \,   \frac{\mu^{2\omega}}{[l^2]^2 \, [(l+p)^2]^\omega} \bigg|_{\omega=0} \, .
\eea
It is easy to see that the contributions in the last line in (\ref{Ch1simple}) cancel
after inserting the explicit value for the $2$-point function in (\ref{Ch1B0general}).

The finite part of the expression is found to be, after removing some additional tadpoles,
\bea
&&
\llangle T_{\mu\nu} \, O \, O  \rrangle(p,q) =
(2\pi)^{3d} \, \delta^{(d)}(k+p+q)\, a\,
\nonumber \\
&\times&
\bigg\{
\frac{C(d/2-1)^3}{d\,(d-2)^2}\,
\bigg[
- 4\, (d-1)\, J_{\mu\nu}(p,-q) - 2 \, (d-1) \, \bigg( (q_\nu - p_\nu)\, J_\mu(p,-q)  + (q_\mu - p_\mu)\, J_\nu(p,-q) \bigg)
\nonumber \\
&&\hspace{25mm}
+\, \bigg(d\,(p_\mu q_\nu + p_\nu q_\mu ) - (d-2)\, (p_\mu p_\nu + q_\mu q_\nu) \bigg) \, J(p,-q) \bigg]
\nonumber \\
&-&
\delta_{\mu\nu}\,
\bigg[
\frac{C(d/2-1)^2}{d\,\pi^{d/2}\,2^{d}\,\Gamma(d/2)} \,
\bigg( \left(\gamma -\log 4 -\psi (d/2)\right)\, \big(I(p) + I(-q) \big) + \big(IL(p,0,-q) + IL(-q,0,p) \big) \bigg)
\nonumber \\
&& \hspace{-2mm}
+ \, \frac{C(d/2-1)}{3\,d\,2^{2d+1}\, \pi^d\,\Gamma(d/2)^2}\,
\bigg(
12\, \left(\gamma - \log 4 - \psi (d/2)\right)\, \big(IL(p,0,0) + IL(-q,0,0)\big)
\nonumber \\
&&\hspace{33mm}
+\, 3 \, \big(ILL(p,p,0,0) + 2\, ILL(p,-q,0) + ILL(-q,-q,0,0) \big)
\bigg)
\bigg]
\bigg\} \, ,
\eea
where now also double logarithmic integrals have appeared. Using the relations (\ref{Ch1B0general}) and (\ref{Ch1IntegralTrick}), 
the terms proportional to $(\gamma - \log 4 - \psi (d/2))$, which are just a remain of the regularization procedure, cancel out,
leaving us with the simplified result
\bea
\label{Ch1toologs}
&&
\llangle T_{\mu\nu} \, O \, O  \rrangle(p,q) =
(2\pi)^{3d} \, \delta^{(d)}(k+p+q)\, a\,
\nonumber \\
&\times&
\bigg\{
\frac{C(d/2-1)^3}{d\,(d-2)^2}\,
\bigg[
- 4\, (d-1)\, J_{\mu\nu}(p,-q) - 2 \, (d-1) \, \bigg( (q_\nu - p_\nu)\, J_\mu(p,-q)  + (q_\mu - p_\mu)\, J_\nu(p,-q) \bigg)
\nonumber \\
&&\hspace{25mm}
+\, \bigg(d\,(p_\mu q_\nu + p_\nu q_\mu ) - (d-2)\, (p_\mu p_\nu + q_\mu q_\nu) \bigg) \, J(p,-q) \bigg]
\nonumber \\
&-&
\delta_{\mu\nu}\, \frac{C(d/2-1)}{d\,(4\pi)^d\,\Gamma(d/2)^2}
\bigg[
(4\pi)^{d/2}\,\Gamma(d/2) \,C(d/2-1)\, \bigg(\big(IL(p,0,-q) + IL(-q,0,p) \big) \bigg)
\nonumber \\
&& \hspace{30mm}
+ \, (d-4)\, \big(ILL(p,p,0) + 2\, ILL(p,-q,0,0) + ILL(-q,-q,0,0) \big)
\bigg]
\bigg\}.
\eea
It is slightly lengthy but quite straightforward to show that (\ref{Ch1toologs}) can be re-expressed in terms of ordinary Feynman integrals. 
This can be obtained by reducing all the tensor integrals (logarithmic and non-logarithmic) to scalar form.  
After the reduction, one can check directly that specific combinations of logarithmic integrals can be expressed in terms of ordinary master
integrals. In this case these relations hold since the integrands of the logarithmic expansion (linear combinations thereof) are equivalent to 
non-logarithmic ones, given the finiteness of the correlators. 
Obviously for a correlator which is not integrable such a correspondence does not exist and the logarithmic integrals cannot be avoided. 
This would be another signal, obviously, that the theory does not have a realization in terms of a local Lagrangian, 
since a Lagrangian field theory has a diagrammatic description only in terms of ordinary Feynman integrals.

We conclude this section with few more remarks concerning the treatment of correlators with more general scaling dimensions.
For instance one could consider correlators of the  generic form
\beq
\llangle O_i(x_i) O_j(x_j)O_k(x_k)\rrangle =
\frac{\lambda_{ijk}}{((x_i-x_j)^2)^{\Delta_i + \Delta_j - \Delta_k} ((x_j - x_k)^2)^{\Delta_j + 
\Delta_k - \Delta_i}((x_k-x_i)^2)^{\Delta_k + \Delta_i - \Delta_j}}.
\eeq

In this case their expression in momentum space can be found by applying Mellin-Barnes methods. They can be reconducted to
integrals in momentum space of the form
\beq
J(\nu_1,\nu_2,\nu_3)=\int \frac{d^d l}{(l^2)^{\nu_1}((l-k)^2)^{\nu_2}((l+p)^2)^{\nu_3}} \, ,
\eeq
\beq
\nu_1=d/2-\Delta_i - \Delta_j + \Delta_k \, ,
\qquad 
\nu_2=d/2-\Delta_j - \Delta_k + \Delta_i \, ,
\qquad 
\nu_3=d/2-\Delta_k - \Delta_i + \Delta_j \, ,
\eeq
which can be expressed \cite{Davydychev:1992xr} in terms of generalized hypergeometric functions $F_4[a,b,c,d;x,y]$
of two variables $(x,y)$, the two ratios of the 3 external momenta. The computation of these integrals  with arbitrary exponents at the 
denominators is by now standard lore in perturbation theory, with recursion relations which allow to relate shifts in the exponents 
in a systematic way. The problem is more involved for correlators which require an intermediate regularization in order to be transformed 
to momentum space. In this case one can show,  in general, that the pole structure (in $1/\omega$) of these can be worked out closely, 
but the finite $O(1)$ contributions involve derivatives of generalized hypergeometric functions respect to their indices $a,b,c,d$.
Only in some cases the latter can be re-expressed in terms of polylogarithmic functions, which are typical and common 
in ordinary perturbation theory.  The possibility to achieve this is essentially related to finding simple expansions of the hypergeometric functions 
around non integer (and not just rational) indicial points.
For integrable correlators, the analysis of Mellin-Barnes methods remains, however, a significant option, which will probably deserve a closer look.

\section{Conclusions}

The work presented in this chapter has the main goal to close the gap between position space and momentum space
analysis of $3$-point CFT's correlators characterized by the presence of one and three EMT's,
and to provide a general method to establish whether correlators built in position space CFT's admit or not a Lagrangian formulation.
We have tried to map position space and momentum space approaches, showing their interrelation, using free field theory
realizations of the general solutions of these correlators in order to establish their expression in momentum space. 

A parallel has been drawn between the approach to renormalization typical of standard perturbation theory and the same
approach based on the solution of the anomalous Ward identities, as discussed in \cite{Osborn:1993cr,Erdmenger:1996yc}.
As a non trivial test of the equivalence of both methods in $4$ dimensions, we have verified that the counterterms predicted
by the general analysis in position space coincide with those obtained from momentum space in the Lagrangian predictions derived from 
$1$-loop free field theory calculations.


We have also discussed a, in the second part, a general algorithm that should prove useful to regulate and 
map correlators from position space to momentum space, and we have illustrated how to perform such a mapping 
in a systematic way with a number of examples. 
The method can be applied to the analysis of more complex correlators. 
The power of the approach has been shown by re-analyisng conformal correlators investigated in the first part, 
offering a complete test of its consistency.

\clearpage{\pagestyle{empty}\cleardoublepage}

\chapter{Dilaton interactions and the anomalous breaking of scale invariance in the Standard Model}\label{Effective}

\section{Introduction} 

In this chapter we discuss the main features of dilaton interactions for fundamental and effective dilaton fields. 
In particular, we elaborate on the various ways in which dilatons can couple to the Standard Model and on the role played by the conformal 
anomaly as a way to characterize their interactions.
In the case of a dilaton derived from a metric compactification (graviscalar), we present the structure of the radiative corrections 
to its decay into two photons, a photon and a $Z$, two $Z$ gauge bosons and two gluons, together with their renormalization 
properties. We prove that, in the electroweak sector, the renormalization of the theory is guaranteed only if the Higgs is 
conformally coupled. For such a dilaton, its coupling to the trace anomaly is quite general, and determines, for instance, 
an enhancement of its decay rates into two photons and two gluons.   
We then turn our attention to theories containing a non-gravitational (effective) dilaton, which, in our perturbative analysis, 
manifests as a pseudo-Nambu Goldstone mode of the dilatation current ($J_D$). The infrared coupling of such a state to the 
$2$-photons and to the $2$-gluons sector, together with the corresponding anomaly enhancements of its decay rates in these channels,
is critically analysed. \\

Dilatons are part of the low energy effective action of several different types of theories, from string theory to theories with compactified extra 
dimensions, but they may appear also in appropriate bottom-up constructions. For instance, in scale invariant extensions of the Standard Model, 
the introduction of a dilaton field allows to recover scale invariance, which is violated by the Higgs potential, by introducing a new, enlarged, 
Lagrangian. This is characterized both by a spontaneous breaking of the conformal and of the electroweak symmetries.  

In this case, one can formulate simple scale invariant extensions of the potential which can accommodate, via spontaneous breaking, 
two separate scales: the electroweak scale ($v$), related to the vev of the Higgs field, and the conformal symmetry breaking scale 
($\Lambda$), related to the vev of a new field $\Sigma=\Lambda + \rho$, with $\rho$ being the dilaton. The second scale can be 
fine-tuned in order to proceed with a direct phenomenological analysis and is, therefore, of utmost relevance in the search for new 
physics at the LHC.

In a bottom-up approach, and this will be one of the main points that we will address in our analysis, the dilaton of the effective 
scale invariant Lagrangian can also be interpreted as a composite scalar, with the dilatation current taking the role of an operator 
which interpolates between this state and the vacuum. We will relate this interpretation to the appearance of an anomaly pole in the 
correlation function involving the dilatation current ($J_D$)
and two neutral currents ($V, V'$) of the Standard Model, providing evidence, in the ordinary perturbative picture, in favour of such 
a statement.   

One of the main issues which sets a difference between the various types of dilatons is, indeed, the contribution coming from the 
anomaly, which is expected to be quite large. Dilatons obtained from compactifications with large extra dimensions and a low gravity 
scale, for instance, carry this coupling, which is phenomenologically relevant. 
The same coupling is present in the case of an effective dilaton, appearing as a Goldstone mode of the dilatation current, with some 
differences that we will specify in a second part of the chapter. The analysis will be carried out in analogy to the pion case, which in 
a perturbative picture is associated with the appearance of an anomaly pole in the $AVV$ diagram (with $A$ being the axial current).

This chapter is organized as follows. In a first part we will characterize the leading $1$-loop interactions of a dilaton derived from a 
Kaluza-Klein compactification of the gravitational metric. The set-up is analogous to that presented in 
\cite{Han:1998sg,Giudice:2000av} for a compactified theory with large extra dimensions and it involves all the neutral currents of 
the Standard  Model. We present also a discussion of the same interaction in the QCD case for off-shell gluons.    

These interactions are obtained by tracing the $TVV$ vertex, with $T$ denoting the (symmetric and conserved) 
energy-momentum tensor (EMT) of the Standard Model. This study is accompanied by an explicit proof of the renormalizability of these 
interactions in the case of a conformally coupled Higgs scalar.

In a second part then we turn our discussion towards models in which dilatons are introduced from the ground up, starting with simple 
examples which should clarify - at least up to operators of dimension 4 - how one can proceed with the formulation of scale invariant 
extensions of the Standard Model.  Some of the more technical material concerning this point has been left to the appendices, where 
we illustrate the nature of the coupling of the dilaton to the mass dependent terms of the corresponding Lagrangian. The goal of 
these technical additions is to clarify that a fundamental (i.e. not a composite) dilaton, in a {\em classical} scale invariant 
extension of a given Lagrangian, does not necessarily couple to the anomaly, but only to massive states, exactly as in the Higgs 
case. For an effective dilaton, instead,  the Lagrangian is derived at tree level on the basis of classical scale invariance, as for 
a fundamental dilaton, but needs to be modified with the addition of an anomalous contribution, due to the composite nature of the 
scalar, in close analogy to the pion case.   

As we are going to show, if the dilaton is a composite state, identified with the anomaly pole of 
the $J_D VV$ correlator, an infrared coupling of this pole (i.e. a non-zero residue) is necessary in order to claim the presence of an 
anomaly enhancement in the $VV$ decay channel, with the $VV$ denoting on-shell physical asymptotic states, in a typical $S$-matrix 
approach. Here our reasoning follows quite closely the chiral anomaly case, where the anomaly pole of the $AVV$ diagram, which 
describes the pion exchange between the axial vector ($A$) and the vector currents, is infrared coupled only if $V$ denote physical 
asymptotic states. 
 
Clearly, our argument relies on a perturbative picture and is, in this respect, admittedly limited, forcing this issue to be resolved 
at experimental level, as in the pion case. We recall that in the pion the enhancement is present in the di-photon channel and not in 
the 2-gluon decay channel. 
 
Perturbation theory, in any case, allows to link the enhancement of a certain dilaton production/decay channel, to the virtuality of 
the gauge currents in the initial or the final state. 
  
We conclude with a discussion of the possible phenomenological implications of our results at the level of anomaly-enhanced dilaton 
decays, after pointing out the difference between the various ways in which the requirement of scale invariance (classical or 
quantum) can be realized in a typical scale invariant extension of the Standard Model Lagrangian. 

\subsection{The energy-momentum tensor}

We start with a brief summary of the structure of the Standard Model interactions with a $4D$ gravitational background, which is 
convenient in order to describe both the coupling of the graviscalar dilaton, emerging from the Kaluza-Klein compactification, and of 
a graviton at tree level and at higher orders.   
In the background metric $g_{\mu\nu}$ the action takes the form 
\beq S = S_G + S_{SM} + S_{I}= -\frac{1}{\kappa^2}\int d^4 x \sqrt{-g}\, R + \int d^4 x
\sqrt{-g}\mathcal{L}_{SM} + \chi \int d^4 x \sqrt{-g}\, R \, H^\dag H      \, ,
\eeq
where $\kappa^2=16 \pi G_N$, with $G_N$ being the four dimensional Newton's constant and $\mathcal H$ is the Higgs doublet.
We recall that the EMT in our conventions is defined as
\beq  T_{\mu\nu}(x)  = \frac{2}{\sqrt{-g(x)}}\frac{\d [S_{SM} + S_I ]}{\d g^{\mu\nu}(x)},
\eeq
or, in terms of the SM Lagrangian, as
\beq \label{TEI spaziocurvo}
\frac{1}{2} \sqrt{-g} T_{\mu\nu}{\equiv} \frac{\pd(\sqrt{-g}\mathcal{L})}
{\pd g^{\mu\nu}} - \frac{\pd}{\pd x^\s}\frac{\pd(\sqrt{-g}\mathcal{L})}{\pd(\pd_\s g^{\mu\nu})}\, ,
\eeq
which is classically covariantly conserved   ($g^{\mu\rho}T_{\mu\nu; \rho}=0$).  In flat spacetime, the covariant derivative is 
replaced by the ordinary derivative, giving the ordinary conservation equation ($ \pd_\mu T^{\mu\nu} = 0$).

We use the convention $\eta_{\mu\nu}=(1,-1,-1,-1)$ for the metric in flat spacetime, parametrizing its deviations 
from the flat case as
\beq\label{QMM} g_{\mu\nu}(x) \equiv \h_{\mu\nu} + \kappa \, h_{\mu\nu}(x)\,,\eeq
with the symmetric rank-2 tensor $h_{\mu\nu}(x)$ accounting for its fluctuations.

 In this limit, the coupling of the Lagrangian to gravity is given by the term
\beq\label{Lgrav} \mathcal{L}_{grav}(x) = -\frac{\kappa}{2}T^{\mu\nu}(x)h_{\mu\nu}(x). \eeq
In the case of theories with extra spacetime dimensions the structure of the corresponding Lagrangian can be found in 
\cite{Han:1998sg,Giudice:2000av}. For instance, in the case of a compactification over a $S_1$ circle of a 5-dimensional theory to 
4D, equation (\ref{Lgrav}) is modified in the form 
\beq
\label{Lgrav1} \mathcal{L}_{grav}(x) = -\frac{\kappa}{2} T^{\mu\nu}(x) \left(h_{\mu\nu}(x) + \rho(x) \,  \eta_{\mu\nu} \right) 
\eeq
which is sufficient in order to describe dilaton $(\rho)$ interactions with the fields of the Standard Model at leading order in 
$\kappa$, as in our case. In this case the graviscalar field $\rho$ is related to the $g_{55}$ component of the 5D metric and 
describes its massless Kaluza-Klein mode. The compactification generates an off-shell coupling of $\rho$ to the trace of the 
symmetric EMT. Notice that in this construction the fermions are 
assumed to live on the 4D brane and their interactions can be described by the ordinary embedding of the fermion Lagrangian of the 
Standard Model to a curved 4D gravitational background.  We use the spin connection $\Omega$ induced by the
curved metric $g_{\mu\nu}$. This allows to define a spinor derivative $\mathcal{D}$ which transforms covariantly under local Lorentz 
transformations. If we denote with $\underline{a},\underline{b}$ the Lorentz indices of a local free-falling frame, and denote with
$\s^{\underline{a}\underline{b}}$ the generators of the Lorentz group in the spinor representation, the spin connection takes 
the form
\beq
 \Omega_\mu(x) = \frac{1}{2}\s^{\underline{a}\underline{b}}V_{\underline{a}}^{\,\nu}(x)V_{\underline{b}\nu;\mu}(x)\, ,
\eeq
where we have introduced the Vielbein $V_{\underline{a}}^\mu(x)$. The covariant derivative of a spinor in a given representation
$(R)$ of the gauge symmetry group, expressed in curved $(\mathcal{D}_{\mu})$ coordinates is then given by
\beq \mathcal{D}_{\mu} = \frac{\pd}{\pd x^\mu} + \Omega_\mu   + A_\mu,\eeq
where $A_\mu\equiv A_\mu^a\, T^{a  (R)}$ are the gauge fields and $T^{a (R)}$ the group generators,
giving a Lagrangian of the form
\bea \mathcal{L}& = & \sqrt{-g} \bigg\{\frac{i}{2}\bigg[\bar\psi\g^\mu(\mathcal{D}_\mu\psi)
 - (\mathcal{D}_\mu\bar\psi)\g^\mu\psi \bigg] - m\bar\psi\psi\bigg\}\, .       
\eea
The derivation of the complete dilaton/gauge/gauge vertex in the Standard Model requires the computation of the trace of the EMT 
${T^\mu}_\mu$ (for the tree-level contributions), and of a large set of 1-loop 3-point functions. 
These are diagrams characterized by the insertion of the trace into 2-point functions of gauge currents. 
The full EMT is given by a minimal tensor $T_{Min}^{\mu\nu}$ (without improvement) and by a term of improvement, 
$T_I^{\mu\nu}$, originating from $S_I$, as
\bea
T^{\mu\nu} = T_{Min}^{\mu\nu} + T_I^{\mu\nu} \,,
\eea
where the minimal tensor is decomposed into gauge, ghost, Higgs, Yukawa and gauge fixing (g.fix.) contributions which can be found in 
\cite{Coriano:2011zk} 
\bea
T_{Min}^{\mu\nu} = 
T_{gauge}^{\mu\nu} + T_{ferm.}^{\mu\nu} + T_{Higgs}^{\mu\nu} + T_{Yukawa}^{\mu\nu} + T_{g.fix.}^{\mu\nu} + T_{ghost}^{\mu\nu}.
\eea
Concerning the structure of the EMT of improvement, we introduce the ordinary parametrizations of the Higgs field  
\beq
\label{Higgsparam}
H = \left(\begin{array}{c} -i \phi^{+} \\ \frac{1}{\sqrt{2}}(v + h + i \phi) \end{array}\right)
\eeq
and of its conjugate $H^\dagger$, expressed in terms of $h$, $\phi$ and $\phi^{\pm}$, corresponding to the physical Higgs and the 
Goldstone bosons of the $Z$ and $W^{\pm}$ respectively. As usual, $v$ denotes the Higgs vacuum expectation value. This expansion 
generates a non-vanishing EMT, induced by $S_I$, given by
\bea
\label{Timpr}
T^{\mu\nu}_I = - 2 \chi (\partial^\mu \partial^\nu - \eta^{\mu \nu} \Box) H^\dag H = 
- 2 \chi (\partial^\mu \partial^\nu - \eta^{\mu\nu} \Box) \left( \frac{h^2}{2} + \frac{\phi^2}{2} + \phi^+ \phi^- + v \, h\right)\, .
\eea
Notice that this term is generated by a Lagrangian which does not survive the flat spacetime limit. We are going to show by an 
explicit computation that $T_I$, if properly included with $\chi=1/6$, guarantees the renormalizability of the model.

\section{One loop electroweak corrections to dilaton-gauge-gauge vertices}

In this section we will present results for the structure of the radiative corrections to the dilaton/gauge/gauge vertices in the 
case of two photons, photon/$Z$ and $Z Z$ gauge currents. We have included in appendix \ref{rules} the list of the relevant tree 
level interactions extracted from the SM Lagrangian  
introduced above and which have been used in the computation of these corrections. We identify three classes of contributions, 
denoted as $\mathcal{A}$, $\Sigma$ and $\Delta$, with the $\mathcal A$-term coming from the conformal anomaly while the $\Sigma$ and 
$\Delta$ terms are related to the exchange of fermions, gauge bosons and scalars (Higgs/Goldstones). The separation between the 
anomaly part and the remaining terms is typical of the $TVV$ interaction. In particular one can check that in a 
mass-independent renormalization scheme, such as dimensional regularization with minimal subtraction, this separation can be verified 
at least at one loop level and provides a realization of the (anomalous) conformal Ward identity 
\beq
\Gamma^{\alpha\beta}(z,x,y) 
\equiv \eta_{\mu\nu} \left\langle T^{\mu\nu}(z) V^{\alpha}(x) V^{\beta}(y) \right\rangle 
= \frac{\delta^2 \mathcal A(z)}{\delta A_{\alpha}(x) \delta A_{\beta}(y)} + \left\langle {T^\mu}_\mu(z) V^{\alpha}(x) V^{\beta}(y) 
\right\rangle,
\label{traceid1}
\eeq
where we have denoted by $\mathcal A(z)$ the anomaly and $A_{\alpha}$ the gauge sources coupled to the current $V^{\alpha}$. Notice 
that in the expression above $\Gamma^{\alpha\beta}$ denotes a generic 
dilaton/gauge/gauge vertex, which is obtained form the $TVV$ vertex by tracing the spacetime indices $\mu\nu$. A simple way to test 
the validity of (\ref{traceid1}) is to compute the renormalized vertex $\langle T^{\mu\nu} V^\alpha V'^\beta\rangle$ (i.e. the 
graviton/gauge/gauge vertex) and perform afterwards its 4-dimensional trace. This allows to identify the left-hand-side of this 
equation. On the other hand, the insertion of the trace of $T^{\mu\nu}$ (i.e. $T^\mu_\mu$ )into a two point function $VV'$, allows to 
identify the second term on the right-hand-side of (\ref{traceid1}),  $\langle T^{\mu}_\mu(z) V^{\alpha}(x) V^{\beta}(y)\rangle$. 
The difference between the two terms so computed can be checked to correspond to the $\mathcal A$-term, obtained by two 
differentiations of the  anomaly functional $\mathcal A$. 
We recall that, in general, when scalars are conformally coupled, this takes the form
\bea \label{TraceAnomaly}
\mathcal A(z)
&=& \sum_{i} \frac{\beta_i}{2 g_i} \, F^{\alpha\beta}_i(z) F^i_{\alpha\beta}(z) +... \,,
\eea
where $\beta_i$ are clearly the mass-independent $\beta$ functions of the gauge fields
and $g_i$ the corresponding coupling constants, while the ellipsis refer to curvature-dependent terms. 
We present explicit results starting for the $\rho VV'$ vertices ($V,V' = \gamma, Z$), denoted as $\Gamma_{VV'}^{\alpha \beta}$, 
which are decomposed in momentum space in the form 
\beq
\Gamma_{VV'}^{\alpha \beta}(k,p,q) = (2\, \pi)^4\, \delta(k-p-q) \frac{i}{\Lambda} 
\left( \mathcal A^{\alpha \beta}(p,q) + \Sigma^{\alpha \beta}(p,q) + \Delta^{\alpha \beta}(p,q)\right),
\eeq
where 
\beq
\mathcal A^{\alpha \beta}(p,q) = \int d^4 x\, d^4 y \, e^{i p \cdot x + i q\cdot y}\, 
\frac{\delta^2 \mathcal A(0)}{\delta A^\alpha(x)\delta A^\beta(y)}
\eeq
and 
\beq
 \Sigma^{\alpha \beta}(p,q) +  \Delta^{\alpha \beta}(p,q) = \int d^4 x\, d^4 y\, e^{ i p \cdot x + i q\cdot y}\, 
\left\langle {T^\mu}_\mu(0) V^\alpha(x) V^\beta(y) \right\rangle \,.
\eeq
We have denoted with $\Sigma^{\alpha \beta}$ the cut vertex contribution to $\Gamma^{\alpha\beta}_{\rho VV'}$, 
while $\Delta^{\alpha \beta}$ includes the dilaton-Higgs mixing on the dilaton line, as shown in fig. \ref{figuremix}.
Notice that $\Sigma^{\alpha \beta}$ and $\Delta^{\alpha \beta}$ take contributions in two cases, specifically if the theory has an 
explicit (mass dependent) breaking and/or if the scalar - which in this case is the Higgs field - is not conformally coupled. 
The $\mathcal A^{\alpha\beta}(p,q)$ represents the conformal anomaly while $\Lambda$ is dilaton interaction scale.

\subsection{The $\rho\gamma\gamma$ vertex}

The interaction between a dilaton and two photons is identified by the diagrams in 
figs. \ref{figuretriangle}, \ref{figuretadpole}, \ref{figuremix} and is summarized by the expression 
\bea
\Gamma_{\gamma \gamma}^{\alpha\beta}(p,q) = \frac{i}{\Lambda} \bigg[ \mathcal A^{\alpha\beta}(p,q) + 
\Sigma^{\alpha\beta}(p,q) + \Delta^{\alpha\beta}(p,q) \bigg] \,,\eea
with the anomaly contribution given by
\bea
\label{Agammagamma}
\mathcal A^{\alpha\beta} =   \frac{\alpha}{\pi}\, \bigg[ -\frac{2}{3}\sum_{f} Q_{f}^2 + \frac{5}{2}  + 6\,\chi\bigg]\,
 u^{\alpha\beta}(p,q) \stackrel{\chi\rightarrow\frac{1}{6}}{=} - 2\, \frac{\beta_e}{e}\, u^{\alpha\beta}(p,q) \, ,
\eea
where
\bea
\label{utensor}
u^{\alpha\beta}(p,q) = (p\cdot q) \eta^{\alpha\beta} - q^{\alpha}p^{\beta} \,,
\eea
and the explicit scale-breaking term $\Sigma^{\alpha \beta}$ which splits into
\bea
\label{Sigmagammagamma}
\Sigma^{\alpha\beta}(p,q) = \Sigma_F^{\alpha\beta}(p,q) +  \Sigma_B^{\alpha\beta}(p,q) +\Sigma_I^{\alpha\beta}(p,q) \,.
\eea
We obtain for the on-shell photon case ($p^2 = q^2 = 0$)
\bea
\Sigma_F^{\alpha\beta}(p,q) 
&=& 
 \frac{\alpha}{\pi}\, \sum_f Q_f^2 m_f^2 \left[ \frac{4}{s} + 2 \left(\frac{4 
m_f^2}{s}-1\right) \mathcal C_0\left(s,0,0,m_f^2,m_f^2,m_f^2 \right)\right]\, u^{\alpha\beta}(p,q) \, , \nn \\
\Sigma_B^{\alpha\beta}(p,q) 
&=& 
 \frac{\alpha}{\pi}\, \left[ 6 M_W^2 \left(1-2\frac{M_W^2}{s}\right)  
\mathcal C_0(s,0,0,M_W^2,M_W^2,M_W^2) - 6 \frac{M_W^2}{s} - 1 \right]\, u^{\alpha\beta}(p,q) \, , \nn \\
\Sigma_I^{\alpha\beta}(p,q) 
&=& 
  \frac{\alpha}{\pi}\, 6 \chi \bigg[ 2 M_W^2 \mathcal C_0\left(s,0,0,M_W^2,M_W^2,M_W^2\right) \,
u^{\alpha\beta}(p,q) \nonumber \\
&-& 
M_W^2\, \frac{s}{2}\, \mathcal C_0(s,0,0,M_W^2,M_W^2,M_W^2) \, \eta^{\alpha \beta} \bigg]\, ,
\eea
while the mixing contributions are given by
\bea
\label{DeltaHgammagamma}
\Delta^{\alpha\beta}(p,q) 
&=&
\frac{\alpha}{\pi (s - M_H^2)} 6 \chi \bigg\{ 2 \sum_f Q_f^2 m_f^2  \bigg[ 2 + (4 m_f^2 -s ) \mathcal C_0(s,0,0,m_f^2,m_f^2,m_f^2) 
\bigg] \nn \\
&+& 
 M_H^2 + 6 M_W^2 + 2 M_W^2 (M_H^2 + 6 M_W^2 -4 s) \mathcal C_0(s,0,0,M_W^2,M_W^2,M_W^2)\bigg\} u^{\alpha \beta}(p,q) \nn \\
&+&  
\frac{\alpha}{\pi} 3 \chi s \, M_W^2 \, \mathcal C_0(s,0,0,M_W^2,M_W^2,M_W^2) \eta^{\alpha \beta} \, ,
\eea
with $\alpha$ the fine structure constant. The scalar integrals are defined in appendix \ref{scalars}.
The $\Sigma$'s  and $\Delta$ terms are the contributions obtained from the insertion on the photon 2-point function of the trace of 
the EMT, ${T^\mu}_\mu$. Notice that $\Sigma_I$ includes all the trace insertions which originate from the terms of improvement $T_I$ 
except for those which are bilinear in the Higgs-dilaton fields and 
which have been collected in $\Delta$. The analysis of the Ward and Slavnov-Taylor identities for the graviton-vector-vector correlators shows that these can be consistently solved only if we include the graviton-Higgs mixing on the graviton line. 

We have included contributions proportional both to fermions ($F$) and boson ($B$) loops, beside the $\Sigma_I$.
A conformal limit on these contributions can be performed by sending to zero all the mass terms, which is equivalent to sending 
the vev $v$ to zero and requiring a conformal coupling of the Higgs $(\chi=1/6)$. 
In the $v\to 0$ limit, but for a generic parameter $\chi$, we obtain 
\beq
\lim_{v\to 0} \left( \Sigma^{\a\b} + \Delta^{\a\b} \right)
= \lim_{v\to 0} \left(\Sigma_{B}^{\a\b}  + \Sigma_{I}^{\a\b}\right) 
=  \frac{\alpha}{\pi} (6 \chi -1) u^{\alpha \beta}(p,q),
\eeq
which, in general, is non-vanishing.
Notice that, among the various contributions, only the exchange of a boson or the term of improvement contribute in this limit and 
their sum vanishes only if the Higgs is conformally coupled $(\chi = \frac{1}{6})$. \\

Finally, we give the decay rate of the dilaton into two on-shell photons in the simplified case in which we remove the term of 
improvement by sending $\chi \to 0$
\bea
\Gamma(\rho \rightarrow \gamma\gamma) 
&=&
\frac{\alpha^2\,m_{\rho}^3}{256\,\Lambda^2\,\pi^3} \, \bigg| \beta_{2} + \beta_{Y} 
-\left[ 2 + 3\, x_W  +3\,x_W\,(2-x_W)\,f(x_W) \right]
+ \frac{8}{3} \, x_t\left[1 + (1-x_t)\,f(x_t) \right] \bigg|^2 \,,
\label{PhiGammaGamma} 
\eea
where the contributions to the decay, beside the anomaly term, come from the $W$ and the fermion (top) loops
and $\beta_2 (= 19/6)$ and $\beta_Y (= -41/6)$ are the $SU(2)$ and $U(1)$ $\beta$ functions respectively.
Here, as well as in the other decay rates evaluated all through the paper, the $x_i$ are defined as
\beq \label{x}
x_i = \frac{4\, m_i^2}{m^2_\rho} \, ,
\eeq
with the index "$i$" labelling the corresponding massive particle, and $x_t$ denoting the contribution from the top quark,
which is the only massive fermion running in the loop.
The function $f(x)$ is given by
\bea
\label{fx}
f(x) = 
\begin{cases}
\arcsin^2(\frac{1}{\sqrt{x}})\, , \quad \mbox{if} \quad \,  x \geq 1 \\ 
-\frac{1}{4}\,\left[ \ln\frac{1+\sqrt{1-x}}{1-\sqrt{1-x}} - i\,\pi \right]^2\, , \quad \mbox{if} \quad \, x < 1.
\end{cases}
\eea
which originates from the scalar $3$-point master integral through the relation 
\beq \label{C03m}
C_0(s,0,0,m^2,m^2,m^2) = - \frac{2}{s} \, f(\frac{4\,m^2}{s}) \, .
\eeq

\subsection{The $\rho\gamma Z$ vertex} 
%
The interaction between a dilaton, a photon and a $Z$ boson is described by the $\Gamma^{\alpha \beta}_{\gamma Z}$ correlation 
function (figs.. \ref{figuretriangle}, \ref{figuretadpole}, \ref{figuremix}). In the on-shell case, with the kinematic defined by
\beq
p^2 = 0 \, \quad q^2 = M_Z^2 \, \quad k^2 = (p+q)^2 = s \,,
\eeq
the vertex $\Gamma^{\alpha \beta}_{\gamma Z}$ is expanded as 
\bea
\Gamma^{\alpha \beta}_{\gamma Z} &=& \frac{i}{\Lambda} \bigg[ \mathcal A^{\alpha\beta}(p,q)
+ \Sigma^{\alpha\beta}(p,q) + \Delta^{\alpha\beta}(p,q) \bigg] \nn \\
&=&
\frac{i}{\Lambda}\, \Bigg\{\,
\left[ \frac{1}{2}\,\left(s - M^2_Z\right)\,\eta^{\alpha\beta} - q^\alpha\,p^\beta\right] \, 
\left( \mathcal A_{\gamma Z} +  \Phi_{\gamma Z}(p,q)\right)
+\eta^{\alpha\beta}\, \Xi_{\gamma Z}(p,q) \Bigg\} \, .
\eea
The anomaly contribution is
\beq
\mathcal A_{\gamma Z} =  \frac{\alpha}{\pi\,s_w c_w}\,\left[-\frac{1}{3}\sum_{f}C^f_v\,Q_f + \frac{1}{12}\,(37-30s_w^2)
+ 3\, \chi\, (c_w^2 - s_w^2) \right]\, ,
\eeq
where $s_w$ and $c_w$ to denote the sine and cosine of the $\theta$-Weinberg angle.
Here $\Delta^{\alpha \beta}$ is the external leg correction on the dilaton line and the form factors $\Phi(p,q)$ and $\Xi(p,q)$ 
are introduced to simplify the computation of the decay rate and decomposed as
\bea
\Phi_{\gamma Z}(p,q)  &=& \Phi^{\Sigma}_{\gamma Z}(p,q) + \Phi^{\Delta}_{\gamma Z}(p,q) \,, \nn \\
\Xi_{\gamma Z}(p,q) &=& \Xi^{\Sigma}_{\gamma Z}(p,q) + \Xi^{\Delta}_{\gamma Z}(p,q) \,,
\eea
in order to distinguish the contributions to the external leg corrections ($\Delta$) from those to the cut vertex ($\Sigma$).
They  are given by
\bea
\Phi^{\Sigma}_{\gamma Z}(p,q)
&=& 
\frac{\alpha}{\pi\, s_w\,c_w}\, \Bigg\{
\sum_f C_v^f \, Q_f \, \Bigg[\frac{2\, m_f^2}{s - M_Z^2} + \frac{2 m_f^2 \, M_Z^2}{(s - M_Z^2)^2}\, \mathcal D_0(s,M_Z^2,m_f^2,m_f^2)
\nonumber \\
&& \hspace{-20mm}
- m_f^2\, \left(1 - \frac{4\, m_f^2}{s - M_Z^2}\right)\, \mathcal C_0(s,0,M_Z^2,m_f^2,m_f^2,m_f^2) \Bigg]
-  \Bigg[ \frac{M_Z^2}{2\,\left(s - M_Z^2\right)}\, (12\, s_w^4 - 24\, s_w^2 + 11)
\nonumber\\
&& \hspace{-20mm}
\frac{M_Z^2}{2\,\left(s - M_Z^2 \right)^2} 
\left[2\, M_Z^2\,\Big(6\, s_w^4 - 11\, s_w^2 + 5 \Big)- 2\, s_w^2\, s + s \right]\, \mathcal D_0(s,M_Z^2,M_W^2,M_W^2)
\nonumber \\
&& \hspace{-20mm}
+ \frac{M_Z^2\, c_w^2}{s - M_Z^2}\,
\Big[2\, M_Z^2\, \left(6\, s_w^4 - 15\, s_w^2 + 8 \right) + s\, \left(6\, s_w^2 - 5\right)\Big]\,
\mathcal C_0(s,0,M_Z^2,M_W^2,M_W^2,M_W^2) \Bigg]
\nonumber \\
&& \hspace{-20mm}
+ \frac{3 \chi \,(c_w^2 - s_w^2)\,}{s - M_Z^2}\,
\bigg[M_Z^2 +  s\, \bigg( 2\, M_W^2 \, \mathcal C_0(s,0,M_Z^2,M_W^2,M_W^2,M_W^2) 
+ \frac{M_Z^2}{s-M_Z^2}\, \mathcal D_0(s,M_Z^2,M_W^2,M_W^2) \bigg) \bigg]
\Bigg\} \, ,
\nonumber
\eea
\bea
\Xi^{\Sigma}_{\gamma Z}(p,q)
&=&
\frac{\alpha}{\pi}
\Bigg\{ - \frac{c_w\, M_Z^2}{ s_w} \mathcal B_0(0,M_W^2,M_W^2) + 3\, s\, \chi \, s_w^2\, M_Z^2 \, 
\mathcal C_0(s,0,M_Z^2,M_W^2,M_W^2,M_W^2) \Bigg\} \, ,  \nn \\
\Phi^{\Delta}_{\gamma Z}(p,q)
&=& 
\frac{3\,\alpha\,s\,\chi}{\pi s_w c_w (s-M_H^2)(s-M_Z^2)} 
\bigg\{ 2 \sum_f m_f^2 C_v^f Q_f \bigg[ 2 + 2 \frac{M_Z^2}{s-M_Z^2} \mathcal D_0(s,M_Z^2,m_f^2,m_f^2)  \nn \\
&& \hspace{-20mm}
+ (4 m_f^2 + M_Z^2 - s) \mathcal C_0(s,0,M_Z^2,m_f^2,m_f^2,m_f^2) \bigg]  + M_H^2(1-2 s_w^2) + 2 M_Z^2 (6 s_w^4 - 11 s_w^2 +5) \nn \\
&& \hspace{-20mm}
+ \frac{M_Z^2}{s-M_Z^2} ( M_H^2 (1-2 s_w^2) + 2 M_Z^2 (6 s_w^4 -11 s_w^2 + 5) ) \mathcal D_0(s,M_Z^2,M_W^2,M_W^2) \nn \\
&& \hspace{-20mm}
+ 2 M_W^2 \mathcal C_0(s,0,M_Z^2,M_W^2,M_W^2,M_W^2) ( M_H^2 (1-2 s_w^2) + 2 M_Z^2 (6 s_w^4 - 15 s_w^2 + 8) + 2 s (4 s_w^2-3)) \bigg\} 
\nn \\
\Xi^{\Delta}_{\gamma Z}(p,q)
&=&
\frac{3\,\alpha\,s\, \chi\, c_w}{\pi\, s_w}\, M_Z^2 \bigg\{ \frac{2}{s-M_H^2} \mathcal B_0(0, M_W^2,M_W^2) 
- s_w^2 \mathcal C_0(s,0,M_Z^2,M_W^2,M_W^2,M_W^2) \bigg\}\, .
\eea

As for the previous case, we give the decay rate in the simplified limit $\chi \to 0$ which is easily found to be
\bea
\Gamma(\rho\rightarrow \gamma Z) 
&=&
\frac{9\,m_{\rho}^3}{1024\,\Lambda^2\,\pi} \, \sqrt{1-x_Z} \, \bigg( |\Phi^{\Sigma}_{\gamma Z}|^2(p,q)\,m_{\rho}^4\,(x_Z-4)^2 + 
48 \, Re\, \left\{\Phi^{\Sigma}_{\gamma Z}(p,q)\,\Xi^{\Sigma\,*}_{\gamma Z}(p,q) \, m_{\rho}^2\,(x_Z-4)\right\}
\nonumber \\
&& \hspace{35mm}
- \, 192 \, |\Xi^{\Sigma}_{\gamma Z}|^2(p,q)  \bigg) \, ,
\label{RateRhoGammaZ}
\eea
where $Re$ is the symbol for the real part.

\subsection{The $\rho Z Z$ vertex}

\begin{figure}[t]
\centering
\subfigure[]{\includegraphics[scale=.7]{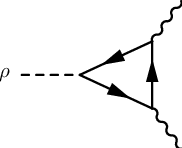}}
\hspace{.2cm}
\subfigure[]{\includegraphics[scale=.7]{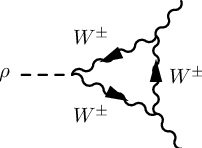}}
\hspace{.2cm}
\subfigure[]{\includegraphics[scale=.7]{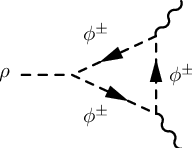}}
\hspace{.2cm}
\subfigure[]{\includegraphics[scale=.7]{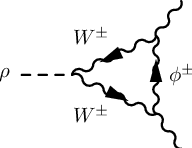}}
\hspace{.2cm}
\subfigure[]{\includegraphics[scale=.7]{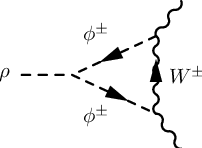}} \\
\hspace{.2cm}
\subfigure[]{\includegraphics[scale=.7]{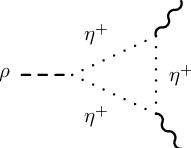}}
\hspace{.2cm}
\subfigure[]{\includegraphics[scale=.7]{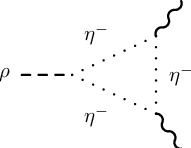}}
\hspace{.2cm}
\subfigure[]{\includegraphics[scale=.7]{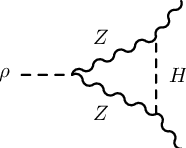}}
\hspace{.2cm}
\subfigure[]{\includegraphics[scale=.7]{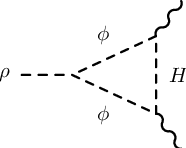}}
\hspace{.2cm}
\subfigure[]{\includegraphics[scale=.7]{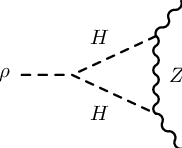}}
\hspace{.2cm}
\subfigure[]{\includegraphics[scale=.7]{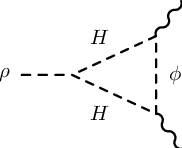}}  
\caption{Amplitudes of triangle topology contributing to the $\rho \gamma\gamma$, $\rho \gamma Z$ and $\rho ZZ$ interactions. They 
include fermion $(F)$, gauge bosons $(B)$ and contributions from the term of improvement (I). Diagrams (a)-(g) contribute to all the 
three channels while (h)-(k) only in the $\rho ZZ$ case.}
\label{figuretriangle}
\end{figure}

\begin{figure}[t]
\centering
\subfigure[]{\includegraphics[scale=.7]{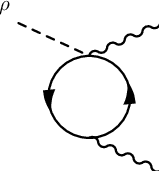}}
\hspace{.2cm}
\subfigure[]{\includegraphics[scale=.7]{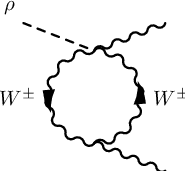}}
\hspace{.2cm}
\subfigure[]{\includegraphics[scale=.7]{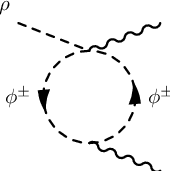}}
\hspace{.2cm}
\subfigure[]{\includegraphics[scale=.7]{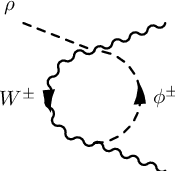}}
\hspace{.2cm}
\subfigure[]{\includegraphics[scale=.7]{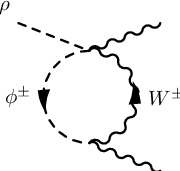}}
\hspace{.2cm}
\subfigure[]{\includegraphics[scale=.7]{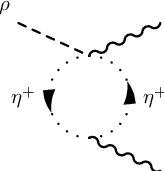}} \\
\subfigure[]{\includegraphics[scale=.7]{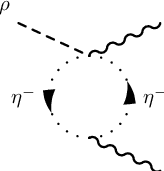}}
\hspace{.2cm}
\subfigure[]{\includegraphics[scale=.7]{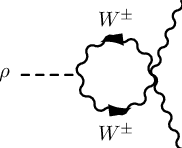}}
\hspace{.2cm}
\subfigure[]{\includegraphics[scale=.7]{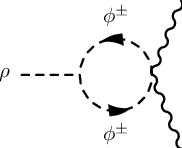}}
\hspace{.2cm}
\subfigure[]{\includegraphics[scale=.7]{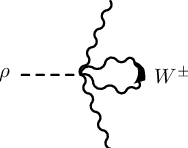}}
\hspace{.2cm}
\subfigure[]{\includegraphics[scale=.7]{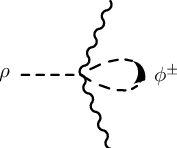}}
\hspace{.2cm}
\subfigure[]{\includegraphics[scale=.7]{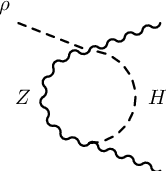}} \\
\hspace{.2cm}
\subfigure[]{\includegraphics[scale=.7]{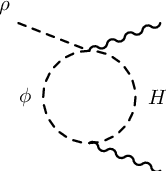}}  
\hspace{.2cm}
\subfigure[]{\includegraphics[scale=.7]{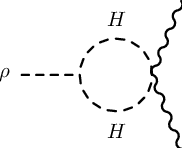}}
\hspace{.2cm}
\subfigure[]{\includegraphics[scale=.7]{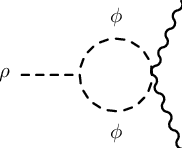}}  
\hspace{.2cm}
\subfigure[]{\includegraphics[scale=.7]{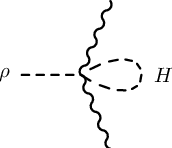}}
\hspace{.2cm}
\subfigure[]{\includegraphics[scale=.7]{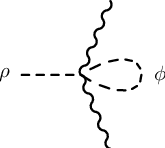}}  
\caption{Bubble and tadpole-like diagrams for $\rho \gamma\gamma$ $\rho \gamma Z $ and $\rho Z Z$. 
Amplitudes (l)-(q) contribute only in the $\rho ZZ$ channel.}
\label{figuretadpole}
\end{figure}

\begin{figure}[t]
\centering
\subfigure[]{\includegraphics[scale=.7]{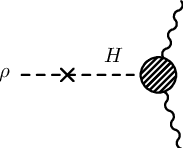}}
\hspace{.2cm}
\subfigure[]{\includegraphics[scale=.7]{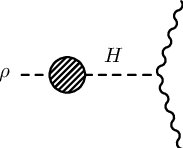}}
\hspace{.2cm}
\subfigure[]{\includegraphics[scale=.7]{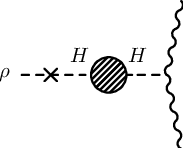}}
\caption{External leg corrections. Diagrams (b) and (c) appear only in the $\rho Z Z$ sector.}
\label{figuremix}
\end{figure}

The expression for the $\Gamma^{\alpha\beta}_{Z Z}$ vertex (figs.. \ref{figuretriangle},\ref{figuretadpole},\ref {figuremix}) defining 
the $\rho ZZ$ interaction is presented here in the kinematic limit given by $k^2 = (p+q)^2 = s$, $p^2 = q^2 = M_Z^2$ with two 
on-shell $Z$ bosons. The completely cut correlator takes contributions from a fermion sector, a $W$ gauge boson sector, a $Z-H$ 
sector together with a term of improvement.
There is also an external leg correction $\Delta^{\alpha \beta}$ on the dilaton line which is much more involved than in the previous 
cases because there are contributions coming from the minimal EMT and from the improven EMT . \\
At one loop order we have 
\bea \label{1loopZZ}
&&
\Gamma_{Z Z}^{\alpha\beta}(p,q)
\equiv 
\frac{i}{\Lambda}\, \left[ \mathcal A^{\alpha\beta}(p,q) + \Sigma^{\alpha\beta}(p,q)+ \Delta^{\alpha\beta}(p,q) \right] \nn \\
&& = 
\frac{i}{\Lambda}\,
\Bigg\{
\left[\left(\frac{s}{2} - M^2_Z \right)\, \eta^{\alpha\beta} - q^\alpha\,p^\beta \right]\,
\left( \mathcal A_{ZZ} + \Phi^{\Sigma}_{ZZ}(p,q) + \Phi^{\Delta}_{ZZ}(p,q) \right)
+ \eta^{\alpha\beta}\, \left(\Xi^{\Sigma}_{ZZ}(p,q) + \Xi^{\Delta}_{ZZ}(p,q) \right) \Bigg\} ,
\eea
where again $\Sigma$ stands for the completely cut vertex and $\Delta$ for the external leg corrections and
we have introduced for convenience the separation
\bea
\Phi^{\Sigma}_{ZZ}(p,q) &=& \Phi^{F}_{ZZ}(p,q) + \Phi^{W}_{ZZ}(p,q) + \Phi^{ZH}_{ZZ}(p,q) + \Phi^{I}_{ZZ}(p,q) \,, \nn \\
\Xi^{\Sigma}_{ZZ}(p,q) &=& \Xi^{F}_{ZZ}(p,q) + \Xi^{W}_{ZZ}(p,q) + \Xi^{ZH}_{ZZ}(p,q) + \Xi^{I}_{ZZ}(p,q)\, .
\eea
The form factors are given in appendix \ref{VZZ}, while here we report only the purely anomalous contribution
\beq
\mathcal A_{ZZ} = \frac{\alpha}{6 \pi c_w^2 s_w^2}\,
\left\{ -\sum_{f} \left({C_a^f}^2+{C_v^f}^2\right) + \frac{60\,s_w^6 - 148\,s_w^2 + 81}{4} - \frac{7}{4} 
+ 18\,\chi\,\left[ 1 - 2\,s_w^2\,c_w^2 \right]\right\}\, .
\eeq
Finally, we give the decay rate expression for the $\rho \to  Z Z$ process. 
At leading order it can be computed from the tree level amplitude
\bea
\mathcal M^{\alpha\beta}(\rho \rightarrow ZZ) 
&=& 
\frac{2}{\Lambda}\,M_Z^2\, \eta^{\alpha\beta} \, , 
\eea
and it is given by  
\bea
\Gamma(\rho \rightarrow ZZ) 
&=&
\frac{ m_{\rho}^3}{32\,\pi \Lambda^2} \, (1-x_Z)^{1/2}\, \left[ 1 - x_Z + \frac{3}{4}\,x_Z^2 \right].
\label{PhiZZ}
\eea
Including the $1$-loop corrections defined in eq. (\ref{1loopZZ}), one gets the decay rate at next-to-leading order
\bea
 \Gamma(\rho\rightarrow ZZ) 
&=&
\frac{m_{\rho}^3}{32\,\pi\,\Lambda^2} \,\sqrt{1-x_Z}\, \bigg\{ 1 - x_Z + \frac{3}{4}\,x_Z^2 
+ \frac{3}{x_Z} \, \bigg[4\, Re\, \{\Phi^{\Sigma}_{ZZ}(p,q)\}(1-x_Z + \frac{3}{4}\,x_Z^2) 
\nn \\
&-& 
Re\, \{\Xi^{\Sigma}_{ZZ}(p,q)\}\,m_{\rho}^2\, \left( \frac{3}{4}\,x_Z^3 - \frac{3}{2}\,x_Z^2 \right) \bigg] \bigg\}\, .
\eea
%
%
%
\subsection{Renormalization of dilaton interactions in the broken electroweak phase}
\label{renorm}

In this section we address the renormalization properties of the correlation functions given above. Although the 
proof is quite cumbersome, one can check, from our previous results, that the 1-loop renormalization of the Standard Model 
Lagrangian is sufficient to cancel all the singularities in the cut 
vertices independently of whether the Higgs is conformally coupled or not. Concerning the uncut vertices, instead, the term of 
improvement plays a significant role in the determination of Green functions which are ultraviolet finite. In particular such a term 
has to appear with $\chi=1/6$ in order to guarantee the cancellation of a singularity present in the 1-loop 2-point function 
describing the Higgs dilaton mixing ($\Sigma_{\rho H}$).  
The problem arises only in the $\Gamma^{\alpha\beta}_{ZZ}$ correlator, where the $\Sigma_{\rho H}$ $2$-point function is present as 
an external leg correction on the dilaton line. 

The finite parts of the counterterms are determined in the on-shell renormalization scheme, which is widely used in the electroweak 
theory. In this scheme the renormalization conditions are fixed in terms of the physical parameters to all orders in perturbation 
theory and the wave-function normalizations of the fields are obtained by requiring a unit residue of the full 2-point functions on 
the physical particle poles.

From the counterterm Lagrangian we compute the corresponding counterterm to the trace of the EMT. As we have already mentioned, 
one can also verify from the explicit computation that the terms of improvement, in the conformally coupled case, are necessary 
to renormalize the vertices containing an intermediate scalar with an external bilinear mixing (dilaton/Higgs).
The counterterm vertices for the correlators with a dilaton insertion are
\bea
\delta [\rho \gamma \gamma]^{\alpha\beta}  &=& 0
\\
\delta [\rho \gamma Z]^{\alpha\beta}  &=&  - \frac{i}{\Lambda}  \delta Z_{Z\gamma} \, M_Z^2  \, \eta^{\alpha\beta}  \, , 
\\
\delta [\rho Z Z]^{\alpha\beta}  &=&  - 2 \frac{i}{\Lambda}  (M_Z^2 \, \delta Z_{ZZ} + \delta M^2_Z)  \, \eta^{\alpha\beta} \, , 
\eea
where the counterterm coefficients are defined in terms of the 2-point functions of the fundamental fields as
\bea
\delta Z_{Z \gamma} = 2 \frac{\Sigma_T^{\gamma Z}(0)}{M_Z^2} \,, \quad \delta Z_{ZZ} = - Re \frac{\partial 
\Sigma_T^{ZZ}(k^2)}{\partial k^2} \bigg |_{k^2 = M_Z^2} \,, \quad
\delta M_Z^2 = Re \, \Sigma_T^{ZZ}(M_Z^2) \,,
\eea
and are defined in appendix \ref{SigmaSM}.
It follows then that the $\rho \gamma \gamma$ interaction must be finite, as one can find by a direct inspection of the 
$\Gamma^{\alpha\beta}_{\gamma \gamma}$ vertex, while the others require the subtraction of their divergences.

These counterterms are sufficient to remove the divergences of the completely cut graphs which do 
not contain a bilinear mixing, once we set on-shell  the external gauge lines. This occurs both for those diagrams which do not 
involve the terms of improvement and for those involving $T_I$. 
Regarding those contributions which involve the bilinear mixing on the external dilaton line, we encounter two different situations. \\
In the $\rho \gamma Z$ vertex the insertion of the bilinear mixing $\rho H$ generates a reducible diagram of the form Higgs/photon/Z 
whose renormalization is guaranteed, within the Standard Model, by the use of the Higgs/photon/Z counterterm
\bea
\delta [H\gamma Z]^{\alpha\beta} = \frac{e \, M_Z}{2 s_w c_w} \delta Z_{Z\gamma}\, \eta^{\alpha\beta} \,.
\eea
As a last case, we discuss the contribution to $\rho ZZ$ coming from the bilinear mixing, already mentioned above. The corrections on 
the dilaton line involve the dilaton/Higgs mixing $\Sigma_{\rho H}$, the Higgs self-energy $\Sigma_{HH}$ and the term of improvement 
$\Delta^{\alpha\beta}_{I\,,HZZ}$, which introduces the Higgs/Z/Z vertex (or $HZZ$) of the Standard Model. 
The Higgs self-energy and the $HZZ$ vertex, in the Standard Model, are renormalized with the usual counterterms
\bea
\delta [HH](k^2) 
&=& 
(\delta Z_H \, k^2 - M_H^2 \delta Z_H - \delta M_H^2) \, , 
\\
\delta [HZZ]^{\alpha\beta} 
&=& 
\frac{e \, M_Z}{s_w \, c_w} \bigg[ 1 + \delta Z_e + \frac{2 s_w^2 
- c_w^2}{c_w^2} \frac{\delta s_w}{s_w} + \frac{1}{2} \frac{\delta M_W^2}{M_W^2} + \frac{1}{2} \delta Z_H + \delta Z_{ZZ}  \bigg] 
\, \eta^{\alpha\beta} \,,
\eea
where
\bea
&& 
\delta Z_H = 
- Re \frac{\partial \Sigma_{HH}(k^2)}{\partial k^2} \bigg|_{k^2=M_H^2}\, ,\quad \delta M_H^2 = Re \Sigma_{HH}(M_H^2) \, , \quad
\delta Z_e = - \frac{1}{2} \delta Z_{\gamma \gamma} + \frac{s_w}{2 c_w} \delta Z_{Z \gamma} \,, \nn \\
&& \delta s_w = - \frac{c_w^2}{2 s_w} \left( \frac{\delta M_W^2}{M_W^2} - \frac{\delta M_Z^2}{M_Z^2} \right) \,, \quad \delta M_W^2 = 
Re \Sigma^{WW}_T(M_W^2) \,, \quad 
\delta Z_{\gamma \gamma} = - \frac{\partial \Sigma_T^{\gamma \gamma}(k^2)}{\partial k^2} \bigg|_{k^2 = 0} \,.
\eea
The self-energy $\Sigma_{\rho H}$ is defined by the minimal contribution generated by ${{T_{Min}}^\mu}_\mu$ 
and by a second term derived from ${{T_I}^\mu}_\mu$.
This second term, with the conformal coupling $\chi = \frac{1}{6}$, is necessary in order to ensure the renormalizability 
of the dilaton/Higgs mixing.
In fact, the use of the minimal EMT in the computation of this self-energy involves a divergence of the form
\bea
\delta [\rho H]_{Min} = - 4 \frac{i}{\Lambda}\, \delta t \, , \label{CThH}
\eea
with $\delta t$ fixed by the condition of cancellation of the Higgs tadpole $T_{ad}$ ($\delta t + T_{ad} = 0$).
A simple analysis of the divergences in $\Sigma_{Min, \, \rho H}$ shows that the counterterm given in eq.  (\ref{CThH}) is not 
sufficient to remove all the singularities of this correlator unless we also include the renormalization constant provided by the term of improvement 
which is given by
\bea
\delta [\rho H]_{I}(k) = - \frac{i}{\Lambda} 6 \, \chi \, v \bigg[ \delta v + \frac{1}{2} \delta Z_H \bigg]  k^2\,, \qquad 
\qquad \text{with} \quad \chi = \frac{1}{6} \,,
\eea
and
\bea
\delta v = v \bigg( \frac{1}{2} \frac{\delta M_W^2}{M_W^2} + \frac{\delta s_w}{s_w} - \delta Z_e \bigg) \,.
\eea
One can show explicitly that this counterterm indeed ensures the finiteness of $\Sigma_{\rho H}$.



\begin{figure}[t]
\centering
\subfigure[]{\includegraphics[scale=.7]{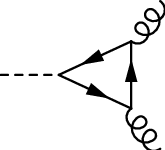}\label{QCD_NLOa}}
\hspace{.5cm}
\subfigure[]{\includegraphics[scale=.7]{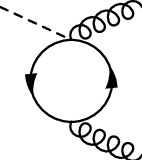}\label{QCD_NLOb}}
\hspace{.5cm}
\subfigure[]{\includegraphics[scale=.7]{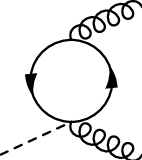}\label{QCD_NLOc}}
\hspace{.5cm}
\subfigure[]{\includegraphics[scale=.7]{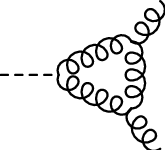}\label{QCD_NLOd}}
\hspace{.5cm}
\subfigure[]{\includegraphics[scale=.7]{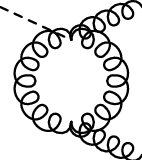}\label{QCD_NLOe}}
\hspace{.5cm}
\subfigure[]{\includegraphics[scale=.7]{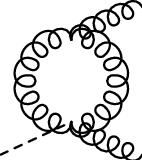}\label{QCD_NLOf}}\\
\hspace{.5cm}
\subfigure[]{\includegraphics[scale=.7]{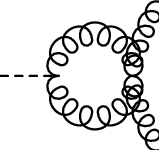}\label{QCD_NLOg}}
\hspace{.5cm}
\subfigure[]{\includegraphics[scale=.7]{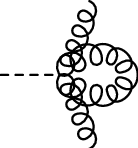}\label{QCD_NLOh}}
\hspace{.5cm}
\subfigure[]{\includegraphics[scale=.7]{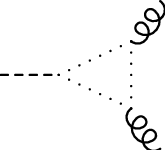}\label{QCD_NLOi}}
\hspace{.5cm}
\subfigure[]{\includegraphics[scale=.7]{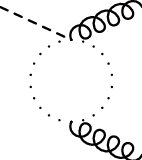}\label{QCD_NLOl}}
\hspace{.5cm}
\subfigure[]{\includegraphics[scale=.7]{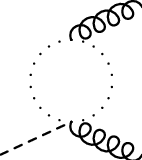}\label{QCD_NLOm}}
\caption{QCD vertices at next-to-leading order. In the on-shell gluon case only diagram (a)contributes.}
\label{QCD_NLO}
\end{figure}
%

\section{The off-shell dilaton-gluon-gluon vertex in QCD}

After a discussion of the leading corrections to the vertices involving one dilaton and two electroweak currents we investigate the 
interaction of a dilaton and two gluons beyond leading order, giving the expression of the full off-shell vertex. The corresponding 
interaction with two on-shell gluons has been computed in \cite{Giudice:2000av} and is simply given by the contributions of the 
anomaly and of the quark loop. We will come back to rediscuss the on-shell case in the second part of this work, where we will 
stress some specific perturbative features of this interaction.  

We show in fig. \ref{QCD_NLO} a list of the NLO QCD contributions to dilaton interactions. As we have just mentioned, in the 
$2$-gluon on-shell case one can show by an explicit computation that each of these contributions vanishes, except for diagram $(a)$, 
which is non-zero when a massive fermion runs in the loop. For this specific reason, in the parton model, the production of the 
dilaton in $pp$ collisions at the LHC is mediated by the diagram of gluon fusion, which involves a top quark in a loop. 

We find convenient to express the result of the off-shell $\Gamma^{\alpha\beta}_{g g}$ vertex in the form 
\bea \label{OffShellQCD}
\Gamma_{g g}^{\alpha \beta}(p,q) = 
\frac{i}{\Lambda}\, \bigg\{ A^{00}(p,q) \eta^{\alpha \beta} + A^{11}(p,q) p^{\alpha} p^{\beta} + A^{22}(p,q) q^{\alpha} q^{\beta} 
+ A^{12}(p,q) p^{\alpha}q^{\beta} + A^{21}(p,q) q^{\alpha}p^{\beta}\bigg\} \, ,
\eea
where $A^{ij}(p,q) = A^{ij}_g(p,q) + A^{ij}_q(p,q)$ which are diagonal ($\propto \delta^{a b}$) in colour space.\\
After an explicit computation, we find
\bea
A^{00}_g(p,q) 
&=& 
-\delta_{ab}\,\frac{g^2 \, N_C}{16 \pi^2} \bigg\{   
2 \left( p^2 +  q^2 + \frac{11}{3} p \cdot q \right) 
+ (p^2-q^2) \bigg[ \mathcal B_0(p^2,0,0) - \mathcal B_0(q^2,0,0)\bigg] 
\nonumber \\
&& 
+ \left(p^4 + q^4 - 2(p^2 + q^2) p \cdot q -6 p^2 q^2 \right) \mathcal C_0((p+q)^2, p^2, q^2,0,0,0)
\bigg\} \, , 
\nonumber 
\eea
\bea
A^{11}_g(p,q) = A^{22}_g(q,p)
&=&   
\delta_{ab} \, \frac{g^2 \, N_C \,}{16\,\pi^2 }  \, 
\bigg\{ 2 + \frac{1}{p \cdot q^2 - p^2 \, q^2} \, 
\bigg[ (p+q)^2 \, p \cdot q \, \mathcal B_0((p+q)^2,0,0)
\nonumber \\
&& 
- p^2 \, (q^2 + p \cdot q) \, \mathcal B_0(p^2,0,0)  - (2 p \cdot q^2 - p^2 \, q^2 + p \cdot q \, q^2) \mathcal B_0(q^2,0,0) 
\nonumber \\
&& 
+ \left( p^2 \, q^2 ( 5 q^2 - p^2) + 2 p \cdot q^2 (p^2 + p \cdot q - 2 \, q^2 ) \right) \, 
\mathcal C_0((p+q)^2, p^2, q^2,0,0,0) \bigg] \bigg\}  \, , 
\nonumber 
\eea
\bea
A^{12}_g(p,q) 
&=& 
\delta_{ab}\, \frac{g^2 \, N_C}{4 \pi^2} \, p \cdot q \, \mathcal C_0((p+q)^2, p^2, q^2,0,0,0)\, ,
\nonumber 
\eea
\bea
A^{21}_g(p,q) 
&=& 
\delta_{ab}\, \frac{g^2 \, N_C}{24\,\pi^2 }  \, 
\bigg\{11 + \frac{3}{2}\,\frac{1}{p \cdot q^2 - p^2 \, q^2} (p^2 + q^2) \, \bigg[(p^2 + p \cdot q)\mathcal B_0(p^2,0,0) 
\nonumber \\
&& 
+ (q^2 + p \cdot q) \mathcal B_0(q^2,0,0) - (p+q)^2  \mathcal B_0((p+q)^2,0,0)  
\nonumber \\
&& 
- \left( p \cdot q (p^2 + 4 p \cdot q + q^2) - 2 \, p^2 \, q^2 \right) \mathcal C_0((p+q)^2, p^2, q^2,0,0,0) \bigg]\bigg\} \, ,
\nonumber 
\eea
\bea
A^{00}_q(p,q) &=&
\delta_{ab} \frac{g^2}{8 \pi^2} \sum_{i=1}^{n_f} \bigg\{ \frac{2}{3} p\cdot q - 2\, m_i^2 
+ \frac{m_i^2}{p \cdot q^2 - p^2 q^2} \bigg[
p^2 \left(p \cdot q +q^2\right)  \mathcal B_0(p^2,m_i^2,m_i^2 ) \nn \\
&+&
q^2 \left(p^2+p \cdot q \right) \mathcal B_0(q^2,m_i^2,m_i^2)  -
\left(p^2 \left(p \cdot q+2 q^2\right) + p \cdot q \, q^2\right) \mathcal B_0( (p+q)^2,m_i^2,m_i^2) \nn \\
&-&
\left(p^2 q^2 \left(p^2+q^2 -4 m_i^2 \right)+4 m_i^2 \, p \cdot q^2 +4 p^2 \, q^2 \, p \cdot q -2 \, p \cdot q^3\right)
  \mathcal C_0 ((p+q)^2,p^2,q^2,m_i^2,m_i^2,m_i^2) 
\bigg] \bigg\} , \nn
\eea
%
\bea
A^{11}_q(p,q) &=& A^{22}_q(q,p) =
\delta_{ab} \frac{g^2}{ 8\,\pi^2} \sum_{i=1}^{n_f} \frac{2 \, m_i^2 \, q^2}{ p \cdot q^2 - p^2 q^2} \bigg\{ -2 + \frac{1}{p \cdot q^2 - p^2 q^2} \bigg[
\left(q^2 \left(p^2+3 \, p \cdot q \right) \right. \nn \\
&+& \left. 2 \, p \cdot q^2\right) \mathcal B_0(q^2,m_i^2,m_i^2)
+ \left(p^2 \left(3 \, p \cdot q+q^2\right)+2 p \cdot q^2\right)  \mathcal B_0(p^2,m_i^2,m_i^2) \nn \\
&-& \left(p^2 \left(3 \, p \cdot q+2 q^2\right) + p \cdot q \left(4 \, p \cdot q+3 q^2\right)\right) \mathcal B_0((p+q)^2,m_i^2,m_i^2) \nn \\
&-& \left(2 \, p \cdot q^2 \left(2 m_i^2+p^2+q^2\right) + p^2 q^2 \left(p^2+q^2-4 m_i^2 \right)  \right.) \nn \\
&+& \left. 4 p^2 q^2 \, p \cdot q +2 \, p \cdot q ^3\right)
   \mathcal C_0((p+q)^2,p^2,q^2,m_i^2,m_i^2,m_i^2)
\bigg] \bigg\} , \nn
\eea
\bea
A^{12}_q(p,q) &=&
\delta_{ab} \frac{g^2}{ 8\,\pi^2} \sum_{i=1}^{n_f} \frac{2 \, m_i^2 \, p \cdot q^2}{ p \cdot q^2 - p^2 q^2} \bigg\{ 2 + \frac{1}{p \cdot q^2 - p^2 q^2} \bigg[
-\left(q^2 \left(p^2+3 \, p \cdot q \right)+ 2 \, p \cdot q^2\right) \mathcal B_0(q^2,m_i^2,m_i^2) \nn \\
&-&
\left(p^2 \left(3 \, p \cdot q +q^2\right)+ 2 \, p \cdot q^2\right) \mathcal B_0(p^2,m_i^2,m_i^2) +
\left(p^2 \left(3 \, p \cdot q +2 q^2\right)+p \cdot q \left(4 \, p \cdot q +3 q^2 \right)\right) \nn \\
&\times& \mathcal B_0( (p+q)^2,m_i^2,m_i^2)+
\left(2 \, p \cdot q^2 \left(2 m_i^2+p^2+q^2\right)+p^2 q^2 \left(p^2+q^2 -4 m_i^2 \right) \right. \nn \\
&+& \left.  4 \, p^2 \, q^2 \, p \cdot q +2\, p \cdot q^3\right)
 \mathcal C_0((p+q)^2,p^2,q^2,m_i^2,m_i^2,m_i^2)
\bigg] \bigg\} , \nn 
\eea
\bea
A^{21}_q(p,q)  &=& \delta_{ab} \frac{g^2}{ 8\,\pi^2} \sum_{i=1}^{n_f} \bigg\{ - \frac{2}{3} + \frac{2 \, m_i^2 \, p \cdot q}{p \cdot q - p^2 q^2} + \frac{m_i^2}{(p \cdot q -p^2 q^2)^2} \bigg[
- p^2 \left(q^2 \left(2 p^2+3 \, p \cdot q \right)+p \cdot q ^2\right)  \nn \\ 
&\times& \mathcal B_0(p^2, m_i^2,m_i^2) 
- q^2 \left(p^2 \left(3 \, p \cdot q+2 q^2\right)+p \cdot q^2\right) \mathcal B_0( q^2, m_i^2,m_i^2) 
+ \left(2 p^4 q^2+p^2 \left(6 \, p \cdot q \, q^2 \right. \right. \nn \\ 
&+& \left. \left. p \cdot q^2+2 q^4\right)+p \cdot q^2 q^2\right) \mathcal B_0( (p+q)^2,m_i^2,m_i^2)
+ p \cdot q \left(p^2 q^2 \left(3 p^2+3 q^2 - 4 m_i^2\right) \right. \nn \\
&+& \left. 4 \,   m_i^2 \, p \cdot q^2
 +8 p^2 \, q^2 \, p \cdot q -2 \, p \cdot q ^3\right)
   \mathcal C_0( (p+q)^2,p^2,q^2,m_i^2,m_i^2,m_i^2)
\bigg] \bigg\}, 
\eea
where $N_C$ is the number of colours, $n_f$ is the number of flavour and $m_i$ the mass of the quark.
In the on-shell gluon case, eq. (\ref{OffShellQCD}) reproduces the same interaction responsible for Higgs production at LHC augmented by 
an anomaly term.
This is given by
\bea \label{OnShellQCD}
\Gamma^{\alpha\beta}_{gg}(p,q) =  
\frac{i}{\Lambda} \Phi(s) \, u^{\alpha\beta}(p,q) \, ,
\eea
with $u^{\alpha\beta}(p,q)$ defined in eq. (\ref{utensor}), and with the gluon/quark contributions included in the $\Phi(s)$ form factor ($s = k^2 = (p+q)^2$) 
\bea \label{OnShellPhi}
\Phi(s) 
= 
- \delta^{ab} \frac{g^2}{24\,\pi^2} \, \bigg\{
 \, \left(11\, N_C - 2\, n_f \right) + 12 \, \sum_{i=1}^{n_f} m_i^2 \, 
\bigg[ \frac{1}{s} \, - \, \frac{1} {2 }\mathcal C_0 (s, 0, 0, m_i^2, m_i^2, m_i^2) \bigg(1-\frac{4 m_i^2}{ s}\bigg) \bigg]
\bigg\} \, ,
\eea
where the first mass independent terms represent the contribution of the anomaly, while the others are the explicit mass corrections. \\ 
The decay rate of a dilaton in two gluons can be evaluated from the on-shell limit in eq. (\ref{OnShellQCD}) and it is given by
\bea
\Gamma(\rho \rightarrow gg) 
&=&
\frac{\alpha_s^2\,m_\rho^3}{32\,\pi^3 \Lambda^2} \, \bigg| \beta_{QCD} + x_t\left[1 + (1-x_t)\,f(x_t) \right] \bigg|^2 \,,
\label{Phigg}
\eea
where we have taken the top quark as the only massive fermion and $x_i$ and $f(x_i)$ are defined in eq.  (\ref{x}) and eq.  (\ref{fx})
respectively. Moreover we have set
$\beta_{QCD} = 11 N_C/3 - 2 \, n_f/3$ for the QCD $\beta$ function.

\section{Non-gravitational dilatons from scale invariant extensions of the Standard Model}
\label{NonGrav} 

As we have pointed out in the introduction, a dilaton may appear in the spectrum of different extensions of the Standard Model not 
only as a result of the compactification of extra spacetime dimensions, but also as an effective state, related to the breaking of
dilatation symmetry. In this respect, notice that in its actual formulation 
the Standard Model is not scale invariant, but can be such, at classical level, if we slightly modify the scalar potential with the 
introduction of a dynamical field $\Sigma$ that 
allows to restore this symmetry and acquires a vacuum expectation value. This task is accomplished by the replacement of every 
dimensionful parameter $m$ according to $m \rightarrow m \frac{\Sigma}{\Lambda}$, where $\Lambda$ is the classical conformal 
breaking scale. 
In the case of the Standard Model, classical scale invariance can be easily accommodated with a simple change of the scalar potential. 

This is defined, obviously, modulo a constant, therefore we may consider, for instance, two equivalent choices 
\bea
V_1(H, H^\dagger)&=& - \mu^2 H^\dagger H +\lambda(H^\dagger H)^2 =
\lambda \left( H^\dagger H - \frac{\mu^2}{2\lambda}\right)^2 - \frac{\mu^4}{4 \lambda}\nonumber \\
V_2(H,H^\dagger)&=&\lambda \left( H^\dagger H - \frac{\mu^2}{2\lambda}\right)^2
\eea
which give two {\em different} scale invariant extensions 
\bea
V_1(H,H^\dagger, \Sigma)&=&- \frac{\mu^2\Sigma^2}{\Lambda^2} H^\dagger H +\lambda(H^\dagger H)^2 \nonumber \\
V_2(H,H^\dagger, \Sigma)&=& \lambda \left( H^\dagger H - \frac{\mu^2\Sigma^2}{2\lambda \Lambda^2}\right)^2 \,,
\eea 
where $H$ is the Higgs doublet, $\lambda$ is its dimensionless coupling constant, while $\mu$ has the dimension of a mass and, 
therefore, is the only term involved in the scale invariant extension. More details of this analysis can be found in 
the next section.\\
The invariance of the potential under the addition of constant terms, typical of any Lagrangian, is lifted once we 
require the presence of a dilatation symmetry. Only the second choice $(V_2)$ guarantees the existence of a stable ground state 
characterized by a spontaneously 
broken phase. In $V_2$ we parameterize the Higgs, as usual, around the electroweak vev $v$ as in eq. (\ref{Higgsparam}), 
and indicate with $\Lambda$ the vev of the dilaton field $\Sigma = \Lambda + \rho$, 
and we have set $\phi^+ = \phi = 0$ in the unitary gauge. \\
The potential $V_2$ has a massless mode due to the existence of a flat direction. 
Performing a diagonalization of the mass matrix we define the two mass eigenstates $\rho_0$ and $h_0$, which are given by 
\beq
 \left( \begin{array}{c}
 {\rho_0}\\
  h_0 \\
  \end{array} \right)
 =\left( \begin{array}{cc}
\cos\alpha & \sin\alpha \\
-\sin\alpha & \cos\alpha  \\
 \end{array} \right)
 \left( \begin{array}{c}
  \rho\\
 {h} \\
  \end{array} \right)
\eeq
with 
\beq
\cos\alpha=\frac{1}{\sqrt{1 + v^2/\Lambda^2}}\qquad \qquad  \sin\alpha=\frac{1}{\sqrt{1 + \Lambda^2/v^2}}.
\eeq
We denote with ${\rho_0}$ the massless dilaton generated by this potential, while 
$h_0$ will describe a massive scalar, interpreted as a new Higgs field, whose mass is given by  
\beq 
m_{h_0}^2= 2\lambda v^2 \left( 1 +\frac{v^2}{\Lambda^2}\right) \qquad \textrm{with} \qquad v^2=\frac{\mu^2}{\lambda},
\eeq
and with $m_h^2=2 \lambda v^2$ being the mass of the Standard Model Higgs.
The Higgs mass, in  this case, is corrected by the new scale of the spontaneous breaking of the dilatation symmetry ($\Lambda$), 
which remains a free parameter. 
 
The vacuum degeneracy of the scale invariant model  can be lifted by the introduction of 
extra (explicit breaking) terms which give a small mass to the dilaton field.
To remove such degeneracy, one can introduce, for instance, the term
\beq
\mathcal{L}_{break} 
= \frac{1}{2} m_{\rho}^2 {\rho}^2 + \frac{1}{3!}\, {m_{\rho}^2} \frac{{\rho}^3}{\Lambda} + \dots \, ,
\eeq
where $m_{\rho}$ represents the dilaton mass.

It is clear that in this approach the coupling of the dilaton to the anomaly has to be added by hand.
The obvious question to address, at this point, is if one can identify in the effective action of the Standard Model 
an effective state which may interpolate between the dilatation current of the same model and the final state with two
neutral currents, for example with two photons. The role of the following sections will be to show 
rigorously that such a state can be identified in ordinary perturbation theory in the form of an anomaly pole.

We will interpret this scalar exchange as a composite state whose interactions with the rest of the Standard Model are 
defined by the conditions of scale and gauge invariance. In this respect, the Standard Model Lagrangian, enlarged by the introduction 
of a potential of the form $V_2(H,H^\dagger,\Sigma)$, which is expected to capture the dynamics of this pseudo-Goldstone mode, could
take the role of a workable model useful for a phenomenological analysis.  
We will show rigorously that this state couples to the conformal anomaly by a direct analysis of the $J_DVV$ correlator, 
in the form of an anomaly pole, with $J_D$ and $V$ being the dilatation and a vector current respectively.
Usual polology arguments support the fact that a pole  in a correlation function is there to indicate that a specific state can be created by 
a field operator in the Lagrangian of the theory, or, alternatively, as a composite particle of the same elementary fields.  

Obviously, a perturbative hint of the existence of such intermediate state does not correspond to a complete 
description of the state, in the same way as the discovery of an anomaly pole in the $AVV$ correlator of QCD (with $A$ being the 
axial current) is not equivalent to a proof of the existence of the pion.

\subsection{A classical scale invariant Lagrangian with a dilaton field}
\label{classical}

In this section we briefly describe the construction of a scale invariant theory to clarify some of the issues
concerning the coupling of a dilaton. In particular, the example has the goal to 
illustrate that in a classical scale invariant extension of a given theory, the dilaton couples only to operators which are mass dependent, 
and thus scale breaking, before the extension. We take the case of a fundamental dilaton field (not a composite) introduced 
in this type of extensions.

A scale invariant extension of a given Lagrangian can be obtained if we promote all the dimensionful constants to dynamical 
fields. 
We illustrate this point in the case of a simple interacting scalar field theory incorporating the Higgs mechanism. 
At a second stage we will derive the structure of the dilaton interaction at order $1/\Lambda$, where 
$\Lambda$ is the scale characterizing the spontaneous breaking of the dilatation symmetry.

Our toy model consists in a real singlet scalar with a potential of the kind of $V_2(\phi)$ introduced in section \ref{NonGrav},
\beq
\label{original}
\mathcal L = \frac{1}{2} (\partial \phi)^2 - V_2(\phi) =
\frac{1}{2} (\partial \phi)^2 + \frac{\mu^2}{2}\, \phi^2 - \lambda\, \frac{\phi^4}{4} - \frac{\mu^4}{4\,\lambda}\, ,
\eeq
obeying the classical equation of motion
\beq \label{scalarEOM}
\square \phi = \mu^2\,\phi - \lambda\, \phi^3\, .
\eeq
Obviously this theory is not scale invariant due to the appearance of the mass term $\mu$. This feature is reflected in the trace of the EMT.
Indeed the canonical EMT of such a theory and its trace are
\bea
T^{\mu\nu}_{c}(\phi) 
&=& 
\partial^\mu \phi\, \partial^\nu \phi 
- \frac{1}{2}\,\eta^{\mu\nu} \bigg[ (\partial \phi)^2 + \mu^2 \,\phi^2 
-  \lambda\,\frac{\phi^4}{2} -  \frac{\mu^4}{2\,\lambda} \bigg] \, ,
\nn \\
{T_{c}^\mu}_\mu(\phi) &=& 
- (\partial\phi)^2 - 2\, \mu^2 \,\phi^2 +\lambda\, \phi^4 + \frac{\mu^4}{\lambda} \, .
\eea
Improving the EMT of the scalar field in such a way as to make its trace proportional only to the scale breaking parameter,
i.e. the mass $\mu$, which is done by adding an extra contribution $T_I^{\mu\nu}(\phi, \chi)$,
\beq
T_I^{\mu\nu}(\phi,\chi)= \chi\, \left(\eta^{\mu\nu} \square \phi^2 - \partial^\mu \partial^\nu \phi^2 \right) \, ,
\eeq
where the parameter $\chi$ is left generic.
The combination of the canonical plus the improvement EMT, 
$T^{\mu\nu} \equiv T_c^{\mu\nu} + T_I^{\mu\nu}$ has the off-shell trace
\beq
{T^\mu}_\mu(\phi,\chi)= 
(\partial\phi)^2\, \left( 6 \chi - 1 \right) - 2\, \mu^2\, \phi^2 
+ \lambda\, \phi^4 + \frac{\mu^4}{\lambda} + 6 \chi \phi\, \square \phi\, .
\eeq
Using the equation of motion (\ref{scalarEOM}) and choosing $\chi=1/6$ the trace relation given above 
becomes proportional uniquely to the scale breaking term $\mu$  
\beq \label{ImprovenTrace}
{T^\mu}_\mu(\phi,1/6) = - \mu^2 \phi^2 + \frac{\mu^4}{\lambda} \, .
\eeq
The scale invariant extension of the Lagrangian given in eq. (\ref{original}) is achieved by promoting the mass terms to dynamical fields by the replacement 
\beq
\label{rep}
\mu \to \frac{\mu}{\Lambda} \, \Sigma,
\eeq
obtaining
\beq
\label{sigmaphi}
\mathcal L = 
\frac{1}{2}\, (\partial \phi)^2 +\frac{1}{2} (\partial \Sigma)^2 
+  \frac{ \mu^2}{2\,\Lambda^2}\, \Sigma^2\, \phi^2 - \lambda \frac{\phi^4}{4}
-  \frac{\mu^4}{4\,\lambda \, \Lambda^4}\, \Sigma^4
\eeq
where we have used eq. (\ref{rep}) and introduced a kinetic term for the dilaton $\Sigma$. 
Obviously, the new Lagrangian is dilatation invariant, as one can see from the trace of the improven EMT 
\beq
{T^{\mu}}_{\mu}(\phi,\Sigma,\chi,\chi') = \left( 6\, \chi - 1 \right)\, (\partial\phi)^2
+ \left( 6 \chi^\prime -1\right)\, (\partial\Sigma)^2 
+ 6 \chi\, \phi\, \square \phi + 6 \chi^\prime\, \Sigma\, \square \Sigma 
- 2\, \frac{\mu^2}{\Lambda^2}\, \Sigma^2\, \phi^2 + \lambda\, \phi^4
+ \frac{1}{\lambda}\,\frac{\mu^4}{\Lambda^4}\,\Sigma^4 \,,
\eeq
which vanishes upon using the equations of motion for the $\Sigma$ and $\phi$ fields,
\bea
\square \phi &=& \frac{\mu^2}{\Lambda^2}\, \Sigma^2\, \phi -  \lambda\, \phi^3\, ,
\nn \\
\square \Sigma &=& \frac{\mu^2}{\Lambda^2}\, \Sigma\, \phi^2 - \frac{1}{\lambda}\, \frac{\mu^4}{\Lambda^4}\,\Sigma^3  \, ,
\eea
and setting the $\chi, \chi'$ parameters at the special value $\chi=\chi^\prime=1/6$. 

As we have already discussed in section \ref{NonGrav}, the scalar potential $V_2$ allows to perform 
the spontaneous breaking of the scale symmetry around a stable minimum point, 
giving the dilaton and the scalar field the vacuum expectation values $\Lambda$ and $v$ respectively
\beq
\Sigma  =  \Lambda + \rho \, , \quad \phi = v + h\, .
\eeq
For our present purposes, it is enough to expand the Lagrangian (\ref{sigmaphi}) around the vev for the dilaton field, 
as we are interested in the structure of the couplings of its fluctuation $\rho$
\beq \label{Manifest}
\mathcal L = \frac{1}{2}\, (\partial\phi)^2 + \frac{1}{2}\, (\partial\rho)^2 + \frac{\mu^2}{2}\, \phi^2 
- \lambda \, \frac{\phi^4}{4} - \frac{\mu^4}{4\,\lambda}
- \frac{\rho}{\Lambda}\,\left(- \mu^2\, \phi^2  + \frac{\mu^4}{\lambda}\right) + \dots\, ,
\eeq
where the ellipsis refer to terms that are higher order in $1/\Lambda$.
It is clear, from (\ref{ImprovenTrace}) and (\ref{Manifest}), that one can write an dilaton Lagrangian
at order $1/\Lambda$, as
\beq \label{RhoInteraction}
\mathcal L_{\rho} = (\partial\rho)^2 - \frac{\rho}{\Lambda}\, {T^{\mu}}_{\mu}(\phi,1/6) + \dots\, ,
\eeq
where the equations of motion have been used in the trace of the energy-momentum tensor.
Expanding the scalar field around $v$ would render the previous equation more complicated and we omit it for definiteness.
We only have to mention that a mixing term $\sim \rho\, h$ shows up and it has to be removed diagonalizing 
the mass matrix, switching from interaction to mass eigenstates exactly in the way we discussed in section \ref{NonGrav}, 
to which we refer for the details. 

It is clear, from this simple analysis, that a dilaton, in general, does not couple to the anomaly,
but only to the sources of explicit breaking of scale invariance, i.e. to the mass terms.
The coupling of a dilaton to an anomaly is, on the other hand, necessary, 
if the state is interpreted as a composite pseudo Nambu-Goldstone mode of the dilatation symmetry.
Thus, this coupling has to be introduced by hand, in strict analogy with the chiral case.

\subsection{The $J_DVV$ and $TVV$ vertices}

This effective degree of freedom emerges both from the spectral analysis of the $TVV$ \cite{Giannotti:2008cv, Armillis:2009pq} and, 
as we are now going to show, of the  $J_D VV$ 
vertices, being the two vertices closely related. 
We recall that the dilatation current can be defined as 
\beq
J_D^\mu(z)= z_\nu T^{\mu \nu}(z) \qquad \textrm{with}  \qquad \partial\cdot J_D = {T^\mu}_\mu. 
\label{def}
\eeq
The $T^{\mu\nu}$ has to be symmetric and on-shell traceless for a classical scale invariant theory, and includes, at
quantum level, the contribution from the trace anomaly together with the additional terms describing the explicit breaking of the 
dilatation symmetry. 
The separation between the anomalous and the explicit contributions to the breaking of dilatation symmetry is present 
in all the analysis that we have performed on the $TVV$ vertex in dimensional regularization. 
In this respect, the analogy between these types of correlators and the $AVV$ diagram of the chiral anomaly goes 
quite far, since in the $AVV$ case such a separation has been shown to hold in the Longitudinal/Transverse 
(L/T) solution of the anomalous Ward identities \cite{Knecht:2003xy, Jegerlehner:2005fs, Armillis:2009sm}. 
This has been verified in perturbation theory in the same scheme.

We recall that the $U(1)_A$ current is characterized by an anomaly pole which describes the interaction between the 
Nambu-Goldstone mode, generated by the breaking of the chiral symmetry, and the gauge currents. 
In momentum space this corresponds to the nonlocal vertex 
\beq
\label{AVVpole}
V_{\textrm{anom}}^{\lambda \mu\nu}(k,p,q)=  \frac{k^\lambda}{k^2}\epsilon^{\mu \nu \alpha \beta}p_\alpha q_\beta +...
\eeq
with $k$ being the momentum of the axial-vector current and $p$ and $q$ the momenta of the two photons.
In the equation above, the ellipsis refer to terms which are suppressed at large energy. 
In this regime, this allows to distinguish the operator accounting for the chiral anomaly (i.e. $\square^{-1}$ in coordinate space)
from the contributions due to mass corrections. 
Polology arguments can be used to relate the appearance of such a pole to the pion state 
around the scale of chiral symmetry breaking. 

To identify the corresponding pole in the dilatation current of the $J_D VV$ correlator at zero momentum transfer, one can follow the 
analysis of \cite{Horejsi:1997yn}, where it is shown that the appearance of the trace anomaly is related to the presence of a
superconvergent sum rule in the spectral density of this correlator. At non-zero momentum transfer the derivation of a similar
behaviour can be obtained by an explicit computation of the spectral density of the 
$TVV$ vertex \cite{Giannotti:2008cv} or of the entire correlator, as done for QED and QCD 
\cite{Armillis:2009pq, Armillis:2010qk} and as we will show next.

Using the relation between $J_D^\mu$ and the EMT $T^{\mu\nu}$ we introduce the $J_DVV$ correlator 
\bea
\Gamma_D^{\mu\alpha\beta}(k,p)
&\equiv& 
\int d^4 z\, d^4 x\, e^{-i k \cdot z + i p \cdot x}\,
\left\langle J^\mu_D(z) V^\alpha(x)V^\beta(0)\right\rangle 
\label{gammagg}
\eea
which can be related to the $TVV$ correlator 
\bea
\Gamma^{\mu\nu\alpha\beta}(k,p)&\equiv& \int d^4 z\, d^4 x\, e^{-i k \cdot z + i p \cdot x}\, 
\left\langle T^{\mu \nu}(z) V^\alpha(x) V^\beta(0)\right\rangle 
\eea
according to
\bea
\Gamma_D^{\mu\alpha\beta}(k,p)&=& 
i \frac{\partial}{\partial k^\nu}\Gamma^{\mu\nu\alpha\beta}(k,p) \,.
\eea
As we have already mentioned, this equation allows us to identify a pole term in the $J_DVV$ diagram from the corresponding pole 
structure in the $TVV$ vertex. 
In the following we will show the emergence of the anomaly poles in the QED and QCD cases. 

\subsection{The dilaton anomaly pole in the QED case}
\label{SecPoleQED}

For definiteness, it is convenient to briefly review the characterization of the $TVV$ vertex in the QED case with a massive fermion 
(see \cite{Armillis:2009pq} for more details).
The full amplitude  can be expanded in a specific basis of $13$ tensors first identified in \cite{Giannotti:2008cv}
\bea
\Gamma^{\mu\nu\alpha\beta}(p,q) =  \, \sum_{i=1}^{13} F_i (s; s_1, s_2,m^2)\phi_i^{\mu\nu\alpha\beta}(p,q)\,,
\label{Gamt}
\eea
where the $13$ invariant amplitudes $F_i$ are functions of the kinematic invariants $s=k^2=(p+q)^2$, \mbox{$s_1=p^2$}, $s_2=q^2$, 
with $p$ and $q$ the momenta of the external photons,  
and of the internal fermion mass $m$. The list of the tensor structures $\phi_i$  can be found in \cite{Giannotti:2008cv}. The number 
of these form factors reduces from $13$ to $3$ in the case of on-shell photons.  For our purposes, being interested in the appearance 
of the anomaly poles, we only need the contributions that generate a non zero trace. These come from the tensors 
$\phi_1^{\mu\nu\alpha\beta}$ and $\phi_2^{\mu\nu\alpha\beta}$ which are
\bea	
\phi_1^{\mu\nu\alpha\beta}&=&\left(k^2 \eta^{\mu\nu} - k^{\mu } k^{\nu}\right) u^{\alpha\beta}(p,q), \nn \\
\phi_2^{\mu\nu\alpha\beta}&=&\left(k^2 \eta^{\mu\nu} - k^{\mu } k^{\nu}\right) w^{\alpha\beta}(p,q),
\eea
where
\bea
&&u^{\alpha\beta}(p,q) \equiv (p\cdot q) \eta^{\alpha\beta} - q^{\alpha}p^{\beta}\,,\nonumber \\
&&w^{\alpha\beta}(p,q) \equiv p^2 q^2 \eta^{\alpha\beta} + (p\cdot q) p^{\alpha}q^{\beta}
- q^2 p^{\alpha}p^{\beta} - p^2 q^{\alpha}q^{\beta}.\,
\label{uwdef}
\eea
For two on-shell final state photons ($s_1=s_2=0$) and a massive fermion we obtain
\beq
\label{oom}
{F_1 (s;\,0,\,0,\,m^2)} = 
F_{1\, pole} \,  + \, \frac{e^2 \,   m^2}{3 \, \pi ^2 \, s^2} \, 
- \frac{e^2 \, m^2}{3 \, \pi^2 \, s}  \, \mathcal C_0 (s, 0, 0, m^2,m^2,m^2) 
\bigg[\frac{1}{2}-\frac{2 \,m^2}{ s}\bigg],  \\
\eeq
where 
\beq F_{1\, pole}=- \frac{e^2 }{18 \, \pi^2 s} 
\eeq
and the scalar $3$-point function $ \mathcal C_0 (s, 0,0,m^2,m^2,m^2) $ is given by 
\beq
\mathcal C_0 (s, 0,0,m^2,m^2,m^2) = \frac{1}{2 s} \, \log^2 \frac{a_3+1}{a_3-1}, \qquad \mbox{with} \quad a_3 = \sqrt {1-4m^2/s} \,.
\eeq
In the massless fermion case two properties of this expansion are noteworthy: 1) the trace anomaly 
takes contribution only from a single tensor structure $(\phi_1)$ and invariant amplitude $(F_1)$ which coincides with the pole term; 
2) the residue of this pole as $s\to 0$ is non-zero, showing that the pole is coupled in the infrared. Notice that the form factor 
$F_2$, which in general gives a non-zero contribution to the trace in the presence of mass terms, is multiplied by a tensor 
structure ($\phi_2$) which {\em vanishes} when the two photons are on-shell.
Therefore, similarly to the case of the chiral anomaly, also in this case the anomaly is {\em entirely} given by the appearance 
of an anomaly pole. 
We stress that this result is found to be exact in dimensional regularization, which is a mass independent scheme:
at perturbative level, the anomalous breaking of the dilatation symmetry, related to an anomaly pole in the spectrum of all
the gauge-invariant correlators studied in this work, is separated from the sources of {\em explicit} breaking. The latter are related to the mass 
parameters and/or to the gauge bosons virtualities $p^2$ and $q^2$. 

\begin{figure}[t]
\begin{center}
\includegraphics[scale=1.2]{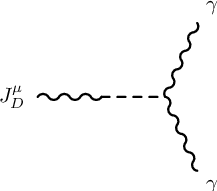}
\end{center}
\caption{Exchange of a dilaton pole mediated by the $J_D V V$ correlator. }
\label{dilatonpole}
\end{figure}

To analyse the implications of the pole behaviour discussed so far for the $TVV$ vertex and its connection with the $J_DVV$
correlator, we limit our attention on the anomalous contribution ($F_1\, \phi_1^{\mu\nu\alpha\beta}$), 
which we rewrite in the form
\beq
\Gamma_{pole}^{\mu\nu\alpha \beta}(k,p)\equiv
- \frac{e^2}{18\pi^2}\frac{1}{k^2}\left( \eta^{\mu \nu} k^2 - k^\mu k^\nu\right) u^{\alpha \beta}(p,q) \, , 
\qquad q = k - p \, .
\label{uref}
\eeq
This implies that the $J_D VV$ correlator acquires a pole as well
\bea
\Gamma_{D\, pole}^{\mu\alpha\beta}
&=& 
- i \frac{e^2}{18 \pi^2}\frac{\partial}{\partial k^\nu}
\left[ \frac{1}{k^2}\, \left( \eta^{\mu\nu } k^2 - k^{\mu} k^{\nu}\right) u^{\alpha \beta}(p,k-p) \right]
\eea
and acting with the derivative on the right hand side we finally obtain 
\beq
\Gamma_{D \, pole}^{\mu\alpha\beta}(k,p)= 
i\, \frac{e^2}{6 \pi^2}\frac{k^\mu}{k^2}u^{\alpha \beta}(p,k-p) - i \frac{e^2}{18 \pi^2}\frac{1}{k^2}
\left( \eta^{\mu\nu } k^2 - k^{\mu} k^{\nu}\right)\frac{\partial}{\partial k_\nu}u^{\alpha \beta}(p,k-p).
\eeq
Notice that the first contribution on the right hand side of the previous equation corresponds to an anomaly pole, 
shown pictorially in fig. \ref{dilatonpole}. In fact, by taking a derivative of 
the dilatation current only this term will contribute to the corresponding Ward identity
\beq
k_\mu \, \Gamma_D^{\mu\alpha\beta}(k,p) = i \frac{e^2}{6 \pi^2}u^{\alpha \beta}(p,k-p),
\label{res}
\eeq
which is the expression in momentum space of the usual relation $\partial J_D\sim FF$, while the second term trivially vanishes.  
Notice that the pole in (\ref{res}) has disappeared, and we are left just with its residue on the r.h.s., or, 
equivalently, the pole is removed in eq.  (\ref{uref}) if we trace the two indices $(\mu,\nu)$.

%

\subsection{The dilaton anomaly pole in the QCD case}

The analysis presented for the dilatation current of QED can be immediately generalized to the case of QCD.
Following a similar reasoning, we expand the on-shell $TVV$ vertex, with $V$ denoting now the gluon, as
\bea
\Gamma^{\mu\nu\alpha\beta}(p,q) = \delta^{a b} \sum_{i = 1}^{3} \Phi_i(s; 0,0) \, t_i^{\, \mu\nu\alpha\beta}(p,q) \qquad \mbox{with} 
\quad p^2 = q^2 = 0 \, ,
\eea
with the tensor basis given by
\bea
t_1^{\, \mu \nu \a \b} (p,q) &=&
(s \, \eta^{\mu\nu} - k^{\mu}k^{\nu}) \, u^{\a \b} (p,q),
\label{widetilde1} \nn \\
t_2^{\, \mu \nu \a \b} (p,q) &=& - 2 \, u^{\a \b} (p,q) \left[ s \, \eta^{\mu \nu} + 2 (p^\mu \, p^\nu + q^\mu \, q^\nu )
- 4 \, (p^\mu \, q^\nu + q^\mu \, p^\nu) \right],
\label{widetilde2} \nn \\
t^{\, \mu \nu \alpha \beta}_{3} (p,q) &=&
\big(p^{\mu} q^{\nu} + p^{\nu} q^{\mu}\big)g^{\alpha\beta}
+ \frac{s}{2} \left(\eta^{\alpha\nu} \eta^{\beta\mu} + \eta^{\alpha\mu} \eta^{\beta\nu}\right) \nn \\
&&
- \eta^{\mu\nu} \left(\frac{s}{2} \eta^{\alpha \beta}- q^{\alpha} p^{\beta}\right)
-\left(\eta^{\beta\nu} p^{\mu} + \eta^{\beta\mu} p^{\nu}\right)q^{\alpha}
- \big (\eta^{\alpha\nu} q^{\mu} + \eta^{\alpha\mu} q^{\nu }\big)p^{\beta},
\label{widetilde3}
\eea
where $\delta^{a b}$ is the diagonal matrix in colour space. 
Again we have $s=k^2 = (p+q)^2$, with the virtualities of the two gluons being $p^2 = q^2 = 0$.
Notice that in the massless fermion limit only the first ($t_1$) of these 3 form factors contributes to the anomaly.
The corresponding on-shell form factors with massive quarks are
\bea
\Phi_{1}(s;0,0) &=& - \frac{g^2}{72 \pi^2 \,s}\left(2 n_f - 11 N_C \right) + \frac{g^2}{6 \pi^2}\sum_{i=1}^{n_f} m_i^2 \, \bigg\{ 
\frac{1}{s^2} \, - \, \frac{1} {2 s}\mathcal C_0 (s, 0, 0, m_i^2,m_i^2,m_i^2)
\bigg[1-\frac{4 m_i^2}{ s}\bigg] \bigg\}, \,
\label{Phi1} \nn \\
\Phi_{2}(s;0,0) &=& - \frac{g^2}{288 \pi^2 \,s}\left(n_f - N_C\right) \nn \\
&-& \frac{g^2}{24 \pi^2} \sum_{i=1}^{n_f} m_i^2 \, \bigg\{ \frac{1}{s^2}
+ \frac{ 3}{ s^2} \mathcal D_0 (s, 0, 0, m_i^2, m_i^2)
+ \frac{ 1}{s } \mathcal C_0(s, 0, 0, m_i^2,m_i^2,m_i^2 )\, \left[ 1 + \frac{2 m_i^2}{s}\right]\bigg\},
\label{Phi2} \nn \\
\Phi_{3}(s;0,0) &=& \frac{g^2}{288 \pi^2}\left(11 n_f - 65 N_C\right) - \frac{g^2 \, N_C}{8 \pi^2} 
\bigg[ \frac{11}{6} \mathcal B_0(s,0,0) - \mathcal B_0(0,0,0) +  s  \,\mathcal C_0(s,0,0,0,0,0) \bigg] \nn \\
&+&   \frac{g^2}{8 \pi^2} \sum_{i=1}^{n_f}\bigg\{  \frac{1}{3}\mathcal B_0(s, m_i^2,m_i^2) + m_i^2 \, \bigg[
\frac{1}{s}
 + \frac{5}{3 s}  \mathcal D_0 (s, 0, 0, m_i^2) + \mathcal C_0 (s, 0,0,m_i^2,m_i^2,m_i^2) \,\left[1 + \frac{2 m_i^2}{s}\right]
\bigg]\bigg\} , \nn \\ 
\label{Phi3}
\eea
where $m_i$ denotes the quark mass, and we have summed over the fermion flavours $(n_f)$, while $N_C$ denotes the number of colours.
Notice the appearance of the $1/s$ pole in $\Phi_1$, which saturates the contribution to the trace 
anomaly in the massless limit which becomes
\beq
\Phi_{1}(s;0,0) = - \frac{g^2}{72 \pi^2 \,s}\left(2 n_f - 11 N_C\right).
\label{polepole}
\eeq
As for the QED case, this is the only invariant amplitude which contributes to the anomalous trace part of the correlator.
The pole completely accounts for the trace anomaly and is clearly inherited by the QCD dilatation current, 
for the same reasonings discussed above.

\subsection{Mass corrections to the dilaton pole}

The discussion of the mass corrections to the massless dilaton can follow quite closely the strategy adopted in 
the pion case using partially conserved axial currents (PCAC) techniques. Also in this case, as for PCAC in the past, 
one can assume a partially conserved dilaton current (PCDC) in order to relate the decay amplitude of the dilaton $f_\rho$ 
to its mass $m_\rho$ and to the vacuum energy. \\
For this goal we define the 1-particle transition amplitudes for the dilatation current and the EMT between the vacuum and a dilaton 
state with momentum $p_\mu$
\bea
\langle 0| J^\mu_D(x) |\rho, p \rangle 
&=& 
- i \, f_\rho \, p^\mu \, e^{-i p \cdot x} 
\nn \\
\langle 0| T^{\mu\nu}(x) |\rho, p \rangle 
&=&
\frac{f_\rho}{3} \, \left( p^\mu p^\nu - \eta^{\mu\nu} \, p^2\right)\, e^{-i p \cdot x},
\eea
both of them giving 
\beq
\label{rel1}
\partial_\mu \langle 0| J^\mu_D(x) |\rho, p\rangle = \eta_{\mu\nu} \langle 0| T^{\mu\nu}(x) |\rho, p \rangle = - f_\rho \, m_\rho^2 \, e^{-i p \cdot x}.
\eeq
We introduce the dilaton interpolating field $\rho(x)$ via a PCDC relation
\beq
\partial_\mu J^\mu_D(x) = - f_\rho \, m_\rho^2 \, \rho(x)
\label{pcdcrel}
\eeq
with 
\beq
\langle 0|\rho(x)|\rho, p\rangle = e^{-i p \cdot x}
\eeq
and the matrix element 
\beq
\mathcal A^\mu(q)= \int d^4 x \, e^{i q \cdot x} \,\langle 0| T \{ J^\mu_D(x) {T^\alpha}_\alpha(0) \} |0 \rangle,
\label{www}
\eeq
where $T\{\ldots\}$ denotes the time ordered product. \\
Using dilaton pole dominance we can rewrite the contraction of 	$q_\mu$ with this correlator as 
\bea
\label{inter}
\lim_{q_\mu \to 0} q_\mu \, \mathcal A^\mu(q)  = f_\rho \,\left\langle \rho, q=0| {T^\alpha}_\alpha(0) |0\right\rangle \,, 
\eea
where the soft limit $q_\mu \to 0$ with $q^2 \gg m_\rho^2 \sim 0$ has been taken. \\  
At the same time the dilatation Ward identity on the amplitude $\mathcal A^{\mu}(q)$ in eq. (\ref{www}) gives 
\bea
\label{WIJD}
q_\mu \mathcal A^\mu(q)
&=&
i \int d^4 x\,e^{i q \cdot x}\, \frac{\partial}{\partial x^\mu}\, 
\left\langle 0 \left| T\{ J^\mu_D(x) {T^\alpha}_\alpha(0) \} \right|0\right\rangle
\nn \\
&=&
i \int d^4 x \, e^{i q \cdot x} \,  
\left\langle 0 \left| T\{ \partial_\mu J^\mu_D(x) {T^\alpha}_\alpha(0)\} \right|0\right\rangle
+ i  \int d^4 x \, e^{i q \cdot x} \, \delta(x_0) \, 
\left\langle 0\left| \left[ J_D^0(x), {T^\alpha}_\alpha(0)\right] \right|0\right\rangle. 
\eea
The commutator of the time component of the dilatation charge density and the trace of the EMT can be rewritten as 
\beq
\left[ J_D^0(0, {\bf x}), {T^\alpha}_\alpha(0)\right] = -i \delta^3 ({\bf x})\left( 
d_T + x\cdot \partial\right){T^\alpha}_\alpha(0) 
\label{derr}
\eeq
where $d_T$ is the canonical dimension of the EMT $(d_T=4)$. 
Inserting eq. (\ref{derr}) in the Ward identity (\ref{WIJD}) and neglecting the first term due to the nearly 
conserved dilatation current ($m_\rho \sim 0 $), we are left with 
\beq
q_\mu \mathcal A^\mu(q)= d_T \, \left\langle 0|{T^\alpha}_\alpha(0) |0\right\rangle.
\label{interdue}
\eeq
In the soft limit, with $q^2 \gg  m_\rho^2$, comparing eq. (\ref{inter}) and eq. (\ref{interdue}) we obtain 
\bea
\lim_{q^\mu\to 0} q^\mu \mathcal{A}_\mu =
f_\rho \, \left\langle \rho, q=0| {T^\alpha}_\alpha(0) |0\right\rangle  = d_T \, \left\langle 0|{T^\alpha}_\alpha(0) |0\right\rangle.
\eea
Introducing the vacuum energy density $\epsilon_{vac}=\left\langle 0|T^0_0 |0\right\rangle = \frac{1}{4}  \left\langle 
0|T^\alpha_\alpha(0) |0\right\rangle$ and using the relation in eq. (\ref{rel1}) we have 
\beq
\left\langle\rho, p=0| {T^\mu}_\mu |0\right\rangle =- f_\rho m_\rho^2 = \frac{d_T}{f_\rho} \epsilon_{vac}
\eeq
from which we finally obtain ($d_T = 4$)
\beq
f_\rho^2 m_\rho^2 = -16 \, \epsilon_{vac}. 
\eeq
This equation fixes the decay amplitude of the dilaton in terms of its mass and the vacuum energy. 
Notice that $\epsilon_{vac}$ can be related both to the anomaly and possibly to explicit contributions of the 
breaking of the dilatation symmetry since 
\beq
\label{beta}
\epsilon_{vac}= \frac{1}{4} \, \left\langle 0 \left| \frac{\beta(g)}{2 g} F_{\mu\nu}F^{\mu\nu} \right|0\right\rangle + ... 
\eeq
where the ellipsis saturate the anomaly equation with extra mass-dependent contributions, which may be far larger in size then 
the anomaly term. 
In (\ref{beta}) we have assumed, for simplicity, the coupling of the pole to a single gauge field, with 
a beta function $\beta(g)$, but obviously, it can be generalized to several gauge fields. 

 In the case of higher dimensional operators we would get 
\beq
\epsilon_{vac} = \frac{1}{4} \, \left\langle 0 \left| \frac{\beta(g)}{2 g} \, F_{\mu\nu}\,F^{\mu\nu} \right|0\right\rangle 
+ \sum_i g_i \, (d_i-4) \, \left\langle 0| O_i | 0 \right\rangle,
\eeq
valid around the scale at which the PCDC approximation holds. Therefore a massless pole can be corrected nonperturbatively according 
to some completion theory, causing its mass to shift. Perturbation theory 
gives indications about the interpolating fields which can couple to it, as we have seen by exploiting the chiral analogy, but not 
more than that. The corrections are model-dependent and can be the subject of additional 
phenomenological searches, but the dilatation current takes the role, with no doubt, of an interpolating field for the propagation of 
such a scalar intermediate state.

\section{The infrared coupling of an anomaly pole and the anomaly enhancement} 

It is easy to figure out from the results of the previous sections that the coupling of a (graviscalar) dilaton to the anomaly 
causes a large enhancement of its $2$-photons and $2$-gluons decays. 
One of the features of the graviscalar interaction is that its coupling includes 
anomalous contributions which are part both of the $2$-photons and of the $2$-gluons cross sections. For this reason, 
if an enhancement with respect to the Standard Model rates is found only in one of these two channels and it is associated 
to the exchange of a spin zero intermediate state, this result could be used to rule out the exchange of a graviscalar.

On the other hand, for an effective dilaton, identified by an anomaly pole in the $J_D VV$ correlator of the Standard Model, the case is more 
subtle, since the coupling of this effective state to the anomaly has to be introduced by hand. This state should be identified, in the perturbative 
picture,  with the corresponding anomaly pole. The situation, here, is closely similar to the pion case: 
in fact, in perturbative QCD, the anomalous $AVV$ diagram is characterized by the presence of an anomaly pole in the variable 
$k^2$, with $k$ denoting the momentum of the axial-vector current, which is explicitly shown in eq. (\ref{AVVpole}). 
It is interesting to note that this structure has a non-vanishing residue for on-shell photons and for massless quarks running
in the loop. In this case the pole is said to be \emph{infrared coupled}.
This feature, supplemented by usual polology arguments, leads to a $\pi\to \gamma \gamma$ decay rate which 
is enhanced with respect to the non-anomalous case.
On the other hand, if the photons are virtual or the quarks are massive the anomaly pole decouples, namely, its residue is zero. 
We refer to \cite{Armillis:2009sm} for more details. 

The same behaviour is shared by the conformally anomalous $TVV$ diagram \cite{Armillis:2009pq}, so
let us illustrate this important point in the QED case by considering the off-shell correlator.
We denote with $s_1$ and  $s_2$ the virtualities of the two final state photons and with $m$ the mass 
of the fermion running in the loops. 
The case with on-shell photons and a massive fermion has already been discussed in section \ref{SecPoleQED}. There we have shown that 
the anomaly pole has a non-vanishing residue only in the conformal limit, when all masses are set to zero. 
Indeed, in the case of a massive fermion, besides the fact that the anomaly pole anyway appears in the corresponding invariant 
amplitude $F_1 (s;\,0,\,0,\,m^2)$, as one can see from eq. (\ref{oom}), it will decouple, showing a zero residue
\bea
\lim_{s\rightarrow0} \, s \, \Gamma^{\mu\nu\a\b}(s;\,0,\,0,\,m^2) =0 \,.
\eea
As for the chiral anomaly case, the absence of the internal fermion masses is not sufficient to guarantee the infrared coupling of 
the anomaly pole. Indeed, if $m=0$ but the photons are taken off-shell, being characterized by non zero virtualities $s_1$ and $s_2$, 
one can check that the entire correlator is completely free from anomaly poles as
\bea
\lim_{s\rightarrow0} \, s \, \Gamma^{\mu\nu\a\b}(s;\,s_1,\,s_2,\,0) =0 \,.
\eea
The computation of this limit needs the explicit results for all the invariant amplitudes $F_i$, which are not given here 
due to their lengthy expressions but can be found in \cite{Armillis:2009pq}. \\

One should be aware of the fact that the same pole is present in the $AVV$ diagram when $VV$ are now the gluons. If the two gluons 
are on-shell, as in the 2-photon case, the perturbative anomaly pole is again infrared coupled. Obviously, such an an enhancement is 
not observable, since the gluons cannot be  on-shell, because of confinement. In the perturbative picture, a 
non-zero virtuality of the two gluons is then sufficient to exclude an infrared coupling of the anomaly pole.

We feel, however, that a simple perturbative analysis may not be completely sufficient to decide whether or not the coupling of such a state to 
the gluon anomaly takes place. On the other hand, there is no doubt, by the same reason, that such a coupling should occur in the 2-photon 
case, being the photons massless asymptotic states. In this case the corresponding anomaly pole of the $J_D \gamma\gamma$ vertex is 
infrared coupled.
  
In general, in the case of an effective dilaton, one is allowed to write down a Lagrangian which is assumed to be scale invariant and, at a second 
stage, introduce the direct coupling of this state to the trace anomaly. The possibility of coupling such a state to the photon and to the gluons or 
just to the photons, for instance, is a delicate issue for which a simple perturbative approach is unable to offer a definite answer. 

\section{Quantum conformal invariance and dilaton couplings at low energy}

Similar enhancements are present in the case of quantum scale invariant extensions of the Standard Model \cite{Goldberger:2007zk},  where 
one assumes that the spectrum of the theory is extended with new massive states in order to set the $\beta$ functions of the gauge couplings to 
vanish. In a quantum scale invariant theory such as the one discussed in \cite{Goldberger:2007zk}, the dilaton couples only to massive states, 
but the heavy mass limit and the condition of the vanishing of the complete $\beta$ functions, leave at low energy a dilaton interaction 
proportional only to the $\beta$ functions of the low energy states. 
The "remnant" low energy interaction is mass-independent and coincides with that due to a typical anomalous coupling, 
although its origin is of different nature, since anomalous contributions are genuinely mass-independent. 

For this reason, the decays of a dilaton produced by such extensions carries anomaly-like enhancements as in the graviscalar case. 
Obviously, such enhancements to the low energy states of the Standard Model would also be typical of the decay of a Higgs field, 
which couples proportionally to the mass of an intermediate state, if quantum scale invariance is combined with the decoupling of a heavy sector.
This, in general, causes an enhancement of the Higgs decay rates into photons and gluons. A partial enhancement only of the di-photon 
channel could be accomplished, in this approach, by limiting the above quantum scale invariant arguments only to the electroweak sector. 

As a second example, we consider the situation in which all the SM fields are embedded in a (quantum) Conformal Field Theory (CFT) 
extension \cite{Goldberger:2007zk} and we discuss the (loop-induced) couplings of the dilaton to the massless gauge bosons. 
At tree level the dilaton of \cite{Goldberger:2007zk} couples to the SM fields only through their masses, as the fundamental dilaton which we 
have discussed previously, and, in this respect, it behaves like the SM Higgs, without scale anomaly contributions. 
For this reason the dilaton interaction with the massless gauge bosons is induced by quantum effects mediated by heavy particles
running in the loops (in this context heavier or lighter is referred to the dilaton mass), and not by anomalous terms.
When the mass $m_i$ of the particle running in the loop is much greater than the dilaton mass, the coupling to the massless 
gauge bosons becomes 
\bea
\mathcal L_{\rho} = \frac{\alpha_s}{8 \pi} \sum_i b_g^i \, \frac{\rho}{\Lambda} (F_{g \, \mu\nu}^a)^2
+ \frac{\alpha_{em}}{8 \pi} \sum_i b_{em}^i \, \frac{\rho}{\Lambda} (F_{\gamma \, \mu\nu})^2 \,,
\eea
where $b_{em}^i$ and $b_{g}^i$ are the contributions of the heavy field $i$ to the $1$-loop $\beta$ function (computed in the 
$\overline{MS}$ scheme) for the electromagnetic and strong coupling constants respectively. The $\beta$ functions are normalized as
\bea
\beta_i = \frac{g^3}{16 \pi^2} b^i \,.
\eea
Note that this result is independent from the heavy mass $m_i$ as one can prove by analysing the structure of the mass corrections 
of the dilaton coupling, which reads as
\beq
\Gamma_{\rho V V}\sim \frac{g^2}{\pi^2 \Lambda} \, m_i^2 \, 
\bigg[ \frac{1}{s}  -  \frac{1} {2 }\mathcal C_0 (s, 0, 0, m_i^2, m_i^2, m_i^2) \bigg(1-\frac{4 m_i^2}{ s}\bigg)\bigg] 
\sim 
\frac{g^2}{\pi^2 \Lambda} \, \frac{1}{6} + O\left( \frac{s}{m_i^2} \right)
\eeq
where $s=m_{\rho}^2$ is fixed at the dilaton mass and we have performed the large mass limit of the amplitude using 
\bea
\mathcal C_0(s, 0, 0, m_i^2, m_i^2, m_i^2) \sim - \frac{1}{2 m_i^2} \left( 1 + \frac{1}{12} \frac{s}{m_i^2} 
+ O(\frac{s^2}{m_i^4} ) \right)
\eea
valid for $m_i^2 \gg s = m_{\rho}^2$. This shows that in the case of heavy fermions, 
the dependence on the fermion mass cancels. 
Obviously, this limit generates an effective coupling which is proportional to the $\beta$ function related to the heavy flavours. 
The same reasonings can be employed to the Higgs case as well. It clear that this coupling to the massless gauge bosons is dependent 
from new heavy states and, therefore, from the UV completion of the SM. This is certainly the case for the Standard Model Higgs whose 
double photon decay is one of the most important decay channel for new physics discoveries. \\ 
For the dilaton case the situation is slightly different. Surely we do not understand the details of the CFT extension, nor its 
particle spectrum, but nevertheless we know that the conformal symmetry is realized at the quantum level. Therefore the complete 
$\beta$ functions, including the contribution from all states, must vanish
\bea
\beta = \frac{g^3}{16 \pi^2} \bigg[ \sum_{i} b^i + \sum_{j} b^j \bigg] = 0 \,,
\eea
where $i$ and $j$ run over the heavy and light states respectively. Exploiting the consequence of the quantum conformal symmetry, the 
dilaton couplings to the massless gauge bosons become
\bea
\mathcal L_{\rho} = 
- \frac{\alpha_s}{8 \pi} \sum_j b_g^j \, \frac{\rho}{\Lambda} (F_{g \, \mu\nu}^a)^2  -  \frac{\alpha_{em}}{8 \pi}
\sum_j b_{em}^j \, \frac{\rho}{\Lambda} (F_{\gamma \, \mu\nu})^2  \,,
\eea
in which the dependence from the $\beta$ functions of the light states is now explicit. We emphasize that the appearance of the light 
states contributions to the $\beta$ functions is a consequence of the vanishing of the complete $\beta$, and, therefore, of the CFT 
extension and not the result of a direct coupling of the dilaton to the anomaly.

\section{Conclusions}

We have presented a general discussion of dilaton interactions with the neutral currents sector of the Standard Model. In the case of a 
fundamental graviscalar as a dilaton, we have presented the complete electroweak corrections to the corresponding interactions and we have 
discussed the renormalization properties of the same vertices. In particular, we have shown that the renormalizability of the dilaton vertices is 
inherited directly from that of the Standard Model only if the Higgs sector is characterized by a conformal coupling ($\chi$) 
fixed at the value $1/6$.

Then we have moved to an analysis of the analytic structure of the $J_D VV$ correlator, showing that it supports an anomaly pole as an 
interpolating state, which indicates that such a state can be interpreted as the Nambu-Goldstone (effective dilaton) mode of the anomalous 
breaking of the dilatation symmetry. 

In fact, the trace anomaly seems to bring in some important information concerning the dynamics of the Standard Model, aspects that we have 
tried to elucidate. For this reason, we have extended a previous analysis of ours of the $TVV$ vertex, performed in the broken electroweak 
phase and in QCD, in order to characterize the dynamical behaviour of the analogous $J_D VV$ correlator. The latter carries relevant 
information on the anomalous breaking of the dilatation symmetry in the Standard Model. In fact, as we move to high energy, far above the 
electroweak scale, the Lagrangian of the Standard Model becomes approximately scale invariant. This approximate dilatation 
symmetry is broken by a quantum anomaly and its signature, as we have shown in our analysis, is in the appearance of 
an anomaly pole in the $J_DVV$ correlator. The same pole might appear in correlators with multiple insertions of $J_D$, but the proof 
of their existence is far more involved and requires further investigations. 
This pole is clearly massless in the perturbative picture, and accounts for the anomalous breaking 
of this approximate scale invariance.

\clearpage{\pagestyle{empty}\cleardoublepage}

\chapter{Higher order dilaton interactions in the nearly conformal limit of the Standard Model}\label{Traced4T}
 
\section{Introduction}

In the previous chater we have elaborated on dilaton interactions with neutral gauge currents in QCD and in the Standard Model
and we have seen that the two main features distinguishing dilaton vertices from those of the Higgs field are respectively the appearance of the 
conformal symmetry breaking scale $\Lambda$ and the anomalous enhancements, which are present both for the fundamental graviscalar and 
the effective, composite dilaton field.
The whole analysis was carried out including all the mass terms in the $1$-loop radiative corrections.

In this chapter we turn to the investigation of dilaton self-interactions in the nearly conformal limit of the Standard Model,
in which conformal symmetry is broken only by the dilatation anomaly, through a hierarchy of anomalous Ward identities for the divergence
of its dilatation current.  In this approximation, the identities allow to extract the coupling of the dilaton to the trace anomaly, 
which we compute up to the quartic order in the conformal breaking scale. Our approach can be easily extended to discuss the 
anomaly contributions to the dilaton effective action to an arbitrarily high order. They allow to make a distinction between 
the Higgs and the dilaton at a phenomenological level.

The possibility that the Standard Model is characterized at high energy by a nearly conformal dynamics has motivated several investigations 
spanning considerable time \cite{Goldberger:2007zk,Buchmuller:1987uc, Buchmuller:1988cj,Shaposhnikov:2008xi, Meissner:2006zh}. 
If not for a quadratic term present in the Higgs potential, the model would in fact enjoy a dilatation symmetry which 
is broken by the vev of the Higgs field in the process of spontaneous symmetry breaking.

A dilaton couples to the trace of the energy-momentum tensor (EMT) ${T^\mu}_\mu$, and the coupling is affected by a trace anomaly. The 
trace anomaly equation plays a key role in characterizing the dynamics of the dilaton interactions, with a breaking of the dilatation symmetry which 
is enforced by two different contributions. 

They can be easily identified from the structure of the corresponding Ward identity satisfied by the 
dilaton $(\rho)VV$ vertex \cite{Coriano:2012nm}, with $V$ denoting a neutral (or charged) vector current, but also of higher vertices, such 
as the cubic ($TTT$) and quartic ($TTTT$) dilaton interactions, which are part of the dilaton effective action as well. 
One specific contribution is the coupling of the dilaton to the anomaly, the second one being related to explicit mass terms generated 
at the electroweak scale.  
In fact, the basic trace anomaly equation which takes the role of the generating functional of all the Ward identities 
satisfied by the dilaton vertices is given by 
\beq
g_{\mu\nu}\left\langle T^{\mu\nu} \right\rangle_s  =  \mathcal A[g]  +  \left\langle {T^\mu}_{\mu} \right\rangle_s \, ,
\label{Ch2TraceAnomaly1}
\eeq
where the anomalous $(\mathcal A[g])$ and the explicit  $(\left\langle T^{\mu\nu} \right\rangle)$ contributions are clearly separated. 
As usual, gravity plays simply an auxiliary role, since one takes the flat limit in all the hierarchical
Ward identities which are obtained from (\ref{Ch2TraceAnomaly1}) after the functional differentiations. 

The goal of this chapter is to stress on some specific features of the dilaton effective action which follow up from (\ref{Ch2TraceAnomaly1}) 
and which are related to the structure of the anomalous contributions. In particular, in a nearly conformal phase of the Standard Model, 
which can be approximated by an exact $SU(3)_C \times SU(2)_L \times U(1)_Y$ gauge theory, 
cubic and quartic contributions to the dilaton dynamics are essentially fixed by the anomaly and can be extracted, with some effort, 
from a diagrammatic analysis of  (\ref{Ch2TraceAnomaly1}) expanded up to the fourth order in the metric.  
This is the approach that we will be following in our case and on which we are going to elaborate. 
In particular, we will present the expressions of such contributions. 
These interactions set a key distinction between a Higgs and a dilaton at every order, being the Higgs not affected by the scale anomaly,  
and can provide the basis for a direct phenomenological analysis of possible dilaton interactions at the LHC.

Their derivation will bypass the direct diagrammatic computation, relying instead on the connection between conformal anomalies
and counterterms in dimensional regularization, which was thoroughly discussed in section \ref{CountAnom}.

\section{Anomalous interactions from the Ward identities}

To illustrate the role of the anomaly in a more direct way and its possible significance in setting a distinction between the 
Higgs and the dilaton, we recall that the interaction of the dilaton $\rho$ with the Standard Model fields is given to first order by
\bea
\label{Ch2Lint}
\mathcal L_{int} = - \frac{1}{\Lambda} \rho \, {T^\mu}_\mu \, ,
\eea
where $\Lambda$ is the conformal breaking scale, which remains a free parameter of the effective action.
For convenience, let us recall the ordinary definition of the EMT of the Standard Model
\beq \label{Ch2EMTCh2}
T^{\mu\nu} = -\frac{2}{\sqrt{g}}\frac{\delta\,\mathcal{S}}{\delta g_{\mu\nu}}  \, ,
\eeq
in terms of the action $\mathcal{ S}$, so that its quantum average in terms of the euclidean generating functional of the theory,
$\mathcal {W}$, depending from the background metric $g_{\mu\nu}(x)$, 
\beq\label{Ch2GeneratingCh2}
\mathcal W = \frac{1}{\mathcal{N}} \, \int \, \mathcal D\Phi \, e^{-\, \mathcal{S}} \, . 
\eeq
is given by
\beq \label{Ch2VEVEMTCh2}
\left\langle T^{\mu\nu} \right\rangle_s = \frac{2}{\sqrt{g}}\frac{\delta\, \mathcal W}{\delta\, g_{\mu\nu}}\, ,
\eeq
where the background fields are kept switched on, as the subscript $s$ indicates.
\begin{figure}[t]
\centering
\includegraphics[scale=0.8]{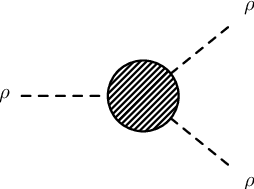}
\hspace{1cm}\includegraphics[scale=0.8]{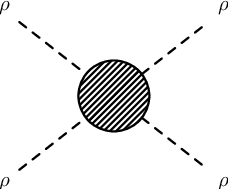}
\caption{Cubic and quartic dilaton interactions. In the nearly conformal limit the computation of the interactions involves virtual 
scalars, spin 1 and fermion exchanges. }
\label{Ch2rgg1}
\end{figure}

The identification of the anomaly contributions to dilaton interactions, on general grounds, requires an analysis of the anomalous Ward identities 
satisfied by the respective correlators. In this chapter we will concentrate on the extraction of the anomalous contribution to the quartic dilaton 
interactions, using as a fundamental scenario the Standard Model in the unbroken phase (i.e. with $v\to 0$). 

In this approximation, explicit trace insertions vanish for on-shell massless final states (i.e. for gauge fields)
and we neglect all the contributions related both to the Higgs and to virtual corrections with a 
massive dilaton in the loops. In the same limit all the mixing contributions related to a possible term of improvement are not present 
\cite{Coriano:2012nm} and the computation of the anomalous terms amounts to the extraction of some finite parts.

Explicit (i.e. non anomalous) corrections, also present in the fundamental Ward identity, are calculable, but they are model-dependent. 
In fact, they require the introduction of some extra potential for the dilaton/Higgs system. 
Its form necessarily has to rely on extra assumptions, such as the specific choice of breaking of the conformal symmetry,
as illustrated in chapter \ref{Effective}. 
They are obtained by inserting the trace (${T^\mu}_\mu$) operator on generic correlators involving all the fields of the 
Standard Model (plus other dilaton lines). In the conformal limit such contributions, which have been discussed in \cite{Coriano:2012nm} 
in the case of a single dilaton, drop out and the computations simplify considerably.

In this approximation the breaking of the dilatation symmetry does not contain any explicit scale-dependent term, and it is only due  to 
the anomaly, which is induced by renormalization. We call this approximation "nearly conformal".

As we have pointed out before, the breaking of the dilatation symmetry shows up, at a perturbative level, 
with the appearance of a massless pole in the $J_D VV$ correlator in the neutral and charged current sectors of the theory, with a residue which is 
proportional to a specific beta-function of the theory, related to the final state. This takes the role of a Nambu-Goldstone mode of the broken 
dilatation symmetry and it has been shown to affect each gauge invariant sector of the dilaton-to-two gauge bosons matrix elements.

To extract the anomalous contributions of the higher order interactions shown in fig. \ref{Ch2rgg1} 
we start from the explicit expression of the anomaly, which  in $d=4$ is given by \cite{Duff:1977ay,Duff:1993wm}
\bea \label{Ch2TraceAnomalyCh2}
 \mathcal A[g]
&=&
\sum_{I=f,s,V}n_I \bigg[\beta_a(I)\, F + \beta_b(I)\, G + \beta_c(I) \, \square R \bigg] \, .
\eea
$\mathcal{A}[g]$ contains the diffeomorphism-invariants built out of the Riemann tensor, ${R^{\alpha}}_{\beta\gamma\delta}$, 
the Ricci tensor $R_{\alpha\beta}$ and the scalar curvature $R$. In particular, $G$ and $F$ in eq.  (\ref{Ch2TraceAnomalyCh2}) 
are the Euler density and the square of the Weyl tensor respectively, given in appendix \ref{Geometrical},  with coefficients $\beta_a$
and $\beta_b$ depending on the field content of the Lagrangian (scalar, fermion,vector) and we have a multiplicity factor $n_I$ 
for each particle species. The form of the trace anomaly in (\ref{Ch2TraceAnomalyCh2}) already contains the 
constraint $\beta_d = 0 $ discussed in chapter \ref{TTTVertex} and we are discarding the gauge sector, 
which is not concerned for the purposes of this chapter.

In addition, the value of $\beta_c$ is regularization-dependent, corresponding to 
the fact that it can be changed by the addition of an arbitrary local $R^2$ term in the effective action. 
The values given in table \ref{Ch1AnomalyCoeff} are those obtained in dimensional regularization, 
in which the relation $\beta_c = -{2}/{3}\,\beta_a$ is found to hold. 
The values of the coefficients for three theories of spin $0, \frac{1}{2}, 1$, that we are going to consider 
are listed in chapter \ref{TTTVertex},  table \ref{Ch1AnomalyCoeff}.

Taking (\ref{Ch2VEVEMTCh2}) into account, in terms of $\mathcal{W}$ the fundamental trace anomaly 
equation can be rewritten in the form
\beq
\label{Ch2TraceAnomalySymmCh2}
2 \, g_{\mu\nu}\frac{\delta \mathcal{W}}{\delta g_{\mu\nu}}= \sqrt{g} \, \mathcal A[g]
\eeq
and plays the role of a generating functional for the anomalous Ward identities of any underlying 
Lagrangian field theory being, therefore, model independent. 
From (\ref{Ch2TraceAnomalySymmCh2}) we can extract several identities satisfied by the anomaly term, for correlators involving $n$ 
insertions of energy-momentum tensors, by performing $n-1$ functional derivatives with respect to the metric of both sides of 
(\ref{Ch2TraceAnomalySymmCh2}) and taking the trace of the result at the very end.

\section{EMT's and Correlators}

In perturbation theory, imposing the conservation Ward identity for the EMT and the Ward identities for the vector currents - whenever these are 
present - is sufficient to obtain the corresponding anomalous term from the complete diagrammatic expansion. 
In particular, in dimensional regularization, the anomaly comes for free at the end of the computations, but this is a demanding job. 

In the case of the $TVV$,
for instance, it is a common practice to perform a direct computation, since only one term ($\sim F^{a\,\mu\nu}(z)\,F^a_{\mu\nu} (z)$)
can appear in the anomaly. We have omitted it in (\ref{Ch2TraceAnomalyCh2}), since our analysis is focused on the anomaly-induced 
radiative corrections to correlators involving only dilaton self-interactions. A general discussion of contributions containing neutral currents 
(the $TVV$ vertex) has been given in \cite{Coriano:2011zk, Coriano:2012nm} in the Standard Model and in 
\cite{Giannotti:2008cv, Armillis:2009pq} for QED. However, things are far more involved for vertices containing multiple insertions of EMT's, 
such as the $TTT$ and $TTTT$, and it is convenient to infer the structure of the anomalous corrections without having to perform a complete 
diagrammatic analysis. In any case, a successful test of the anomalous Ward identities is crucial in order to secure the correctness of the result of 
the computation.

As mentioned above, in the nearly conformal limit of the Standard Model we will need to consider a scalar, a fermion and an abelian vector theory 
coupled to a background gravitational field. In fact the non abelian character of the theory is not essential in the study of the higher order terms
to the dilaton effective action. In this case we can reconstruct the entire contribution to the anomaly from the abelian case 
by correcting the result just by one extra multiplicity factor.

We recall here the EMT's for the theories that we consider, i.e.
\bea
T^{\mu\nu}_{\phi}
&=&
\nabla^\mu \phi \, \nabla^\nu\phi - \frac{1}{2} \, g^{\mu\nu}\,g^{\alpha\beta}\,\nabla_\alpha \phi \, \nabla_\beta \phi
+ \chi \bigg[g^{\mu\nu} \Box - \nabla^\mu\,\nabla^\nu - \frac{1}{2}\,g^{\mu\nu}\,R + R^{\mu\nu} \bigg]\, \phi^2 
\label{Ch2ScalarEMT}\\
T^{\mu\nu}_{\psi}
&=&
\frac{1}{4} \,
\bigg[ g^{\mu\lambda}\,{V_\alpha}^\nu + g^{\nu\lambda}\,{V_\alpha}^\mu - 2\,g^{\mu\nu}\,{V_\alpha}^\lambda \bigg]
\bigg[\bar{\psi} \, \gamma^a \, \left(\mathcal{D}_\lambda \,\psi\right) -
\left(\mathcal{D}_\lambda \, \bar{\psi}\right) \, \gamma^a \, \psi \bigg] \, ,
\label{Ch2FermionEMT} \\
T^{\mu\nu}_V
&=&
F^{\mu\a}{F^\nu}_{\a}  - \frac{1}{4}g^{\mu\nu}F^{\a\b}F_{\a\b} \, ,
\label{Ch2MaxwellEMT}
\eea
where ${V_\alpha}^\nu$ is the \emph{vierbein} needed to embed the fermion in the gravitational background
and the corresponding covariant derivative is $ \mathcal{D}_\mu = \pd_\mu + \Gamma_\mu =
\pd_\mu + \frac{1}{2} \, \Sigma^{\alpha\beta} \, {V_\alpha}^\sigma \, \nabla_\mu\,V_{\beta\sigma} \, $
where the $\Sigma^{\alpha\beta}$ are the generators of the Lorentz group  in the case of a spin $1/2$-field. \\

It is convenient to define the correlation functions with $n$ external insertions of EMT's, which can be effectively thought as 
gravitons, as functional derivatives of order $n$ of $\mathcal W$, evaluated in the flat limit
\bea \label{Ch2NPFCh2}
\left\langle T^{\mu_1\nu_1}(x_1)...T^{\mu_n\nu_n}(x_n) \right\rangle 
&=&
\bigg[\frac{2}{\sqrt{g_{x_1}}}...\frac{2}{\sqrt{g_{x_n}}} \,
\frac{\delta^n \mathcal{W}}{\delta g_{\mu_1\nu_1}(x_1)...\delta g_{\mu_n\nu_n}(x_n)}\bigg]
\bigg|_{g_{\mu\nu} = \delta_{\mu\nu}} \nonumber \\ 
&=&  
2^n\, \frac{\delta^n \mathcal{W}}{\delta g_{\mu_1\nu_1}(x_1)...\delta g_{\mu_n\nu_n}(x_n)}\bigg|_{g_{\mu\nu} = 
\delta_{\mu\nu}} \, .
\eea
%
%
%
for any functional (or function) $\mathcal F$ which depends on the background field $g_{\mu\nu}(x)$. 
Denoting with
\beq
\left\langle \mathcal{O} \right\rangle = \int\, \mathcal{D}\Phi\, \mathcal{O}\, e^{-S} \, 
\eeq
the vacuum expectation values of each operator, with $\mathcal{S}$ the generic action, we obtain 
\bea
\left\langle T^{\mu\nu}(x)\right\rangle
&=&
- 2 \, \left\langle \left[\mathcal S\right]^{\mu\nu}(x) \right\rangle \label{Ch21PF} \\
\left\langle T^{\muu\nuu}(x_1)T^{\mud\nud}(x_2)\right\rangle
&=&
4 \, \bigg[ 
\left\langle \left[\mathcal S\right]^{\muu\nuu}(\xu)\, \left[\mathcal S\right]^{\mud\nud}(\xd) \right\rangle 
- \, \left\langle \left[\mathcal S\right]^{\muu\nuu\mud\nud}(\xu,\xd)\right\rangle \bigg] \, , \label{Ch22PFCh2}\\
\left\langle T^{\muu\nuu}(x_1)T^{\mud\nud}(x_2)T^{\mut\nut}(x_3)\right\rangle
&=&
8 \, \bigg[ - \left\langle \left[\mathcal S \right]^{\muu\nuu}(\xu)\left[\mathcal S \right]^{\mud\nud}(\xd)
\left[\mathcal S \right]^{\mut\nut}(\xt) \right\rangle 
\nn \\
&& \hspace{-50mm}
+ \,\bigg( \left\langle \left[\mathcal S\right]^{\muu\nuu\mud\nud}(\xu,\xd)\,\left[\mathcal S\right]^{\mut\nut}(\xt) \right\rangle
+ 2\, \text{perm.} \bigg)
- \,\left\langle  \left[\mathcal S\right]^{\muu\nuu\mud\nud\mut\nut}(\xu,\xd,\xt)\right\rangle \bigg] \label{Ch23PFCh2} \, ,
\eea
\bea
\left\langle T^{\muu\nuu}(x_1)T^{\mud\nud}(x_2)T^{\mut\nut}(x_3)T^{\muq\nuq}(x_4)\right\rangle
&=&
16 \, \bigg[
\left\langle \left[\mathcal S \right]^{\muu\nuu}(\xu)\left[\mathcal S \right]^{\mud\nud}(\xd)
\left[\mathcal S \right]^{\mut\nut}(\xt)\left[\mathcal S \right]^{\muq\nuq}(\xq) \right\rangle 
\nn \\
&& \hspace{-80 mm}
- \,\bigg( \left\langle \left[\mathcal S\right]^{\muu\nuu\mud\nud}(\xu,\xd)\left[\mathcal S\right]^{\mut\nut}(\xt)
\left[\mathcal S\right]^{\muq\nuq}(\xq) \right\rangle + 5 \, \text{perm.} \bigg) 
+ \bigg( \left\langle \left[\mathcal S\right]^{\muu\nuu\mud\nud}(\xu,\xd)\left[\mathcal 
S\right]^{\mut\nut\muq\nuq}(\xt,\xq)\right\rangle + 2 \, \text{perm.} \bigg)
\nn \\
&& \hspace{-80mm}
+ \,\bigg( \left\langle \left[\mathcal S\right]^{\muu\nuu\mud\nud\mut\nut}(\xu,\xd,\xt)\left[\mathcal 
S\right]^{\muq\nuq}(\xq)\right\rangle + 3 \, \text{perm.} \bigg)
- \, \left\langle \left[\mathcal S\right]^{\muu\nuu\mud\nud\mut\nut\muq\nuq}(\xu,\xd,\xt,\xq) \right\rangle \bigg] \, .
\label{Ch24PF}
\eea
Notice that in dimensional regularization 
\beq
\left\langle \left[\mathcal S\right]^{\mu\nu}(x) \right\rangle= \left\langle \left[\mathcal 
S\right]^{\muu\nuu\mud\nud}(\xu,\xd)\right\rangle
=\left\langle  \left[\mathcal S\right]^{\muu\nuu\mud\nud\mut\nut}(\xu,\xd,\xt)\right\rangle
= \left\langle \left[\mathcal S\right]^{\muu\nuu\mud\nud\mut\nut\muq\nuq}(\xu,\xd,\xt,\xq) \right\rangle=0
\eeq
being proportional to massless tadpoles. 
In particular, this implies that, to perform a perturbative computation of a correlator of order $n$,
one would be needing interaction vertices with at most $n-1$ gravitons.
Concerning the diagrammatic structure of each contribution, the correlator
\beq
\left\langle \left[\mathcal S \right]^{\muu\nuu}(\xu)\left[\mathcal S \right]^{\mud\nud}(\xd)
      \left[\mathcal S \right]^{\mut\nut}(\xt)\left[\mathcal S \right]^{\muq\nuq}(\xq) \right\rangle 
\eeq 
has a box topology;
\beq     
\left\langle \left[\mathcal S \right]^{\muu\nuu}(\xu)\left[\mathcal S \right]^{\mud\nud}(\xd)
\left[\mathcal S \right]^{\mut\nut}(\xt) \right\rangle 
\eeq
which is the first contribution to the graviton $3$-point function, and
\beq 
\left\langle \left[\mathcal S\right]^{\muu\nuu\mud\nud}(\xu,\xd) 
\left[\mathcal S\right]^{\mut\nut}(\xt)\left[\mathcal S\right]^{\muq\nuq}(\xq) \right\rangle \, ,
\eeq 
which corresponds to a contact term in $\left\langle TTTT \right\rangle$, are represented by triangles. \\
The remaining contributions, 
\beq
\left\langle \left[\mathcal S\right]^{\muu\nuu}(\xu)\, \left[\mathcal S\right]^{\mud\nud}(\xd) \right\rangle \, ,
\eeq 
the contact terms $\left\langle TTT \right\rangle$, which are 
\beq
\left\langle \left[\mathcal S\right]^{\muu\nuu\mud\nud}(\xu,\xd)\,\left[\mathcal S\right]^{\mut\nut}(\xt) \right\rangle
\eeq 
and the two remaining types of diagrams which enter into $\left\langle TTTT \right\rangle$,
\beq     
\left\langle \left[\mathcal S\right]^{\muu\nuu\mud\nud}(\xu,\xd)\left[\mathcal S\right]^{\mut\nut\muq\nuq}(\xt,\xq)\right\rangle
\eeq
and
\beq
\left\langle \left[\mathcal S\right]^{\muu\nuu\mud\nud\mut\nut}(\xu,\xd,\xt)\left[\mathcal S\right]^{\muq\nuq}(\xq)\right\rangle \, ,
\eeq
have the topologies of 2-point functions.
Our conventions for the choice of the external momenta, which are taken to be all incoming, are defined via the Fourier transform
\bea
&&
\int \, d^4\xu\,d^4\xd\,d^4\xt\,d^4\xq\, 
\left\langle T^{\muu\nuu}(\xu)T^{\mud\nud}(\xd)T^{\mut\nut}(\xt)T^{\muq\nuq}(\xq)\right\rangle \,
e^{-i(\ku\cdot \xu + \kd\cdot \xd + \kt\cdot \xt + \kq \cdot \xq)} = 
\nn \\
&& \hspace{40mm}
(2\pi)^4\,\delta^{(4)}(\ku + \kd + \kt + \kq)\, 
\left\langle T^{\muu\nuu}T^{\mud\nud}T^{\mut\nut}T^{\muq\nuq}\right\rangle(\kd,\kt,\kq) \, , 
\label{Ch24PFMom}
\eea
and similar for the $3$- and $2$-point functions.

\section{Ward identities}

We start from the analysis of the general covariance Ward identities, which partially overlaps with the 
discussion in chapter \ref{TTTVertex}, which is limited to the $TTT$ vertex.  \\
The identities we look for  are obtained from the functional relation
\beq\label{Ch2masterWI0Ch2}
\nabla_{\nuu} \left\langle T^{\muu\nuu}(x_1) \right\rangle
= \nabla_{\nuu} \bigg(\frac{2}{\sqrt{g_{x_1}}}\frac{\delta\mathcal W}{\delta g_{\muu\nuu}(x_1)}\bigg)= 0 \, 
\eeq
which, after an expansion, becomes
\bea \label{Ch2InterWard}
&&
\frac{2}{\sqrt{g_{x_1}}}\bigg(\pd_{\nuu}\frac{\delta\mathcal W}{\delta g_{\muu\nuu}(x_1)}
- \Gamma^\lambda_{\lambda\nuu}(x_1)\frac{\delta\mathcal W}{\delta g_{\muu\nuu}(x_1)}
+ \Gamma^{\muu}_{\kappa\nuu}(x_1)\frac{\delta\mathcal W}{\delta g_{\kappa\nuu}(x_1)}
+ \Gamma^{\nuu}_{\kappa\nuu}(x_1)\frac{\delta\mathcal W}{\delta g_{\muu\kappa}(x_1)}\bigg) = 0. \nn\\
\eea
Cancelling the second and fourth terms in parentheses, (\ref{Ch2InterWard}) takes the simpler form
\bea \label{Ch2masterWICh2}
&& 
2 \, \bigg(\pd_{\nuu}\frac{\delta\mathcal W}{\delta g_{\muu\nuu}(x_1)}
+ \Gamma^{\muu}_{\kappa\nuu}(x_1)\frac{\delta\mathcal W}{\delta g_{\kappa\nuu}(x_1)}\bigg)
= 0\, .
\eea
The Ward identities we are interested in are obtained by functional differentiation of (\ref{Ch2masterWICh2}) and give
\bea
&&
4\, \bigg[ \pd_{\nuu} \frac{\delta^2\mathcal W}{\delta g_{\mud\nud}(x_2)\delta g_{\muu\nuu}(x_1)}
+ \frac{\delta \Gamma^{\muu}_{\kappa\nuu}(x_1)}{\delta g_{\mud\nud}(x_2)}
\frac{\delta\mathcal W}{\delta g_{\kappa\nuu}(x_1)}
+ \Gamma^{\muu}_{\kappa\nuu}(x_1) 
\frac{\delta^2 \mathcal W}{\delta g_{\mud\nud}(x_2)\delta g_{\kappa\nuu}(x_1)}\bigg] = 0 \, ,
\label{Ch2WI2PFCoordinateCh2} 
\eea
for single
\bea
&&
8\, \bigg[\pd_{\nuu}
\frac{\delta^3\mathcal W}{\delta g_{\mut\nut}(x_3)\delta g_{\mud\nud}(x_2)\delta g_{\muu\nuu}(x_1)}
+ \frac{\delta \Gamma^{\muu}_{\kappa\nuu}(x_1)}{\delta g_{\mud\nud}(x_2)}
\frac{\delta^2 \mathcal W}{\delta g_{\mut\nut}(x_3)\delta g_{\kappa\nuu}(x_1)}
+ \frac{\delta \Gamma^{\muu}_{\kappa\nuu}(x_1)}{\delta g_{\mut\nut}(x_3)}
\frac{\delta^2 \mathcal W}{\delta g_{\mud\nud}(x_2)\delta g_{\kappa\nuu}(x_3)} \nn\\
&&
+ \frac{\delta^2\Gamma^{\muu}_{\kappa\nuu}(x_1)}{\delta g_{\mud\nud}(x_2)\delta g_{\mut\nut}(x_3)}
\frac{\delta\mathcal W}{\delta g_{\kappa\nuu}(x_1)}
+ \Gamma^{\muu}_{\kappa\nuu}(x_1)
\frac{\delta^3\mathcal W}{\delta g_{\mud\nud}(x_2)\delta g_{\mut\nut}(x_2)\delta g_{\kappa\nuu}(x_1)}\bigg] = 0 \, ,
\label{Ch2WI3PFCoordinate} 
\eea
double, and 
\bea
&&
16\, \bigg[\pd_{\nuu}\frac{\delta^4\mathcal W}
{\delta g_{\muq\nuq}(x_4)\delta g_{\mut\nut}(x_3)\delta g_{\mud\nud}(x_2)\delta g_{\muu\nuu}(x_1)}
+ \bigg(\frac{\delta \Gamma^{\muu}_{\kappa\nuu}(x_1)}{\delta g_{\mud\nud}(x_2)}
\frac{\delta^3 \mathcal W}{\delta g_{\muq\nuq}(x_4) \delta g_{\mut\nut}(x_3)\delta g_{\kappa\nuu}(x_1)}
\nn \\
&&
+ \big( 2 \leftrightarrow 4, 2 \leftrightarrow 3\big) \bigg)
+ \bigg( \frac{\delta^2 \Gamma^{\muu}_{\kappa\nuu}(x_1)}{\delta g_{\mut\nut}(x_3) \delta g_{\mud\nud}(x_2)}
\frac{\delta^2 \mathcal W}{\delta g_{\muq\nuq}(x_4)\delta g_{\kappa\nuu}(x_1)} + 
\big( 2 \leftrightarrow 4, 3 \leftrightarrow 4 \big) \bigg)
\nn \\
&&
+ \frac{\delta^3 \Gamma^{\muu}_{\kappa\nuu}(x_1)}
{\delta g_{\muq\nuq}(x_4) \delta g_{\mut\nut}(x_3) \delta g_{\mud\nud}(x_2)}
\frac{\delta \mathcal W}{\delta g_{\kappa\nuu}(x_1)}
+ \Gamma^{\muu}_{\kappa\nuu}(x_1) \frac{\delta^4 \mathcal W}
{\delta g_{\muq\nuq}(x_4)\delta g_{\mut\nut}(x_3)\delta g_{\mud\nud}(x_2)\delta g_{\kappa\nuu}(x_1)}=0
\label{Ch2WI4PFCoordinate} 
\eea
triple differentiations of the master equation (\ref{Ch2masterWI0Ch2}).
In the Ward identity satisfied by the $4$-point function we have left implicit the contributions obtained by permuting the action of
the functional derivatives. 

To move to the flat spacetime limit of (\ref{Ch2WI2PFCoordinateCh2}) and (\ref{Ch2WI3PFCoordinate}), 
we use the notations in (\ref{Ch1Flat}) and set to zero the contributions from the massless tadpoles, obtaining  
\bea
\pd_{\nuu} \left\langle T^{\muu\nuu}(x_1) T^{\mud\nud}(x_2) \right\rangle &=& 0 \, , \label{Ch2WI2PFCoordinateFlat} \\
\pd_{\nuu} \left\langle T^{\muu\nuu}(x_1)T^{\mud\nud}(x_2)T^{\mut\nut}(x_3) \right\rangle 
&=& 
- 2\, \left[\Gamma^{\muu}_{\kappa\nuu}(x_1)\right]^{\mud\nud}(x_2) 
\left\langle T^{\kappa\nuu}(x_1)T^{\mut\nut}(x_3)\right\rangle
\nn \\
&&
- 2\, \left[\Gamma^{\muu}_{\kappa\nuu}(x_1)\right]^{\mut\nut}(x_3) 
\left\langle T^{\kappa\nuu}(x_1)T^{\mut\nut}(x_3)\right\rangle \, , 
\label{Ch2WI3PFCoordinateFlat} \\
\pd_{\nuu} \left\langle T^{\kappa\nuu}(x_1)T^{\mud\nud}(x_2)T^{\mut\nut}(x_3)T^{\muq\nuq}(x_4) \right\rangle 
&=&
- 2\, \bigg(\left[\Gamma^{\muu}_{\kappa\nuu}(x_1)\right]^{\mud\nud}(x_2)
\left\langle T^{\kappa\nuu}(x_1) T^{\mut\nut}(x_3)T^{\muq\nuq}(x_4) \right\rangle 
\nn \\
&& \hspace{-40mm}
+ \big( 2 \leftrightarrow 3, 2 \leftrightarrow 4 \big) \bigg)
- 4\, \bigg( \left[\Gamma^{\muu}_{\kappa\nuu}(x_1)\right]^{\mud\nud\mut\nut}(x_2,x_3)
\left\langle T^{\kappa\nuu}(x_1)T^{\muq\nuq}(x_4) \right\rangle 
+ \big( 2 \leftrightarrow 4, 3 \leftrightarrow 4 \big) \bigg) \, , \label{Ch2WI4PFCoordinateFlat} 
\nn \\
\eea
which after some manipulations give the transversality constraint for the $2$-point functions and
\bea
k_{1\,\nuu} \left\langle T^{\muu\nuu}T^{\mut\nut}T^{\mud\nud} \right\rangle(\kd,\kt) 
&=&
- \kt^{\muu} \left\langle T^{\mut\nut}T^{\mud\nud}\right\rangle(\kd) 
- \kd^{\muu} \left\langle T^{\mud\nud}T^{\mut\nut}\right\rangle(\kt) \nn \\
&+&
k_{3\, \nuu} \bigg[\delta^{\muu\nut} \left\langle T^{\nuu\mut}T^{\mud\nud} \right\rangle(\kd)
+ \delta^{\muu\mut}\left\langle T^{\nuu\nut}T^{\mud\nud}  \right\rangle(\kd)\bigg] 
\nn \\
&+& 
k_{2\, \nuu} \bigg[\delta^{\muu\nud}\left\langle T^{\nuu\mud}T^{\mut\nut}  \right\rangle(\kt)
+ \delta^{\muu\mud}  \left\langle T^{\nuu\nud}T^{\mut\nut}\right\rangle(\kt)\bigg]	\, .
\nn \\
\label{Ch2WI3PFCh2}
\eea
Similarly, in the case of the $4$-point function $TTTT$, using (\ref{Ch24PFMom}) we obtain
\bea
&&
k_{1\,\nuu} \, \left\langle T^{\muu\nuu}T^{\mut\nut}T^{\mud\nud}T^{\muq\nuq} \right\rangle(\kd,\kt,\kq) =
\bigg[ - \kd^{\muu} \, \left\langle T^{\mud\nud}T^{\mut\nut}T^{\muq\nuq}\right\rangle(\kt,\kq)
\nn \\
&+&
 k_{2\,\nuu}\, \bigg( \delta^{\muu\nud}\, \left\langle T^{\nuu\mud}T^{\mut\nut}T^{\muq\nuq}\right\rangle(\kt,\kq)
+ \delta^{\muu\mud}\, \left\langle T^{\nuu\nud}T^{\mut\nut}T^{\muq\nuq}\right\rangle(\kt,\kq) \bigg)
+ \big( 2 \leftrightarrow 3, 2 \leftrightarrow 4\big) \bigg]
\nn \\
&+&
 \bigg[ 2\, k_{2\,\nuu}\, \bigg( 
\left[g^{\muu\mud}\right]^{\mut\nut}\, \left\langle T^{\nuu\nud} T^{\muq\nuq}\right\rangle(\kq) + 
\left[g^{\muu\nud}\right]^{\mut\nut}\, \left\langle T^{\nuu\mud}T^{\muq\nuq} \right\rangle(\kq) \bigg)
\nn \\
&+& \hspace{2mm}
 2\, k_{3\,\nuu}\, \bigg(
\left[g^{\muu\mut}\right]^{\mud\nud}\, \left\langle T^{\nuu\nut} T^{\muq\nuq} \right\rangle(\kq) + 
\left[g^{\muu\nut}\right]^{\mud\nud}\, \left\langle T^{\nuu\mut} T^{\muq\nuq} \right\rangle(\kq) \bigg)
\nn \\
&+&
 \bigg( k_2^{\nut} \delta^{\muu\mut} + k_2^{\mut} \delta^{\muu\nut}\bigg)\, 
\left\langle T^{\mud\nud}T^{\mut\nut}\right\rangle(\kq)
+ \bigg( k_3^{\nud} \delta^{\muu\mud} + k_3^{\mud} \delta^{\muu\nud}\bigg)\, 
\left\langle T^{\mut\nut}T^{\mud\nud}\right\rangle(\kq) + \big( 2 \leftrightarrow 4, 3 \leftrightarrow 4 \big) \bigg] \, .
\nn \\
\label{Ch2WI4PF}
\eea
Similar identities are obtained for the momenta of the other external gravitons. 

\section{Counterterms}

Coming to a discussion of the counterterms to the $4$-dilaton amplitude, these are obtained from the $1$-loop Lagrangian
which accounts for the gravitational counterterms to pure graviton amplitudes in the $\overline{MS}$ scheme,
\beq\label{Ch2CounterActionCh2}
S_{counter} = - \frac{\mu^{-\epsilon}}{\bar\epsilon}
\sum_{I=s,f,V}n_I \int d^d x \sqrt{g}\,  \bigg( \beta_a(I) \, F + \beta_b(I) \, G \bigg)\, .
\eeq
Again, the dimensional parameter is $\epsilon= 4 - d$.

In the case of the $4$-graviton vertex the counterterm action (\ref{Ch2CounterActionCh2}) generates the vertex
\beq
-\frac{\mu^{-\epsilon}}{\bar\epsilon}\, \bigg(
\beta_a\, D_F^{\muu\nuu\mud\nud\mut\nut\muq\nuq}(x_1,x_2,x_3,x_4) + 
\beta_b\, D_G^{\muu\nuu\mud\nud\mut\nut\muq\nuq}(x_1,x_2,x_3,x_4)\bigg)\, ,
\eeq
where
\bea
D_F^{\muu\nuu\mud\nud\mut\nut\muq\nuq}(x_1,x_2,x_3,x_4)
&=&
2^4 \, \frac{\delta^4}
{\delta g_{\muu\nuu}(x_1)\delta g_{\mud\nud}(x_2)\delta g_{\mut\nut}(x_3)\delta g_{\muq\nuq}(x_4)}
\int\,d^d w\,\sqrt{g}\, F\, ,
\label{Ch2DFCh2}\\
D_G^{\muu\nuu\mud\nud\mut\nut\muq\nuq}(x_1,x_2,_3,x_4)
&=&
2^4 \, \frac{\delta^4}
{\delta g_{\muu\nuu}(x_1)\delta g_{\mud\nud}(x_2)\delta g_{\mut\nut}(x_3)\delta g_{\muq\nuq}(x_4)}
\int\,d^d w\,\sqrt{g}\, G 
\label{Ch2DGCh2}\,, 
\eea
and similarly for the $2$- and $3$-point correlators.

Using these expressions, the fully renormalized $2$-, $3$- and $4$-point correlators in momentum space can be written down as
\bea
\left\langle T^{\muu\nuu}T^{\mud\nud} \right\rangle_{ren}(\kd) 
&=&
\left\langle T^{\muu\nuu}T^{\mud\nud} \right\rangle_{bare}(\kd) -
\frac{\mu^{-\epsilon}}{\bar\epsilon}\, \beta_a\, D_F^{\muu\nuu\mud\nud}(\kd) \, ,
\label{Ch2Ren2PF}
\nn \\
\left\langle T^{\muu\nuu}T^{\mud\nud}T^{\mut\nut} \right\rangle_{ren}(\kd,\kt)
&=&
\left\langle T^{\muu\nuu}T^{\mud\nud}T^{\mut\nut} \right\rangle_{bare}(\kd,\kt)
\nn \\
&-&
\frac{\mu^{-\epsilon}}{\bar\epsilon} \, \bigg(
\beta_a\, D_F^{\muu\nuu\mud\nud\mut\nut}(\kd,\kt) + \beta_b\, D_G^{\muu\nuu\mud\nud\mut\nut}(\kd,\kt) \bigg)\, ,
\label{Ch2Ren3PFCh2}
\nn \\
\left\langle T^{\muu\nuu}T^{\mud\nud}T^{\mut\nut}T^{\muq\nuq} \right\rangle_{ren}(\kd,\kt,\kq)
&=&
\left\langle T^{\muu\nuu}T^{\mud\nud}T^{\mut\nut}T^{\muq\nuq} \right\rangle_{bare}(\kd,\kt,\kq)
\nn \\
&-&
\frac{\mu^{-\epsilon}}{\bar\epsilon}\,\bigg(
\beta_a\, D_F^{\muu\nuu\mud\nud\mut\nut\muq\nuq}(\kd,\kt,\kq) + 
\beta_b\, D_G^{\muu\nuu\mud\nud\mut\nut\muq\nuq}(\kd,\kt,\kq) \bigg)\,  .
\nn \\
\label{Ch2Ren4PF}
\eea
From these relations and from (\ref{Ch2WI3PFCh2}), (\ref{Ch2WI4PF}) it is clear that counterterms must be related by the same 
general covariance Ward identities which relate the bare correlators.  One can also separately check these identites for F- and G- counterterms just 
by writing them down and equating the coefficients of $\beta_a$ and $\beta_b$.
We omit the explicit forms of the counterterms, which are necessary in order to test all these constraints.
We have checked all of them with a symbolic calculus program.

A second, powerful constraint on the counterterms comes from the anomalous Ward identities for the Green functions at hand,
which are obtained through functional derivation of (\ref{Ch2TraceAnomalySymmCh2}),
passing to the flat space limit and using the definition (\ref{Ch2NPFCh2}). 
A direct computation gives the equations
\bea 
\delta_{\muu\nuu}\, \left\langle T^{\muu\nuu}T^{\mud\nud}\right\rangle(\kd)
&=&
2 \, \left[\sqrt{g}\, \mathcal{A}\right]^{\mud\nud}(\kd) = 2 \, \left[ \Box R\right]^{\mud\nud}(\kd)\, , \label{Ch2AWard2PF} \\
\delta_{\muu\nuu}\, \left\langle T^{\muu\nuu}T^{\mud\nud}T^{\mut\nut} \right\rangle(\kd,\kt)
&=&
4 \, \left[\sqrt{g}\, \mathcal{A}\right]^{\mud\nud\mut\nut}(\kd,\kt)
- 2 \, \left\langle T^{\mud\nud}T^{\mut\nut} \right\rangle(\kd) 
- 2 \, \left\langle T^{\mud\nud}T^{\mut\nut} \right\rangle(\kt)\nn\\
&& \hspace{-40mm}
=\, 4 \, \bigg[ \beta_a\,\bigg(\left[F\right]^{\mut\nut\mud\nud}(\kd,\kt)
- \frac{2}{3} \left[\sqrt{g}\, \Box R\right]^{\mut\nut\mud\nud}(\kd,\kt)\bigg)
+ \beta_b\, \left[G\right]^{\mut\nut\mud\nud}(\kd,\kt) \bigg]\nn\\
&-&
  2 \, \left\langle T^{\mud\nud}T^{\mut\nut} \right\rangle(\kd) 
- 2 \, \left\langle T^{\mud\nud}T^{\mut\nut} \right\rangle(\kt) \, , \label{Ch2AWard3PF} \\
\delta_{\muu\nuu}\, \left\langle T^{\muu\nuu}T^{\mud\nud}T^{\mut\nut}T^{\muq\nuq} \right\rangle(\kd,\kt,\kq)
&=&
8 \, \left[ \sqrt{g}\, \mathcal{A}\right]^{\mud\nud\mut\nut\muq\nuq}(\kd,\kt,\kq)
\nn \\
&& \hspace{-60mm}
-\,  2 \, \left\langle T^{\mud\nud}T^{\mut\nut}T^{\muq\nuq} \right\rangle(\kd,\kt)
- 2 \, \left\langle T^{\mud\nud}T^{\mut\nut}T^{\muq\nuq} \right\rangle(\kd,\kq)
- 2 \, \left\langle T^{\mud\nud}T^{\mut\nut}T^{\muq\nuq} \right\rangle(\kt,\kq) 
\nn\\
&& \hspace{-70mm}
=\, 8 \, \bigg[ \beta_a\, \bigg(\left[\sqrt{g}\,F\right]^{\mud\nud\mut\nut\muq\nuq}(\kd,\kt,\kq)
- \frac{2}{3} \left[\sqrt{g}\, \Box R\right]^{\mud\nud\mut\nut\muq\nuq}(\kd,\kt,\kq)\bigg)
+ \beta_b\, \left[\sqrt{g}\, G\right]^{\mud\nud\mut\nut\muq\nuq}(\kd,\kt,\kq) \bigg] 
\nn\\
&& \hspace{-70mm}
- 2 \, \left\langle T^{\mud\nud}T^{\mut\nut}T^{\muq\nuq} \right\rangle(\kd,\kt)
- 2 \, \left\langle T^{\mud\nud}T^{\mut\nut}T^{\muq\nuq} \right\rangle(\kd,\kq)
- 2 \, \left\langle T^{\mud\nud}T^{\mut\nut}T^{\muq\nuq} \right\rangle(\kt,\kq) \, . \label{Ch2AWard4PF}
\eea
The explicit expressions of the multiple functional derivatives of the various operators in square bracket $([\,\,])$ 
are very lengthy and we omit them.

At this stage, we can extract from (\ref{Ch2AWard4PF}) four trace identities
(one for each graviton) for the counterterms of the $4$-point functions  (\ref{Ch2DFCh2}) and (\ref{Ch2DGCh2}), 
relating them to the corresponding $2$- and $3$-point ones. 
As these counterterms have been independently tested through general covariance Ward identities, this provides a 
useful test of the anomaly contributions to the $4$-point function as well, which is used to deduce the form of the quartic dilaton interactions. 
The identity, involving traces in $d$ dimensions, is a direct consequence of eq.  (\ref{dTraceCt}), which was extensively discussed, 
and its form is
\bea
\delta^{d}_{\muu\nuu}\, D_{F}^{\muu\nuu\mud\nud\mut\nut\muq\nuq}(\kd,\kt,\kq) 
&=& 
- \frac{\epsilon}{2}\, \left(\left[\sqrt{g}\, F\right]^{\mud\nud\mut\nut\muq\nuq}(\kd,\kt,\kq) 
- \frac{2}{3}\,  \left[\sqrt{g}\,\Box R \right]^{\mud\nud\mut\nut\muq\nuq}(\kd,\kt,\kq)\right)
\nn \\
&-& 
2\, D_{F}^{\mud\nud\mut\nut\muq\nuq}(\kt,\kq) - 2\, D_{F}^{\mud\nud\mut\nut\muq\nuq}(\kd,\kq)
- 2\, D_{F}^{\mud\nud\mut\nut\muq\nuq}(\kd,\kt) \, , 
\nn \\
\delta^{d}_{\muu\nuu}\, D_{G}^{\muu\nuu\mud\nud\mut\nut\muq\nuq}(\kd,\kt,\kq) 
&=& 
- \frac{\epsilon}{2}\, \left[\sqrt{g}\, G \right]^{\mud\nud\mut\nut\muq\nuq}(\kd,\kt,\kq)
\nn\\
&-& 
2\, D_{G}^{\mud\nud\mut\nut\muq\nuq}(\kt,\kq) - 2\, D_{G}^{\mud\nud\mut\nut\muq\nuq}(\kd,\kq)
- 2\, D_{G}^{\mud\nud\mut\nut\muq\nuq}(\kd,\kt) \, .
\nn \\
\label{TraceConstraint}
\eea
The superscript $d$ in (\ref{TraceConstraint}) indicates that the trace has to be taken in $d = 4 - \epsilon$ dimensions
and the $3$-point function counterterms were computed and tested in \cite{Coriano:2012wp}, 
as discussed in chapter \ref{TTTVertex}.
The correctness of the counterterms computed for the $4$-point function was already put to a test by the general covariance
Ward identities. Now we can understand why it was worth computing them. In fact, the identities (\ref{TraceConstraint})
relate these counterterms to the functional derivatives of the trace anomaly, whose computation is not less difficult.
Nevertheless, with well checked expressions for the counterterms, we can use eq.  (\ref{TraceConstraint}) to test
our functional derivatives of the trace anomaly. The test is highly non trivial, given the complexity of the expressions involved.
We succesfully performed it.

From the discussion in section \ref{CountAnom}, it follows that the traced $3$- and $4$- point correlators of the EMT exactly coincide
with the traces of their counterterms in dimensional regularization. These correlators are expected to be the constituents of the
dilaton perturbative effective action, which we indicate as $\Gamma[\rho]$ and can be tentatively written as the functional series expansion
\bea
\Gamma[\rho] 
&=& 
\dots + \int d^4\xu d^4\xd d^4\xt \, 
\left\langle {T^{\muu}}_{\muu}(\xu){T^{\mud}}_{\mud}(\xd){T^{\mut}}_{\mut} (\xt) \right\rangle\,
\rho(\xu)\rho(\xd)\rho(\xt) \nn \\
&&
+\, \int d^4\xu d^4\xd d^4\xt d^4\xq \, 
\left\langle {T^{\muu}}_{\muu}(\xu){T^{\mud}}_{\mud}(\xd)
{T^{\mut}}_{\mut} (\xt){T^{\muq}}_{\muq} (\xq) \right\rangle\, \rho(\xu)\rho(\xd)\rho(\xt)\rho(\xq)
+ \dots
\label{EffectiveDilaton}
\eea
We will see in the next chapter that the solution of the anomalous constraints generates an expression which is 
more involved than (\ref{EffectiveDilaton}). 
Nevertheless, the traced correlators of the EMT are the basic constituents of both actions and, in the on-shell limit, they exactly coincide.

\section{Three and four dilaton interactions from the trace anomaly}

From (\ref{Ch2AWard2PF})-(\ref{Ch2AWard4PF}) and from the knowledge of the trace anomalies therein, that we have explicitly 
computed, one can get the form of the off-shell 3- and $4$-dilaton ($\rho$) interactions, which are found to be
\bea \label{Ch23DVertex}
\mathcal V_{\rho\rho\rho}^{\phi}
&=&
- \frac{1}{\Lambda^3}\, \frac{3\,\left( \kd^4+ \kt^4\right) + 6 \, \kd\cdot \kt \, \left( \kd^2 + \kt^2 \right) 
+ 4 \,(\kd\cdot \kt)^2 + 5 \, \kd^2\,\kt^2}{360\,\pi^2} 
\, , \nonumber \\
\mathcal V_{\rho\rho\rho}^{\psi}
&=&
- \frac{1}{\Lambda^3}\, \frac{18\,\left( \kd^4+ \kt^4\right) + 36 \, \kd\cdot \kt \, \left( \kd^2 + \kt^2 \right) 
+ 29 \,(\kd\cdot \kt)^2 + 25 \, \kd^2\,\kt^2}{360\, \pi^2} \, , 
\nonumber \\
\mathcal V_{\rho\rho\rho}^V
&=&
- \frac{1}{\Lambda^3}\, \frac{18\,\left( \kd^4+ \kt^4\right) + 36 \, \kd\cdot \kt \, \left( \kd^2 + \kt^2 \right) 
+ 49 \,(\kd\cdot \kt)^2 + 5 \, \kd^2\,\kt^2}{180\,\pi^2} \, ,
\eea
together with the new quartic dilaton interactions 
\bea \label{Ch24DVertex}
\mathcal V_{\rho\rho\rho\rho}^{\phi}
&=&
- \frac{1}{\Lambda^4}\,\frac{1}{60\,\pi^2}\, \bigg[
3\, \left((\kd^2)^2 + (\kt^2)^2 + (\kq^2)^2 \right) + 
6\, \left( \kq^2\, \kq \cdot (\kd +\kt) +  \kt^2\, \kt \cdot(\kd + \kq) +  \kd^2\, \kd\cdot(\kt+\kq) \right)
\nn \\
&& \hspace{-10mm}
+ \, 4\, \left( ( \kd \cdot \kq)^2 + ( \kd \cdot \kt)^2 + ( \kt \cdot \kq)^2 \right) + 
6\, \left( \kd \cdot \kt\, \kd \cdot \kq + \kt \cdot \kd\, \kt \cdot \kq +  \kq \cdot \kd\, \kq \cdot \kt  \right)
\nn \\
&& \hspace{-10mm}
+\, 5\, \left( \kd^2\, \kt^2 + \kt^2\, \kq^2 +  \kd^2\, \kq^2 \right)
+ 5\, \left( \kd^2\, \kt \cdot \kq + \kt^2\, \kd \cdot \kq +  \kq^2\, \kd \cdot \kt \right) \bigg] \, , 
\nonumber \\
\mathcal V_{\rho\rho\rho\rho}^{\psi}
&=&
- \frac{1}{\Lambda^4}\, \frac{1}{120\,\pi^2}\,
\, \bigg[
36\, \left((\kd^2)^2 + (\kt^2)^2 + (\kq^2)^2 \right) + 
72\, \left( \kq^2\, \kq \cdot (\kd +\kt) +  \kt^2\, \kt \cdot(\kd + \kq) +  \kd^2\, \kd\cdot(\kt+\kq) \right)
\nn \\
&& \hspace{-10mm}
+ \, 58\, \left( ( \kd \cdot \kq)^2 + ( \kd \cdot \kt)^2 + ( \kt \cdot \kq)^2 \right) + 
82\, \left( \kd \cdot \kt\, \kd \cdot \kq + \kt \cdot \kd\, \kt \cdot \kq +  \kq \cdot \kd\, \kq \cdot \kt  \right)
\nn \\
&& \hspace{-10mm}
+\, 50\, \left( \kd^2\, \kt^2 + \kt^2\, \kq^2 +  \kd^2\, \kq^2 \right)
+ 55\, \left( \kd^2\, \kt \cdot \kq + \kt^2\, \kd \cdot \kq +  \kq^2\, \kd \cdot \kt \right) \bigg] \, , 
\nonumber \\
\mathcal V_{\rho\rho\rho\rho}^V
&=&
- \frac{1}{\Lambda^4}\, \frac{1}{60\,\pi^2}\,
\, \bigg[
36\, \left((\kd^2)^2 + (\kt^2)^2 + (\kq^2)^2 \right) + 
72\, \left( \kq^2\, \kq \cdot (\kd +\kt) +  \kt^2\, \kt \cdot(\kd + \kq) +  \kd^2\, \kd\cdot(\kt+\kq) \right)
\nn \\
&& \hspace{-10mm}
+ \,98 \, \left( ( \kd \cdot \kq)^2 + ( \kd \cdot \kt)^2 + ( \kt \cdot \kq)^2 \right) + 
122\, \left( \kd \cdot \kt\, \kd \cdot \kq + \kt \cdot \kd\, \kt \cdot \kq +  \kq \cdot \kd\, \kq \cdot \kt  \right)
\nn \\
&& \hspace{-10mm}
+\,10 \, \left( \kd^2\, \kt^2 + \kt^2\, \kq^2 +  \kd^2\, \kq^2 \right)
+ 35 \, \left( \kd^2\, \kt \cdot \kq + \kt^2\, \kd \cdot \kq +  \kq^2\, \kd \cdot \kt \right) \bigg] \, .
\eea
Both the cubic and the quartic terms can be easily modified to account for all the contributions
generated in the nearly conformal limit of the Standard Model, in the form

\bea
\mathcal V_{\rho\rho\rho / \rho \rho\rho\rho} = \sum_{i=s,f,V} N_i \, V^i_{\rho\rho\rho / \rho \rho\rho\rho}
\eea
where $N_V = 8+3+1 = 12$ is the number of gauge fields, $N_f = 3 \times 6 + 3 + 3/2 = 45/2$ is the number of Dirac fermions, where 
the factor 1/2 is due to the fermion chirality, and $N_s = 4$ counts the real scalars of the Higgs doublet. 
These corrections, as we have already remarked, are typical of the dilaton interactions and can be derived 
without any explicit diagrammatic computation. They provide the starting ground for an analysis of the dilaton effective action, 
and characterize the terms which allows to differentiate between the Higgs and the dilaton at the radiative level.

\section{Conclusions}

The analysis of dilaton interactions and of their role in the context of the electroweak symmetry 
breaking is particularly interesting at phenomenological level.

In fact, the Standard Model, in the limit in which we drop the Higgs vev, is conformally invariant at high energy, 
with a breaking of scale invariance, in this limit, which is related only to the trace anomaly.
The anomalous coupling of the dilaton is responsible for setting a remarkable difference between this state and the Higgs, a property
which remains valid - with no distinction - even if the dilaton is assumed to be a fundamental or a composite scalar or a graviscalar. 
We have shown that the anomalous corrections at any order to the dilaton effective action, in the conformal limit, can be extracted 
from a general (and model independent) analysis of the Ward identities, with no further input. 
We have illustrated the approach up to the quartic order.

\clearpage{\pagestyle{empty}\cleardoublepage}

\chapter{Conformal Trace Relations from the Dilaton Wess-Zumino Action}\label{Recursive}

\section{Introduction} 

Anomaly-induced actions play a considerable role among effective field theories.
Simple instances of these types of actions are theories with chiral fermions in the presence of anomalous abelian symmetries 
\cite{Wess:1971yu,Treiman:1986ep,Faddeev:1984jp,Babelon:1986sv}, other examples involve conformal 
\cite{Duff:1977ay,Deser:1993yx,Deser:1999zv,Polyakov:1981re} and superconformal anomalies  \cite{Chaichian:2000wr}.

Direct computations of these actions can be performed in ordinary perturbation theory by the usual Feynman expansion at 1 loop, 
but alternative approaches are also possible. In fact, an action which reproduces the same anomaly at low energy can be constructed 
quite directly, just as a variational solution of the anomaly condition, without any reference to the diagrammatic expansion.
In gravity, typical examples are anomaly actions such as the Riegert action \cite{Riegert:1987kt}, or the 
Wess-Zumino (WZ from now on) dilaton action \cite{Antoniadis:1992xu}, 
which reproduce the anomaly either with a non-local (Riegert) or with a local 
(WZ) effective operator, using a dilaton field in the latter case \cite{Codello:2012sn}. These types of actions are not unique, 
since possible contributions which are conformally invariant are not identified by the variational procedure.
It should also be mentioned that a prolonged interest in these actions has been and is linked to the study of the irreversibility of the
Renormalization Group (RG) flow in various dimensions
 \cite{Jack:1990eb,Cappelli:1990yc, Anselmi:1997ys, Komargodski:2011vj,Yonekura:2012kb} 
and of the trace anomaly matching \cite{Schwimmer:2010za}, 
since Zamolodchikov's proof of his $c$-theorem for $d=2$ \cite{Zamolodchikov:1986gt}.
 
A salient feature of some of these anomaly actions, if formulated in a local form, as in the WZ case, is the inclusion 
of extra degrees of freedom compared to the original tree-level action. In the case of the chiral anomaly this additional degree of 
freedom is the axion $(\theta(x))$, which is linearly coupled to the anomaly functional in the form of a $(\theta/M) F\tilde{F}$ term
- the anomaly coupling - with $F$ and $\tilde{F}$ denoting the field strength of the gauge field and its dual respectively. 
The anomaly interaction is accompanied by a new scale ($M$). This is the scale at which the anomalous symmetry starts to play a role 
in the effective theory. A large value of $M$, for instance, is then associated with a decoupling of the anomaly in the low energy theory. 
In the $1$-particle irreducible (1PI) effective action this is obtained - in the chiral case - by allowing the mass of the fermions ($\sim M$)
that run in the anomaly loops to grow large. The underlying idea of keeping the anomaly interaction in the form of a local operator at low energy - 
such as the $(\theta/M) F\tilde{F}$ term - while removing part of the physical spectrum, is important in the study of 
the renormalization group (RG) flows of large classes of theories, both for chiral and for conformal anomalies.

For conformal anomalies \cite{Duff:1977ay}, which is the case of interest in this chapter, the pattern is similar to the 
chiral case, with the introduction of a dilaton field in place of the axion in order to identify the structure of the corresponding WZ action, 
and the inclusion of a conformal scale ($\Lambda$).  As in the chiral case, one of the significant features of the WZ conformal anomaly 
action is the presence of a linear coupling of the Goldstone mode of the broken symmetry (the dilaton) to the anomaly functional, 
but with a significant variant. In this case, in fact, this linear term has to be corrected by additional contributions, due to the non invariance 
of the anomaly functional under a Weyl transformation. 

This procedure, which allows to identify the structure of WZ action, goes under the name of the {\em Noether method} (see for instance
\cite{Coriano:2013xua,Elvang:2012st}) and has to be iterated several times, due to the structure of the anomaly functional, 
before reaching an end. Given the fact that anomaly functional takes a different form in each spacetime dimension, the anomaly action 
will involve interactions of the dilaton field of different orders in each dimension.

In \cite{Coriano:2013xua} we have investigated an alternative approach, useful for the computation of this action, 
which exploits the structure of the counterterms in dimensional regularization and their {\em Weyl-gauging}, bypassing altogether the 
Noether procedure. This approach has been discussed in $d=4$ by several authors
\cite{Antoniadis:1992xu,Tomboulis:1988gw}, and in a cohomological context in \cite{Mazur:2001aa}.
This construction in dimensions higher than $2$ or $4$ is interesting for several reasons. 
The WZ action was been used in \cite{Elvang:2012st} in the attempt to generalize the proof of the weak a-theorem provided in 
\cite{Komargodski:2011vj}, although some additional work is required, due to the complexity of dispersive analysis of scattering
amplitudes of more than $4$ particles.
At the same time it plays an equally important role in the study of the AdS/CFT correspondence. An example is the investigation of the 
anomaly matching between conformal tensor multiplets on the six dimensional boundary and a stack of M5 branes of $AdS_7$ supergravity in 
the bulk \cite{Bastianelli:1999ab, Bastianelli:2000hi}. We will present, as an application of our formalism, the expression of the WZ action for 
this specific CFT realization in $d=6$. \\
 
In the two previous chapters, we have focused our attention on interactions concerning a dilaton field, featuring a coupling to the trace anomaly, 
which we have interpreted as an effective, low-energy signature of a possible hidden conformal sector that could be investigated in future 
data analysis at hadron colliders, maybe at the LHC. \\
In this final chapter, we turn to a different kind of analysis, switching from the phenomenological aspects to a more formal
application of dilaton interactions. We look at our dilaton as at a WZ Goldstone boson for CFT's, which
means that its self-interactions stem only from the trace anomaly.
These self-interactions are related, in turn, to definite combinations of traced EMT correlation functions, which coincide with
those analysed in chapter \ref{Traced4T} in the on-shell limit.

We will start our investigation by introducing our conventions for the anomalous equations and 
the structure of correlation functions of traces of the EMT for a generic CFT.
Then we give a brief account of the method of Weyl-gauging, which was already extensively discussed, for scale invariant theories,
in chapter \ref{Weyl}. This gives us the chance to introduce the quadratic kinetic term for the dilaton field.
We also comment on the Weyl-gauging for theories containing dimensionful constants, as the dilaton might actually be a massive state.
This way, we can account for a mass term preserving Weyl symmetry.

In the past, the gauging has been discussed in various ways both in the context of extensions of the Standard Model 
\cite{Buchmuller:1987uc,Buchmuller:1988cj} and in cosmology, where it has been shown that the introduction of an extra scalar brings 
to a dynamical adjustment of the cosmological constant \cite{Tomboulis:1988gw}. 
Recent discussions of the role of the dilaton in quantum gravity can be found in \cite{Henz:2013oxa, Percacci:2011uf}. 
This review gives us the chance to classify the possible kinetic terms for the dilaton, on the ground of conformal symmetry requirements
and limiting ourselves to terms that are at most marginal in a wilsonian sense.

Then we turn to the original part of this chapter and apply the method of Weyl-gauging to the counterterms of a CFT in order
to determine the WZ conformal anomaly action and show that in any even ($d = 2 k$) dimensions all the hierarchy of correlation 
functions involving traces of the EMT of a CFT is determined in terms of those of lower orders, up to $2 k$,
which are, in turn, completely fixed by the conformal anomaly. 
For every even dimension $d$, it is the highest order of dilaton anomalous self-interactions which corresponds to the maximum 
order of the independent traced correlators that are necessary to fix the entire hierarchy.
This order, in turn, equals the space dimension, as discussed in \cite{Coriano:2013xua,Coriano:2013nja}.
It turns out that all the correlators which are order $2k+1$ or higher are recursively generated by the first $2k$, 
through a simple algorithm that is discussed in detail.
The method also allows to compute the first $2k$ traced correlation functions of the EMT just by knowing the structure of the WZ action,
thus providing an alternative way to the counterterm approach used in chapter \ref{Traced4T}.

We work out explicitly the cases $d=4$ and $d=6$, while the case in $2$ dimensions is left to appendix \ref{Ch42D}.
Both in the $4$- and in the $6$-dimensional case, we derive the dilaton effective action by taking into account the counterterms 
required within a minimal subtraction scheme, in a sense that will be made clearer below.

We mention that  general results on the structure of the WZ action in any even dimensions have been presented in
\cite{Baume:2013ika}, using the general form of the Euler density, which is sufficient to identify the 
nonlocal structure of the anomaly in a specific scheme. However, the identification of the contributions related to the 
so called \emph{local anomalies} requires a separate effort, that we undertake here.
This more general approach allows us to set a distinction between the nonlocal and local contributions to the anomalous effective action. 

As an application in $6$ dimensions, we present the WZ action for the (2,0) tensor multiplet, which has been investigated in the past in the 
context of  the $AdS_7/CFT_6$ holographic anomaly matching.
Of the first $6$ independent correlators, we give the explicit expressions in the most general scheme of the first $4$.
We have computed the order $5$ and $6$ Green functions, but they are lengthy and add nothing essential to our discussion. \\

The gauging procedure, especially in $6$ dimensions, in a general scheme, is quite demanding from the technical side.
We set a distinction between the operators in the anomaly that are responsible for the universal (scheme-independent)
contributions, i.e. the Euler density and the Weyl invariants, and the operators which are responsible for the scheme-dependence,
such as the $\square R $ term in $4$ dimensions, establishing clearly the relation between their contribution to the WZ
effective action and the counterterms in the minimal scheme. This clarifies the difference with the prescription of \cite{Baume:2013ika}.
We then move to the analysis of the structure of the traced correlators and of their hierarchy, showing how to solve it in terms of the 
first  $d$ correlation functions. We have left to appendix \ref{Geometrical} a discussion of some of the more technical steps.
appendix \ref{Ch42D} includes the consistency checks of the recursion relations satisfied by the traced correlators 
in $2$ dimensions, presenting the expressions of the first traced Green functions up to rank $6$ in this case.

\section{Conventions}\label{RecConv}

For practical reasons, we recollect here all the basic definitions necessary for the purpose of this chapter.
In a generic euclidean field theory, defining the generating functional of the theory $\mathcal{W}$  as
\beq
\mathcal{W}[g] = \int \mathcal D \Phi\, e^{- \mathcal S}\, ,
\eeq
where $\mathcal{S}$ is the generic euclidean action depending on the set of all the quantum fields ($\Phi$) and on the 
background metric ($g$), the vev of the EMT is given by
\beq \label{Ch3EMTvev}
\left\langle T^{\mu\nu}(x) \right\rangle_s
= \frac{2}{\sqrt{g_x}}\, \frac{\delta \mathcal{W}[g]}{\delta g_{\mu\nu}(x)}
= \frac{2}{\sqrt{g_x}}\, \frac{\delta}{\delta g_{\mu\nu}(x)} \int\, \mathcal{D}\Phi\,  e^{-S}\, , 
\eeq
and contains the response to the metric fluctuations keeping the background sources turned on, as the superscript $s$ indicates; 
for the purpose of this chapter, the only background field is the metric tensor $g_{\mu\nu}$,
whereas $g_x\equiv \left|g_{\mu\nu}(x)\right|$ is its determinant.

For CFT's in a even dimensional space which are coupled to a background metric and neither scalar fields nor vector currents,
the trace anomaly condition takes the general symbolic form
\bea \label{Ch3TraceAnomaly}
g_{\mu\nu} \langle T^{\mu\nu} \rangle_s = \mathcal{A}[g] \, ,
\eea
An anomalous relation of the form (\ref{Ch3TraceAnomaly}) holds in any even dimension, where $\mathcal{A}[g]$ is a 
scalar functional depending only on the metric tensor, which can be written, in complete generality, 
as \cite{Duff:1977ay,Duff:1993wm,Deser:1993yx}
\beq
\label{Ch4anom}
\mathcal{A}[g] = 
\sum_{i} c_i\, \left( I_i + \nabla_\mu J^\mu_i \right) - (-1)^{d/2}\, a\, E_d \, ,
\eeq
where $\sqrt{g}\, I_i$ are the conformal invariants available in $d$ dimensions, whose number rapidly increases with $d$,
whereas $E_d$ is the Euler density in $d$ dimensions. They are both defined in appendix \ref{Geometrical}. \\
The contribution coming from the Euler density is usually denoted as the $A$ part of the anomaly, while the rest is called the $B$ part.
For every even value of the space dimension, say $2k$, the $I_i$'s are $2k$-dimensional combinations of the Weyl tensor for $2k$ 
dimensions and covariant derivatives thereof. 
The total derivative terms $\nabla_\mu J_i^\mu$ are known under the name of \emph{local anomaly contributions} and are sometimes omitted,
as they are scheme-dependent. This scheme dependence has already been discussed in section \ref{Ch1Renormalization} 
for the case $d=4$ and will be further explored in the following in $d=4$ and $d=6$.
The total derivatives can be removed by adding proper local counterterms to the action, as we thoroughly show in sections
\ref{WZ4} and \ref{WZ6}. 

Let us provide the explicit expressions of the trace anomaly in $4$ and $6$ dimensions. In the first case, in order to stick to the notations 
of the previous chapters, we set $- a = \beta_b$ and $ c = \beta_a$ and have the anomaly functional
\beq
\mathcal{A}[g] =  \beta_a\, \left( F - \frac{2}{3}\, \Box R \right)  + \beta_b\, G \, ,
\label{4DANOMALY}
\eeq
where $F$ is the squared $4$-dimensional Weyl tensor and $G$ is the Euler density, both defined appendix \ref{Geometrical}
and we are omitting the distinction between the various kinds of contributions that can go into the $\beta$ coefficients.

The specific expression of (\ref{Ch4anom}) for $d=6$ is instead
\beq \label{Ch4anom6D}
\mathcal{A}[g] = 
\sum_{i=1}^3 c_i\, \left( I_i + \nabla_\mu J^\mu_i \right) + a\, E_6\, , 
\eeq
where $\sqrt{g}\, I_i\, , (i=1,2,3)$, are the three Weyl invariants available in $6$ dimensions.

Our goal will be to determine the structure of the dilaton WZ action in the most general case for both $d=4$ and $d=6$,
with the inclusion of the contributions related to the total derivative terms and to completely clarify the relation of the latter to the
various possible choices of the counterterms.

As already emphasized, multiple stress-energy tensor correlators can be defined in various ways, 
differing by contact terms. These depend on the positions of the $g^{-\frac{1}{2}}$ factors entering in the definition 
of the EMT respect to the functional derivatives.
We choose to define the Green function of $n$ EMT's in flat space in the completely symmetric fashion as 
\beq \label{Ch3NPF}
\langle T^{\mu_1\nu_1}(x_1)\ldots T^{\mu_n\nu_n}(x_n)\rangle 
\equiv
\frac{2^n}{\sqrt{g_{\xu}}\ldots \sqrt{g_{x_n}}}
\frac{\delta^n \mathcal{W}[g]}{\delta g_{\mu_1\nu_1}(\xu)\ldots \ldots\delta g_{\mu_n\nu_n}(x_n)}
\bigg|_{g_{\mu\nu}=\delta_{\mu\nu}} \, .
\eeq
It is also useful to recall some notation to denote the functional derivatives 
with respect to the metric of generic functionals in the limit of a flat background
\bea \label{Ch3funcder}
\left[f(x)\right]^{\muu\nuu\dots\mu_{n}\nu_{n}}(\xu,\dots,x_n) 
\equiv
\frac{\delta^n\, f(x)}{\delta g_{\mu_n\nu_n}(x_{n}) \, \ldots\, \delta g_{\muu\nuu}(\xu)}
\bigg|_{g_{\mu\nu}=\delta_{\mu\nu}} 
\eea
and the corresponding expression with traced indices
\beq
\left[f(x)\right]^{\muu\dots\mu_n}_{\spu\muu\dots\nu_n}\left(\xu,\xd,\dots,x_n\right)
\equiv \delta_{\muu\nuu}\dots\delta_{\mu_{n}\nu_{n}}\,
\left[f(x)\right]^{\muu\nuu\dots\mu_{n}\nu_{n}}\left(\xu,\dots,x_n\right)\, ,
\eeq
where the curved euclidean metric $g_{\mu\nu}$ is replaced by $\delta_{\mu\nu}$.

It is clear that, in any CFT in even dimensions, the only object which plays a role in the
determination of the traces of these correlators is the anomaly functional, as one can realize by a direct computation. 
Specifically, from (\ref{Ch3TraceAnomaly}) one can derive trace identities for the $n$-point correlation functions.
In fact, in momentum space the entire hierarchy, which is generated by functional differentiation of (\ref{Ch3TraceAnomaly}),
takes the form
\bea
\label{Ch3hier}
\left\langle T(\ku) \, \dots \, T(k_{n+1})\right\rangle
&=&
2^n\, \left[\sqrt{g}\, \mathcal A \right]^{\muu\dots\mu_n}_{\spu\muu\dots\nu_n}\left(\ku,\dots,k_{n+1}\right)
\nn \\
&&
-\, 2 \sum_{i=1}^{n} \left\langle T(\ku)\dots T(k_{i-1})T(k_{i+1})\dots T(k_{n+1}+k_i) \right\rangle \, .
\eea
In the expression above we have introduced the notation $T \equiv {T^\mu}_\mu$ to denote the trace of the EMT. All 
the momenta characterizing the vertex are taken as incoming. \\

The identity (\ref{Ch3hier}) relates a $n$-point correlator to correlators of order $n-1$, together with the completely 
traced derivatives of the anomaly functionals. In $4$ dimensions, these are 
$\sqrt{g}\,F, \sqrt{g}\,G$ and $\sqrt{g}\,\square R$. 
For $\sqrt{g}\,F$, which is a conformal invariant, the completely traced functional derivatives are identically zero.
For $\sqrt{g}\,G$ these are non vanishing at any arbitrary order $n \geq 2$, i.e. for the $TTT$ and higher order functions,
whereas $\sqrt{g}\,\Box R$ contributes also to the trace of the $2$-point function. \\


In order to characterize the expansion of the scalars appearing in the trace anomaly equation for $d=6$ ,
we introduce the basis of dimension $6$ scalars obtained from the Riemann tensor, its contractions and derivatives, which is given 
by the $15$ terms in table \ref{Ktable}, according to the conventions of \cite{Henningson:1998gx}
\begin{table}[h]
$$
\begin{array}{lll}
 K_1    = R^3  &  
 K_2    = R\,R^{\mu\nu}\,R_{\mu\nu}  &  
 K_3    = R\,R^{\mu\nu\lambda\kappa}\,R_{\mu\nu\lambda\kappa}   \\
 K_4    = {R_\mu}^\nu\, {R_\nu}^\alpha\, {R_\alpha}^\mu   &  
 K_5    = R^{\mu\nu}\, R^{\lambda\kappa}\, R_{\mu\lambda\kappa\nu}  &  
 K_6    = R_{\mu\nu}\, R^{\mu\alpha\lambda\kappa}\, {R^\nu}_{\alpha\lambda\kappa}    \\
 K_7    = R_{\mu\nu\lambda\kappa}\, R^{\mu\nu\alpha\beta}\, {R^{\lambda\kappa}}_{\alpha\beta}  &
 K_8    = R_{\mu\nu\lambda\kappa}\, R^{\mu\alpha\beta\kappa}\, {{R^\nu}_{\alpha\beta}}^\lambda  &
 K_9    = R\square R    \\
 K_{10} = R_{\mu\nu}\square R^{\mu\nu} & 
 K_{11} = R_{\mu\nu\lambda\kappa}\square R^{\mu\nu\lambda\kappa} &  
 K_{12} = \pd_{\mu}R\, \pd^{\mu}R   \\
 K_{13} = \nabla_{\rho}R_{\mu\nu}\, \nabla^{\rho}R^{\mu\nu} & 
 K_{14} = \nabla_{\rho}R_{\mu\nu\alpha\beta}\, \nabla^{\rho}R^{\mu\nu\alpha\beta} &  
 K_{15} = \nabla_{\lambda}R_{\mu\kappa}\, \nabla^{\kappa}R^{\mu\lambda}   \\
\end{array}
$$
\caption{Basis of dimension-$6$ scalars on which the Euler density and the conformal invariants in $6$ dimensions are expanded.}
\label{Ktable}
\end{table}
\\
In terms of such basis, the Euler density takes the form 
\beq \label{Ch4Euler}
E_6 = K_1 - 12\, K_2 + 3\, K_3 + 16\, K_4 - 24\, K_5 - 24\, K_6 + 4\, K_7 + 8\, K_8  \, .
\eeq

Defining a Weyl transformation of the metric as
\beq
g_{\mu\nu}(x) \to  e^{2\, \sigma(x)}\, g_{\mu\nu}(x),
\label{Ch4WeylT}
\eeq
the three Weyl invariants (modulo a $\sqrt{g}$ factor) in $d=6$, restricted to operators of dimension $6$, {$I_i\, , i=1,2,3$},  
are given by the expressions (see appendix \ref{Geometrical} for their definitions in terms of the Weyl and Riemann tensors)
\bea \label{Ch4WeylInv6}
I_1
&=&
\frac{19}{800}\, K_1 - \frac{57}{160}\, K_2 + \frac{3}{40}\, K_3 + \frac{7}{16}\, K_4 
- \frac{9}{8}\, K_5 - \frac{3}{4}\, K_6 + K_8
\, ,\nn \\
I_2
&=&
\frac{9}{200}\, K_1 - \frac{27}{40}\, K_2 + \frac{3}{10}\, K_3 + \frac{5}{4}\, K_4 - \frac{3}{2}\, K_5 - 3\, K_6 + K_7
\, ,
\nn \\
I_3 &=&
\frac{1}{25}\, K_1 - \frac{2}{5}\, K_2 + \frac{2}{5}\, K_3 + \frac{1}{5}\, K_9 - 2\, K_{10} + 2\, K_{11} 
+ K_{13} + K_{14} - 2\,K_{15} \, .
\eea
It is easy to prove that for the three scalars defined above the products $\sqrt{g}\, I_i$ are Weyl invariant in $6$ dimensions, i.e.,
denoting with $\delta_W$ the operator implementing an infinitesimal Weyl transformation,
\beq
\delta_{W} I_i = -6 \, \sigma I_{i}\, .
\eeq
The origin of the derivative terms $\nabla_\mu J^\mu_i$ in eq.  (\ref{Ch4anom6D}) is discussed in section \ref{WZ6}.
Their explicit expressions are
\bea
\nabla_\mu J_1^\mu 
&=& 
- \frac{3}{800}\, \nabla_\mu \bigg[
- 5\, \bigg( 44\, R^{\lambda\kappa}\, \nabla^\mu R_{\lambda\kappa} - 50\, R_{\lambda\kappa}\, \nabla^\sigma R^{\mu\rho} 
- 3\, R^{\mu\nu}\, \pd_\nu R - 4\, R_{\nu\lambda\kappa\alpha }\, \nabla^\mu R^{\nu\lambda\kappa\alpha} 
\nn \\
&& \hspace{20mm}
+\, 40\, R^{\mu\lambda\nu\kappa}\, \nabla_\nu R_{\lambda\kappa} \bigg) + 19\, R\, \pd^\mu R  \bigg] \, ,
\nn \\
\nabla_\mu J_2^\mu 
&=& 
- \frac{3}{200}\, \nabla_\mu \bigg[
- 5\, \bigg( 4\, R^{\lambda\kappa}\, \nabla^\mu R_{\lambda\kappa} 
+ 10\, R_{\lambda\kappa}\, \nabla^\kappa R^{\mu\lambda} + 7\, R^{\mu\nu}\, \pd_\nu R 
- 4\, R_{\nu\lambda\kappa\alpha }\, \nabla^\mu R^{\nu\lambda\kappa\alpha}
\nn \\
&& \hspace{20mm}
-\, 40\, R^{\mu\lambda\nu\kappa}\, \nabla_\nu R_{\lambda\kappa}\,  \bigg) + 9 R\, \pd^\mu R  \bigg] \, ,
\nn \\
\nabla_\mu J_3^\mu
&=& 
\frac{1}{25}\, \nabla_\mu \bigg[ 
10\,  \bigg( 2\,\pd^\mu \square R - 5\, \nabla_\nu \square R^{\mu\nu} + R_{\nu\rho}\, \nabla^\mu R^{\nu\rho}
-2\, R^{\mu\nu}\, \pd_\nu R - R_{\nu\lambda\kappa\alpha }\, \nabla^\mu R^{\nu\lambda\kappa\alpha}
\nn \\
&& \hspace{15mm}
-\, 10\, R^{\mu\lambda\nu\kappa}\, \nabla_\nu R_{\lambda\kappa}\,  \bigg) - 3\, R\, \pd^\mu R  \bigg] \, .
\eea
For about the completely traced derivatives of the anomaly functionals in $d=6$,
$\sqrt{g}\,I_{i}$, $\sqrt{g}\,E_6$ and $\sqrt{g}\,\nabla_\mu J^\mu_i$,
those of $\sqrt{g}\,I_i$, which are conformal invariants, are identically zero.
Concerning $\sqrt{g}\,E_6$, which is cubic in the Riemann tensor, these contributions are non vanishing at any arbitrary order $n \geq 3$.
Finally, $\sqrt{g}\,\nabla_\mu J^\mu_i$ contribute to lower order functions as well.
In particular, being $\nabla_\mu J^\mu_{1}$ and $\nabla_\mu J^\mu_{2}$ 
at least quadratic in the Riemann tensor, they give non-vanishing contributions 
from order $3$ onwards, whereas $\nabla_\mu J^\mu_3$ contains a term which is linear in $R$ 
and thus contributes a non-vanishing trace to the $2$-point function as well.

\section{Overview of Weyl-gauging }

The goal of this section is to extend the method of Weyl-gauging, already discussed in chapter \ref{Weyl}, 
to non scale invariant theories.
We first give a quick overview of the method for classical theories and later discuss its application to the renormalized quantum 
effective action of a CFT, so as to provide the basics for our derivation of the WZ action for conformal anomalies.
We also comment on possible kinetic terms for the dilaton, which obviously cannot be determined by the anomaly equation.
Their possible form is suggested on the sole requirement of conformal invariance.

\subsection{Weyl-gauging for scale invariant theories}

Scale invariance in flat space is equivalent, once the Lagrangian has been merged in a gravitational background,
to global Weyl invariance.
The equivalence is shown by rewriting a scale transformation acting on the coordinates of flat space and the fields $\Phi$,
\bea \label{Ch4FlatScaling}
x^\mu &\to& {x'}^{\mu}=e^\sigma x^\mu \, , \nn\\
\Phi(x)&\to& \Phi'(x')=e^{-d_\Phi \sigma}\Phi(x) \, ,
\eea
in terms of a rescaling of the metric tensor, the Vielbein and the matter fields
\bea \label{Ch4CurvedScaling}
g_{\mu\nu}(x) &\to&  e^{2\sigma}\, g_{\mu\nu}(x)\, , \nn\\
{V}_{a\,\rho}(x) &\to&  e^{\sigma}\, V_{a\,\rho}(x)\, , \nn\\
\Phi(x)&\to& e^{-d_\Phi \sigma}\Phi(x)\, ,
\eea
leaving the coordinates $x$ unchanged, where $d_\Phi$ is the canonical scaling dimension of the field $\Phi$.
In a curved metric background,  we can promote $\sigma$ to a local function, and thus turn the global scale transformations
(\ref{Ch4CurvedScaling}) into
\bea 
\label{Ch4WeylTransf}
{g'}_{\mu\nu}(x)  &=&  e^{2\sigma(x)}\, g_{\mu\nu}(x)\, , \nn \\
{V'}_{a\,\rho}(x)  &=&  e^{\sigma(x)}\, V_{a\,\rho}(x)\, , \nn \\
\Phi'(x)           &=&  e^{-d_{\Phi}\, \sigma(x)}\, \Phi(x)\, ,
\eea
leave the fundamental Lagrangian invariant.  \\
Following closely the analogy with quantum electrodynamics, derivative terms are modified according to the prescription
\beq \label{Ch4DerTransf}
\pd_\mu \rightarrow \pd^W_\mu = \pd_\mu - d_{\phi}\, W_{\mu}\, ,
\eeq
where $W_{\mu}$ is the Weyl vector gauge field, that shifts under Weyl scaling as
\beq \label{Ch4WeylGaugeField}
W_{\mu} \rightarrow W_{\mu} - \pd_\mu \sigma\, ,
\eeq
just as for a gauge transformation of the vector potential.
In the case of higher spin fields, e.g. a vector field $v_\mu$, the Weyl-gauging has to be supplemented
with a prescription to render the general covariant derivative Weyl invariant, which is to add to (\ref{Ch4DerTransf}) the modified 
Christoffel connection
\beq \label{Ch4ModChristoffel}
\hat\Gamma^\lambda_{\mu\nu} =
\Gamma^\lambda_{\mu\nu} + {\delta_\mu}^\lambda\, W_\nu + {\delta_\nu}^\lambda\, W_\mu - g_{\mu\nu}\, W^\lambda\, .
\eeq
This hatted Christoffel symbol is Weyl invariant, so that we can define the Weyl covariant derivatives acting on vector fields as
\bea \label{Ch4WeylChristoffel}
\nabla^W_\mu v_\nu 
&=& 
\pd_\mu v_\nu - d_v\, W_\mu v_\nu - \hat\Gamma^\lambda_{\mu\nu} v_\lambda \, ,
\nn\\
\nabla^W_\mu v_\nu  
&\rightarrow& 
e^{- d_v\sigma(x)}\, \nabla^W_\mu v_\nu \, ,
\eea
which is obviously generalized to tensors of arbitrary rank. \\
In order to include fermions, we can define the covariant derivative
\beq \label{Ch4WeylSpinConnection}
\nabla_{\mu} \rightarrow \nabla^W_\mu = \nabla_{\mu} - d_\psi\, W_{\mu} + 2\, {\Sigma_\mu}^\nu\, W_\nu\, , \quad 
\Sigma^{\mu\nu} \equiv {V_a}^\mu\, {V_b}^\nu \Sigma_{ab}\, ,
\eeq
where $\Sigma_{ab}$ are the spinor generators of the Lorentz group.
More details on the Weyl-gauging for scale invariant theories are given in chapter \ref{Weyl}.

Here we want to discuss kinetic terms for the additional degree of freedom that is introduced when a theory is Weyl-gauged.
For instance, the Weyl vector field $W_{\mu}$ can be rendered dynamical by the inclusion 
of a kinetic term built out of an appropriate field strength 
\beq
F^W_{\mu\nu}\equiv\partial_{\mu} W_{\nu} - \partial_{\nu} W_{\mu} \, ,
\label{Ch4FieldStrength}
\eeq
which is manifestly Weyl invariant.
 
A second possibility is to maintain the expression of $W_\mu$ identifying it with the gradient of a dilaton field, 
 \beq
 \label{Ch4dil}
 W_\mu(x) = \frac{\partial_{\mu} \rho(x)}{\Lambda}\, . 
 \eeq
As we will shortly point out below, this second choice offers an interesting physical interpretation, in connection with the breaking
of the conformal symmetry, which is related to the conformal scale $\Lambda$.  Notice that in this second case the 
$\Omega$ term generates non trivial cubic and quartic interactions between
the original scalar and the dilaton 
\beq \label{Ch4Omegatau}
\Omega_{\mu\nu}\left(\frac{\pd\rho}{\Lambda}\right) = 
\frac{\nabla_\mu \pd_\nu\rho}{\Lambda}  - \frac{\pd_\mu\rho\,\pd_\nu\rho}{\Lambda^2}
+ \frac{1}{2}\,g_{\mu\nu}\, \frac{\left(\pd\rho \right)^2}{\Lambda^2},
\eeq
which bring the Weyl-gauged action for the scalar field (\ref{ChIntgauged}) to the form
\beq
\mathcal S_{\phi,\pd\rho} = 
\frac{1}{2}\, \int d^dx\, \sqrt{g}\,  g^{\mu\nu}\, \left(
\pd_\mu \phi\, \pd_\nu\phi  + \frac{d-2}{2}\, \phi^2\, \frac{\Box\rho}{\Lambda} 
+ \left(\frac{d-2}{2}\right)^2 \, \phi^2\,  \frac{\left(\pd\rho\right)^2}{\Lambda^2} \right) \, .
\label{Ch4gaugedtau}
\eeq
As the field strength $F^W_{\mu\nu}$ in (\ref{Ch4FieldStrength}), on account of (\ref{Ch4dil}) is obviously zero, the standard way 
to give a kinetic term to the dilaton is by introducing  a conformally coupled scalar field $\chi$ and imposing the field redefinition
\beq
\chi(\rho) \equiv \Lambda^{\frac{d-2}{2}}\, e^{-\frac{(d-2)\,\rho}{2\Lambda}}.
 \label{Ch4chitau}
 \eeq
It is clear that, with the choice (\ref{Ch4dil}), we are no longer gauging the Weyl group, but we are just using a compensation 
procedure to introduce an additional scalar field which we later render dynamical via the kinetic term for $\chi$.
As we shall see in the next section, this is mandatory if one wants to achieve Weyl invariance for a theory containing dimensional
parameters.
At this point, the dynamics of the combined scalar/dilaton/graviton system is described by the Weyl invariant action
 \beq
 \mathcal{S}= S_{\chi(\rho), imp} + S_{\phi,\partial \rho} \, ,
 \eeq
having combined (\ref{ChIntgauged}), where $\phi$ is replaced by $\chi$, and (\ref{Ch4gaugedtau}). 
The  kinetic action for $\chi$, $S_{\chi(\rho), imp}$, takes the form
\beq \label{Ch4DilatonKinetic}
\mathcal S_{\chi(\rho), imp} = \frac{\Lambda^{d-2}}{2}\, \int d^dx\, \sqrt{g}\, e^{-\frac{(d-2)\,\rho}{\Lambda}}\, \bigg( 
\frac{(d-2)^2}{4\,\Lambda^2}\, g^{\mu\nu}\,\pd_\mu \rho\, \pd_\nu\rho - \frac{1}{4}\, \frac{d-2}{d-1}\, R \bigg) \, .
\eeq
The Weyl-gauging procedure, as we have described it so far, is possible only when we take as a starting point a scale invariant Lagrangian, 
with dimensionless constants.
Things are different when an action is not scale invariant in flat space, and in that case the same gauging requires some extra steps. 
We illustrate this point below and discuss the modification of the procedure outlined above, by considering again a scalar theory as an example.
This approach exemplifies a situation which is typical in theories with spontaneous breaking of the ordinary gauge symmetry, 
such as the Standard Model. 
 
 \subsection{Weyl-gauging for non scale invariant theories } 

We consider a free scalar theory with a mass term 
\beq
\label{Ch4tre}
\mathcal{S}_2  = \frac{1}{2}\, \int d^d x \sqrt{g}\, \left(g^{\mu\nu}\,\partial_{\mu}\phi\, \partial_\nu \phi + m^2\, \phi^2 \right) \, .
\eeq
Scale invariance is lost, but it can be recovered. 
There are two ways to promote this action to a scale invariant one. The first is simply to render the mass term dynamical 
\beq
 \label{Ch4redef}
 m\to m\, \frac{\Sigma}{\Lambda}\, ,
 \eeq
using a second scalar field, $\Sigma$. The action (\ref{Ch4tre}), 
with the replacement (\ref{Ch4redef}), can be extended with the inclusion of the 
kinetic term for $\Sigma$.  The inclusion of $\Sigma$ and the addition of two conformal couplings (i.e. of two Ricci gaugings) both for 
$\phi$ and $\Sigma$ brings to the new action
\beq
\label{Ch4tre1}
\mathcal{S}^{\Sigma}_2  = \int d^d x \sqrt{g}\, \left[ \frac{1}{2}\,g^{\mu\nu}\, 
\bigg(\partial_{\mu}\phi\, \partial_\nu \phi + \partial_{\mu}\Sigma \, \partial_\nu \Sigma \bigg)
+ \frac{1}{2}\, m^2\, \frac{\Sigma^2}{\Lambda^2}\, \phi^2
+ \frac{1}{4}\, \frac{d-2}{d-1}\,R\, \bigg(\phi^2 + \Sigma^2\bigg) \right]\, ,
\eeq
which is Weyl invariant in curved space. 
These types of actions play a role in the context of Higgs-dilaton mixing in conformal invariant extension of the Standard 
Model, where $\phi$ is replaced by the Higgs doublet and $\Sigma$ is assumed to acquire a vacuum expectation value (vev)
which coincides with the conformal breaking scale $\Lambda$  ($\langle \Sigma \rangle=\Lambda$) 
(see for instance \cite{Coriano:2012nm}). 
The mixing is induced by a simple extension of (\ref{Ch4tre1}), where the mass term is generated via the scale invariant potential 
\beq
\mathcal{S}_{pot}= 
\lambda\, \int d^4 x\, \sqrt{g}\, \left( \phi^2 - \frac{\mu^2}{2\,\lambda}\frac{\Sigma^2}{\Lambda^2} \right)^2 \,  
\label{Ch4example}
\eeq
(with $m=\mu$). This choice provides a clear example of a Weyl invariant Lagrangian that allows a spontanous breaking of the 
$Z_2$ symmetry of the scalar sector $\phi$, following the breaking of the conformal symmetry
($\langle\Sigma \rangle=\Lambda, \textrm{with} \langle \rho\rangle=0$). 
The theory is obviously Weyl invariant, but the contributions proportional to the Ricci scalar $R$ do not survive in the flat limit.  \\

We have to mention that the approach to Weyl-gauging discussed so far for scale invariant and non scale invariant theories is not unique. 
In fact, a second alternative in the construction of a Weyl invariant Lagrangian in curved space, 
starting from (\ref{Ch4tre}), using the compensation procedure, which amounts to the replacements 
\bea \label{Ch4Compensate}
 m &\to& m \, e^{- \rho/\Lambda}\, ,\nn \\
 g_{\mu\nu}&\to& \hat{g}_{\mu\nu}\equiv g_{\mu\nu}\, e^{-2 {\rho/\Lambda }} \, \nn \\
 \phi&\to & \hat{\phi}\equiv \phi \, e^{\rho/\Lambda} \, , \nn\\
 \partial_\mu \phi &\to& \partial_\mu \hat{\phi}= e^{\rho/\Lambda}\, \partial^W_\mu\, \phi  \, .
 \eea
It is immediately seen that, for instance, that the application of the replacements (\ref{Ch4Compensate}) 
to the scalar field action (\ref{Ch4tre}) give
\beq
\hat{\mathcal{S}}_2 
\equiv
\mathcal{S}_2(\hat{g},\hat{\phi})= 
\frac{1}{2}\, \int d^4 x\,  \sqrt{g}\,  \bigg[ g^{\mu\nu}\,\partial_{\mu}\phi\, \partial_\nu \phi
+ g^{\mu\nu}\,\Omega_{\mu\nu}\left(\frac{\partial \rho}{\Lambda}\right)\phi^2 + m^2\, e^{-2\,\rho/\Lambda}\, \phi^2 
\bigg] \, ,
\label{Ch4newflat}
\eeq
where $\Omega(\pd\rho/\Lambda)$ was defined in (\ref{Ch4Omegatau}). 
Also in this case, the compensator $\rho$ ca be promoted to a dynamical field by adding to
$\hat{\mathcal{S}}_2$ the kinetic contribution of a conformally coupled scalar (\ref{Ch4DilatonKinetic}), 
thereby obtaining the total action
\beq 
\mathcal{S}_T \equiv \hat{\mathcal{S}}_2 + \mathcal{S}_{\chi(\rho), imp}\, . 
\eeq
Notice that in this case we choose not to require the Ricci gauging of the $\Omega\left(\pd\rho/\Lambda \right)$ 
term in $\hat{\mathcal{S}}_2$, but we leave it as it is, thereby generating additional interactions between the dilaton and the scalar $\phi$ 
in flat space. Obviously, also following this second route, we can incorporate spontaneous breaking of the $Z_2$ symmetry of the 
$\phi$ field after the breaking of conformal invariance (with $\langle\Sigma\rangle=\Lambda)$. This is obtained, as before, by the inclusion of 
the potential (\ref{Ch4example}).

In this second approach the $\Omega(\partial \rho/\Lambda)$ terms are essential in order to differentiate between the two residual dilaton 
interactions in flat space. In the context of Weyl invariant extensions of the Standard Model, such terms are naturally present in the analysis  
of \cite{Buchmuller:1988cj}.

\subsection{The dynamical dilaton}\label{DynDil}

So far, we have reviewed how to build standard kinetic terms for a dilaton  field, essentially by requiring Weyl invariance to be preserved.
Nevertheless, a general Lagrangian describing the dilaton dynamics could contain, in principle, higher derivative contributions.

It is trivial that any diffeomorphism-invariant functional of the field-enlarged metric $\hat{g}_{\mu\nu}$ in (\ref{Ch4Compensate})
is Weyl invariant. Thus, by applying the Weyl-gauging procedure to all the infinite set of diffeomorphism invariant functionals which can be built
out of the metric tensor and of increasing mass dimension, one can identify the homogeneous terms of the anomaly action.
Beside, there will be the anomalous contributions, which are accounted for by the WZ action. The latter can be added in 
order to identify a consistent anomaly action.
As we will see, these terms play no role in the determination of the hierarchy (\ref{Ch3hier}), so that the ambiguity intrinsic to their choice 
is no harm for our results.

The Weyl invariant terms may take the form of any scalar contraction of $\hat R_{\mu\nu\rho\sigma}$, $\hat R_{\mu\nu}$, $\hat R$ù
and Weyl covariant derivatives thereof and can be classified by their mass dimension. Typical examples are
\beq
\mathcal{J}_n\sim \frac{1}{\Lambda^{2(n-2)}}\int d^4 x \sqrt{\hat{g}}\hat{R}^n \, .
\eeq
In principle, all these terms can be included into $\Gamma_0[\hat{g}]\equiv \Gamma_0[g,\rho]$ which describes 
the non anomalous part of the renormalized action 
\beq
\Gamma_0[\hat{g}]\sim \sum_n \mathcal J_n[\hat{g}]\, .
\eeq
Here we recall the structure of the operators that are at most marginal from the Renormalization Group viewpoint.
The first term that can be included is trivial, corresponding to a cosmological constant contribution 
\beq
\mathcal S^{(0)}_\rho =  \Lambda^d\, \int d^dx\, \sqrt{\hat{g}}=  
\Lambda^d\, \int d^dx\, \sqrt{g}\,e^{-\frac{d\,\rho}{\Lambda}} \, .
\eeq
Here the superscript number in round brackets in $\mathcal{S}^{(2 n)}$ denotes the order of the contribution in the derivative expansion, 
so to distinguish the scaling behaviour of the various terms under the variation of the length scale. 
These terms can be included into the effective action of the theory, that we name $\Gamma_0[\hat{g}]\equiv \Gamma_0[g,\rho]$, 
which describes the non anomalous part of the interactions,
\beq
\Gamma_0[\hat{g}]\sim \sum_n \mathcal J_n[\hat{g}] \, .
\eeq
The next contribution to $\Gamma_0$ is the kinetic term for the dilaton, which can be obtained in two ways.
The first method is to consider the Weyl-gauged Einstein-Hilbert term
\bea
\int d^dx\, \sqrt{\hat g}\, \hat R 
&=& 
\int d^dx\, \sqrt{g}\, e^{\frac{(2-d)\,\rho}{\Lambda}}\, \bigg[
R - 2\, \left(d-1\right)\, \frac{\Box\rho}{\Lambda} +
\left(d-1\right)\,\left(d-2\right)\, \frac{\left(\pd\rho\right)^2}{\Lambda^2} \bigg] \nn \\
&=&
\int d^dx\, \sqrt{g}\, e^{\frac{(2-d)\,\rho}{\Lambda}}\, \bigg[R - 
\left(d-1\right)\,\left(d-2\right)\, \frac{\left(\pd\rho\right)^2}{\Lambda^2} \bigg]\, ,
\eea
with the inclusion of an appropriate normalization 
\bea \label{Ch3EH}
\mathcal S^{(2)}_{\rho} = - \frac{\Lambda^{d-2}\, \left(d-2\right)}{8\,\left(d-1\right)}\, \int d^dx\, \sqrt{\hat g}\, \hat R  \, ,
\eea
which reverts the sign in front of the Einstein term. We recall that the extraction of a conformal factor ($\tilde{\sigma}$) 
from the Einstein-Hilbert term from a fiducial metric $\bar{g}_{\mu\nu}$ ($g_{\mu\nu}=\bar{g}_{\mu\nu}e^{\tilde{\sigma}}$)
generates a kinetic term for ($\tilde{\sigma}$) which is ghost-like. In this case the non-local anomaly action, which in perturbation theory 
takes Riegert's form \cite{Riegert:1984kt}, can be rewritten in the WZ form but at the cost of sacrificing covariance, due 
to the specific choice of the fiducial metric. 

An alternative method consists in writing down the usual conformal invariant action for a scalar field $\chi$ 
in a curved background
\bea \label{Ch3ScalarImprovenChi}
\mathcal S^{(2)}_{\chi} = \frac{1}{2}\, \int d^dx\, \sqrt{g}\, \bigg( 
g^{\mu\nu}\,\pd_\mu \chi\, \pd_\nu\chi - \frac{1}{4}\,\frac{d-2}{d-1}\, R\,\chi^2 \bigg) \, .
\eea
By the field redefinition $\chi\equiv\Lambda^{\frac{d-2}{2}}\, e^{-\frac{(d-2)\,\rho}{2\Lambda}}$ 
eq.  (\ref{Ch3ScalarImprovenChi}) becomes
\beq \label{Ch3DilatonKinetic}
\mathcal S^{(2)}_{\rho} = \frac{\Lambda^{d-2}}{2}\, \int d^dx\, \sqrt{g}\, e^{-\frac{(d-2)\,\rho}{\Lambda}}\, \bigg( 
\frac{(d-2)^2}{4\,\Lambda^2}\, g^{\mu\nu}\,\pd_\mu \rho\, \pd_\nu\rho - \frac{1}{4}\, \frac{d-2}{d-1}\, R \bigg) \, ,
\eeq
which, for $d=4$, reduces to the familiar form
\beq \label{Ch3KinTau}
\mathcal S^{(2)}_{\rho} = \frac{1}{2}\, \int d^4 x\, \sqrt{g}\, e^{-\frac{2\,\rho}{\Lambda}}\, \bigg(
g^{\mu\nu}\,\pd_\mu \rho\, \pd_\nu\rho - \frac{\Lambda^2}{6}\, R \bigg)\, 
\eeq
and coincides with the previous expression (\ref{Ch3EH}), obtained from the formal Weyl invariant construction. \\

In four dimensions we can build the following possible subleading contributions (in $1/\Lambda$) to the effective action which, when gauged, 
can contribute to the fourth order dilaton action 
\beq
S^{(4)}_\rho = \int d^4x\, \sqrt{g}\, \bigg( \alpha\, R^{\mu\nu\rho\sigma}\, R_{\mu\nu\rho\sigma} 
                             + \beta\, R^{\mu\nu}\, R_{\mu\nu} + \gamma\, R^2 + \delta\, \Box R\bigg)\, .
\eeq
The fourth term ($\sim \Box R$) is just a total divergence, whereas two of the remaining three terms can be 
traded for the squared Weyl tensor $F$ and the Euler density $G$. 
As $\sqrt{g}\, F$ is Weyl invariant and $G$ is a topological term, neither of them contributes, 
when gauged according to (\ref{Ch4Compensate}), so that the only non vanishing $4$-derivative term
in the dilaton effective action in four dimensions is
\beq \label{Ch3UpToMarginal}
S^{(4)}_{\rho} = \gamma\, \int d^4x\, \sqrt{\hat{g}}\, \hat{R}^2 = 
\gamma\, \int d^4x\, \sqrt{g}\, 
\bigg[ R - 6\, \bigg( \frac{\Box\rho}{\Lambda} - \frac{\left(\pd\rho\right)^2}{\Lambda^2} \bigg) \bigg]^2 \, .
\eeq
with $\gamma$ a dimensionless constant. 
Thus, we have got the final form of the dilaton effective action in $d=4$ up to order four in the derivatives of the metric tensor
\bea
\label{Ch3tot}
S_{\rho} &=& 
S^{(0)}_{\rho} + S^{(2)}_{\rho} + S^{(4)}_{\rho} + \dots =
\int d^4x\, \sqrt{\hat{g}}\, \bigg\{ \alpha
- \frac{\Lambda^{d-2}\, \left(d-2\right)}{8\,\left(d-1\right)}\, \hat{R} + \gamma\, \hat{R}^2\, 
\bigg\} + \dots\, ,
\eea
where the ellipsis refer to additional operators which are suppressed in $1/\Lambda$. 
In flat space ($g_{\mu\nu}\rightarrow \delta_{\mu\nu}$), (\ref{Ch3UpToMarginal}) becomes
\beq
\mathcal S_{\rho} = \int d^4x\, \bigg[ e^{-\frac{4\,\rho}{\Lambda}}\, \alpha + 
\frac{1}{2}\, e^{-\frac{2\,\rho}{\Lambda}} \, \left(\pd\rho\right)^2 +
36\,\gamma\, \bigg( \frac{\Box\rho}{\Lambda} - \frac{\left(\pd\rho\right)^2}{\Lambda^2} \bigg) \bigg] + \dots
\eeq
where the ellipsis refer to higher dimensional contributions.
In general we can identify $\Gamma_0[\hat{g}]$ with $S_{\rho}$ as given in (\ref{Ch3tot}), thereby fixing the Weyl invariant 
contribution to $\Gamma_{\textrm{ren}}$. \\


Now we consider the case of $6$ dimensions, where operators are marginal up to dimension $6$.
From (\ref{Ch3DilatonKinetic}), the kinetic term in $6$ dimensions is just
\beq \label{KinRho6}
\mathcal S_{\chi(\tau), imp} = 
\int d^6 x\, \sqrt{g}\, e^{-\frac{4\,\tau}{\Lambda}}\, 
\bigg( 2\, \Lambda^2\, g^{\mu\nu}\, \pd_\mu \tau\, \pd_\nu\tau - \frac{\Lambda^4}{10}\, R \bigg)\, .
\eeq
The possible $4$-derivative terms ($n=2$) are
\beq
\label{Ch44der}
\int d^6x\, \sqrt{\hat g}\, \bigg( \alpha\, \hat R^{\mu\nu\lambda\kappa}\, \hat R_{\mu\nu\lambda\kappa} 
+ \beta\, \hat R^{\mu\nu}\, \hat R_{\mu\nu}+ \gamma\, \hat R^2 + \delta\, \hat \Box \hat R \bigg)\, .
\eeq
The $\square R$ contribution in this expression can be obviously omitted, being a total derivative.
We can also replace the Riemann tensor with the Weyl tensor squared  and remain with only two 
(as $\sqrt{\hat g}\,\hat C^{\mu\nu\lambda\kappa}\, \hat C_{\mu\nu\lambda\kappa}= 
\sqrt{g}\,C^{\mu\nu\lambda\kappa}\, C_{\mu\nu\lambda\kappa}\, e^{\frac{2\rho}{\Lambda}} $)
non trivial contributions, $\hat R^{\mu\nu}\, \hat R_{\mu\nu}$ and $\hat R^2$.
We present here the expression of (\ref{Ch44der}) for a conformally flat metric, while the result for a general gravitational background 
can be computed exploiting the Weyl-gauged tensors given in appendix \ref{Geometrical},
\bea
S^{(4)}_\rho 
&=& 
\int d^6x\, \sqrt{\hat g}\, \bigg( \alpha\, \hat R^{\mu\nu}\, \hat R_{\mu\nu} + \beta\, \hat R^2\bigg)
\nn \\
&=& 
\int d^6x\, e^{-\frac{2\,\rho}{\Lambda}}\, \bigg[ 
100\, \alpha\, \bigg( \frac{\Box\rho}{\Lambda} - 2\,\frac{\left(\pd\rho\right)^2}{\Lambda^2} \bigg)^2 
 + 2\, \beta\, \bigg( 15\,\frac{\left(\Box\rho\right)^2}{\Lambda^2}  
- \frac{68}{\Lambda^3}\, \Box\rho\, \left(\pd\rho\right)^2 + 72\, \frac{\left(\pd\rho\right)^4}{\Lambda^4}  \bigg)
\bigg]\, .
\eea
The last contributions that are significant down to the infrared regime are the marginal ones, i.e. the 6-derivative operators. 
To derive them we follow the analysis in \cite{Elvang:2012st}. We use the basis of diffeomorphic invariants which are order 6 in the derivatives,
on which the $I_i's$ are expanded (see eq.  (\ref{Ch4WeylInv6})). It is made of 11 elements, 6 of which contain the Riemann tensor, that 
can be traded for a combination of the Weyl tensor and the Ricci tensor and scalar, so that we are left with only the 5 terms in 
$(K_1 - K_{11})$   (see table \ref{Ktable}) that do not contain the Riemann tensor.
As we are going to write down the result only in the flat limit, we can exploit two additional constraints.
Indeed in \cite{Anselmi:1999uk} it was shown that, in this case, the integral of
\beq
R^3 - 11\, R\,R^{\mu\nu}\, R_{\mu\nu} + 30\,  {R_\mu}^\nu\,{R_\nu}^\alpha\,{R_\alpha}^\mu
- 6\, R\Box R + 20\, R^{\mu\nu}\Box R_{\mu\nu}
\eeq
vanishes, so that we can use this result to eliminate $R^{\mu\nu}\Box R_{\mu\nu}$. \\
Then, as the Euler density can be written in the form
\bea
E_6 
&=&
\frac{21}{100}\, R^3 - \frac{27}{20}\, R\, R^{\mu\nu}\, R_{\mu\nu}
+ \frac{3}{2}\, {R_\mu}^\nu\,{R_\nu}^\alpha\, {R_\alpha}^\mu
+ 4\, C_{\mu\nu\lambda\kappa}\,{C^{\mu\nu}} _{\alpha\beta}\, C^{\lambda\kappa\alpha\beta}
\nn \\
&&
-\, 8\, C_{\mu\nu\lambda\kappa}\,C^{\mu\alpha\lambda\beta}\, {{{C^\nu}_\alpha}^\kappa}_\beta
-6\, R_{\mu\nu}\, C^{\mu\alpha\lambda\kappa}\, {C^\nu}_{\alpha\lambda\kappa}
+\frac{6}{5}\, R\, C^{\mu\nu\lambda\kappa}\, C_{\mu\nu\lambda\kappa}
- 3\, R^{\mu\nu}\, R^{\lambda\kappa}\, C_{\mu\lambda\kappa\nu} \, ,
\eea
it is apparent that only the first three terms are non vanishing on a conformally flat metric.
Now, as in the effective action these contributions are integrated and the Euler density is a total derivative,
one can thereby replace ${R_\mu}^\nu\,{R_\nu}^\alpha\,{R_\alpha}^\mu$ for $R^3$ and $R\,R^{\mu\nu}\,R_{\mu\nu}$.
In the end, Weyl-gauging $R^3$, $R^{\mu\nu}\,R_{\mu\nu}$ and $R\Box R$ is sufficient to account for all the possible 
6-derivative terms of the dilaton effective action which do not vanish in the flat space limit.
After some integrations by parts, one can write the overall contribution as
\bea
S^{(6)}_\rho 
&=&
\int d^6x\, \sqrt{\hat{g}}\, \bigg[  \gamma\, \hat{R}^3 
+ \delta \hat{R}\,\hat R^{\mu\nu}\,\hat R_{\mu\nu}\, + \zeta\, \hat R \hat \Box \hat R  \bigg] \nn \\
&=&
\int d^6x\, 20\, \bigg[ \frac{1}{\Lambda^3}\,
\bigg( 5\,\zeta\, \Box^2\rho\, \Box\rho -(50\, \gamma + 7\, \delta - 30\, \zeta)\, (\Box\rho)^3 
- 8 \,(\delta + 5\,\zeta)\, \Box\rho\, (\pd\pd\rho)^2  \bigg)\nn \\
&& \hspace{14mm}
+\,  \frac{1}{\Lambda^4}\, \bigg( 50\, (6\,\gamma + \delta - 2\,\zeta)\, (\Box\rho)^2\, (\pd\rho)^2   
- 16\, (\delta + 5\, \zeta)\, \Box\rho\, \pd_\mu \pd_\nu\rho\, \pd^\mu\rho\, \pd^\nu\rho 
\nn \\
&& \hspace{14mm}
+\, 8\,(2\,\delta + 5\,\zeta)\, (\pd\rho)^2 (\pd\pd\rho)^2  \bigg)
- \frac{120}{\Lambda^5}\,(5\,\gamma + \delta - \zeta)\, \Box\rho\,(\pd\rho)^4 
+   \frac{80}{\Lambda^6}\,(5\,\gamma + \delta - \zeta)\,(\pd\rho)^6 
\bigg].
\eea
We have introduced the compact notation $\left(\pd\rho\right)^n \equiv 
\left( \pd_\lambda\rho\,\pd^\lambda\rho \right)^{n/2}\, , \left(\pd\pd\rho\right)^2 \equiv 
\pd_\mu\pd_\nu\rho\, \pd^\mu\pd^\nu\rho $ to denote multiple derivatives of the dilaton field. 
The Weyl invariant part of the dilaton effective action is then given by
\beq
\Gamma_0[g,\rho] =  
\mathcal{S}^{(0)}_\rho + \mathcal{S}^{(2)}_\rho + \mathcal{S}^{(4)}_\rho + \mathcal{S}^{(6)}_\rho
+ \dots \, ,
\eeq
where the ellipsis denote all the possible higher-order, irrelevant terms.

\subsection{ Weyl-gauging of the renormalized effective action}

The starting relation of our argument is the cocycle condition satisfied by the WZ anomaly-induced action.
We recall that a WZ action is constructed by solving the constraints coming from the conformal anomaly and 
differs from the effective action computed using perturbation theory and integrating out the matter fields.

For instance, direct computations of several correlators \cite{Giannotti:2008cv,Armillis:2009pq} have shown that these 
are in agreement with the expression predicted by the non-local anomaly action proposed by Riegert \cite{Riegert:1984kt}.
In this respect the WZ and the Riegert's action show significantly different features. 

A WZ form of the non-local anomaly action is regained from Riegert's expression only at the cost of 
sacrificing covariance, by the choice of a fiducial metric \cite{Antoniadis:1991fa}. However, the WZ action becomes generally 
covariant at the price of introducing one extra field, the dilaton, which plays a key role in order to extract information on some significant 
implications of the anomaly, as we are going to show.

We will derive this action from the Weyl-gauging of the counterterms for CFT's in dimensional regularization \cite{Mazur:2001aa}. 
We will be using the term {\em renormalized action} to denote the anomaly-induced action which is given by the sum 
of the Weyl-invariant (non anomalous) terms, denoted by $\Gamma_0$, and of the counterterms 
$\Gamma_{\textrm{Ct}}$ which one extracts in ordinary perturbation theory \cite{Duff:1977ay}. Explicitly

\beq
\label{oneloop}
\Gamma_{\textrm{ren}}[g, \rho]=\Gamma_0[g,\rho] + \Gamma_{\textrm{Ct}}[g],
\eeq
where the dependence on the dilaton $\rho$ in $\Gamma_0$ is generated by the Weyl-gauging of diffeomorphism invariant functionals of 
the metric, discussed in the previous section. 
This action correctly reproduces the anomaly, which is generated by the Weyl variation of  $\Gamma_{\textrm{Ct}}$. 

The change in the notation with respect to (\ref{RenW}) requires some comment.
In (\ref{RenW}) the generating functional $\mathcal{W}$ was used, whereas here we are replacing it by the effective 
action $\Gamma$. This notation is more general and better suits the conventions that are met in the literature. 
In fact, even though we are dealing only with exactly conformal field theories, we know that CFT's are believed to be 
both the ultraviolet and infrared limits of any RG flow; but despite the fact that in the two phases the mathematics is the same, 
yet the physics is quite different, because down in the infrared regime a lot of the microspcopic degrees of freedom 
showing up at high energies have been integrated out.
This is why we choose to switch from the rather generic generating functional $\mathcal W$ to the 
more common $\Gamma$, which typically indicates effective actions. 
Of course, the different numbers of degrees of freedom in the two CFT's show up in the different values of the scalar coefficients 
$c$ and $a$ in the trace anomaly equation (\ref{Ch4anom}).

We have also hidden the $1/\epsilon$ pole of the counterterm inside $\Gamma_{Ct}$, just for convenience, so that the following relation
holds (see eq.  (\ref{RenW}))
\beq
\Gamma_{Ct}[g] = -\frac{\mu^{-\epsilon}}{\bar\epsilon}\, \mathcal{W}_{Ct}[g]\, .
\eeq
The cocycle condition summarizes the response of the functional $\Gamma$ under a Weyl-gauging of the metric, which is
given by (see also eq.  (\ref{Ch4Compensate}))
\beq
\label{fet}
g_{\mu\nu} \to \hat{g}_{\mu\nu} = g_{\mu\nu}\, e^{- 2\, \rho/\Lambda} \, .
\eeq
In particular, we can define the Weyl-gauged renormalized effective action,
\beq
\hat\Gamma_{\textrm{ren}}[g,\rho] \equiv \Gamma_{0}[g,\rho] + \Gamma_{\textrm{Ct}}[\hat{g}]\, ,
\label{HattedGamma}
\eeq
and the WZ action, $\Gamma_{WZ}$, is identified from the relation
\beq
\label{coc}
\hat\Gamma_{\textrm{ren}}[g,\rho] = \Gamma_{\textrm{ren}}[g,\rho] - \Gamma_{\textrm{WZ}}[g,\rho] \, .
\eeq
We recall that the Weyl transformation of the field-enlarged system is realised by
\bea
g'_{\mu\nu} &=& g_{\mu\nu}\, e^{2\sigma}\, , \nn \\
\rho' &=& \rho + \sigma\, .
\label{WeylVar}
\eea
The defining condition of the WZ action is that its variation equals the trace anomaly, i.e.
\beq
\delta_W \Gamma_{WZ}[g,\rho] = \int d^dx\, \sqrt{g}\, \sigma\, \mathcal{A}[g]\, .
\label{WZVar}
\eeq
%
We will exploit explicitly the relation (\ref{coc}) in $4$ and $6$ dimensions in the main text and for $d=2$ in appendix \ref{Ch42D}.
It is possible to prove it in complete generality for arbitrary even dimensions, under very general hypotheses.
The proof combines cohomological arguments and dimensional regularization, explicitly deriving the algorithm to be used below,
showing that it automatically generates effective actions satifying the WZ consistency condition for the Weyl group.
The proof lies beyond the goal of this chapter, so that we refer to the original paper by Mazur and Mottola for the details \cite{Mazur:2001aa}. 

Here we limit to a sketchy derivation of the result. First, we observe that the Weyl-gauged the effective action (\ref{HattedGamma})
is naturally expected to consist of the non gauged action plus a function of the dilaton and the metric, say $\Gamma_{1}[g,\rho]$
\beq
\Gamma_{ren}[\hat{g}] = \Gamma_{ren}[g] + \Gamma_{1}[g,\rho] \, .
\eeq
Recalling the discussion of section \ref{CountAnom}, where it was shown that the Weyl variation of the renormalized generating functional
in dimensional regularization is the trace anomaly, we can write this condition,
with the generating functional replaced by the effective action $\Gamma$, in the equivalent integral form
\beq
\delta_W \Gamma_{ren}[g] = \int d^dx\, \sqrt{g}\, \mathcal{A}[g]\, ,
\label{DeltaGammaRen}
\eeq
which is the same constraint satisfied by the WZ action, (\ref{WZVar}).
Observing that any functional of the hatted metric $\hat{g}_{\mu\nu}$ is Weyl invariant by construction\footnote{This observation
is definitely true for functionals that are defined in the physical dimensions. For the counterterms in dimensional regularization 
some care is needed. For details, see \cite{Mazur:2001aa}}, it follows that the Weyl variation of the hatted renormalized effective
action (\ref{HattedGamma}) is expected to vanish, which, taking (\ref{DeltaGammaRen}) into account, is 
explicitly written as
\beq
\int d^dx\, \sqrt{g}\, \mathcal{A}[g] + \delta_W \Gamma_{1}[g,\rho] = 0\, .
\label{WanishWeylGamma}
\eeq
Considering (\ref{WanishWeylGamma}) together with (\ref{WZVar}), it is natural to identify $\Gamma_{1}$ 
with the WZ effective action,
\beq
\Gamma_{1}[g,\rho] \equiv - \Gamma_{WZ}[g,\rho]\, ,
\label{IdentifyWZ}
\eeq
thus obtaining (\ref{coc}).
Of course, our argument is of variational nature, so we are allowed to write (\ref{IdentifyWZ}) modulo Weyl invariant terms. 
Nevertheless, these can always be absorbed into the part of the effective action giving the kinetic terms for the dilaton, 
which were reviewed above.

\section{The WZ effective action for $d=4$}\label{WZ4}

Having reviewed the structure of the Weyl invariant operators in the dilaton effective action $\Gamma_{\textrm{ren}}$ and the cocycle 
condition, we move on to the hard core of this chapter and construct the WZ effective actions, which are the mean we will exploit
to derive the recursive algorithm for the computation of traced EMT correlation functions of arbitrary order.

\subsection{The counterterms in $4$ dimensions} \label{Ch3Counterterms}

One standard approach followed in the derivation of the WZ anomaly action is the Noether method, in which $\rho$ is linearly coupled 
to the anomaly. Further terms are then introduced in order to correct for the non invariance under Weyl transformations of the 
anomaly functional itself. This approach, reviewed in appendix \ref{Ch3WessZumino}, does not make transparent the functional 
dependence of the WZ action on the Weyl invariant metric $\hat{g}_{\mu\nu}$, which motivates our analysis. 
Here, instead, we proceed with a construction of the same effective action by applying the Weyl-gauging procedure to the renormalized 
effective action. For definiteness, in the following we briefly review the discussion of the connection of the counterterms in to the trace anomaly 
and the scheme dependence of the latter.

Following the discussion in \cite{Duff:1977ay}, we start by introducing the counterterm action
\beq\label{Ch3CounterAction4}
{\Gamma}_{\textrm{Ct}}[g] = 
- \frac{\mu^{-\epsilon}}{\epsilon}\int d^d x\, \sqrt{g}\, \bigg( \beta_a F + \beta_b G\bigg) \, ,  \quad \epsilon = 4 - d \, ,
\eeq
where $\mu$ is a regularization scale. It is this form of ${\Gamma}_{\textrm{Ct}}$, which is part of 
$\Gamma_{\textrm{ren}}$  to induce the anomaly condition 
\beq
\label{Ch3qt}
\frac{2}{\sqrt{g}}g_{\mu\nu}\frac{\delta{\Gamma_{\textrm{ren}}}[g]}{\delta g_{\mu\nu}}\bigg|_{d\rightarrow 4} =
\frac{2}{\sqrt{g}}g_{\mu\nu}\frac{\delta{\Gamma_{\textrm{Ct}}}[g]}{\delta g_{\mu\nu}}\bigg|_{d\rightarrow 4} = 
\mathcal{A}[g].
\eeq
In (\ref{Ch3qt}) we have exploited the Weyl invariance of the non anomalous action $\Gamma_0[g]$
\beq \label{Ch3BareActionInvariance}
g_{\mu\nu}\frac{\delta{\Gamma}_0[g]}{\delta g_{\mu\nu}}\bigg|_{d\rightarrow 4} = 0  \, , 
\eeq
with the anomaly generated entirely by the counterterm action ${\Gamma}_{\textrm{Ct}}[g]$. This follows from the well known relations
\bea \label{Ch3NTracesCTF}
\frac{2}{\sqrt{g}}\, g_{\mu\nu}\, \frac{\delta}{\delta g_{\mu\nu}}\, \int d^d x\,\sqrt{g}\, F 
&=&
-\epsilon \, \left(F - \frac{2}{3}\, \Box R\right)\, , 
\\
\label{Ch3NTracesCTG}
\frac{2}{\sqrt{g}}\, g_{\mu\nu}\, \frac{\delta}{\delta g_{\mu\nu}}\, \int d^d x\, \sqrt{g}\, G 
&=& 
-\epsilon \, G \,,
\eea
which give 
\beq \label{Ch3TraceAnomaly4d}
\left\langle T \right\rangle = 
\frac{2}{\sqrt{g}}\, g_{\mu\nu}\, \frac{\delta{\Gamma}_{\textrm{Ct}}[g]}{\delta g_{\mu\nu}}\bigg|_{d\rightarrow 4} =  
\beta_a\, \left( F - \frac{2}{3}\, \Box R\right) + \beta_b\, G   \, .
\eeq
The $\Box R$ term in eq. ~(\ref{Ch3NTracesCTF}) is prescription dependent and can be avoided if the 
$F$-counterterm is chosen to be conformal invariant in $d$ dimensions, i.e. using the square $F_d$ of the Weyl tensor in $d$ 
dimensions (see appendix (\ref{Geometrical})),
\beq \label{Ch3FdCounterterm}
{\Gamma}^{d}_{\textrm{Ct}}[g] = 
- \frac{\mu^{-\epsilon}}{\epsilon}\, \int d^d x\, \sqrt{g}\, \bigg( \beta_a F_d + \beta_b G\bigg)\, .
\eeq
In fact, expanding (\ref{Ch3FdCounterterm}) around $d=4$ and computing the $O(\epsilon)$ 
contribution to the vev of the traced EMT we find
\bea
\int d^d x\,\sqrt{g}\, F_d 
&=& \int d^d x\,\sqrt{g}\, \bigg[ F - \epsilon\, \bigg( R^{\alpha\beta}R_{\alpha\beta} - \frac{5}{18}\, R^2 \bigg)
+ O\big(\epsilon^2\big) \bigg] \, , \\
\frac{2}{3}\, \Box R 
&=&
\frac{2}{\sqrt{g}}\, g_{\mu\nu}\, \frac{\delta}{\delta g_{\mu\nu}}\, 
\int d^4 x\, \sqrt{g}\, \left( R^{\alpha\beta} R_{\alpha\beta} - \frac{5}{18} R^2 \right) \, .
\eea
These formulae, combined with (\ref{Ch3NTracesCTF}), give
\beq \label{Ch3NTracesCT2}
\frac{2}{\sqrt{g}}\, g_{\mu\nu}\, \frac{\delta}{\delta g_{\mu\nu}}\, \int d^d x\,\sqrt{g}\, F_d
= - \epsilon\, F + O\big(\epsilon^2\big) \, ,
\eeq
in which the $\square R$ term is now absent.

In general, one may want to vary arbitrarily the coefficient in front of the $\Box R$ anomaly in (\ref{Ch3TraceAnomaly}).
This can be obtained by the inclusion of the counterterm
\beq \label{Ch3ll}
\beta_{\textrm{fin}}\, \int d^4x\, \sqrt{g}\, R^2\, ,
\eeq
where $\beta_{\textrm{fin}}$ is an arbitrary parameter, and the subscript $\textrm{fin}$ stands for "finite",
given that (\ref{Ch3ll}) is just a finite, prescription-dependent contribution.
In fact, the relation
\beq \label{Ch3LocalAnomaly}
\frac{2}{\sqrt{g}}\, g_{\mu\nu}\, \frac{\delta}{\delta g_{\mu\nu}}\, \int d^4 x\, \sqrt{g}\, R^2 = 12\, \Box R
\eeq
allows to modify at will the coefficient in front of $\square R$ in the anomaly functional. This is obtained by adding the finite 
contribution (\ref{Ch3ll}) to the action of the theory and by tuning appropriately the coefficient $\beta_{\textrm{fin}}$.
When (\ref{Ch3ll}) is present, the overall counterterm is
\beq
{\Gamma}_{\textrm{Ct}}[g] + \beta_{\textrm{fin}}\, \int d^4x\, \sqrt{g}\, R^2
\eeq
and the modified trace anomaly equation becomes
\beq
\left\langle T \right\rangle = 
\beta_a\, F + \beta_b\, G -\frac{2}{3}\, \bigg( \beta_a -18\, \beta_{\textrm{fin}} \bigg)\, \Box R \, .
\label{Ch3ModifiedTraceAnomalies}
\eeq
Nevertheless, the contribution (\ref{Ch3ll}) breaks the conformal symmetry of the theory. 
This implies that the only modification of the effective action which modifies the coefficient of $\square R$ in the trace anomaly 
and is, at the same time, consistent with conformal symmetry is the replacement of $F$ with $F_d$, 
removing the $\Box R$ anomaly altogether.

\subsection{Weyl-gauging of the counterterms in $4$ dimensions}\label{Ch3GaugeCount}

At this point we illustrate the practical implementation of the gauging procedure on the renormalized effective action.

It is natural to expand the gauged counterterms in a double power series with respect to $\epsilon = 4-d$
and $\kappa_{\Lambda}\equiv1/\Lambda$ around $(\epsilon , \kappa_\Lambda )=( 0,0)$. 
Their formal expansions are
\bea
- \frac{1}{\epsilon}\,\int d^dx \, \sqrt{\hat{g}} \, \hat{F}\, 
&=& 
- \frac{1}{\epsilon}\, \int d^dx \sum_{i,j=0}^{\infty}\frac{1}{i!j!}\, \epsilon^i\,\left( \kappa_\Lambda\right)^j\,
\frac{\pd^{i+j}\left(\sqrt{\hat g}\, \hat F\, \right)}{\pd \epsilon^i\,\pd \kappa_\Lambda^{j}}\, ,
\nn \\
- \frac{1}{\epsilon}\,\int d^dx \, \sqrt{\hat{g}} \, \hat{G}
&=& 
- \frac{1}{\epsilon}\, \int d^dx \sum_{i,j=0}^{\infty}\frac{1}{i!j!}\, \epsilon^i\,\left( \kappa_\Lambda\right)^j\,
\frac{\pd^{i+j}\left(\sqrt{\hat g}\, \hat G \right)}{\pd \epsilon^i\,\pd \kappa_\Lambda^{j}}\, .
 \label{Ch3PreGauging}
\eea
It is clear that only the $O(\epsilon)$ contributions are significant, as every higher order term would still be $O(\epsilon)$ at least
after the division by $\epsilon$, therefore vanishing for $d \rightarrow 4$. \\
On the other hand, we observe that in the gauged Riemann tensor there are no more than two dilaton fields (see appendix \ref{Geometrical}).
But $\kappa_\Lambda$ can appear in the gauged counterterms only through the dilaton, so that the conditions
\beq
\frac{\pd^{n} \left(\sqrt{\hat{g}} \, \hat{F}\right)}{\pd\kappa_\Lambda^{n}} = O(\epsilon^2)\, , 
\qquad
\frac{\pd^{n} \left(\sqrt{\hat{g}} \, \hat{G}\right)}{\pd\kappa_\Lambda^{n}} = O(\epsilon^2) \, ,
\quad n \geq 5
\eeq
are found to hold. 

Finally, in (\ref{Ch3PreGauging}) there are terms which are $O(1/\epsilon)$ and deserve special attention. 
Taking, for example, the first of (\ref{Ch3PreGauging}), they are $F/\epsilon$ plus something more.
But everything differing from $F/\epsilon$ is found to vanish after proper integrations by parts and the same holds for the other
gauged counterterm.

Thus we have obtained the intermediate results
\bea
\label{Ch3effe}
- \frac{\mu^{-\epsilon}}{\epsilon}\,\int d^dx \, \sqrt{\hat{g}} \, \hat{F} \, 
&=&
- \frac{\mu^{-\epsilon}}{\epsilon}\,\int d^dx \, \sqrt{g}\, F
+\int d^4x \, \sqrt{g}\, \bigg\{\frac{1}{\Lambda}\, 
\bigg( - \rho\, F - \frac{4}{3}\, R\,\Box\rho + 4\, R^{\alpha\beta}\, \nabla_\alpha\pd_\beta\rho \bigg)
\nn \\
&&
+\, \frac{2}{\Lambda^2}\, \bigg( 2\, R^{\alpha\beta}\,\pd_\alpha\rho\,\pd_\beta\rho - \frac{R}{3}\, \left(\pd\rho\right)^2
+ \left(\Box\rho\right)^2 - 2\, \nabla_\beta\pd_\alpha\rho\, \nabla^\beta\pd^\alpha\rho
\bigg) 
\nn \\
&&
-\, \frac{8}{\Lambda^3}\, \pd^\alpha\rho\,\pd^\beta\rho\, \nabla_\beta\pd_\alpha\rho
- \frac{2}{\Lambda^4}\, \left(\left(\pd\rho\right)^2\right)^2\, \bigg\}\, , 
\eea
\bea
\label{Ch3gi}
- \frac{\mu^{-\epsilon}}{\epsilon}\,\int d^dx \, \sqrt{\hat{g}} \, \hat{G} \,
&=&
- \frac{\mu^{-\epsilon}}{\epsilon}\,\int d^dx \, \sqrt{g}\, G
+ \int d^4x \, \sqrt{g}\, \bigg\{ \frac{1}{\Lambda}\, 
\bigg( - \rho\, G - 4\, R\, \Box \rho + 8\, R^{\alpha\beta}\, \nabla_\beta \pd_\alpha \rho \bigg)
\nn \\
&& \hspace{-5mm}
+\, \frac{2}{\Lambda^2}\, \bigg[ 2\, R\, \rho\,\Box\rho + R\, \left(\pd\rho\right)^2 
+ 4\, R^{\alpha\beta}\, \bigg(\pd_\alpha\rho\,\pd_\beta\rho - \rho\,\nabla_\beta \pd_\alpha\rho \bigg)
+ 6\, \left( \Box\rho \right)^2 - 6\, \nabla^\beta\pd^\alpha\rho\,\nabla_\beta\pd_\alpha\rho  \bigg]
\nn \\
&& \hspace{-5mm}
-\, \frac{4}{\Lambda^3}\, \bigg( 2\, R^{\alpha\beta}\,\rho\, \pd_\alpha\rho\, \pd_\beta\rho + 2\, \rho\, \left(\Box\rho\right)^2 +
5\, \left(\pd\rho\right)^2\, \Box\rho + 6\,\pd^\alpha\rho\,\pd^\beta\rho\,\nabla_\beta\pd_\alpha\rho - 
2\, \rho\, \nabla^\beta\pd^\alpha\rho\, \nabla_\beta\pd_\alpha\rho \bigg) 
\nn \\
&& \hspace{-5mm}
+\, \frac{2}{\Lambda^4}\, \bigg(
4\, \rho\, \left(\pd\rho\right)^2\, \Box\rho + 
3\, \left( \left(\pd\rho\right)^2 \right)^2 + 
8\, \rho\, \pd^\alpha\rho\,\pd^\beta\rho\, \nabla_\beta\pd_\alpha \rho \bigg)\bigg\} \, .
\eea
%
The expressions above can be simplified using integrations by parts and the identity 
for the commutator of covariant derivatives of a vector,
\beq
[\nabla_\mu,\nabla_\nu]\, v_\rho = {R^\lambda}_{\rho\mu\nu}\,v_\lambda \, .
\eeq
After these manipulations we find that  the Weyl-gauging of the counterterms gives
\bea
- \frac{\mu^{-\epsilon}}{\epsilon}\,\int d^dx \, \sqrt{\hat{g}} \, \hat{F} \, 
&=&  
- \frac{\mu^{-\epsilon}}{\epsilon}\, \int d^dx \, \sqrt{g} \, F +
\int d^4 x \, \sqrt{g}\,  \bigg[ - \frac{\rho}{\Lambda}\, \bigg( F - \frac{2}{3} \Box R \bigg)
- \frac{2}{\Lambda^2}\, \bigg( \frac{R}{3}\, \left(\pd\rho\right)^2 + \left(\Box \rho \right)^2  \bigg)
\nn \\ 
&& \hspace{50mm}
+ \frac{4}{\Lambda^3}\, \left(\pd\rho\right)^2\,\Box\rho \, 
- \frac{2}{\Lambda^4}\, \left(\left(\pd\rho\right)^2\right)^2 \bigg] \, ,
\label{Ch3GaugingF} \\
- \frac{\mu^{-\epsilon}}{\epsilon}\,\int d^dx \, \sqrt{\hat{g}} \, \hat{G}
&=&
- \frac{\mu^{-\epsilon}}{\epsilon}\, \int d^dx \, \sqrt{g} \, G  + 
\int d^4 x \, \sqrt{g}\, \bigg[ -  \frac{\rho}{\Lambda}\, G 
+ \frac{4}{\Lambda^2}\, \left( R^{\alpha\beta} - \frac{R}{2}\,g^{\alpha\beta} \right)\, \pd_\alpha\rho\,\pd_\beta\rho
\nn \\
&& \hspace{50mm}
+ \, \frac{4}{\Lambda^3}\, \left(\pd\rho\right)^2\, \Box \rho
- \frac{2}{\Lambda^4}\, \left(\left(\pd\rho\right)^2 \right)^2 \bigg]\, .
\label{Ch3GaugingG}
\eea  
As a consistency check, one can apply a Weyl transformation to (\ref{Ch3GaugingF}) and (\ref{Ch3GaugingG}), 
and see that it gives zero, which is the basic requirement for a WZ action. 
Recalling the defining relation for the WZ action, (\ref{coc}), we can finally write
\bea
\Gamma_{WZ}[g,\rho] 
&=&
\int d^4x\, \sqrt{g}\, \bigg\{ \beta_a\, \bigg[ \frac{\rho}{\Lambda}\, \bigg( F - \frac{2}{3} \Box R \bigg) + 
\frac{2}{\Lambda^2}\, \bigg( \frac{R}{3}\, \left(\pd\rho\right)^2 + \left( \Box \rho \right)^2 \bigg) - 
\frac{4}{\Lambda^3}\, \left(\pd\rho\right)^2\,\Box \rho + 
\frac{2}{\Lambda^4}\,  \left(\pd\rho\right)^4 \bigg] \nn \\
&& 
\hspace{15mm}
+\, \beta_{b}\, \bigg[ \frac{\rho}{\Lambda}\,G  - 
\frac{4}{\Lambda^2}\, \bigg( R^{\alpha\beta} - \frac{R}{2}\,g^{\alpha\beta} \bigg)\, \pd_\alpha\rho\, \pd_\beta\rho -
\frac{4}{\Lambda^3} \, \left(\pd\rho\right)^2\,\Box \rho + 
\frac{2}{\Lambda^4}\, \left( \left(\pd\rho\right)^2 \right)^2 \bigg]\bigg\} \, .
\label{Ch3Effective4d}
\eea
Notice that the ambiguity in the choice of the Weyl tensor discussed above - i.e. between F and $F_d$ - implies that no dilaton 
vertex is expected to emerge from the gauging of the $F_d$-counterterm, as it is conformal invariant.
This is indeed the case and we find the relation
\bea \label{Ch3GaugingFd}
&&  \hspace{-8 mm}
-\frac{\mu^{-\epsilon}}{\epsilon}\,\int d^d x \, \sqrt{\hat{g}} \, \hat{F}_d \, = 
-\frac{\mu^{-\epsilon}}{\epsilon}\,\int d^d x \, \sqrt{g} \, F_{d} 
- \int d^4 x \,\sqrt{g}\, \frac{\rho}{\Lambda}\, F\, ,
\eea
that modifies the structure of the WZ action eliminating from (\ref{Ch3Effective4d}) all the terms multiplying $\beta_a$
but $\left(\rho/\Lambda\right)\, F$.

Finally we remark that, in the case in which a finite counterterm of the kind (\ref{Ch3ll}) is present,
the formulae of this section are modified according to the simple prescription (see eq.  (\ref{Ch3ModifiedTraceAnomalies})),
\beq \label{Ch3LocalToBeta}
\beta_a \rightarrow \beta_a - 18\, \beta_{\textrm{fin}}\, ,
\eeq
as it is possibile to render all the quantum effective action Weyl invariant. This is obtained, as discussed above, 
by the Weyl-gauging of the complete counterterm
\beq
\Gamma_{\textrm{Ct}}[g] = 
-\frac{\mu^{-\epsilon}}{\epsilon}\, \int d^dx\, \sqrt{g}\, \bigg( \beta_a\,F + \beta_b\,G \bigg) + 
\beta_{\textrm{fin}}\, \int d^4x\, \sqrt{g}\, R^2\, .
\eeq
In this case the compensating WZ action for $\int d^4x\, \sqrt{g}\, R^2$ can be generated by the relation
\bea
\int d^4x\, \sqrt{\hat g}\, \hat{R}^2 
&=& 
\int d^4x\, \sqrt{g}\, R^2 +
18\, \int d^4 x \, \sqrt{g}\,  \bigg[ - \frac{2}{3}\, \frac{\rho}{\Lambda}\, \Box R
+ \frac{2}{\Lambda^2}\, \bigg( \frac{R}{3}\, \left(\pd\rho\right)^2 + \left(\Box \rho \right)^2  \bigg)
\nn \\ 
&& \hspace{47mm}
- \frac{4}{\Lambda^3}\, \left(\pd\rho\right)^2\,\Box\rho \, 
+ \frac{2}{\Lambda^4}\, \left(\left(\pd\rho\right)^2\right)^2 \bigg] \, .
\eea
Comparing the result given above with (\ref{Ch3GaugingF}), eq.  (\ref{Ch3LocalToBeta}) follows immediately.

\section{The WZ effective action for $d=6$}\label{WZ6}

We repeat step by step the procedure outlined for the $4$-dimensional case, discussing the possible choices of the counterterms
in $6$ dimensions in full generality, highlighting the details of the difference between the choice of the Weyl invariants 
$\sqrt{g}\, I_i$ and their $d$-dimensional counterparts, $\sqrt{g}\, I^d_i$, in particular deriving the finite renormalization
distinguishing them. These results are then used in the study of the various possible WZ actions.

\subsection{The counterterms in $6$ dimensions} \label{Ch4Counterterms}

As we have discussed above, we construct the effective action by applying the Weyl-gauging procedure to the renormalized effective action, 
which breaks scale invariance via the anomaly. First we must introduce the $1$-loop counterterm action, 
which is given by the integrals of all the possible Weyl invariants and of the Euler density, each continued to $d$ dimensions,
\beq\label{Ch4CounterAction6}
{\Gamma}_{\textrm{Ct}}[g] = 
- \frac{\mu^{-\epsilon}}{\epsilon}\int d^d x\, \sqrt{g}\, \bigg( \sum_{i=1}^{3} c_i\, I_i + a E_6 \bigg) \, ,  
\quad \epsilon = 6 - d \, ,
\eeq
where $\mu$ is a regularization scale. 
This form of ${\Gamma}_{\textrm{Ct}}$ induces the anomaly relation 
\beq \label{Ch4AnomalyCondition}
\frac{2}{\sqrt{g}}g_{\mu\nu}\frac{\delta{\Gamma_{\textrm{ren}}}[g]}{\delta g_{\mu\nu}}\bigg|_{d\rightarrow 6} =
\frac{2}{\sqrt{g}}g_{\mu\nu}\frac{\delta{\Gamma_{\textrm{Ct}}}[g]}{\delta g_{\mu\nu}}\bigg|_{d\rightarrow 6}= 
\mathcal{A}[g] \, .
\eeq
where we have exploited once again the Weyl invariance of the non anomalous action $\Gamma_0[g]$ in $6$ dimensions
\beq \label{Ch4BareActionInvariance}
g_{\mu\nu}\frac{\delta{\Gamma}_0[g]}{\delta g_{\mu\nu}}\bigg|_{d\rightarrow 6} = 0,  \, 
\eeq
with the anomaly generated entirely by the counterterm action ${\Gamma}_{\textrm{Ct}}[g]$, due to the relations
\bea \label{Ch4NTracesCTI}
\frac{2}{\sqrt{g}}\, g_{\mu\nu}\, \frac{\delta}{\delta g_{\mu\nu}}\, \int d^d x\,\sqrt{g}\, I_i 
&=&
-\epsilon \, \bigg( I_i + \nabla_\mu J^\mu_i \bigg) \, , 
\\
\label{Ch4NTracesCTE}
\frac{2}{\sqrt{g}}\, g_{\mu\nu}\, \frac{\delta}{\delta g_{\mu\nu}}\, \int d^d x\, \sqrt{g}\, E_6
&=& 
-\epsilon \, E_6 \, ,
\eea
so that from (\ref{Ch4AnomalyCondition}) we find
\beq \label{Ch4TraceAnomaly6d}
\left\langle T \right\rangle = 
\frac{2}{\sqrt{g}}\, g_{\mu\nu}\, \frac{\delta{\Gamma}_{\textrm{Ct}}[g]}{\delta g_{\mu\nu}}\bigg|_{d\rightarrow 4} =  
\sum_{i=1}^3 c_i\, \left( I_i + \nabla_\mu J_i^\mu \right) + a E_6   \, .
\eeq
The explicit expressions of the derivative terms was given in section \ref{RecConv}
and can be obtained through the functional variations listed in appendix \ref{Ch4Geometrical3}.
These terms above are renormalization prescription dependent and are not present if, instead of the counterterms
$\sqrt{g}\, I_{i}$,  one chooses scalars  that are conformal invariant in $d$ dimensions, i.e. the $I^d_i$'s defined in 
appendix \ref{Geometrical}. Notice that the inclusion of $d$-dimensional counterterms simplifies considerably the computation of the 
dilaton WZ action, as shown in \cite{Baume:2013ika}. In fact, in this scheme, the contribution of the $I^d_i$'s to the same action 
is just linear in the dilaton field and can be derived from the counterterm
\beq \label{Ch4IdCounterterm}
{\Gamma}^{d}_{\textrm{Ct}}[g] = - \frac{\mu^{-\epsilon}}{\epsilon}\, \int d^d x\, \sqrt{g}\, 
\bigg( \sum_{i=1}^{3} c_i\, I^d_i + a E_6 \bigg). 
\eeq
It can be explicitly checked that by expanding (\ref{Ch4IdCounterterm}) around $d=6$ and computing the order $O(\epsilon)$ contribution 
to the vev of the traced EMT one obtains the relation 
\beq \label{Ch4NTracesCT1}
\frac{2}{\sqrt{g}}\, g_{\mu\nu}\, \frac{\delta}{\delta g_{\mu\nu}}\, \int d^d x\,\sqrt{g}\, I^d_i = 
- \epsilon\, I_i + O\big( \epsilon^2 \big). \,
\eeq
In this simplified scheme, it is possible to give the structure of the WZ action in any even dimension \cite{Baume:2013ika}, 
just by adding to the contribution of such invariants the one coming from the Euler density $E_d$, being the 
total derivative terms $\nabla_\mu J^\mu_i$ absent. 

\subsection{General scheme-dependence of the trace anomaly in $6$ dimensions}\label{Ch4SchemDep}

In this section we establish a connection between the two renormalization schemes used to derive the dilaton WZ action, with the inclusion of 
invariant counterterms of $B$ type which are either $d$ or $6$-dimensional, in close analogy with the $4$-dimensional case.
For $d=6$ we proceed in a similar way. We expand the $d$-dimensional counterterms around $d=6$ to identify the finite contributions as
\beq \label{Ch4SeriesEps}
I_i^d = I_i + (d-6)\, \frac{\pd I^d_i}{\pd d}\bigg|_{d=6} = I_i - \epsilon\, \frac{\pd I^d_i}{\pd d}\bigg|_{d=6}\, .
\eeq
Using (\ref{Ch4SeriesEps}) in the $d$-dimensional counterterms, we have
\beq
-\frac{1}{\epsilon}\, \int d^d x\, \sqrt{g}\, I^d_i = 
-\frac{1}{\epsilon} \int d^d x\, \sqrt{g}\, I_i + \int d^dx\, \sqrt{g}\, \frac{\pd I^d_i}{\pd d}\bigg|_{d=6}\, .
\eeq
This implies, due to (\ref{Ch4NTracesCTI}) and (\ref{Ch4NTracesCT1}), that 
\bea
-\frac{1}{\epsilon}\, \frac{2}{\sqrt{g}}\, g_{\mu\nu}\, \frac{\delta}{\delta g_{\mu\nu}}\, \int d^d x\,\sqrt{g}\, I_i 
&=&
I_i -  \frac{2}{\sqrt{g}}\, g_{\mu\nu}\, 
\frac{\delta}{\delta g_{\mu\nu}}\, \int d^dx\, \sqrt{g}\, \frac{\pd I^d_i}{\pd d}\bigg|_{d=6}
\eea
and hence we conclude that a finite counterterms which can account for the $i$-th total derivative term in the trace anomaly 
\bea
\frac{2}{\sqrt{g}}\, g_{\mu\nu}\, \frac{\delta}{\delta g_{\mu\nu}}\, 
\int d^dx\, \sqrt{g}\, \frac{\pd I^d_i}{\pd d}\bigg|_{d=6}
&=& 
- \nabla_\mu J^\mu_i\, .
\eea
This clearly identifies the terms that we can add to (\ref{Ch4CounterAction6}) in order to arbitrarily vary the coefficients $c_i$ 
in (\ref{Ch4TraceAnomaly6d}). They are given by the derivatives of the $d$-dimensional terms $I^d_i$ evaluated at $d=6$, 
linearly combined with arbitrary coefficients $c'_i$
\beq
\Gamma'_{\textrm{Ct}}[g] = 
- \frac{\mu^{-\epsilon}}{\epsilon}\int d^d x\, \sqrt{g}\, \bigg( \sum_{i=1}^{3} c_i\, I_i + a E_6 \bigg)
+ \int d^6x\, \sqrt{g}\, \sum_{i=1}^3 c'_i\, \frac{\pd I_i^d}{\pd d}\bigg|_{d=6}
\label{Ch4CounterAction6Mod}
\eeq
which gives the modified trace anomaly relation
\bea
\left\langle T' \right\rangle \equiv
\frac{2}{\sqrt{g}}\, g_{\mu\nu}\, \frac{\delta \Gamma'_{\textrm{Ct}}[g]}{\delta g_{\mu\nu}}\bigg|_{d\rightarrow 4}= 
\sum_{i=1}^3 c_i\, I_i  + a E_6 + \sum_{i=1}^3 \left( c_i - c'_i \right)\, \nabla_\mu J_i^\mu   \, .
\label{Ch4CounterAction6Mod1}
\eea
Then, the choice $c'_i=c_i$ in (\ref{Ch4CounterAction6Mod})  allows to move back to the scheme in which the local anomaly 
contribution is not present. 

We list the three local counterterms of (\ref{Ch4CounterAction6Mod}),
\bea
\frac{\pd I^d_1}{\pd d}\bigg|_{d=6}
&=& 
\frac{1}{16000}\, \bigg(-307\, K_1 + 3465\, K_2 - 540\, K_3 - 3750\, K_4 + 6000\, K_5 + 3000\, K_6\bigg) \, , \nn \\
\frac{\pd I^d_2}{\pd d}\bigg|_{d=6} 
&=& 
\frac{1}{4000}\, \bigg( -167\, K_1 + 1965\, K_2 - 540\, K_3 - 2750\, K_4 + 3000\, K_5 + 3000\, K_6\bigg) \, , \nn \\
\frac{\pd I^d_3}{\pd d}\bigg|_{d=6} 
&=& 
\frac{1}{500}\, \bigg(-18\,K_1 + 140\,K_2 - 90\, K_3 - 70\,K_9 + 500\,K_{10} - 250\,K_{11} + 25\,K_{12} 
\nn \\ 
&& \hspace{11mm}
-\,  625\,K_{13} + 750\,K_{15}\bigg) \, .\nn\\
\eea
Finally, in general one might also be interested to generate an anomaly functional in which 
the derivative terms appear in combinations that are different from those in eq.  (\ref{Ch4anom6D}). 
For this goal, one should use proper linear combinations of the $K_i$ according to the relations listed in (\ref{Ch4Geometrical3}).

\subsection{Weyl-gauging of the counterterms in $6$ dimensions}

At this point we have to Weyl-gauge the renormalized effective action.

Again, we expand the gauged counterterms in a double power series with respect to $\epsilon = 6-d$
and $\kappa_{\Lambda}\equiv1/\Lambda$ around $(\epsilon , \kappa_\Lambda )=( 0,0)$. 
Denoting generically with $A$ either the Euler density $E_6$ or the three invariants $I_i$' s, the expansion takes the form

\beq
- \frac{1}{\epsilon}\,\int d^dx \, \sqrt{\hat{g}} \, \hat{A}  = 
- \frac{1}{\epsilon}\, \int d^dx \sum_{i,j=0}^{\infty}\frac{1}{i!j!}\, \epsilon^i\,\left( \kappa_\Lambda\right)^j\,
\frac{\pd^{i+j}\left(\sqrt{\hat g}\, \hat A\,   \right)}{\pd \epsilon^i\,\pd \kappa_\Lambda^{j}}\, ,
\label{Ch4PreGauging}
\eeq
only the $O(\epsilon)$ contributions are significant, due to the $1/\epsilon$ factor in front of the counterterms. 
On the other hand, similarly to the case in $4$ dimensions, the condition
\beq
\frac{\pd^{n} \left(\sqrt{\hat{g}} \, \hat A\right)}{\pd\kappa_\Lambda^{n}} = O(\epsilon^2) \, , 
\quad n \geq 7 \, 
\eeq
holds, as the Euler density and the three conformal invariants are at most cubic in the Riemann tensor
and in its double covariant derivatives and, besides, there are no terms with more than two dilatons in the gauged Riemann tensor.
All the terms which are of $O(1/\epsilon)$ in (\ref{Ch4PreGauging}) and are different from $I_i$' s or, respectively, $E_6$
are found to vanish after some integrations by parts. Therefore, after gauging the counterterms we end up with the general result
\beq
- \frac{\mu^{-\epsilon}}{\epsilon}\,\int d^dx \, \sqrt{\hat{g}} \, \hat{A} \, =
- \frac{\mu^{-\epsilon}}{\epsilon}\, \int d^dx \, \sqrt{g} \, A +  \Sigma_{A} + O(\epsilon)  \, .
\label{Ch4ExpandCT}
\eeq
where each $\Sigma_{A}$ term is related to the corresponding specific invariant $A$. 
For instance, if $A=I_1$, then the corresponding $\Sigma$ term is $\Sigma_1$, and so on for each of the $I_i$'s, 
whereas for the contribution of the Euler density we have $A=E_6$  and $\Sigma_A=\Sigma_a$. \\
Their explicit expressions are
\bea
&&
\Sigma_1 =  \int d^6 x \, \sqrt{g}\,  \bigg\{- \frac{\rho}{\Lambda}\, \bigg( I_1 + \nabla_\mu J_1^\mu \bigg) \nn \\ 
&&
+\, \frac{1}{\Lambda^2}\, \bigg[ \frac{3}{4}\, R^{\mu\lambda\kappa\alpha}\, 
{R^\nu}_{\lambda\kappa\alpha}\, \pd_\mu \rho\, \pd_\nu \rho
- \frac{3}{40}\, R^{\mu\nu\lambda\kappa}\, R_{\mu\nu\lambda\kappa}\, \left(\pd \rho\right)^2 
- \frac{3}{10}\, R\, \left(\nabla \pd \rho\right)^2 \nn \\ 
&& 
+\, \frac{9}{4}\, R^{\mu\lambda\kappa\nu}\, R_{\lambda\kappa}\, \pd_\mu \rho\, \pd_\nu \rho
- 3\, R^{\mu\nu\lambda\kappa}\, \nabla_\nu \pd_\lambda \rho\, \nabla_\mu \pd_\kappa \rho
-\frac{57}{800}\, R^2\, \left(\pd \rho\right)^2 \nn \\
&& 
-\, \frac{21}{16}\, R^{\mu\lambda}\, {R_\lambda}^\nu\, \pd_\mu \rho\, \pd_\nu \rho  
- \frac{9}{4}\, R^{\mu\nu}\, \Box \rho\,\nabla_\mu \pd_\nu \rho
+ \frac{57}{160}\, R^{\mu\nu}\,R_{\mu\nu}\, \left(\pd \rho\right)^2 \nn \\
&& 
+\, \frac{3}{2}\, R^{\mu\nu} \nabla^\lambda \pd_\mu \rho \, \nabla_\lambda \pd_\nu \rho 
+ \frac{57}{80}\, R\, R^{\mu\nu} \pd_\mu \rho \, \pd_\nu \rho 
+ \frac{57}{160}\, R\, \left(\Box\rho\right)^2  \bigg] \nn \\
&& 
+\, \frac{1}{\Lambda^3}\, \bigg[
- \frac{7}{16}\, \left(\Box \rho \right)^3
+ \frac{3}{2}\, \left(\nabla \pd \rho \right)^2 \, \Box \rho
- 6\, R^{\mu\nu\lambda\kappa}\, \pd_\rho \rho\, \pd_\nu \rho\, \nabla_\mu \pd_\sigma \rho \nn \\ 
&& 
+ 3\, R^{\mu\nu}\, \nabla_\lambda \pd_\nu \rho \, \pd_\mu \rho\, \pd^\lambda \rho 
-\frac{9}{4}\, R^{\mu\nu}\, \pd_\mu \rho \,\pd_\nu \rho \,\Box \rho
- \frac{3}{5}\, R\, \pd^\mu \rho\, \pd^\nu \rho\, \nabla_\mu \pd_\nu \rho \bigg] \nn \\
&& 
+\, \frac{1}{\Lambda^4}\, \bigg[
- \frac{3}{2}\, \left(\pd \rho \right)^2\, \left(\nabla \pd \rho \right)^2
-\, \frac{3}{8}\,  \left(\pd \rho \right)^2\, \left(\Box \rho\right)^2
+ \frac{3}{4}\, \pd^\mu \left( \pd\rho \right)^2\, \pd_\mu \left( \pd\rho \right)^2
-\frac{3}{20}\, R\,\left(\pd \rho\right)^4 
\bigg] \nn \\
&& 
+\, \frac{1}{\Lambda^5}\, \frac{3}{2}\, \left(\pd \rho\right)^4\, \Box \rho
- \frac{\left(\pd \rho\right)^6}{\Lambda^6}\,     \bigg\},
\label{Ch4WZI1}
\eea
for $I_1$, 
\bea
&&
\Sigma_2 = \int d^6 x \, \sqrt{g}\,  \bigg\{- \frac{\rho}{\Lambda}\, \bigg( I_2 + \nabla_\mu J_2^\mu \bigg) \nn \\ 
&& 
+ \frac{1}{\Lambda^2}\, \bigg[
3\, R^{\mu\lambda\kappa\alpha}\, {R^\nu}_{\lambda\kappa\alpha}\, \pd_\mu \rho\, \pd_\nu \rho 
+ \frac{27}{40}\, R\,  \left(\Box\rho\right)^2 
- \frac{6}{5}\,  R\, \left(\nabla \pd \rho\right)^2  - \frac{27}{200}\, R^2\, \left(\pd \rho \right)^2 \nn \\
&& 
-\, \frac{3}{10}\, R^{\mu\nu\lambda\kappa}\, R_{\mu\nu\lambda\kappa}\, \left(\pd  \rho\right)^2 
+ 3\, R^{\mu\lambda\kappa\nu}\, R_{\lambda\kappa}\, \pd_\mu \rho\, \pd_\nu \rho
- \frac{15}{4}\, R^{\mu\lambda}\, {R_\lambda}^\nu\, \pd_\mu \rho\, \pd_\nu\rho \nn \\ 
&& 
-\, 3\, R^{\mu\nu}\, \nabla_\nu \pd_\mu \rho\, \Box\rho + \frac{27}{40}\, R^{\mu\nu}\, R_{\mu\nu}\, \left(\pd\rho\right)^2
+ 6\, R^{\mu\nu}\, \nabla^\lambda \pd_\mu \rho\, \nabla_\lambda  \pd_\nu \rho
+\frac{27}{20}\, R\, R^{\mu\nu}\, \pd_\mu\rho\, \pd_\nu \rho  \bigg]  \nn
\eea
\bea
&& 
+\, \frac{1}{\Lambda^3}\, \bigg[ 
\frac{11}{4}\, \left(\Box \rho \right)^3 - 6\, \left(\nabla \pd \rho \right)^2\, \Box \rho
- 8\,  R^{\mu\nu}\, \nabla_\lambda\pd_\nu \rho \, \pd_\mu\rho \, \pd^\lambda \rho
- 6\,  R^{\mu\nu}\, \nabla_\mu \pd_\nu \rho \left(\pd\rho\right)^2
\nn \\
&& 
+\, 8\, R^{\mu\nu\lambda\kappa}\, \pd_\nu\rho\, \pd_\lambda\rho\, \nabla_\mu\pd_\kappa\rho 
+ 5\, R^{\mu\nu}\, \pd_\mu \rho \, \pd_\nu \rho\, \Box \rho
+ \frac{18}{5}\, R\, \pd^\mu \rho\, \pd^\nu \rho\, \nabla_\mu \pd_\nu \rho  
+ 3\, R\,  \left(\pd \rho\right)^2\, \Box \rho \bigg]  \nn \\
&& 
+\, \frac{1}{\Lambda^4}\, \bigg[ 
6\, \left(\pd \rho\right)^2\, \left(\nabla \pd \rho\right)^2 - \frac{9}{2}\, \left(\pd \rho\right)^2\, \left(\Box \rho\right)^2 
- 3\, \pd^\mu \left( \pd\rho \right)^2\, \pd_\mu \left( \pd\rho \right)^2 - \frac{3}{5}\, R\, \left(\pd \rho\right)^4
\bigg]  \nn \\ 
&& 
+\, \frac{6}{\Lambda^5}\, \left(\pd \rho\right)^4\, \Box \rho - \frac{4}{\Lambda^6}\, \left(\pd \rho\right)^6
\bigg\}, 
\label{Ch4WZI2}
\eea
for the second invariant $I_2$ and
\bea 
&&
\Sigma_3 =
\int d^6 x \, \sqrt{g}\,  \bigg\{ - \frac{\rho}{\Lambda}\, \bigg( I_3 + \nabla_\mu J_3^\mu \bigg) 
\nn \\
&& 
+\, \frac{1}{\Lambda^2}\, \bigg[ -\frac{3}{25} R^2 \left(\pd \rho \right)^2  
+ \frac{13}{10}R^{\mu\nu}R\, \pd_\mu \rho\, \pd_\nu \rho
-\frac{2}{5}\, R^{\mu\nu\lambda\kappa}\, R_{\mu\nu\lambda\kappa}\, \left(\pd \rho \right)^2 \nn \\ 
&&
+\frac{9}{10}\, R\,  \left(\Box \rho\right)^2  - \frac{3}{10}\, R\, \pd^\mu \rho \,\Box \pd_\mu \rho  
- \frac{12}{5}\, R\, \left(\nabla \pd \rho \right)^2 - 5\, R^{\mu\lambda}\, {R_\lambda}^\nu\, \pd_\mu \rho\, \pd_\nu \rho \nn \\ 
&& 
+ 7\, R^{\mu\nu}\, \nabla_\mu \pd_\nu \rho\, \Box \rho  - 9\, R^{\mu\nu}\, \pd_\mu \rho \,\Box \pd_\nu \rho\,
- \Box \rho \, \Box^2 \rho + \frac{2}{5}\, R^{\mu\nu}\, R_{\mu\nu}\, \left(\pd \rho \right)^2 \nn \\ 
&& 
+ 8\, R^{\mu\nu} \,\nabla_\lambda \nabla_\nu \pd_\mu \rho\, \pd^\lambda \rho  
+ 16\, R^{\mu\nu\lambda\kappa}\,\nabla_\nu \pd_\lambda \rho \, \nabla_\mu \pd_\kappa \rho  \bigg]
\nn \\
&&
+\, \frac{1}{\Lambda^3}\, \bigg[
2\, \left(\Box \rho \right)^3 - 8\, \left(\nabla \pd \rho \right)^2\, \Box\rho
- \frac{16}{5}\, R\,\pd^\mu \rho\, \pd^\nu \rho \, \nabla_\mu \pd_\nu \rho
+ 8\, R^{\mu\nu}\, \pd_\mu \rho\, \pd_\nu \rho\, \Box \rho
\nn \\
&&
+\, 32\, R^{\mu\nu\lambda\kappa}\,\pd_\nu \rho \, \pd_\rho \rho \, \nabla_\mu \pd_\sigma \rho
\bigg] \nn \\
&&
+\, \frac{1}{\Lambda^4}\, \bigg[
- 4\, \left(\pd \rho\right)^2\, \left(\Box \rho \right)^2  - 4\, \pd^\mu \left( \pd\rho \right)^2\, \pd_\mu \left( \pd\rho \right)^2
+ 16\, \left(\pd \rho\right)^2\, \left(\nabla \pd \rho\right)^2  - \frac{4}{5}\, R\, \left(\pd \rho\right)^4 \bigg]
\bigg\}
\label{Ch4WZI3}
\eea
for the third invariant $I_3$, while the contribution from the integrated Euler density is
\bea
&&
\Sigma_a =
\int d^6 x \, \sqrt{g}\,  \bigg\{- \frac{\rho}{\Lambda}\,  E_6 
+\, \frac{1}{\Lambda^2}\, \bigg[ 
12\, R^{\mu\lambda\kappa\alpha}\, {R^\nu}_{\lambda\kappa\alpha}\, \pd_\mu \rho\, \pd_\nu \rho   
- 3\,R^{\mu\nu\lambda\kappa}\, R_{\mu\nu\lambda\kappa}\,\left(\pd \rho \right)^2  \nn \\
&&
+\, 24\, R^{\mu\lambda\kappa\nu}\, R_{\lambda\kappa}\, \pd_\mu \rho\, \pd_\nu \rho
+ 12\, R^{\mu\nu}\, R_{\mu\nu}\,\left(\pd \rho  \right)^2 
- 24\,R^{\mu\lambda}\,{R^\nu}_\lambda \, \pd_\mu \rho\, \pd_\nu \rho
+ 12\, R\, R^{\mu\nu}\, \pd_\mu \rho\, \pd_\nu \rho -  3\, R^2\,\left(\pd\rho \right)^2  \bigg] \nn \\
&& 
+\, \frac{1}{\Lambda^3}\, \bigg[ 16\, R^{\mu\nu\lambda\kappa}\, \pd_\nu \rho \, \pd_\lambda \rho\, \nabla_\mu \pd_\kappa \rho
- 16\, R^{\mu\nu}\,\nabla_\mu \pd_\nu \rho\, \left(\pd \rho\right)^2 
+ 32\, R^{\mu\nu}\,\nabla_\mu \pd_\lambda \rho\, \nabla^\lambda \pd_\nu \rho \nn \\
&&
-\, 8\, R\, \pd^\mu \rho\, \pd^\nu \rho\, \nabla_\mu \pd_\nu \rho 
+   8\, R\, \left( \pd \rho \right)^2\, \Box \rho - 16\, R^{\mu\nu}\, \pd_\mu \rho\, \pd_\nu \rho \bigg] \nn \\
&&
+\, \frac{1}{\Lambda^4}\, \bigg[ 24\, \left( \pd \rho \right)^2\, \left(\nabla \pd \rho\right)^2 
- 24\, \left(\pd \rho \right)^2\, \left(\Box \rho \right)^2 - 6\, R\, \left(\pd \rho \right)^4 \bigg]
+ \frac{36}{\Lambda^5}\, \Box \rho\, \left(\pd \rho \right)^4 - \frac{24}{\Lambda^6}\, \left(\pd \rho\right)^6
\bigg\}. \, 
\label{Ch4WZE6}
\eea
The derivation of (\ref{Ch4WZI1})-(\ref{Ch4WZE6}) is very involved
and we have used several integration by parts to get to the previous expressions.
The WZ effective action is then obtained from (\ref{coc}) and, in a general gravitational background, 
it is just given by the combination of (\ref{Ch4WZI1})-(\ref{Ch4WZE6}) with the proper coefficients, up to a minus sign, i.e.
\beq
\label{Ch4WZfinal}
\Gamma_{WZ}[g,\rho]  = - \bigg( \sum_{i=1}^3 c_i\, \Sigma_i  + a\, \Sigma_a \bigg)\, .
\eeq
Before presenting the expression of the dilaton WZ action, we pause for a comment. It is clear from our analysis that the form of this action is not 
unique, due to the renormalization scheme dependence of the counterterms which are chosen before performing the Weyl-gauging. As we have 
seen in section \ref{Ch4SchemDep}, this ambiguity manifests in the coefficients $c'_i$ which parametrizes the local terms of the anomaly, 
proportional to the derivatives of the currents $J^{\mu}_i$.  
Obviously, it is preferable to be able to characterize this ambiguity in a more complete way, 
and the analysis of the relation between counterterms in $6$ and in $d$ dimensions serves this purpose. 
In fact, this allows to identify the functionals whose variation generates the local anomaly terms. 
By proceeding in this way one is able to identify a 3-parameter class of renormalization schemes, related to the coefficients $c'_i$, which 
become free parameters in the anomaly action. 
For this purpose we exploit the relations 
\bea
- \frac{\mu^{-\epsilon}}{\epsilon}\, \int d^dx\,\sqrt{g}\, \hat I^d_i 
&=& 
- \frac{\mu^{-\epsilon}}{\epsilon}\, \int d^dx\,\sqrt{g}\, \hat I_i + \int d^6x\,\sqrt{g}\, \frac{\pd \hat I^d_i}{\pd d}\bigg|_{d=6}
\nn \\
&=&
- \frac{\mu^{-\epsilon}}{\epsilon}\, \int d^dx\,\sqrt{g}\, \hat I^d_i + \int d^6 x\, \sqrt{g}\, \frac{\rho}{\Lambda}\, I_i\, ,
\label{Ch4Additional}
\eea
where the last line follows from the transformations properties of $I^d_i$ in $d$ dimensions under Weyl scaling.
From (\ref{Ch4ExpandCT}) and (\ref{Ch4Additional}) we infer that
\bea
\int d^6x\,\sqrt{g}\, \frac{\pd \hat I_i}{\pd d}\bigg|_{d=6} = 
-\Sigma_{i} + \int d^6x\, \sqrt{g}\, \frac{\rho}{\Lambda}\, I_i \, .
\label{Ch4ModifyWZ}
\eea
Then can immediately use (\ref{Ch4ModifyWZ}) to infer that the WZ action corresponding to the modified effective action
(\ref{Ch4CounterAction6Mod}) is given by
\beq
\label{Ch4WZfinalMod}
\Gamma_{WZ}[g,\rho]  = - \bigg( \sum_{i=1}^3 (c_i-c'_i)\, \Sigma_i  + a\, \Sigma_a 
                                                        + \sum_{i=1}^3 c'_i\, \int d^6x\, \sqrt{g}\, \frac{\rho}{\Lambda}\, I_i     \bigg)\, .
\eeq
In particular, it is clear that, choosing the counterterms with $I_i \rightarrow I^d_i$, as in eq.  (\ref{Ch4IdCounterterm}),
we get just the so-called universal terms, as done in \cite{Baume:2013ika} for general even dimensions.
\beq
\label{Ch4WZfinalUniv}
\Gamma_{WZ}[g,\rho]  = - \bigg( \sum_{i=1}^3 c_i\, \int d^6x\, \sqrt{g}\, \frac{\rho}{\Lambda}\, I_i + a\, \Sigma_a  \bigg) \, .
\eeq

In the flat space limit $(g_{\mu\nu}\to \delta_{\mu\nu})$ 
there are obvious simplifications and (\ref{Ch4WZfinal})takes the form
\bea
\Gamma_{WZ}[\delta,\rho] 
&=&
- \int d^6x\, \sqrt{g}\, \bigg\{ 
- \frac{c_3}{\Lambda^2}\, \Box \rho\, \Box^2\rho
+ \frac{1}{\Lambda^3}\, \bigg[ 
\bigg( - \frac{7}{16}\, c_1 + \frac{11}{4}\, c_2 + 2\, c_3 \bigg)\, \left(\Box\rho\right)^3 \nn \\
&&
+ \bigg( \frac{3}{2}\, c_1 - 6\, c_2 - 8\, c_3 \bigg)\, \left(\pd \pd \rho \right)^2\, \Box\rho \bigg]
+ \frac{1}{\Lambda^4}\, \bigg[ 
  \bigg( -\frac{3}{2}\, c_1 + 6\, c_2 + 16\, c_3 + 24\, a\bigg)\, \left(\pd\rho\right)^2\, \left(\pd\pd\rho\right)^2 
\nn \\
&&
-\, 
\bigg( \frac{3}{8}\, c_1 + \frac{9}{2}\, c_2 + 4\, c_3 + 24\, a \bigg)\, \left(\pd\rho\right)^2\, \left(\Box\rho\right)^2
+ \bigg( \frac{3}{4}\, c_1 - 3\, c_2 - 4\, c_3 \bigg)\, \pd^\mu\left(\pd\rho\right)^2\, \pd_\mu\left(\pd\rho\right)^2
\bigg] \nn \\
&&
\frac{1}{\Lambda^5}\, \bigg( \frac{3}{2}\, c_1 + 6\, c_2 + 36\, a \bigg)\, \left(\pd\rho\right)^4\, \Box\rho
- \frac{1}{\Lambda^6}\, \bigg( c_1 +4\, c_2 + 24\, a \bigg)\, \left(\pd\rho\right)^6
\bigg\}\, .
\label{Ch4Effective6dFlat}
\eea
The structures of the flat space limits of (\ref{Ch4WZfinalMod}) and (\ref{Ch4WZfinalUniv}) follow trivially.
Having obtained the most general form for the WZ action for conformal anomalies in 6 dimensions, 
we now turn to discuss one specific example in $d=6$, previously studied within the AdS/CFT correspondence. 
This provides an application of the results of the previous sections.

\subsection{The WZ action action for a free CFT: the $(2,0)$ tensor multiplet}

In this section we are going to determine the coefficients of the WZ action for the (2,0) tensor multiplet in $d=6$, which has been investigated 
in the past in the context of the $AdS_7/CFT_6$ holographic anomaly matching.

Free field realizations of CFT's are particularly useful in the analysis of the anomalies and their matching between theories 
in regimes of strong and weak coupling, allowing to relate free and interacting theories of these types. 
In this respect, the analysis of correlation functions which can be uniquely fixed by the symmetry is crucial in order to compute the anomaly for 
theories characterized by different field contents in general spacetime dimensions. This is the preliminary step in order to investigate the
matching with other realizations which share the same anomaly content. These are correlation functions which contain up to 3 EMT's and 
that can be determined uniquely, in any dimensions, modulo a set of 
coefficients, such as the number of fermions, scalars and/or spin 1, 
which can be fixed within a specific field theory realization \cite{Osborn:1993cr, Erdmenger:1996yc}

While in $d=4$ these correlation functions can be completely identified by considering a generic theory which combines free scalar, fermions and 
gauge fields \cite{Osborn:1993cr, Erdmenger:1996yc,Coriano:2012wp}, in $d$ dimensions scalars and fermions need to be 
accompanied not by a spin 1 (a $1$-form) but by  a $\kappa$ -form ($d=2 \kappa +2$). 
In $d=6$ this is a 2-form, $B_{\mu\nu}$  \cite{Bastianelli:2000hi}.  

Coming to specific realizations and use of CFT's in $d=6$, we mention that, for instance, the dynamics of a single M5 brane is described by a 
free $\mathcal N=(2,0)$ tensor multiplet which contains 5 scalars, 2 Weyl fermions and a 2-form whose strength is anti-selfdual. For $N$ 
coincident $M5$ branes, at large $N$ values, the anomaly matching between the free field theories realizations and the interacting $(2,0)$ 
CFT's, investigated in the $AdS_7\times S^4$ supergravity description, has served as an interesting test of the correspondence between the $A$
 and $B$ parts of the anomalies in both theories \cite{Bastianelli:1999ab,Bastianelli:2000hi}. 
 
We have summarized in table \ref{Ch4AnomalyCoeff} the coefficients 
of the WZ anomaly action in the case of a scalar, a fermion and a non-chiral $B_{\mu\nu}$  form, which are the fields appearing 
in the (2,2)CFT. Anomalies in the (2,0) and the (2,2) theories are related just by a factor 1/2, after neglecting the gravitational 
anomalies related to the imaginary parts of the (2,0) multiplet \cite{Bastianelli:2000hi}.  

We have extracted the anomaly coefficients in table \ref{Ch4AnomalyCoeff} 
from \cite{Bastianelli:2000hi}, having performed a redefinition of the third invariant $I_3$ 
in the structure of the anomaly functional (\ref{Ch4anom6D}). We choose to denote with $\tilde I_i, \tilde J_i$ and $\tilde c_i$ the anomaly 
operators and coefficients in \cite{Bastianelli:2000hi} 
\beq \label{Ch4Correspondence}
\tilde I_1 = I_1\, , \quad \tilde I_2 = I_2\, , \quad \tilde I_3 = 3\, I_3 + 8\, I_1 -2\, I_2 \, .
\eeq
Actually in \cite{Bastianelli:2000hi} the third conformal invariant, whose implicit expression can be found in \cite{Arakelian:1995ye},
is given by
\beq \label{Ch4I3tilde}
\tilde I_3 \equiv
C^{\alpha\gamma\lambda\kappa} \bigg(
{\delta_\alpha}^\beta\, \Box - 4\, {R_\alpha}^\beta 
+ \frac{6}{5}\, {\delta_\alpha}^\beta \,R \bigg)\, C_{\beta\gamma\lambda\kappa}
+ \bigg( 8\, {\delta_\alpha}^\kappa\, {\delta_\beta}^\lambda - \frac{1}{2}\, g_{\alpha\beta}\, g^{\kappa\lambda} \bigg)\, 
\nabla_\kappa\nabla_\lambda C^{\alpha\gamma\lambda\kappa}\, {C^\beta}_{\gamma\lambda\kappa}\, ,
\eeq
which differs from our choice, reported in appendix \ref{Geometrical}.
The relation in (\ref{Ch4Correspondence}) between the third invariant $\tilde{I}_3$ and $I_3$ can be derived
expanding (\ref{Ch4I3tilde}) on the basis of the $K$-scalars given in table \ref{Ktable}
and comparing it to the third of (\ref{Ch4WeylInv6}).

In light of (\ref{Ch4Correspondence}), as the conformal anomalies depend only on the field content of the theory, i.e.
\beq \label{Ch4Matching}
\mathcal{A}[g]=\sum_{i=1}^3 c_i\, \left( I_i + \nabla_\mu J^\mu_i \right)=
\sum_{i=1}^3 \tilde c_i\, \left( \tilde I_i + \nabla_\mu \tilde J^\mu_i \right) \, ,
\eeq
by replacing (\ref{Ch4Correspondence}) on the r.h.s. of (\ref{Ch4Matching}), 
we conclude that the relations between the anomaly coefficients $\tilde c_i$ and $c_i$ are
\beq \label{Ch4MatchCoefficients}
c_1 = \tilde c_1 + 8\, \tilde c_3\, , \quad 
c_2 = \tilde c_2 - 2\, \tilde c_3\, , \quad 
c_3 = 3\, \tilde c_3 \, .
\eeq
The WZ action can be derived from eq.  (\ref{Ch4WZfinal}) by inserting the expressions of the $c_i$'s and $a$ extracted from 
table \ref{Ch4AnomalyCoeff}.
These can be specialized to the scalar (S), fermion (F) and to the 2-form (B) cases, thereby generating via (\ref{Ch4Matching}) the 
corresponding anomaly functionals. For the $(2,0)$ tensor multiplet this is obtained from the relation
\beq
\mathcal A^{T}[g] = \frac{1}{2}\, \bigg( 10\, \mathcal A^{S}[g] + 2\, \mathcal A^{F}[g] + \mathcal A^{B}[g] \bigg).
\eeq

\begin{table}
$$
\begin{array}{|c|c|c|c|c|}\hline
I & c_1\times 7!\,(4\,\pi)^3 & c_2 \times 7!\,(4\,\pi)^3 & c_3 \times 7!\,(4\,\pi)^3 & a \times 7!\,(4\,\pi)^3
\\ \hline\hline
S & \frac{20}{3} & -\frac{7}{3} & 6 & -\frac{5}{72}
\\ \hline
F & \frac{64}{3} & -112 & 120  & -\frac{191}{72}
\\ \hline
B & - \frac{3688}{3} & -\frac{3458}{3} & 540 & -\frac{221}{4}
\\ \hline
T & -560 & -700 & 420 & -\frac{245}{8} 
\\
\hline
\end{array}
$$
\caption{Anomaly coefficients for a conformally coupled scalar (S), a Dirac Fermion (F), a 2-form field (B)
and the chiral $(2,0)$ tensor multiplet (T), to be normalized by an overall $1/ (7!\,(4\pi)^3)$}
\label{Ch4AnomalyCoeff}
\end{table}

\section{Dilaton interactions and constraints from $\Gamma_{WZ}$}

So far, we have extracted the structure of the WZ action and thus of the anomaly-related dilaton interactions via
the Weyl-gauging of the effective action, adding to the results which can be found in the literature all the scheme-dependent terms.
We now turn to analysing the Weyl-gauged effective action of the CTF's with a perturbative approach.
This analysis will generate a new expression of the WZ effective action, in terms of traced correlators of the EMT.
We wiil then require this expression to match the form obtained in the previous sections, for consistency.
From this the recursive algorithm will follow. \\

We proceed with a Taylor expansion in $\kappa_{\Lambda}$ of the gauged metric which is given by 
\beq \label{Ch3SeriesInG}
\hat{g}_{\mu\nu} = g_{\mu\nu}\, e^{-2\,\kappa_{\Lambda}\rho} =
\bigg(\d_{\mu\nu} + \kappa\, h_{\mu\nu} \bigg)\, e^{-2\,\kappa_{\Lambda}\rho} =
\bigg(\delta_{\mu\nu} + \kappa\, h_{\mu\nu} \bigg)\,
\sum_{n=0}^{\infty} \frac{(-2)^n}{n!}\,(\kappa_{\Lambda}\,\rho)^n \, ,
\eeq
where $\kappa = \sqrt{16\,\pi\,G_{N}}$, with $G_{N}$ the Newton constant in $4$ ($6$) dimensions.
As we are considering only the dilaton contributions, we focus on the functional expansion of the renormalized and Weyl-gauged effective 
action $\hat\Gamma_{\textrm{ren}}[g,\rho]$ with respect to $\kappa_{\Lambda}$.
This is easily done using the rule for the derivation of composite functionals,
\beq \label{Ch3CompositeDiff}
\frac{\pd\hat\Gamma_{\textrm{ren}}[g,\rho]}{\pd\kappa_{\Lambda}} = 
\int d^dy\, \frac{\delta\hat\Gamma_{\textrm{ren}}[g,\rho]}{\delta\hat{g}_{\mu\nu}(x)}
            \frac{\pd\hat{g}_{\mu\nu}(x)}{\pd\kappa_{\Lambda}}\, .
\eeq
Applying (\ref{Ch3CompositeDiff}) repeatedly and taking (\ref{Ch3SeriesInG}) into account, the perturbative series takes the form
\bea \label{Ch3Expansion}
\hat\Gamma_{\textrm{ren}}[g,\rho]
&=&
\Gamma_{\textrm{ren}}[g,\rho]
+\, \frac{1}{2!\,\Lambda^2}\, \int d^d \xu d^d \xd\, 
\frac{\delta^2\hat\Gamma_{\textrm{ren}}[g,\rho]}{\delta\hat{g}_{\muu\nuu}(\xu)\delta\hat{g}_{\mud\nud}(\xd)}
\frac{\pd \hat{g}_{\muu \nuu}(\xu)}{\pd \kappa_\Lambda}\frac{\pd \hat{g}_{\mud\nud}(\xd)}{\pd\kappa_\Lambda}
\nn \\
&& \hspace{10mm}
+\, \frac{1}{3!\, \Lambda^3}\, \bigg(\int d^d x_1 d^d \xd d^d \xt\,
\frac{\delta^3\hat\Gamma_{\textrm{ren}}[g,\rho]}
{\delta\hat{g}_{\muu\nuu}(\xu)\delta\hat{g}_{\mud\nud}(\xd)\delta\hat{g}_{\mut\nut}(\xt)}
\frac{\pd \hat{g}_{\muu \nuu}(\xu)}{\pd \kappa_\Lambda}\frac{\pd \hat{g}_{\mud \nud}(\xd)}{\pd \kappa_\Lambda}
\frac{\pd \hat{g}_{\mut \nut}(\xt)}{\pd \kappa_\Lambda}
\nn \\
&& \hspace{23mm}
+\, 3\, \int d^d \xu d^d \xd\, 
\frac{\delta^2\hat\Gamma_{\textrm{ren}}[g,\rho]}{\delta\hat{g}_{\muu\nuu}(\xu)\delta\hat{g}_{\mud\nud}(\xd)}
\frac{\pd^2 \hat{g}_{\muu\nuu}(\xu)}{\pd \kappa_\Lambda^2}\frac{\pd\hat{g}_{\mud\nud}(\xd)}{\pd \kappa_\Lambda} 
\bigg) +\ldots
\eea
Until now we have always defined the Green functions of the EMT in terms of the ordinary generating functional $\mathcal W$, 
accorging to (\ref{Ch3NPF}). It goes without saying that, either at UV or IR fixed points of the RG flow, this definition is
generalized through the simple replacement $\mathcal{W} \rightarrow \Gamma$.
As we are interested in the flat space limit of the dilaton action, we can write (\ref{Ch3Expansion}) 
by taking the limit of a conformally flat background metric 
$(\hat{g}_{\mu\nu}\rightarrow \hat \delta_{\mu\nu} \equiv\delta_{\mu\nu}\, e^{- 2\,\kappa_{\Lambda}\rho})$ 
obtaining
\bea \label{Ch3FinalExp}
\hat\Gamma_{\textrm{ren}}[\delta,\rho]
&=&
\Gamma_{\textrm{ren}}[\delta,\rho] + 
\frac{1}{2!\,\Lambda^2}\,  
\int d^d \xu d^d \xd\, \langle T(\xu) T(\xd)\rangle\, \rho(\xu)\rho(\xd)
\nn \\
&& 
-\, \frac{1}{3!\,\Lambda^3}\, \bigg[ \int d^d \xu d^d \xd d^d \xt\,\langle T(\xu) T(\xd) T(\xt)\rangle\, \rho(\xu)\rho(\xd)\rho(\xt)
\nn \\
&& \hspace{20mm}
+\, 6\,\int d^d \xu d^d \xd\, \langle T(\xu) T(\xd)\rangle\, (\rho(\xu))^2\rho(\xd) \bigg] + \ldots \, ,
\eea
where we have used eq.  (\ref{Ch3NPF}) in the definition of the EMT' s correlators and the obvious relation
\beq \label{Ch3MetricDil}
\frac{\pd^n \hat{g}_{\mu\nu}(x)}{\pd \kappa_{\Lambda}^n}
\bigg|_{g_{\mu\nu}=\delta_{\mu\nu},\kappa_{\Lambda}= 0} = 
\left(-2\right)^n \, \left(\rho(x)\right)^n\, \delta_{\mu\nu}\, .
\eeq
From (\ref{Ch3FinalExp}) one may identify the expression of 
$\Gamma_{WZ}= \Gamma_{\textrm{ren}}[\delta,\rho] - \hat\Gamma_{\textrm{ren}}[\delta,\rho]$
written in terms of the traced n-point correlators of stress-energy tensors. 
This has to coincide with eq.  (\ref{Ch3Effective4d}) evaluated in the conformally flat limit and given by
\beq \label{Ch3FlatWZ}
- \Gamma_{WZ}[\delta,\rho] = 
- \int d^4x\,  \bigg[
\frac{2\,\beta_a}{\Lambda^2}\, \left( \Box \rho \right)^2
+ \left(\beta_a + \beta_b\right)\, \bigg( - \frac{4}{\Lambda^3}\, \left(\pd\rho\right)^2\,\Box \rho
+ \frac{2}{\Lambda^4}\, \left(\pd\rho\right)^4 \bigg) \bigg ] \, .
\eeq
At this point, a comparison between the dilaton vertices extracted from (\ref{Ch3FinalExp}) and (\ref{Ch3FlatWZ}) allows to establish a 
consistency condition between the first four of such vertices and a relation among the entire hierarchy of the traced correlators,
which is the key result of this chapter.

For this purpose we denote by $\mathcal I_n(\xu,\dots,x_n)$ the dilaton vertices obtained by functional differentiation of 
$\Gamma_{\textrm{ren}}[\hat{\delta}]$,
\beq \label{Ch3FuncDiffWZ}
\mathcal I_{n}(\xu,\dots,x_n) = 
\frac{\delta^n \left(\hat\Gamma_{\textrm{ren}}[\delta,\rho]-\Gamma_{\textrm{ren}}[\delta,\rho]\right)}
{\delta\rho(\xu)\dots\delta\rho(x_n)} 
= -  \frac{\delta^n \Gamma_{WZ}[\delta,\rho]}{\delta\rho(\xu)\dots\delta\rho(x_n)} 
\eeq
in coordinate space, which we can promptly transformed to momentum space. 

The expressions of such vertices for the first six orders in $\kappa_{\Lambda}$ are given by
\bea
{\mathcal I}_2(\ku,-\ku) 
&=& 
\frac{1}{\Lambda^2}\, \langle T(\ku) T(-\ku)\rangle \, ,
\nn \\
{\mathcal I}_3(\ku,\kd,\kt)
&=& 
-\frac{1}{\Lambda^3}\, \bigg[ 
\langle T(\ku) T(\kd) T(\kt) \rangle
+\, 2\, \bigg( \langle T(\ku) T(-\ku)\rangle + \langle T(\kd) T(-\kd)\rangle + \langle T(\kt) T(-\kt) \rangle \bigg) \bigg]\, ,
\nn \\ 
{\mathcal I}_4(\ku,\kd,\kt,\kq)
&=&
\frac{1}{\Lambda^4}\, \Biggl[
\langle T(\ku) T(\kd) T(\kt) T(\kq) \rangle
+\, 2\,\sum_{\mathcal T\left\{4,(k_{i_1},k_{i_2})\right\}}
\langle T(k_{i_1}) T(k_{i_2}) T(-k_{i_1}-k_{i_2})\rangle
\nn \\
&& \hspace{6mm}
+\, 2\, \sum_{\mathcal T\left\{4,(k_{i_1},k_{i_2})\right\}} 
\langle T(k_{i_1}+k_{i_2}) T(-k_{i_1}-k_{i_2})\rangle
+ 4\, \sum_{i=1}^{4} \langle T(k_i) T(-k_i)\rangle \Biggr]\, ,
\nn \\
{\mathcal I}_5(\ku,\kd,\kt,\kq,\kc) 
&=&
- \frac{1}{\Lambda^5}\, \Biggl[
\langle T(\ku) T(\kd) T(\kt) T(\kq) T(\kc) \rangle 
\nn \\
&&
+\, 2\, \sum_{\mathcal T\left\{5,(k_{i_1},k_{i_2},k_{i_3})\right\}}
\langle T(k_{i_1}) T(k_{i_2}) T(k_{i_3}) T(-k_{i_1}-k_{i_2}-k_{i_3}) \rangle
\nn \\
&&
+ \,4\, \Biggl(
\sum_{\mathcal T\left\{5,(k_{i_1},k_{i_2})\right\}}
\langle T(k_{i_1}) T(k_{i_2}) T(-k_{i_1}-k_{i_2}) \rangle 
\nn \\
&&
+\, \sum_{\mathcal T\left\{5,[(k_{i_1},k_{i_2}),(k_{i_3},k_{i_4})]\right\}}
\langle T(k_{i_1}+k_{i_2}) T(k_{i_3}+k_{i_4}) T(-k_{i_1}-k_{i_2}-k_{i_3}-k_{i_4}) \rangle \Biggr)
\nn \\
&&
+\, 8\, \Biggl(
\sum_{\mathcal T\left\{5,(k_{i_1},k_{i_2})\right\}} \langle T(k_{i_1}+k_{i_2}) T(-k_{i_1}-k_{i_2}) \rangle +
\sum_{i=1}^{5} \langle T(k_i) T(-k_i)\rangle \Biggr)
\Biggr]\, ,
\label{Ch3DilIntStructure} \\
{\mathcal I}_6(\ku,\kd,\kt,\kq,\kc,\ks)
&=&
\kappa_{\Lambda}^6\, \Biggl[
\langle T(\ku) T(\kd) T(\kt) T(\kq) T(\kc) T(\ks) \rangle 
\nn \\
&&
+\, 2\, \sum_{\mathcal T\left\{6,(k_{i_1},k_{i_2})\right\}}
\langle T(k_{i_1}+k_{i_2}) T(k_{i_3}) T(k_{i_4}) T(k_{i_5}) T(k_{i_6}) \rangle
\nn \\
&&
+ \,4\, \Biggl( \sum_{\mathcal T\left\{6,(k_{i_1},k_{i_2},k_{i_3})\right\}}
\langle T(k_{i_1}+k_{i_2}+k_{i_3}) T(k_{i_4}) T(k_{i_5}) T(k_{i_6}) \rangle
\nn \\
&&
+ \, \sum_{\mathcal T\left\{6,[(k_{i_1},k_{i_2}),(k_{i_3},k_{i_4})]\right\}}
\langle T(k_{i_1}+k_{i_2}) T(k_{i_3}+k_{i_4}) T(k_{i_5}) T(k_{i_6}) \rangle 
\Biggr) \nn \\
&&
+\, 8\, \Biggl( 
\sum_{\mathcal T\left\{6,(k_{i_1},k_{i_2},k_{i_3},k_{i_4})\right\}}
\langle T(k_{i_1}+k_{i_2}+k_{i_3}+k_{i_4}) T(k_{i_5}) T(k_{i_6}) \rangle 
\nn
\eea
\bea
&&
+\, \sum_{\mathcal T\left\{6,[(k_{i_1},k_{i_2},k_{i_3}),(k_{i_4},k_{i_5})]\right\}}
\langle T(k_{i_1}+k_{i_2}+k_{i_3}) T(k_{i_4}+k_{i_5}) T(k_{i_6}) \rangle
\nn \\
&&
+\, \sum_{\mathcal T\left\{6,[(k_{i_1},k_{i_2}),(k_{i_3},k_{i_4})]\right\}}
\langle T(k_{i_1}+k_{i_2}) T(k_{i_3}+k_{i_4}) T((k_{i_5}+k_{i_6})) \rangle\Biggr)
\nn \\
&&
+\, 16\, \Biggl(
\frac{1}{2}\, \sum_{\mathcal T\left\{6,(k_{i_1},k_{i_2},k_{i_3})\right\}} 
\langle T(k_{i_1}+k_{i_2}+k_{i_3}) T(-k_{i_1}-k_{i_2}-k_{i_3}) \rangle
\nn \\
&&
+\, \, \sum_{\mathcal T\left\{6,(k_{i_1},k_{i_2})\right\}} \langle T(k_{i_1}+k_{i_2}) T(-k_{i_1}-k_{i_2}) \rangle
+ \sum_{i=1}^{6} \langle T(k_i) T(-k_i)\rangle
\Biggr)
\Biggr].
\label{Ch4DilIntStructure6}
\eea
These results can be easily extended to any higher order. 
The recipe, in this respect, is really simple: \\
in order to construct the vertex at order $n$ one has to sum to the $n$-point function
all the lower order functions in the hierarchy, down to $n=2$, partitioning the momenta in all the possible ways
and symmetrising each single contribution. 
The normalization factor in front of the correlator of order-$k$ is always $2^{n-k}$, while the factor in front of the 
vertex of order $n$ is $(-\kappa_\Lambda)^n$. Notice that, for $n$ even, we have an additional $1/2$ factor in front of the contributions from 
the $2$-point functions in which each EMT carries $n/2$ momenta, to avoid double counting. 

Notice that the expressions of the dilaton interactions in (\ref{Ch4DilIntStructure6}) have been derived withour any reference to the
dimensions of space, so that they hold for any even dimension. As such, they can be thoroughly tested in the simplest case, i.e. $d=2$.
We did check them in $2$ dimensions, as illustrated in appendix \ref{Ch42D}.

We pause for a moment to clarify the notation used in 
(\ref{Ch3DilIntStructure}) for the organization of the momenta and the meaning of the symbol $\mathcal T$.

For example $\mathcal T\left\{4,(k_{i_1},k_{i_2})\right\} $ denotes the six pairs of distinct momenta in the case of the four point 
functions
\beq
\mathcal T\left\{4,(k_{i_1},k_{i_2})\right\} = 
\left\{(\ku,\kd),(\ku,\kt),(\ku,\kq),(\kd,\kt),(\kd,\kq),(\kt,\kq) \right\}   \, ,
\eeq
where we are combining the 4 momenta $k_1,...k_4$ into all the possible pairs, for a total of $\binom{4}{2}$ terms.
With five momenta $\left(\ku,\kd,\kt,\kq,\kc\right)$ the available pairs are
\bea
\mathcal T\left\{5,(k_{i_1},k_{i_2})\right\}
&=& 
\left\{(\ku,\kd),(\ku,\kt),(\ku,\kq),(\ku,\kc),(\kd,\kt),(\kd,\kq),(\kd,\kc),(\kt,\kq),(\kt,\kc),(\kq,\kc) \right\}  \, \nn\\
\eea
while the possible triples are 
\bea
\mathcal T\left\{5,(k_{i_1},k_{i_2},k_{i_3})\right\}
&=& 
\left\{(\ku,\kd,\kt),(\ku,\kd,\kq),(\ku,\kd,\kc),(\ku,\kt,\kq),(\ku,\kt,\kc),
\right.
\nn \\
&&
\left.
\hspace{1mm}
(\ku,\kq,\kc),(\kd,\kt,\kq),(\kd,\kt,\kc),(\kd,\kq,\kc),(\kt,\kq,\kc) \right\}\, .
\eea
As we move to higher orders, the description of the momentum dependence gets slightly more involved and we need to distribute the external 
momenta into two pairs. The notation $\mathcal T\left\{5, [(k_{i_1},k_{i_2}),(k_{i_3},k_{i_4})]\right\}$ denotes the set of 
independent paired couples which can be generated out of $5$ momenta. 
Their number is $15$ and they are given by
\bea
\mathcal T\left\{5,[(k_{i_1},k_{i_2}),(k_{i_3},k_{i_4})]\right\}
&=& 
\left\{[(\ku,\kd),(\kt,\kq)],[(\ku,\kd),(\kt,\kc)],[(\ku,\kd),(\kq,\kc)] 
\right. 
\nn \\
&& \hspace{-55mm}
\left.     
[(\ku,\kt),(\kd,\kq)],[(\ku,\kt),(\kd,\kc)],[(\ku,\kt),(\kq,\kc)],[(\ku,\kq),(\kd,\kt)],
[(\ku,\kq),(\kd,\kc)],[(\ku,\kq),(\kt,\kc)], 
\right.
\nn \\
&& \hspace{-55mm}    
\left.
[(\ku,\kc),(\kd,\kt)],[(\ku,\kc),(\kd,\kq)],[(\ku,\kc),(\kt,\kq)],[(\kd,\kt),(\kq,\kc)],
[(\kd,\kq),(\kt,\kc)],[(\kd,\kc),(\kt,\kq)] \right\}\, .
\eea
%

\subsection{The recursive relation in $4$ dimensions}

It is obvious that a direct computation of $\mathcal{I}_2, \mathcal{I}_3$ and $\mathcal{I}_4$ 
from the anomaly action (\ref{Ch3FlatWZ}) allows to extract the explicit structure of these vertices in momentum space
\bea \label{Ch3DilatonInt}
{{\mathcal I}}_2(\ku,-\ku) 
&=&
-\frac{4}{\Lambda^2}\, \beta_a\, {\ku}^4\, , \nn \\
{{\mathcal I}}_3(\ku,\kd,\kt) 
&=& 
\frac{8}{\Lambda^3} \, \bigg(\beta_a + \beta_b \bigg)\, \bigg( 
\ku^2\, \kd\cdot\kt + \kd^2\, \ku\cdot\kt + \kt^2\, \ku\cdot\kd \bigg) 
\nn \\
{{\mathcal I}}_4(\ku,\kd,\kt,\kq)
&=&
- \frac{16}{\Lambda^4}\,\bigg(\beta_a + \beta_b \bigg)\, \bigg( 
\ku\cdot\kd\, \kt\cdot\kq + \ku\cdot\kt\, \kd\cdot\kq + \ku\cdot\kq\, \kd\cdot\kt  \bigg)\, ,
\eea
(with $k_{i}^n \equiv (k_i^2)^{n/2}$). These relations can be used together with (\ref{Ch3FinalExp}) in order to extract the 
structure of the $2$- $3$- and $4$-point functions of the traced correlators, solving an elementary linear system.
Their expressions are very easily found to be
\bea \label{Ch3BuildingBlocks}
\langle T(\ku) T(-\ku) \rangle 
&=& 
- 4\, \beta_a\, {\ku}^4 \, , 
\nn \\
\langle T(\ku) T(\kd) T(\kt) \rangle
&=& 
8 \bigg[ 
- \bigg( \beta_a+\beta_b \bigg)\,\bigg( f_{3}(\ku,\kd,\kt)+f_{3}(\kd,\ku,\kt)+f_{3}(\kt,\ku,\kd)\bigg) 
+        \beta_a\, \sum_{i=1}^{3} k_i^4 \bigg]\, ,
\nn \\
\langle T(\ku) T(\kd) T(\kt) T(\kq)\rangle 
&=&
8\, \bigg\{ 
6\, \bigg( \beta_a + \beta_b \bigg)\, \bigg[
\sum_{\mathcal T\left\{4,[(k_{i_1},k_{i_2}),(k_{i_3},k_{i_4})]\right\}} 
k_{i_i}\cdot k_{i_2}\, k_{i_3}\cdot k_{i_4}
\nn \\
&&
+\, f_{4}(\ku\,\kd,\kt,\kq) + f_{4}(\kd\,\ku,\kt,\kq) + f_{4}(\kt\,\ku,\kd,\kq) + f_{4}(\kq\,\ku,\kd,\kt) \bigg]
\nn \\
&&
-\, \beta_a\, 
\bigg( \sum_{\mathcal T\left\{4,(k_{i_1},k_{i_2})\right\}}(k_{i_1} + k_{i_2})^4 
+ 4\, \sum_{i=1}^{4} k_{i}^4 \bigg)
\bigg\}\, ,
\eea
where we have introduced the compact notations
\bea
f_{3}(k_a,k_b,k_c)
&=&
k_a^2\, k_b \cdot k_c \, ,
\nn \\
f_{4}(k_a,k_b,k_c,k_d)
&=&
k_a^2\, \left( k_b \cdot k_c + k_b \cdot k_d + k_c \cdot k_d \right)\, .
\eea
The third and fourth order results, in particular, were established in \cite{Coriano:2012dg}
via the explicit computation of the first three functional derivatives of the anomaly $\mathcal A[g]$
and exploiting recursively the hierarchical relations (\ref{Ch3hier}).

here comes the second significant result: it is quite immediate to realize that the hierarchy in eq.  (\ref{Ch3hier})
can be entirely re-expressed in terms of the first four traced correlators. 
For this purpose, one has just to notice that $\Gamma_{WZ}[\hat{\delta}]$ is quartic in $\rho$, with  $\mathcal{I}_n=0$, for $n\ge 5$. 
Therefore, for instance, the absence of vertices with $5$ dilaton external lines, which sets $\mathcal{I}_5=0$, 
combined with the 4 fundamental traces in (\ref{Ch3BuildingBlocks}), are sufficient to completely fix the structure of the 5-point 
function, which takes the form 
\bea
&&
\langle T(\ku) T(\kd) T(\kt) T(\kq) T(\kc) \rangle =
16\, \Biggl\{
-24\,\bigg(\beta_a + \beta_b\bigg)\, \Biggl[
\sum_{\mathcal T\left\{5,[(k_{i_1},k_{i_2}),(k_{i_3},k_{i_4})]\right\}}
k_{i_1}\cdot k_{i_2}\,k_{i_3}\cdot k_{i_4}
\nn \\
&& \hspace{-10mm}
+\, f_{5}(\ku,\kd,\kt,\kq,\kc) + f_{5}(\kd,\ku,\kt,\kq,\kc) + f_{5}(\kt,\ku,\kd,\kq,\kc) 
+   f_{5}(\kq,\ku,\kd,\kt,\kc) + f_{5}(\kc,\ku,\kd,\kt,\kq) \Biggr]
\nn \\
&& \hspace{-10mm}
+\, \beta_a\, \Biggl[
                     \sum_{\mathcal T\left\{5,(k_{i_1},k_{i_2},k_{i_3})\right\}} 
										\left(k_{i_1} + k_{i_2} + k_{i_3} \right)^4
                     + 3\, \sum_{\mathcal T\left\{5,(k_{i_1},k_{i_2})\right\}} \left(k_{i_1} + k_{i_2} \right)^4
                     + 12\, \sum_{i=1}^{5} k_{i}^4 
\Biggr]
\Biggr\}\, ,
\label{Ch35T}
\eea
where $f_5$ is defined as
\beq
f_{5}(k_a,k_b,k_c,k_d,k_e) = 
k_a^2\, \left( k_b \cdot k_c + k_b \cdot k_d + k_b \cdot k_e + k_c \cdot k_d + k_c \cdot k_e + k_d \cdot k_e  \right)\, .
\eeq
The construction that we have outlined can be extended to any arbitrary traced $n$-point function of the EMT: 
it just takes one to apply the simple recipe to express the $n$-dilaton interactions, which is given under eq.  (\ref{Ch4DilIntStructure6}).
These relations can be compared, for consistency, with their equivalent expression obtained directly from the hierarchy (\ref{Ch3hier}).
In general, this requires the computation of functional derivatives of the anomaly functional $\mathcal{A}$ up to the relevant order.
One can check by a direct computation using (\ref{Ch3hier}) the agreement with (\ref{Ch35T}) up to the $5$-th order,
as we explicitly did to test the correctness of our result.

All the results given in this section can be easily generalized with the inclusion of a counterterm (\ref{Ch3ll}), 
using the prescription (\ref{Ch3LocalToBeta}), as discussed above. 

\subsection{The recursive relation in $6$ dimensions}

At this point we can move on to the evaluation of dilaton interactions in $6$ dimensions and, consequently, 
of the first $6$ traced correlators, being clear from (\ref{Ch4DilIntStructure6}) that a direct computation of 
$\mathcal{I}_2 - \mathcal{I}_6$ from the anomaly action (\ref{Ch4Effective6dFlat}) allows to extract the structure of these 
Green functions.

The dilaton interactions are straightforwarly computed,
\bea \label{Ch4DilatonInt}
{\mathcal I}_2(\ku,-\ku) 
&=&
\frac{2}{\Lambda^2}\, c_3\, k_1^6\, ,  \nn \\
{\mathcal I}_3(\ku,\kd,\kt) 
&=& 
\frac{1}{\Lambda^3}\, \bigg[
\bigg( \frac{21}{8}\, c_1 - \frac{33}{2}\,c_2 - 12\,c_3 \bigg)\, k_1^2\,k_2^2\,k_3^2  \nn \\
&& \hspace{4mm}
+\, \bigg( - 3\, c_1 + 12\, c_2 + 16\,c_3 \bigg)\, \bigg( k_1^2\, \left(k_2\cdot k_3\right)^2 
+ k_2^2\, \left(k_1\cdot k_3\right)^2 + k_3^2\, \left(k_1\cdot k_2\right)^2 \bigg) \bigg] \, ,
\nn \\
{\mathcal I}_4(\ku,\kd,\kt,\kq)
&=&
\frac{1}{\Lambda^4}\, \bigg[
\bigg( 6\, c_1 - 24\,c_2 - 64\,c_3 - 96 \,a\bigg)\,
\sum_{\mathcal T\left\{4,(k_{i_1},k_{i_2})\right\}} k_{i_1}\cdot k_{i_2}\, \left(k_{i_3}\cdot k_{i_4}\right)^2  \nn \\
&& \hspace{-10mm}
+\, \bigg(\frac{3}{2}\, c_1 + 18\, c_2 + 16 \,c_3 + 96\,a \bigg)\,
\sum_{\mathcal T\left\{4,(k_{i_1},k_{i_2})\right\}} k_{i_1}\cdot k_{i_2}\, k_{i_3}^2\, k_{i_4}^2
\nn \\
&& \hspace{-10mm}
+\, \bigg( - 6\, c_1 + 24\,c_2 + 32\,c_3 \bigg)\,
\sum_{\mathcal T\left\{4,\left[(k_{i_1},k_{i_2}),(k_{i_3},k_{i_4})\right]\right\}} 
(k_{i_1}+k_{i_2})\cdot(k_{i_3}+k_{i_4})\, k_{i_1}\cdot k_{i_2}\, k_{i_3}\cdot k_{i_4} \bigg] \, ,
\nn
\eea
\bea
{\mathcal I}_5(\ku,\kd,\kt,\kq,\kc)
&=&
- \frac{12}{\Lambda^5}\, \bigg(c_1 + 4\,c_2 + 24\,a \bigg) \nn \\ 
&&
\times \sum_{\mathcal T\left\{5,(k_{i_1},k_{i_2},k_{i_3},k_{i_4})\right\}}
k_{i_5}^2\, \left( k_{i_1} \cdot k_{i_2} \, k_{i_3} \cdot k_{i_4} + k_{i_1} \cdot k_{i_3} \, k_{i_2} \cdot k_{i_4}
                                         + k_{i_1} \cdot k_{i_4} \, k_{i_2} \cdot k_{i_3} \right)\, , \nn \\
{\mathcal I}_6(\ku,\kd,\kt,\kq,\kc,\ks)
&=&
\frac{48}{\Lambda^6}\, \bigg(c_1 + 4\, c_2 + 24\,a\bigg)\, 
\sum_{\mathcal T\left\{6,\left[(k_{i_1},k_{i_2}),(k_{i_3},k_{i_4}),(k_{i_5},k_{i_6})\right]\right\}}
k_{i_1}\cdot k_{i_2}\, k_{i_3}\cdot k_{i_4}\, k_{i_5}\cdot k_{i_6} \, , \nn \\
\eea
(with $k_{i}^n \equiv (k_i^2)^{n/2}$).

These vertices can be used together with the relations (\ref{Ch4DilIntStructure6}) in order to extract the structure of the traced correlators.
Solving the linear system, we find that the first two of them are given by
\bea \label{Ch4BuildingBlocks23}
\langle T(\ku) T(-\ku) \rangle 
&=& 
2\, c_3\, \ku^6 \, , 
\nn \\
\langle T(\ku) T(\kd) T(\kt) \rangle
&=& 
\bigg( 3\, c_1 - 12\,c_2 - 16\,c_3 \bigg)\, 
\bigg( \ku^2\,\left( \kd\cdot\kt \right)^2 + \kd^2\,\left( \ku\cdot\kt \right)^2 + \kt^2\,\left( \ku\cdot\kd \right)^2 \bigg)
\nn \\
&&
- \bigg( \frac{21}{8}\, c_1 - \frac{33}{2}\, c_2 - 12\, c_3 \bigg)\, \ku^2\,\kd^2\,\kt^2 
-\, 4\,c_3\,\bigg( \ku^6 + \kd^6 + \kt^6  \bigg) \, .
\eea
The structure of the $4$-point Green function is much more complicated and can be expressed in a compact notation through the expression
\bea
\langle T(\ku) T(\kd) T(\kt) T(\kq)\rangle 
&=&
\bigg[ 4\, c_3\, \bigg(7\, f^{2i,2i,2i} + 6\, f^{2i,2i,ij} + 3\, f^{2i,2i,2j}
+ 12\, f^{2i,ij,ij} + 12\, f^{2i,2j,ij} + 8\, f^{ij,ij,ij} \bigg) 
\nn \\
&&
+\, \bigg(-18\, c_1 + 72\, c_2 + 96\, c_3 \bigg)\, f^{2i,jk,jk}
+ 4\, \bigg(24\, a + 3\, c_1 - 12\, c_2 - 8\, c_3\bigg)\, f^{2i,2j,kl}
\nn \\
&&
-\, 6\, \bigg(16\, a + c_1 - 4\,c_2\bigg) f^{ij,kl,kl} 
+ \bigg(\frac{63}{4}\, c_1 - 99\, c_2 - 72\, c_3 \bigg)\, f^{2i,2j,2k}
\nn \\
&&
+\, \bigg(-6\,c_1 + 24\,c_2 + 32\,c_3 \bigg)\, \bigg( 2\, \, f^{2i,jk,jl} +  f^{ij,ik,jl} \bigg)
\bigg]\, .
\label{Ch4BuildingBlocks4}
\eea
Here we have introduced a compact notation for the basis of the $12$ scalar functions $f^{\dots}(\ku,\kd,\kt,\kq)$
on which the correlator is expanded, leaving their dependence on the momenta implicit not to make the formula clumsy.
As dimensional analysis forces every term in the Green function to be the product of three scalar products of momenta,
the role of the tree superscripts on each of the $f$'s is to specify the way in which the momenta are distributed.
We present below the expressions of the first four scalar $f$'s, from which it should be clear how to derive the explicit forms of all the 
others. We obtain
\bea
f^{2i,2i,2i}(\ku,\kd,\kt,\kq)
&=&
\sum_{i=1}^{4} (k_i)^6 \, , \nn \\
f^{2i,2i,ij}(\ku,\kd,\kt,\kq)
&=&
\sum_{i=1}^{4} (k_i)^4\, \sum_{j\neq i} k_i \cdot k_j\, , \nn \\
f^{2i,2i,2j}(\ku,\kd,\kt,\kq)
&=&
\sum_{i=1}^{4} (k_i)^4\, \sum_{j\neq i} k_j^2\, , \nn \\
f^{2i,ij,ij}(\ku,\kd,\kt,\kq)
&=&
\sum_{i=1}^{4} (k_i)^2\, \sum_{j\neq i} \left(k_i \cdot k_j\right)^2\, .
\eea
Notice that each one of the $f$'s is completely symmetric with respect to any permutation of the momenta, as for the whole correlator. 
The structure of the $5$- and $6$-point functions is essentially similar to (\ref{Ch4BuildingBlocks23}), 
although they require much broader bases of scalar functions to account for all their terms and we do not report them explicitly.

Again, it is clear that the hierarchy in eq.  (\ref{Ch3hier}) can be entirely re-expressed in terms of the first $6$ 
traced correlators. In fact, one notices that $\Gamma_{WZ}[\hat{\delta}]$ is at most of order $6$ in $\rho$, which implies
\beq \label{Ch4NoInt}
\mathcal I_n(\xu,\dots,x_n) = 0\, , \quad n \geq 7 \, .
\eeq
Therefore, for instance, the absence of vertices with $7$ dilaton external lines, which sets $\mathcal{I}_7=0$
and the knowledge of the first $6$ fundamental Green functions are sufficient to completely fix the structure of the $7$-point function, 
and so for the vertices of higher orders. In this way one can determine all the others recursively, up to the desired order.
The consistency of these relations could be checked, in principle, by a direct comparison with their expression obtained directly
from the hierarchy (\ref{Ch3hier}). This requires the explicit computation of functional derivatives of the anomaly functional $\mathcal{A}$ 
up to the relevant order, which is a much more time-consuming task.

\section{Conclusions}

Our analysis has had the goal of showing that the infinite hierarchy of fully traced correlation functions generated by the
anomaly constraint in a generic CFT in even dimensions has as fundamental building blocks, in $d$ dimensions, only the first $d$ 
correlators. For instance, in $d=4$ only correlators with $2, 3$ and $4$ traces are necessary to identify the entire hierarchy. 
This result can be simply derived from the structure of the WZ action, which only contains dilaton 
interactions up to the quartic order. Non anomalous terms, which are homogeneous under Weyl transformations and can be 
of arbitrarily higher orders in $\rho$, do not play any role in this construction.
The WZ action can also be determined, in general, by the Noether method, where the dilaton is coupled directly to the anomaly 
and corrections are included in order to take care of the Weyl non-invariance of the functional. 
Alternatively, the same action is fixed by the cocycle condition, which shows that its functional dependence 
on the dilaton field takes place via the Weyl-gauging of the metric tensor. In our analysis we have introduced an expression of the 
anomaly-induced action in which the anomaly contribution is generated directly by the counterterms, evaluated in dimensional
regularization.

The WZ conformal anomaly actions that we have derived include all the contributions related to the local part of the anomaly.
This result adds full generality to the analysis of dilaton effective actions, which carry an intrinsic regularization scheme dependence, 
due to the appearance in the anomaly functional of terms that are different from the Euler density and the $d$-dimensional Weyl invariants 
in the specific space dimension. 
In general, the extraction of these extra contributions, as one can figure out from our study, is very involved, with a level of difficulty that grows 
with the dimensionality of the space in which the underlying CFT is formulated. Our main results for WZ dilaton actions, especially in $d=6$,
are remarkably simplified in flat space. 
Comparing our results with those of the previous literature \cite{Bastianelli:1999ab}, we have given the form of the WZ action 
in the case of the CFT of the $(2,0)$ tensor multiplet, which, in the past, has found application in the $AdS_7/CFT_6$ correspondence.

\clearpage{\pagestyle{empty}\cleardoublepage}

\begin{appendix}

\chapter{Appendix}\label{Ch1ComputeTTT}

\section{Sign conventions}\label{Sign}

 The definition of the Fourier transform of a $n$-EMT's correlation functions, which holds for any other $n$-point function as well, 
 is given by
\beq
\int \, d^d\xu\, \dots d^d x_n\, \left\langle T^{\muu\nuu}(\xu)\dots T^{\mu_n\nu_n}(x_n)\right\rangle \,
e^{-i(\ku\cdot \xu + \dots + k_n \cdot x_n)} = (2\pi)^d\,
\delta^{(d)}\left( \sum_{i=1}^n k_i \right)\,\left\langle T^{\muu\nuu}(\ku)\dots T^{\mu_n\nu_n}(k_n)\right\rangle \, ,
\label{Ch3NPFMom}
\eeq
where all the momenta are conventionally taken to be incoming. \\

The covariant derivatives of a contravariant vector $A^\mu$ and of a covariant one $B_\mu$ are respectively
\bea
\nabla_{\nu} A^\mu \equiv \pd_\nu A^\mu + \Gamma^\mu_{\nu\r}A^\r\, ,\\
\nabla_{\nu} B_\mu \equiv \pd_\nu B_\mu - \Gamma^\r_{\nu\mu}B_\r\, ,
\eea
with the Christoffel symbols defined as
\beq\label{Ch1Christoffel}
\Gamma^{\a}_{\b\g} = \frac{1}{2}g^{\a\k}\left[-\pd_\k g_{\b\g} + \pd_\b g_{\k\g} + \pd_\g g_{\k\b} \right]\, .
\eeq
Our definition of the Riemann tensor is
\bea \label{Ch1Tensors}
{R^\lambda}_{\mu\kappa\nu}
&=&
\pd_\nu \Gamma^\lambda_{\mu\kappa} - \pd_\kappa \Gamma^\lambda_{\mu\nu}
+ \Gamma^\lambda_{\nu\eta}\Gamma^\eta_{\mu\kappa} - \Gamma^\lambda_{\kappa\eta}\Gamma^\eta_{\mu\nu}.
\eea
The Ricci tensor is defined by the contraction $R_{\mu\nu} = {R^{\lambda}}_{\mu\lambda\nu}$ 
and the scalar curvature by $R = g^{\mu\nu}R_{\mu\nu}$.\\

The functional variations with respect to the metric tensor are computed using the relations
\bea\label{Ch1Tricks}
\delta \sqrt{g} = -\frac{1}{2} \sqrt{g}\, g_{\a\b}\,\delta g^{\a \b}\quad &&
\delta \sqrt{g} = \frac{1}{2} \sqrt{g}\, g^{\a\b}\,\delta g_{\a \b}  \nonumber \\
\delta g_{\mu\nu} = - g_{\mu\a} g_{\nu\b}\, \delta g^{\a\b} \quad&&
\delta g^{\mu\nu} = - g^{\mu\a} g^{\nu\b}\, \delta g_{\a\b}\,
\eea
The structure $s^{\a\b\g\delta} $ has been repeatedly used throughout the calculations: it comes from
\bea\label{Ch1Tricks2} 
- \frac{\d g^{\a\b}(z)}{\d g_{\g\d}(x)}\bigg|_{g_{\mu\nu}=\delta_{\mu\nu}} = 
\frac{1}{2}\left[\delta^{\a\g}\delta^{\b\delta} + \delta^{\a\delta}\delta^{\b\g}\right]\, \delta^{(4)}(z-x)
= s^{\mu\nu\alpha\beta}\, \delta^{(4)}(z-x)\, .
\eea
The variations of the Christoffel symbols are tensors themselves and their expression is
\bea \label{Ch3deltaChristoffel}
\delta \Gamma^\alpha_{\beta\gamma}
&=&
\frac{1}{2}\,g^{\alpha\lambda}\big[ 
- \nabla_{\lambda}(\delta g_{\beta\gamma}) + \nabla_{\gamma}(\delta g_{\beta\lambda}) + \nabla_{\beta}(\delta g_{
\gamma\lambda})
\big]\, ,
\nn\\
\nabla_\rho \delta\Gamma^\alpha_{\beta\gamma}
&=&
\frac{1}{2}\,g^{\alpha\lambda}\big[ - 
\nabla_{\rho}\nabla_{\lambda}(\delta g_{\beta\gamma}) + 
\nabla_{\rho}\nabla_{\gamma} (\delta g_{\beta\lambda}) 
+ \nabla_{\rho}\nabla_{\beta}(\delta g_{\gamma\lambda}) \big]\, .
\eea

\section{Results for Weyl-gauging}

In the Weyl-gauging of the counterterms, we use the following relations 
\bea
{\hat{\Gamma}}^\alpha_{\beta\gamma} 
&=& 
\Gamma^\alpha_{\beta\gamma} + \frac{1}{\Lambda}\, \bigg( {\delta_\beta}^\alpha\, \nabla_\gamma\rho 
+ {\delta_\gamma}^\alpha\, \nabla_\beta\rho - g_{\beta\gamma}\, \nabla^\alpha\rho \bigg) \, ,
\nn \\
\hat {R^\mu}_{\nu\lambda\sigma}
&=& 
{R^\mu}_{\nu\lambda\sigma}
+ g_{\nu\lambda}\,   \bigg( \frac{\nabla_{\sigma}\pd^\mu\rho}{\Lambda}  + \frac{\pd^\mu\rho\, \pd_\sigma\rho}{\Lambda^2} \bigg)
- g_{\nu\sigma}\, \bigg( \frac{\nabla_{\lambda}\pd^\mu\rho}{\Lambda}    + \frac{\pd^\mu\rho\, \pd_\lambda\rho}{\Lambda^2}   \bigg)
\nn \\
&& +\, 
{\delta^\mu}_\sigma\,\bigg( \frac{\nabla_{\lambda}\pd_\nu\rho}{\Lambda}   + \frac{\pd_\nu\rho\,\pd_\lambda\rho}{\Lambda^2} \bigg) -
{\delta^\mu}_\lambda\,  \bigg( \frac{\nabla_{\sigma}\pd_\nu\rho}{\Lambda} + \frac{\pd_\nu\rho\,\pd_\sigma\rho}{\Lambda^2} \bigg) +
\bigg( {\delta^\mu}_\lambda\, g_{\nu\sigma} - {\delta^\mu}_\sigma\, g_{\nu\lambda} \bigg)\,
\frac{(\pd\rho)^2}{\Lambda^2} \, , 
\nn \\
\hat R_{\mu\nu}
&=& 
R_{\mu\nu} - g_{\mu\nu}\, \bigg( \frac{\Box\rho}{\Lambda} 
- (d-2)\,\frac{(\pd\rho)^2}{\Lambda^2}\bigg) 
- (d-2)\, \bigg( \frac{\nabla_\mu \pd_\nu\rho}{\Lambda} + \frac{\pd_\mu\rho\,\pd_\nu\rho}{\Lambda^2} \bigg)\, ,
\nn \\
%
\hat R
&\equiv&
\hat g^{\mu\nu}\, \hat R_{\mu\nu} =
e^{\frac{2\,\rho}{\Lambda}}\bigg[ R - 2\, (d-1)\, \frac{\Box \rho}{\Lambda} 
+ (d-1)\,(d-2)\, \frac{(\pd\rho)^2}{\Lambda^2} \bigg]\, .
\label{Ch3GaugeRiemann}
\eea
When specialized to the case of Weyl transformations, for which $ \delta_{W} g_{\mu\nu} = 2 \sigma g_{\mu\nu}$,
the variations of the Christoffel symbols and their covariant derivatives (\ref{Ch3deltaChristoffel})
are
\bea \label{Ch3deltaWeylChristoffel}
\delta_{W} \Gamma^\alpha_{\beta\gamma}
&=&
- g_{\beta\gamma}\, \pd^\alpha \sigma + {\delta_\beta}^\alpha\, \pd_\gamma \sigma + {\delta_\gamma}^\alpha\, \pd_\beta\sigma
\quad \Rightarrow \quad \delta_{W} \Gamma^\alpha_{\alpha\gamma} = d\, \pd_\gamma \sigma \, , \nn \\
\nabla_\lambda \delta_{W} \Gamma^\alpha_{\beta\gamma}
&=&
- g_{\beta\gamma}\, \nabla_\lambda\pd^\alpha\sigma + {\delta_\beta}^\alpha\, \nabla_\lambda\pd_\gamma\sigma
+ {\delta_\gamma}^\alpha\, \nabla_\lambda\pd_\beta\sigma
\quad \Rightarrow \quad \delta_{W} \nabla_{\lambda}\Gamma^\alpha_{\alpha\gamma} = d\, \nabla_\lambda \pd_\gamma\sigma \, .
\eea
Using the Palatini identity
\beq \label{Ch3Palatini}
\delta {R^\alpha}_{\beta\gamma\lambda} =
\nabla_{\lambda}(\delta\Gamma^\alpha_{\beta\gamma}) - \nabla_{\gamma}(\delta\Gamma^\alpha_{\beta\lambda})
\quad \Rightarrow \quad
\delta R_{\alpha\beta} =
\nabla_{\alpha} (\delta\Gamma^\lambda_{\beta\lambda}) - \nabla_{\lambda}(\delta\Gamma^\lambda_{\alpha\beta})
\eeq
we obtain the expressions for the Weyl variations of the Riemann and Ricci tensors
\bea \label{Ch3deltaWeylRiemann}
\delta_{W} {R^\alpha}_{\beta\gamma\delta} 
&=& 
  g_{\beta\delta}\, \nabla_{\gamma}\pd^\alpha\sigma 
- g_{\beta\gamma}\, \nabla_{\delta}\pd^\alpha \sigma
+ {\delta_{\gamma}}^{\alpha}\, \nabla_{delta}\pd_\beta\sigma 
- {\delta_{delta}}^{\alpha}\, \nabla_{\gamma}\pd_\beta\sigma \, , \nn \\
\delta_{W} R_{\alpha\beta} 
&=& 
g_{\alpha\beta}\, \Box \sigma + (d-2)\, \nabla_{\alpha}\pd_\beta\sigma \, .
\eea
It is also customary to introduce the Cotton tensor,
\beq \label{Ch4Cotton}
\tilde{C}_{\alpha\beta\gamma} = \nabla_{\gamma} K_{\alpha\beta} - \nabla_{\beta} K_{\alpha\gamma}\, , 
\quad \text{where}\, \quad
K_{\alpha\beta} = \frac {1}{d-2}\, \bigg( R_{\alpha\beta} - \frac{g_{\alpha\beta}}{2\,(d-1)}\, R \bigg)\, .
\eeq
Using (\ref{Ch3deltaWeylChristoffel})-(\ref{Ch3deltaWeylRiemann}) 
one can easily show that the variation of the Cotton tensor is simply given by
\beq \label{Ch4deltaWeylCotton}
\delta_{W} \tilde{C}_{\alpha\beta\gamma} = - \pd_\lambda\sigma\, {C^\lambda}_{\alpha\beta\gamma}\, ,
\eeq
which is expressed in terms of the Weyl tensor.

\section{Weyl invariants and Euler densities in $2$, $4$ and $6$ dimensions}\label{Geometrical}

It is well known that the object one has to deal with in order to construct Weyl-invariant objects for general dimensions $d$
is the traceless part of the Riemann tensor, called the Weyl tensor, defined by
\beq \label{WeyldDef}
C_{\alpha\beta\gamma\delta} = R_{\alpha\beta\gamma\delta} -
\frac{1}{d-2}( g_{\alpha\gamma} \, R_{\delta\beta} + g_{\alpha\delta} \, R_{\gamma\beta}
- g_{\beta\gamma} \, R_{\delta\alpha} - g_{\beta\delta} \, R_{\gamma\alpha} ) +
\frac{1}{(d-1)(d-2)} \, ( g_{\alpha\gamma} \, g_{\delta\beta} + g_{\alpha\delta} \, g_{\gamma\beta}) R\, .
\eeq
This object enjoys the same symmetry properties of the Riemann tensor, i.e.
\beq
C_{\alpha\beta\gamma\delta} = - C_{\b\a\g\d}=  C_{\b\a\d\g} = C_{\d\g\b\a} \, ,
\eeq
and, moreover, is traceless with respect to any couple of its indices. 
It is invariant under Weyl scalings of the metric
\beq
\delta_W {C^{\a}}_{\b\g\d} = 0 \, .
\eeq
It is apparent from (\ref{WeyldDef}), it is not defined for $d=2$,so that no Weyl-invariant object depending only on the metric
can be built out of it. 

In $4$ dimensions, the only quantity which is Weyl invariant, when multiplied by $\sqrt{-g}$, 
is the Weyl tensor squared, which is given by
\beq \label{Ch1Weyl}
F \equiv
C^{\alpha\beta\gamma\delta}C_{\alpha\beta\gamma\delta} =
R^{\alpha\beta\gamma\delta}R_{\alpha\beta\gamma\delta} - 2\, R^{\alpha\beta}R_{\alpha\beta} + \frac{1}{3}R^2 \, .
\eeq
Its d-dimensional version, which we call $F_d$, is instead
\beq \label{Ch1Weyld}
F_d \equiv
C^{\alpha\beta\gamma\delta}C_{\alpha\beta\gamma\delta} =
R^{\alpha\beta\gamma\delta}R_{\alpha\beta\gamma\delta} - \frac{4}{d-2}R^{\alpha\beta}R_{\alpha\beta}
+ \frac{2}{(d-2)(d-1)}R^2
\eeq

There are three dimension-6 scalars that are Weyl invariant when multiplied by $\sqrt{g}$.
Their choice is not unique at all, as one can always take linear combinations of them.
In particular, the literature is full of different choices for the third one, which involves differential operators.
In this work we adopt the definition of $I_3$ given, for general dimensions, in \cite{Parker:1987}
\bea \label{Ch4WeylInvd}
I^d_1 &\equiv& 
C_{\mu\nu\alpha\beta}\, C^{\mu\rho\sigma\beta}\, {{C^\nu}_{\rho\sigma}}^\alpha  =
\frac{d^2+d-4}{(d-1)^2\,(d-2)^3}\, K_1 - \frac{3\,(d^2+d-4)}{(d-1)\,(d-2)^3}\, K_2 + \frac{3}{2\,d^2-6\,d+4}\, K_3 
\nn \\
&& \hspace{35mm}
+\, \frac{6\,d-8}{(d-2)^3}\, K_4 - \frac{3\,d}{(d-2)^2}\, K_5 - \frac{3}{d-2}\, K_6 + K_8 
\nn \\
I^d_2 &\equiv& 
C_{\mu\nu\alpha\beta}\, C^{\alpha\beta\rho\sigma}\, {C^{\mu\nu}}_{\rho\sigma} =
\frac{8\,(2\,d-3)}{(d-1)^2\,(d-2)^3}\, K_1 + \frac{72-48\,d}{(d-1)\,(d-2)^3}\, K_2 + \frac{6}{d^2-3\,d+2}\, K3
\nn \\
&& \hspace{35mm}
+\, \frac{16\,(d-1)}{(d-2)^2}\, K_4 - \frac{24}{(d-2)^2}\, K_5 - \frac{12}{d-2}\, K_6 + K_7
\nn \\
I^d_3
&\equiv&
\frac{d-10}{d-2}\, \bigg( \nabla^{\alpha}C^{\beta\gamma\rho\sigma}\, \nabla_{\alpha}C_{\beta\gamma\rho\sigma} 
- 4\, (d-2)\, \tilde{C}^{\gamma\rho\sigma}\, \tilde{C}_{\gamma\rho\sigma}\bigg) +
\frac{4}{d-2}\, \bigg(\square + \frac{2}{(d-1)}\, R\bigg)\,C^{\alpha\beta\rho\sigma}\,C_{\alpha\beta\rho\sigma}
\nn \\ 
&=&
\frac{16}{(d^2-3d+2)^2}\, K_{1} - \frac{32}{(d-1)\,(d-2)^2}\, K_{2} + 
\frac{8}{d^2-3\,d+2} K_{3} + \frac{16}{(d-1)\,(d-2)^2}\, K_{9}
\nn \\
&& \hspace{-5mm}
-\, \frac{32}{(d-2)^2}\, K_{10} + \frac{8}{d-2}\, K_{11} + \frac{4\,(d-6)}{(d-1)\,(d-2)^2}\, K_{12} 
+ \frac{88-12\,d}{(d-2)^2}\, K_{13} + K_{14} + \frac{8\,(d-10)}{(d-2)^2}\, K_{15}\, . \nn \\
\eea
We mention that the general problem of constructing all the possible Weyl-invariant coordinate scalars 
depending on the metric tensor was solved in full generality in \cite{Fefferman:2007rka}.

The other quantity with which one can construct an integral which is Weyl invariant (in fact, a constant) for general dimensions
is the Euler density, defined, for a general even number of dimensions $d=2k$, as
\beq \label{EulerdDef}
E_{2k} = \frac{1}{2^k}\, \delta_{\mu_1 a_1\nu_1b_1\dots\mu_k a_k\nu_k b_k}\, 
R^{\mu_1\nu_1\lambda_1 \kappa_1}\dots R^{\mu_k \nu_k a_k b_k}\, .
\eeq
The antisymmetric Kronecker symbol is defined by
\beq
\delta_{\nu_1 a_1 \nu_2 a_2\dots \nu_n a_n } = 
n!\, \sum_{\mathcal P(a_1,\dots,a_n)}(-1)^{T_{\mathcal P}}\,  
g_{\nu_1 \mathcal P(a_1)}\dots g_{\nu_n \mathcal P(a_n)}\, ,
\eeq
where $T_{\mathcal P}$ \`{e} is the number of inversions 
the permutation $\mathcal P$ of the $n$ numbers $a_1, \dots a_n$ is made of.

By applying the general definition \ref{EulerdDef}, we find that in $2$, $4$ and $6$ dimensions respectively, it is given by the expressions
\bea
E_2 &=& R \, , \nn \\
E_4 &\equiv&G =
R^{\alpha\beta\gamma\delta}R_{\alpha\beta\gamma\delta} - 4\,R^{\alpha\beta}R_{\alpha\beta} + R^2\, , \nn \\
E_6 
&=&
\frac{21}{100}\, R^3 - \frac{27}{20}\, R\, R^{\mu\nu}\, R_{\mu\nu}
+ \frac{3}{2}\, {R_\mu}^\nu\,{R_\nu}^\alpha\, {R_\alpha}^\mu
+ 4\, C_{\mu\nu\rho\sigma}\,{C^{\mu\nu}} _{\alpha\beta}\, C^{\rho\sigma\alpha\beta}
\nn \\
&&
-\, 8\, C_{\mu\nu\rho\sigma}\,C^{\mu\alpha\rho\beta}\, {{{C^\nu}_\alpha}^\sigma}_\beta
-6\, R_{\mu\nu}\, C^{\mu\alpha\rho\sigma}\, {C^\nu}_{\alpha\rho\sigma}
+\frac{6}{5}\, R\, C^{\mu\nu\rho\sigma}\, C_{\mu\nu\rho\sigma}
- 3\, R^{\mu\nu}\, R^{\rho\sigma}\, C_{\mu\rho\sigma\nu} \, .
\label{VariousEuler}
\eea
Of course, in the last expression $C$ is the Weyl tensor in $6$ dimensions.

\section{Functional derivation of Riemann-quadratic integrals} \label{Ch1FunctionalIntegral}

In this appendix we explicitly show how to evaluate the functional variation of the most general integral
which is quadratic in the Riemann tensor and its contractions, which we call $\mathcal{I}(a,b,c)$,
\beq
\mathcal{I}(a,b,c)
\equiv
\int\,d^d x\,\sqrt{g}\, K\,
\equiv
\int\,d^d x\,\sqrt{g}\,
\big(a\,R^{\alpha\beta\gamma\delta}R_{\alpha\beta\gamma\delta} + b\,R^{\alpha\beta}R_{\alpha\beta} + c\, R^2 \big)\, ,
\eeq
needed to compute the counterterms found in section \ref{Ch1Renormalization}. \\
The same techniques apply to integrals which are higher order in the Riemann tensor,
as the one listed below for the case of $6$ dimensions.

Our index conventions for the Riemann and Ricci tensors are those in (\ref{Ch1Tensors}).
We have
\bea
\delta (R^{\alpha\beta\gamma\delta}R_{\alpha\beta\gamma\delta})
&=&
\delta(g_{\alpha\sigma}g^{\beta\eta}g^{\gamma\zeta}g^{\delta\rho}
{R^\alpha}_{\beta\gamma\delta}{R^\sigma}_{\eta\zeta\rho}) \nn\\
&=&
\delta (g_{\alpha\sigma}g^{\beta\eta}g^{\gamma\zeta}g^{\delta\rho})
{R^\alpha}_{\beta\gamma\delta}{R^\sigma}_{\eta\zeta\rho}
+ g_{\alpha\sigma}g^{\beta\eta}g^{\gamma\zeta}g^{\delta\rho}\delta 
({R^\alpha}_{\beta\gamma\delta}{R^\sigma}_{\eta\zeta\rho}) \nn\\
&=&
\delta (g_{\alpha\sigma}g^{\beta\eta}g^{\gamma\zeta}g^{\delta\rho})
{R^\alpha}_{\beta\gamma\delta}{R^\sigma}_{\eta\zeta\rho} 
+  2\,\delta ({R^\alpha}_{\beta\gamma\delta}){R_\alpha}^{\beta\gamma\delta}\, ,
\eea
Using (\ref{Ch1Tricks}) and (\ref{Ch1Tricks2}) and the product rule for derivatives one easily finds out that
the variation can be written at first as
\bea
\delta \mathcal I(a,b,c)
&=& \int\,d^dx\,\sqrt{g}\,\bigg\{ \bigg[ \frac{1}{2}g^{\mu\nu}K - 2a\, R^{\mu\alpha\beta\gamma}{R^\nu}_{\alpha\beta\gamma}
   - 2b\,R^{\mu\alpha}{R^\nu}_\alpha - 2c\,R R^{\mu\nu}\bigg]\delta g_{\mu\nu}\nn\\
&&\hspace{20mm}
+ \, 2a\, {R_\alpha}^{\beta\gamma\delta}\delta {R^\alpha}_{\beta\gamma\delta}
+ 2b\, R^{\alpha\beta}\delta R_{\alpha\beta} + 2c\, R\, g^{\alpha\beta}\, \delta R_{\alpha\beta}\bigg\}\, .
\eea
Now we have to exploit the Palatini identities
\beq \label{Ch1Palatini}
\delta {R^\alpha}_{\beta\gamma\delta}
=
(\delta\Gamma^\alpha_{\beta\gamma})_{;\delta} - (\delta\Gamma^a_{\beta\delta})_{;\gamma} \quad \Rightarrow \quad
\delta R_{\beta\delta}
=
(\delta\Gamma^\lambda_{\beta\lambda})_{;\delta} - (\delta\Gamma^\lambda_{\beta\delta})_{;\lambda}\, ,
\eeq
and the Bianchi identities,
\bea\label{Ch1Bianchi}
R_{\alpha\beta\gamma\delta;\eta} + R_{\alpha\beta\eta\gamma;\delta} + R_{\alpha\beta\delta\eta;\gamma}
&=& 0
\quad \Rightarrow \quad
R_{\beta\delta;\eta} - R_{\beta\eta;\delta} + {R^\gamma}_{\beta\delta\eta;\gamma} =  0 \nn \\
\Rightarrow \quad
R_{;\delta}
&=&
2\,{R^\alpha}_{\delta;\alpha}
\quad \Leftrightarrow \quad
\big( R^{\alpha\beta} - \frac{1}{2}g^{\alpha\beta} R \big)_{;\beta} = 0 \, .
\eea
After an integration by parts and a reshuffling of indices we get
\bea\label{Ch1deltaISecond}
\delta \mathcal I(a,b,c)
&=&
\int\,d^dx\,\sqrt{g}\,\bigg\{\bigg[ \frac{1}{2}g^{\mu\nu}K - 2\big( a\, R^{\mu\alpha\beta\gamma}{R^\nu}_{\alpha\beta\gamma}
                                     + b\,R^{\mu\alpha}{R^\nu}_\alpha + c\,R R^{\mu\nu}\big)\bigg]\delta g_{\mu\nu}\nn\\
&&
+ \,\left[ 4a\,g_{\beta\delta}\,g^{\gamma\eta}\,(\delta\Gamma^\delta_{\alpha\gamma})_{;\eta}
- (4a+2b)\,(\delta \Gamma^\gamma_{\alpha\beta})_{;\gamma}
+ (4c+2b)\, (\delta \Gamma^\lambda_{\alpha\lambda})_{;\beta}
- 4c\, g^{\eta\delta}\,g_{\gamma\alpha}\,(\delta \Gamma^\gamma_{\eta\delta})_{;\beta}\right]\,R^{\alpha\beta}  \bigg\}.\nn\\
\eea
The variations of the Christoffel symbols and of their covariant derivatives
in terms of covariant derivatives of the metric tensors variations are
\bea\label{Ch1deltaChristoffel}
\delta \Gamma^\alpha_{\beta\gamma}
&=&
\frac{1}{2}\,g^{\alpha\delta}\big[-(\delta g_{\beta\gamma})_{;\delta}
+ (\delta g_{\beta\delta})_{;\gamma} +(\delta g_{\gamma\delta})_{;\beta} \big]\, ,\nn\\
(\delta\Gamma^\alpha_{\beta\gamma})_{;\delta}
&=&
\frac{1}{2}\,g^{\alpha\eta}\big[-(\delta g_{\beta\gamma})_{;\eta;\delta} + (\delta g_{\beta\eta})_{;\gamma;\delta}
                                + (\delta g_{\gamma\eta})_{;\beta;\delta} \big]\, .
\eea
Now we use them to rewrite (\ref{Ch1deltaISecond}) as
\bea\label{Ch1Bivio}
\delta \mathcal I(a,b,c)
&=&
\int\,d^dx\,\sqrt{g}\bigg\{\bigg[ \frac{1}{2}g^{\mu\nu}K - 2\big( a\, R^{\mu\alpha\beta\gamma}{R^\nu}_{\alpha\beta\gamma}
                                  + b\,R^{\mu\alpha}{R^\nu}_\alpha + c\,R R^{\mu\nu}\big)\bigg]\delta g_{\mu\nu}\nn\\
&&\hspace{18,5mm}
+ \, \bigg[2a\,\big[-(\delta g_{\alpha\delta})_ {;\beta;\gamma} +(\delta g_{\alpha\beta})_ {;\gamma;\delta}
+ (\delta g_{\beta\delta})_ {;\alpha;\gamma} \big]\nn\\
&& \hspace{19mm}
- \, (2a + b)\,\big[- (\delta g_{\alpha\beta})_ {;\delta;\gamma} +(\delta g_{\alpha\delta})_ {;\beta;\gamma}
+ (\delta g_{\beta\delta})_ {;\alpha;\gamma} \big] + (2c + b)\, (\delta g_{\gamma\delta})_ {;\alpha;\beta}\nn\\
&& \hspace{19mm}
- \, 2c\, \big[- (\delta g_{\gamma\delta})_ {;\alpha;\beta} +(\delta g_{\alpha\delta})_ {;\gamma;\beta}
+ (\delta g_{\alpha\gamma})_ {;\delta;\beta} \big]\bigg]g^{\gamma\delta}\,R^{\alpha\beta} \bigg\}\, .
\eea
The presence of the factor $g^{cd}R^{ab}$ imposes two symmetry constraints on the terms in the last contribution in square brackets. By adding and subtracting  $-(4a+2b)\,(\delta g_{ac})_{;d;b}$ we obtain the expression
\bea
\delta \mathcal I(a,b,c)
&=& \int\,d^dx\,\sqrt{g}\,\bigg\{\bigg[\frac{1}{2}g^{\mu\nu}K
- 2\big( a\, R^{\mu\alpha\beta\gamma}{R^\nu}_{\alpha\beta\gamma}
+ b \, R^{\mu\alpha}{R^\nu}_\alpha + c\,R R^{\mu\nu}\big)\bigg]\delta g_{\mu\nu}\nn\\
&&\hspace{18,5mm}
+ \, \bigg[(4a+2b)\,\big[(\delta g_{\alpha\gamma})_ {;\beta;\delta}
- (\delta g_{\alpha\gamma})_ {;\delta;\beta}\big]+
(4a + b)(\delta g_{\alpha\beta})_ {;\gamma;\delta} + (4c + b)\, (\delta g_{\gamma\delta})_ {;\alpha;\beta}\nn\\
&& \hspace{18,5mm}
- \, (4a+2b+4c)\, (\delta g_{\alpha\gamma})_ {;\delta;\beta}\bigg]g^{\gamma\delta}\,R^{\alpha\beta} \bigg\}\, .
\label{Ch1HalfFirstFunctional}
\eea
The commutation of covariant derivatives allows us to write
\bea
g^{\gamma\delta} \big[ (\delta g_{\alpha\gamma})_{;\beta;\delta}
- (\delta g_{\alpha\gamma})_{;\delta;\beta} \big]R^{\alpha\beta}
&=&
g^{\gamma\delta} \big[ -\delta g_{\alpha\sigma} {R^\sigma}_{\gamma\delta\beta}
- \delta g_{\gamma\sigma} {R^\sigma}_{\alpha\beta\delta} \big]R^{\alpha\beta}\nn\\
&=&
g^{\gamma\delta} \big[-s^{\mu\nu}_{\alpha\sigma}{R^\sigma}_{\gamma\beta\delta}
- s^{\mu\nu}_{c\sigma}{R^\sigma}_{\alpha\beta\delta}\big]R^{\alpha\beta} \,
\delta g_{\mu\nu}\nn\\
&=&
(- R^{\mu\alpha}{R^\nu}_\alpha + R^{\mu\alpha\nu\beta}R_{\alpha\beta}) \delta g_{\mu\nu}\, .
\eea
Inserting this back into (\ref{Ch1HalfFirstFunctional}) we get
\bea
\delta \mathcal I(a,b,c) =
\nonumber\\
&& \hspace{-20mm}
\int\,d^dx\,\sqrt{g}\,\bigg\{\bigg[ \frac{1}{2}g^{\mu\nu}K
- 2a\, R^{\mu\alpha\beta\gamma}{R^\nu}_{\alpha\beta\gamma} + 4a\,R^{\mu\alpha}{R^\nu}_\alpha
-(4a+2b)\, R^{\mu\alpha\nu\beta}R_{\alpha\beta} - 2c \, R R^{\mu\nu}\bigg]\delta g_{\mu\nu}\nn\\
&&
+ \, \bigg[(4a + b)(\delta g_{\alpha\beta})_ {;\gamma;\delta}
+ (4c + b)\, (\delta g_{\gamma\delta})_ {;\alpha;\beta}
- (4a+2b+4c)\, (\delta g_{\alpha\gamma})_ {;\delta;\beta}\bigg]g^{\gamma\delta}\,R^{\alpha\beta}\bigg\}\, .
\nonumber\\
\eea
If the coefficients are $a = c = 1$ and $b=-4$, i.e. if the integrand is the Euler density,
the last three terms are zero. \\
All that is left to do is a double integration by parts for each one of the last three terms,
to factor out $\delta g_{\mu\nu}$.
This is easily performed and the final result can be written as
\bea  \label{Ch1Magic}
\frac{\delta}{\delta g_{\mu\nu}} \mathcal I(a,b,c)
&=&
\frac{\delta}{\delta g_{\mu\nu}} \int\,d^d x\,\sqrt{g\,}
\big( a\,R^{\alpha\beta\gamma\delta}R_{\alpha\beta\gamma\delta} + b\,R^{\alpha\beta}R_{\alpha\beta}
+ c\,R^2 \big)\nn\\
&=&
\sqrt{g}\, \bigg\{\frac{1}{2}g^{\mu\nu}K
- 2a\, R^{\mu\alpha\beta\gamma}{R^\nu}_{\alpha\beta\gamma}
+ 4a\,R^{\mu\alpha}{R^\nu}_\alpha -(4a+2b)\, R^{\mu\alpha\nu\beta}R_{\alpha\beta} - 2c \, R R^{\mu\nu}\nn\\
&& \hspace{8mm}
+ \, (4a + b)\,\Box{R^{\mu\nu}} + (4c + b)\,g^{\mu\nu}{R^{\alpha\beta}}_{;\alpha;\beta}
- (4a+2b+4c){{R^{\nu\beta}}_{;\beta}}^{;\mu}\bigg\}\, .
\eea
%

\section{Functional variations in $6$ dimensions}\label{Ch4Geometrical3}

The results for the trace anomaly in $6$ dimensions, presented in section \ref{Ch4Counterterms},
are obtained by computing the functional variations of the integrals of the $K_i$ in dimensional regularization,
which can be obtained with the same techniques employed in appendix \ref{Ch1FunctionalIntegral}.

A simple counting of the metric tensors needed to contract all the indices for any $K_i$ shows that
\beq
\delta_{W} \int d^dx\, \sqrt{g}\, K_i = \int d^dx\, \sqrt{g}\, \left[- \epsilon\, K_i + D(K_i) \right]\, \sigma
\, , \quad \epsilon = 6-d\, ,
\eeq
where the second term on the right hand side, $D(K_i)$, is a total derivative contribution.
We give the complete list of these terms below. We obtain
\bea \label{Ch4FuncVar}
D(K_1) &=& 12\, (d-1)\,\nabla_\mu \left(R\,\pd^\mu R \right) \nn \\
D(K_2) &=& \nabla_\mu \bigg[ 4\,(d-1)\, R_{\nu\lambda}\, \nabla^\mu R^{\nu\lambda}
           + 2\,\left(d-2\right)\, R^{\mu\nu}\, \pd_\nu R + \left(d+2\right)\, R\,\pd^\mu R \bigg]\nn \\
D(K_3) &=& 4\, \nabla_\mu \bigg[ R\, \pd^\mu R + 2\, R^{\mu\nu}\, \pd_\nu R
           +\left(d-1\right)\, R_{\nu\lambda\kappa\alpha}\, \nabla^\mu R^{\nu\lambda\kappa\alpha} \bigg] \nn \\
D(K_4) &=& 3\, \nabla_\mu \bigg[ \frac{d-2}{2}\,R^{\mu\nu}\, \pd_\nu R 
              +\left(d-2\right) R_{\nu\lambda}\, \nabla^\lambda R^{\mu\nu}
              +2\, R_{\nu\rho}\, \nabla^\mu R^{\nu\rho} \bigg] \nn \\
D(K_5) &=& \nabla_\mu \bigg[ -R\, \pd^\mu R - R^{\mu\nu}\, \pd_\nu R
           +2\, \left(d-1\right)\, R_{\nu\lambda}\, \nabla^\lambda R^{\mu\nu}
           -2\,d\, R_{\nu\lambda}\,\nabla^\mu R^{\nu\lambda} 
           + 2\,\left(d-2\right)\, R^{\mu\lambda\nu\kappa}\, \nabla_\kappa R_{\nu\lambda} \bigg] \nn \\
D(K_6) &=& \nabla_\mu \bigg[ 2\, R^{\mu\nu}\, \pd_\nu R + 4\, R_{\nu\lambda}\, \nabla^\mu R^{\nu\lambda}
           +\frac{d+2}{2}\, R_{\nu\lambda\kappa\alpha}\, \nabla^\mu R^{\nu\lambda\kappa\alpha}
           - 2\,d\, R^{\mu\lambda\nu\kappa}\, \nabla_\kappa R_{\nu\lambda} \bigg] \nn \\
D(K_7) &=& 6\, \nabla_\mu \bigg[ R_{\nu\lambda\kappa\alpha}\, \nabla^\mu R^{\nu\lambda\kappa\alpha} 
                            +4\, R^{\nu\lambda\kappa\mu}\, \nabla_\lambda R_{\nu\kappa}      \bigg] \nn \\
D(K_8) &=& 3\, \nabla_\mu \bigg[ \frac{1}{2}\,R_{\nu\lambda\kappa\alpha}\, \nabla^\mu R^{\nu\lambda\kappa\alpha}
                                 + 2\, R^{\nu\lambda}\, \left( \nabla_\lambda R_{\mu\nu} - \nabla_\mu R_{\nu\lambda} \right)  \bigg] \nn \\
D(K_9) &=& \nabla_\mu \bigg[ 4\,\left(d-1\right)\, \pd^\mu \square R - \left(d-2\right)\, R\,\pd^\mu R \bigg] \nn \\
D(K_{10}) &=& \nabla_\mu \bigg[ 2\,\pd^\mu \square R + 2\,\left(d-2\right)\,\nabla_\nu \square R^{\mu\nu}
                               +2\, R^{\mu\nu}\, \pd_\nu R - 4\, R_{\nu\lambda}\, \nabla^\lambda R^{\mu\nu}
                               -\left(d-2\right)\, R_{\nu\lambda}\,\nabla^\mu R^{\nu\lambda}  \bigg]  \nn \\
D(K_{11}) &=& \nabla_\mu \bigg[ 8\, \nabla_\nu \square R^{\mu\nu} 
             - \left(d+2\right)\, R_{\nu\lambda\kappa\alpha}\, \nabla^{\mu}R^{\nu\lambda\kappa\alpha}
             - 16\, R^{\mu\lambda\nu\kappa}\,\nabla_\kappa R_{\nu\lambda} \bigg] \nn \\
D(K_{12}) &=& 4\, \nabla_\mu \bigg[ R\,\pd^\mu R - \left(d-1\right)\, \pd^\mu \square R \bigg]  \nn \\
D(K_{13}) &=& 2\,\ \nabla_\mu \bigg[ - \pd^\mu\square R  - \left(d-2\right)\, \nabla_\nu \square R^{\mu\nu} 
             - R^{\mu\nu}\, \pd_\nu R + 2\, R_{\nu\lambda}\, \nabla^{\lambda} R^{\mu\nu} 
             + 2\, R_{\nu\lambda}\, \nabla^\mu R^{\nu\lambda}   \bigg] \nn \\
D(K_{14}) &=& 8\, \nabla_\mu \bigg[ - \nabla_\nu \square R^{\mu\nu} 
              +R_{\nu\lambda\kappa\alpha}\, \nabla^{\mu} R^{\nu\lambda\kappa\alpha} 
              + 2\, R^{\mu\lambda\nu\kappa}\,\nabla_\kappa R_{\nu\lambda}  \bigg] \nn \\
D(K_{15}) &=& \nabla_\mu \bigg[ - \pd^\mu\square R - 2\,\left(d-2\right)\, \nabla_\nu \square R^{\mu\nu}
              -3\, R^{\mu\nu}\, \pd_\nu R + 6\, R_{\nu\lambda}\, \nabla^\lambda R^{\mu\nu} \nn \\
&& \hspace{10mm}
+ 2 \, R_{\nu\lambda}\, \nabla^\mu R^{\nu\lambda} 
- 2 \, \left(d-2\right)\, R^{\mu\lambda\nu\kappa}\,\nabla_\kappa R_{\nu\lambda} \, \bigg]\, .
\eea
%

\section{ List of functional derivatives}
\label{Ch1Functionals}

We list here the contributions to the trace anomalies for three point function
coming from the elementary quadratic objects. They are given by
\bea
\big[R_{\lambda\mu\kappa\nu}R^{\lambda\mu\kappa\nu}\big]^{\alpha\beta\rho\sigma}(p,q)
&=& p \cdot q\, \big[p \cdot q \big(\delta^{\alpha\rho}\delta^{\beta\sigma} + \delta^{\alpha\sigma}\delta^{\beta\rho}\big)
- \big(\delta^{\alpha\rho}p^{\sigma}q^{\beta} + \delta^{\alpha\sigma}p^{\rho}q^{\beta}\nn\\
&+& \delta^{\beta\rho}p^{\sigma}q^{\alpha} + \delta^{\beta\sigma}p^{\rho}q^{\alpha}\big)\big]
+ 2 \, p^{\rho}p^{\sigma}q^{\alpha}q^{\beta}\, , \nn
\eea
\bea
\big[R_{\mu\nu}R^{\mu\nu}\big]^{\alpha\beta\rho\sigma}(p,q)
&=& \frac{1}{4} p \cdot q \big(\delta^{\alpha\rho}p^\beta q^\sigma + \delta^{\alpha\sigma}p^\beta q^\rho
+ \delta^{\beta\rho}p^\alpha q^\sigma + \delta^{\beta\sigma}p^\alpha q^\rho \big)\nn\\
&+&
\frac{1}{2}(p \cdot q)^2 \delta^{\alpha\beta}\delta^{\rho\sigma}
+ \frac{1}{4}p^2 q^2\big(\delta^{\alpha\rho}\delta^{\beta\sigma} + \delta^{\alpha\sigma}\delta^{\beta\rho}\big)\nn\\
&-&
\bigg[\frac{1}{4} p^2\big(q^\alpha q^\rho \delta^{\beta\sigma}+ q^\alpha q^\sigma \delta^{\beta\rho}
+ q^\beta q^\rho \delta^{\alpha\sigma}+ q^\beta q^\sigma \delta^{\alpha\rho}\big)\nn\\
&+&
\frac{1}{2}\delta^{\alpha\beta}\big( p\cdot q\,(p^\rho q^\sigma + p^\sigma q^\rho) - q^2 p^\rho p^\sigma \big)
+ (\alpha,\beta,p)\leftrightarrow(\rho,\sigma,q)\bigg]\, ,\nn
\eea
\bea
\big[R^2\big]^{\alpha\beta\rho\sigma}(p,q)
&=& 2\big(p^\alpha p^\beta q^\rho q^\sigma - p^2 q^\rho q^\sigma \delta^{\alpha\beta}
- q^2 p^\alpha p^\beta \delta^{\rho\sigma} + p^2 \, q^2 \delta^{\alpha\beta}\delta^{\rho\sigma} \big)\, , \nn
\eea
\bea\label{Ch1QuadraticFunctionals}
\big[\Box\,R\big]^{\alpha\beta\rho\sigma}(p,q)
&=&(p+q)^2\bigg\{-\frac{1}{2}\delta^{\alpha\beta}\big(p^\rho q^\sigma + p^\sigma q^\rho + 2\,p^\rho p^\sigma\big)
- \frac{1}{2}\delta^{\rho\sigma}\big(q^\alpha p^\beta + q^\beta q^\alpha + 2\,q^\alpha q^\beta \big)\nn\\
&+&
\frac{1}{2} p \cdot q \, \delta^{\alpha\beta}\delta^{\rho\sigma}
+\frac{1}{4}\big(p^\rho q^\beta \delta^{\alpha\sigma} + p^\rho q^\alpha \delta^{\beta\sigma}
+ p^\sigma q^\beta \delta^{\alpha\rho} + p^\sigma q^\alpha \delta^{\beta\rho}\big)\nn\\
&+&
\frac{1}{2}\bigg[\big(q^\rho p^\beta \delta^{\alpha\sigma} + q^\rho p^\alpha \delta^{\beta\sigma}
+ q^\sigma p^\beta \delta^{\alpha\rho} + q^\sigma p^\alpha \delta^{\beta\rho}\big)\nn\\
&+&
\delta^{\alpha\rho}\big(p^\beta p^\sigma + q^\beta q^\sigma \big)
 + \delta^{\alpha\sigma}\big(p^\beta p^\rho + q^\beta q^\rho \big)
 +\delta^{\beta\rho}\big(p^\alpha p^\sigma + q^\alpha q^\sigma \big)\nn\\
&+&
\delta^{\beta\sigma}\big(p^\alpha p^\rho + q^\alpha q^\rho \big)
-\big(\delta^{\alpha\sigma}\delta^{\beta\rho} + \delta^{\alpha\rho}\delta^{\beta\sigma}\big)
\big(p^2 +q^2 + \frac{3}{2}p \cdot q\big)\bigg]\bigg\}\nn\\
&+&
\frac{1}{2}\big(p^2 \delta^{\alpha\beta} - p^\alpha p^\beta\big)\big(p \cdot q \, \delta^{\rho\sigma}
- (p^\rho q^\sigma + p^\sigma q^\rho) - 2\,p^\rho p^\sigma \big)\nn\\
&+& \frac{1}{2}\big(q^2 \delta^{\rho\sigma} - q^\sigma q^\rho\big)\big(p \cdot q \, \delta^{\alpha\beta}
- (p^\alpha q^\beta + p^\beta q^\alpha) - 2\,q^\alpha q^\beta \big)\, ,
\eea

\section{ Graviton interaction vertices}\label{Ch1Vertices}

Here we list the vertices which are needed for the momentum space computation of the $TTT$ correlator.
Notice that they are computed differentiating the first and second functional derivatives of the action, 
because this allows to keep multi-graviton correlators symmetric (see \ref{Ch13PF}).
\begin{itemize}
\item{graviton - scalar - scalar vertex}
\\ \\
\begin{minipage}{95pt}
\includegraphics[scale=0.7]{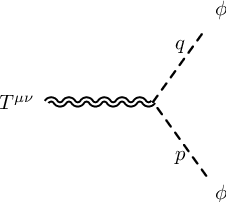}
\end{minipage}
\begin{minipage}{70pt}
\bea
&\equiv& V_{\mathcal{S}\phi\phi}^{\mu\nu}(p,q)=
\frac{1}{2}\,p_\alpha \, q_\beta \, C^{\mu\nu\alpha\beta}
+\chi \bigg( \delta^{\mu\nu} \left( p + q \right)^2 - \left( p^\mu + q^\mu \right)\,\left( p^\nu + q^\nu \right) \bigg)\, ,
 \nn
\eea
\end{minipage}
\item{graviton - fermion - fermion vertex}
\\ \\
\begin{minipage}{95pt}
\includegraphics[scale=0.7]{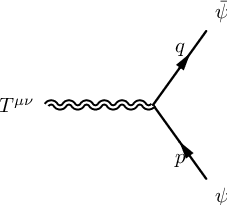}
\end{minipage}
\begin{minipage}{70pt}
\bea
&\equiv& V_{\mathcal{S} \bar{\psi}\psi}^{\mu\nu}(p,q) =
\frac{1}{8} \, A^{\mu\nu\alpha\lambda}\, \gamma_\alpha \,\left(p_\lambda - q_\lambda \right)\, , \nn
\eea
\end{minipage}
\item{graviton - photon - photon vertex}
\\ \\
\begin{minipage}{95pt}
\includegraphics[scale=0.7]{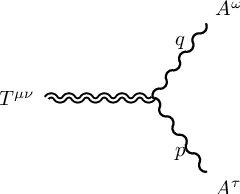}
\end{minipage}
\begin{minipage}{70pt}
\bea
&\equiv& V_{\mathcal{S}AA}^{\mu\nu\tau\omega}(p,q) 
= \frac{1}{2}\,\bigg[ p \cdot q\, C^{\mu\nu\tau\omega} + D^{\mu\nu\tau\omega}(p,q) 
+ \frac{1}{\xi}E^{\mu\nu\tau\omega}(p,q) \bigg] \, , \nn
\eea
\end{minipage}
\item{graviton - ghost - ghost vertex}
\\ \\
\begin{minipage}{95pt}
\includegraphics[scale=0.7]{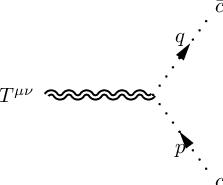}
\end{minipage}
\begin{minipage}{70pt}
\bea
&\equiv& V_{\mathcal{S}\bar{c}c}^{\mu\nu}(p,q) = - V_{\mathcal{S}\phi\phi}^{\mu\nu}(p,q)\bigg|_{\chi=0} \, , \nn
\eea
\end{minipage}
\item{graviton - graviton - scalar - scalar vertex}
\\ \\
\begin{minipage}{95pt}
\includegraphics[scale=0.7]{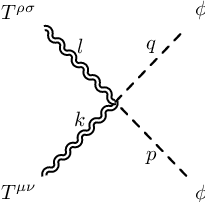}
\end{minipage}
\begin{minipage}{70pt}
\bea
\equiv V_{\mathcal{S}\mathcal{S}\phi\phi}^{\mu\nu\rho\sigma}(p,q,l) &=&
\frac{1}{2}\, p\cdot q\, s^{\mu\nu\rho\sigma} - \frac{1}{4}\, G^{\mu\nu\rho\sigma}(p,q)
+ \frac{1}{4}\, \delta^{\rho\sigma}\, p_\alpha\, q_\beta\,  C^{\mu\nu\alpha\beta} \nn \\
&& \hspace{-10mm}
+\, \chi \, \bigg\{\bigg[  \bigg(\delta^{\mu\lambda}\, \delta^{\alpha\kappa}\, \delta^{\nu\beta}
+ \delta^{\mu\alpha}\,\delta^{\nu\kappa}\,\delta^{\beta\lambda} - \delta^{\mu\kappa}\,\delta^{\nu\lambda}\,\delta^{\alpha\beta}
- \delta^{\mu\nu}\,\delta^{\alpha\lambda}\,\delta^{\beta\kappa}\bigg)\, s^{\rho\sigma}_{\lambda\kappa} \nn \\
&& \hspace{4mm}
+\, \frac{1}{2} \, \delta^{\rho\sigma} \, \bigg(\delta^{\mu\alpha}\,\delta^{\nu\beta} - 
\delta^{\mu\nu}\, \delta^{\alpha\beta}\bigg)\bigg]
\left(p_\alpha \, q_\beta + p_\beta \, q_\alpha + p_\alpha \, p_\beta + q_\alpha \, q_\beta \right)\nn \\
&&
+\, \bigg[\bigg(\delta^{\mu\nu}\,\delta^{\alpha\beta} - \delta^{\mu\alpha}\,\delta^{\nu\beta} \bigg)
\big[\Gamma^\lambda_{\alpha\beta}\big]^{\rho\sigma}(l)\, i\, \left( \, p_\lambda +  q_\lambda\right) \nn \\
&&
+\, \bigg(\delta^{\mu\alpha}\,\delta^{\nu\beta} - \frac{1}{2}\,\delta^{\mu\nu}\,\delta^{\alpha\beta}\bigg)\,
  \big[R_{\alpha\beta}\big]^{\rho\sigma}(l) \,  \bigg]\bigg\}\, , \nn
\eea
\end{minipage}
\item{graviton - graviton - fermion - fermion vertex}
\\ \\
\begin{minipage}{95pt}
\includegraphics[scale=0.7]{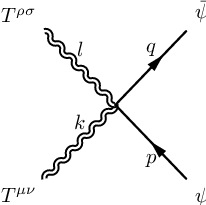}
\end{minipage}
\begin{minipage}{70pt}
\bea
\equiv 
V_{\mathcal{S}\mathcal{S} \bar{\psi}\psi}^{\mu\nu\rho\sigma}(p,q) 
&=&
\frac{1}{16} \bigg[- 4 \, s^{\mu\nu\rho\sigma} - 2 \, \delta^{\mu\nu}\,s^{\alpha\lambda\rho\sigma}
+ 2 \, \delta^{\alpha\mu}\,s^{\nu\lambda\rho\sigma} + 2\, \delta^{\alpha\nu}\,s^{\mu\lambda\rho\sigma} \nn \\
&& \hspace{10mm}
+\, \delta^{\mu\lambda}\,s^{\alpha\nu\rho\sigma} + \delta^{\nu\lambda}\,s^{\alpha\mu\rho\sigma} 
+ \delta^{\rho\sigma}\, A^{\mu\nu\alpha\lambda} \bigg]\,\gamma_{\alpha}\,(p_\lambda - q_\lambda) \, , \nn
\eea
\end{minipage}
\item{graviton - graviton - photon - photon vertex}
\\ \\
\begin{minipage}{95pt}
\includegraphics[scale=0.7]{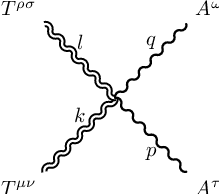}
\end{minipage}
\begin{minipage}{70pt}
\bea
\equiv V_{\mathcal{S}\mathcal{S} AA}^{\mu\nu\rho\sigma\tau\omega}(p,q,l) 
&=&
\frac{1}{2} \, \bigg\{
\bigg[ B^{\alpha\mu\rho\sigma\beta\lambda\gamma\nu} + \frac{1}{4}\, B^{\mu\nu\rho\sigma\alpha\lambda\gamma\beta} \bigg]\,
{F_{\alpha\beta\gamma\lambda}}^{\tau\omega} (p,q) \nn \\
&&
+\, \frac{1}{\xi}\bigg( H^{\mu\nu\rho\sigma\tau\omega}(p,q,l) + I^{\mu\nu\rho\sigma\tau\omega}(p,q,l) \bigg) \bigg\} \nn \\
&&
+\, \frac{1}{4}\, \delta^{\rho\sigma}\, \bigg[ p \cdot q \,  C^{\mu\nu\tau\omega}
 + D^{\mu\nu\tau\omega}(p,q) + \frac{1}{\xi}E^{\mu\nu\tau\omega}(p,q) \bigg] \, , \nn
\eea
\end{minipage}
\item{graviton - graviton - ghost - ghost vertex}
\\ \\
\begin{minipage}{95pt}
\includegraphics[scale=0.7]{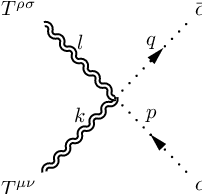}
\end{minipage}
\begin{minipage}{70pt}
\bea
\equiv  V_{\mathcal{S}\mathcal{S}\bar{c}c}^{\mu\nu\rho\sigma}(p,q,l) =  
  - V_{\mathcal{S}\mathcal{S}\phi\phi}^{\mu\nu\rho\sigma}(p,q,l)\bigg|_{\chi=0}\, .
	\nn
\eea
\end{minipage}
\end{itemize}
%

\vspace{0.5cm}%

We have simplified the notation by introducing, for convenience, the tensor structures
\bea
A^{\mu\nu\alpha\lambda} = 2\,\delta^{\mu\nu}\,\delta^{\alpha\lambda}
- \delta^{\alpha\mu}\,\delta^{\lambda\nu} - \delta^{\alpha\nu}\,\delta^{\lambda\mu}
\nonumber
\eea
\bea
B^{\alpha\mu\rho\sigma\beta\lambda\gamma\nu}
&=&
  s^{\alpha\mu\rho\sigma}    \, \delta^{\beta\lambda} \, \delta^{\gamma\nu}
+ s^{\beta\lambda\rho\sigma} \, \delta^{\alpha\mu}    \, \delta^{\gamma\nu}
+ s^{\gamma\nu\rho\sigma}    \, \delta^{\alpha\mu}    \, \delta^{\beta\lambda}
\nonumber
\eea
\bea
C^{\mu\nu\alpha\beta}
&=&
\delta^{\mu\alpha} \delta^{\nu\beta} + \delta^{\mu\beta} \delta^{\nu\alpha} - \delta^{\mu\nu} \delta^{\alpha\beta}\, ,
\nonumber
\eea
\bea
D^{\mu\nu\rho\sigma} (p,q)
&=&
\delta^{\mu\nu} p^\sigma q^\rho + \delta^{\rho\sigma} \big(p^\mu q^\nu + p^\nu q^\mu\big)
- \delta^{\mu\sigma} p^\nu q^\rho - \delta^{\mu\rho} p^\sigma q^\nu
- \delta^{\nu\sigma}p^\mu q^\rho - \delta^{\nu\rho} p^\sigma q^\mu
\nonumber
\eea
\bea
E^{\mu\nu\rho\sigma} (p, q)
&=&
\delta^{\mu\nu}\, \big[p^\rho p^\sigma + q^\rho q^\sigma + p^\rho q^\sigma \big]
- \big[\delta^{\nu\sigma} p^\mu p^\r + \delta^{\nu\rho}
q^\mu q^\sigma + \delta^{\mu\sigma} p^\nu p^\rho + \delta^{\mu\rho} q^\nu q^\sigma \big]\, ,
\nonumber
\eea
\bea
F^{\mu\nu\rho\sigma\tau\omega} (p, q)
&=&
-  \delta^{\tau\rho}    \delta^{\omega\mu} p^{\sigma} q^{\nu} + \delta^{\tau\rho} \delta^{\omega\nu} p^{\sigma} q^{\mu}
+  \delta^{\tau\sigma}  \delta^{\omega\mu} p^{\rho} q^{\nu} - \delta^{\tau\sigma} \delta^{\omega\nu} p^{\rho} q^{\mu}
+ (\tau,p) \leftrightarrow (\omega,q)
\nonumber
\eea
\bea
G^{\mu\nu\rho\sigma}(p,q)
&=&
  \delta^{\mu\sigma} \big[p^{\rho} q^{\nu}  +  q^{\rho}   p^{\nu}\big]
+ \delta^{\nu\sigma}\big[p^{\rho}  q^{\mu}  +  q^{\rho}   p^{\mu}\big]+
  \delta^{\mu\rho}\big[p^{\sigma}  q^{\nu}  +  q^{\sigma} p^{\nu}\big]
+ \delta^{\nu\rho}\big[p^{\sigma}  q^{\mu}  +  q^{\sigma} p^{\mu}\big]
\nonumber\\
&-&
\delta^{\mu\nu}\big[p^{\rho} q^{\sigma} + q^{\rho} p^{\sigma} \big]
\nonumber
\eea
\bea
H^{\mu\nu\rho\sigma\tau\omega}(p,q,l)
&=&
\bigg[
\bigg(s^{\mu\omega\rho\sigma}\, \delta^{\nu\lambda} + s^{\nu\lambda\rho\sigma}\, \delta^{\mu\omega} \bigg)\, p_\lambda\, p^\tau
+  \delta^{\mu\omega}\,\bigg(s^{\lambda\tau\rho\sigma}\, l^\nu + s^{\lambda\tau\rho\sigma}\, p^\nu\bigg)\, p_\lambda
\nonumber \\
&+&
 \frac{1}{2} \delta^{\mu\omega}\,
\left(p + l\right)^\nu\, \bigg(- l^{\tau}\, \delta^{\rho\sigma} + 2\, l_\lambda\, s^{\tau\lambda\rho\sigma}\bigg)
+ (\mu \leftrightarrow \nu)\bigg] + (\tau,p) \leftrightarrow (\omega,q)
\nonumber
\eea
\bea
I^{\mu\nu\rho\sigma\tau\omega}(p,q,l)
&=&
\delta^{\mu\nu}\,\bigg\{
\frac{1}{2}\, \delta^{\rho\sigma}\, l^\tau\, \left(p + q + l\right)^\omega
-  s^{\lambda\tau\rho\sigma}   \, \bigg[q^\omega\, p_\lambda + l_\lambda\, \left( p + q + l \right)^\omega \bigg]
-  s^{\lambda\omega\rho\sigma} \, \bigg[p^\tau  \, p_\lambda + q_\lambda\, \left( q + l     \right)^\tau   \bigg]
\bigg\}
\nonumber \\
&-&
s^{\mu \nu \rho \sigma}\, \bigg(p^\omega\, p^\tau + q^\omega\, p^\tau\bigg) + (\tau,p) \leftrightarrow (\omega,q).
\nonumber
\eea
We have performed all our computations in the Feynman gauge ($\xi=1$)
The euclidean propagators of the fields in this case are
\bea
\left\langle\phi \, \phi \right\rangle (p)
&=&
\frac{1}{p^2}
\nonumber\\
\left\langle \bar{\psi}\, \psi \right\rangle (p)
&=&
\frac{p\cdot\gamma}{p^2} \, .
\nonumber \\
\left\langle A^\mu \, A^\nu \right\rangle (p)
&=&
\frac{\delta^{\mu\nu}}{p^2} \, ,
\nonumber\\
\left\langle \bar{c}\, c \right\rangle (p)
&=&
- \frac{1}{p^2} \, .
\eea
%

\section{ Comments on the inverse mapping} \label{Ch1InverseTTT}

In this appendix we offer some calculational details in the derivation of the expression of the $TTT$
correlator in position space via the inverse mapping procedure. The remarks apply as well to any other correlator.\\
For example,  eq.  (\ref{Ch1ScalarTriangle}) refers to the contribution coming from the triangle diagram shown in fig.
\ref{Ch1Fig.diagramsTTT}. We assign the loop momentum $l$ to flow from the upper external point
($x_3$) to the lower one ($x_2$) on the right, the other two flows being determined by momentum conservation.
We denote the third external point as $x_1$.
For the scalar case, for instance, the complete $1$-loop triangle diagram is
\beq\label{Ch1TriangleLoop}
8\, \int\, \frac{d^dl}{(2\pi)^d} \,
\frac{V^{\mu\nu}_{\mathcal{S}\phi\phi}(l-q,-l-p)V^{\rho\sigma}_{\mathcal{S}\phi\phi}(l,-l+q)
V^{\alpha\beta}_{\mathcal{S}\phi\phi}(l+p,-l)}{l^2\,(l-q)^2\,(l+p)^2}
\eeq
The vertices are defined in appendix \ref{Ch1Vertices}.
The first argument in each vertex denotes the momentum of the incoming particle, the second argument
is the momentum of the outgoing one.
A typical term appearing in the loop integral is then
\beq\label{Ch1TypicalInverse3}
I\equiv\int\, \frac{d^dl}{(2\pi)^d} \,
\frac{l^{\mu}\,l^{\nu}\,(l+p)^{\rho}\,(l+p)^{\sigma}(l-q)^{\alpha}\,(l-q)^{\beta}}{l^2\,(l-q)^2\,(l+p)^2}\, .
\eeq
From (\ref{Ch1fund}) the propagators in configuration space are
\beq
\frac{1}{l^2\,(l-q)^2\,(l+p)^2} =
C(1)^3 \, \int\,d^d x_{12}\,d^d x_{23}\,d^d x_{31}\,
\frac{e^{i\,[l\cdot x_{23}+(l-q)\cdot x_{12}+(l+p)\cdot x_{31}]}}{(x^2_{12})^{d/2-1}\,(x^2_{23})^{d/2-1}\,(x^2_{31})^{d/2-1}}
\, ,
\eeq
where $C\alpha)$ has been defined in (\ref{Ch1fund}).
It is straightforward to see that (\ref{Ch1TypicalInverse3}) is given by
\bea\label{Ch1TypicalInverse3Coord}
\int\, \frac{d^dl}{(2\pi)^d} \,
\frac{l^{\mu}\,l^{\nu}\,(l+p)^{\rho}\,(l+p)^{\sigma}(l-q)^{\alpha}\,(l-q)^{\beta}}{l^2\,(l-q)^2\,(l+p)^2}
&=&
\nonumber\\
&& \hspace {-70mm}
C(1)^3\,\int\, \frac{d^dl}{(2\pi)^d}\,d^d x_{12}\,d^d x_{23}\,d^d x_{31}\,
\frac{(-i)^6\,\partial^\mu_{23}\,\partial^\nu_{23}\,\partial^\rho_{31}\,\partial^\sigma_{31}\,
\partial^\alpha_{12}\,\partial^\beta_{12} \,e^{i\,[l\cdot x_{23}+(l-q)\cdot x_{12}+(l+p)\cdot x_{31}]}}
{(x^2_{12})^{d/2-1}\,(x^2_{23})^{d/2-1}\,(x^2_{31})^{d/2-1}} \nn \\
\, .
\eea
We can now integrate by parts moving the derivatives onto the propagators, getting
\bea
I&=&
C(1)^3\,\int\, \frac{d^dl}{(2\pi)^d}\,d^d x_{12}\,d^d x_{23}\,d^d x_{31}\,
e^{i\,[l\cdot x_{23}+(l-q)\cdot x_{12}+(l+p)\cdot x_{31}]}
\nonumber\\
&\times&
i^6\,\partial^\mu_{23}\,\partial^\nu_{23}\,\partial^\rho_{31}\,\partial^\sigma_{31}\,
\partial^\alpha_{12}\,\partial^\beta_{12}\,\frac{1}{(x^2_{12})^{d/2-1}\,(x^2_{23})^{d/2-1}\,(x^2_{31})^{d/2-1}} \, .
\eea
The second line is immediately identified with the coordinate space Green's function.\\
This can be done for each term of (\ref{Ch1TriangleLoop}), justifying the rule quoted
in section \ref{Ch1InverseMappingTTT}, that we have used for all the inverse mappings of the work.
According to this the correlators in coordinate space can be obtained replacing the momenta in the vertices with ``$i$" times
the respective derivative which then act directly on the propagators after a partial integration.

The same arguments could be applied to the bubbles. Nevertheless, we have seen in \ref{Ch1InverseMappingTTT}
that derivatives of delta functions appear in the scalar case.
These are generated by the dependence of the $V^{\mu\nu\rho\sigma}_{\mathcal{S}\mathcal{S}\phi\phi}(p,q,l)$ 
from the momentum $l$ of the graviton bringing the pair of indices $\rho\sigma$ (see appendix \ref{Ch1Vertices}).
They are due to coupling of the scalar with derivatives of the metric through the Ricci scalar $R$ in
the improvement term (see eq.  (\ref{Ch1scalarAction})) and state that the graviton feels the metric gradient.
We discuss this below, showing how to inverse-map the third bubble in fig. (\ref{Ch1Fig.diagramsTTT}), getting
(\ref{Ch1ScalarKBubble}). \\
This bubble can be seen as the ($x_2\rightarrow x_3$) limit of the triangle
and its diagrammatic momentum-space expression at $1$-loop is
\beq\label{Ch1KBubbleLoop}
\int\, \frac{d^dl}{(2\pi)^d} \,
\frac{V^{\mu\nu}_{\mathcal{S}\phi\phi}(l-q,-l-p)
V^{\alpha\beta\rho\sigma}_{\mathcal{S}\mathcal{S}\phi\phi}(l+p,-l+q,-q)}{(l-q)^2\,(l+p)^2}\, .
\eeq
As the two propagators are expressed by
\beq
\frac{1}{(l+q)^2\,(l+p)^2} =
C(1)^2 \, \int\,d^d x_{12}\,d^d x_{31}\,
\frac{e^{i\,[(l-q)\cdot x_{12}+(l+p)\cdot x_{31}]}}{(x^2_{12})^{d/2-1}\,(x^2_{31})^{d/2-1}} \, ,
\eeq
the dependence of the second vertex on $p$ cannot be ascribed to neither of them.\\
Two typical terms encountered in (\ref{Ch1KBubbleLoop}) are
\bea\label{Ch1TypicalInverse2}
\int\, \frac{d^dl}{(2\pi)^d} \,
\frac{(l+p)^{\rho}\,(l+p)^{\sigma}(l-q)^{\alpha}\,(l-q)^{\beta}}{(l-q)^2\,(l+p)^2} \, ,
\nonumber\\
\int\, \frac{d^dl}{(2\pi)^d} \,
\frac{(l+p)^{\rho}\,(l+p)^{\sigma}(l-q)^{\alpha}\,p^{\beta}}{(l-q)^2\,(l+p)^2} \, .
\eea
The first one is treated at once restricting the procedure used for the three point function to the case of two
propagators.\\
For the second one, the following relation is immediately checked:
\bea\label{Ch1TypicalInverse2Coord}
\int\, \frac{d^dl}{(2\pi)^d} \,
\frac{(l+p)^{\rho}\,(l+p)^{\sigma}(l-q)^{\alpha}\,p^{\beta}}{(l-q)^2\,(l+p)^2}
&=&
\nonumber\\
&& \hspace {-70mm}
C(1)^2\,\int\, \frac{d^dl}{(2\pi)^d}\,d^d x_{12}\,d^d x_{23}\,d^d x_{31}\,
\delta^{(d)}(x_{23})\,\frac{(-i)^4\,\partial^\rho_{31}\,\partial^\sigma_{31}\,\partial^\alpha_{12}
\,(\partial_{31}-\partial_{23})^\beta\,
e^{i\,[l\cdot x_{23}+(l-q)\cdot x_{12}+(l+p)\cdot x_{31}]}}
{(x^2_{12})^{d/2-1}\,(x^2_{31})^{d/2-1}} \, . \nn \\
\eea
Notice that an integration by parts brings in a derivative on the delta functions giving
\bea
C(1)^2\,\int\, \frac{d^dl}{(2\pi)^d}\,d^d x_{12}\,d^d x_{23}\,d^d x_{31}\,
e^{i\,[l\cdot x_{23}+(l-q)\cdot x_{12}+(l+p)\cdot x_{31}]}\,
(i)^4\,\partial^\rho_{31}\,\partial^\sigma_{31}\,\partial^\alpha_{12}\,(\partial_{31}-\partial_{23})^\beta\,
\frac{\delta^{d}(x_{23})}{(x^2_{12})^{d/2-1}\,(x^2_{31})^{d/2-1}} \, . \nn \\
\eea
This approach has been followed in all the derivations of the expressions given in (\ref{Ch1InverseMappingTTT}).\\
The integration on $l$ brings about a $\delta^{(d)}(x_{12}+x_{23}+x_{31})$, so that it is natural to chose
the parameterization
\beq
x_{12} = x_1 - x_2\, , \quad x_{23} = x_2 - x_3\, , \quad x_{31} = x_3 - x_1\, .
\eeq
A more inviolved example is the 4-particle vertex. For instance the 
$V_{\mathcal{S}\mathcal{S}\phi\phi}(i\, \partial_{31},- i\, \partial_{12},i \, (\partial_{12}-\partial_{23}))$
is obtained from $V_{\mathcal{S}\mathcal{S}\phi\phi}(p,q,l)$ with the functional replacements 
\beq
p\to \hat{p}=i\, \partial_{31},\qquad  q \to \hat{q}=- i\, \partial_{12}\qquad l\to \hat{l}=i \, (\partial_{12}-\partial_{23})
\eeq
giving
\bea
V_{\mathcal{S}\mathcal{S}\phi\phi}^{\mu\nu\rho\sigma}
(i\, \partial_{31},- i\, \partial_{12},i \, (\partial_{12}-\partial_{23})) =
&&
\nonumber\\
&& \hspace{-60mm} \frac{1}{2}\, i\, \partial_{31}\cdot (-i)\,\partial_{12} s^{\mu\nu\rho\sigma} 
  - \frac{1}{4}\, G^{\mu\nu\rho\sigma}(i\,\partial_{31},-i\,\partial_{12})
+ \frac{1}{4}\, \delta^{\rho\sigma}\, i\,\partial_{31\,\alpha} \, (-i)\,\partial_{12\,\beta}\, C^{\mu\nu\alpha\beta}
\nonumber\\
&&\hspace{-60mm}
+ \, \chi \, \bigg\{\bigg[  \bigg(\delta^{\mu\lambda}\,\delta^{\alpha\kappa}\,\delta^{\nu\beta}
+ \delta^{\mu\alpha}\,\delta^{\nu\kappa}\,\delta^{\beta\lambda}
- \delta^{\mu\kappa}\,\delta^{\nu\lambda}\,\delta^{\alpha\beta}
- \delta^{\mu\nu}\,\delta^{\alpha\lambda}\,\delta^{\beta\kappa}\bigg)
s^{\rho\sigma}_{\lambda\kappa}
\nonumber \\
&& \hspace{-60mm}
+\, \frac{1}{2} \, \delta^{\rho\sigma} \, \bigg(\delta^{\mu\alpha}\,\delta^{\nu\beta} -
\delta^{\mu\nu}\,\delta^{\alpha\beta}\bigg)\bigg]
\left(i\,\partial_{31\,\alpha} \, (-i)\partial_{12\,\beta} + i\,\partial_{31\,\beta} \,(-i)\partial_{12\,\alpha} 
+ i\,\partial_{31\,\alpha} \, i\,\partial_{31\,\beta} + (-i)\partial_{12\,\alpha} \,(-i)\partial_{12\,\beta} \right)
\nonumber \\
&&\hspace{-60mm}
- \, \bigg[\bigg(\delta^{\mu\alpha}\,\delta^{\nu\beta} - \delta^{\mu\nu}\,\delta^{\alpha\beta}\bigg)
\bigg(\big[\Gamma^\lambda_{\alpha\beta}\big]^{\rho\sigma}\left(i\,(\partial_{12}-\partial_{23})\right)\bigg)\, 
(-i) \, \left( i \,  \partial_{31\,\lambda} + \, (-i)\, \partial_{12\,\lambda}\right)
\nonumber	\\
&& \hspace{-60mm} 
+ \, \frac{1}{2} \, \bigg(\delta^{\mu\alpha}\,\delta^{\nu\beta} - \frac{1}{2}\,\delta^{\mu\nu}\,\delta^{\alpha\beta}\bigg)\,
 \bigg( \big[R_{\alpha\beta}\big]^{\rho\sigma}\left(i\,(\partial_{12}-\partial_{23})\right)\bigg) \,  \bigg]\bigg\} \, .
\eea

\section{ Regularizations and distributional identities} \label{Ch1Distributional}

We add few more comments and examples which illustrate the regularization that we have applied in the
computation of the various correlators.

The computation of the logarithmic integrals requires some care due to the distributional nature of some of these formulas.
As an example we consider the integrals
\bea
H_1 = \int d^d l\, e^{i l\cdot x}\,  \frac{\mu^{2 \omega}}{[l^2]^{1 + \omega}} \qquad
H_2 = \int d^d l\, e^{i l\cdot x}\,  \frac{\mu^{2 \omega}}{[l^2]^{ \omega}} \qquad
H_3 = \int d^d l\, e^{i l\cdot x}\,  \log\left(\frac{l^2}{\mu^2}\right)
\eea
We can relate them in the form
\beq
H_3 = -\frac{\partial}{\partial \omega}H_2 \bigg|_{\omega=0}
    = \Box \left(\frac{\partial}{\partial \omega}H_1 \bigg|_{\omega=0}\right)
\eeq
In the two cases we get, using (\ref{Ch1fund})
\beq
-\frac{\partial}{\partial \omega}H_2 \bigg|_{\omega=0} = -  \frac{(4\,\pi)^{d/2} \, \Gamma(d/2)}{(x^2)^{d/2}}
\eeq
and
\beq
\frac{\partial}{\partial \omega}H_1 \bigg|_{\omega=0}
= \frac{2^{d-2} \pi^{d/2} \Gamma(d/2-1)}{[x^2]^{d/2-1}}
  \bigg( \log(x^2 \mu^2) + \gamma - \log 4  - \psi \left( \frac{d-2}{2}\right) \bigg)
\label{Ch1intermed}
\eeq
By redefining the regularization scale $\mu$ with eq.  (\ref{Ch1massscale}) we clearly obtain from (\ref{Ch1intermed})
\beq
\int d^d l \frac{\log( l^2/\mu^2) e^{i l\cdot x}}{l^2} =
2^{d-2}\pi^{d/2} \Gamma(d/2-1) \frac{\log x^2 \bar{\mu}^2}{[x^2]^{d/2-1}}
\eeq
and
\beq \label{Ch1First}
H_3 = \Box \left(\frac{\pd}{\pd\omega} H_1\bigg|_{\omega=0} \right) =
2^{d - 2} \pi^{d/2} \Gamma(d/2-1) \Box \left( \frac{\log x^2 \bar{\mu}^2}{[x^2]^{d/2-1}}\right)
\eeq
The use of $H_2$ instead gives
\beq
H_3 = -\frac{\partial}{\partial \omega}H_2 \bigg|_{\omega=0}
    = - \frac{2^d \pi^{d/2} \Gamma(d/2)}{[x^2]^{d/2}}
\eeq
Notice that this second relation coincides with (\ref{Ch1First}) away from the point $x=0$,
but differs from it right on the singularity, since
\beq
\Box \frac{\log x^2\mu^2}{[x^2]^{d/2-1}}
= - 2 \, (d-2) \bigg( \frac{\pi^{d/2}}{\Gamma(d/2)} \, \log( x^2\mu^2) \,  \delta^d (x)
  + \frac{1}{[x^2]^{d/2}} \bigg)
\eeq
For this reason we take (\ref{Ch1First}) as the regularized expression of $H_3$, in agreement
with the standard approach of differential regularization. \\

The direct method discussed in the second part of the work, though very general and applicable to any correlator, introduces in 
momentum space some logarithmic integrals which are more difficult to handle. They take the role of the ordinary master integrals of 
perturbation theory. The scalar integrals needed for the tensor reduction of the logarithmic contributions in the text are defined
in (\ref{Ch1StandInt}).
After a shift of the momentum in the argument of the logarithm,  a standard tensor reduction gives
\bea
IL_\mu(0,p_1,p_2)
&=&
{CL}_1(p_1,p_2)\, p_{1\,\mu} + {CL}_2(p_1,p_2)\, p_{2\,\mu} \, ,
\nonumber\\
{CL}_1(p_1,p_2)
&=&
\frac{({p_1}^2 - p_1\cdot p_2){p_2}^2\, IL(0,p_1,p_2) + ({p_2}^2 - p_1\cdot p_2)\, {IL^\mu}_\mu(0,p_1,p_2)}
{2\, (p_1\cdot p_2)^2 - {p_1}^2\, {p_2}^2}\, ,
\nonumber \\
{CL}_2(p_1,p_2)
&=&
\frac{({p_2}^2 - p_1\cdot p_2){p_1}^2\, IL(0,p_1,p_2) + ({p_1}^2 - p_1\cdot p_2)\, {IL^\mu}_\mu(0,p_1,p_2)}
{2\, (p_1\cdot p_2)^2 - {p_1}^2\, {p_2}^2}\, .
\eea
To complete the computation of the $VVV$ correlator we need the explicit form of the logarithmic integrals
in terms of ordinary logarithmic and polylogarithmic functions.
We define
\beq
\mathcal I \equiv \int d^d l \frac{\log \left(l^2 / \mu^2\right)}{(l+p_1)^2 (l-p_2)^2} =
- \frac{\partial}{\partial \lambda}
\int d^dl\,  \frac{\mu^{2 \lambda}}{(l^2)^\lambda\,(l+p_1)^2\,(l-p_2)^2}_{\lambda = 0} \, .
\eeq
The logarithmic integral is identified from the term of O($\lambda$) in the series expansion of the previous expression.
Because the coefficient in front of the parametric integral starts at this order, we just need to know the zeroth
order expansion of the integrand, which we separate into two terms. The first one is integrable
\bea
I_1 = \int_0^1 d t  \frac{t^{-\epsilon}  (y t)^{1-\epsilon-\lambda}}{A(t)^{1-\epsilon}} =
\int_0^1 d t  \frac{t^{-\epsilon}  (y t)^{1-\epsilon}}{A(t)^{1-\epsilon}}  + O(\lambda) \equiv  I_1^{(0)}  + O(\lambda)\, ,
\eea
while the last term has a singularity in $t=0$ which must be factored out and re-expressed in terms of a pole in $\lambda$
\bea
I_2 &=& - \int_0^1 d t  \frac{t^{-\epsilon}  (x / t)^{1-\epsilon-\lambda}}{A(t)^{1-\epsilon}} =
- \frac{ x^{1-\epsilon-\lambda}}{\lambda} \int_0^1 d t \frac{1}{A(t)^{1-\epsilon}} \frac{d }{dt} t^\lambda \nn \\
&=&
- \frac{ x^{1-\epsilon-\lambda}}{\lambda} \bigg[ 1 - (\epsilon -1) \int_0^1 d t \frac{t^\lambda}{A(t)^{1-\epsilon}}
\left(\frac{1}{t-t_1} + \frac{1}{t-t_2}\right)\bigg] \nn \\
&=&
\frac{x^{1-\epsilon} }{\lambda} \bigg\{ -1 +  (\epsilon -1) \int_0^1 d t \frac{1}{A(t)^{1-\epsilon}}
\left(\frac{1}{t-t_1} + \frac{1}{t-t_2}\right)\bigg] \bigg\} \nn \\
&+&
x^{1-\epsilon} \bigg[ \log x + (\epsilon -1) \int_0^1 d t \frac{\log \left(t/x\right)}{A(t)^{1-\epsilon}}
\left(\frac{1}{t-t_1} + \frac{1}{t-t_2}\right)\bigg] + O(\lambda)  \equiv \frac{1}{\lambda} I_2^{(-1)} +  I_2^{(0)}
+ O(\lambda)\, , \nonumber \\
\eea
where $t_1$ and $t_2$ are the two roots of $A(t) = y t^2 + (1-x-y)t + x$.
We are now able to write down the full $\lambda$-expansion of $J(1,1,\lambda)$ and to extract the logarithmic
integral $\mathcal I$
\bea
\mathcal I = -  \frac{\pi^{2-\epsilon} i^{1+2\epsilon}}{(p_3^2)^{\epsilon}} \frac{ \Gamma(1-\epsilon) \Gamma(2-\epsilon)
\Gamma(\epsilon )}{ \Gamma(2-2\epsilon)} \frac{1}{\epsilon-1} \bigg\{ I_1^{(0)} + I_2^{(0)} \bigg\} \,.
\eea
The previous expression can be expanded in $d=4-2\epsilon$ dimensions in which it manifests a $1/\epsilon$
pole of ultraviolet origin %
\bea
\mathcal I =
\frac{\pi^{2-\epsilon} i^{1+2\epsilon}}{(p_3^2)^{\epsilon}} \left( -\frac{1}{\epsilon}
+ \gamma \right)\bigg[ A(x,y) + \epsilon \, B(x,y) \bigg]  + O(\epsilon) \,,
\eea
where $A(x,y)$ and $B(x,y)$ are defined from the $\epsilon$-expansion of the two integrals $I_1^{(0)}$ and $I_2^{(0)}$ as
\bea
A(x,y) &=&  x \log x + \int_0^1  \frac{dt}{A(t)} \bigg[ y t - x \log \left( t/x \right)
\left(\frac{1}{t-t_1} + \frac{1}{t-t_2}\right)\bigg] \,, \\
B(x,y) &=& - x \log^2 x + \int_0^1  \frac{dt}{A(t)}  \bigg[ y t \left( \log\left(t-t_1 \right)
+ \log\left(t-t_2 \right) - 2 \log t \right) \nn \\
&-& x \log\left(t/x \right) \left(\frac{1}{t-t_1} + \frac{1}{t-t_2}\right)
\left( \log\left(t-t_1 \right) + \log\left(t-t_2 \right) -  \log \left( x/y\right) -1\right) \bigg].
\eea

We introduce here a systematic short-hand notation to denote the momentum-space integrals.
We define
\bea \label{Ch1StandInt}
I_{\mu_1,\dots,\mu_n}(p)
&=&
\int d^d l \, \frac{l_{\mu_1}\, \dots \, l_{\mu_n}}{l^2\,(l+p)^2}\, ,
\nonumber\\
J_{\mu_1,\dots,\mu_n}(p_1,p_2)
&=&
\int d^d l \, \frac{l_{\mu_1}\, \dots \, l_{\mu_n}}{l^2\,(l+p_1)^2(l+p_2)^2}\, ,
\nonumber \\
IL_{\mu_{1}\dots\mu_{n}}(p_1,p_2,p_3)
&=&
\int d^dl \, \frac { l_{\mu_1}\, \dots l_{\mu_n} \, \log \left((l+p_1)^2/\mu^2\right)}{(l+p_2)^2(l+p_3)^2} \, ,
\nonumber\\
ILL_{\mu_{1}\dots\mu_{n}}(p_1,p_2,p_3,p_4)
&=&
\int d^d l \frac{l_{\mu_1}\, \dots l_{\mu_n} \,
\log \left((l+p_1)^2/\mu^2\right)\log \, \left((l+p_2)^2/\mu^2\right)}{(l+p_3)^2(l+p_4)^2}\, .
\eea
For correlators which are finite, the double logarithmic contributions will appear in combinations
that can be re-expressed in terms of ordinary Feynman integrals.

\section{ The Wess-Zumino action in $4$ dimensions by the Noether method}\label{Ch3WessZumino}

The dilaton effective action that we have obtained by the Weyl-gauging of the counterterms in chapter \ref{Recursive}
can be recovered also through an iterative technique, that we briefly review.

In this second approach one begins by requiring that the variation of the dilaton effective action under the Weyl transformations
be equal to the anomaly
\bea
\delta_W \Gamma_{WZ}[g,\tau]
&=&
\int d^4x\, \sqrt{g}\, \sigma\, \bigg[ \beta_{a}\, \left( F - \frac{2}{3}\, \Box R \right) + \beta_b\, G \bigg]\, .
\label{Ch3VarWZ}
\eea
It is natural to start with the ansatz
\beq \label{Ch3ansatz1}
\Gamma^{(1)}_{WZ}[g,\tau] = 
\int d^4 x\, \sqrt{g}\, \frac{\tau}{\Lambda} \, \bigg[ \beta_{a}\, \left( F - \frac{2}{3}\, \Box R \right) + \beta_b\, G \bigg] \, .
\eeq
As $\sqrt{g}\, F$ is Weyl invariant, the variation of $\tau$ saturates the $F$-contribution in (\ref{Ch3VarWZ}).
But $\sqrt{g}\, G$ and $\sqrt{g}\, \Box R$ are not conformally invariant. Their variations under Weyl scalings
introduce additional terms that must be taken into account. The general strategy is to compute the infinitesimal variation
of these terms and to add terms which are quadratic in the derivatives of the dilaton and that cancel this extra contributions. 
But these, when Weyl-transformed, will generate additional terms which must be compensated in turn.
The iteration stops at the fourth order in the dilaton field.
We go through all the computation of the WZ action in some detail.
The piece of (\ref{Ch3ansatz1}) whose first Weyl variation is most easily worked out is the contribution $\sqrt{g}\,\tau\, \Box R$. 
We integrate it by parts twice and find that
\bea
\delta_W \int d^4x\, \sqrt{g}\, \frac{\tau}{\Lambda}\, \Box R
&=& 
\delta_W \int d^4x\, \sqrt{g}\,g^{\mu\nu}\,g^{\rho\sigma}\, R_{\mu\nu}\, \frac{1}{\Lambda}\, \nabla_\sigma\pd_\rho\tau\, , 
\nn \\
&=&
\int d^4x\, \sqrt{g} \bigg( \Box\sigma\, R - \frac{1}{\Lambda}\, g^{\rho\sigma}\,\delta_W \Gamma^\lambda_{\rho\sigma}\, 
\pd_\lambda\tau\, R + \frac{1}{\Lambda}\,\Box \tau\, g^{\mu\nu}\, \delta_W R_{\mu\nu}  \bigg)\, .
\eea
Using (\ref{Ch3deltaWeylChristoffel}) and (\ref{Ch3deltaWeylRiemann}) this turns into
\beq \label{Ch3FirstWeylBoxR}
\delta_W \int d^4x\, \sqrt{g}\, \frac{\tau}{\Lambda}\, \Box R = 
\int d^4 x\, \sqrt{g}\, \bigg( \sigma\, \Box R + \frac{2}{\Lambda}\, R\, \pd_\lambda\tau\, \pd^\lambda\sigma 
+ \frac{6}{\Lambda}\, \Box\tau\Box\sigma \bigg) \, .
\eeq

Now we perform an infinitesimal Weyl variation of the contribution $\sqrt{g}\,\tau\,G$ and we find
\bea
\delta_{W} \int d^4x\, \sqrt{g}\, \frac{\tau}{\Lambda}\, G
&=& 
\int d^4 x\, \sqrt{g}\, \bigg\{ \sigma\, G  +
\frac{2}{\Lambda}\,\tau \bigg[ {R_{\alpha}}^{\beta\gamma\delta}\delta_W {R^{\alpha}}_{\beta\gamma\delta}
- \left(4\,R^{\alpha\beta} - g^{\alpha\beta}R \right) \delta_W R_{\alpha\beta} \bigg] \bigg\}\, .
\eea
After using the algebraic symmetries of the Riemann and Ricci tensors and relabeling the indices we obtain
\bea
\delta_{W} \int d^4x\, \sqrt{g}\, \frac{\tau}{\Lambda} \, G
&=& 
\int d^4 x\, \sqrt{g}\, \bigg\{ 
\sigma\, G + \frac{\tau}{\Lambda}\, \bigg[ 4\,R^{\lambda\gamma\beta\alpha}
- 4\,\bigg( g^{\alpha\gamma}R^{\beta\lambda} + g^{\beta\lambda} R^{\alpha\gamma} 
- 2\,g^{\lambda\gamma}\,R^{\alpha\beta} \bigg) \nn \\
&& \hspace{35mm} 
+\, 2\, \bigg( g^{\alpha\gamma}g^{\beta\lambda} - g^{\alpha\beta}g^{\gamma\lambda} \bigg)\, R \bigg]\,
\nabla_\lambda\nabla_\beta(\delta_W g_{\alpha\gamma}) 
\bigg\}\, . \nn \\
\eea
Now we set $\delta_W g_{\alpha\gamma} = 2\, \sigma g_{\alpha\gamma}$ and after a double integration by parts we obtain
\beq \label{Ch3FirstWeylG}
\delta_{W} \int d^4x\, \sqrt{g}\, \frac{\tau}{\Lambda} \, G = 
\int d^4 x\, \sqrt{g}\, \bigg\{ \sigma\, G
+ \frac{8}{\Lambda}\,\bigg[ \bigg( R^{\alpha\beta}- \frac{1}{2}\, g^{\alpha\beta}\,R \bigg)\, \pd_\alpha\sigma\, \pd_\beta\tau
\bigg] \bigg\} \, .
\eeq
Combined together, eqs.  (\ref{Ch3FirstWeylBoxR}) and (\ref{Ch3FirstWeylG}) give the Weyl variation of the first ansatz (\ref{Ch3ansatz1})
\bea \label{Ch3Var1}
\delta_{W} \Gamma^{(1)}_{WZ}[g,\tau] 
&=&
\int d^4 x\, \sqrt{g}\, \bigg\{
\sigma\, \bigg[\beta_a\, \left(F - \frac{2}{3}\, \Box R  \right) + \beta_b\, G \bigg]
\nn \\
&& \hspace{14mm}
+ \frac{1}{\Lambda}\, \bigg[ \beta_a\, 
\left( \frac{4}{3}\,R\, \pd^\lambda\tau\, \pd_\lambda\sigma + 4\, \Box \tau\, \Box \sigma \right)
+\, 8\, \beta_b\, \pd_\alpha\sigma\, \pd_\beta\tau\, \left( R^{\alpha\beta} - \frac{g^{\alpha\beta}}{2}\, R  \right) \bigg]
\bigg\}\, .
\eea
In order to cancel second line in the integrand in (\ref{Ch3Var1}) we correct with the second ansatz
\bea \label{Ch3ansatz2}
\Gamma^{(2)}_{WZ}[g,\tau] = \Gamma^{(1)}_{WZ}[g,\tau]
+ \frac{1}{\Lambda^2}\,\int d^4 x\, \sqrt{g}\, \bigg\{
\beta_a\, \bigg(\frac{2}{3}\,R\, \left(\pd\tau\right)^2 + 2\, \left(\Box\tau\right)^2 \bigg)
- 4\, \beta_b\, \bigg( R^{\alpha\beta} - \frac{g^{\alpha\beta}}{2}\, R \bigg)\, \pd_\alpha\tau\, \pd_\beta\tau \bigg\}\, . \nn \\
\eea
We then find that the variation of this second ansatz is given by
\bea \label{Ch3Var2}
\delta_{W} \Gamma^{(2)}_{WZ}[g,\tau] 
&=&
\int d^4 x\, \sqrt{g}\, \bigg\{
\sigma\, \bigg[\beta_a\, \left(F - \frac{2}{3}\, \Box R  \right) + \beta_b\, G \bigg] 
+ \frac{1}{\Lambda^2}\, \bigg[ \beta_a\, \left( 8\, \Box \tau\, \pd^\lambda\tau\, \pd_\lambda\sigma
+ 4\, \left(\pd\tau\right)^2\, \Box \sigma \right) 
\nn \\
&& \hspace{16mm}
+\, 8\, \beta_b\, \left( \left(\pd\tau\right)^2 \,\Box \sigma 
- \pd_\alpha\tau\,\pd_\beta\tau\, \nabla^\beta\pd^\alpha\sigma \right) 
\bigg]
\bigg\}\, .
\eea
It is then necessary to compensate for terms which are cubic in the dilaton. 
The structure of the spurious contributions in (\ref{Ch3Var2}) suggests the third ansatz
\bea \label{Ch3ansatz3}
\Gamma^{(3)}_{WZ}[g,\tau] = \Gamma^{(2)}_{WZ}[g,\tau]
- \frac{4}{\Lambda^3}\,\int d^4 x\, \sqrt{g}\,
\left( \beta_a + \beta_b \right)\, \left(\pd\tau\right)^2\,\Box \tau \, .
\eea
The variation of (\ref{Ch3ansatz3}) is
\bea \label{Ch3Var3}
\delta_{W} \Gamma^{(3)}_{WZ}[g,\tau] 
&=&
\int d^4 x\, \sqrt{g}\, \bigg\{ 
\sigma\, \bigg[\beta_a\, \left(F - \frac{2}{3}\, \Box R  \right) + \beta_b\, G \bigg]
+ \frac{4}{\Lambda^2}\, \beta_b\, \left( 2\, \nabla_\beta\pd_\alpha\tau\, \pd^\beta\tau\, \pd^\alpha\sigma
+ \left(\pd\tau\right)^2\, \Box \sigma \right) \nn \\
&& \hspace{22mm}
-\, \frac{8}{\Lambda^3}\, \left(\beta_a + \beta_b\right)\, \pd^\alpha\tau\,\pd_\alpha\tau\, \pd^\beta\tau\, \pd_\beta\sigma
\bigg\} \nn \\
&=&
\int d^4 x\, \sqrt{g}\, \bigg\{
\sigma\, \bigg[\beta_a\, \left(F - \frac{2}{3}\, \Box R  \right) + \beta_b\, G \bigg]
-\, \frac{8}{\Lambda^3}\, \left(\beta_a + \beta_b\right)\,
\pd^\alpha\tau\,\pd_\alpha\tau\, \pd^\beta\tau\,\pd_\beta\sigma \bigg\}\, ,
\eea
which finally allows to infer the structure of the complete WZ action 
\bea \label{Ch3ansatz4}
\Gamma_{WZ}[g,\tau] = \Gamma^{(3)}_{WZ}[g,\tau] 
+ \frac{2}{\Lambda^4}\, \int d^4 x\, \sqrt{g}\, \left( \beta_a + \beta_b \right)\,
\left(\left(\pd\tau\right)^2\right)^2 \, .
\eea
eq.  (\ref{Ch3ansatz4}) coincides with (\ref{Ch3Effective4d}) and it is easy to check that no more terms are needed to ensure
that (\ref{Ch3VarWZ}) holds.
%

\section{ The case $d=2$ as a direct check of the recursive formulae}\label{Ch42D}

Here we discuss how to cross-check the relations (\ref{Ch4DilIntStructure6}), 
elaborating on the $2$-dimensional case, which is the easiest one of course. \\

It is clear that the expressions of the dilaton vertices $\mathcal{I}_n$ given in (\ref{Ch4DilIntStructure6}) do not depend on the working 
dimensions. Therefore we take $d=2$ and check the agreement between the correlators that result from (\ref{Ch4DilIntStructure6}) 
and those found by a direct functional differentiation of the anomaly via the hierarchy (\ref{Ch3hier}). This provides a strong
check of the correctness of (\ref{Ch4DilIntStructure6}). In fact, the equation of the trace anomaly in 2 dimensions takes the form
\beq \label{Ch4TraceAnomaly2D}
\langle T \rangle = - \frac{c}{24\,\pi}\, R\, ,
\eeq
where $c = n_s + n_f$, with $n_s$ and $n_f$ being the numbers of free scalar and fermion fields respectively. 
It is derived from the counterterm
\beq\label{Ch4Counterterm2D}
\Gamma_{\textrm{Ct}}[g] = - \frac{\mu^{\epsilon}}{\epsilon}\, \frac{c}{24\,\pi} \,\int d^d x\, \sqrt{g}\, R\, , 
\quad \epsilon = d - 2\, .
\eeq
The Weyl-gauging procedure  for the integral of the scalar curvature gives
\beq \label{Ch4GaugeCT2D}
-\frac{\mu^{\epsilon}}{\epsilon}\, \int d^dx \, \sqrt{\hat g}\, \hat R =
- \frac{\mu^{\epsilon}}{\epsilon}\, \int d^dx \, \sqrt{g} \, R + 
\int d^2 x \, \sqrt{g}\, \left[ \frac{\tau}{\Lambda}\, R + \frac{1}{\Lambda^2}\, \left(\pd\tau\right)^2 \right]\, .
\eeq
The second term in (\ref{Ch4GaugeCT2D}) is, modulo a constant, the Wess-Zumino action in $2$ dimensions,
\beq
\Gamma_{WZ}[g,\tau] = - \frac{c}{24\,\pi}\, \int d^2x\, \sqrt{g}\, \left[
\frac{\tau}{\Lambda}\, R  + \frac{1}{\Lambda^2}\, \left(\pd\tau\right)^2 \right] \, ,
\eeq
from which we can extract the 2-dilaton amplitude according to (\ref{Ch3FuncDiffWZ})
\beq
\mathcal I_{2}(\ku\,-\ku) = \frac{1}{\Lambda^2}\, \langle  T(\ku) T(-\ku) \rangle = \frac{c}{12\,\pi}\, \ku^2 \, .
\eeq
Starting from the 2-dilaton vertex, which is the only non-vanishing one, 
exploiting (\ref{Ch4NoInt}) and inverting the remaining relations, we get the Green functions
\bea
\langle T(\ku) T(\kd) T(\kt) \rangle 
&=& 
- \frac{c}{6\,\pi}\, \left( \ku^2 + \kd^2 + \kt^2 \right)\, ,
\nn \\
\langle T(\ku) T(\kd) T(\kt) T(\kq) \rangle 
&=&
\frac{c}{\pi}\, \left( \ku^2 + \kd^2 + \kt^2 + \kq^2 \right)\, , \nn \\
\langle T(\ku) T(\kd) T(\kt) T(\kq) T(\kc) \rangle 
&=&
-\frac{8\,c}{\pi}\, \left( \ku^2 + \kd^2 + \kt^2 + \kq^2 + \kc^2 \right)\, , \nn \\
\langle T(\ku) T(\kd) T(\kt) T(\kq) T(\kc) T(\ks) \rangle 
&=&
\frac{80\,c}{\pi}\, \left( \ku^2 + \kd^2 + \kt^2 + \kq^2 + \kc^2 + \ks^2 \right)\, .
\eea
These results exactly agree with the combinations of completely traced multiple functional derivatives of the anomaly
(\ref{Ch4TraceAnomaly2D}) that one derives from (\ref{Ch3hier}), providing a consistency check of our recursive formulas
(\ref{Ch4DilIntStructure6}). \\

We also report the expression we obtain of the order-six correlator in $d=4$. 
\bea
&&
\langle T(\ku) T(\kd) T(\kt) T(\kq) T(\kc) T(\ks) \rangle =
32\, \Biggl\{
120\,\bigg(\beta_a + \beta_b\bigg)\, \Biggl[
\sum_{\mathcal T\left\{6,[(k_{i_1},k_{i_2}),(k_{i_3},k_{i_4})]\right\}} 
k_{i_1}\cdot k_{i_2}\,k_{i_3}\cdot k_{i_4} \nn \\
&& 
+\, f_{6}(\ku,\kd,\kt,\kq,\kc,\ks) + f_{6}(\kd,\ku,\kt,\kq,\kc,\ks) + f_{6}(\kt,\ku,\kd,\kq,\kc,\ks)  \nn \\
&&
+\,  f_{6}(\kq,\ku,\kd,\kt,\kc,\ks) + f_{6}(\kc,\ku,\kd,\kt,\kq,\ks) + f_{6}(\ks,\ku,\kd,\kt,\kq,\kc) \Biggr] \nn \\
&& \hspace{-10mm}
-\, \beta_a\, \Biggl[ 
\sum_{\mathcal T\left\{6,(k_{i_1},k_{i_2},k_{i_3},k_{i_4})\right\}} 
\left(k_{i_1} + k_{i_2} + k_{i_3}+ k_{i_4} \right)^4
+  4\, \sum_{\mathcal T\left\{6,(k_{i_1},k_{i_2},k_{i_3})\right\}} \left(k_{i_1} + k_{i_2} + k_{i_3} \right)^4
\nn \\
&&
+\,  11\, \sum_{\mathcal T\left\{6,(k_{i_1},k_{i_2})\right\}} \left(k_{i_1} + k_{i_2} \right)^4
+ 48\, \sum_{i=1}^{5} k_{i}^4 
\Biggr]
\Biggr\}\, ,
\label{Ch4BuildingBlocks}
\eea
where we have introduced the compact notation
\bea
f_{6}(k_a,k_b,k_c,k_d,k_e,k_f)
&=& 
k_a^2\, \left( k_b \cdot k_c + k_b \cdot k_d + k_b \cdot k_e + k_b \cdot k_f + k_c \cdot k_d + k_c \cdot k_e 
\right.
\nn \\
&& \hspace{5mm}
\left.
+\, k_c \cdot k_f + k_d \cdot k_e + k_d \cdot k_f + k_e \cdot k_f  \right)\, .
\eea


\chapter{Appendix}\label{Effective dilaton}

\section{Dilaton interaction vertices}
\label{rules}
The Feynman rules used throughout the paper are collected here. We have 
\begin{itemize}
%
\item{ dilaton - gauge boson - gauge boson vertex}
\\ \\
\begin{minipage}{95pt}
\includegraphics[scale=1.0]{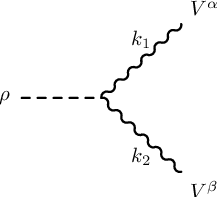}
\end{minipage}
\begin{minipage}{70pt}
\bea
= V^{\alpha\beta}_{\rho VV}(k_1,k_2) =
- \frac{2\, i}{\Lambda} \bigg\{M_V^2 \, \eta^{\alpha\beta} 
- \frac{1}{\xi}\, \left( k_1^\alpha\, k_1^\beta + k_2^\alpha\, k_2^\beta + 2\, k_1^\alpha\, k_2^\beta \right) \bigg\}
\nn
\eea
\end{minipage}
\bea
\label{FRdilVV}
\eea
where $V$ stands for the gluons or for the vector gauge bosons $A, Z$ and $W^{\pm}$ and,
if the gauge bosons are gluons, a color-conserving $\delta_{ab}$ matrix must be included.
\\ \\
\item{dilaton - fermion - fermion vertex}
\\ \\
\begin{minipage}{95pt}
\includegraphics[scale=1.0]{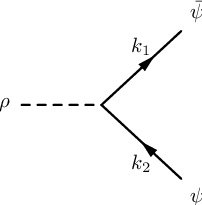}
\end{minipage}
\begin{minipage}{70pt}
\bea
&=& 
V_{\rho \bar\psi\psi}(k_1,k_2) = 
- \frac{i}{2\,\Lambda} \, \bigg\{ 3 \, \left( \ksl_1 - \ksl_2 \right) + 8 \, m_f \bigg\}
\nn
\eea
\end{minipage}
\bea
\label{FRdilFF}
\eea
If the fermions are quarks, the vertex must be multiplied by the identity color matrix $\delta_{a b}$. \\ \\
\item{dilaton - ghost - ghost vertex }
\\ \\
\begin{minipage}{95pt}
\includegraphics[scale=1.0]{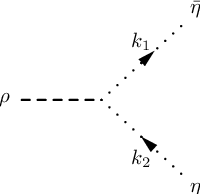}
\end{minipage}
\begin{minipage}{70pt}
\bea
&=& 
V_{\rho \bar c c}(k_1,k_2)
= - \frac{2 \, i}{\Lambda} \bigg\{ k_1 \cdot k_2 + 2 \, M_{\eta}^2\bigg\}
\nn
\eea
\end{minipage}
\bea
\label{FRdilUU}
\eea
where $\eta$ denotes both the QCD ghost fields $c^{a}$ or the electroweak ghost fields $\eta^{+}$, $\eta^{-}$ ed $\eta^Z$. In the QCD 
case one must include a color-conserving $\delta_{ab}$ matrix. 
\\ \\
\item{dilaton - scalar - scalar vertex}
\\ \\
\begin{minipage}{95pt}
\includegraphics[scale=1.0]{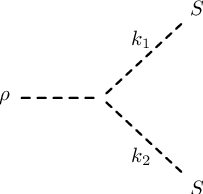}
\end{minipage}
\begin{minipage}{70pt}
\bea
=  
V_{\rho SS}(k_1,k_2)
&=& - \frac{2 \, i}{\Lambda} \bigg\{ k_1 \cdot k_2 + 2 \, M_S^2  \bigg\} 
\nn \\
&=&  
\frac{i}{\Lambda} \,6\,\chi\, (k_1+k_2)^2 \nn
\eea
\end{minipage}
\bea
\label{FRdilSS}
\eea
where $S$ stands for the Higgs $H$ and the Goldstones $\phi$ and  $\phi^{\pm}$. 
The first expression is the contribution coming from 
the minimal energy-momentum tensor while the second is due to the term of improvement.
\\ \\
\item{dilaton - Higgs vertex}
\\ \\
\begin{minipage}{95pt}
\includegraphics[scale=1.0]{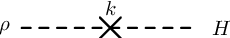}
\end{minipage}
\begin{minipage}{70pt}
\bea
\qquad 
&=& 
V_{I,\, \rho H }(k)
= - \frac{i}{\Lambda} \frac{12\,\chi\, s_w M_W}{e} \, k^2 \nn
\eea
\end{minipage}
\bea
\label{FRdilH}
\eea
This vertex is derived from the term of improvement of the energy-momentum tensor and it is a feature of the electroweak symmetry 
breaking because it is proportional to the Higgs vev.
\\ \\
\item{ dilaton - three gauge boson vertex}
\\ \\
\begin{minipage}{95pt}
\includegraphics[scale=1.0]{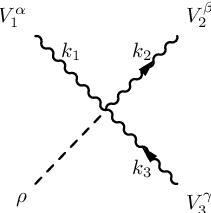}
\end{minipage}
\begin{minipage}{70pt}
\bea
&=&  
V^{\alpha\beta\gamma}_{\rho V V V}
= 0
\nn
\eea
\end{minipage}
\bea
\label{FRdilVVV}
\eea
where $V_1$, $V_2$, $V_3$ stand for gluon, photon, $Z$ and $W^{\pm}$.
\\ \\
\item{dilaton - gauge boson - scalar - scalar vertex }
\\ \\
\begin{minipage}{95pt}
\includegraphics[scale=1.0]{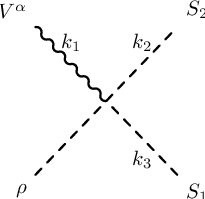}
\end{minipage}
\begin{minipage}{70pt}
\bea
&=&
V^\alpha_{\rho VSS}(k_2,k_3)
=  - \frac{2\,i}{\Lambda}\, e \, \mathcal C_{V S_1 S_2} \, \left( k_2^\alpha - k_3^\alpha \right)
\nn
\eea
\end{minipage}
\bea
\label{FRdilVSS}
\eea
with $\mathcal C_{V S_1 S_2}$ given by
\bea
\mathcal C_{A\phi^{+}\phi^{-}} = 1 \qquad
\mathcal C_{Z\phi^{+}\phi^{-}} = \frac{c_w^2 - s_w^2}{2 s_w \, c_w} \qquad
\mathcal C_{Z H \phi} =  \frac{i}{2 s_w \, c_w}. \nn
\eea
\\ \\
\item{dilaton - gauge boson - ghost - ghost vertex}
\\ \\
\begin{minipage}{95pt}
\includegraphics[scale=1.0]{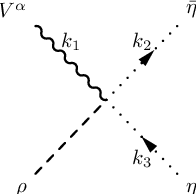}
\end{minipage}
\begin{minipage}{70pt}
\bea
&=&
V^\alpha_{\rho V \bar\eta\eta}(k_2)
= - \frac{2\,i}{\Lambda} \, \mathcal C_{V \eta} \, k_2^\alpha
\nn
\eea
\end{minipage}
\bea
\label{FRdilVUU}
\eea
where $V$ denotes the $g^a$, $A$, $Z$ gauge bosons and $\eta$ the ghosts $c^b$, $\eta^{+}$, $\eta^{-}$.
The coefficients $\mathcal C$ are defined as
\bea
\mathcal C_{g^a  c^b} = f^{abc}\, g \qquad
\mathcal C_{A \eta^{+}} = e \qquad
\mathcal C_{A \eta^{-}} = -e \qquad
\mathcal C_{Z \eta^{+}} =  e \, \frac{c_w}{s_w} \qquad
\mathcal C_{Z \eta^{-}} =  - e \, \frac{c_w}{s_w}. \nn
\eea
%

\item{dilaton - gauge boson - gauge boson - scalar vertex}
\\ \\
\begin{minipage}{95pt}
\includegraphics[scale=1.0]{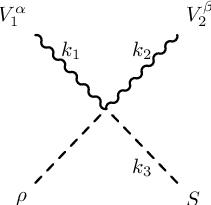}
\end{minipage}
\begin{minipage}{70pt}
\bea
&=&
V^{\alpha\beta}_{\rho VVS}
= - \frac{2}{\Lambda} \, e \, \mathcal C_{V_1 V_2 S} \, M_W \, \eta^{\alpha\beta}
\nn
\eea
\end{minipage}
\bea
\label{FRdilVVS}
\eea
where $V$ stands for $A$, $Z$ and $W^{\pm}$ and $S$ for $\phi^{\pm}$
and $H$. The coefficients are defined as
\bea
\mathcal C_{A W^{+} \phi^{-}} = 1 \qquad
\mathcal C_{A W^{-} \phi^{+}} = -1 \qquad
\mathcal C_{Z W^{+} \phi^{-}} = - \frac{s_w}{c_w} \qquad \nn \\
\mathcal C_{Z W^{-} \phi^{+}} = \frac{s_w}{c_w} \qquad
\mathcal C_{Z Z H} = - \frac{i}{s_w \, c_w^2} \qquad
\mathcal C_{W^{+} W^{-} H} = - \frac{i}{c_w}. \nn
\eea
\\ \\
\item{dilaton - scalar - ghost - ghost vertex}
\\ \\
\begin{minipage}{95pt}
\includegraphics[scale=1.0]{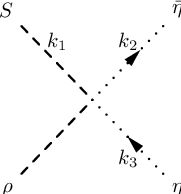}
\end{minipage}
\begin{minipage}{70pt}
\bea
&=&
V{\rho S\bar\eta\eta}
= - \frac{4 \, i}{\Lambda}\, e \, \mathcal C_{S \eta} \, M_W
\nn
\eea
\end{minipage}
\bea
\label{FRdilSUU}
\eea
where $S = H$ and $\eta$ denotes $\eta^{+}$, $\eta^{-}$ and $\eta^{z}$. The vertex is defined with the coefficients
\bea
\mathcal C_{H \eta^{+}} = \mathcal C_{H \eta^{-}} = \frac{1}{2 s_w} \qquad \mathcal C_{H \eta^{z}} = \frac{1}{2 s_w \, c_w}. 
\nn
\eea
%

\item{dilaton - three scalar vertex}
\\ \\
\begin{minipage}{95pt}
\includegraphics[scale=1.0]{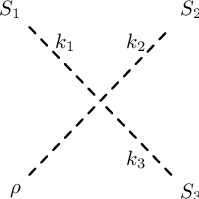}
\end{minipage}
\begin{minipage}{70pt}
\bea
&=&
V_{\rho SSS}
= - \frac{4 \, i}{\Lambda} \, e \, \mathcal C_{S_1 S_2 S_3}
\nn
\eea
\end{minipage}
\bea
\label{FRdilSSS}
\eea
with $S$ denoting $H$, $\phi$ and $\phi^{\pm}$. We have defined the coefficients
\bea
\mathcal C_{H \phi \phi} = \mathcal C_{H \phi^{+} \phi^{-}} = \frac{1}{2 s_w \, c_w}
\frac{M_H^2}{M_Z} \qquad \mathcal C_{H H H} = \frac{3}{2 s_w \, c_w} \frac{M_H^2}{M_Z}. 
\nn
\eea
\\ \\
\item{dilaton - scalar - fermion - fermion vertex}
\\ \\
\begin{minipage}{95pt}
\includegraphics[scale=1.0]{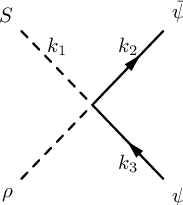}
\end{minipage}
\begin{minipage}{70pt}
\bea
&=&
V_{\rho S\bar\psi\psi}
= - \frac{2\,i}{\Lambda} \, \frac{e}{s_w \, c_w} \frac{m_f}{M_Z}
\nn
\eea
\end{minipage}
\bea
\label{FRdilSFF}
\eea
where  $S$ is only the Higgs scalar $H$.
\\ \\
\item{dilaton - gluon - fermion - fermion vertex}
\\ \\
\begin{minipage}{95pt}
\includegraphics[scale=1.0]{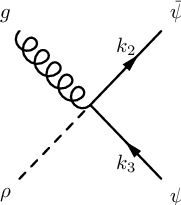}
\end{minipage}
\begin{minipage}{70pt}
\bea
&=&
V^{a\,\alpha}_{\rho g\bar\psi\psi}
= \frac{3\, i}{\Lambda}\, g \, T^a \, \gamma^\alpha\, .
\nn
\eea
\end{minipage}
\bea
\label{FRdilgFF}
\eea
\\ \\
\item{dilaton - photon - fermion - fermion vertex}
\\ \\
\begin{minipage}{95pt}
\includegraphics[scale=1.0]{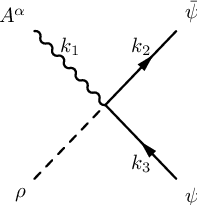}
\end{minipage}
\begin{minipage}{70pt}
\bea
&=&
V^{\alpha}_{\rho \gamma\bar\psi\psi}
= Q_f \, e \frac{3\,i}{\Lambda}\, \gamma^\alpha
\nn
\eea
\end{minipage}
\bea
\label{FRdilAFF}
\eea
where $Q_f$ is the fermion charge expressed in units of $e$.
\\ \\
\item{dilaton - Z - fermion - fermion vertex}
\\ \\
\begin{minipage}{95pt}
\includegraphics[scale=1.0]{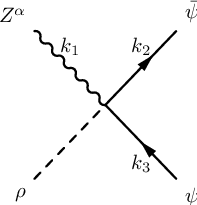}
\end{minipage}
\begin{minipage}{70pt}
\bea
&=&
V^{\alpha}_{\rho Z\bar\psi\psi}
= \frac{3 \,i }{2\, \Lambda s_w \, c_w} \, e \, (C_v^f - C_a^f \gamma^5) \, \gamma^\alpha
\nn
\eea
\end{minipage}
\bea
\label{FRdilZFF}
\eea
where $C_v^f$ and $C_a^f$ are the vector and axial-vector couplings of the $Z$ gauge boson
to the fermion ($f$). Their expressions are
\bea
C_v^f = I^f_3 - 2 s_w^2 \, Q^f \qquad \qquad C_a^f = I^f_3. \nn
\eea
$I^f_3$ denotes the 3rd component of the isospin.
\\ \\
\item{dilaton - four gauge bosons vertex}
\\ \\
\begin{minipage}{95pt}
\includegraphics[scale=1.0]{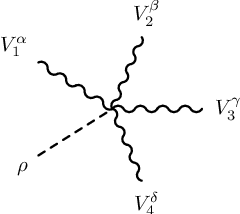}
\end{minipage}
\begin{minipage}{70pt}
\bea
&=&
V^{\alpha\beta\gamma\delta}_{\rho VVVV}
\qquad \qquad = 0 \nn
\eea
\end{minipage}
\bea
\label{FRdilVVVV}
\eea
where $V_1$, $V_2$, $V_3$ and $V_4$ denote $g$, $A$, $Z$ or $W^{\pm}$.
\\ \\
\item{dilaton - gauge boson - gauge boson - scalar - scalar vertex}
\\ \\
\begin{minipage}{95pt}
\includegraphics[scale=1.0]{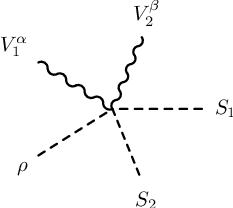}
\end{minipage}
\begin{minipage}{70pt}
\bea
&=&
V^{\alpha\beta}_{\rho VVSS}
\qquad = \frac{2 \, i}{\Lambda} \, e^2 \, \mathcal C_{V_1 V_2 S_1 S_2} \, \eta^{\alpha\beta}
\nn
\eea
\end{minipage}
\bea
\label{FRdilVVSS}
\eea
where $V_1$ and $V_2$ denote the neutral gauge bosons $A$ and $Z$, while the possible scalars are
$\phi$, $\phi^{\pm}$ and $H$. The coefficients are
\bea
\mathcal C_{A A \phi^+ \phi^-} = 2
\qquad
\mathcal C_{A Z \phi^+ \phi^-} = \frac{c_w^2 - s_w^2}{s_w \, c_w}\qquad \mathcal C_{Z Z \phi^+ \phi^-} = 
\frac{\left( c_w^2 - s_w^2 \right)^2}{2 s_w^2 \,c_w^2}
\qquad
\mathcal C_{Z Z \phi \phi} =  \mathcal C_{Z Z H H} = \frac{1}{2 s_w^2 \, c_w^2} . \nn
\eea
\end{itemize}

\section{The scalar integrals}
\label{scalars}

We collect in this appendix the definition of the scalar integrals appearing in the computation of the correlators.
$1$-, $2$- and $3$-point functions are denoted, respectively as $\mathcal A_0$, $\mathcal B_0$ and $\mathcal C_0$, with
\bea
\mathcal A_0 (m_0^2) &=& \frac{1}{i \pi^2}\int d^n l \, \frac{1}{l^2 - m_0^2} \,, \nn\\
\mathcal B_0 (k^2, m_0^2,m_1^2) &=&  \frac{1}{i \pi^2} \int d^n l \, \frac{1}{(l^2 - m_0^2) \, ((l + k )^2 - m_1^2 )} \,, \nn \\
\mathcal C_0 ((p+q)^2, p^2, q^2, m_0^2,m_1^2,m_2^2) &=& \frac{1}{i \pi^2} \int d^n l \, \frac{1}{(l^2 - m_0^2) \, ((l + p )^2 - m_1^2 ) \, ((l -q )^2 - m_2^2 ) } \,.
\eea
We have also used the finite combination of $2$-point scalar integrals
\bea
\mathcal D_0 (p^2, q^2, m_0^2,m_1^2) = \mathcal B_0 (p^2, m_0^2,m_1^2) - \mathcal B_0 (q^2, m_0^2,m_1^2) \,.
\eea
The explicit expressions of $\mathcal A_0$, $\mathcal B_0$ and $\mathcal C_0$ can be found in \cite{Denner:1991kt}.

\section{Contributions to $\mathcal V_{\rho Z Z }$}
\label{VZZ}

\begin{itemize}
\item {\bf The fermion sector}
%
\bea
&&
\Phi^{F}_{ZZ}(p,q)
=\sum_f \bigg\{
\frac{ \alpha  \, m_f^2}{\pi\,  s c_w^2 s_w^2 \left(s-4 M_Z^2\right)} \left(s-2 M_Z^2\right)
\left(C_a^{f \, 2} + C_v^{f\, 2}\right)
\nn \\
&&
+ \frac{2\, \alpha  \, m_f^2}{ \pi\, s c_w^2 \left(s-4 M_Z^2\right)^2 s_w^2}
\left[\left(2 M_Z^4-3 s M_Z^2+s^2\right) C_a^{f \, 2}+C_v^{f \, 2} M_Z^2 \left(2 M_Z^2 + s \right)\right]
\mathcal D_0(s, M_Z^2,m_f^2,m_f^2)
\nn \\
&&
+ \frac{\alpha \, m_f^2}{2 \,\pi \, s c_w^2 \left(s - 4 M_Z^2\right){}^2 s_w^2}
\left(s-2 M_Z^2\right) \big\{\left[4 M_Z^4-2 \left(8 m_f^2+s \right) M_Z^2+s \left(4 m_f^2+s \right)\right] C_a^{f\,2}
\nn \\
&&
+ C_v^{f \, 2} \left[4 M_Z^4+2 \left(3 s - 8 m_f^2\right) M_Z^2-s \left(s - 4 m_f^2\right) \right] \big\}
\vphantom{\frac{m_f^2}{\left(s-4 M_Z^2\right)}}\, \mathcal C_0(s, M_Z^2,M_Z^2 ,m_f^2,m_f^2,m_f^2)\bigg\}
\nn \\
&& \Xi^{F}_{ZZ}(p,q) =
%
\sum_f \bigg\{ - \frac{\alpha\, m_f^2}{\pi \,  s c_w^2 \left(s-4 M_Z^2\right) s_w^2} 
\left[\left(2 M_Z^4-4 s M_Z^2+s^2\right) C_a^{f \, 2}+2 C_v^{f \, 2} M_Z^4 \right]
\nn \\
&&
- \frac{\alpha\,m_f^2}{ \pi\, c_w^2 s_w^2}C_a^{f\,2}\mathcal B_0(s,m_f^2,m_f^2)
- \frac{2\,\alpha\,  m_f^2 \,M_Z^2}{ \pi \, s c_w^2 \left(s-4 M_Z^2\right){}^2 s_w^2} 
\big[s^2 C_a^{f \, 2} - 2 \left( C_a^{f \, 2} + C_v^{f \, 2} \right) M_Z^4
\nn \\
&&
+ 2 s \left(C_v^{f \, 2}-C_a^{f \, 2}\right) M_Z^2 \big]\, \mathcal D_0(s, M_Z^2, m_f^2,m_f^2) 
- \frac{\alpha \, m_f^2}{ \pi\, s c_w^2 \left(s - 4 M_Z^2\right){}^2 s_w^2}
\bigg[ \left[ 4 M_Z^8 
\right.
\nn \\
&&
\left.
-2 \left(8 m_f^2 + 5 s \right) M_Z^6 + 3 s \left(12 m_f^2+s \right) M_Z^4 - 16 s^2 m_f^2 M_Z^2 
+ 2 s^3 m_f^2 \right] C_a^{f \,2} \nn \\ 
&&
+ C_v^{f \, 2} M_Z^4 \left[4 M_Z^4-2 \left(8 m_f^2+s \right) M_Z^2  + s \left(4 m_f^2+s \right)\right] \bigg]\, 
\mathcal C_0(s, M_Z^2,M_Z^2 ,m_f^2,m_f^2,m_f^2) \vphantom{\frac{m_f^2}{\left(s-4 M_Z^2\right)}} \bigg\} \nn 
\eea
%

\item{\bf The $W$ boson sector}
%
\bea
&&
\Phi^{W}_{ZZ}(p,q)=
\frac{\alpha}{ s_w^2 \, c_w^2 \, \pi }
\bigg[ 
\frac{M_Z^2}{2\,s\left(s-4 M_Z^2\right)} \big[ 2 M_Z^2 \left(-12 s_w^6+32 s_w^4-29 s_w^2+9\right) 
+ s \left(12 s_w^6-36 s_w^4+33 s_w^2-10\right)\big] \bigg]  
\nn
\eea
\bea
&&
+ \frac{\alpha \, M_Z^2}{ 2\, s_w^2 \, c_w^2 \,\pi\,s\,(s-4 M_Z)^2} 
[ 4 M_Z^4 (12 s_w^6 -32 s_w^4 +29 s_w^2 - 9)
\nn\\
&&
+ 2 M_Z^2 s (s_w^2 - 2)(12 s_w^4 - 12 s_w^2 +1) + 
s^2 (-4 s_w^4+8s_w^2-5) ] \mathcal D_0(s, M_Z^2, M_W^2,M_W^2) 
\nn\\
&&
+ \frac{\alpha \, M_Z^2}{2\,\pi \, s_w^2 \, c_w^2 \, s \, (s-4 M_Z)^2}
\bigg[-4 M_Z^6(s_w^2-1)(4 s_w^2-3)(12 s_w^4 - 20s_w^2 + 9) 
\nn \\
&&
 + 2 M_Z^4 s (18 s_w^4-34s_w^2 + 15)(4(s_w^2-3)s_w^2+7) - 2M_Z^2 s^2 (12 s_w^8-96s_w^6 +201s_w^4-157s_w^2+41)+ 
\nn \\
&&
 s^3(-12 s_w^6+32s_w^4-27s_w^2+7)  \bigg]\, \mathcal C_0(s, M_Z^2,M_Z^2 ,M_W^2,M_W^2,M_W^2)
\nn  \\ 
&&
\Xi^{W}_{ZZ}(p,q) = 
%
\frac{\alpha\,M_Z^2}{2 \,\pi\, s_w^2\,c_w^2\,s (s-4M_Z^2)}
\bigg[2 M_Z^4 \left(-12 s_w^6+32 s_w^4-29 s_w^2+9\right)
\nn\\
&&
+ s M_Z^2 \left[4\left(s_w^4+ s_w^2\right)-7\right]- 2 s^2 \left(s_w^2-1\right) \bigg]
+ \frac{\alpha M_Z^2}{\pi \,s_w^2c_w^2 }(-2s_w^4 + 3s_w^2 - 1)\, \mathcal B_0(s,M_W^2,M_W^2)
\nn \\
&&
+ \frac{\alpha \, M_Z^2}{\pi\, s_w^2 \, c_w^2\, s (s-M_Z^2)^2} 
\big[\vphantom{\left(s_w^2-1\right)^2}2 M_Z^6 \left(12 s_w^6-32 s_w^4+29 s_w^2-9\right)
+ s M_Z^4 \left(-24 s_w^6 + 92 s_w^4 - 110 s_w^2 + 41 \right)
\nn\\
&&
+ s^2 M_Z^2 \left(-12 s_w^4+26 s_w^2-13\right)+ 2 s^3 \left(s_w^2-1\right)^2\big]\mathcal D_0(s, M_Z^2, M_W^2,M_W^2)
\nn \\
&&
+ \frac{\alpha\,M_Z^2}{4 \,\pi \,s_w^2\,c_w^2\,s(s-M_Z^2)^2} 
\big[ -8 M_Z^8 \left(s_w^2 - 1\right) \left(4 s_w^2-3\right) \left(12 s_w^4-20 s_w^2+9\right)
\nn \\
&&
+ 4 s M_Z^6 \left(24 s_w^8-60 s_w^6+30 s_w^4+25 s_w^2 - 18 \right)
+ 2 s^2 M_Z^4 (-20 s_w^6+76 s_w^4  - 103 s_w^2+46) 
\nn \\
&&
+ s^3 M_Z^2 \left(-4 s_w^4+24 s_w^2-19\right) - 2 s^4 \left(s_w^2-1\right)\big]\, 
\mathcal C_0(s, M_Z^2,M_Z^2 ,M_W^2,M_W^2,M_W^2)\vphantom{\frac{M_Z^2}{(s-4M_Z^2)}} \nn 
\eea
%
\item{\bf The $(Z,H)$ sector}
%
\bea
&&
\Phi^{ZH}_{ZZ}(p,q) 
= 
\frac{- \alpha}{4\,\pi\,s\,c_w^2 s_w^2\, \left( s - 4 M_Z^2 \right)}
\left\{\vphantom{\frac{M_H^2}{\left(s-4 M_Z^2\right)}} \left[M_H^2 \left(s-2 M_Z^2\right)+3 s M_Z^2- 2 M_Z^4\right] 
\right.
\nn\\
&&
\left.
+ 2 \left(M_Z^2-M_H^2\right)\left(\mathcal A_0(M_Z^2)-\mathcal A_0(M_H^2)\right)
\right.
\nn\\
&&
\left.
+ \frac{1}{\left(s-4 M_Z^2\right)} \left[2 M_H^2 \left(s M_Z^2-2 M_Z^4+s^2\right)+3 s^2 M_Z^2-14 s M_Z^4+8 M_Z^6\right]
\mathcal B_0(s, M_Z^2,M_Z^2)
\right.
\nn\\
&&\left. 
- \frac{1}{\left(s-4 M_Z^2\right)}
\left(2 M_H^2+s\right) \left[2 M_H^2 \left(s-M_Z^2\right)-3 s M_Z^2\right]
\mathcal B_0(s, M_H^2,M_H^2)
\right.
\nn\\
&&\left.
+ \frac{2}{\left(s-4 M_Z^2\right)} \left[s M_H^4+6 \left(s-M_H^2\right) M_Z^4+\left(2 M_H^4-3 s M_H^2-3 s^2\right) 
M_Z^2\right]\mathcal B_0(M_Z^2, M_Z^2,M_H^2)
\right.
\nn\\
&&
\left. 
+ \frac{\left(2 M_H^2+s\right)}{\left(s-4 M_Z^2\right)}  \bigg[M_Z^2 \left(-8 s M_H^2-2 M_H^4+s^2\right) 
+ 2 M_Z^4 \left(4 M_H^2 + s\right) + 2 s M_H^4\bigg]\mathcal C_0(s, M_Z^2,M_Z^2 ,M_Z^2,M_H^2,M_H^2) 
\right.
\nn\\
&&
\left.
+ \frac{M_H^2}{\left(s-4 M_Z^2\right)} 
\left[2 M_H^2 \left(s M_Z^2-2 M_Z^4+s^2\right) - 20 s M_Z^4+16 M_Z^6+s^3\right]
\mathcal C_0(s, M_Z^2,M_Z^2 ,M_H^2,M_Z^2,M_Z^2) \right\}  
\nn \\
&& \Xi^{ZH}_{ZZ}(p,q)=
-  \frac{\alpha}{8\, \pi \, s\,c_w^2s_w^2 \left(s-4 M_Z^2\right)}
\bigg\{- 4 M_Z^2 \left(M_Z^4+M_H^2 M_Z^2-3 s M_Z^2+s^2\right)
\nn\\
&&
+ 2 \left(M_H^2-M_Z^2 \right)\left(s-2 M_Z^2\right)\left(\mathcal A_0(M_Z^2)-\mathcal A_0(M_H^2)\right)
\nn\\
&& 
- \frac{1}{\left(s-4 M_Z^2\right)} \left[\left(8 M_Z^6+s^3\right) M_H^2+M_Z^2 \left(s-4 M_Z^2\right) \left(s-2 
M_Z^2\right)\left(3 s-2 M_Z^2\right)\right]\mathcal B_0(s, M_Z^2,M_Z^2)  \nn\\
&&
+ \frac{1}{\left(s-4 M_Z^2\right)} \left[2 \left(4 M_H^4-s^2\right) M_Z^4 
- s \left(2 M_H^2+s \right){}^2 M_Z^2+s^2 M_H^2\left(2 M_H^2+s \right)\right]\, \mathcal B_0(s, M_H^2,M_H^2)
\nn\\
&&
+ \frac{8 M_Z^2}{\left(s-4 M_Z^2\right)} \left[-s M_H^4-\left(3 M_H^2+5 s \right) M_Z^4
+ \left(M_H^2+s \right)\left(M_H^2+2 s \right) M_Z^2\right]\mathcal B_0(M_Z^2, M_Z^2,M_H^2)
\nn\\
&& 
- \frac{\left(2 M_H^2+s \right)}{\left(s-4 M_Z^2\right)} 
\left[4 \left(7 s-4 M_H^2\right) M_Z^6+4 \left(M_H^2-s \right) \left(M_H^2+3 s \right) M_Z^4
\right.
\nn\\
&&
\left. 
+ 2 s \left(-M_H^4-2 s M_H^2+s^2\right) M_Z^2+s^2 M_H^4\right]\mathcal C_0(s, M_Z^2,M_Z^2 ,M_Z^2,M_H^2,M_H^2)
\nn
\eea
\bea
&&
- \frac{1}{\left(s-4 M_Z^2\right)} \left[\left(8 M_Z^6+s^3\right) M_H^4
+ 4 M_Z^2 \left(s-4 M_Z^2\right) \left(2 M_Z^4-s M_Z^2+s^2\right) M_H^2 
\right.
\nn \\
&&
\left.
+ 4 s M_Z^4 \left(s-4 M_Z^2\right){}^2\right]\mathcal C_0(s, M_Z^2,M_Z^2 ,M_H^2,M_Z^2,M_Z^2)
\bigg\}  \nn 
\eea
\item{\bf Term of improvement}

\bea
&&
\Phi^{I}_{ZZ}(p,q)
= \frac{3\,\chi\,\alpha}{2\, \pi\, s_w^2\, c_w^2\,(s-4M_Z^2)^2}
\bigg\{ (c_w^2 - s_w^2)^2 \bigg[ 2 M_Z^2 s - 8 M_Z^4 
\nn \\
&&
+ 2 M_Z^2 (s + 2 M_Z^2) \, \mathcal D_0 \left( s, M_Z^2 , M_W^2, M_W^2 \right) 
+ 2 \left( c_w^2 M_Z^2 (8 M_Z^4 - 6 M_Z^2 s + s^2) - 2 M_Z^6 + 2 M_Z^4 s \right)
\nn \\
&& 
\times \, \mathcal C_0 \left( s, M_Z^2, M_Z^2, M_W^2,M_W^2,M_W^2 \right) \bigg]  
+  2 M_Z^2 s - 8 M_Z^4  + 2 M_Z^2 (s + 2\, M_Z^2) \big[
\mathcal B_0 \left(s, M_Z^2, M_Z^2 \right)  \nn \\
&& 
-  \mathcal B_0\left(M_Z^2, M_Z^2, M_H^2 \right)  \big] 
+ \left( 3 M_Z^2 s - 2 M_H^2 (s - M_Z^2) \right) \big[ \mathcal B_0 
\left(s, M_H^2, M_H^2 \right) -  \mathcal B_0 \left(s, M_Z^2, M_Z^2 \right)\big] 
\nn \\
&& 
+ M_H^2 \left( 2 M_H^2 (s-M_Z^2) + 8 M_Z^4 - 6 M_Z^2 s + 
s^2 \right) \mathcal C_0\left( s, M_Z^2,M_Z^2, M_H^2,M_Z^2,M_Z^2\right) 
\nn \\
&& 
+ \left( 2 M_H^2 (M_H^2 - 4 M_Z^2)(s-M_Z^2) + s M_Z^2 (s+ 2 M_Z^2)\right) \mathcal C_0\left( s, M_Z^2,M_Z^2, 
M_Z^2,M_H^2,M_H^2\right) \bigg\} \nn 
\eea
\bea
&& \Xi^{I}_{ZZ}(p,q) =
- \frac{3\,\chi\,\alpha\,s}{8\,\pi\, s_w^2 \, c_w^2 (s-4 M_Z^2)^2} 
\, \bigg\{ (c_w^2 -s_w^2)^2 \bigg[ 4 M_Z^4 (s - 4 M_Z^2)
+ 8 M_Z^4 (s-M_Z^2) \, \mathcal D_0 \left( s, M_Z^2, M_W^2, M_W^2\right)
\nn \\
&&  
+ 2 M_Z^4 \left[ s^2 - 2 M_Z^2 s + 4 M_Z^4 + 4 c_w^2 M_Z^2 (s - 4 M_Z^2) \right]\,\mathcal C_0 \left( s, M_Z^2, M_Z^2, 
M_W^2,M_W^2,M_W^2 \right) \bigg] 
\nn \\
&&  
+ 4 M_Z^2 s_w^4 c_w^2 s (s - 4 M_Z^2)^2 \mathcal C_0 \left( s, M_Z^2, M_Z^2, M_W^2,M_W^2,M_W^2 \right) 
+ 4 M_Z^2 (s-4 M_Z^2)
\nn \\
&&
+ \left[M_Z^2 s (s + 2 M_Z^2) - M_H^2 (s^2 - 2 M_Z^2 s + 4 M_Z^4)\right] 
\big[ \mathcal B_0 \left(s, M_H^2, M_H^2 \right) -  \mathcal B_0 \left(s, M_Z^2, M_Z^2 \right) \big] 
\nn \\
&& 
+ 8 M_Z^4 (s - M_Z^2) \big[ \mathcal B_0 \left(s, M_Z^2, M_Z^2 \right) -  \mathcal B_0 \left(M_Z^2, M_Z^2, M_H^2 \right)  \big] 
+ M_H^2 \left[ 4 M_Z^4 (s-4 M_Z^2)
\right. 
\nn \\
&&
\left. 
+ M_H^2 (s^2 - 2 MZ^2 s + 4 M_Z^4)\right] \mathcal C_0\left( s, M_Z^2,M_Z^2, M_H^2,M_Z^2,M_Z^2\right) 
+ \left[ M_H^2(M_H^2 - 4 M_Z^2) (s^2 - 2 M_Z^2 s + 4 M_Z^4) 
\right.
\nn \\
&&
\left. 
+ 2 M_Z^2 s (s^2 - 6 M_Z^2 s + 14 M_Z^4)\right] \mathcal C_0\left( s, M_Z^2,M_Z^2, M_Z^2,M_H^2,M_H^2 \right)
\bigg\} \, . \nn 
\eea

\item{\bf External leg corrections}
\end{itemize}
%
The $\Delta^{\alpha\beta}(p,q)$ correlator is decomposed as
\bea
\Delta^{\alpha\beta}(p,q) 
&=& 
\bigg[ \Sigma_{Min, \, \rho H}(k^2) + \Sigma_{I,\,\rho H}(k^2) \bigg] 
\frac{1}{s - M_H^2} V_{HZZ}^{\alpha\beta} + \left( \frac{\Lambda}{i} \right) V_{I, \, \rho H}(k) \frac{1}{s - M_H^2} 
\Sigma_{HH}(k^2) \frac{1}{s - M_H^2} 
V_{HZZ}^{\alpha\beta}  
\nn \\
&+& 
\Delta^{\alpha\beta}_{I, \, HZZ}(p,q) \nn 
\eea
where $\Sigma_{HH}(k^2)$ is the Higgs self-energy, $V_{HZZ}^{\alpha\beta}$ and
$ V_{I, \, \rho H}$ are tree level vertices defined in appendix (\ref{rules}) and $\Delta^{\alpha\beta}_{I, \, HZZ}(p,q)$ 
is expanded into the three contributions of improvement as
\bea
\Delta^{\alpha\beta}_{I, \, HZZ}
&=&  
\Delta^{\alpha\beta}_{(F), \, HZZ}(p,q) + \Delta^{\alpha\beta}_{(W), \, HZZ}(p,q) + 
\Delta^{\alpha\beta}_{(Z,H), \, HZZ}(p,q) \nn  \\
&=&
\left[ \left( \frac{s}{2} - M_Z^2 \right) \eta^{\alpha\beta} - q^\alpha p^\beta \right]\, \Phi^{\Delta}_{ZZ}(p,q) +
\eta^{\alpha\beta}\, \Xi^{\Delta}_{ZZ}(p,q)
\nn \\
&=& 
\left[ \left( \frac{s}{2} - M_Z^2 \right) \eta^{\alpha\beta} - q^\alpha p^\beta \right]
\left(\Phi^{\Delta\,F}_{ZZ}(p,q) + \Phi^{\Delta\,W}_{ZZ}(p,q) + \Phi^{\Delta\,W}_{ZZ}(p,q)\right)  
\nn \\
&& \hspace{35mm}
+\, \eta^{\alpha\beta}\, \left(\Xi^{\Delta\,F}_{ZZ}(p,q) + \Xi^{\Delta\,W}_{ZZ}(p,q) + \Xi^{\Delta\,W}_{ZZ}(p,q) \right)
\, . \nn  
\eea
These are given by
\bea
&&
\Phi^{\Delta\,F}_{ZZ}(p,q) =
%
- \sum_f \frac{6 \, \alpha \,\chi \, m_f^2}{\pi\, s_w^2 \, c_w^2 \, (s-M_H^2)(s - 4 M_Z^2)}
\bigg\{
( C_v^{f\, 2} + C_a^{f\, 2}) (s-2 M_Z^2)  \nn \\
&& 
+  \frac{2}{s-4 M_Z^2} [ M_Z^2 ( C_v^{f\, 2} + C_a^{f\, 2}) (s +2 M_Z^2) + C_a^{f \, 2} (s-4M_Z^2) s ] \mathcal 
D_0(s,M_Z^2,m_f^2,m_f^2) \nn \\
&&
+ \frac{s - 2 M_Z^2}{2(s-4 M_Z^2)} 
[ ( C_v^{f\, 2} + C_a^{f\, 2}) (4 m_f^2 (s-4 M_Z^2)  + 4 M_Z^4 + 6 M_Z^2 s - s^2 ) + 2 C_a^{f \, 2} s (s-4 M_Z^2) ] 
\times \nn \\
&&
\mathcal C_0(s, M_Z^2,M_Z^2,m_f^2,m_f^2,m_f^2) 
\bigg\} \, , \nn  \\
%
&& \Xi^{\Delta\,F}_{ZZ}(p,q)
= \sum_f \frac{18\,\alpha\,s\,\chi \, m_f^2}{\pi \,  s_w^2 \, c_w^2 \, (s - M_H^2)(s -4 M_Z^2) } \, 
\eta^{\alpha \beta} 
\bigg\{
2 M_Z^2 ( C_v^{f\, 2} + C_a^{f\, 2}) \nn \\
&&
+ 2 s (s- 4 M_Z^2) C_a^{f \, 2} \mathcal B_0(s, m_f^2, m_f^2) 
+ \frac{2}{s- 4M_Z^2} [ 2 M_Z^4 ( C_v^{f\, 2} + C_a^{f\, 2}) (s - M_Z^2) 
+ C_a^{f \, 2} M_Z^2 (s-4M_Z^2) s ] \times \nn \\
&&
\mathcal D_0(s,M_Z^2,m_f^2,m_f^2)
+ \frac{1}{s -4 M_Z^2} [  ( C_v^{f\, 2} + C_a^{f\, 2}) M_Z^4 ( 4 m_f^2 (s -4 M_Z^2) + 4 M_Z^4 - 2 M_Z^2 s + s^2) \nn \\
&&
+ 2 C_a^{f \, 2} s (s - 4 M_Z^2) (m_f^2 (s- 4 M_Z^2) + M_Z^4)] \mathcal C_0(s, M_Z^2,M_Z^2,m_f^2,m_f^2,m_f^2) 
\bigg\} \nn \\
&&
\Phi^{\Delta\,W}_{ZZ}(p,q) = 
%
\frac{3\,\alpha \, \chi }{\pi\, s_w^2 \, c_w^2 \, (s-M_H^2)(s-4 M_Z^2)} 
\bigg\{
\frac{s - 2 M_Z^2}{2} [ M_H^2 (1- 2 s_w^2)^2 \nn \\
&&
+ 2 M_Z^2 (-12 s_w^6 + 32 s_w^4 - 29 s_w^2 + 9)]
+ \frac{M_Z^2}{s - 4 M_Z^2} [ M_H^2 (1 - 2 s_w^2)^2 (s + 2 M_Z^2) \nn \\
&&
- 2 (s_w^2 -1) ( 2 M_Z^4 (12 s_w^4 - 20 s_w^2 + 9) + s M_Z^2 (12 s_w^4 - 20 s_w^2 +1 ) + 2 s^2) ]
\mathcal D_0(s, M_Z^2, M_W^2,M_W^2) \nn \\
&&
+ \frac{M_Z^2}{s - 4 M_Z^2} [ 2 (s_w^2 -1) ( 2 M_Z^6 (4 s_w^2-3)(12 s_w^4 - 20 s_w^2 + 9) 
+ 2 M_Z^4 s (- 36 s_w^6 + 148 s_w^4 - 163 s_w^2 + 54) \nn \\
&&
+ M_Z^2 s^2 (12 s_w^6 - 96 s_w^4 + 125 s_w^2 - 43) + 4 s^3 ( 2 s_w^4 - 3 s_w^2 + 1) ) 
- M_H^2 (1- 2 s_w^2)^2 (M_Z^4 (8 s_w^2 -6) \nn  \\
&&
+ 2 M_Z^2 s (2 - 3 s_w^2) + s^2 (s_w^2-1))  ] \mathcal C_0(s,M_Z^2,M_Z^2,M_W^2,M_W^2,M_W^2) \bigg\}\, ,  \nn   \\
%
&& \Xi^{\Delta\,W}_{ZZ}(p,q)
%
=\frac{3\,\alpha\, \chi \, M_Z^2}{\pi\,s_w^2 \, c_w^2 \, (s-M_H^2)(s-4 M_Z^2)}
\bigg\{
- M_Z^2 [ M_H^2 (1- 2 s_w^2)^2 + 2 M_Z^2 (-12 s_w^6 + 32 s_w^4 - 29 s_w^2 + 9)] \nn \\
&&
+ \frac{s (s- 4 M_Z^2)}{2} (8 s_w^4 - 13 s_w^2 + 5) \mathcal B_0(s, M_W^2,M_W^2) 
+ \frac{2}{s - 4 M_Z^2} [ M_H^2 M_Z^2 (1-2s_w^2)^2 (M_Z^2- s ) \nn \\
&&
- 2 (s_w^2 -1)(M_Z^6 (12 s_w^4 - 20 s_w^2 + 9)  
- 3 M_Z^4 s ( 4 (s_w^2 -3)s_w^2 + 7) + M_Z^2 s^2 (7 - 8 s_w^2) + s^3 (s_w^2 -1)  )] 
\times \nn \\ 
&&
\mathcal D_0(s, M_Z^2, M_W^2,M_W^2)
+ \frac{1}{2 (s-4 M_Z^2)} [  M_H^2 (- 4 M_Z^6 (1 -2 s_w^2 )^2 (4 s_w^2-3) 
+ 2 M_Z^4 s (24 s_w^6 - 28 s_w^4 + 6 s_w^2 - 1)  \nn \\
&&
+ M_Z^2 s^2 (- 16 s^6 + 12 s_w^4 + 4 s_w^2 -1) + 2 s^3 s_w^4 (s_w^2 -1)  )  
+ 2 (s_w^2 -1) ( 4 M_Z^8 ( 4 s_w^2 -3)(12 s_w^4 - 20 s_w^2 + 9) \nn \\
&&
- 2 M_Z^6 s (24 s_w^6 - 52 s_w^4 + 6 s_w^2 + 15) + M_Z^4 s^2 (45 - 4 s_w^2 (s_w^2+ 13)) 
+ 2 M_Z^2 s^3 (4 s_w^4 + 2 s_w^2 -5) \nn \\
&&
- s^4(s_w^4 -1) ) ]  \mathcal C_0(s,M_Z^2,M_Z^2,M_W^2,M_W^2,M_W^2) 
\bigg\}\nn \\
&& \Phi^{\Delta\,ZH}_{ZZ}(p,q) =  
%
\frac{\alpha \, \chi}{\pi\,s_w^2\,c_w^2\,(s - M_H^2)(s- 4 M_Z^2)}\,
\bigg\{
(2 M_H^2 + M_Z^2) (s - 2 M_Z^2) \nn \\
&&
- 2 (M_H^2 - M_Z^2) (\mathcal A_0(M_Z^2) - \mathcal A_0(M_H^2)) 
+ \frac{1}{s - 4 M_Z^2} [2 M_H^4 (s - M_Z^2) + 3 M_H^2 M_Z^2 s \nn \\
&&
+ 2 M_Z^2 (4 M_Z^4 - 9 M_Z^2 s + 2 s^2)] \mathcal B_0(s, M_Z^2, M_Z^2) 
- \frac{3}{s - 4 M_Z^2} [2 M_H^4(s - M_Z^2)- 3 M_H^2 M_Z^2 s] 
\mathcal B_0(s, M_H^2,M_H^2)  \nn \\
&&
- \frac{2}{s-4 M_Z^2} [ M_H^2 (s + 2 M_Z^2)(4 M_Z^2 - M_H^2) 
+ 2 M_Z^2 s (s - 4 M_Z^2)] \mathcal B_0(M_Z^2, M_Z^2, M_H^2) \nn
\eea
\bea
&&
- \frac{3 M_H^2}{s- 4 M_Z^2} [2 M_H^2 (s-M_Z^2)(4 M_Z^2 - M_H^2) 
- M_Z^2 s (s + 2 M_Z^2)] \mathcal C_0(s,M_Z^2,M_Z^2,M_Z^2,M_H^2,M_H^2)  
+ \frac{M_H^2}{s - 4 M_Z^2} \times \nn \\
&&
[ 2 M_H^4 (s-M_Z^2) + M_H^2 (4 M_Z^4 - 2 M_Z^2 s + s^2) + 2 M_Z^2 (8 M_Z^4 - 14 M_Z^2 s + 3 s^2)] \mathcal 
C_0(s,M_Z^2,M_Z^2,M_H^2,M_Z^2,M_Z^2)
\bigg\} \nn \\
%
&& \Xi^{\Delta\,ZH}_{ZZ}(p,q)
%
= - \frac{3\,\alpha\,\chi}{\pi\,s_w^2\,c_w^2\,(s- M_H^2)(s -4 M_Z^2)}\, \bigg\{ M_Z^4 (2 M_H^2 + M_Z^2) 
\nn \\
&&
+ \frac{1}{2} (M_Z^2 - M_H^2) (s - 2 M_Z^2)  (\mathcal A_0(M_Z^2) - \mathcal A_0(M_H^2))
+ \frac{1}{4 (s - 4 M_Z^2)}\, [ M_H^4 (4 M_Z^4 - 2 M_Z^2 s + s^2) 
\nn \\
&& 
+ M_H^2 M_Z^2 s (s + 2 M_Z^2) - M_Z^2 (16 M_Z^6 - 28 M_Z^4 s + 18 
M_Z^2 s^2 - 3 s^3)] \, \mathcal B_0(s, M_Z^2, M_Z^2) 
\nn \\
&&
- \frac{3 M_H^2}{4(s-4 M_Z^2)} [M_H^2 (4 M_Z^4 - 2 M_Z^2 s + s^2) - M_Z^2 s (s + 2 M_Z^2)] \mathcal B_0 
(s, M_H^2, M_H^2) 
\nn \\
&&
+ \frac{2}{s- 4 M_Z^2} [ M_Z^2 s (M_H^2 - 2 M_Z^2)^2 - M_H^2 M_Z^4 (M_H^2 - 4 M_Z^2) 
- M_Z^4 s^2] \mathcal B_0(M_Z^2, M_Z^2, M_H^2)
\nn \\
 &&
+ \frac{3 M_H^2}{4 (s- 4 M_Z^2)} \times
[ M_H^2 (4 M_Z^4 - 2 M_Z^2 s + s^2) (M_H^2 - 4 M_Z^2) + 2 M_Z^2 s (16 M_Z^4 - 6 M_Z^2 s + s^2)]\,\times 
\nn \\
&&
\mathcal C_0(s, M_Z^2, M_Z^2, M_Z^2, M_H^2, M_H^2)
+ \frac{1}{4(s - 4 M_Z^2)} [ M_H^6 (4 M_Z^4 - 2 M_Z^2 s + s^2) + 2 M_H^4 M_Z^2 (s^2 - 4 M_Z^4) \nn \\
&&
- 4 M_H^2 M_Z^2 (8 M_Z^6 - 10 M_Z^4 s + 6 M_Z^2 s^2 - s^3)
+ 4 M_Z^4 s (s - 4 M_Z^2)^2 ] \mathcal C_0(s, M_Z^2, M_Z^2, M_H^2, M_Z^2, M_Z^2)
\bigg\} \nn
\eea
\bea
\Sigma_{Min, \, \rho H}(k^2) 
&=& 
\frac{e }{48 \pi^2 \, s_w \, c_w \, M_Z}  \bigg\{ \sum_f m_f^2 \bigg[ 3 (4 m_f^2 - k^2) \mathcal 
B_0(k^2,m_f^2,m_f^2) + 12 \mathcal A_0(m_f^2) -2 k^2 + 12 m_f^2 \bigg] \nn \\
&-& 
\frac{1}{2} \bigg[ 3 \left( k^2 (M_H^2 -6 M_W^2)  + 2 M_W^2 (M_H^2 + 6 M_W^2) \right)
\mathcal B_0(k^2, M_W^2,M_W^2) + 6 (M_H^2 + 6 M_W^2) \mathcal A_0(M_W^2) 
\nn \\
&& 
- k^2 (M_H^2 + 18 M_W^2) + 6 M_W^2 (M_H^2 - 2 M_W^2) \bigg] 
- \frac{1}{4} \bigg[ 3 \left(M_H^2 (2 M_Z^2 + s) + 12 M_Z^4 - 6 M_Z^2 s \right) \nn \\
&\times& 
\mathcal B_0(s, M_Z^2,M_Z^2) + 9 M_H^2 (2 M_H^2 + s) \mathcal B_0(s, M_H^2,M_H^2) 
+ 6 (M_H^2 + 6 M_Z^2) \mathcal A_0(M_Z^2) + 18 M_H^2 \mathcal A_0(M_H^2) \nn \\
&&  
+ 2 (9 M_H^4 + M_H^2 (3 M_Z^2 -2 s) - 6 M_Z^4 - 9 M_Z^2 s ) \bigg]
\bigg\} \nn
\eea
\bea
\Sigma_{I, \, \rho H}(k^2)
&=& 
\frac{3 e}{16 \pi^2 \, s_w \, c_w} \frac{k^2 \, M_H^2}{M_Z^2}\chi \, \bigg[ \mathcal B_0(k^2, M_W^2, M_W^2)+ 
\frac{3}{2} \mathcal B_0(k^2,M_H^2,M_H^2)  + \frac{1}{2} \mathcal B_0(k^2, M_Z^2, M_Z^2) \bigg] \nn
\eea

\section{Standard Model self-energies}
\label{SigmaSM}

We report here the expressions of the self-energies appearing in Section \ref{renorm} which define the renormalization conditions. 
They are given by
\bea
\Sigma^{\gamma\gamma}_T(p^2)
&=& 
- \frac{\alpha}{4 \pi} \bigg\{ \frac{2}{3} \sum_f \, N_C^f 2 Q_f^2 \bigg[
-(p^2 + 2 m_f^2)B_0(p^2, m_f^2, m_f^2) + 2 m_f^2 B_0(0, m_f^2, m_f^2) + \frac{1}{3}p^2 \bigg] \nn \\
&+& \bigg[ (3 p^2 + 4 M_W^2 ) B_0(p^2, M_W^2, M_W^2) - 4 M_W^2 B_0(0, M_W^2, M_W^2)\bigg]\bigg\}
\nn
\eea
\bea
\Sigma^{ZZ}_T(p^2)
&=& 
- \frac{\alpha}{4 \pi} \bigg\{ \frac{2}{3} \sum_f \,
N_C^f \bigg[ \frac{C_V^{f \, 2} + C_A^{f \, 2}}{2 s_w^2 c_w^2}
\bigg( -(p^2 + 2m_f^2) B_0(p^2, m_f^2, m_f^2)  
+ 2 m_f^2 B_0(0, m_f^2, m_f^2) + \frac{1}{3}p^2 \bigg) \nn\\
&& 
+ \frac{3}{4 s_w^2 c_w^2} m_f^2 B_0(p^2,m_f^2, m_f^2) \bigg]
 + \frac{1}{6 s_w^2 c_w^2}\bigg[ \bigg( (18 c_w^4 + 2 c_w^2 -\frac{1}{2})p^2 
+ (24 c_w^4 + 16 c_w^2 -10)M_W^2 \bigg) \nn \\
&& 
\times B_0(p^2, M_W^2, M_W^2) - (24 c_w^4 - 8 c_w^2 + 2)M_W^2 B_0(0, M_W^2, M_W^2) 
+ (4 c_w^2-1)\frac{p^2}{3} \bigg] \nn \\
&& 
+  \frac{1}{12 s_w^2 c_w^2} \bigg[ (2 M_H^2 -10 M_Z^2 - p^2) B_0(p^2, M_Z^2, M_H^2) 
-  2 M_Z^2 B_0(0, M_Z^2, M_Z^2) - 2 M_H^2 B_0(0, M_H^2, M_H^2) 
\nn \\
&& 
- \frac{(M_Z^2 - M_H^2)^2}{p^2}\left( B_0(p^2, M_Z^2, M_H^2) - B_0(0, M_Z^2, M_H^2) \right) -
\frac{2}{3} p^2  \bigg] \bigg\} \nn
\eea
\bea
\Sigma^{\gamma Z}_T(p^2)
&=& 
\frac{\alpha}{4 \pi \, s_w \, c_w} \bigg\{ \frac{2}{3} \sum_f \, N_C^f \,
Q_f \, C_V^f \bigg[ (p^2 + 2m_f^2) B_0(p^2, m_f^2, m_f^2) - 2 m_f^2 B_0(0, m_f^2, m_f^2) -\frac{1}{3}p^2 \bigg] \nn \\
&& 
- \frac{1}{3} \bigg[ \left( (9 c_w^2 + \frac{1}{2})p^2 + (12 c_w^2 + 4)M_W^2 \right) B_0(p^2, M_W^2, M_W^2) 
- (12 c_w^2 -2)M_W^2 B_0(0, M_W^2, M_W^2) + \frac{1}{3}p^2 \bigg]\bigg\} \nn
\eea
\bea
\Sigma_{HH}(p^2) 
&=& 
- \frac{\alpha}{4 \pi} \bigg\{ \sum_{f} N_C^f \frac{m_f^2}{2 s_w^2 M_W^2}
\bigg[ 2 \mathcal A_0\left( m_f^2 \right) + (4 m_f^2 - p^2) \mathcal B_0 \left( p^2, m_f^2,m_f^2\right) \bigg]  \nn \\
&& 
- \frac{1}{2 s_w^2} \bigg[ \left(6 M_W^2 - 2p^2 
+ \frac{M_H^4}{2 M_W^2} \right) \mathcal B_0 \left( p^2, M_W^2, M_W^2 \right) 
+ \left( 3 + \frac{M_H^2}{2 M_W^2} \right) \mathcal A_0 \left( M_W^2 \right) - 6 M_W^2 \bigg]  \nn \\
&& 
- \frac{1}{4 s_w^2 \, c_w^2} \bigg[ \left(6 M_Z^2 - 2p^2 
+ \frac{M_H^4}{2 M_Z^2} \right) \mathcal B_0 \left( p^2, M_Z^2, M_Z^2 \right) 
+ \left( 3 + \frac{M_H^2}{2 M_Z^2} \right) \mathcal A_0 \left( M_Z^2 \right) - 6 M_Z^2 \bigg] \nn \\
&-& \frac{3}{8 s_w^2} \bigg[ 3 \frac{M_H^4}{M_W^2} \mathcal B_0 \left( p^2, M_H^2,M_H^2 \right) 
+ \frac{M_H^2}{M_W^2} \mathcal A_0 \left( M_H^2 \right) \bigg] \bigg\} \nn
\eea
\bea
\Sigma^{WW}_T(p^2)
 &=& 
-\frac{\alpha}{4 \pi} \bigg\{ \frac{1}{3 s_w^2} 
\sum_i \bigg[ \left( \frac{m_{l,i}^2}{2} - p^2 \right)\mathcal B_0\left(p^2, 0, m_{l,i}^2 \right) 
+ \frac{p^2}{3} + m_{l,i}^2 \mathcal B_0 \left( 0, m_{l,i}^2,m_{l,i}^2 \right)   \nn \\
&& 
+ \frac{m_{l,i}^4}{2 p^2} \left( \mathcal B_0\left(p^2, 0, m_{l,i}^2 \right) 
- \mathcal B_0\left(0, 0, m_{l,i}^2 \right)\right) \bigg] + \frac{1}{s_w^2} \sum_{i,j}|V_{ij}|^2 \bigg[ 
\left( \frac{m_{u,i}^2 + m_{d,j}^2}{2} - p^2\right) \mathcal B_0 \left( p^2, m_{u,i}^2, m_{d,j}^2\right) \nn \\
&& 
+ \frac{p^2}{3} + m_{u,i}^2 \mathcal B_0 \left( 0, m_{u,i}^2, m_{u,i}^2\right) 
+ m_{d,j}^2 \mathcal B_0 \left( 0, m_{d,j}^2, m_{d,j}^2\right) 
+ \frac{(m_{u,i}^2 - m_{d,j}^2)^2}{2 p^2} \big( \mathcal B_0 \left( p^2, m_{u,i}^2, m_{d,j}^2\right) \nn \\
&&
- \mathcal B_0 \left( 0, m_{u,i}^2, m_{d,j}^2\right)\big) \bigg]  
+ \frac{2}{3} \bigg[ (2 M_W^2 + 5 p^2) \mathcal B_0 \left( p^2, 
M_W^2, \lambda^2 \right) - 2 M_W^2 \mathcal B_0 \left( 0, M_W^2, M_W^2\right) \nn \\
&&
- \frac{M_W^4}{p^2} \big( \mathcal B_0\left( p^2, M_W^2, \lambda^2 \right) - \mathcal B_0 \left( 0,M_W^2, \lambda^2 \right) \big) 
+ \frac{p^2}{3} \bigg] + \frac{1}{12 s_w^2} \bigg[ \big( (40 c_w^2 -1)p^2 \nn \\
&&
+(16 c_w^2 + 54 - 10 c_w^{-2}) M_W^2 \big) \mathcal B_0 \left(p^2, M_W^2, M_Z^2 \right) 
- (16 c_w^2 + 2) \big( M_W^2 \mathcal B_0 \left( 0,M_W^2,M_W^2\right) \nn \\
&&
+ M_Z^2 \mathcal B_0 \left( 0, M_Z^2, M_Z^2\right) \big) + (4 c_w^2 -1) \frac{2 p^2}{3} 
- (8 c_w^2 +1) \frac{(M_W^2 - M_Z^2)^2}{p^2} \big( \mathcal B_0 \left( p^2, M_W^2,M_Z^2\right) \nn \\
&&
- \mathcal B_0 \left(0, M_W^2,M_Z^2 \right) \big) \bigg] + \frac{1}{12 s_w^2} \bigg[ (2 M_H^2 - 10 M_W^2 - p^2) \mathcal B_0 
\left(p^2, M_W^2,M_H^2 \right) - 2 M_W^2 \mathcal B_0 \left(0,M_W^2,M_W^2 \right) \nn \\
&& 
- 2 M_H^2 \mathcal B_0 \left( 0, M_H^2,M_H^2\right) - \frac{(M_W^2 -M_H^2)^2}{p^2} \big( \mathcal B_0 \left( p^2, M_W^2, 
M_H^2\right) - \mathcal B_0 \left( 0,M_W^2,M_H^2\right) \big) - \frac{2 p^2}{3}\bigg] \bigg\}\, , \nn
\eea
where the subscripts $l$, $u$ and $d$ stand for "leptons", "up" and "down" (quarks) respectively.
The sum runs over the three generations and $\lambda$ is the photon mass introduced to regularize the infrared
divergences.

\end{appendix}

\bibliography{ThesisBib}
\bibliographystyle{h-physrev5}

\end{document}